\documentclass[
11pt, 
english, 
singlespacing, 
headsepline, 
]{MastersDoctoralThesis} 

\usepackage[utf8]{inputenc} 
\usepackage[T1]{fontenc} 
\usepackage[resetlabels,labeled]{multibib}
\newcites{K}{Published work}

\usepackage{mathpazo} 
\usepackage{amsmath}
\usepackage{amsfonts}
\usepackage[backref=page]{hyperref}
\usepackage{amssymb}
\usepackage{graphicx}
\usepackage{subfig}
\usepackage{physics}
\usepackage{array}
\usepackage{tikz-feynman}
\renewcommand*{\backref}[1]{}
\renewcommand*{\backrefalt}[4]{%
    \ifcase #1 (Not cited.)%
    \or        (Cited on page~#2.)%
    \else      (Cited on pages~#2.)%
    \fi}
\usepackage{nicefrac}

\newcommand{\mpl}{M_\mathrm{P}}
\newcommand{\sch}{Schwarzschild }
\newcommand{\aaa}{\alpha}
\newcommand{\bbb}{\beta}
\newcommand{\be}{\begin{equation}}
\newcommand{\ee}{\end{equation}}

\usepackage{pdfpages}

\thesistitle{Testing gravity with the two-body problem} 
\supervisor{Dr. Federico \textsc{Piazza}} 
\examiner{} 
\degree{Thèse de doctorat} 
\author{Adrien \textsc{Kuntz}} 
\addresses{} 

\subject{Theoretical Physics} 
\university{\href{https://www.univ-amu.fr/}{Aix-Marseille Université}} 
\department{\href{http://department.university.com}{}} 
\group{\href{http://researchgroup.university.com}{Research Group Name}} 
\faculty{\href{http://www.cpt.univ-mrs.fr/}{Centre de Physique Théorique}} 

\AtBeginDocument{
\hypersetup{pdftitle=\ttitle} 
\hypersetup{pdfauthor=\authorname} 
\hypersetup{pdfkeywords=\keywordnames} 
\hypersetup{hypertexnames=true}
}

\begin{document}

\frontmatter 

\pagestyle{plain} 

\includepdf{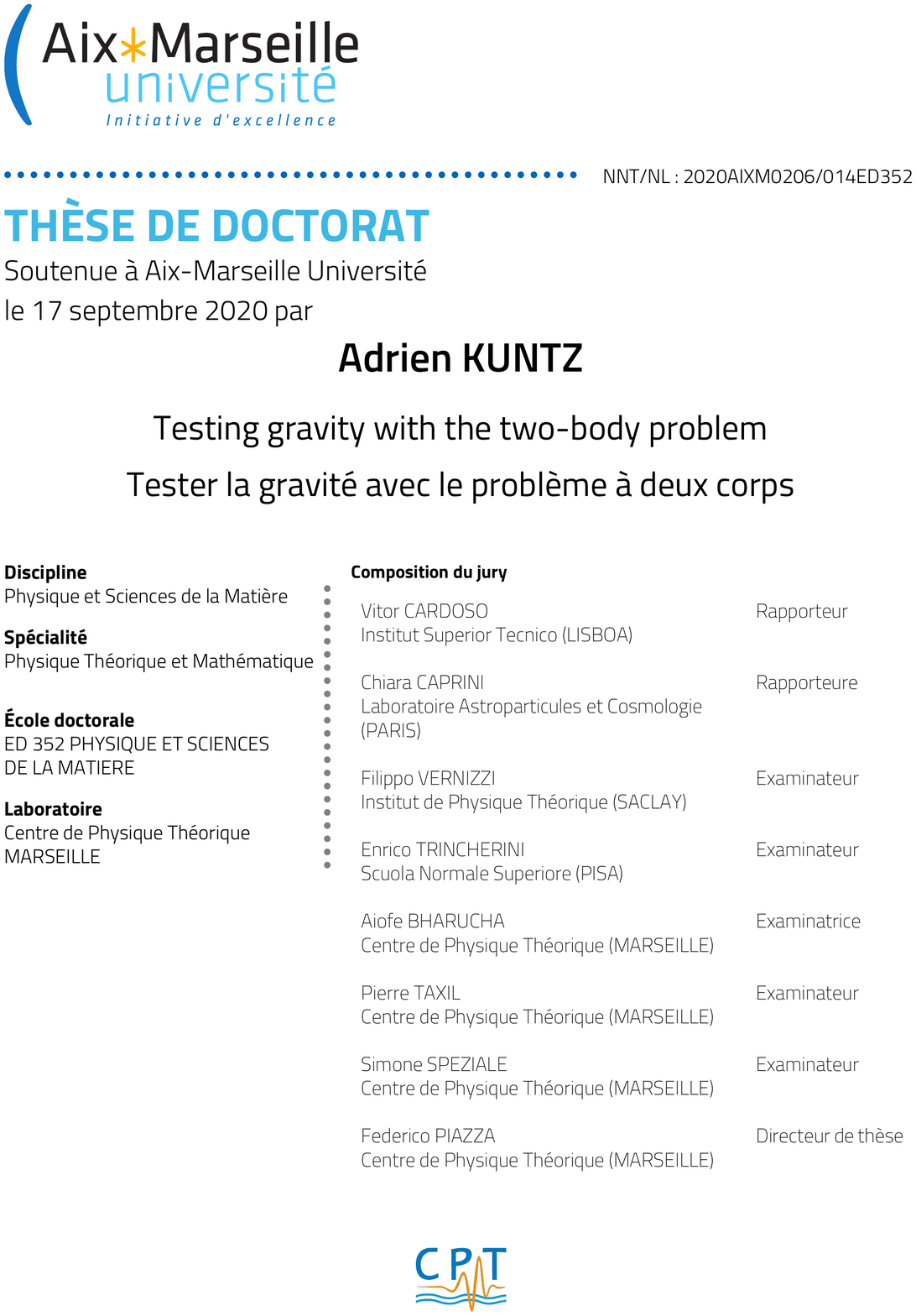}


\begin{titlepage}
\begin{center}

\vspace*{.06\textheight}
{\scshape\LARGE \univname\par}\vspace{1.5cm} 
\textsc{\Large Doctoral Thesis}\\[0.5cm] 

\HRule \\[0.4cm] 
{\huge \bfseries \ttitle\par}\vspace{0.4cm} 
\HRule \\[1.5cm] 
 
\begin{minipage}[t]{0.4\textwidth}
\begin{flushleft} \large
\emph{Author:}\\
\href{http://adrien.kuntz.perso.luminy.univ-amu.fr/index.html}{\authorname} 
\end{flushleft}
\end{minipage}
\begin{minipage}[t]{0.4\textwidth}
\begin{flushright} \large
\emph{Supervisor:} \\
\supname 
\end{flushright}
\end{minipage}\\[3cm]
 
\vfill

 
\vfill

{\large \today}\\[4cm] 

\vfill
\end{center}
\end{titlepage}


\chapter*{Déclaration sur l'honneur}

Je soussigné, Adrien Kuntz, 
    déclare par la présente que le travail présenté dans ce manuscrit est mon propre travail, réalisé sous la direction scientifique de Federico Piazza, 
    dans le respect des principes d’honnêteté, d'intégrité et de responsabilité inhérents à la mission de recherche. Les travaux de recherche et la rédaction de ce manuscrit ont été réalisées dans le respect à la fois de la charte nationale de déontologie des métiers de la recherche et de la charte d’Aix-Marseille Université relative à la lutte contre le plagiat.
    
    Ce travail n'a pas été précédemment soumis en France ou à l'étranger dans une version identique ou similaire à un organisme examinateur.\\
    
    Fait à Marseille le 9/06/2020
    
    \begin{flushright}\includegraphics[width=120px,height=40px]{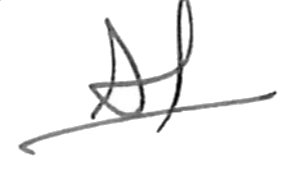}\end{flushright}

\clearpage
%
%
%
%


\vspace*{0.2\textheight}

\noindent{\itshape Douter de tout ou tout croire, ce sont les deux solutions également commodes qui l'une et l'autre nous dispensent de réfléchir.}\bigbreak

\hfill Henri \textsc{Poincaré}


\chapter*{Abstract}
\addchaptertocentry{\abstractname} 
Gravitational waves provide a new probe into the strong-field regime of gravity. It is thus essential to identify the predictions of General Relativity on the nature of the two-body problem, and to contrast them to alternative theories. This thesis aims at comparing the predictions of General Relativity and scalar-tensor theories on gravitational observables using Effective Field Theory techniques. In a first part, we show how simple scalar-tensor theories can be embedded in the Non-Relativistic General Relativity approach to the two-body problem and highlight their essential features. Furthermore, we study the effects of a disformal coupling of the scalar on the two-body dynamics and introduce a resummation technique. This new Non-Relativistic Scalar-Tensor formalism will provide a basis for the study of the Vainshtein mechanism in two-body configurations which is at the core of the second part of this thesis. Finally, in a last part we devise an Effective Field Theory formalism adapted to gravitational wave generation in theories featuring scalar hair, in the extreme mass ratio regime.

\vspace{0.5cm}
Keywords: Effective Field Theories, Two-Body Problem, Scalar-Tensor Theories, Gravitational Waves

\chapter*{Résumé}
Les ondes gravitationnelles sont une nouvelle sonde pour le régime de champ fort de la gravité. Il est donc essentiel d'identifier les prédictions de la Relativité Générale sur la nature du problème à deux corps et de les confronter à des théories alternatives. Cette thèse se propose de comparer les prédictions de la Relativité Générale et des théories tenseur-scalaires sur les observables gravitationnelles en utilisant des techniques de théories effectives des champs. Dans une première partie, on montrera comment de simples théories tenseur-scalaire peuvent être intégrées dans l'approche "Relativité Générale Non Relativiste" du problème à deux corps, et on soulignera leurs propriétés essentielles. De plus, l'effet d'un couplage difforme du scalaire sur la dynamique à deux corps sera étudié et une technique de resommation sera introduite. Ce nouveau formalisme "Tenseur-Scalaire Non Relativiste" formera les bases de l'étude du mécanisme de Vainshtein dans le cadre du problème à deux corps qui est au coeur de la seconde partie de cette thèse. Enfin, dans une dernière partie on introduira une nouvelle théorie effective des champs adaptée à l'étude de la génération d'ondes gravitationnelles dans les théories prédisant des "cheveux scalaires" autour de trous noirs, dans l'approximation d'un rapport de masse extrême.

\vspace{0.5cm}
Mots clés: Théories Effectives des Champs, Problème à Deux Corps, Théories Tenseur-Scalaires, Ondes Gravitationnelles


\begin{acknowledgements}
\addchaptertocentry{\acknowledgementname} 
Il est toujours délicat de remercier en si peu de lignes des personnes qui ont tant contribué au succès de cette thèse. En premier lieu, je voudrais bien sûr remercier mon directeur de thèse Federico, qui a donné sans compter son temps pour m'aider à avancer sur les diverses problématiques de ma thèse. Aller voir Federico dans son bureau pour discuter, c'est ressortir avec de nombreuses réponses mais généralement toujours plus de nouvelles questions, ce qui est indubitablement la marque d'un grand physicien ! Je tiens ensuite à remercier ma famille pour tous les bons moments passés ensemble au cours de cette thèse, et en particulier ma mère pour son soutien indéfectible et ses conseils avisés (que je n'écoute toujours que d'une oreille, en bon fils qui se respecte). Merci aussi à Morgane et Martin pour leur indéfectible enthousiasme et bonne humeur, à Hélène et Brigitte pour les nombreux moments de détente ensemble. Je pense très fort à mes grands-parents qui resteront toujours le noyau dur de notre famille, même s'ils nous ont quittés tandis que j'écrivais ces lignes.

Mes remerciements vont ensuite à tous mes amis et en particulier à tous les 'gros' avec qui j'ai tant en commun. Merci à Ségolène pour ces moments de joie que nous avons eu, merci à Alexandre pour ces vacances au grand air qu'elles soient dans les Dolomites, les Pyrénées, un chalet perdu des Alpes de haute-Provence ou sur un bateau : je ne pourrais tout simplement pas vivre sans ces moments de liberté ! Merci à Nicolas, Elie, Alice, Guillaume, Jean, Manuel pour ces vacances et diverses randonnées que nous avons faites ensemble. Merci à Assaf avec qui j'ai pu partager ma passion pour la gastronomie de haut niveau. Je veux ensuite remercier tous les autres doctorants marseillais avec qui j'ai passé tant de bons moments, et en particulier Philippe pour ces séjours magiques en Corse. Merci à Sylvain, Marie, Adrian, Romain, Loïc pour les diverses soirées et autres escapades à Sugiton. Merci aux habitués du club oeno (en particulier Lorenzo et Théo) qui ont égayé la plupart de mes jeudi soirs à Marseille, et merci aux 'bons vivants' Hichem et Romain pour les repas gargantuesques que nous avons partagé. Un grand merci à Pierre, Jeremy, Robin et tous les organisateurs de l'Agape : je souhaite une longue vie à cette école si riche en rencontres et discussions !

Enfin, je souhaite remercier vivement tous les collègues avec qui j'ai pu travailler durant cette thèse, en particulier Filippo, Philippe et Riccardo : j'ai retiré de nombreux enseignements de nos collaborations. Un grand merci à Vitor pour le séjour extraordinaire que j'ai pu passer à Lisbonne, ainsi qu'à Lavinia pour l'invitation à Zurich, à Matthew pour m'avoir permis de découvrir l'intitut Périmètre et à Riccardo pour son invitation à Pittsburgh. Je souhaite remercier les membres de mon jury de thèse qui ont tous accepté mon invitation avec un enthousiasme qui m'a fait chaud au coeur : Vitor, Chiara, Filippo, Enrico, Pierre, Simone, et bien sûr Federico. Merci à tous les collègues professeurs avec qui j'ai eu le plaisir d'enseigner durant ces trois années : Frédéric, André, Jean-Marc, Nikolas, William, Guillaume ; ainsi qu'à tous les chercheurs du CPT avec qui j'ai pu avoir de nombreuses et agréables discussions. Finalement, mention spéciale pour le pôle administratif du CPT qui m'a assisté sans une seule turbulence au cours de ces trois années riches en voyages.


\end{acknowledgements}


\tableofcontents 

\listoffigures 

\listoftables 


\begin{abbreviations}{ll} 
\textbf{$\Lambda$CDM} & \textbf{$\Lambda$} \textbf{C}old \textbf{D}ark \textbf{M}atter \hfill \\
\textbf{(P)PN} & (\textbf{P}arametrized) \textbf{P}ost \textbf{N}ewtonian \\
\textbf{(S/W)EP} & (\textbf{S}trong/\textbf{W}eak) \textbf{E}quivalence \textbf{P}rinciple \\
\textbf{ADM} & \textbf{A}rnowitt \textbf{D}eser \textbf{M}isner \\
\textbf{BAO} & \textbf{B}aryon \textbf{A}coustic \textbf{O}scillations \\
\textbf{BBN} & \textbf{B}ig \textbf{B}ang \textbf{N}ucleosynthesis \\
\textbf{BD(T)} & \textbf{B}rans-\textbf{D}icke (\textbf{T}ype) \\
\textbf{BH(s)} & \textbf{B}lack \textbf{H}ole(s) \\
\textbf{BOSS} & \textbf{B}aryon \textbf{O}scillations \textbf{S}pectroscopic \textbf{S}urvey \\
\textbf{CMB} & \textbf{C}osmic \textbf{M}icrowave \textbf{B}ackground \\
\textbf{CS} & \textbf{C}hern-\textbf{S}imons \\
\textbf{DECIGO} & \textbf{D}eci-Hertz interferometer \textbf{G}ravitational-wave \textbf{O}bservatory \\
\textbf{DE} & \textbf{D}ark \textbf{E}nergy \\
\textbf{DGP} & \textbf{D}vali \textbf{G}abadadze \textbf{P}orrati \\
\textbf{DHOST} & \textbf{D}egenerate \textbf{H}igher \textbf{O}rder \textbf{S}calar \textbf{T}ensor \\
\textbf{DM} & \textbf{D}ark \textbf{M}atter \\
\textbf{EFT} & \textbf{E}ffective \textbf{F}ield \textbf{T}heory \\
\textbf{EH} & \textbf{E}instein-\textbf{H}ilbert \\
\textbf{EMRI} & \textbf{E}xtreme \textbf{M}ass \textbf{R}atio \textbf{I}nspiral \\
\textbf{EOB} & \textbf{E}ffective \textbf{O}ne-\textbf{B}ody \\
\textbf{EOM} & \textbf{E}quations \textbf{O}f \textbf{M}otion \\
\textbf{EPTA} & \textbf{E}uropean \textbf{P}ulsar \textbf{T}iming \textbf{A}rray \\
\textbf{ET} & \textbf{E}instein \textbf{T}elescope \\
\textbf{FLRW} & \textbf{F}riedmann, \textbf{L}emaître, \textbf{R}obertson, \textbf{W}alker \\
\textbf{GB} & \textbf{G}auss-\textbf{B}onnet \\
\textbf{GR} & \textbf{G}eneral \textbf{R}elativity \hfill \\
\textbf{GSF} & \textbf{G}ravitational \textbf{S}elf-\textbf{F}orce \\
\textbf{GW(s)} & \textbf{G}ravitational \textbf{W}ave(s) \\
\textbf{I(S)CO} & \textbf{I}nnermost (\textbf{S}table) \textbf{C}ircular \textbf{O}rbit \\
\textbf{IR} & \textbf{I}nfra-\textbf{R}ed \\
\textbf{KAGRA} & \textbf{Ka}miokia \textbf{Gra}vitational-wave detector \\
\textbf{LIGO} & \textbf{L}aser \textbf{I}nterferometer \textbf{G}ravitational-wave \textbf{O}bservatory \\
\textbf{LISA} & \textbf{L}aser \textbf{I}nterferometer \textbf{S}pace \textbf{A}ntenna \\
\textbf{LLR} & \textbf{L}unar \textbf{L}aser \textbf{R}anging \\
\textbf{MACHO} & \textbf{M}assive \textbf{A}strophysical \textbf{C}ompact \textbf{H}alo \textbf{O}bject \\
\textbf{MOND} & \textbf{Mo}dified \textbf{N}ewton \textbf{G}ravity \\
\textbf{NRGR} & \textbf{N}on-\textbf{R}elativistic \textbf{G}eneral \textbf{R}elativity \\
\textbf{NR} & \textbf{N}umerical \textbf{R}ealtivity \\
\textbf{NS(s)} & \textbf{N}eutron \textbf{S}tar(s) \\
\textbf{ODE} & \textbf{O}rdinary \textbf{D}ifferential \textbf{E}quation \\
\textbf{PDE} & \textbf{P}artial \textbf{D}ifferential \textbf{E}quation \\
\textbf{PM} & \textbf{P}ost-\textbf{M}inkowskian \\
\textbf{QFT} & \textbf{Q}uantum \textbf{F}ield \textbf{T}heory \\
\textbf{QNM(s)} & \textbf{Q}uasi-\textbf{N}ormal \textbf{M}ode(s) \\
\textbf{RW} & \textbf{R}egge-\textbf{W}heeler \\
\textbf{ST} & \textbf{S}calar-\textbf{T}ensor \\
\textbf{TeVeS} & \textbf{Te}nsor \textbf{Ve}ctor \textbf{S}calar \\
\textbf{TT} & \textbf{T}ransverse-\textbf{T}raceless \\
\textbf{UV} & \textbf{U}ltra-\textbf{V}iolet \\
\textbf{vDVZ} & \textbf{v}an \textbf{D}am, \textbf{V}eltman, \textbf{Z}akharov \\
\textbf{VLBI} & \textbf{V}ery \textbf{L}ong \textbf{B}aseline \textbf{I}nterferometry \\
\textbf{WIMP} & \textbf{W}eakly \textbf{I}nteracting \textbf{M}assive \textbf{P}article \\
\textbf{WKB} & \textbf{W}entzel \textbf{K}ramers \textbf{B}rillouin \\

\end{abbreviations}

\chapter*{Conventions}

In this thesis, we will use units in which $\hbar = c = 1$. Planck's mass is related to the Newton constant \textit{via} $\mpl^2 = 1/(8 \pi G)$, and the metric convention is 'mostly plus'. Unless specified otherwise, we adopt the following convention for the Fourier transform,
\begin{equation*}
f(t,\mathbf{x}) = \int \frac{\mathrm{d}\omega}{2 \pi} \frac{\mathrm{d}^3 k}{(2 \pi)^3} e^{-i \omega t} e^{i \mathbf{k} \cdot \mathbf{x}} \tilde f(\omega, \mathbf{k})
\end{equation*}

{ \begingroup
\renewcommand*{\backref}[1]{}
\renewcommand*{\backrefalt}[4]{%
    \ifcase #1 %
    \or        %
    \else      %
    \fi}
        \bibliographystyleK{utphys}
		\bibliographyK{moi.bib}

    \endgroup
}

%
%
%
%
%

%
%
%
%
%


\dedicatory{A mes grands-parents, vous qui m'avez transmis votre infatiguable curiosité. J'aurais aimé pouvoir vous parler de ce manuscrit.} 


{
\ihead{Foreword}
\chapter*{Foreword}
\addchaptertocentry{Foreword}

Physics is an endless quest for simplicity. Looking back in time, it is obvious that the emergence of physics as a science took place with celestial mechanics. Even if the world was full of complicated phenomenons that no one could explain at that time (Whats causes tides? Why are there storms? Why is there life?), the first astronomers remarked that up in the sky the mechanisms seemed very pristine and quite simple. It is the numerous observations of these periodic phenomenons which led to the first ever exactly solved problem in physics: the two-body problem. It is beyond doubt that, if the sky on Earth had been cloudy on every possible night, the development of physics would have been quite delayed!

It might come as a surprise that the very first academic problem of physics, the two-body motion, is still of interest today. In Einstein's theory of General Relativity (GR) as well as in the other competing theories which we will consider in this thesis, there is no exact solution to the two-body trajectories; it is however possible to resort to approximation methods. The very first and historic approximation to the two-body motion in GR consists in expanding the conserved energy of the system in powers of $v \ll 1$ (in this thesis, we use units in which $c=1$). One thus obtains the first relativistic correction to the two-body energy; this Einstein-Infeld-Hoffmann Lagrangian will be computed in the more general context of scalar-tensor (ST) theories in Chapter~\ref{Chapter4}. This approximation, which yielded one of the first experimental test of GR with the anomalous perihelion precession of Mercury which we will recall in Chapter~\ref{Chapter1}, has evolved today in the post-Newtonian formalism. A complete presentation of all the recent developments in this formalism is beyond the scope of this thesis; we will content ourselves with a short description of its essential building blocks in Chapters~\ref{Chapter1} and~\ref{Chapter3}. 

Solving as accurately as possible the two-body problem is of importance today for the ermergent field of gravitational wave (GW) astronomy. Indeed, observable GW are mainly produced by binary systems of very compact objects, namely black holes (BHs) or neutron stars (NS), in a close-by trajectory. Because of the emission of GW, the two objects lose energy, getting closer and closer up to the point where they merge together. The main steps of this physical process will be recalled in Chapter \ref{Chapter3}.

At the core of this thesis is the attempt to answer one simple question: 'What can we learn about gravity by studying the two-body problem?'. More precisely, do the two-body orbits that we observe in Nature correspond to our best theory of gravity up to now, namely GR, or are they described by another fundamental theory? If the two-body problem agrees with GR up to the observable precision, how much are the competing theories constrained? Giving a precise answer to these questions would greatly improve our knowledge about Nature. Indeed, cosmological observations may suggest that General Relativity could be replaced by another fundamental theory, as we will recall in Chapter~\ref{Chapter2}. However, this putative theory should also satisfy to all the variety of stringent tests that GR passed with flying colors.
 In this respect, two different approaches can be followed \cite{damour1996gravitation}. To illustrate them, we will again take the example of Mercury's perihelion precession.

The first obvious approach is simply to compare the experimental predictions of different theories, and to falsify them if they do not correspond to experiment. There, experiments have a negative value, telling us something about a theory only when it is wrong. For example, the perihelion precession of Mercury $\dot \omega$ is compatible with Einstein's theory, but not with Newton's. The essential limitation of this approach is that it tells us nothing about which part of the theory is actually being tested. Let's say that two theories agree with experimental data: Occam's razor criterion dictates that, on the absence of any other possibility to distinguish among theories, we should content ourselves with the simplest one. However, the comparing approach did not tell us if the experiment tested all the aspects of the underlying theory, nor did it tell
if there would exist other theories agreeing with data but differing in some aspect from the simplest theory. To answer these questions, it is often more useful to adopt a contrasting approach.

To contrast (rather than compare) theories, one can embed different theories within a finite set of parameters. The Eddington parameters $\beta$ and $\gamma$ are a prototypical example of such a parametrization: we will present them in Chapter \ref{Chapter1}, so let us just state here that they encode the simplest possible deviations from GR in a weak-field metric. Experiments place constraints on the parameters $\beta$ and $\gamma$ and theories can be selected on the basis that they satisfy these constraints. However, the full version of the parametrized post-Newtonian (PPN) formalism which we will present in Chapter \ref{Chapter1} contains many more parameters which are not constrained by the perihelion data. Thus, it is essential to list all aspects in which gravity theories could differ in the Solar system; this could suggest new experimental tests of gravity. The contrasting approach is essential here in the sense that it points towards directions in which one could improve our global knowledge on gravity.

This contrasting approach finds a natural embedding into the Effective Field Theories (EFT) ideas which are at the heart of this thesis. The EFT line of reasoning was born in the field of particle physics, but it finds many different and useful applications to gravity. One of the very first and simplest application of EFT ideas to gravity is the construction of the Fierz-Pauli action, showing that linearized GR can simply be recovered by imposing gauge-invariance to a spin-2 field theory. But the unifying power of EFT in gravity is more apparent in both the PPN framework and the EFT of Dark Energy (EFT of DE), which we will present in the two first Chapters of this thesis. Together with the third Chapter containing a basic introduction to GW science and observations, they form the first introductory part of this thesis designed to present all the essential tools which are underlying the work produced during this PhD.

The usefulness of effective formalisms lie in the fact that they define a 'bottom-up' approach to theory space: all valid theories are considered on an equal footing when it comes to comparison with observations.
This is especially convenient in the cosmological setup because of the somewhat overwhelming number of new theories which emerged in the last decades. Indeed, cosmological observations come with some striking and yet unsolved issues, among which the most prominent are the nature of dark energy and dark matter. Even if dark energy can be simply accounted for by a mere cosmological constant, the solution is not really satisfying from a field theory viewpoint; as for dark matter, the unobservability of any putative candidate up to now have motivated theorists to explore new solutions to the problem, including modifications of gravity. Furthermore, the growing Hubble tension between local and distant observations of the universe could be a sign of new physics. Since it provides a strong motivation to many of the theories which we will examine in more details in this thesis, and even if they are not directly related to the two-body problem at the core of this thesis, we will devote Chapter~\ref{Chapter2} to cosmological issues and to some of the main theories developed in recent years for cosmological applications.

A second motivation to dedicate a full Chapter to cosmology is to present in some details the EFT of Dark Energy (EFT of DE) as an example of application of EFT ideas to gravity. The formalism and tools explored here will be useful in other parts of the thesis, especially for Chapter~\ref{Chapter10} which uses the same construction. The EFT of DE can be used to group all scalar-tensor modifications of gravity and compare them efficiently to cosmological observations. 
As much as the PPN formalism, it acts as a sort of 'proxy' between fundamental theories and observational parameters. Thus, it is one of the best tools to use for analyzing large-scale surveys such as Euclid or LSST.

But if one changes gravity to account for cosmological observations, one is also likely to have modifications to small-scale gravitational observables like the motion of planets in the Solar system or the production of GW by compact objects, even if these are wildly separated scales. Indeed, the modifications of gravity considered in this thesis proceed by the addition of a scalar degree of freedom which participates in \textit{all} gravitational phenomenon. Thus, in Chapters~\ref{Chapter1} and~\ref{Chapter3} we will introduce the reader to small-scale tests of gravity. Since the early historical tests of GR were Solar system experiments, we will devote the opening Chapter of this thesis to them. This will also allow us to introduce the first scalar-tensor modification of gravity in its historical context, namely Brans-Dicke theory, and comment on some of its observational consequences. But the main objective of Chapter~\ref{Chapter1} is to present in details the PPN formalism which classifies all the possible predictions of gravity theories in the Solar system. To date, the Solar system measurements are among the most accurate constraints on GR and many of its alternatives; the PPN formalism and its associated physics will be a source of inspiration for many developments presented in this thesis, notably when studying the two-body problem with Vainshtein screening in Chapter~\ref{Chapter7}. As already emphasized before, the PPN formalism is particularly useful in the sense that it relates the different gravity theories on the one hand and the experimental data on the other hand. Thus, when we will derive the PPN parameters for Brans-Dicke type theories in Chapter~\ref{Chapter4}, we will immediately be able to impose experimental constraints on the theory.

Solar system tests only probe the weak-field regime of gravity. On the other hand, GW observations allow for a direct investigation of the strong-field regime. Chapter~\ref{Chapter3} contains an introduction to GW science today. We will present the essentials of GW theory which will underline many of the results of the next three parts: propagation of GW in flat space, the quadrupole formula and its beautiful confirmation by the Hulse-Taylor pulsar. However, with the current an planned laser interferometers used for direct detection, the physics of GW goes well beyond the quadrupolar emission. We will present the matched filtering technique used in data analysis and the necessity to dispose of an accurate theoretical waveform template in order to detect a signal. The templates used in today's interferometers are provided by the post-Newtonian (PN) formalism of which we will give a short summary, together with some other tools used in the two-body problem: gravitational self-force (GSF), post-Minkowskian (PM) expansion, effective one-body (EOB) approach. While the subject of Chapter~\ref{Chapter4} will be to offer an alternative view to the PN formalism in scalar-tensor theories, we will often use ideas from these other approaches in the following Chapters. Thus, the PM philosophy will be adopted in Chapter~\ref{Chapter6}, while the EOB approach will motivate the developments presented in Chapter~\ref{Chapter10}. Finally, Chapter~\ref{Chapter3} contains the constraints on gravity theories placed by GW observations, the most striking one being the 2017 measurement of the speed of gravitational waves (compared to that of light) which happened at the beginning of this PhD. As we will recall there, many cosmological models were disfavored by this observation. One of the major objectives of this thesis will be to propose new tests of GR using GW data, thus completing the last section of this Chapter.

Beyond this first introductory part, the rest of this thesis is constituted of the new results obtained during this PhD and aim towards an exploration of the two-body problem in theories differing from GR. The second part of the thesis is constituted of Chapters \ref{Chapter4}, \ref{Chapter5} and \ref{Chapter6}, which are based on the published works \citeK{Kuntz:2019zef}, \citeK{Brax:2019tcy} and \citeK{Kuntz:2020gan} respectively. In Chapter \ref{Chapter4}, we will adopt an EFT viewpoint on the two-body problem in a simple class of scalar-tensor theories generalizing GR, namely Brans-Dicke type theories. This Chapter will serve as a theoretical toolbox introducing several ideas which will be used afterwards.  Perhaps one of the most useful field theory tool introduced there is the concept of effective action as the saddle point of a path integral. It allows to recast the two-body problem in the following way: starting with your favorite gravity action (we take GR here as an example), supplemented with two point-particles,
\begin{equation}\label{eq:GR_2body_intro}
S = \frac{1}{16 \pi G} \int \mathrm{d}^4 x \; \sqrt{-g} R - m_1 \int \mathrm{d}t \sqrt{-g_{\mu \nu} v_1^\mu v_1^\nu} - m_2 \int \mathrm{d}t \sqrt{-g_{\mu \nu} v_2^\mu v_2^\nu} \; ,
\end{equation}
where $v_A^\mu = \mathrm{d} x_A^\mu / \mathrm{d} t = (1, \mathbf{v}_A)$ is the four-velocity of each particle $A=1,2$ and $m_A$ is its mass, the effective action ruling the entire dynamics of the two point-particles can be found by integrating out the gravitational field,
\begin{equation}
\exp \big(i S_\mathrm{eff}[x_1^\mu, x_2^\mu] \big) = \int \mathcal{D}g_{\mu \nu} \exp \big( i S[x_1^\mu, x_2^\mu, g_{\mu \nu}] \big)
\end{equation}
In practice, the path integral is computed as a series of Feynman diagrams in a perturbative expansion around Minkowski space. The complete procedure will be detailed in Chapter~\ref{Chapter4} for Brans-Dicke type theories. These kind of theories have been studied for a while so that the results obtained in this Chapter will only recover some well-known facts, however the new EFT approach that we will adopt will allow us to have a viewpoint complementary to the usual PN approach. Notably, we will show that violations of the strong equivalence principle typical of scalar-tensor theories are linked to the renormalization of scalar charge.

The formalism developed in Chapter~\ref{Chapter4} will be directly used in Chapter \ref{Chapter5} to study the effects of a particular type of scalar coupling (namely the \textit{disformal} coupling) on the two-body problem. We will establish a theorem showing that the disformal coupling does not contribute to the binary dynamics in the case of circular orbits. However, elliptic orbits such as the ones observed in binary pulsars will allow us to constrain the magnitude of a disformal coupling. Finally,
 in Chapter \ref{Chapter6} we will develop a resummation technique highlighting some new aspects of the two-body problem in GR. More precisely, we will show that by introducing two parameters along the wordline of the two bodies, one can resum a particular class of Feynman diagrams contributing to the two-body energy. These parameters (generically denoted by $e$ in this thesis) are defined from the point-particle action $S_\mathrm{pp}$ already introduced in Eq.~\eqref{eq:GR_2body_intro}:
\begin{equation}
S_\mathrm{pp} = - m \int \mathrm{d}t \sqrt{-g_{\mu \nu} v^\mu v^\nu} \quad \Leftrightarrow \quad S_\mathrm{pp} = - m \int \mathrm{d}t \left[ e - \frac{g_{\mu \nu} v^\mu v^\nu}{e} \right] \; .
\end{equation}
The equivalence can be seen by varying the action with respect to the additional parameter $e$, yielding $e = \sqrt{-g_{\mu \nu} v^\mu v^\nu}$ and then replacing $e$ in the action by its equation of motion.
The interest of the worldline parameters lie in the fact that the point-particle action is now \textit{linear} in the gravitational field, allowing for an exact computation of the effective action. Thus, we will be able to resum a particular a particular class of Feynman diagrams, as is illustrated in Figure~\ref{fig:worldline_resummation}. Furthermore, the worldline parameters are no longer well-defined when $g_{\mu \nu} v^\mu v^\nu$ changes sign and we will show that this is related to strong-field nonperturbative effects such as the location of an 'effective two-body horizon'. This will allow us to generalize the notion of innermost circular orbit to the two-body problem.
\begin{figure}
\centering
\includegraphics[scale=0.6]{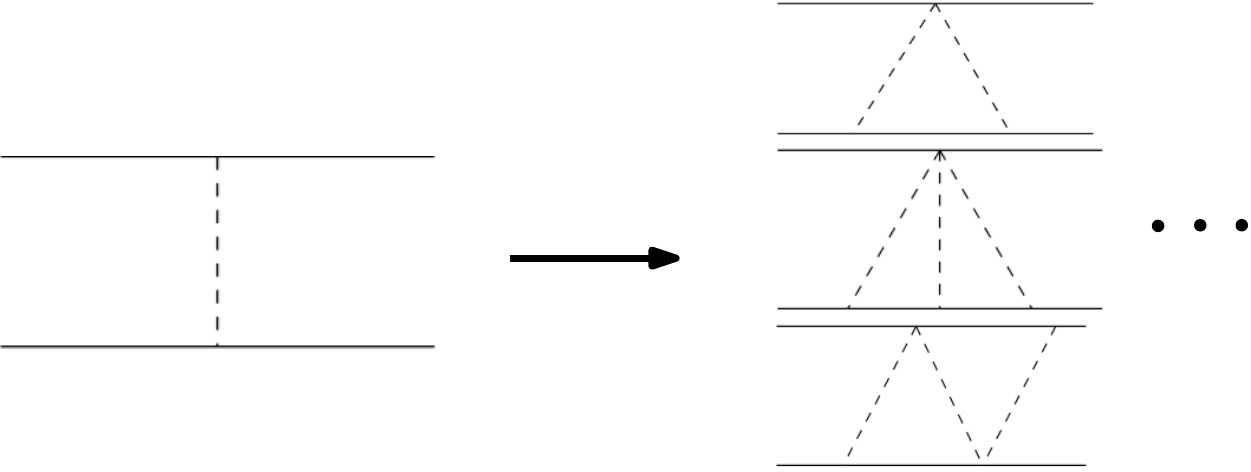}
\caption[Worldline resummation]{Resummation defined by the worldline parameters. A single linear Feynman diagram contains infinitely many nonlinear couplings.}
\label{fig:worldline_resummation}
\end{figure}

The third part of the thesis is dedicated to the study of the Vainshtein screening mechanism typical of a class of scalar-tensor theories. The essence of this mechanism is to allow the scalar field to take appreciable values at large, cosmological scales while at the same time satisfying the stringent Solar system constraints presented in Chapter~\ref{Chapter1}. In Chapter \ref{Chapter7}, based on the PRD paper \citeK{Kuntz:2019plo}, we will present a simple approach to the full two-body problem in these kind of theories. We will show how the Effective One-Body formalism allows for a convenient packaging of the two-body energy, and we will carry out a numerical simulation in order to extract the behavior of this energy in the static case. Essentially, the results of this Chapter can be illustrated by the simple formula giving the two-body energy of a theory equipped with Vainshtein screening in the small-scale approximation,
\begin{equation}
E = - \frac{G m_1 m_2}{r} \left[ 1 + h\left( \nu \right) \left( \frac{r}{r_*} \right)^n \right] \; .
\end{equation}
In this equation, $r$ is the distance between the two point-particles, $r_*$ is a scale named the Vainshtein radius and the function $h$ depends on the symmetric mass ratio $\nu = m_1 m_2/(m_1+m_2)^2$. In the small-scale limit, $r \ll r_*$ so that the modification to Newton's law is small. However, one of the new results of this thesis is that 
the nontrivial mass dependence of $h$ implies a direct violation of the Weak Equivalence Principle (WEP): thus we will show how one can obtain new constraints on Vainshtein screened theories by using Lunar Laser Ranging data which are very sensitive to any violation of the WEP. 

After having discussed the two-body problem in the weak-field regime of Vainshtein screened theories, we will move towards the exploration of the gravitational strong-field regime and study GW generation in a cubic Galileon in Chapter~\ref{Chapter8} (based on the JCAP paper \citeK{Brax:2020ujo}). There, we will study the possibility of observing a signal characteristic of these theories using GW data. Since the conventional post-Newtonian tools are not applicable to Vainhstein screening, we will rather consider a highly asymmetric system composed of a giant supermassive Black Hole (BH) orbited by a much smaller one. This will allow us to set up a perturbative computation using the smallness of the mass ratio. We will then derive the perturbation of two-body orbits caused by the Vainshtein field and its associated modification to the GW signal.

Finally, the last part of this thesis aims at constructing an EFT formalism adapted to GW generation in a large class of theories, namely the ones predicting the existence of a nontrivial scalar profile (or 'hair') around a BH. Chapter~\ref{Chapter9} is an introduction to no-hair theorems, and as such does not contain any new result. We will show how the formation of any scalar profile around a BH is forbidden in a large class of scalar-tensor theories, and then critically examine the assumptions lying behind this derivation. It will clearly appear that some perfectly viable scalar-tensor theories evade the no-hair theorem, among which the most well-known is the Gauss-Bonnet coupling which will serve as an example in the following Chapter. 
We will then adopt a pragmatic approach in Chapter~\ref{Chapter10} (based on the JCAP paper \citeK{Kuntz:2020yow}) by assuming that scalar 'hair' can indeed exist and derive the observational predictions stemming from this postulate. In the extreme mass ratio limit, a small object orbiting a supermassive BH can be thought of a small perturbation on top of a background profile. Then,  the existence of a nontrivial background scalar field allows to choose a gauge where the perturbations of the scalar vanish everywhere: this is called the 'unitary gauge' by analogy with particle physics. We can then use the methodology described in Chapter~\ref{Chapter2} to construct the EFT of DE, time being replaced by the radial variable. We end up with the most generic action describing a BH endowed with scalar hair, given by\footnote{We are restricting to the odd sector in this work, see Chapter~\ref{Chapter10} for further details.}
\begin{align} \label{eq:effective_action_intro}
& S =  \int \mathrm{d}^4x \, \sqrt{-g} \bigg[
\frac{1}{2}M^2_1(r) R -\Lambda(r) - f(r)g^{rr} - \alpha(r)\bar K^\mu_{\ \nu} K^\nu_{\ \mu} \nonumber
\\
& \qquad \qquad \qquad \qquad \qquad + M_{10}^2(r)\delta K^\mu_{\ \nu}\delta K^\nu_{\ \mu}  + M_{12}(r)  \bar K^\mu_{\ \nu} \delta K^\nu_{\ \rho} \delta K^\rho_{\ \mu} \bigg] . 
\end{align}
In this equation, $r$ is the distance to the supermassive BH, $R$ is the Ricci scalar, $K^\mu_{\ \nu} = \bar K^\mu_{\ \nu} + \delta K^\mu_{\ \nu}$ is the extrinsinc curvature of constant-$r$ hypersurfaces decomposed as a background value and a perturbation, and $M_1$, $M_{10}$, $M_{12}$, $\Lambda$, $f$ and $\alpha$ are arbitrary functions of $r$. In the post-Newtonian regime, one can expand these functions for large radius, thus leaving an action depending on some parameters describing deviations from GR. The usefulness of such an action is that one can easily identify the effective theory with a fundamental one depending on a few parameters only; at the same time, its form is ready to use for practical computations involving gravitational observables. Thus, in Chapter~\ref{Chapter10} we will derive a Regge-Wheeler equation from the effective action~\eqref{eq:effective_action_intro} and extract from it the power dissipated in GW, following a classic computation by Poisson and Sasaki.
Consequently, we will be able to predict the GW waveform signal in \textit{any} theory predicting scalar hair around static, spherically symmetric and asymptotically flat BH, and in the extreme mass ratio regime. The EFT approach advocated here will allow a straightforward comparison of theory with data by providing a bank of waveform templates in several theories differing from GR in the strong-field regime. We will end the thesis with this Chapter opening the door to many interesting developments.
}

\mainmatter 

\pagestyle{thesis} 

\part{Introduction} \label{part1}

\chapter{First tests of General Relativity and the Parametrized Post-Newtonian formalism} 

\label{Chapter1} 

\begin{quotation}
For over half a century, the general theory of relativity has stood as a monument to the genius of Albert Einstein. It has altered forever our view of the nature of space and time, and has forced us to grapple with the question of the birth and fate of the universe. Yet, despite its subsequently great influence on scientific thought, general relativity was supported initially by very meager observational evidence. It has only been in the last two decades that a technological revolution has brought about a confrontation between general relativity and experiment at unprecedented levels of accuracy. It is not unusual to attain precise measurements within a fraction of a percent (and better) of the minuscule effects predicted by general relativity for the solar system.
To keep pace with these technological advances, gravitation theorists have developed a variety of mathematical tools to analyze the new high- precision results, and to develop new suggestions for future experiments made possible by further technological advances. The same tools are used to compare and contrast general relativity with its many competing theories of gravitation, to classify gravitational theories, and to under- stand the physical and observable consequences of such theories.
The first such mathematical tool to be thoroughly developed was a "theory of metric theories of gravity" known as the Parametrized Post-Newtonian (PPN) formalism, which was suited ideally to analyzing solar system tests of gravitational theories.
\end{quotation} \hfill \textsc{Clifford Will}, \textit{Theory and experiments in gravitational physics}

\vspace{2em}

This Chapter aims at presenting the basic tenets of General Relativity together with the experimental tests that the theory has passed with flying colors in the Solar system, from the famous Mercury anomalous perihelion precession to contemporary tests like the Gravity Probe B experiment, aimed at measuring the Lense-Thirring effect, or the MICROSCOPE satellite \cite{Touboul:2017grn} designed to test the Equivalence Principle with a relative precision of $10^{-14}$. We will follow an historical route by first recalling what led Einstein to build his theory, then presenting the first competitor of General Relativity (namely Brans-Dicke theory) and finally expanding on the modern version of the parametrized post-Newtonian (PPN) formalism, which condenses in a few parameters the conceivable deviations from GR in the Solar system. We will draw heavily from the work by Will \cite{Will_2014,Will:1993ns}, to which we refer the reader for a thorough presentation of the PPN formalism.

\section{Two classical tests of General Relativity}

When deriving General Relativity, Einstein was not concerned by the desire to shed light on unexplained experimental results. Instead, he was willing to construct a theory of gravity which would incorporate special relativity and Newton's laws. Nonetheless, soon after the theory was published, two experimental results remarkably confirmed its correctness: the anomalous perihelion precession of Mercury, and the bending of light by the Sun. We briefly review in what they consist below.

\subsection{Anomalous perihelion precession} \label{subsec:perihelion}

At this time, it had been known for half a century that the perihelion of Mercury presented a small anomaly in its precession. The french astronomer Urbain Le Verrier calculated precisely the perihelion precession caused by other planets and found a discrepancy of the order of 1\% with the observed value. Le Verrier immediately proposed that it would imply the existence of a small yet unobserved planet orbiting near to the Sun, which he named Vulcain. Indeed, it is by observing the anomalous prihelion precession of Uranus that Le Verrier successfully predicted the existence of Neptune a few years ago.

However, the proposed planet was never observed. Other attempts to account for this supplementary precession were all shown to be inconsistent with observation \cite{Will_2014}. Einstein sucessfully predicted a supplementary precession of $43$ arcsecond per century without conflicting with other observations. This prediction was obtained by deriving what is now known as the Einstein-Infeld-Hoffmann Lagrangian which is the first relativistic correction to the Newtonian potential. We will give its full expression in Chapter \ref{Chapter3}, Eq. \eqref{eq:LEIH}.

\subsection{Bending of light by the Sun} \label{subsec:light_bending}

\begin{figure}[th]
\centering
\includegraphics[scale=0.5]{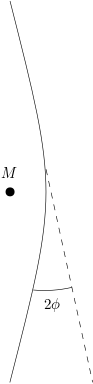}
\caption[Bending of light by the Sun]{Bending of light by the Sun. In Newtonian mechanics, the photon is assumed to follow an hyperbola.}
\label{fig:bending}
\end{figure}

As soon as 1801, Johann Georg von Soldner, a German physicist, calculated the bending that light should experience when passing close to the Sun, using only Newtonian mechanics. To this aim, he treated the photon as a massive particle (the mass of the photon will not influence its trajectory by the Equivalence Principle). We can quickly rederive his result by observing that the Newtonian orbits are parametrized by
\begin{equation}
r = \frac{p}{1+e \cos(\theta)}, \quad p=(m+M) \frac{L^2}{G M^2 m^2} \; ,
\end{equation}
where $m$ is the photon mass, $M$ is the Sun mass, $L$ is the angular momentum of the system, $e$ the eccentricity, and $\theta$ the true anomaly. The orbit is assumed to be hyperbolic, and we would like to obtain the deviation angle (see Fig. \ref{fig:bending}). The total (kinetic and potential) conserved energy of the system is found by evaluating it at periastron, where $r = p/(1+e)$,
\begin{align}
\begin{split}
E &= \frac{m v^2}{2} - \frac{GmM}{r} \\
&= \frac{(GM)^2 m^3}{2L^2} (e^2 -1) \; .
\end{split}
\end{align}
If we assume that the velocity of the photon at periastron is $c = 1$ in our units (note that it implies that its velocity at other points of the trajectory will differ from $c$ !), then the eccentricity is
\begin{align}
\begin{split}
e &= \sqrt{1+2\frac{L^2 E}{G^2 M^2 m^3}} \\
&= \frac{ r_p}{GM} - 1 \; ,
\end{split}
\end{align}
where $r_p$ is the periastron distance to the Sun. The eccentricity is simply related to the bending angle $2 \phi$ by the formula $\phi = \mathrm{Arcsin}(1/e)$. Assuming a very eccentric trajectory $e \gg 1$ (i.e, a small deflection), the bending angle is found to be
\begin{equation}
2\phi \simeq 2 \frac{GM}{ r_p} \; .
\end{equation}

Einstein first discovered the same result using special relativity. But, just after the end of World War I and using general relativity, he found that the predicted angle is twice the Newtonian value. Let us derive it using simple arguments. In isotropic coordinates, the Schwarzschild metric describing the weak-field Solar system is
\begin{equation}
ds^2 = - \left(1+\frac{2GM}{r} \right) dt^2 + \left(1-\frac{2GM}{r} \right) (dx^2+dy^2+dz^2) \; ,
\end{equation}
We have ignored terms of order $\mathcal{O}(2GM/r)$ or higher. For a light ray, setting $ds^2=0$ the trajectory can be described by
\begin{equation}
\frac{dl}{dt} = 1 - \frac{2GM}{r} \; ,
\end{equation}
where $dl = (dx^2+dy^2+dz^2)^{1/2}$. This equation is reminiscent of propagation in a material of refractive index $n = 1/(1 - 2GM/r) \simeq 1 + 2GM/r$. We then use the eikonal equation,
\begin{equation}
\frac{d(n \mathbf{u})}{ds} = \mathbf{\nabla} n \; ,
\end{equation}
where $\mathbf{u}$ is the vector tangent to the trajectory, and $s$ is the curvilinear abscissa. Integrating from $s=-\infty$ to $s=\infty$, the path of the photon is found to be curved according to
\begin{equation}
\mathbf{u}(\infty) - \mathbf{u}(-\infty) = \int ds \mathbf{\nabla} n \; .
\end{equation}
We can insert as a first-order approximation that the trajectory is along the $x$ axis. Then $ds = dx$, $y=\mathrm{constant}=r_p$ and the integral is
\begin{align}
\begin{split}
\mathbf{u}(\infty) - \mathbf{u}(-\infty) &= - 2 GM \int_{-\infty}^\infty dx \frac{y}{(x^2+y^2)^{3/2}} \hat{y} \\
&= - 4 \frac{GM}{r_s} \hat{y}
\end{split}
\end{align}
It appears that the deflection angle is now $4 GM / r_s$. In June 1918, two expeditions were organized, the first one led by Arthur Eddington and Frank Dyson, the second one by Andrew Crommelin and Charles Rundle Davidson. The aim was to measure the angular displacement of distant stars passing close to the Sun during an eclipse in May 1919, in Principe and Sobral. They confirmed Einstein's result, thus providing a strong check of general relativity.

\section{Mach's principle and Brans-Dicke theory} \label{sec:BD}

During the genesis of General Relativity, Einstein was concerned by an idea which was quite popular at that time, namely Mach's principle. This is a quite vague statement about the fact that the large-scale distribution of matter should determine the local inertial frames (Hermann Bondi and Joseph Samuel established a list of eleven different formulations of Mach's principle \cite{Bondi_1997}). Weinberg expresses it as
\begin{quote}
You are standing in a field looking at the stars. Your arms are resting freely at your side, and you see that the distant stars are not moving. Now start spinning. The stars are whirling around you and your arms are pulled away from your body. Why should your arms be pulled away when the stars are whirling? Why should they be dangling freely when the stars don't move?~\cite{Weinberg1972-WEIGAC}
\end{quote}

Einstein strongly believed that Mach's principle should be incorporated at the roots of a consistent theory of gravity. He stated as being on an equal footing three principles on which a satisfactory theory of gravitation should rest:
\begin{itemize}
\item The principle of relativity expressed by general covariance
\item The principle of equivalence
\item Mach's principle: that the $g_{\mu \nu}$ are completely determined by the mass of bodies, more generally by $T_{\mu \nu}$~\cite{Ebison_1983}.
\end{itemize}

Later on, in the 1960's, Jordan, Fierz, Brans and Dicke initiated a work on a class of gravitational theories which would incorporate more deeply the Mach's principle \cite{PhysRev.124.925, Fierz:1956zz, Jordan:1959eg}. The idea is to assume that even the gravitational "constant", $G$, should be a function of the distribution of matter in the universe. It is then natural to assume that $G$ becomes a new fundamental field which we denote by $G = 1/\phi$. The gravity action then looks like
\begin{equation} \label{eq:action_BD}
S = \int \mathrm{d}^4x \sqrt{-g} \left(\phi R + L_m[\psi_i, g_{\mu \nu}] + L_\phi \right) \; ,
\end{equation}
where $R$ is the Ricci scalar and $L_m[\psi_i, g_{\mu \nu}]$ the matter Lagrangian which is minimally coupled through $g_{\mu \nu}$ to some matter fields $\psi_i$. Since $\phi$ is now promoted to be a dynamical field, we have to provide a Lagrangian $L_\phi$ for it. The simplest possibility is to introduce a standard scalar kinetic term,
\begin{equation}
L_\phi = - \omega g^{\mu \nu} \partial_\mu \phi \partial_\nu \phi \; .
\end{equation}
However, since $\phi$ has a mass dimension of two, the coupling constant $\omega$ would have to bear the same dimension as $G$, which brings us back to the original problem. The second simplest possibility is then
\begin{equation}
L_\phi = - \frac{\omega}{\phi} g^{\mu \nu} \partial_\mu \phi \partial_\nu \phi \; .
\end{equation}
This choice leads to what is known as the Jordan-Fierz-Brans-Dicke (or Brans-Dicke for short) action. One can be even more "Machian" and assume that the dimensionless coupling constant depends itself on $\phi$, $\omega = \omega(\phi)$ (as the experiments usually take place in weak-field situations, we can Taylor expand the coupling function and get predictions depending on a small number of parameters). Indeed, this theory was proposed by Bergmann, Wagoner and Nordtvedt in the late 60's \cite{Bergmann:1968ve, Wagoner:1970vr, Nordtvedt:1970uv} and is now accepted as the most simple playground for departures from GR under the name of scalar-tensor theory. We will mention it again in Chapter \ref{Chapter2}, Section \ref{subsec:BDT}.

Let us come back to Brans-Dicke theory. Since the matter Lagrangian is not modified compared to GR, test particles would still follow geodesics and the Weak Equivalence Principle (WEP) is preserved. However, as we will demonstrate in Chapter \ref{Chapter4}, in Brans-Dicke (and more generally scalar-tensor) theory, the Strong Equivalence Principle (SEP) is not obeyed and we should see a strongly self-gravitating body like a neutron star falling in a different way than a test particle.
 
Adding another scalar component to the gravitational action implies that the gravitational field is now sourced by the total (matter plus scalar) Lagrangian. The scalar, being not directly coupled to matter, is determined by the gravitational field. We will analyze in more details the interaction between two massive bodies in a scalar-tensor theory in Chapter \ref{Chapter4}. 

Before closing this Section, let us make a comment about the change of frames. The action \eqref{eq:action_BD} is said to be in the "Jordan frame", where matter is minimally coupled to the gravitational field. However, we are free to perform a conformal redefinition of the metric when the field $\phi$ is not vanishing, $\tilde{g}_{\mu \nu} = \phi g_{\mu \nu}$, and the field redefinition $\phi = e^{\alpha}$. The action in terms of this new variable is
\begin{equation}
\int \mathrm{d}^4x \sqrt{- \tilde g} \left[ \tilde R - \left(\omega + \frac{3}{2} \right) \tilde g^{\mu \nu} \partial_\mu \alpha \partial_\nu \alpha + e^{-2\alpha} L_m[\psi_i, e^{-\alpha} \tilde g_{\mu \nu}] \right] \; .
\end{equation}
Thus is this "Einstein frame" the action is the one of Einstein gravity plus a massless scalar non-minimally coupled to matter through the Lagrangian $\tilde L_m = e^{-2\alpha} L_m[\psi_i, e^{-\alpha} \tilde g_{\mu \nu}]$. The physical predictions in the two frames are of course identical since we are just dealing with field redefinitions.

Brans-Dicke theory is an ideal playground for analyzing modifications of gravity because it is the simplest theory that one can think of beyond GR, yet giving quantitatively different predictions. It triggered ideas to test GR in new regimes, notably Nordtvedt test of the SEP. In the limit $\omega \rightarrow \infty$, one recovers General Relativity. From the weak bound $\omega > 5$ consistent with observations in the 70's, the current limit on the Brans-Dicke parameter has now greatly improved to $\omega > 40000$ from the Cassini spacecraft (see Section \ref{subsec:expe_ppn}).

\section{Parametrized post-Newtonian formalism}

Brans-Dicke theory pioneered an intense theoretical activity in devising new "metric" theories of gravity (interpreting gravitation as a phenomenon of curved spacetime) which were differing in some way from Einstein's GR. Among them, we can cite vector-tensor and scalar-vector-tensor theories, quadratic gravity and Chern-Simons theories, and massive gravity (we will present them in Chapter \ref{Chapter2}). Due to the proliferation of ideas, there was an urgent need to devise a formalism that would encompass them all when confronting theory and Solar system experiments, in order to bypass the tedious computation of all observational tests for each specific theory.

This was achieved by what is known as the parametrized post-Newtonian (PPN) formalism, pioneered by Nordtvedt and Will \cite{Nordtvedt:1968qs, 1971ApJ...163..611W, 1972ApJ...177..757W}. The post-Newtonian formalism itself will be presented in details in Chapter \ref{Chapter3}, Section \ref{subsec:pn_formalism}; loosely speaking, it consists in expanding the Lagrangian of a N-body system in powers of $v$, where $v$ is the typical velocity of the consituents which is assumed to be small compared to the speed of light, $v \ll 1$. In the Solar system, the first PN order expansion of GR is given by the Einstein-Infeld-Hoffman Lagrangian which is at the origin of the prediction for the perihelion precession of Mercury mentioned in Section \ref{subsec:perihelion}, and which will be derived in Chapter \ref{Chapter4}. However, for more strongly gravitating systems like black holes one has to go to higher perturbative orders in the PN expansion and this will be presented in Chapter \ref{Chapter3}. For the time being though, we will not need such refinements and will content ourselves with this general picture.
 Before giving the full form of the PPN formalism, let us define the class of theories to which it will apply (the metric theories of gravity) and give a simple historical toy model devised by Eddington.

\subsection{Metric theories of gravity}

The theories cited above all share the common property that they can be expressed in a "Jordan frame" where matter is minimally coupled to, and only to, the gravitational field $g_{\mu \nu}$. The supplementary fields can contribute to generate the spacetime curvature associated with the metric, but matter and non-gravitational fields obey equations of motion only sourced by the metric. This means that, in order to get equations of motion for objects in the Solar system, we can make a huge simplification by saying that we need only to parametrize the metric $g_{\mu \nu}$ in a low velocity, weak-field (or post-Newtonian) expansion about Minkowski space $\eta_{\mu \nu}$. Let us now give a concrete example of such an expansion in a simplified case.

\subsection{Eddington parameters} \label{sec:Eddington}

A toy model of the PPN expansion parameters is given by the Eddington-Robertson-Schiff parameters. Starting from the expression of the \sch metric in isotropic coordinates,
\begin{align}
\begin{split}
\mathrm{d}s^2 &= - \left( \frac{1-GM/2r}{1+GM/2r}\right) \mathrm{d}t^2  + \left(1+\frac{GM}{2r} \right)^4 (\mathrm{d}x^2 + \mathrm{d}y^2 + \mathrm{d}z^2) \\
&= - \left(1 - \frac{2GM}{r} + \frac{2G^2M^2}{r^2} + \dots \right) \mathrm{d}t^2 + \left(1+\frac{2GM}{r} + \dots \right) (\mathrm{d}x^2 + \mathrm{d}y^2 + \mathrm{d}z^2) \; ,
\end{split}
\end{align}
Eddington proposed to parametrize the metric of an unknown theory (possibly differing from GR) as
\begin{equation}
\mathrm{d}s^2 = - \left(1 - \frac{2GM}{r} + \beta \frac{2G^2M^2}{r^2} + \dots \right) \mathrm{d}t^2 + \left(1+ \gamma \frac{2GM}{r} + \dots \right) (\mathrm{d}x^2 + \mathrm{d}y^2 + \mathrm{d}z^2) \; .
\end{equation}
In the former metric we ensured that asymptotically $g_{\mu \nu} \sim \eta_{\mu \nu}$ and we did not introduce any parameter $\alpha$ for the first term of the expansion since it can be absorbed into a definition of the (ADM) mass $M$. In GR, $\beta = \gamma = 1$. In Brans-Dicke theory, $\beta = 1$ and $\gamma = \frac{1+\omega}{2+\omega}$~\cite{Will_2014}.

The convenient and simple form of the Eddington metric allows to derive observational signatures straightforwardly. For example, the light bending by the Sun explained in Section \ref{subsec:light_bending} in this new metric yields a deflection angle
\begin{equation} \label{eq:bending_gamma}
\theta = \left(\frac{1+\gamma}{2} \right) \frac{4GM}{r_p}
\end{equation}
which can be used to put constraint on $\gamma$ if no deviation is measured from GR. The historical measurement of Eddington yielded an accuracy of about 30\%; successive experiments based on the same principle did not result on much improved accuracy because of the inherent imprecision of the method. However, the development of Very Long Baseline Interferometry (VLBI) allowed to put much more precise constraints on the parameter $\gamma$, of the order of $10^{-4}$ (see Section \ref{subsec:expe_ppn}). The PPN parameters $\gamma$ and $\beta$ encode the most important deviations from GR inside the Solar system, even if they have been supplemented with several other parameters corresponding to distinct physical effects which we will present in the next Section.

Let us end this Section by quoting an audacious declaration from Einstein concerning the measurement of the bending angle: being asked what he would have done if the measurement of Eddington did not agree with GR, he said that he would "feel sorry for the dear Lord, for the theory is correct!". Since we just showed that a theory even more Machian than GR (thus abiding with Einstein's view on the subject) predicts a different angle, we can surely affirm that Einstein was quite lucky that GR conformed with the observation!

\subsection{Parametrized Post-Newtonian formalism}

The PPN formalism is a generalization of the Eddington metric we introduced in the last Section. It relies on the fact that, inside the Solar system, the velocities and internal energies of all bodies are small,
\begin{equation}
U \sim \frac{\delta \rho}{\rho} \sim \frac{p}{\rho} \sim v^2 \ll 1 \; ,
\end{equation}
where $U$ is the Newtonian potential, $p$ and $\rho$ are the pressure and rest-mass density of matter (as measured in a local freely falling frame momentarily comoving with the matter) and $\delta \rho$ is the perturbed energy density (that is, all forms of internal energy which are not rest-mass and gravitational: e.g., energy of compression and thermal energy). We wish to expand all the metric quantities to the desired order and obtain equations of motion for planetary bodies generalizing the Newtonian ones by including the first-order general relativity correction. We treat each body as a point-particle, neglecting finite-size (tidal) effects. The correct order to expand metric quantities can be seen by expanding the point-particle action encapsulating the equations of motion of each body,
\begin{align}
\begin{split}
S_m &= -m \int dt \sqrt{-g_{\mu \nu} v^\mu v^\nu} \; , \\
&= -m \int dt \sqrt{- g_{00} - 2 g_{0i}v^i - g_{ij}v^i v^j} \; ,
\end{split}
\end{align}
where $v^\mu = \frac{dx^\mu}{dt}$ is the four-velocity of the pointlike body. The point-particle action contains
\begin{itemize}
\item To $\mathcal{O}(v^0)$ only the rest-mass of the body, 
\item To $\mathcal{O}(v^2)$ the Newtonian energy, 
\item To $\mathcal{O}(v^4)$ the post-Newtonian energy (1PN) 
\end{itemize}
(terms containing an odd number of powers of $v$, representing energy dissipation by the system, arise only at a higher order in known metric theories of gravity). We should thus expand $g_{00}$ to $\mathcal{O}(v^4)$, $g_{0i}$ to $\mathcal{O}(v^3)$ and $g_{ij}$ to $\mathcal{O}(v^2)$ in order to get the first post-Newtonian term. 

Let us now model the collection of point-particle objects forming the Solar system as a perfect fluid. Its energy-momentum tensor is given by
\begin{equation}
T^{\mu \nu} = (\rho + \delta \rho + p)v^\mu v^\nu + p g^{\mu \nu} \; .
\end{equation}
With such an energy-momentum tensor one can construct e.g the Newtonian potential,
\begin{equation}
U(\mathbf{x}, t) =G \int \mathrm{d}^3 x' \frac{ \rho(\mathbf{x}', t)}{\vert \mathbf{x} - \mathbf{x}' \vert} \; ,
\end{equation}
which will enter the parametrized metric along with other quantities (which we will describe shortly). We take the point of view of Eulerian fluid dynamics, where a velocity $\mathbf{v}$ is associated to each position in space $\mathbf{x}$. The density $\rho$ obeys the continuity equation,
\begin{equation}
\frac{\partial \rho}{\partial t} + \mathbf{\nabla} (\rho v) = \mathcal{O}(v^2) \; ,
\end{equation}
the $\mathcal{O}(v^2)$ term containing relativistic corrections which we will not need at 1PN. We will use this equation to simplify the 1PN potentials in what follows.

The form of the expanded metric can be restricted using the symmetries of spacetime which are still obeyed in the expansion (the idea of using symmetries to restrict the form of terms entering an expansion is reminiscent of Effective Field Theories (EFT) introduced in particle physics, and will be developed further in Chapter \ref{Chapter2} when presenting the Effective Field Theory of Dark Energy). In this case, while time translations and boosts mix together different components of the metric, we can use that $g_{00}$, $g_{0i}$ and $g_{ij}$ should still be separately covariant under space rotations. Imposing a few other restrictions on the metric (see \cite{Will:1993ns}), one can show that the only quantities allowed in the expansion of the metric at the desired order are:
\begin{itemize}
\item $g_{ij}$ to $\mathcal{O}(v^2)$ (should behave as a tensor under rotations): $U \delta_{ij}$ and $U_{ij}$ where
\begin{equation}
U_{ij} = G \int \mathrm{d}^3 x' \frac{ \rho(\mathbf{x}', t) (x - x')_i (x - x')_j}{\vert \mathbf{x} - \mathbf{x}' \vert^3} \; .
\end{equation}

\item $g_{0i}$ to $\mathcal{O}(v^3)$ (should behave as a vector under rotations): $V_i$, $W_i$ where
\begin{equation}
V_i = G \int \mathrm{d}^3 x' \frac{ \rho(\mathbf{x}', t) v'_i}{\vert \mathbf{x} - \mathbf{x}' \vert} \; , \quad W_i = G \int \mathrm{d}^3 x' \frac{ \rho(\mathbf{x}', t) (x - x')_i \mathbf{v}' \cdot (\mathbf{x} - \mathbf{x}')}{\vert \mathbf{x} - \mathbf{x}' \vert^3} \; .
\end{equation}

\item $g_{00}$ to $\mathcal{O}(v^4)$ (should behave as a scalar under rotations): $U^2$, $\Phi_W$, $\Phi_1$, $\Phi_2$, $\Phi_3$, $\Phi_4$, $\mathcal{A}$, $\mathcal{B}$ where
\begin{align}
\begin{split}
\Phi_W &= G \int \mathrm{d}^3 x' \mathrm{d}^3 x'' \frac{ \rho(\mathbf{x}', t) \rho(\mathbf{x}'', t) (\mathbf{x} - \mathbf{x}')}{\vert \mathbf{x} - \mathbf{x}' \vert^3} \cdot \left( \frac{\mathbf{x}' - \mathbf{x}''}{\vert \mathbf{x} - \mathbf{x}'' \vert} - \frac{\mathbf{x} - \mathbf{x}''}{\vert \mathbf{x}' - \mathbf{x}'' \vert} \right) \; , \\
\Phi_1 &= G \int \mathrm{d}^3 x' \frac{ \rho(\mathbf{x}', t) v'^2}{\vert \mathbf{x} - \mathbf{x}' \vert} \; , \quad \Phi_2 = G \int \mathrm{d}^3 x' \frac{ \rho(\mathbf{x}', t) U(\mathbf{x}', t)}{\vert \mathbf{x} - \mathbf{x}' \vert} \; , \\
\Phi_3 &= G \int \mathrm{d}^3 x' \frac{\delta \rho(\mathbf{x}', t)}{\vert \mathbf{x} - \mathbf{x}' \vert} \; , \quad \Phi_4 = G \int \mathrm{d}^3 x' \frac{p(\mathbf{x}', t)}{\vert \mathbf{x} - \mathbf{x}' \vert} \; , \\
\mathcal{A} &= G \int \mathrm{d}^3 x' \frac{ \rho(\mathbf{x}', t) [\mathbf{v}' \cdot (\mathbf{x} - \mathbf{x}')]^2}{\vert \mathbf{x} - \mathbf{x}' \vert^3} \; , \quad \mathcal{B} = G \int \mathrm{d}^3 x' \frac{ \rho(\mathbf{x}', t)}{\vert \mathbf{x} - \mathbf{x}' \vert} (\mathbf{x} - \mathbf{x}') \cdot \frac{\mathrm{d} \mathbf{v}'}{\mathrm{d}t} \; . 
\end{split}
\end{align}
\end{itemize}
In addition, the expansion can also involve the velocity of the Solar system with respect to the mean rest-frame of the universe, $w^i$. This variable involves preferred-frame effects present when fundamental vector fields participate in the gravitational dynamics.

There just remains to give a name to each parameter multiplying the twelve functions given above. However, before to do this we must choose the gauge. Indeed, with a small coordinate transformation $x^\mu \rightarrow x^\mu + \xi^\mu$ one can generate a 1PN perturbation of the metric $g_{\mu \nu} \rightarrow g_{\mu \nu} + \partial_\mu \xi_\nu + \partial_\nu \xi_\mu$  which has no physical significance. Let us take as an example the spatial part of the transformation to be
\begin{equation}
\xi_i = \lambda \partial_i \chi, \quad \chi = G \int \mathrm{d}^3 x'  \rho(\mathbf{x}', t) \vert \mathbf{x} - \mathbf{x}' \vert
\end{equation} 
Then one easily see that the spatial part of the metric transforms as
\begin{equation}
g_{ij} \rightarrow g_{ij} + \lambda (U \delta_{ij} - U_{ij})
\end{equation}
Consequently, with a gauge transformation one can set $U_{ij}$ to zero. It is straightforward to show that with a similar temporal part of the gauge transformation $\xi_0$ one can set $\mathcal{B} = 0$~\cite{Will:1993ns}. Thus, there remains ten functions to be described by ten PPN parameters. It is then just a matter of convention to assign a letter to each parameter. In the current version of the PPN expansion, the metric writes as
\begin{align}
\begin{split}
g_{ij} &= (1+2 \gamma U) \delta_{ij} + \mathcal{O}(v^4) \; , \\
g_{0i} &= - \frac{1}{2} (4 \gamma + 3 + \alpha_1 - \alpha_2 + \zeta_1 - 2 \xi)V_i - \frac{1}{2}(1+\alpha_2-\zeta_1+2\xi)W_i \\
&-\frac{1}{2}(\alpha_1-2\alpha_2)w_i U - \alpha_2 w^J U_{ij} + \mathcal{O}(v^5) \; , \\
g_{00} &= -1 + 2U - 2\beta U^2 - 2 \xi \Phi_W + (2 \gamma + 2 + \alpha_3 + \zeta_1-2\xi) \Phi_1 \\
&+ 2(3 \gamma - 2 \beta + 1 + \zeta_2 + \xi) \Phi_2 + 2(1+\zeta_3) \Phi_3 + 2(3 \gamma + 3 \zeta_4 - 2 \xi)\Phi_4 \\
& - (\zeta_1 - 2 \xi) \mathcal{A} - (\alpha_1 - \alpha_2 - \alpha_3) w^2 U - \alpha_2 w^i w^j U_{ij} + (2\alpha_3 - \alpha_1) w^i V_i +\mathcal{O}(v^6) \; .
\end{split}
\end{align}

Because in all known theories of gravity the parameter $\zeta_4$ is related to the other PPN parameters via
\begin{equation}
6 \zeta_4 = 3 \alpha_3 + 2 \zeta_1 - 3 \zeta_3 \; ,
\end{equation}
we will not consider it to be an independent parameter, thus reducing the total number of parameters to 9. The total list of PPN parameters together with their physical interpretation and their current constraints can be found in Table \ref{table:ppn}

\begin{table}
\begin{center}
\caption{Experimental constraints on the PPN parameters. Table taken from Ref.~\cite{Will_2014} \label{table:ppn}}
\begin{tabular}{cccc}
\hline
Parameter & Effect & Limit & Experiment \\
\hline
$\gamma - 1$ & Time delay & $2.3 \times 10^{-5}$ & Cassini  \\
$\beta - 1$  & Perihelion shift & $8 \times 10^{-5}$ & Messenger \\
$\xi$  & Spin precession & $4 \times 10^{-9}$ & Millisecond pulsars \\
$\alpha_1$  & Orbital polarization & $4 \times 10^{-5}$ & PSR J1738+0333 \\
$\alpha_2$ & Spin precession & $2 \times 10^{-9}$ & Millisecond pulsars \\
$\alpha_3$  & Pulsar acceleration & $4 \times 10^{-20}$ & Pulsar $\dot P$ statistics \\
$\zeta_1$  & - & $2 \times 10^{-2}$ & Combined PPN bounds \\
$\zeta_2$  & Binary acceleration & $4 \times 10^{-5}$ & PSR 1913+16 \\
$\zeta_3$  & Newton's third law & $10^{-8}$ & Lunar acceleration \\
\hline
\end{tabular}
\end{center}
\end{table}

\subsection{Experimental constraints on the PPN parameters} \label{subsec:expe_ppn}

General Relativity is now tested on small scales with great accuracy and this translates into constraints on the PPN parameters. In this Section we will briefly recap some of the most important experimental tests of GR, but the reader interested in a detailed account should refer to \cite{Will_2014}.

\begin{itemize}
\item \textit{Light deflection}: We have mentioned above that the bending angle of a photon passing near to the Sun depends on the parameter $\gamma$, cf Eq. \eqref{eq:bending_gamma}. By looking at the deviation angle between a distant reference luminous source and another source passing nearby the Sun, one can have access to the deviation angle. The most precise of these measurements are obtained with Very Long Baseline radio Interferometry (VLBI), by looking at distant radio quasars: these techniques can measure angular separations up to 100 microarcseconds. Analyses that incorporated data through 2010 yielded \cite{Lambert_2009, Lambert_2011}:
\begin{equation}
\gamma - 1 = (-0.8\pm 1.2) \times 10^{-4} \; .
\end{equation}

\item \textit{Shapiro time delay}: General Relativity predicts that a photon will take an additional time (with respect to flat space) to make a round trip between the Earth and a planet or a satellite passing close to the Sun (from the Earth's point of view). This effect was discovered by Shapiro in 1964 \cite{Shapiro:1964aa}. The total travel time is \cite{Will_2014}
\begin{equation}
\Delta t = 2 \vert \mathbf{x}_\oplus - \mathbf{x}_p \vert + 2(1+\gamma) m_\odot \ln\left( \frac{4 r_\oplus r_p}{d^2} \right) \; ,
\end{equation}
where, when using a coordinate system centered on the Sun, $\mathbf{x}_\oplus$ is the position of the Earth, $\mathbf{x}_p$ is the position of the planet or the satellite on the other side of the Sun, $r_\oplus = \vert \mathbf{x}_\oplus \vert$, $r_p = \vert \mathbf{x}_p \vert$ and $d$ is the distance of closest approach to the Sun. In practice, experiments measure the time delay between a light ray passing far from the Sun and another one passing close by. We will recover a similar formula when studying black holes with hair in Part \ref{part4}.

The most precise measurement of this delay was made by the Cassini spacecraft on its way to Saturn, the distance of closest approach being $d \simeq 1.6 R_\oplus$. This yielded the most precise constraint on $\gamma$ up to now \cite{Bertotti:2003rm}:
\begin{equation} \label{eq:constraint_gamma}
\gamma - 1 = (2.1 \pm 2.3) \times 10^{-5} \; .
\end{equation}


\item \textit{Geodetic precession:} A gyroscope moving through curved spacetime suffers a precession of its spin axis $\mathbf{S}$ given by \cite{Will_2014}
\begin{equation}
\frac{d\mathbf{S}}{d\tau} = \mathbf \Omega_G \times \mathbf S, \quad \mathbf \Omega_G = \left( \gamma + \frac{1}{2} \right) \mathbf v \times \nabla U \; ,
\end{equation}
where $\mathbf v$ is the velocity of the gyroscope and $U$ is an external Newtonian potential. The Earth-Moon system can be seen as a gyroscope moving in the field of the Sun: this effect has been measured to $0.6$ percent accuracy. Another measurement is from the Gravity Probe B experiment \cite{PhysRevLett.106.221101} placing gyroscopes in orbit around the Earth, giving a precision of $0.3$ percent still not competitive with the one of the Cassini Spacecraft for measuring $\gamma$.

\item \textit{Perihelion precession of Mercury}: In terms of the PPN parameters, the perihelion precession discussed in Section \ref{subsec:perihelion} is \cite{Will_2014}
\begin{equation}
\dot \omega = 42''98 \left( \frac{1}{3}(2+2\gamma-\beta) + 3 \times 10^{-4} \frac{J_2}{10^{-7}} \right) \; ,
\end{equation}
where $\dot \omega$ is the rate of perihelion shift in arcseconds per century, and $J_2$ is the quadrupole moment of the Sun which is poorly known and only estimated as $J_2 \sim 10^{-7}$~\cite{Rozelot_2011}. The Messenger spacecraft yielded the most precise measurement of the perihelion precession, with an uncertainty of $1.5 \times 10^{-3}$ arcseconds per century. Using the Cassini bound on $\gamma$, this translates into a constraint on $\beta$~\cite{Park_2017}:
\begin{equation}
\beta - 1 = (-2.7 \pm 3.9) \times 10^{-5}
\end{equation}

\item \textit{Nordtvedt effect and Lunar Laser Ranging}: Even if the Weak Equivalence Principle is satisfied in numerous alternative gravity theories, there could be violations of the Strong Equivalence Principle (SEP) caused by the fact that the gravitational energy of a body depends on the field configuration and so on its composition \cite{Nordtvedt_1968}. We will derive this effect in a scalar-tensor theory in Chapter \ref{Chapter4}, and show that it can be identified as a renormalization of the scalar charge. The net result of this effect is that the acceleration of a body is related to the gradient of an external gravitational potential \textit{via}
\begin{align}
\begin{split}
\mathbf{a} &= \frac{m_p}{m} \nabla U \; , \\
\frac{m_p}{m} &= 1 - \eta \frac{E_g}{m} \; , \\
\eta &= 4 \beta - \gamma - 3 - \frac{10}{3} \xi - \alpha_1 + \frac{2}{3} \alpha_2 - \frac{2}{3} \zeta_1 - \frac{1}{3} \zeta_2 \; ,
\end{split}
\end{align}
where $E_g$ is the (absolute value of the) gravitational energy of the body. This effect is absent in GR ($\eta = 0$) but generically present in other theories of gravity.

A consequence of this violation of the SEP is that the orbit of the Moon around the Earth would be perturbed with an amplitude \cite{Will_2014}
\begin{equation}
\delta r = 13.1 \eta \cos(\omega_0 - \omega_s)t \; [\mathrm{m}] \; ,
\end{equation}
where $\omega_0$ and $\omega_s$ are the angular frequencies of the orbits of the Moon and Sun around the Earth. We will rederive a similar effect for theories equipped with a screening mechanism in Chapter \ref{Chapter7}.

Since August 1969, the position of the Moon is measured with an incredible accuracy by shining a laser on the retroreflector set by the Apollo 11 mission on the Moon. The time travel is estimated with an accuracy approaching $5$ps ($1$mm). Several effects must be accounted for, such as perturbations due to the Sun or other planets, tidal interactions, the librations of the Moon, the orientation of the Earth, the location of the observatories, and atmospheric effects on the signal propagation. The final result concerning the post-Newtonian parameters is a bound on $\eta$~\cite{Merkowitz_2010}:
\begin{equation}
\vert \eta \vert = (4.4 \pm 4.5) \times 10^{-4}
\end{equation}

Finally, let us mention the MICROSCOPE mission aiming at testing the Equivalence Principle by comparing the motion of two test masses (one of platinum and the other of titanium) embarked on a satellite in orbit around the Earth \cite{Touboul:2017grn,Berge:2015dya,Touboul:2012ui}. The differential relative acceleration of the two bodies has been constrained to be less that $10^{-14}$, which is an order of magnitude better than the LLR experiment (however, in Brans-Dicke type theories, only the Strong Equivalence Principle is violated so that there is no observable deviation from GR concerning the motion of test-masses).

\item \textit{Gravitomagnetism}: A rotating body produces a gravitational field analogous to a magnetic field. In particular, a gyroscope orbiting around this rotating body will experience a precession of its spin $\mathbf S$ different from the geodetic precession previously discussed. This is the Lense-Thirring effect \cite{Will_2014}:
\begin{equation}
\frac{d\mathbf S}{d\tau} = \mathbf \Omega_\mathrm{LT} \times \mathbf S, \quad \mathbf \Omega_\mathrm{LT} = - \frac{1}{2}\left(1+\gamma+\frac{1}{4}\alpha_1\right) \frac{\mathbf J - 3 \mathbf n(\mathbf n \cdot \mathbf J)}{r^3} \; ,
\end{equation}
where $\mathbf J$ is the angular momentum of the central body, $r$ is the distance of the gyroscope to this body and $\mathbf n$ is its unit vector. In $\mathbf \Omega_\mathrm{LT}$ one recognizes the expression of the magnetic field created by a magnetic moment.

The Gravity Probe B experiment measured this effect with an accuracy of 20 percent \cite{Will_2014}.

\item \textit{Temporal variation of Newton's "constant":}: A typical prediction of Brans-Dicke and scalar-tensor gravity is that Newton's "constant", which is replaced by a scalar field, should vary with time. Cosmological boundary conditions imply that this rate of variation should be proportional to the expansion rate of the universe, i.e $\dot G/G = \sigma H_0$ where $H_0 \simeq 7 \times 10^{-11} \mathrm{yr}^{-1}$ is the Hubble parameter today (in units adapted to Solar system measurements), and $\sigma$ is a constant depending on the theory considered. For Brans-Dicke theory, $\sigma = -3 q_0 (\omega+2)^{-1}$ where $q_0$ is the cosmological deceleration parameter \cite{Will_2014, PhysRevLett.106.221101}.

The best constraints of $\dot G/G$ come from the ephemeris of Mars using several satellites \cite{KONOPLIV2011401}, and from Lunar Laser Ranging. The effect can be found by simply replacing $G = G_0 + \dot G_0(t-t_0)$ in Newton's equations of motion and analyzing the perturbations on the orbits of the planets and the Moon. This yields \cite{Will_2014}:
\begin{equation}
\frac{\dot G}{G} = (0.1 \pm 1.6) \times 10^{-13} \mathrm{yr}^{-1} \; .
\end{equation}
It is to be noted that models of Big-Bang nucleosynthesis indicate that $G$ was within 20 percent of its value today. Assuming a power-law variation of $G \sim t^{-\alpha}$ yields $\dot G / G = (0 \pm 4) \times 10^{-13} \mathrm{yr}^{-1}$ \cite{Bambi_2005, Copi_2004}.

\item \textit{Preferred-frame and preferred-location effects:} Some theories of gravity violate SEP by predicting that the outcomes of local gravitational experiments may depend on the velocity of the laboratory relative to the mean rest frame of the universe (preferred-frame effects) or on the location of the laboratory relative to a nearby gravitating body (preferred-location effects). In the post-Newtonian limit, preferred-frame effects are governed by the values of the PPN parameters $\alpha_1$, $\alpha_2$, and $\alpha_3$, and some preferred-location effects are governed by $\xi$~\cite{Will_2014}. 

These violations of the SEP manifest themselves in anomalous orbits of planets and pulsars and anomalous torques on rotating stars or pulsars. The best bound on the PPN parameters come from pulsars observations and are summed up in Table \ref{table:ppn}.

\item \textit{Tests of post-Newtonian conservation laws:} The four PPN parameters $\zeta_1, \zeta_2, \zeta_3$ and $\alpha_3$ measure possible violations of conservation of total momentum or of Newton's third law. This can be constrained using LLR data since there is an asymmetry in the distribution of aluminium and iron in the Moon. This translates into a direct constraint on $\zeta_3$~\cite{Will_2014}:
\begin{equation}
\zeta_3 < 1 \times 10^{-8} \; .
\end{equation}
Another consequence of the violation of conservation of momentum is the self-acceleration of the center of mass of a binary system, proportional to $\zeta_2 + \alpha_3$. This can be constrained using observations of binary pulsars. Since $\alpha_3$ is already very constrained by preferred-frames effects (see Table \ref{table:ppn}), this translates into a bound on $\zeta_2$~\cite{Will_1992}:
\begin{equation}
\zeta_2 < 4 \times 10^{-5} \; .
\end{equation}

\end{itemize}



\chapter{Cosmological tests of gravity, and the Effective Field Theory of Dark Energy} 

\label{Chapter2}

\begin{quotation}
There are two aspects of cosmology today that make it more alluring than ever. First, there is an enormous amount of data. To give just one example of how rapidly our knowledge of the structure of the universe is advancing, consider galaxy surveys which map the sky. In 1985, the state-of-the-art survey was the one carried out by the Center for Astrophysics; it consisted of the positions of 1100 galaxies. Today, the Sloan Digital Sky Survey and the Two Degree Field between them have recorded the 3D positions of half a million galaxies.
The other aspect of modern cosmology which distinguishes it from previous efforts to understand the universe is that we have developed a consistent theoretical framework which agrees quantitatively with the data. These two features are the secret of the excitement in modern cosmology: we have a theory which makes predictions, and these predictions can be tested by observations.
\end{quotation} \hfill Scott \textsc{Dodelson}, \textit{Modern Cosmology}, 2003

\vspace{2em}

This quotation from one of the most well-known cosmology textbook has never been so true.  The Euclid mission is going to measure the shapes of over a billion galaxies \cite{laureijs2011euclid}, increasing the size of galaxy surveys by a thousand since the time Dodelson wrote his introduction. This incredible amount of data fits remarkably well into the picture of a (perturbed) homogeneous and isotropic universe as first described by Friedmann, Lemaître, Robertson and Walker (FLRW), in which the background metric of the universe contains one single free function, the expansion factor $a(t)$:
\begin{equation}
\mathrm{d}s^2 = - \mathrm{d}t^2 + a(t)^2 \left[ \frac{\mathrm{d}r^2}{1-k r^2} + r^2 d\theta^2 + r^2 \sin^2 \theta \mathrm{d} \phi^2 \right] \; ,
\end{equation}
where the metric has been written in polar comoving coordinates and physical time, and $k$ is the normalized spatial curvature of constant-time hypersurfaces which can take the values $0$, $1$ or $-1$. The physical coordinates $\mathbf{x}_\mathrm{phys}$ are related to the comoving ones via $\mathbf{x}_\mathrm{phys} = a(t) \mathbf{x}$, and the redshift of a photon is defined as $1+z = 1/a$ where we have normalized the present time scale factor to one, $a_0 = 1$.
 Einstein's equations then link the geometry of spacetime, here encoded in the sole scale factor $a(t)$, to the content of matter in the universe contained in the energy-momentum tensor $T_{\mu \nu}$. Assuming a  perfect fluid for the matter content,
\begin{equation}
T_{00} = \rho \; , \quad T_{i0} = 0 \; , \quad T_{ij} = a^2 p \delta_{ij} \; ,
\end{equation}
where $\rho$ and $p$ are the proper energy density and pressure, the temporal and spatial parts of the Einstein equations give respectively
\begin{align}
H^2 &\equiv \left( \frac{\dot a}{a} \right)^2 = \frac{8\pi G}{3} \rho - \frac{k}{a^2} + \frac{\Lambda}{3} \\
\frac{\ddot a}{a} &= - \frac{4\pi G}{3}(\rho + 3p) + \frac{\Lambda}{3}
\end{align}
where $\Lambda$ is the cosmological constant, sometimes included in the matter sector with an energy density $\rho_\Lambda = \Lambda/8\pi G$. From the second equation it is clear that the effect of the cosmological constant is to make the universe accelerate, $\ddot a >0$ for large enough $\Lambda$. This is what observations pointed out starting in 1998, as we review in Section~\ref{sec:open_issues}. Assuming that the universe is filled with matter and radiation, whose energy density $\rho_m$ and $\rho_r$ decay respectively as $a^{-3}$ and $a^{-4}$ as the universe expands (the supplementary suppression of the radiation energy density compared to the simple $a^{-3}$ volume suppression comes from the dilution of the energy of photons which goes as $a^{-1}$), one can rewrite the first equation as
\begin{equation}
\frac{H^2}{H_0^2} = \Omega_r a^{-4} + \Omega_m a^{-3} + \Omega_k a^{-2} + \Omega_\Lambda \; ,
\end{equation}
where the $\Omega$ parameters represent the fraction of energy of the universe in each component
\begin{equation} \label{eq:defOmega}
\Omega_r = \frac{8 \pi G \rho_r}{3 H_0^2} \; , \quad \Omega_m = \frac{8 \pi G \rho_m}{3 H_0^2} \; , \quad \Omega_k = - \frac{k}{H_0^2} \; , \quad \Omega_\Lambda = \frac{\Lambda}{3 H_0^2} \; .
\end{equation}
These parameters obey $\Omega_\Lambda + \Omega_k +\Omega_m +\Omega_r = 1$. This is the core of the $\Lambda$CDM model.


The Planck mission has given (among others) the most precise measurement of $H_0$ and the density parameters \cite{Aghanim:2018eyx}, when combined with Baryon Acoustic Oscillations (BAO) from the Baryon Oscillation Spectroscopic Survey (BOSS) \cite{Anderson:2013zyy}:
\begin{align} \label{eq:Planck+BAO}
\begin{split}
H_0 &= (67.36 \pm 0.54) \text{km} . \text{s}^{-1} .  \text{Mpc}^{-1} \; , \quad \Omega_k = 0.001 \pm 0.002 \; , \\
\Omega_m &= 0.3166 \pm 0.0084 \; , \quad \Omega_\Lambda = 0.6847 \pm 0.0073 \; .
\end{split}
\end{align}

 Planck measures the temperature of photons emitted at the last scattering surface when the universe became transparent to light (a time named recombination), while BOSS measures fluctuations in galaxies and quasars situated at respectively $z \lesssim 1$ and $z \lesssim 3$. These fluctuations are due to acoustic waves in the primordial plasma composed of photons, baryons and dark matter: since photons and matter were tightly coupled together, acoustic waves were able to propagate in this plasma at a velocity of around half the speed of light. However, at the decoupling time photons ceased to be coupled to matter so that acoustic waves in the matter distribution were 'frozen' at this epoch. It is the relics of these wave that BOSS measures in the distribution of galaxies, at a much later time. An important point to remember is that, although recombination and decoupling happen at a similar redshift, they are close but distinct phenomena. Recombination happens at $z \simeq 1089$ when the photons can freely stream on the universe. On the other hand, baryons decouple from photons at $z \simeq 1059$ when they cease to feel the drag force from photons.
 
As we can see from the Planck+BAO measurements~\eqref{eq:Planck+BAO}, the curvature energy density is negligible, so that in the remaining of this thesis we will consider a flat universe $k=0$; on the other hand, dark energy is the dominant component of our universe. Explaining such a value for the cosmological constant is currently a challenge for physics, see Section~\ref{sec:open_issues}. This is one of the issues we will discuss in the next Section, another one being the incompatibility of the $H_0$ measurement of Planck-BAO with local measurements with supernovae. 

The growing number of tensions and theoretical issues with the standard model of cosmology has driven a increasing interest to explore modifications of gravity. It is impossible to make justice to all of these new developments in a concise way, so that we will only review a very partial list of them in Section~\ref{sec:theory}. This will be our starting point to introduce effective field theory ideas in cosmology. Indeed, as in the years preceding the post-Newtonian formalism, there was an urgent need of an effective and unifying formalism which would encompass a large class of modifications of gravity in a small set of measurable parameters. This has been achieved by the Effective Field Theory of Dark Energy (EFT of DE), which we will present in Section~\ref{sec:EFTDE} as well as the constraints from cosmology on the EFT parameters in Section~\ref{sec:expe_EFTDE}.

\section{Open issues in cosmology} \label{sec:open_issues}

\subsection{Evidence for an accelerated expansion}

In the introduction above, we have seen that the combination of Planck and BAO data constrains DE to constitute approximately 70 \% of the energy budget of the universe, so that the scale factor accelerates ($\ddot a >0$), contrary to the case of a matter-filled universe ($\ddot a <0$). The first historical proof of the acceleration of the expansion has been given in 1998 by the Supernova Cosmology Project and the High-Z Supernova Search Team led by respectively Saul Perlmutter and Adam Riess \cite{Riess:1998cb, Perlmutter:1998np}. By analyzing supernovae whose intrinsinc luminosity is known (see however the caveats of this methodology in Section~\ref{subsec:H0}), they were able to infer the redshift of the sources from their luminosity distance. Their data showed a nonzero $\Omega_\Lambda$ at the 4$\sigma$ level.

\begin{figure}[ht]
\centering
\includegraphics[scale=0.5]{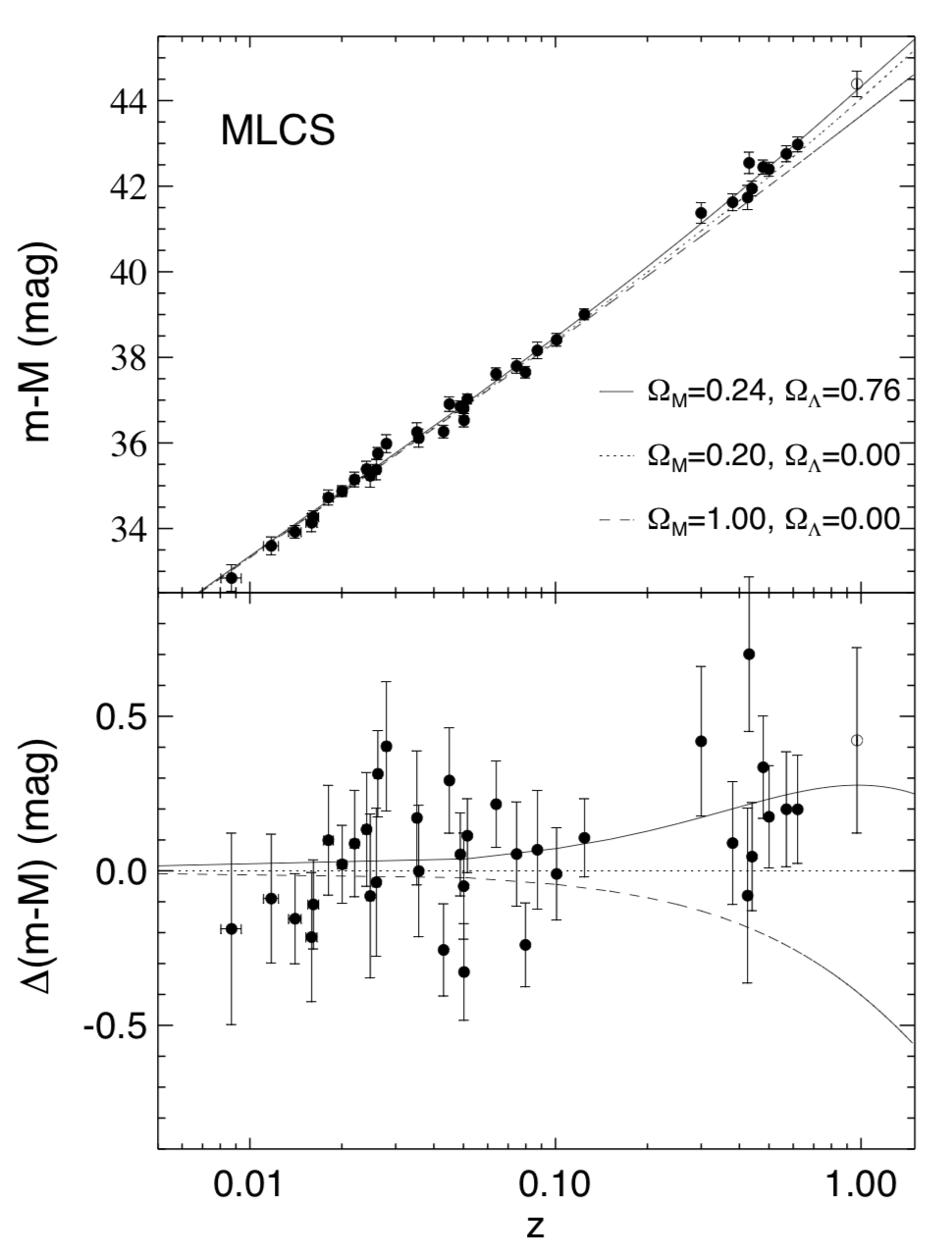}
\caption[Evidence for accelerated expansion]{Data from the High-Z Supernova Search Team \cite{Riess:1998cb} showing an evidence for $\Omega_\Lambda >0$. The $x$ axis is the redshift of the source, while the $y$ axis is the difference between the absolute and relative magnitudes which is proportional to the logarithm of the luminosity distance.}
\label{fig:Riess}
\end{figure}

 However, a number of puzzles  remain associated with the cosmological constant. The most prominent one is its unnaturally small value : if dark energy is really a vacuum energy, then a naive QFT estimate predicts a value which is 120 orders of magnitude too high ! Let us elaborate a bit on this point. For a clear account of the cosmological constant problem, we refer the reader to the Weinberg paper \cite{Weinberg:1988cp} which, even if it dates back to 2000, is still remarkably up to date.

 In QFT, the vacuum energy can be obtained by summing the zero-point energies of the modes of some massive field of mass $m$. Introducing a high-energy cutoff  $\tilde \Lambda$ which can be seen as the highest energy scale at which the QFT we consider is valid, we write for $\tilde \Lambda$ high enough
\begin{equation}
\rho_\Lambda \sim \int^{\tilde \Lambda} \frac{d^3k}{(2 \pi)^2} \frac{1}{2} \sqrt{k^2+m^2} \simeq \frac{\tilde \Lambda^4}{16 \pi^2} \; .
\end{equation}
 If we trust our quantum theories up to the Planck scale so that $\tilde \Lambda \sim \mpl$, then one obtains a value 120 orders of magnitude too important since by Eq.~\eqref{eq:defOmega} one has $\rho_\Lambda \sim H_0^2 \mpl^2$ from observations. One could argue that this result is regularization dependent and as such unphysical. Indeed, in dimensional regularization where power-law divergences disappear one would obtain
\begin{equation}
 \rho_\Lambda \sim \frac{m^4}{64 \pi^2}  \ln \left( \frac{m^2}{\mu^2} \right) \; ,
\end{equation}
where $\mu$ is the substraction scale of dimensional regularization. In this case, taking the Higgs or top quark mass as giving the physical mass entering the bare dark energy density gives a result 'only' 52 orders of magnitudes too high. 

\begin{figure}[ht]
\centering
\includegraphics[width=0.6\textwidth]{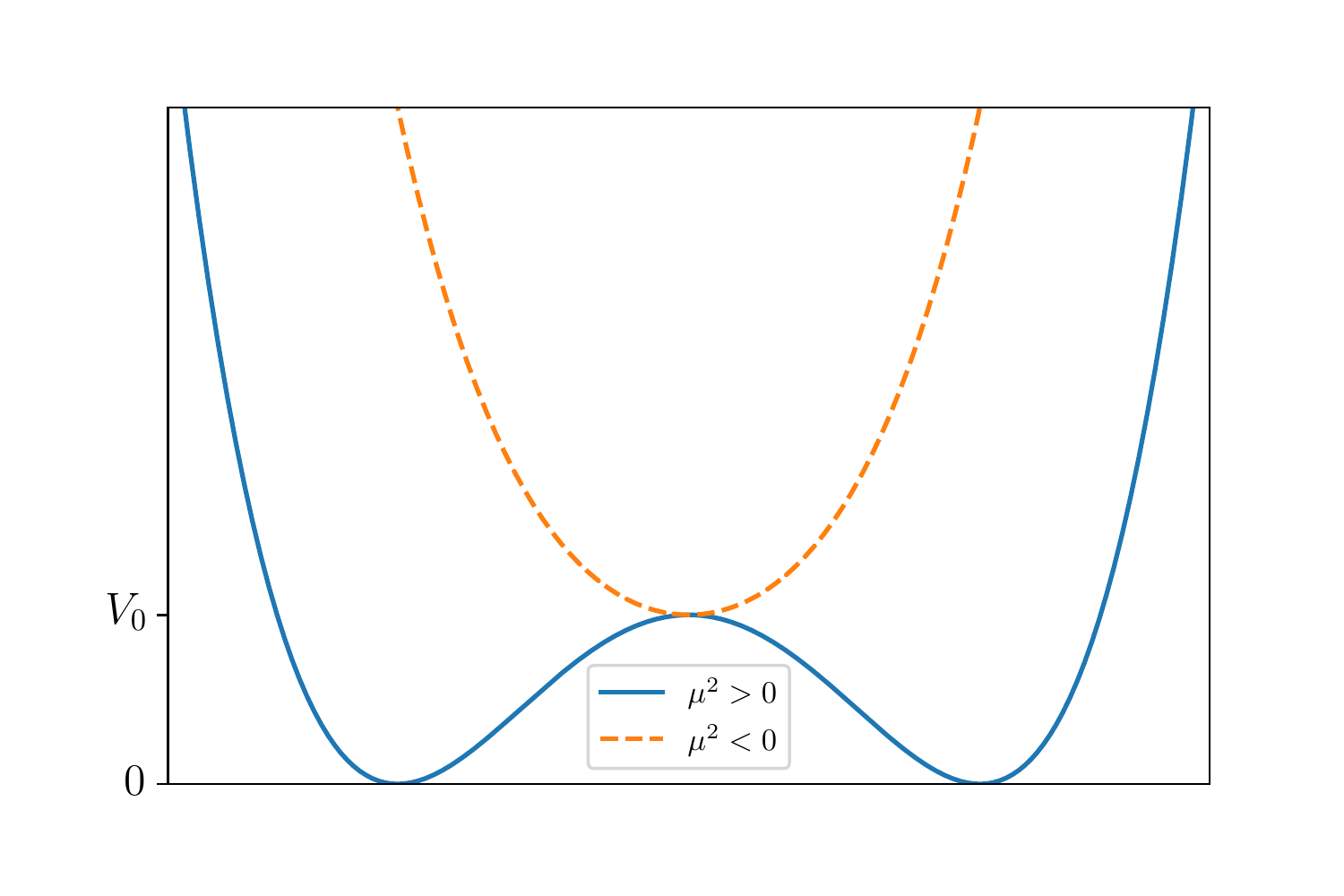}
\caption[Electroweak phase transition]{Plot showing the Higgs potential~\eqref{eq:HiggsPot} for positive and negative $\mu^2$. If we insist on the fact that the minimum should be zero in the symmetry-breaking state, then the value $V_0$ of the cosmological constant before the phase transition should be extremely fine-tuned }
\label{fig:electroweak}
\end{figure}

Even if we disregard this calculation by assuming that it will be solved by a candidate for a quantum theory of gravity,  there remains similar issues related to the electroweak phase transition. This is because the Higgs potential takes the well-known 'mexican hat' form,
\begin{equation} \label{eq:HiggsPot}
V(\varphi) = V_0 - \mu^2 \varphi^\dagger \varphi + g (\varphi^\dagger \varphi)^2 \; ,
\end{equation}
where $\varphi$ is the Higgs doublet. For temperatures high enough, like it was the case in the primordial universe, one has $\mu^2 < 0$; however as the temperature is lowered a phase transition happens and $\mu^2$ becomes negative. This is illustrated in Figure~\ref{fig:electroweak}. In the primordial universe, the Higgs was still in its symmetry-preserving state so that the minimum of $V$ was $V_0$ for $\varphi=0$. We can associate this value of the Higgs potential to its contribution to the cosmological constant of the early universe. However, when the universe began to cool down the minimum of the potential shifted to 
 $V_0 - \mu^4/4g$. This contributes to our present day cosmological constant so that we should set this to (nearly) zero (the value of $\mu^4/4g$ is overwhelmingly larger than the present day cosmological constant). But this is an incredible fine-tuning for the value of $V_0$ : why should the cosmological constant of the early universe be equal to $\mu^4/4g$ with an accuracy of 50 digits ? Notice that it is a fine-tuning issue between two physical and measurable quantities, not between a 'bare' and an 'observable' cosmological constant; moreover, if $g \ll 1$ quantum loops would correct this result by a small amount only so that it is a purely classical phenomenon !

Given the dramatic failure of QFT to predict a consistent cosmological constant value, it would be simpler to assume that some mechanism sets its value to $\Lambda = 0$ exactly. But the supernova observations give an small but nonzero value to $\Lambda$, which is then very difficult to explain !  Let us finish this subsection by drawing an historical parallel: at the beginning of the $20^\mathrm{th}$ century, the ultraviolet catastrophe (predicting that a blackbody would radiate at arbitrarily high energies) was solved by Planck by the introduction of quanta which then gave rise to quantum mechanics; the solution to the cosmological constant problem could maybe be a decisive step towards a quantum theory of gravity.

\subsection{The $H_0$ tension} \label{subsec:H0}

Perhaps one of the most pressing issues in modern cosmology is the tension between different measurements of the Hubble rate $H_0 = (\dot a / a)_0$. We have seen in Eq.~\eqref{eq:Planck+BAO} that the Planck mission together with BAO data gives $H_0 \simeq 67 \text{km} . \text{s}^{-1} .  \text{Mpc}^{-1}$. Note that this value has been obtained by assuming a $\Lambda$CDM cosmology and would be quite sensitive to modifications of the model at large redshift since it depends on the whole path of photons from the last scattering surface until the detector.

\begin{figure}[ht]
\centering
\includegraphics[width=0.5\textwidth]{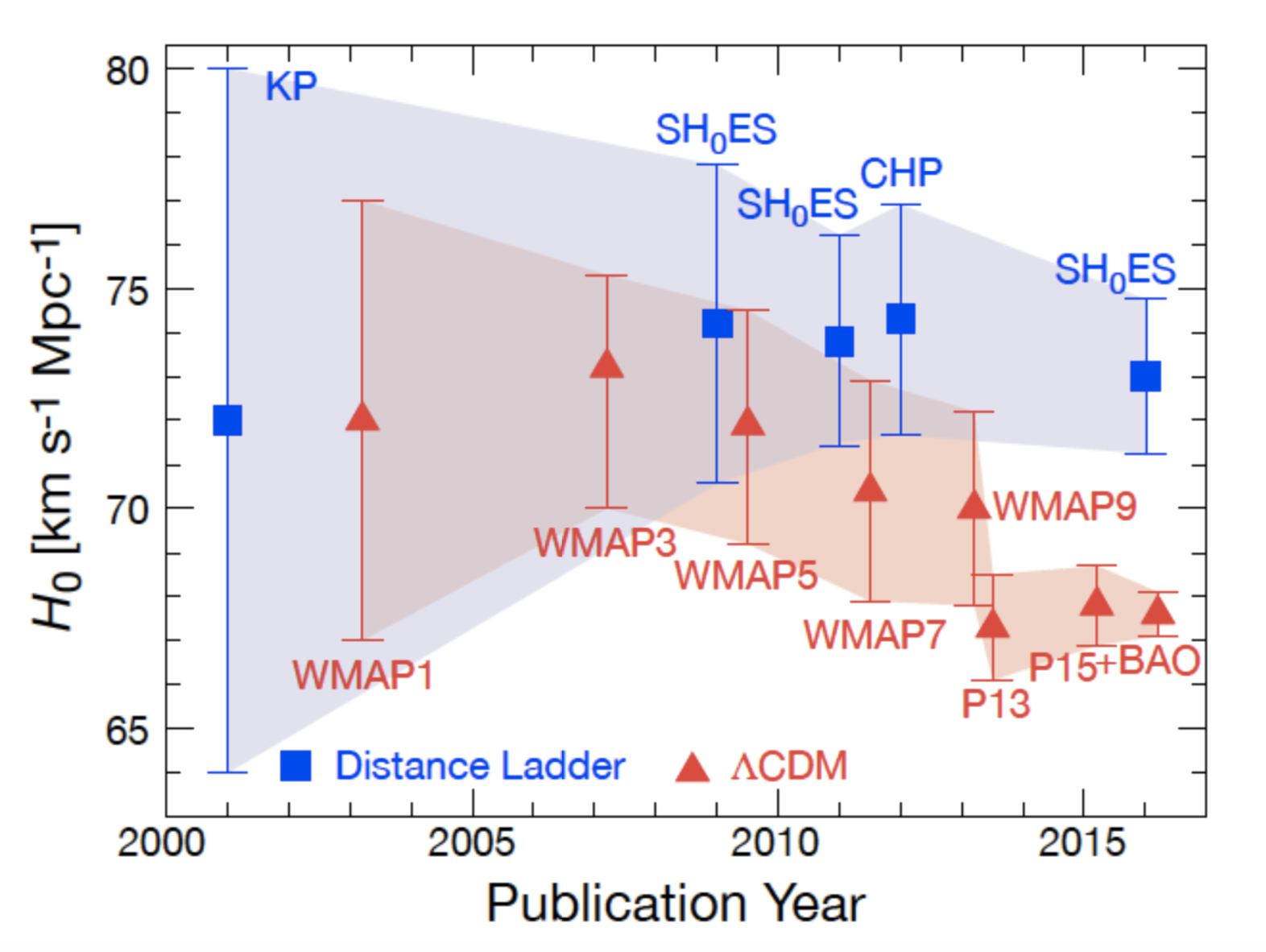}
\caption[$H_0$ tension]{The emergence of the $H_0$ tension over the years. The triangles represent WMAP and Planck points (the distant universe), while the squares are the supernovae measurements (the local universe). Figure taken from Freedman \cite{Freedman:2017yms}   }
\label{fig:H0Tension}
\end{figure}

On the other hand, measurement of $H_0$ in the close universe with supernovae (by the same team who discovered the accelerated expansion)  give a much different value for $H_0$, closer to $ 74 \text{km} . \text{s}^{-1} .  \text{Mpc}^{-1}$. The tension between the two measurements is now more than the $4 \sigma$ level, as illustrated in Figure~\ref{fig:H0Tension}.

There are multiple avenues to try to solve this tension. We can approximately divide them in two groups: astrophysical effects which would bias the supernovae distance ladder, and modifications of gravity at large redshift which would influence the CMB data. Let us briefly comment on these two aspects.

\begin{figure}[ht]
\centering
\includegraphics[width=0.8\textwidth]{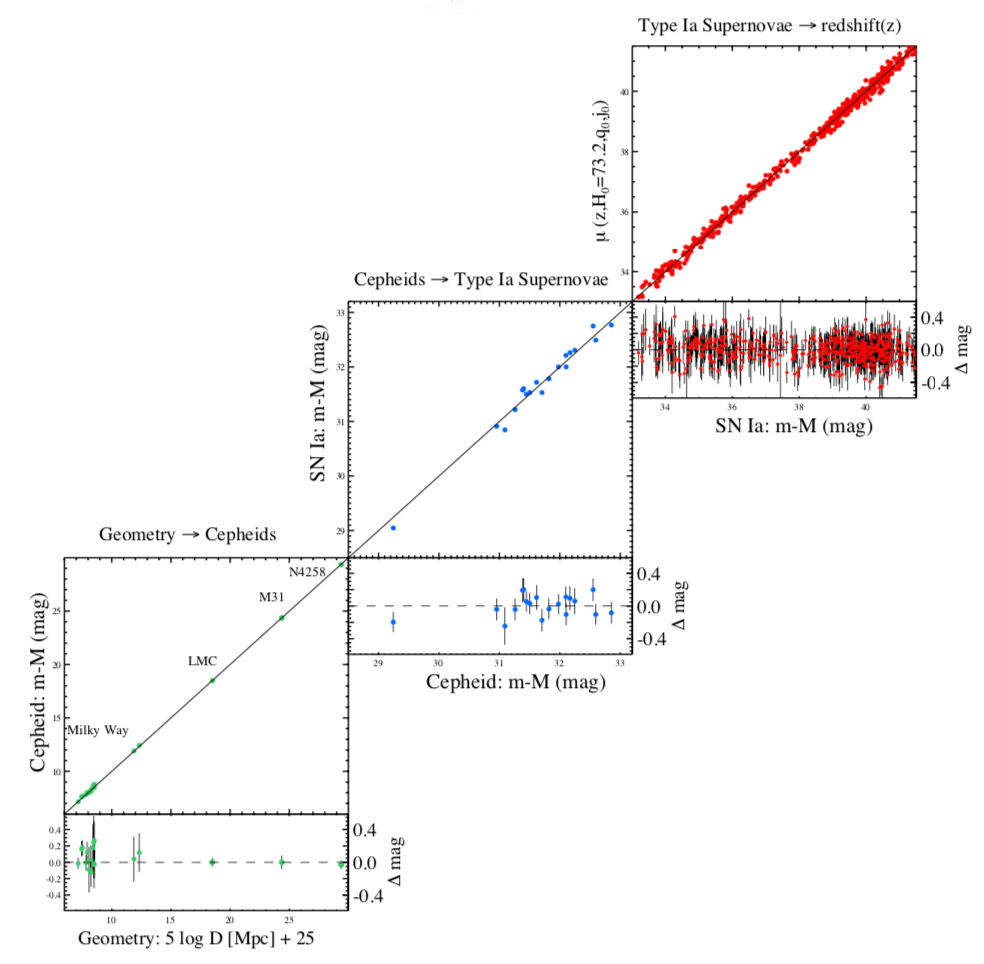}
\caption[Calibration of the supernova distance ladder]{The calibration of the distance ladder for the supernovae measurement of $H_0$. The lower graph represents the calibration of the period-luminosity relation of cepheids through the parallax of close stars; the middle graph is the calibration of supernovae through the cepheids; and the last step is the measurement of $H_0$ from supernovae.  Figure taken from Riess et al. \cite{Riess:2016jrr}   }
\label{fig:RiessLadder}
\end{figure}

\paragraph{Astrophysical effects:} We often refer to supernovae as 'standard candles'; the proper term should be 'standardizable candles'. Despite their crucial cosmological importance, type Ia supernovae are still very much a mystery. A SN Ia is a thermonuclear explosion that completely destroys a carbon/oxygen white
dwarf as it approaches the Chandrasekhar limit of $1.4 M_\odot$.
This is the reason SNe Ia can be
calibrated to be good standard candles. The first challenge to overcome when using SNe Ia
as cosmological standard candles is properly incorporating the intrinsic scatter in SN Ia
peak luminosity. Indeed, this luminosity can depend on the metallicity of the environment and this has to be accounted for. However, there could be other parameters influencing the peak luminosity of SN Ia. For example, Rigault at al. claim that SN Ia could be fainter in locally start-forming environment, which could alleviate the $H_0$ tension \cite{Rigault:2014kaa}. Another source of concern is the use of a distance ladder to estimate $H_0$ from the SN Ia: one first calibrates the period-luminosity relation of variable stars known as cepheids with the parallax of close stars, and then calibrates the SN Ia using cepheids. As illustrated in Figure~\ref{fig:RiessLadder}, there are only five cepheids calibrated with the parallax, which could lead to bias due to an insufficient number of data points; furthermore, the possibility exists that the distant Type Ia supernovae have different properties than nearby Type Ia supernovae. The debate on this issue is not yet settled.

\paragraph{Modifications of gravity: } The other main path to alleviate the $H_0$ tension is to consider that the $\Lambda$CDM model does not describe correctly the universe, so that the Planck+BAO measurement of $H_0$ is biased. The way Planck measures $H_0$ is through an angular scale measurement, mainly the sound horizon at last scattering which is obtained from the peak spacing of the CMB power spectrum. A way to increase $H_0$ while keeping this angular scale fixed is either to modify the content of the universe after recombination ('late-universe solution'), or to modify physics before and at recombination (see \cite{2019arXiv190803663K} for a discussion on this). The last possibility could include an early dark energy component through a supplementary scalar field, see \cite{2019arXiv190806995S}.
\begin{figure}[ht]
\centering
\includegraphics[width=0.5\textwidth]{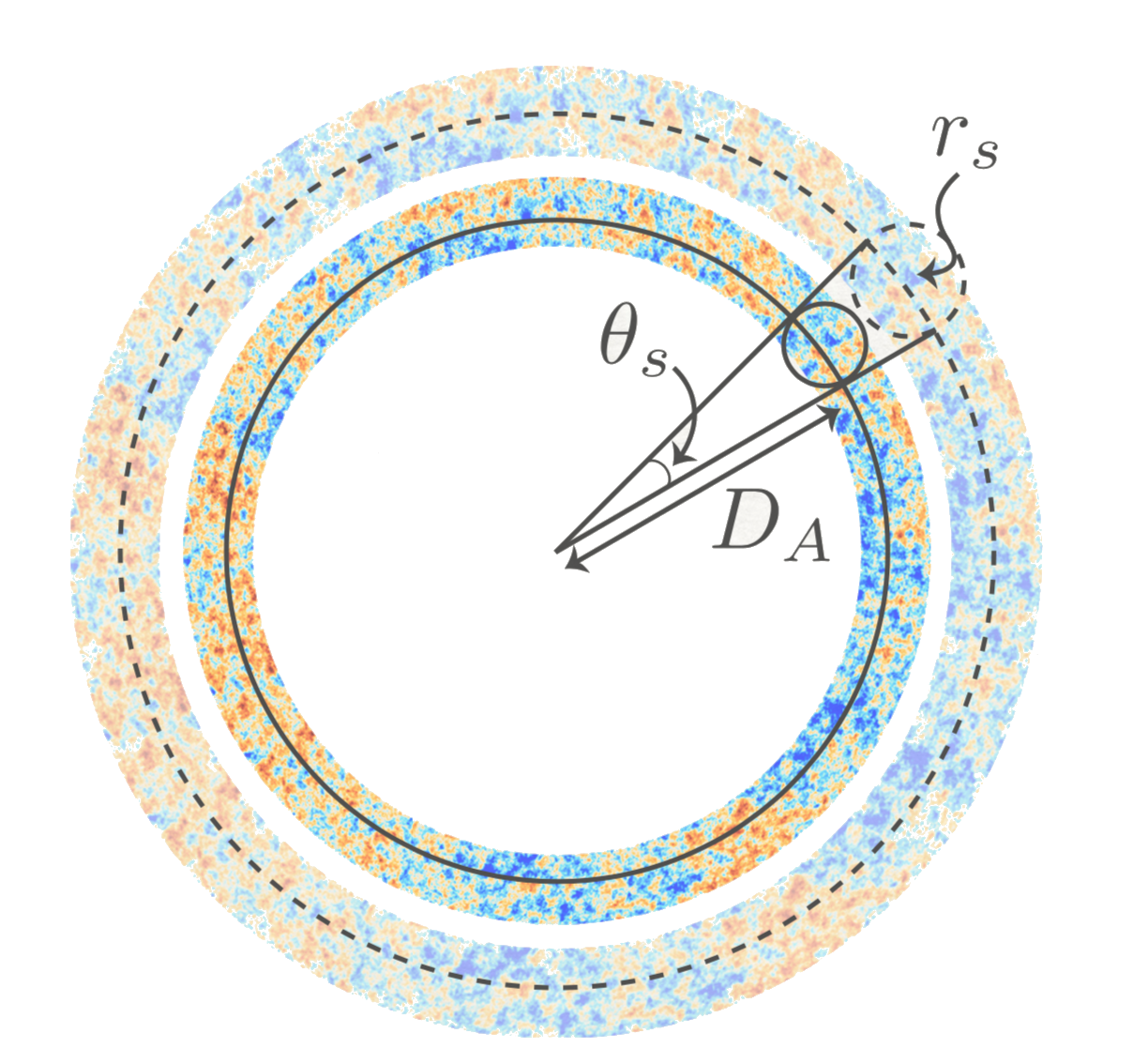}
\caption[Iplication of the $H_0$ tension on the CMB]{Schematic representation of the impact of a higher $H_0$ on CMB physics: a fixed angle $\theta_s$ with higher $H_0$ could imply that the CMB is closer to us than expected and that the size of the sound horizon at recombination $r_s$ is smaller. Figure taken from \href{https://indico.in2p3.fr/event/18900/contributions/75009/attachments/56066/74168/ColloqueDarkEnergy.pdf}{ Vivian Poulin}  }
\label{fig:H0FromPlanck}
\end{figure}

\subsection{Dark matter}

In 1933, Fritz Zwicky inferred the existence of invisible mass by applying the virial theorem to infer the distribution of mass from the velocities of galaxies inside the Coma cluster. He thus coined the term 'dark matter' to refer to this invisible mass which acts to hold the galaxies inside the cluster together. In the sixties, Vera Rubin and Kent Ford observed the same phenomenon when looking at rotation curves of galaxies themselves: the virial theorem predicted that there should be a lot of invisible mass in the outskirts of galaxies. More recently, the CMB observations were also found to be consistent with a universe mainly made of dark matter. Moreover, if dark matter was not present then ordinary matter would not be able to sufficiently clump into dense objects so that we would not be able to explain the large scale structures that we observe today in galaxy surveys. It seems that these four distinct sets of observations point towards the existence of dark matter, which is why dark matter is at the core of the current cosmological model. There are also other approaches, like MOND and TeVeS \cite{Bekenstein:2004ne}, which aim at explaining the rotations curves of galaxies by a modification of Newton's law. However, they can not explain simultaneously the rotation curves and the cosmological observations.

If dark matter exists, what is then its nature ? Several possibilities exist, let us mention a few of them:
\begin{itemize}
\item Weakly Interacting Massive Particle (WIMP): This is one of the favorite dark matter model. These particles interact only through the weak nuclear force and gravity (so that they would be invisible to electromagnetic observations), and have a large mass compared to standard particles (so that they would be slow moving thus they would clump together). The so-called 'WIMP miracle' is that to obtain the correct abundance of dark matter today via thermal production requires a self-annihilation cross section compatible with a mass of $100$GeV, which is the mass range in which we would expect new particles beyond the standard model.
\item Axions: The axion was a particle introduced to solve the strong CP problem. It is expected to be a light particle (with a mass between the $\mu$eV and the meV), it would have no electric charge, and its interaction cross-section for strong and weak forces would be very low thus making a perfect candidate for dark matter.
\item Massive astrophysical compact halo object (MACHO): these are objects like (primordial) black holes or neutron stars as well as brown dwarfs and unassociated planets. Since they are not luminous they would be difficult to detect by other channels than the gravitational one. However, different observations put robust constraints on the abundance of such candidates.
\end{itemize}
The interested reader is encouraged to refer to the abundant literature for more details, in particular the experimental searches for DM candidates.

\subsection{Other problems}

On top of the three main issues discussed above, there are other unsolved problems in the current cosmological model. Let us mention two of them:

\paragraph{The lithium problem: } Big Bang Nucleosynthesis (BBN) correctly predicts the observed abundances of hydrogen and helium in the universe. The predictions are based on a system of coupled Boltzmann equations which govern the temporal evolution of each species. However, the actual amount of observed lithium is 3-4 times less than the one predicted by BBN \cite{Hou_2017}. This seems to indicate that some non-standard physics happened during BBN.

\paragraph{Galaxy formation and the small-scale power spectrum: } In the current cosmological model, galaxy formation proceeds by the accumulation of dark matter in dense halos originating from the primordial adiabatic fluctuations. Gravitational instability leads to the formation of the first dwarf galaxies and the small halos hierarchically merge together to form bigger structures. However, it was realized that supernova and supermassive black holes actively participate in the formation of galaxies \cite{Dekek:1986gu, Silk:1997xw}.  These effects are very small-scale physics and are quite difficult to take into account at cosmological scales, both for theory and numerical simulations. One could e.g. adopt an EFT viewpoint to 'integrate out' this high-energy physics as has been advocated in the EFT of large-scale structures \cite{Carrasco_2012}. This allows to push theoretical predictions up to a 'midly non-linear' scale of $\sim 1$MPc (the linear perturbation theory breaks down at $\sim 50$MPc); however a space mission as Euclid \cite{Amendola_2018} will be able to have access to scales up to $0.1$MPc, where no reliable theoretical prediction is available ! 

\section{A wealth of theoretical ideas} \label{sec:theory}

The tensions and contradictions in the current cosmological model, and in particular the cosmological constant problem, have motivated a growing number of physicists to explore theories in which gravity is different than in GR. One can try to solve one or several of the above mysteries by modifying gravity; more generally, exploring well-behaved theories generalizing GR can provide new insights on the way to test gravity. For example, as already mentioned in Chapter~\ref{Chapter1}, scalar-tensor theories generically predict violations of the strong equivalence principle; consequently, one can test their existence with Lunar Laser Ranging. In this Section we will introduce some of the most well-known modified gravity theories appeared in the last decades. Since in this thesis we will be mostly interested in the small-scale (i.e, solar system size) behavior of modified gravity theories, we will particularly emphasize this aspect rather than their cosmological implications. 

We can broadly divide the modifications of gravity into three classes: the addition of a new scalar, vector or tensor degree of freedom. We will explore separately each of these three options, although in a general setting a mixture of them is perfectly conceivable. However, in the remaining of this thesis we will mainly concentrate on (single) scalar-tensor (ST) theories. Indeed, they are one of the simplest alternative to GR; moreover, scalars can be used in widely different contexts (to model inflation and dark energy as we will discuss, but also dark matter \cite{Berezhiani_2015}; to describe phase transition such as in the Laundau mean-field model; they are also used in condensed matter to describe excitations of solids (phonons) \cite{leutwyler1996phonons}, fluids \cite{Endlich_2011} or superfluids \cite{son2002lowenergy}; last but not least, we have now observed the first fundamental scalar particle with the Higgs boson \cite{Chatrchyan_2012, Aad_2012}). Consequently, in most of this thesis scalars will serve as a 'toolbox' to explore well-motivated and theoretically sound modifications to GR.

\subsection{Brans-Dicke type theories} \label{subsec:BDT}
Brans-Dicke type (BDT) theories are the generalization of the Brans-Dicke theory presented in Section~\ref{sec:BD}, where the parameter $\omega(\phi)$ can itself depend on the scalar. In the literature, they are often called scalar-tensor theories. However, if one defines a ST theory as a theory containing a single scalar degree of freedom on top of the usual graviton, then the ST class can be much broader than the one presented in this Subsection (the full (beyond) Horndeski class, see~\ref{subsec:Horndeski}, belongs to ST theories). Therefore, to avoid confusion we will refer in this thesis to the class of theories introduced in this Subsection by the name Brans-Dicke type theories. The reader should be aware that this denomination is not a standard one.

BDT theories are defined in the Jordan frame by
\begin{equation}
S = \frac{\mpl^2}{2} \int \mathrm{d}^4x \sqrt{-g} \left[ \phi R - \frac{\omega(\phi)}{\phi} g^{\mu \nu} \partial_\mu \phi \partial_\nu \phi - V(\phi) \right] + S_m \; ,
\end{equation}
where $g$ is the determinant of the metric $g_{\mu \nu}$, $R$ is the Ricci scalar and $S_m$ is the matter action, assumed to be minimally coupled to the metric $g_{\mu \nu}$. Alternatively, with a field redefinition $\phi \rightarrow \varphi$ similar to the one in Section~\ref{sec:BD} they can be brought to the following, more convenient form
\begin{equation} \label{eq:ActionBDT}
S = \int \mathrm{d}^4 x \sqrt{-g} \bigg[\frac{\mpl^2}{2} R - \frac{1}{2} g^{\mu \nu} \partial_\mu \varphi \partial_\nu \varphi - \tilde V(\varphi)  \bigg] + \tilde S_m \; ,
\end{equation}
In this frame, matter is coupled to a \textit{Jordan frame} metric $\tilde g_{\mu \nu}$ which can differ from the \textit{Einstein frame} one $g_{\mu \nu}$ by a factor depending on the scalar,
\begin{equation} \label{eq:JordanFrameMetric}
\tilde{g}_{\mu \nu} = A^2(\varphi) g_{\mu \nu} \; ,
\end{equation}
where $A$ is the conformal coupling of the scalar. Since a scalar naturally leads to an accelerated expansion as in inflation, these theories were the first candidates for an alternative to the cosmological constant (in cosmology, they go under the name of Quintessence): an interesting and generic prediction of Quintessence is that the equation of state of dark energy, $w = p / \rho$, can be different from the cosmological constant value $w = -1$, so that one of the main objectives of future cosmological surveys like Euclid is to accurately measure $w$.

However, solar system tests presented in Chapter~\ref{Chapter1} usually put strong constraints on the coupling of such a scalar to matter \textit{via} the conformal coupling $A^2(\varphi)$. For example, the first derivative of $A$ with respect to the scalar is related to the PPN parameter $\gamma$ (we will have more to say concerning this point in Chapter~\ref{Chapter4}, eq.~\eqref{eq:ppn_params}) which is tightly constrained by the Cassini bound~\eqref{eq:constraint_gamma}. More precisely, if one defines the scalar coupling $\beta$ by
\begin{equation}
\beta = \mpl \frac{\partial}{\partial \varphi} \ln A \; ,
\end{equation}
then the Cassini bound~\eqref{eq:constraint_gamma} imposes $\beta \lesssim 10^{-2}$.
 From an effective field theory viewpoint, such a small value of the scalar coupling seems quite unnatural. However, it has been argued~\cite{Damour:1994zq} that cosmological evolution naturally drives $\varphi$ towards a minimum of $A$ so that the scalar coupling to matter (nearly) vanishes. Another way to have a naturally small coupling to matter is to invoke the presence of a screening mechanism such as the ones we will present later in this Section.

Finally, there exists a popular theory in cosmology closely related to BDT theories, namely $f(R)$ theories. These are defined by allowing the action to be an arbitrary function of the Ricci scalar $R$,
\begin{equation} \label{eq:FRtheory}
S = \frac{\mpl^2}{2} \int \mathrm{d}^4 x \sqrt{-g} f(R)  + S_m \; .
\end{equation}
At first sight, these theories look quite different than the BDT theories we introduced above. However, they secretely propagate an additional scalar degree of freedom. To see this, let us consider the gravitational action of a BDT theory in some generic frame,
\begin{equation} \label{eq:scalarFR}
S = \frac{\mpl^2}{2} \int \mathrm{d}^4 x \sqrt{-g} \bigg[f'(\varphi)(R-\varphi) + f(\varphi) \bigg] \; ,
\end{equation}
and integrate out the scalar field $\varphi$. This means that we should replace it using its equation of motion,
\begin{equation}
f''(\varphi) (R-\varphi) = 0
\end{equation}
Assuming $f''(\varphi) \neq 0$, we see that the scalar action~\eqref{eq:scalarFR} is perfectly equivalent to an $f(R)$ theory. This scalar action is a particular BDT theory in the Jordan frame, with no kinetic term. However, when going to the Einstein frame the field acquires a kinetic term so that the theory is well-behaved. The final translation of an $f(R)$ theory in terms of a BDT theory in the Einstein frame is
\begin{equation}
V(\varphi) = \frac{\mpl^2}{2} \frac{\varphi f'-f}{f'^2} \; , \quad A(\varphi) = e^{\varphi/\sqrt{6} \mpl} \; ,
\end{equation}
where $V$ is the potential appearing in Eq.~\eqref{eq:ActionBDT} and $A$ is the conformal coupling in the Jordan frame metric~\eqref{eq:JordanFrameMetric}. In particular, the conformal coupling $\beta = 1/\sqrt{6}$ severely violates solar system constraints, so that $f(R)$ theories would be ruled out without a particular screening mechanism known as the Chameleon mechanism~\cite{Khoury:2003aq}. Let us describe briefly in what it consists.

\begin{figure}[ht]
\centering
\includegraphics[width=0.5\textwidth]{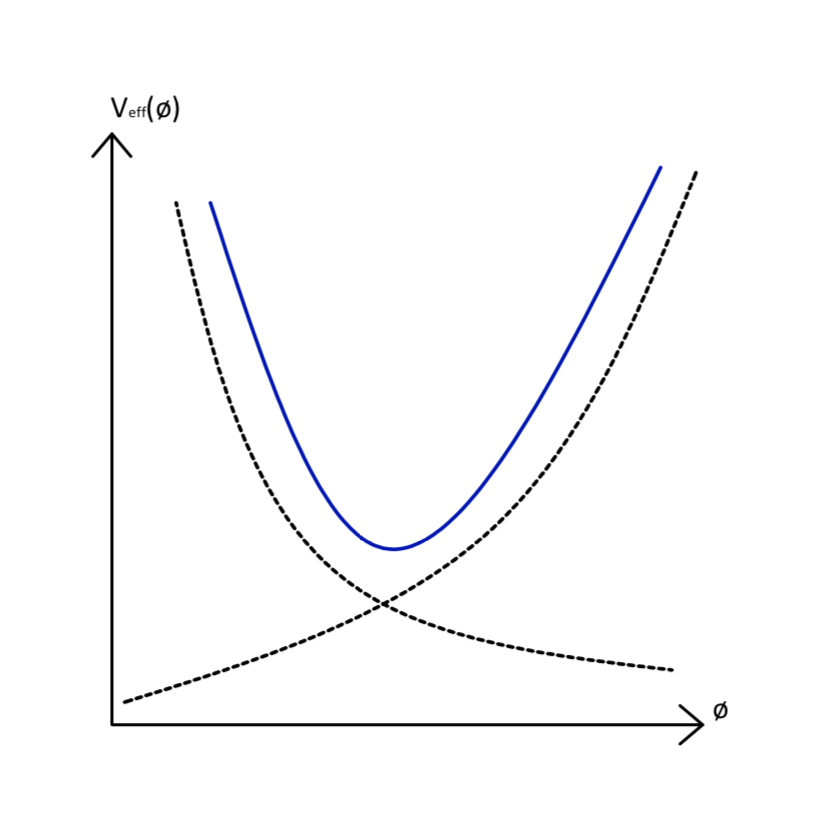}
\caption[Chameleon mechanism]{The effective potential in the chameleon screening mechanism. The monotonically decreasing curve is the potential $V(\varphi)$, while the increasing one is the contribution from matter in Eq.~\eqref{eq:KGChameleon}. The effective potential is the sum of the two curves and exhibits a minimum.  }
\label{fig:Chameleon}
\end{figure}

We consider the class of BDT theories in which $A(\varphi) = e^{\beta \varphi / \mpl}$ where $\beta$ is a constant. Assuming that matter is made of a perfect fluid of nonrelativistic matter so that its energy-momentum tensor is $T^{\mu \nu} = \mathrm{diag}(\rho,0,0,0)$, the equation of motion for the scalar reads
\begin{equation} \label{eq:KGChameleon}
\square \varphi = V'(\varphi) + \frac{\beta}{\mpl} \rho e^{\beta \varphi / \mpl} \; .
\end{equation}
This is equivalent to the dynamics of a scalar field in an effective potential $V_\mathrm{eff}(\varphi) = V(\varphi) + \rho e^{\beta \varphi / \mpl}$. Usually, in cosmological models we are interested by a monotonically decreasing $V(\varphi)$ so that the fields rolls down the potential and produces a behavior similar to inflation. Thus, $V(\varphi)$ does not have a minimum for finite values of $\varphi$. But the \textit{effective} potential can be quite different, as illustrated in Figure~\ref{fig:Chameleon}: when the matter density $\rho$ is large enough, the rhs in Eq.~\eqref{eq:KGChameleon} can dominate the scalar potential so that there exists a minimum at the point where $V(\varphi)$ is counter-balanced by the matter-dependent term. If the field settles to this minimum $\varphi_0$, then it oscillates with a mass
\begin{equation}
m_\varphi^2 = V''(\varphi_0) + \frac{\beta^2}{\mpl^2}  \rho^2 e^{2 \beta \varphi_0 / \mpl} \; .
\end{equation}
Since a massive field gives rise to a Yukawa attractive potential between massive objects $\propto e^{-m_\varphi r}/r$, one can see that for a large enough mass $m_\varphi$ the deviations from Newton's law becomes exponentially suppressed and GR is recovered in the solar system.

\subsection{Bekenstein's disformal coupling} \label{subsec:intro_disf}

Bekenstein showed that Eq.~\eqref{eq:JordanFrameMetric} is not the most general coupling that one can consider in ST theories. Indeed, the Jordan frame metric can be augmented to~\cite{Bekenstein:1992pj}
\begin{equation} \label{eq:JordanFrameMetric_disf}
\tilde{g}_{\mu \nu} = A^2(\varphi, X) g_{\mu \nu} + B(\varphi, X) \partial_\mu \varphi \partial_\nu \varphi \; ,
\end{equation}
where $X = \partial_\mu \varphi \partial^\mu \varphi$. The part proportional to the fields gradients is named the \textit{disformal} transformation. Since matter is coupled to a metric, such a conformal/disformal transformation respects causality and the weak equivalence principle~\cite{Bekenstein:1992pj}. The geometry is 'Finslerian', i.e it is a geometry for matter couplings  in which the squared line element is homogeneous of second
degree in the coordinate variations. By contrast with a conformal transformation which merely stretches all spacetime directions equally,
a disformal transformation also adds a translation along the directions in which the scalar is changing. 

One could legitimately ask if such disformal term spoils the causal structure of the original metric. Indeed, the line element gets modified by such a transformation as
\begin{equation}
\mathrm{d}s^2 = g_{\mu \nu} \mathrm{d}x^\mu \mathrm{d}x^\nu \rightarrow \mathrm{d}\tilde s^2 = A^2(\varphi, X) \mathrm{d}s^2 + B(\varphi, X) \big( \partial_\mu \varphi \mathrm{d}x^\mu \big)^2 \; .
\end{equation}
Thus, $B$ stretches the light-cone. In order to ensure causal behavior for massive particles we require that $B < 0$ everywhere so that $\mathrm{d}\tilde s^2 < 0$. This means that a null vector for $g_{\mu \nu}$ is now a timelike vector for $\tilde g_{\mu \nu}$.

Other features of the original metric that we want to preserve is that it should have a Lorentzian signature and it should be invertible with nonsingular volume element. This can be shown to be the case under the condition $A^2 + BX > 0$~\cite{Bekenstein:1992pj}.

Disformal couplings have been used to predict anomalous light bending~\cite{Bekenstein:1993fs}, in models of inflation~\cite{Creminelli_2014, Kaloper_2004} or dark energy~\cite{de_Bruck_2015, Zumalac_rregui_2010, koivisto2008disformal}. They are naturally embedded inside Horndeski theories (see Section~\ref{subsec:Horndeski}). Different experimental constraints apply on the disformal parameter, see e.g~\cite{Sakstein_2014} for a nice review. In Chapter~\ref{Chapter5}, we will compute the effect of a disformal coupling on the dynamics of a binary system of neutron stars.

Before moving on, let us emphasize a point which is often not put forward in the modified gravity literature: a conformal or disformal \textit{coupling} refers to a coupling of matter to a Jordan frame metric as we explained above. This nonminimal coupling can have observable effects on the bending of light or the speed of photons compared to gravitational waves (see Section~\ref{sec:disformal_speed}). On the other hand, a conformal/disformal \textit{transformation} refers to a change of variables: in this case Eq.~\eqref{eq:JordanFrameMetric_disf} is interpreted as defining a new metric $\tilde g _{\mu \nu}$. Expressing the action and the equations of motion in terms of $g _{\mu \nu}$ (i.e, in the Einstein frame) or $\tilde g _{\mu \nu}$ (i.e, in the Jordan frame) should therefore give the same physical results since we are just dealing with a field redefinition.

\subsection{Galileons} \label{subsec:Galileons}


Galileons in flat space were first introduced in~\cite{Nicolis:2008in}. They are defined as the most generic scalar theories which enjoys 'Galilean' invariance, i.e their Lagrangian is built out of a scalar $\pi$ which is invariant up to a total derivative term under the transformation
\begin{equation}
\pi(x) \rightarrow \pi(x) + b_\mu x^\mu + c \; ,
\end{equation}
where $b_\mu$ and $c$ are constants. Futhermore, it is asked that their equations of motion is still of second order in derivatives. Higher derivatives in the equations of motion usually lead to what is known as an 'Ortrogradski ghost', i.e a supplementary degree of freedom which renders the Hamiltonian unbounded from below so that the theory is sick~\cite{mancarella2017effective}. In four dimensions, there are only five types of Lagrangians which satisfy these two properties, namely
\begin{align} \label{eq:GalileonLagrangians}
L_1 &= \pi \; , \\
L_2 &= -\frac{1}{2} (\partial \pi)^2 \; , \\
L_3 &=  (\partial \pi)^2 \big[ \Pi \big] \; , \\
L_4 &=  (\partial \pi)^2 \big( \big[ \Pi \big]^2 - \big[ \Pi^2 \big] \big) \; , \\
L_5 &=  (\partial \pi)^2 \big( \big[ \Pi \big]^3 - 3 \big[ \Pi \big]  \big[ \Pi^2 \big] + \big[ \Pi^3 \big] \big) \; ,
\end{align}
where $\Pi$ is a shorthand notation for $\partial_\mu \partial_\nu \pi$ and the bracket means to take the trace.

Since the discovery  of the Galileon interactions there has been a flurry of works  related to Galileon cosmology~\cite{Chow:2009fm,DeFelice:2010nf,deRham:2011by,deRham:2010tw}, inflation~\cite{Creminelli:2010ba,LevasseurPerreault:2011mw,Kobayashi:2010cm,Burrage:2010cu}, laboratory tests~\cite{Brax:2011sv}, BHs~\cite{Babichev:2016fbg,Babichev:2015rva}, lensing~\cite{Wyman:2011mp}, superluminal propagation around compact sources~\cite{Goon:2010xh,deFromont:2013iwa}.
There are two main properties that make Galileons attractive from a field theory point of view. First, they enjoy a non-renormalization theorem stating that quantum loops will not renormalize any operator of the form~\eqref{eq:GalileonLagrangians}~\cite{Luty_2003, Hinterbichler:2010xn, Goon_2016}. Second, they give rise to the Vainshtein mechanism which screens the field around massive sources~\cite{Vainshtein:1972sx,Babichev:2013usa}. 
The Vainshtein mechanism was first proposed as a possible solution to the vDVZ discontinuity of massive gravity~\cite{1970NuPhB..22..397V, 1970JETPL..12..312Z} in which GR is not recovered as the limit $m \rightarrow 0$ is taken in Fierz-Pauli massive gravity. The idea is that non-linearities in the scalar action become dominant close to a massive source so that the field does not have the usual Newtonian behavior. Part~\ref{part3} will be dedicated to the study of this mechanism in different regimes, however we will give here a short preview of its essence.

To illustrate the idea, let us take a cubic Galileon coupled to a massive source,
\begin{equation} \label{eq:cubicGal}
S = \int \mathrm{d}^4 x \bigg[ - \frac{1}{2} (\partial \varphi)^2 - \frac{1}{2 \Lambda^3} (\partial \varphi)^2 \square \varphi \bigg] + \frac{\varphi T}{\mpl} \; ,
\end{equation}
where $\Lambda$ is the strong coupling energy scale of the cubic Galileon and $T$ is the trace of the energy-momentum tensor. Assuming that $T = -m \delta^3(\mathbf{x}) = -m \delta(r)/4\pi r^2$ represents a single static point-particle, one can use spherical symmetry and derive the following equation of motion for the scalar profile $\varphi_0$,
\begin{equation}
\frac{1}{r^2} \partial_r \left[ r^3 \left( \frac{\varphi_0'}{r} + \frac{1}{2\Lambda^3} \left( \frac{\varphi_0'}{r} \right)^2 \right) \right] = \frac{m}{4 \pi \mpl r^2} \delta(r) \; ,
\end{equation}
so that integrating both sides of this equation gives
\begin{equation} \label{eq:Vain_cubicGal}
\varphi_0' + \frac{1}{2\Lambda^3} \frac{\varphi_0'\hphantom{}^2}{r} = \frac{m}{4 \pi \mpl r^2} \; .
\end{equation}
We can define the Vainshtein or strong coupling radius $r_*$ as
\begin{equation}
r_* = \left(\frac{m}{8 \pi \mpl \Lambda^3} \right)^{1/3} \; ,
\end{equation}
so that a large distances compared to that Vainshtein radius the linear term in~\eqref{eq:Vain_cubicGal} dominates while the interactions dominate at distances shorter than $r_*$,
\begin{align}
\begin{split}
\varphi_0'(r) &\simeq \frac{m}{4 \pi \mpl r^2} \quad \text{for } r \gg r_* \; , \\
\varphi_0'(r) &\simeq \frac{m}{8 \pi \mpl r_*^{3/2} r^{1/2}} \quad \text{for } r \ll r_* \; ,
\end{split}
\end{align}
So at large distances one recovers Newton's law for the scalar (which adds up to the usual gravitational force, so that the modification of gravity can be quite important), but on distances smaller than the Vainshtein radius the force is quite suppressed,
\begin{equation}
\frac{F_\varphi}{F_\mathrm{Newt}} \sim \left( \frac{r}{r_*} \right)^{3/2} \ll 1 \quad \text{for } r \ll r_* \; .
\end{equation}
On the other hand, in order for the cubic Galileon to give rise to a part of the accelerated expansion, one should relate the strong coupling scale $\Lambda$ to the Hubble parameter. This can be seen from the Galileon operator~\eqref{eq:cubicGal}: setting it equal to a cosmological constant contribution $\mpl^2 H^2$ for values $\varphi_0/\mpl \sim 1$ imposes $\Lambda^3 \sim H^2 \mpl$. The associated Vainshtein radius is huge: a few parsecs for a solar mass object!

 This means that the Galileon force on the earth is suppressed by 12 orders of magnitude compared to the gravitational force, so that it would be nearly invisible. However, as we have seen in Chapter~\ref{Chapter1} tests of gravity in the solar system are extremely accurate: the precise constraints on a Galileon field from solar system measurements will be the subject of Chapter~\ref{Chapter7}. Notably, we will see how one can go beyond the simple spherically symmetric situation of the one-body case.
 
 Finally, we should mention that the existence of a Vainshtein mechanism around massive source also implies that small scalar fluctuations can propagate superluminally. Whether or not this is a true pathology of the theory is still actively discussed~\cite{deRham:2014zqa}. We will illustrate the relation between screening and superluminality in the simpler case of a K-Mouflage theory in the next Section.

\subsection{K-essence, ghost condensate, DBI} \label{subsec:KEssence}

K-essence fields are a generalization of quintessence (i.e, BDT theories) in which the potential can depend on the kinetic term $X = g^{\mu \nu} \partial_\mu \varphi \partial_\nu \varphi$,
\begin{equation}
S = \int \mathrm{d}^4x \sqrt{-g} P(\varphi, X) \; .
\end{equation}
As the Galileons, this type of theory does not possess an Ostrogradsky ghost. An example of such a function $P$ is given by the DBI Lagrangian used in inflation~\cite{Alishahiha:2004eh},
\begin{equation}
P(\varphi, X) = - \frac{1}{f(\varphi)} \sqrt{1+f(\varphi) g^{\mu \nu} \partial_\mu \varphi \partial_\nu \varphi} \; ,
\end{equation}
and which describes the motion of a five-dimensional brane: $\varphi$ is the position of the brane along the fifth dimension and the Lagrangian is thus proportional to its Lorentz factor.

These models are used to describe inflation or dark energy. They also model superfluids~\cite{son2002lowenergy} and as such are used in models where dark matter is made of a superfluid~\cite{Berezhiani_2015}. They appear in a phenomenon similar to the Higgs mechanism known as the ghost condensate~\cite{Hamed_2004}. They possess several properties that make them distinct from quintessence. First, they can feature a dark energy equation of state (EoS) $w = p/\rho < -1$, contrary to quintessence. This behavior is usually referred to as phantom dark energy; measuring precisely the dark energy EoS is one of the most straightforward way to discriminate among dark energy models. Second, they possess a screening mechanism very similar to the Vainshtein screening, dubbed K-Mouflage~\cite{Babichev:2009ee}. Third, they have a speed of sound different from unity, i.e small scalar perturbations propagate with a speed different from that of light.

Let us elaborate on the last two points. For simplicity we focus on the case where the function $P(X)$ only depends on the kinetic term. Adding a coupling to the trace of the energy momentum-tensor as in the Galileon case~\eqref{eq:cubicGal}, the equation of motion for the scalar around an isolated point-particle of mass $m$ is
\begin{equation} \label{eq:EOM_KMouflage}
\varphi_0' P'(\varphi_0'\hphantom{}^2) = - \frac{m}{8 \pi \mpl r^2} \; .
\end{equation}
Let us assume that, on top of the usual kinetic term, the function $P$ is a simple power-law:
\begin{equation} \label{eq:simple_PX}
P(X) = - \frac{X}{2} + c \Lambda^{4-4\alpha} X^\alpha \; ,
\end{equation}
where $c$ and $\alpha$ are constants, and $\Lambda$ is an energy scale. The equation of motion~\eqref{eq:EOM_KMouflage} is equivalent to
\begin{equation}
\varphi_0' \left(- \frac{1}{2} + c \alpha \Lambda^{4-4\alpha} \varphi_0'\hphantom{}^{2\alpha-2}  \right) =  - \frac{m}{8 \pi \mpl r^2}
\end{equation}
For large $r$ one should recover a Newtonian potential for $\varphi$ so that it imposes $\alpha > 1$. On the other hand,
 if the power-law term dominates the kinetic term in the small-scale regime (as in the Vainshtein mechanism) one has:
\begin{equation}
\varphi_0'^{2 \alpha - 1} \sim \Lambda^{4\alpha-4} \frac{m}{\mpl r^2} \; .
\end{equation}
 The previous equation can also be rewritten as
\begin{equation}
\frac{\varphi_0'}{\varphi'_N} \sim \left( \frac{r}{r_*} \right)^{\frac{4\alpha-4}{2\alpha-1}}
\end{equation}
where we have introduced the Newtonian potential $\varphi'_N \sim m/(M_P r^2)$ and the Vainshtein radius $r_*^2 = m/(M_P \Lambda^2)$. Then for $\alpha > 1$ the scalar field is negligible compared to Newtonian gravity.

We will now show that for $\alpha > 1$ the radial sound speed is greater than the speed of light. Let us consider a small fluctuation of the scalar of the form
\begin{align}
\begin{split}
\varphi &= \varphi_0 + \delta \varphi \; , \\
X &= \varphi_0'^2 + 2 \varphi_0' \partial_r \delta \varphi + (\partial \delta \varphi)^2 \; .
\end{split}
\end{align}

It is then easy to find the quadratic action for the perturbations:
\begin{equation}
S_\mathrm{quad} = \int \mathrm{d}^4x \left[P'(\varphi_0'^2) (\partial \delta \varphi)^2 + 2 \varphi_0'^2 P''(\varphi_0'^2) (\partial_r \delta \varphi)^2 \right] \; ,
\end{equation}
from which we read the angular and radial velocities
\begin{align}
\begin{split} \label{eq:cr_px}
c_\Omega^2 &= 1 \\
c_r^2 &= 1 + 2 \varphi_0'^2 \frac{P''(\varphi_0'^2)}{P'(\varphi_0'^2)}
\end{split}
\end{align}
When the nonlinear term in~\eqref{eq:simple_PX} dominates, the radial sound speed is $c_r^2 = 1 + 2(\alpha-1) > 1$ for $\alpha > 1$. More generically, this superluminal propagation has been shown to be present for any function $P(X)$ if we simultaneously demand that the field is screened on small scales~\cite{Barreira_2015}. One could thus be led to think that these theories are pathological, however it has been argued that in spite of the superluminal propagation the causal paradoxes do not arise in these theories and in this respect they are not less safe than General Relativity~\cite{Babichev:2007dw}.

\subsection{Horndeski and beyond} \label{subsec:Horndeski}

Horndeski theories is a wide class of scalar-tensor theories which encompasses all the models mentioned above. The Horndeski Lagrangian is defined as the most generic scalar-tensor theory with both second-order dynamics for both the metric and the scalar~\cite{Horndeski:1974wa} (note that the fact that the Einstein-Hilbert action is the only theory involving second-order dynamics for the metric was proved by Weyl and Cartan~\cite{damour1996gravitation}). The key idea
is that one can admit higher derivatives in the Lagrangian, provided that its variation
gives only second order EOM both for the scalar field and for the metric. The most
general Lagrangian satisfying the above property amounts to the four terms
\begin{align}\label{eq:L_horndeski}
\begin{split}
L_2^\mathrm{H} &= G_2(\varphi, X) \; , \quad L_3^\mathrm{H} = G_3(\varphi, X) \square \varphi \; , \\
L_4^\mathrm{H} &= G_4(\varphi, X) R - 2 G_{4,X}(\varphi, X) \big(\square \varphi^2 - \varphi^{\mu \nu} \varphi_{\mu \nu} \big) \; , \\
L_5^\mathrm{H} &= G_5(\varphi, X) G_{\mu \nu} \varphi^{\mu \nu} + \frac{1}{3} G_{5,X}(\varphi, X) \big(\square \varphi^3 - 3 \square \varphi \varphi_{\mu \nu} \varphi^{\mu \nu} + 2 \varphi_{\mu \nu} \varphi^{\mu \sigma} \varphi^\nu_\sigma \big) \; ,
\end{split}
\end{align}
where the $G_i$'s are arbitrary functions, $R$ is the Ricci tensor and $G_{\mu \nu}$ is the Einstein tensor, and we have introduced the notation
\begin{equation}
\varphi_{\mu \nu} = \nabla_\mu \nabla_\nu \varphi \; , \quad \square \varphi = g^{\mu \nu} \varphi_{\mu \nu} \; .
\end{equation}
We thus see that k-essence theories are described by the function $G_2$, while Galileons are recovered by taking $G_3=X$. The structure of Galileons and Horndeski Lagrangians is very similar; Horndeski theories can be seen as a curved-space generalization of the Galileons~\cite{Deffayet:2009mn}.

A generalization of Horndeski theories which goes under the name of Gleyzes-Langlois-Piazza-Vernizzi (GLPV) or beyond Horndeski theories~\cite{Zumalacarregui:2013pma,Gleyzes:2014dya,Gleyzes:2014qga,Langlois:2015cwa,BenAchour:2016fzp} can be obtained by considering that the coupling to matter is a mixture of conformal and disformal couplings, see Eq.~\eqref{eq:JordanFrameMetric_disf}. When going in the Jordan frame where matter is minimally coupled to the metric, one obtains two supplementary Lagrangians in the action,
\begin{align}\label{eq:L_beyond_horndeski}
L_4^\mathrm{bH} &= F_4(\varphi, X) \epsilon^{\mu \nu \rho}_{\hphantom{\mu \nu \rho}\sigma} \epsilon^{\mu' \nu' \rho' \sigma} \varphi_\mu \varphi_{\mu'} \varphi_{\nu \nu'} \varphi_{\rho \rho'} \\
L_5^\mathrm{bH} &= F_5(\varphi, X) \epsilon^{\mu \nu \rho \sigma} \epsilon^{\mu' \nu' \rho' \sigma'} \varphi_\mu \varphi_{\mu'} \varphi_{\nu \nu'} \varphi_{\rho \rho'} \varphi_{\sigma \sigma'}
\end{align}
Although these terms produce equations of motion higher than second order, it can be shown that they propagate a single scalar degree of freedom and are free of the Ostrogradsky ghost. This is because of a special degeneracy condition on the Lagrangian, which has been further generalized in what is known as the Degenerate Higher-Order Scalar-Tensor (DHOST) theories~\cite{mancarella2017effective, langlois_degenerate_2017}.

(Beyond) Horndeski theories naturally encompass both the Vainshtein and the K-Mouflage mechanism. Their cosmological relevance will be discussed in Section~\ref{sec:EFTDE}. One of the most dramatic constraint on these theories was obtained by the recent measurement of the speed of gravitational waves compared to that of light: essentially, this removed most of the complexity and freedom of the above Lagrangians in a cosmological setup. This will be discussed in Chapter~\ref{Chapter3}.

\subsection{Gauss-Bonnet and Chern-Simons} \label{subsec:GB_CS}

Gauss-Bonnet (GB)~\cite{Sotiriou_2014} and Chern-Simons (CS)~\cite{Jackiw_2003} gravity are examples of scalar-tensor theories in which the scalar couples to a total divergence (so that if the scalar is constant, one recovers GR). Their respective actions are given by
\begin{align}
S_\mathrm{GB} &= \int \mathrm{d}^4 x \sqrt{-g} \bigg[ \frac{\mpl^2}{2} R - \frac{1}{2} g^{\mu \nu} \partial_\mu \varphi \partial_\nu \varphi + f(\varphi) \mathcal{G}  \bigg] \; , \\
S_\mathrm{CS} &= \int \mathrm{d}^4 x \sqrt{-g} \bigg[ \frac{\mpl^2}{2} R - \frac{1}{2} g^{\mu \nu} \partial_\mu \varphi \partial_\nu \varphi + \frac{\gamma}{4} \varphi \;^*R R  \bigg] \; ,
\end{align}
where $f$ is an arbitrary function, $\alpha$ is a constant and $\mathcal{G}$ and $\;^*R R$ are respectively the Gauss-Bonnet and Pontryagin topological invariants,
\begin{align}
\mathcal{G} &= R^{\mu \nu \rho \sigma} R_{\mu \nu \rho \sigma} - 4 R^{\mu \nu} R_{\mu \nu} + R^2 \; , \\
\;^*R R &= \frac{1}{2} R_{\mu \nu \rho \sigma} \epsilon^{\mu \nu \alpha \beta} R^{\rho \sigma}_{\alpha \beta} \; .
\end{align}
The CS coupling naturally arise in compactifications of string theory; one of its interesting features is that it has a characteristic observational signature, which could
allow one to discriminate an effect of this theory from other phenomena. Indeed, the CS term violates parity, and thus
it mainly affects the axial-parity component of the gravitational field. On the other hand, the GB coupling give rise to \textit{hairy} black holes, i.e black holes in which the scalar solution is nontrivial~\cite{Antoniou_2018, Sotiriou_2014} (we will have much more to say on this subject in Part~\ref{part4}). Furthermore, these hairy solutions may even be dynamically preferred over GR~\cite{Silva:2017uqg}.

These theories fit into the beyond Horndeski class, although it is not evident from the form of their action. Indeed, it can be shown that the Gauss-Bonnet coupling is equivalent to the following Horndeski functions~\cite{Kobayashi:2011nu}:
\begin{align} \label{eq:translation_GB_Horndeski}
\begin{split}
G_5 &= - 4 f' \ln X \; , \\
G_4 &= 4 f'' X (2 - \ln X) \; , \\
G_3 &= 4 f^{(3)} X (7 - 3 \ln X) \; , \\
G_2 &= 8 f^{(4)} X^2 (3 - \ln X) \; .
\end{split}
\end{align}

One may be tempted to think that, because the GB coupling is related to the Horndeski $G_4$ and $G_5$ functions, it should be ruled out by the GW observations which we will describe in Section~\ref{sec:tests_GW}. However, this is only the case if we assume that it should be cosmologically relevant. To be definite, let us consider a linear GB coupling $f(\varphi) = \alpha \varphi$ which translates in a pure $G_5$ function from Eq.~\eqref{eq:translation_GB_Horndeski}. On a cosmological background, it induces a shift in the GW speed~\cite{Tattersall_2018}
\begin{equation}
\Delta c_\mathrm{GW} = 4 \frac{(H \dot \varphi - \ddot \varphi) \alpha}{\mpl^2 - 4 H \dot \varphi \alpha} \; .
\end{equation}
If the scalar is assumed to be cosmologically relevant, then the $G_5$ function must participate in the Friedmann equations leading to a scaling of $\alpha$~\cite{Bellini_2014}
\begin{equation}
\alpha \sim \frac{\mpl}{H^2} \; .
\end{equation}
and this indeed induces a shift $\Delta c_\mathrm{GW} \sim 1$. However, interesting physics also comes when one considers the GB coupling to be of the order of the size of a solar-mass BH, so that the scalar hair induces sizable deviations from GR around a BH. In this case, the coupling has to satisfy~\cite{Sotiriou_2014}
\begin{equation}
\alpha \sim \frac{M_\odot^2}{\mpl^3} \; .
\end{equation}
In this case it is easily checked that the deviation of the speed of GW from unity is even more negligible on a cosmological background than the restrictive experimental bound.

\subsection{Generalized Proca}

We now turn towards theories modifying GR by the addition of a new spin-1, i.e vector, degree of freedom in a way similar to Horndeski for the scalar case. The simplest example of this is a Proca field, i.e a massive vector field with action
\begin{equation}
S = \int \mathrm{d}^4 x \sqrt{-g} \bigg[ - \frac{1}{4} F_{\mu \nu} F^{\mu \nu} + m^2 A^2 \bigg] \; ,
\end{equation}
where $F_{\mu \nu} = \partial_\mu A_\nu - \partial_\nu A_\mu$ is the usual electromagnetic field strength tensor. The mass term breaks the $U(1)$ gauge invariance of the theory $A_\mu \rightarrow A_\mu + \partial_\mu \varphi$, but it is well-known that one can restore it using the Stuckelberg trick, i.e by splitting the vector field $A_\mu = A_\mu^\mathrm{T} + \partial_\mu \pi$ into a transverse one $A_\mu^\mathrm{T}$ and a scalar which transforms as $\pi \rightarrow \pi - \varphi$ under a gauge transformation. Such a Lagrangian is generalizable in the way identical to the construction of Horndeski Lagrangians, so that the most general theory giving rise to second order equations of motion for the vector field (and without any dynamics of the temporal component of the vector field) are~\cite{Heisenberg:2017mzp}
\begin{align}
\begin{split}
L_2 &= f_2(X,F,Y) \; , \\
L_3 &= f_3(X) \partial_\mu A^\mu \; , \\
L_4 &= f_4(X) \big[ (\partial_\mu A^\mu)^2 - \partial_\mu A_\nu \partial^\nu A^\mu \big] \; , \\
L_5 &= f_5(X) \bigg[ (\partial_\mu A^\mu)^3 - 3 \partial_\mu A^\mu \partial_\rho A_\sigma \partial^\sigma A^\rho + 2 \partial_\mu A_\nu \partial^\rho A^\mu \partial^\nu A_\rho   \bigg] + f_5(A^2) \tilde F^{\alpha \mu} \tilde{F}^\beta_\mu \partial_\alpha A_\beta \; , \\
L_6 &= f_6(X) \tilde F^{\alpha \beta} \tilde F^{\mu \nu} \partial_\alpha A_\mu \partial_\beta A_\nu \; ,
\end{split}
\end{align}
where $\tilde F$ is the dual of the field strength tensor, and $X = A^2$, $F = - F_{\mu \nu}F^{\mu \nu}/4$ and $Y=A^\mu A^\nu F_\mu^\alpha F_{\nu \alpha}$.

One particular case is Einstein-Aether theory, in which the vector field is forced to take a unit timelike norm through a constraint $\lambda(A^2 + 1)$ in the action. Then, the theory leads to a \textit{spontaneous} breaking of Lorentz invariance in that the vector field is forced to select a particular spacetime direction as vacuum expectation value. This can be tested through PPN parameters related to preferred-frame effects, see Chapter~\ref{Chapter1} and Ref.~\cite{Will_2014}.

\subsection{Massive gravity}

Massive gravity is a vast and growing subject, so we will only give here a very short and partial account of what it consists on. Massive gravity theories attempt to give the putative “graviton” a mass. The simplest attempt to implement this in a ghost-free manner, the Fierz-Pauli Lagrangian~\cite{Fierz:1939ix}, suffers from the so-called van Dam–Veltman–Zakharov (vDVZ) discontinuity~\cite{1970NuPhB..22..397V, 1970JETPL..12..312Z}. Because of the additional helicity states of a massive spin-2 graviton, the limit of small graviton mass does not coincide with pure GR, and the predicted perihelion advance (for example) violates experiment. The resolution behind that puzzle was provided by Vainshtein two years later~\cite{Vainshtein:1972sx}: he found that the extra degree of freedom responsible for the vDVZ discontinuity gets screened by its own interactions which dominate over the linear terms in the massless limit. We illustrated this mechanism when studying the Galileon in Section \ref{subsec:Galileons}.

A second element of concern when dealing with a theory of massive gravity is the realization that most non-linear extensions of Fierz–Pauli massive gravity are plagued with a ghost, known as the Boulware–Deser (BD) ghost~\cite{PhysRevD.6.3368}. The past decade has seen a revival of interest in massive gravity with the realization that this BD ghost could be avoided in a specific ghost-free realization of massive gravity known as dRGT~\cite{deRham:2010kj}.
The reader is encouraged to consult the excellent review~\cite{deRham:2014zqa} for an exhaustive presentation of massive gravity.

\subsection{Nonlocal gravity}

Finally, let us mention a modification of gravity sensibly different than the other approaches since it does not consist in adding a new degree of freedom to GR. Rather, the idea is to consider the \textit{quantum effective action} of gravity obtained by integrating out the quantum fluctuations of fields, and which yields the equations of motion for the vacuum expectation values of all fields. In such an action, nonlocal terms will unavoidably be generated even if the fundamental action is local. In particular, one such term gives rise to an interesting phenomenology in a cosmological setup, namely the 'RR' term~\cite{Belgacem_2018}:
\begin{equation}
\Gamma_\mathrm{RR} = - \int \mathrm{d}^4 x \sqrt{-g} R \frac{\Lambda_\mathrm{RR}^4}{\square^2} R \; ,
\end{equation}
where the nonlocality is contained in the inverse of the d'Alembertian operator $\square$. The RR term reproduces a (modified) FLRW evolution with well-behaved perturbations. It fits the cosmological data as well as the $\Lambda$CDM model and could even reduce the $H_0$ tension. We refer the reader to Ref.~\cite{Belgacem_2018} for a complete review of this model.

\section{One theory to rule them all: the Effective Field Theory of Dark Energy} \label{sec:EFTDE}

The short summary of the last Section should have given to the reader an idea of the high number of theoretical ideas when it comes to modifying gravity. In the cosmological community, there was an urgent need of a formalism generic enough to encompass several modified gravity ideas, and which would at the same time provide an effective way to compare these theories to observations, in a similar spirit to the post-Newtonian formalism. 
 This is the purpose of the Effective Field Theory of Dark Energy (EFT of DE)~\cite{Creminelli:2008wc,Gubitosi:2012hu,Bloomfield:2012ff,Piazza:2013coa} which we will describe in this Section. The underlying idea was already used in inflation~\cite{Creminelli:2006xe,Cheung:2007st}, and is applicable to all models containing a supplementary scalar degree of freedom on top of the GR spin-2 graviton (and thus to the Horndeski class and its generalizations). The approach is 'bottom-up' and theory agnostic in the sense that one parametrizes the modification of gravity in terms of observable parameters without referring to any specific theory. On the other hand, one should not be too libertarian on the parametrization as this could hide some pathology in the theory. The EFT of DE manages the difficult task to provide a proxy between theories and observations, in the sense that each parameter can be related to specific theories while at the same time being ready to use in cosmological observations.

The essential idea of this effective field theory is the following. In the alternatives to $\Lambda$CDM that we are considering, the accelerated expansion is caused by a single scalar which takes a time-dependent vacuum expectation value $\bar \varphi(t)$. In the very same way as the Higgs mechanism of the standard model of particle physics, this breaks a fundamental symmetry which is time reparametrization invariance. This analogy with gauge theories suggests that there will be Goldstone bosons, i.e massless excitations describing the low-energy dynamics. These modes are fluctuations of the scalar degree of freedom, $  \varphi(t, \mathbf{x}) = \bar \varphi(t) + \delta \varphi(t, \mathbf{x})$. In the \textit{unitary gauge}, one sets $\delta \varphi = 0$; this can be achieved by a time diffeomorphism $t \rightarrow t - \delta \varphi / \dot \varphi$. Thus, the scalar degree of freedom has been 'eaten' by the metric; one can build an action principle by considering the most general action built out of the metric and compatible with the residual diffeomorphism, i.e spatial reparametrizations. In this action one can allow operators breaking time diffeomorphism invariance. The coefficients of these operators will be functions of time and can be constrained by observations. The formalism, being constructed to describe fluctuations around a $\Lambda$CDM background, is very convenient to allow comparison with cosmological data; on the other hand, for any covariant theory such as the ones discussed in Section~\ref{sec:theory} it is possible to relate the fundamental parameters to the effective description. We will perform such a \textit{matching} computation in Section~\ref{subsec:matching}.

Before entering into the details of the construction of the EFT of DE, let us mention that the last part of this thesis will be devoted to an analogue construction adapted to gravitational wave observations. There, the observable is the signal of inspiralling black holes or neutron stars which is usually tedious to derive in theories generalizing GR; a unifying formalism relating theory and observations would be of great importance to improve our knowledge about gravity.

\subsection{Construction of the EFT of DE} \label{subsec:construction_EFTDE}

\paragraph{Building blocks: }
Let us now show how to write an action based on the above ideas. The fact that the scalar field has a background value $\bar \varphi(t)$ defines
a preferred foliation of spacetime, given by the hypersurfaces of constant $\varphi$. In
a cosmological context, the usual assumption is that the scalar field gradient is
timelike, $(\partial \varphi)^2 < 0$, so these hypersurfaces are spacelike. To adapt to this preferred
foliation, we can choose the background value of the scalar field as a “clock”, such
that constant time hypersurfaces correspond to constant $\varphi$ ones. This choice of
the time coordinate is called the unitary gauge. More precisely, in a general (perturbed) FLRW universe, $\varphi(t, \mathbf{x}) = \bar \varphi(t) + \delta \varphi(t, \mathbf x)$.
By choosing the coordinate $t$ to be a function of $\varphi$, $t=t(\varphi)$, we thus simply have $\delta \varphi = 0$.
Therefore the DE action written in this gauge only displays metric degrees of
freedom. The building blocks of the action should be invariant under the residual space diffs but can break time diffs. One can therefore build various terms\footnote{The following discussion is taken from~\cite{Cheung:2007st}}:
\begin{itemize}
\item Terms which are invariant under all diffeomorphisms. These are just functions of the Riemann tensor $R_{\mu \nu \rho \sigma}$ and its covariant derivatives, contracted to give a scalar
\footnote{The metric and the completely antisymmetric tensor $(-g)^{-1/2} \epsilon^{\mu \nu \rho \sigma}$ can be used in order to contract indices.}
\item A generic function of $\varphi$ becomes $f(t)$ in the unitary gauge. We are therefore free to use arbitrary functions of time in front of any term in the Lagrangian.
\item The gradient $\partial_\mu \varphi$ becomes a kronecker $\delta_\mu^0$ in unitary gauge. Thus in every tensor we can always have a free upper $0$ index. For example we can use $g^{00}$ (and functions of it) or the component of the Ricci tensor $R^{00}$ in the unitary gauge Lagrangian.
\item It is convenient to define a unit vector perpendicular to surfaces of constant $\varphi$
\begin{equation}
n_\mu = \frac{\partial_\mu \varphi}{\sqrt{-g^{\mu \nu} \partial_\mu \varphi \partial_\nu \varphi}} \; .
\end{equation}
This allows to define the induced spatial metric on surfaces of constant $\varphi$: $h_{\mu \nu} = g_{\mu \nu} + n_\mu n_\nu$.
Every tensor can be projected on these surfaces using $h_{\mu \nu}$ . In particular we can use in our action the Riemann tensor of the induced three-dimensional metric $\hphantom{}^{(3)} R_{\alpha \beta \gamma \delta}$ and covariant derivatives with respect to the 3D metric.

\item New terms also come from the covariant derivatives of $\partial_\mu \varphi$. Notice that we can as well look at covariant derivatives of $n_\mu$: the derivative acting on the normalization factor just gives terms like $\partial_\mu g^{00}$ which are themselves covariant and can be used in the unitary gauge Lagrangian. The covariant derivative of $n_\mu$ projected on the surfaces of constant $\varphi$ defines the extrinsic curvature of these surfaces
\begin{equation}
K_{\mu \nu} = h_\mu^\sigma \nabla_\sigma n_\nu \; .
\end{equation}
The index $\nu$ is already projected on the surface because $n^\nu \nabla_\sigma n_\nu=\nabla_\sigma(n^\nu n_\nu)/2$. The covariant derivative of $n_\nu$ perpendicular to the surface can be expressed as
\begin{equation}
n^\sigma \nabla_\sigma n_\nu = - \frac{1}{2} (-g^{00})^{-1} h_\nu^\mu \partial_\mu\big(-g^{00}\big) \; ,
\end{equation}
so that it does not give rise to new terms beyond the ones we already consider. Therefore all covariant derivatives of $n_\mu$ can be expressed using the extrinsic curvature $K_{\mu \nu}$ (and its covariant derivatives) and derivatives of $g^{00}$.
\item  Notice that using both the Riemann tensor of the induced 3D metric and the
extrinsic curvature is redundant since $\hphantom{}^{(3)} R_{\alpha \beta \gamma \delta}$ can be rewritten with the Gauss-Codazzi relation~\cite{gourgoulhon200731} as
\begin{equation}
\hphantom{}^{(3)} R_{\alpha \beta \gamma \delta} = h_\alpha^\mu h_\beta^\nu h_\gamma^\rho h_\delta^\sigma R_{\mu \nu \rho \sigma} - K_{\alpha \gamma} K_{\beta \delta} + K_{\beta \gamma} K_{\alpha \delta} \; .
\end{equation}
Thus one can forget about the 3D Riemann tensor in the action. We can also avoid referring to the
induced metric $h_{\alpha \beta}$ explicitly, as writing it in terms of the 4D metric and $n_\mu$ one only gets terms already discussed above. Finally, the covariant derivatives with respect to the induced 3D metric can also be avoided: the 3D covariant derivative of a projected tensor can be obtained as the projection of the 4D covariant derivative~\cite{gourgoulhon200731}.

\end{itemize}

We conclude that the most generic action in unitary gauge is given by the arbitrary function
\begin{equation} \label{eq:GenericActionEFT}
S = \int \mathrm{d}^4x \sqrt{-g} L(g_{\mu \nu}, \epsilon^{\mu \nu \rho \sigma}, R_{\mu \nu \rho \sigma}, g^{00}, K_{\mu \nu}, \nabla_\mu, t) \; ,
\end{equation}
where all the free indices inside the Lagrangian $L$ must be upper $0$’s. On top of this, one should add a matter action usually modeled by a perfect fluid of nonrelativistic matter. An important point is that we consider matter to be minimally coupled, i.e the metric $g_{\mu \nu}$ is the Jordan frame metric (so that we will allow for arbitrary functions in front of the usual Einstein-Hilbert term).

\paragraph{Link with the ADM decomposition: } The splitting in '3+1' quantities, with time set apart from space, is remindful of the ADM decomposition. We quickly recall here the link between this decomposition and the EFT building blocks in the unitary gauge. In the ADM formalism, the metric is decomposed as
\begin{equation}
\mathrm{d}s^2 = - N^2 \mathrm{d}t^2 + h_{ij} (\mathrm{d}x^i + N^i \mathrm{d}t ) (\mathrm{d}x^j + N^j \mathrm{d}t ) 
\end{equation}
where $N$ is called the \textit{lapse} and $N^i$ is called the \textit{shift}. Physically, the lapse determines the rate in proper time at which
one progresses from one slice of spacetime to the next, while the shift
vector basically quantifies how much the spatial coordinates change between foliations.
In matrix form, one has
\begin{equation}
g_{\mu \nu} = \begin{pmatrix}
-N^2 + N_i N^i & N_i \\ N_j & h_{ij}
\end{pmatrix} \; , \quad g^{\mu \nu} = \begin{pmatrix}
- \frac{1}{N^2} & \frac{N^i}{N^2} \\ \frac{N^j}{N^2} & h^{ij} - \frac{N^i N^j}{N^2}
\end{pmatrix} \; ,
\end{equation}
where the spatial indices are raised with the induced metric $h_{ij}$. In unitary gauge, $n_\mu = \delta_\mu^0 / \sqrt{-g^{00}}$ so that $n_0 = N$ and $n_i=0$. Therefore, $g_{0i} = h_{0i} = N_i$ and $h_{00}=N^i N_i$. Moreover, the spatial components of the extrinsic curvature $K_{ij}$ can be conveniently written as
\begin{equation} \label{eq:KijADM}
K_{ij} = \nabla_i n_j = \frac{1}{2N} \bigg( \dot{h}_{ij} - \nabla_i N_j - \nabla_j N_i \bigg) \; .
\end{equation}

\paragraph{Background and perturbations: } We now split the different building blocks between background values in an FLRW universe (which we denote with an upper index $0$) and perturbations. A nice feature of the FLRW universe is that due to its high degree of symmetry, every tensor evaluated on the background ($K_{\mu \nu}^{(0)}$, $R_{\mu \nu \rho \sigma}^{(0)}$, $(\nabla_\alpha R_{\mu \nu \rho \sigma})^{(0)}$...) can be written just in terms of $g_{\mu \nu}$, $n_{\mu}$ and functions of time \footnote{This will not be true for the black hole case which we will consider in Part~\ref{part4}}. For example,
\begin{align}
K_{\mu \nu}^{(0)} &= H h_{\mu \nu} \\
R_{\mu \nu \rho \sigma}^{(0)} &= 2  \bigg(H^2 + \frac{k}{a^2} \bigg)  h_{\mu [\rho} h_{\sigma ] \nu} - \frac{1}{4} \left[ (H^2 + \dot H) h_{\mu \sigma} n_{\nu} n_{\rho} + \mathrm{perm} \right] \; ,
\end{align}
where we recall that $H$ is the Hubble parameter and $k=-1,0,1$ is the spatial curvature of the universe. This allows to define \textit{covariant} perturbed operators as $\delta K_{\mu \nu} = K_{\mu \nu} - K_{\mu \nu}^{(0)}$, $\delta R_{\mu \nu \rho \sigma} = R_{\mu \nu \rho \sigma} - R_{\mu \nu \rho \sigma}^{(0)}$, and so on. One can then expand the generic action in~\eqref{eq:GenericActionEFT} in powers of perturbations. Ref~\cite{Gleyzes:2013ooa} showed that up to quadratic order in perturbations the action~\eqref{eq:GenericActionEFT} which does not generate derivatives higher than second order in the linear equations of motion for perturbations amounts to
\begin{align} \label{eq:ActionEFTQuadratic}
\begin{split}
S &= \int \mathrm{d}^4x \sqrt{-g} \bigg[ \frac{\mpl^2}{2} f(t) R - \Lambda(t) - c(t) g^{00}\\
 &+ \frac{M_2^4(t)}{2} (\delta g^{00})^2 - \frac{\bar m_3^3(t)}{2} \delta K \delta g^{00} - m_4^2(t) \big( \delta K^2 - \delta K^{\mu \nu} \delta K_{\mu \nu} \big) + \frac{\tilde m_4^2(t)}{2} \delta R \delta g^{00} \bigg] \; ,
 \end{split}
\end{align}
where $K$ is the trace of $K_{\mu \nu}$ and the different functions $m_i$, $M_i$ depend on time (Notice that $m_i^2$
can have either signs, they are written as a square
just to keep track of dimensions). These unknown functions should be fixed by observations, in an EFT spirit. In the first line of this equation we have written all the \textit{tadpole} operators which are fixed by the background cosmology. Indeed, varying the first line of the action with respect to $g_{\mu \nu}$ fixes the two functions $\lambda$ and $c$ in terms of $f$, the Hubble rate and the matter content:
\begin{align}
c &= \mpl^2 f \bigg(-\dot H + \frac{k}{a^2} - \frac{1}{2} \frac{\ddot f}{f} + \frac{H}{2} \frac{\dot f}{f} \bigg) - \frac{1}{2}(\rho_m + p_m) \; , \\
\Lambda &= \mpl^2 f \bigg(\dot H + 3 H^2 + 2 \frac{k}{a^2} + \frac{1}{2}  \frac{\ddot f}{f} + \frac{5H}{2} \frac{\dot f}{f} \bigg)  - \frac{1}{2}(\rho_m - p_m) \; ,
\end{align}
where $\rho_m$ and $p_m$ are the energy density and pressure of matter respectively. The function $f$ is still arbitrary and corresponds to the freedom of the conformal coupling to matter of BDT theories, cf. Section~\ref{subsec:BDT}.

On the other hand, the second line of Eq.~\eqref{eq:ActionEFTQuadratic} contains operators quadratic in the fluctuations and which contribute to modifications of gravity on linear scales.
Ref.~\cite{Gleyzes:2013ooa} also provided the dictionary between the Horndeski class of theories and the action~\eqref{eq:ActionEFTQuadratic}. It turns out that all Horndeski Lagrangians give rise to operators with $m_4^2 = \tilde m_4^2$ so that this action is a bit more general than Horndeski. Furthermore, there can also be operators at the quadratic level which generate higher spatial derivatives (and so correspond to the generalizations of Horndeski theories):
\begin{equation}
- \bar m_4^2(t) \delta K^2 + \frac{\bar m_5(t)}{2} \delta R \delta K + \frac{\bar \lambda(t)}{2} \delta R^2 \; .
\end{equation}
However, one has
to keep in mind that additional spatial derivatives increase the scaling
dimension of an operator and the higher the derivatives an operator
contains, the less it becomes relevant on large linear scales. Similarly, one could consider operators up to cubic order in perturbations such as $(\delta g^{00})^3$, $\delta g^{00} \delta K$ etc; these kind of operators become relevant on smaller scales when one enters in the nonlinear regime of perturbations~\cite{Cusin_2018}. 

Before moving on, let us make a remark about the operators included in the action~\eqref{eq:ActionEFTQuadratic}. One could naively think that operators such as $K$ and $R^{00}$ are also allowed in the tadpole terms. However, since $K = \nabla_\mu n^\mu$, one can get rid of it using
\begin{equation}
\int \mathrm{d}^4 x \sqrt{-g} f(t) K = - \int \mathrm{d}^4 x \sqrt{-g} n^\mu \partial_\mu f = \int \mathrm{d}^4 x \sqrt{-g} \sqrt{-g^{00}} \dot f \; .
\end{equation}
The same is true for $R^{00}$ or other quadratic operators like $R_{\mu \nu} K^{\mu \nu}$ which can be written in terms of the other operators already included in the action.


\subsection{Matching with theories} \label{subsec:matching}

 Let us give a few examples of the translations of theories presented in Section~\ref{sec:theory} in the EFT language:
\begin{itemize}
\item \textit{Quintessence: } Matching the EFT parameters with a quintessence model~\ref{subsec:BDT} is quite straightforward, as in the unitary gauge
\begin{equation}
- \frac{1}{2} (\partial \varphi)^2 - V(\varphi) \rightarrow - \frac{1}{2} \dot{\bar \varphi}^2(t) g^{00} - V(t) \; ,
\end{equation}
so that $c(t) = - \dot{\bar \varphi}^2(t) / 2$ and $\Lambda(t) = V(t)$.
\item \textit{k-essence: } These models are also very simple to translate in the EFT formalism. It is sufficient to expand the function $P(\bar \varphi, \dot{\bar \varphi}^2(t) g^{00})$ appearing in their Lagrangian~\ref{subsec:KEssence} in powers of $\delta g^{00}$ to obtain
\begin{equation}
\Lambda = \dot{\bar \varphi}^2 \left. \frac{\partial P}{\partial X} \right\vert_{X=\dot{\bar \varphi}^2} - P \; , \quad c = \dot{\bar \varphi}^2 \left. \frac{\partial P}{\partial X} \right\vert_{X=\dot{\bar \varphi}^2} \; , \quad M_2^4 = \dot{\bar \varphi}^4 \left. \frac{\partial^2 P}{\partial X^2} \right\vert_{X=\dot{\bar \varphi}^2} \; .
\end{equation}
\item \textit{Galileons: } More complicated structures such as the Galileon are not so straightforward to translate in the EFT language. For example, Ref.~\cite{Gleyzes:2013ooa} showed that a cubic Galileon $(\partial \varphi)^2 \square \varphi$ gives rise to the following operators
\begin{equation}
\Lambda = M_2^4 = \dot{\bar \varphi}^2( \ddot{\bar \varphi} + 3 H \dot{\bar \varphi}) \; , \quad c = \dot{\bar \varphi}^2( -\ddot{\bar \varphi} + 3 H \dot{\bar \varphi}) \; , \quad m_3^3 = 2  \dot{\bar \varphi}^3 \; .
\end{equation}
\end{itemize}

The translation of the Horndeski (and beyond) Lagrangians in the EFT formalism can be found in Refs.~\cite{Gleyzes:2013ooa, mancarella2017effective}.



\subsection{Experimental constraints on the EFT parameters}\label{sec:expe_EFTDE}

\begin{figure}[ht]
\centering
\includegraphics[width=0.9\textwidth]{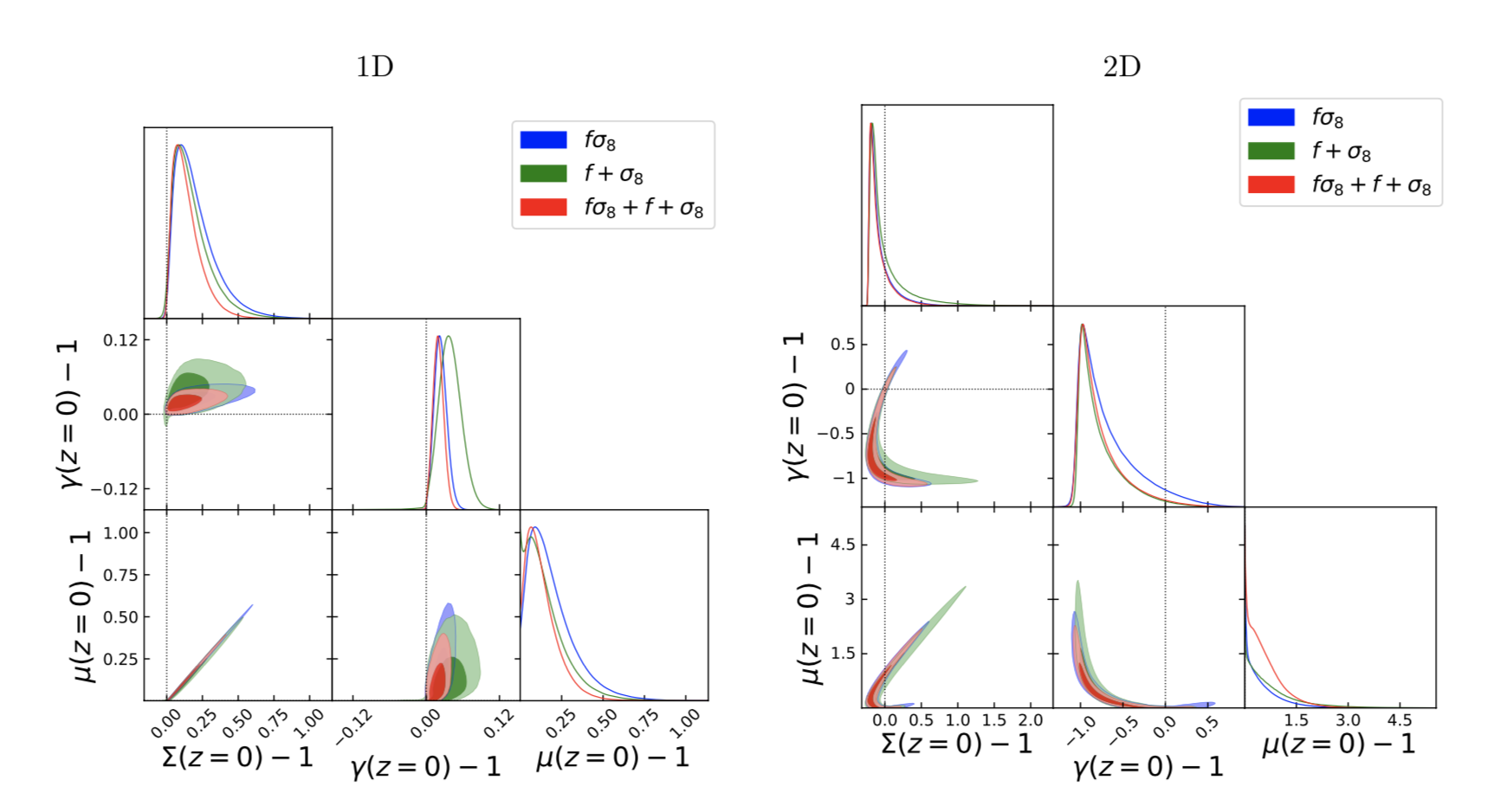}
\caption[Measurement of the EFT coefficients from LSS data]{ Posterior distributions of the observational parameters $\Sigma$, $\gamma$ and $\mu$ for a combination of different LSS data. The '1D' and '2D' graphs refer to different parametrizations of the EFT functions entering in~\eqref{eq:ActionEFTQuadratic}. Figure taken from~\cite{Perenon:2019dpc} }
\label{fig:LSSEFT}
\end{figure}

The most dramatic constraint on the EFT of DE parameters came from the recent observation that the speed of GW is nearly equal to that of light~\cite{Creminelli:2017sry}. However, this will be the subject of the next Chapter. In this Section we will comment on the main cosmological observables allowing to determine the EFT parameters.
 These
observables can be schematically split into two types, the ones linked to the growth of matter perturbations and the ones sensitive to the gravitational potentials. Let us briefly present
them, for which we have adopted the following convention for the perturbed metric in Newtonian
gauge \footnote{The following discussion is mainly based on~\cite{Perenon:2016blf}}:
\begin{equation}
\mathrm{d}s^2 = - (1+2\Phi)\mathrm{d}t^2 + a^2(1-2\Psi) \delta_{ij} dx^i dx^j
\end{equation}
\begin{itemize}
\item \textit{Effective gravitational coupling $\mu$: } In most modified gravity theories it is possible to compile a part
of the modifications of gravity in an observer-friendly quantity, an effective gravitational
coupling $\mu$. It is defined through the Poisson equation,
\begin{equation}
- \frac{k^2}{a^2} \Psi = 4 \pi \mu G \bar \rho \delta \; ,
\end{equation}
where $\bar \rho$ is the mean energy density of matter, and $\delta = (\rho- \bar \rho)/ \bar \rho$ is the density contrast (i.e, the fluctuation in density in the universe). In GR, $\mu=1$. The EFT parameters can be straightforwardly related to $\mu$, see~\cite{P_renon_2015}. Physically, $\mu$ represents the strength of gravity on large scales (on small scales the scalar force can be screened as in Section~\ref{subsec:Galileons}). Large scale-structure data seem to favor a value $\mu>1$, suggesting modifications of gravity on large scales~\cite{Perenon:2019dpc}.
\item \textit{Gravitational slip parameter $\gamma$: } It is defined through the ratio of the two lensing potentials, $\gamma = \Psi/\Phi$, so that $\gamma = 1$ in GR \footnote{ This is exactly the Eddington parameter $\gamma$ which we introduced in Section~\ref{sec:Eddington}. }. However, $\gamma$ itself is of difficult observability; it is easier to constrain the light deflection parameter $\Sigma$ defined by the equation
\begin{equation}
- \frac{k^2}{a^2}(\Phi+ \Psi) = 8 \pi \Sigma G \bar \rho \delta \; .
\end{equation}
Weak lensing measurements which are sensitive to light bending by foreground galaxies can constrain $\Sigma$. The fact that the deflection angle is sensitive to the sum of the gravitational potentials can already be seen from Section~\ref{sec:Eddington}, where it was shown that with the Eddington parametrization of the metric the bending angle was proportional to $1 + \gamma$. However, even if $\gamma$ is strongly constrained by Solar system measurements, one could still invoke a screening mechanism such as the one discussed in~\ref{sec:theory} so that it is important to measure the value of $\gamma$ on large scales. In this respect, the cosmological constraints are much weaker than the Solar system ones, see e.g~\cite{Perenon:2016blf}.
\item \textit{Growth function $f \sigma_8$: }  The effective gravitational constant is naturally part of the
source term in the evolution of the linear density perturbations of matter $\delta$:
\begin{equation}
\ddot \delta + 2H \dot \delta - 4 \pi \mu G \bar \rho \delta = 0 \; .
\end{equation}
The $\delta$ variable is of difficult observability. However its second statistical moment, the
rms of linear density fluctuations on the characteristic scale $R=8$ MPc/h, $\sigma_8$, and its
logarithmic derivative with respect to the scale factor of the universe, the linear growth
rate $f$, can be combined in an observable quantity ($f \sigma_8$) which is minimally affected my
observational biases.
\end{itemize}

The relation between the EFT parameters defined in~\eqref{eq:ActionEFTQuadratic} and the observable parameters discussed above is straightforward but rather lengthy. While referring the reader to~\cite{P_renon_2015,Perenon:2016blf} for further details, here we give their value in the simple case where the quadratic action~\eqref{eq:ActionEFTQuadratic} is restricted to the background terms $f(t)$, $\Lambda(t)$ and $c(t)$:
\begin{align}
\mu &= \frac{1}{f^3(t)} \left(\frac{2 c(t) f(t)/\mpl^2 + 4 \dot f^2}{2 c(t) f(t)/\mpl^2 + 3 \dot f^2} \right) \\
\gamma &= \frac{c(t) f(t) / \mpl^2 + \dot f^2}{c(t) f(t) / \mpl^2 + 2 \dot f^2}
\end{align}

We will finish this chapter by mentioning the future space mission Euclid designed to investigate in details the properties of DE. Through extensive weak lensing and BAO surveys, Euclid will be able to constrain the equation of state of DE and the growth factor at the percent level~\cite{laureijs2011euclid}. The satellite should be launched in 2023.


\chapter{Gravitational waves: an introduction} 


\label{Chapter3}

Between Einstein's first hesitations about the physical nature of GW and their direct detection by the LIGO/Virgo collaboration in 2015 \cite{Abbott:2016blz}, a century has elapsed. Nowadays, the importance of the field lies in the exciting comparison of the theory with astrophysical observations. The two pillars of GW experimental science are binary pulsars like the historical Hulse–Taylor pulsar PSR 1913+16 \cite{Hulse:1974eb} on the one hand, and gravitational waves produced by massive and rapidly evolving systems such as inspiralling compact binaries on the other hand. These begin to be routinely detected on Earth by the network of large-scale laser interferometers which we will present in Section~\ref{subsec:current_planned_inter}. In the future, the space-based laser interferometer LISA should be able to detect supermassive black-hole binaries at cosmological distances.
To prepare these experiments, the required theoretical work consists of carrying out a sufficiently general solution of the Einstein field equations for inspiralling binary systems, by describing the physical processes of the emission and propagation of the gravitational waves from the source to the distant detector, as well as their back-reaction onto the source.

In this Chapter, we will recall some basic facts about the theoretical and experimental knowledge on GW. A more systematic discussion of the emission process will be the subject of Part~\ref{part2}, where we will discuss the effective field theory approach to the two-body problem. However, we find it useful to recall in this Chapter the traditional post-Newtonian approach of the problem and its relation with GW detection. The post-Newtonian formalism remains the most powerful tool to analyze the dissipative dynamics of binary systems \cite{Blanchet:2013haa} and as such it provides the current templates for data analysis. Finally, we will finish the Chapter by highlighting the consistency tests of GR which are carried out using GW data. Since the experimental GW science is a brand new field, this last Section will be quite limited in scope. The remaining of this thesis will be devoted to a more detailed investigation of GW signals in other theories than General Relativity.

\section{GW in flat space} 

In this Section we will closely follow the first volume of Michele Maggiore's book on GW, to which we refer the reader for more details \cite{Maggiore:1900zz}. The gravitational action of GR is constituted of the well-known Einstein-Hilbert action,
\begin{equation}
S = \frac{\mpl^2}{2} \int \mathrm{d}^4 x \sqrt{-g} R \; ,
\end{equation}
where $R$ is the Ricci scalar built out of the metric $g_{\mu \nu}$. To this gravitational action we should supplement a matter action $S_m$ whose functional derivative is the energy-momentum tensor $T^{\mu \nu}$,
\begin{equation}
\frac{\delta S_m}{\delta g_{\mu \nu}} = \frac{\sqrt{-g}}{2} T^{\mu \nu} \; .
\end{equation}
Taking the variation of the total action with respect to $g_{\mu \nu}$ one finds the Einstein equations
\begin{equation} \label{eq:EinsteinEqs}
G^{\mu \nu} = R^{\mu \nu} - \frac{1}{2} g^{\mu \nu} R = 8 \pi G T^{\mu \nu} \; .
\end{equation}
They form a system of ten second-order partial differential equations obeyed by the metric. Among these ten equations, four govern, via the contracted Bianchi identity, the evolution of the matter system
\begin{equation}
\nabla_\mu G^{\mu \nu} = 0 \quad \Rightarrow \quad \nabla_\mu T^{\mu \nu} = 0 \; .
\end{equation}
The matter equations can also be obtained by varying the matter action with respect to the matter fields. The space-time geometry is constrained by the six remaining equations, which place six independent constraints on the ten components of the metric $g_{\mu \nu}$, leaving four of them to be fixed by a choice of the coordinate system.

We will now linearize the Einstein equations for small deviations from flat space. This can be achieved by writing
\begin{equation} \label{eq:g=eta+h}
g_{\mu \nu} = \eta_{\mu \nu} + h_{\mu \nu} \; , 
\end{equation}
where $\eta_{\mu \nu}$ is the Minkowski metric and $h_{\mu \nu}$ is a small fluctuation.
After a bit of algebra, one finds that the linearization of the Einstein equations~\eqref{eq:EinsteinEqs} is
\begin{equation} \label{eq:linearizedEinstein}
\square \bar h_{\mu \nu} + \eta_{\mu \nu} \partial^\rho \partial^\sigma \bar h_{\rho \sigma} - \partial^\rho \partial_\nu \bar h_{\mu \rho} - \partial^\rho \partial_\mu \bar h_{\nu \rho} = - 16 \pi G T_{\mu \nu} \; ,
\end{equation}
where we have defined the barred $h$ variable by
\begin{equation}
\bar h_{\mu \nu} = h_{\mu \nu} - \frac{1}{2} \eta_{\mu \nu} h \; , \quad h = \eta^{\mu \nu} h_{\mu \nu} \; .
\end{equation}

We now use the gauge freedom to choose the \textit{harmonic} or De Donder gauge. At the full nonlinear level, it is defined by the three equivalent conditions
\begin{equation} \label{eq:Harmonic_GF}
g^{\nu \rho} \nabla_\nu \nabla_\rho x^\mu = 0 \quad \Leftrightarrow \quad g^{\nu \rho} \Gamma^\mu_{\nu \rho} = 0 \quad \Leftrightarrow \quad \partial_\nu \big( \sqrt{-g} g^{\mu \nu} \big) = 0 \; ,
\end{equation}
where $\Gamma^\mu_{\nu \rho}$ is the Christoffel symbol for the metric $g_{\mu \nu}$. The first condition express that the coordinates $x^\mu$ are harmonic, i.e they satisfy the harmonic condition: their d'Alembertian vanishes. The second and third conditions are trivially obtained from the first using the definition of the covariant derivative. Linearizing this equation with respect to the metric, one obtains the simple condition on $\bar h_{\mu \nu}$
\begin{equation} \label{eq:linearizedHarmonic}
\partial^\nu \bar h_{\mu \nu} = 0 \; .
\end{equation}
These four equations fix the linearized harmonic gauge. The linearized Einstein equations in this gauge simply amount to a wave equation with source term,
\begin{equation} \label{eq:linearizedEinsteinHarmonic}
\square \bar h_{\mu \nu} = - 16 \pi G T_{\mu \nu} \; .
\end{equation}
Eqs~\eqref{eq:linearizedHarmonic} and~\eqref{eq:linearizedEinsteinHarmonic} together imply for consistency
\begin{equation}
\partial^\mu T_{\mu \nu} = 0 \; ,
\end{equation}
which is the conservation of energy-momentum in the linearized theory. Outside the source, the metric perturbation thus obey the wave equation
\begin{equation}
\square \bar h_{\mu \nu} = 0 \; .
\end{equation}
We can further simplify the form of the metric by noticing that Eq.~\eqref{eq:linearizedHarmonic} does not fix the gauge completely. Indeed, under a small shift of coordinates $x^\mu \rightarrow x'\hphantom{}^\mu = x^\mu + \xi^\mu$ so that $h_{\mu \nu} \rightarrow h'_{\mu \nu} = h_{\mu \nu} - \partial_\mu \xi_\nu - \partial_\nu \xi_\mu$, the harmonic gauge condition~\eqref{eq:linearizedHarmonic} becomes
\begin{equation}
\partial^\nu \bar h_{\mu \nu} \rightarrow (\partial^\nu \bar h_{\mu \nu})' = \partial^\nu \bar h_{\mu \nu} - \square \xi^\mu \; ,
\end{equation}
so that a coordinate change satisfying $\square \xi^\mu = 0$ (i.e, a vector-like wave) remains inside the harmonic gauge. We can use this freedom to simplify the form of $\bar h_{\mu \nu}$: by choosing an appropriate $\xi^0$ we impose that the trace $\bar h = 0$ (so that $\bar h_{\mu \nu} = h_{\mu \nu}$), and by choosing $\xi^i$ we impose $h^{0i} = 0$. The $\mu=0$ part of the harmonic gauge condition~\eqref{eq:linearizedHarmonic} thus becomes $\partial^0 h_{00} = 0$, i.e $h_{00}$ is a constant in time. We simply set it to zero as we are interested by a wavelike behavior. In conclusion, we have set
\begin{equation} \label{eq:TTgauge}
h^{0 \mu} = 0 \; , \quad h^i_i = 0 \; , \quad \partial^j h_{ij} = 0 \; .
\end{equation}
This defines the \textit{transverse-traceless gauge} or TT gauge. By imposing the harmonic gauge we have reduced the 10 degrees of freedom in $h_{\mu \nu}$ to six; the residual gauge freedom allowed us to further reduce the number of degrees of freedom to just two. We will denote the metric in the TT gauge by $h_{ij}^\mathrm{TT}$.

Let us consider a single plane wave $\omega$ propagating in the direction $\hat{ \mathbf n} = \hat{ \mathbf z}$. The three conditions above~\eqref{eq:TTgauge} imply that $h_{ij}^\mathrm{TT}$ can be decomposed on two basis tensors,
\begin{equation} \label{eq:decomp_hijtt}
h_{ij}^\mathrm{TT} = \begin{pmatrix}
h_+ & h_\times & 0 \\ h_\times & -h_+ & 0 \\ 0&0&0
\end{pmatrix}_{ij} e^{i \omega (t - z)} = h_+ e_{ij}^+ e^{i \omega (t - z)} + h_\times e_{ij}^\times e^{i \omega (t - z)} \; ,
\end{equation}
where the tensors $e_{ij}^+$ and $e_{ij}^\times$ are defined by
\begin{equation}\label{eq:def_eij+x}
e_{ij}^+ = \hat{\mathbf x}_i \hat{\mathbf x}_j - \hat{\mathbf y}_i \hat{\mathbf y}_j = \begin{pmatrix}
1 & 0 & 0 \\ 0 & -1 & 0 \\ 0&0&0
\end{pmatrix}_{ij} \; , \quad e_{ij}^\times = \hat{\mathbf x}_i \hat{\mathbf y}_j + \hat{\mathbf y}_i \hat{\mathbf x}_j = \begin{pmatrix}
0 & 1 & 0 \\ 1 & 0 & 0 \\ 0&0&0
\end{pmatrix}_{ij} \; ,
\end{equation}
where $\hat{\mathbf x}$ and $\hat{\mathbf y}$ are the unit vectors orthogonal to $\hat{\mathbf z}$. $h_+$ and $h_\times$ are characteristic amplitudes and are called the GW \textit{strain}. More generally, a wave which is not in the TT gauge can be projected on it with the help of the Lambda tensor $\Lambda_{ij;kl}$. It is a tensor symmetric under the simultaneous exchange $(i,j) \leftrightarrow (k,l)$ and such that
\begin{equation}
\Lambda_{ij;kl} \Lambda_{kl;mn} = \Lambda_{ij;mn} \; , \quad n^i \Lambda_{ij;kl} = n^j \Lambda_{ij;kl} = 0 \; , \quad \Lambda_{ii;kl} = 0 \; ,
\end{equation}
where $\hat{ \mathbf n}$ is the direction of propagation of the GW.
The first condition ensure that $\Lambda_{ij;kl}$ is a projector, the second expresses its transverse property, and the last one sets its trace to zero. These conditions select a unique tensor,
\begin{equation} \label{eq:def_lambda}
\Lambda_{ij;kl} = P_{ik} P_{jl} - \frac{1}{2} P_{ij} P_{kl} \; , \quad P_{ij} = \delta_{ij} - n_i n_j \; .
\end{equation}

We are now ready to derive the effect that a GW has on physical objects such as detectors, which will be the subject of the next Section.

\section{Laser interferometers}

The first historical GW detector was built by Joseph Weber at the university of Maryland; he was the first physicist to seriously consider experimental detection of GW. The 'Weber bars' were massive aluminium cylinders designed to be set in a resonant motion by gravitational waves. Weber announced that he had detected GW but it is now commonly accepted that his detector did not reach the required accuracy. Nowadays, the laser interferometers LIGO and Virgo have detected dozens of GW using a giant Fabry-Perot cavity (the reader interested by the total number of detections can consult the \href{https://www.ligo.org/detections.php}{LIGO} page which is constantly updated). We will begin this Section by recalling the detection principle as well as the interferometer design. Then, we will present the data analysis techniques used to extract the GW signal from the noise. We finish by a short description of the peculiarities of each current and planned GW detector.

\subsection{Detection principle} \label{subsec:detection_principle}

\begin{figure}[ht]
\centering
\includegraphics[scale=0.5]{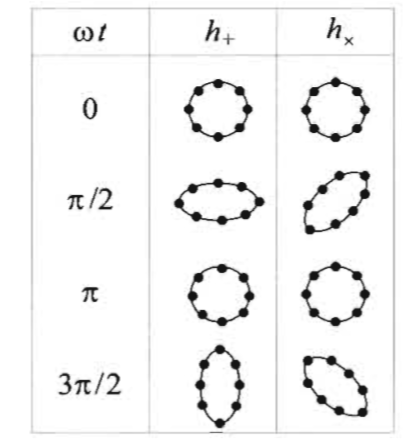}
\caption[Deformation of a ring of test-masses by a GW]{The deformation of a ring of test masses situated in the $(x,y)$ plane by the passage of a GW in the direction $z$. Figure taken from Maggiore \cite{Maggiore:1900zz}.   }
\label{fig:testmasses}
\end{figure}

Let us consider an object (e.g a rod whose orientation is denoted by $\mathbf{L}$) of size $L$ small compared to the spatial variation of the GW, $L \ll \lambda$ where $\lambda$ is the GW wavelength. Then as a GW passes its proper length varies as
\begin{equation}
(\mathbf{L} + \delta \mathbf{L})^2 = (1 + h_+ e^{i \omega t}) L_x^2 + (1 - h_+ e^{i \omega t}) L_y^2 + 2 h_\times e^{i \omega t} L_x L_y \; ,
\end{equation}
where we recall that we consider a GW propagating in the $\hat{\mathbf z}$ direction.
Since the proper length is a physical quantity, we see that we could hope to detect a GW by monitoring the variation of length of this object. A purely plus polarization would induce an oscillation
\begin{align} \label{eq:oscill_h+}
\begin{split}
\delta L_x &= \frac{h_+}{2} L_x e^{i \omega t} \; , \\
\delta L_y &= - \frac{h_+}{2} L_y e^{i \omega t} \; , 
\end{split}
\end{align}
where we have expanded for small deformations. On the other hand, a cross polarization would provoke a change
\begin{align} \label{eq:oscill_hcross}
\begin{split}
\delta L_x &= \frac{h_\times}{2} L_y e^{i \omega t} \; , \\
\delta L_y &= \frac{h_\times}{2} L_x e^{i \omega t} \; .
\end{split}
\end{align}
The effect of a GW on a ring of test masses located in the $(x,y)$ plane is shown in Figure~\ref{fig:testmasses}.

We have given a quick way to derive the effect of a GW on a physical object, but it can be recovered using other tools. In particular, we can use the geodesic \textit{deviation} equation expressing how nearby objects evolve under a gravitational field.
 We denote by $\xi_i$ the separation between two  test masses---for instance the mirrors of a  detector---located at a distance shorter than the typical spatial variation of a gravitational wave. In the proper detector frame, i.e.~choosing coordinates such that the spacetime metric is flat up to tidal effects even during the passage of a gravitational wave, the acceleration between the two masses is given by
 (see e.g.~\cite{Maggiore:1900zz})
\begin{equation}
\ddot{\xi}_i = - R_{i0j0}    \xi_j \;,
\label{eq:motion_riemann}
\end{equation}
where 
\be
R_{i0j0} = - \frac{1}{2}  \ddot{ {h}}_{ij}^\mathrm{TT}  \;.
\ee 
is the corresponding linearized component of the Riemann tensor. Then, using the expression of $h_{ij}^\mathrm{TT}$ given in Eq.~\eqref{eq:decomp_hijtt}, one recovers the oscillations~\eqref{eq:oscill_h+}~-~\eqref{eq:oscill_hcross}.

An order-of-magnitude estimate (we will present it in the next Section) gives that for a black hole merger in our galaxy $h_+ \sim h_\times \sim 10^{-22}$. This is utterly small! To enhance the signal, we can see from Eq.~\eqref{eq:oscill_h+} that we should choose $L$ to be the biggest possible.

This is the strategy adopted in laser interferometers such as LIGO or Virgo: the principle is to build a giant Michelson interferometer whose arms are of the kilometer size. A small variation of the length of the arms would result in interference fringes between the two beams. Furthermore, a Fabry-Pérot cavity is placed at each arm so that the effective length is multiplied by $10^3$ (one can picture this cavity as having the effect of making the laser 'bounce' between two mirrors); a power-recycling mirror is also added to enhance the power of the laser so that interference fringes are more contrasted.

\begin{figure}[ht]
\centering
\includegraphics[width=0.6\textwidth]{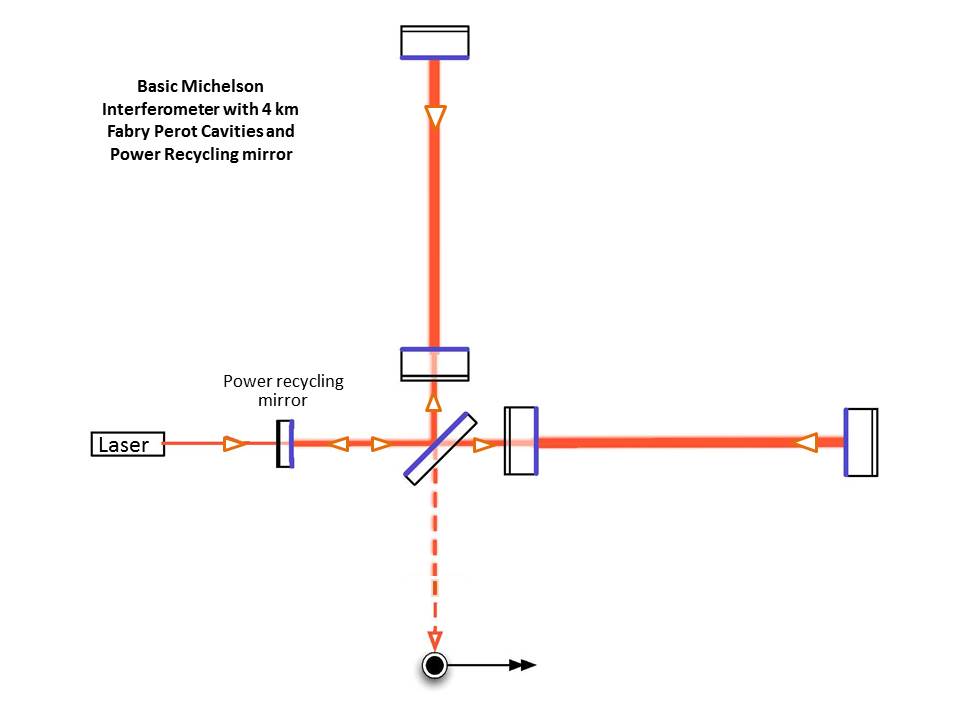}
\caption[Schematic representation of the laser interferometer LIGO]{Schematic representation of the laser interferometer LIGO. Figure taken from LIGO \href{https://www.ligo.caltech.edu/page/ligos-ifo}{website}   }
\label{fig:LIGO_detector}
\end{figure}

Since it is not immediately clear that a laser interferometer measures the proper length variations described in Eqs.~\eqref{eq:oscill_h+}-\eqref{eq:oscill_hcross}, let us elaborate on this point. We consider a GW propagating in the $\hat{\mathbf z}$ direction and containing only a plus polarization for simplicity, and we will work in the TT gauge. For a photon, the spacetime interval is
\begin{equation}
0 = \mathrm{d}s^2 = - \mathrm{d}t^2 + \left[ 1 + h_+ e^{i \omega t} \right] \mathrm{d}x^2 + \left[ 1 - h_+ e^{i \omega t} \right] \mathrm{d}y^2 + \mathrm{d}z^2 \; .
\end{equation}
Let us also choose our axes so that the $x$ direction coincides with one arm of the interferometer (of length $L_x$), while the other arm is aligned with the $y$ axis. Then, for the $x$ arm to first order in $h_+$,
\begin{equation}
\mathrm{d}x = \pm \mathrm{d}t \left[ 1 -  \frac{h_+}{2} e^{i \omega t} \right] \; ,
\end{equation}
where the sign indicates the direction of propagation of the photon. Integrating over the path of the photon for a round trip and considering a large-wavelength GW, $\omega L_x \ll 1$, one gets that the time elapsed is
\begin{equation}
t_1 - t_0 = 2 L_x + L_x h_+ e^{i \omega (t_0 + L_x)} \; ,
\end{equation}
where $t_0$ is the initial time and $t_1$ the time at which the photon finishes its round trip. A similar calculation gives for the $y$ arm
\begin{equation}
t_1 - t_0 = 2 L_y - L_y h_+ e^{i \omega (t_0 + L_y)} \; .
\end{equation}
Thus, by making the $x$ and $y$ photons interfere as it is the case in a Michelson interferometer, one gets a phase difference
\begin{equation}
\Delta \phi \simeq 2 \omega_\mathrm{laser} L h_+ e^{i \omega (t_0 + L)} \; ,
\end{equation}
where $\omega_\mathrm{laser}$ is the frequency of the laser, $L = (L_x+L_y)/2$ and we have assumed $L_x \simeq L_y$. Thus, the passage of a GW can be recorded using the interference fringes of a giant interferometer.

\subsection{Data analysis techniques} \label{subsec:data_analysis}

Even with their effective $10^3$ km size, the current interferometers would be completely unable to see a GW signal if there were no data analysis. Indeed, the signal is 100 times fainter than the detector noise (mainly driven by the seismic noise at low frequencies under 10 Hz, by the thermal noise of the mirrors from 10 Hz to a few hundred Hz, and by the laser shot noise above a few hundred Hz \cite{2015, Abbott_2020}). To overcome this difficulty, the idea is to use a matched filtering analysis. Let us sketch briefly in what it consists.

The total signal $s(t)$ at the detector output can be modeled as
\begin{equation}
s(t) = h(t) + n(t) \; ,
\end{equation}
where $h(t)$ is the expected signal and $n(t)$ is the noise. Let us assume that we perfectly know the form of the signal $h(t)$ (because we have a bank of GW templates theoretically predicted by the PN formalism as we will present in Section~\ref{sec:GWModeling}), and that the total signal duration is $T$. We can average the signal against the template,
\begin{equation}
\frac{1}{T} \int_0^T \mathrm{d}t \; s(t) h(t) = \frac{1}{T} \int_0^T \mathrm{d}t \; h(t)^2 + \frac{1}{T} \int_0^T \mathrm{d}t \; n(t) h(t) \; .
\end{equation}
$h(t)$ is close to a cosine so that the average of $h^2$ is of order one in $T$,
\begin{equation}
\frac{1}{T} \int_0^T \mathrm{d}t \; h(t)^2 \sim h_0^2 \; ,
\end{equation}
where $h_0$ is the amplitude of $h$. On the other hand, the noise is decorrelated from the signal so that the second integral should average to zero. More precisely, as expected for systems performing a random walk, one should have the scaling
\begin{equation}
\frac{1}{T} \int_0^T \mathrm{d}t \; n(t) h(t) \sim n_0 h_0 \left( \frac{\tau}{T} \right)^{1/2} \; ,
\end{equation}
where $n_0$ is the typical amplitude of the noise and $\tau$ is a typical characteristic time such as the period of $h(t)$. Typically, one can achieve $\left( \tau / T \right)^{1/2} \lesssim 10^{-2}$ by monitoring the signal during a sufficient amount of GW cycles so that the observability of the event is greatly enhanced.

\subsection{Current and planned interferometers} \label{subsec:current_planned_inter}

Finally, let us mention the characteristics of the current as well as of the future GW detectors:

\begin{itemize}
\item \textit{(Advanced) Laser Interferometer Gravitational-Wave Observatory (LIGO) \cite{2015}}: this is currently the most performant GW detector. The LIGO team has detectors in two locations, one in Livingston, Louisiana and the other in Handford near Richland, Washington. These sites are separated by 3000 km.  As described above, the principle of the two observatories relies on a giant Michelson interferometer with Fabry-Pérot cavities. Actually, since 2010 the LIGO observatory is in its 'advanced' configuration which means that sophisticated techniques such as cryogenic mirrors and the injection of squeezed vacuum are being used. These techniques allowed for the first GW detection the 14 September 2015. The GW sources aimed by LIGO are mostly neutrons stars up to 100 MPc in distance and solar-size black hole collisions up to 1 GPc. LIGO is also monitoring for unmodeled signals such as from the corecollpase of massive stars.
\begin{figure}[ht]
\centering
\includegraphics[width=0.8\textwidth]{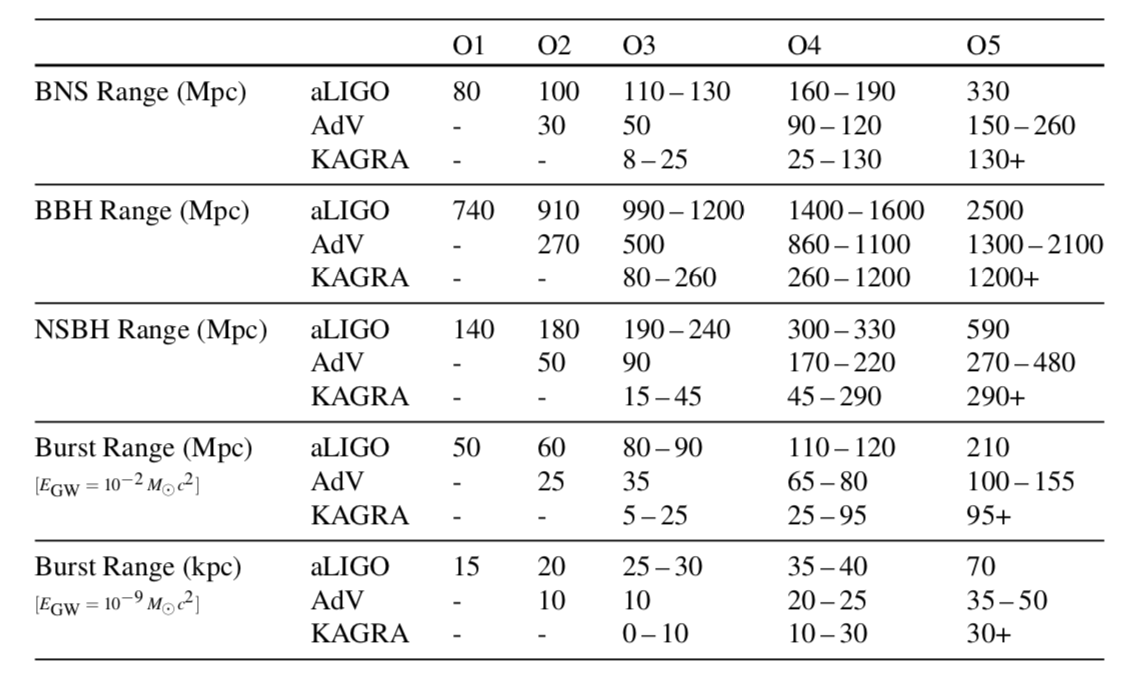}
\caption[Average maximal distances for GW detection]{Average maximal distances for GW detection in the three projects (Advanced) LIGO and Virgo and KAGRA. BNS stands for binary neutron stars, BBH for binary black holes, and NSBH for neutron star-black hole. The two burst signals are derived for different peak luminosities consistent with core-collapse supernovae models. Observation sessions are divided in 'runs'; we are currently in the 'O3' run; the 'O4' is expected to begin in early 2022, and the 'O5'      in 2025. Figure taken from \cite{collaboration2013prospects}. }
\label{fig:detectors_sensitivity}
\end{figure}

\item \textit{(Advanced) Virgo \cite{Acernese_2014}}: This observatory is similar to LIGO except that it is an european project situated near to Pisa, Italy. Note that having several GW observatories is useful because it allows for a more precise estimation of the direction of the event (a single detector sees event from all direction without being able to discriminate among them). The first detection of GW of the Virgo collaboration has been announced on 27 september 2017.

\item \textit{Kamioka Gravitational Wave Detector (KAGRA) \cite{Akutsu:2018axf}:} The Japanese LIGO-like detector, said to be a '2.5' generation project (a little bit more advanced than the second-generation project like Advanced Virgo and LIGO since it is located underground to lower seismic noise, but less powerful than the third-generation instruments which we will present shortly). The observation run started in February 2020.

\item \textit{Einstein Telescope (ET) \cite{punturo:hal-00629986}: } This European 'third-generation' project is expected to come in the 2030's. It should be constituted of a triangular shape interferometer of 10km long situated underground (which should lower the seismic noise). Also, cryogenic cooling of the mirrors will allow to reduce the thermal vibration of the test masses. The expected sensitivity is expect to be a factor of 10 better than the current GW observatories. This improved sensitivity will allow for a more precise mass and spin estimation; furthermore the lower limiting frequency will allow to detect intermediate mass black hole binaries with component masses up to a few thousand solar masses. ET should provide an all-sky survey of such objects up to redshifts of 2 and more. ET should allow to detect a stochastic background of gravitational waves up to $\Omega_\mathrm{GW} \sim 10^{-12}$; although
ET’s sensitivity is a few orders of magnitude poorer than that required to detect
backgrounds predicted by inflationary Universe models, there is the possibility
that phase transitions in the early Universe and other processes could give rise
to a detectable background. Finally, ET could observe normal modes in neutron stars which is the best way to probe neutron stars interiors and to understand the equation of state of matter under extreme conditions of density, pressure, temperature and magnetic fields.

\item \textit{Cosmic Explorer~\cite{reitze2019cosmic}: } It is the planned U.S. contribution to the global third-generation ground-based gravitational-wave detector network. It is expected to be a 40km L-shaped interferometer located on the surface. Its targets are the same than ET and it should be built at the same time.

\item \textit{ Laser Interferometer Space Antenna (LISA) \cite{AmaroSeoane:2012km}: } The LISA underlying idea is that an interferometer based in space should be free from seismic noise, thus allowing for detections at much lower frequencies (down to $10^{-4}$ Hz). Planned to be launched in 2034, the LISA mission will consist in a triangular interferometer with sides 2.5 millions km long, flying along an Earth-like heliocentric orbit. Potential sources for signals are merging massive black holes at the centre of galaxies (LISA is expected to detect \textit{all} of them, all the way back to their earliest formation around $z \simeq 15$), massive black holes orbited by small compact objects up to $z \simeq 4$, known as extreme mass ratio inspirals, binaries of compact stars in our Galaxy (LISA is expected to detect and resolve around 25,000 galactic compact binaries!), and possibly other sources of cosmological origin, such as the very early phase of the Big Bang, and speculative astrophysical objects like cosmic strings and domain boundaries.

\item \textit{Deci-Hertz Interferometer Gravitational wave Observatory (DECIGO) \cite{Kawamura:2006up}: }  DECIGO is a proposed Japanese, space-based, gravitational wave observatory. The laser interferometric gravitational wave detector is so named because it is to be most sensitive in the frequency band between 0.1 and 10 Hz, filling in the gap between the sensitive bands of LIGO and LISA. If funding can be found, its designers hope to launch it in 2027.
The design is similar to LISA, with three zero-drag satellites in a triangular arrangement, but using a smaller separation of only 1000 km.

\item \textit{European Pulsar Timing Array (EPTA) \cite{2013CQGra..30v4009K}: } This project is clearly different from the other ones since it does not consist in an interferometer. Rather, the idea is to monitor millisecond pulsars which serve as very precise clocks distributed all over the galaxy. If a GW passes between us and the pulsars, there would be an additional time delay - typically tens of ns! - which we can measure. On could thus measure low-frequency gravitational waves, with a frequency of $10^{-9}$ to $10^{-6}$ hertz; the expected astrophysical sources of such gravitational waves are massive black hole binaries in the centres of merging galaxies, where tens of millions of solar masses are in orbit with a period between months and a few years. One could also hope to detect stochastic GW background originating from inflation or from cosmic strings. The European Pulsar Timing Array uses five European telescopes. These are the Westerbork Synthesis Radio Telescope, the Effelsberg Radio Telescope, the Lovell Telescope, the Nançay Radio Telescope and the Sardinia Radio Telescope.

\end{itemize}

\begin{figure}[ht]
\centering
\includegraphics[width=0.8\textwidth]{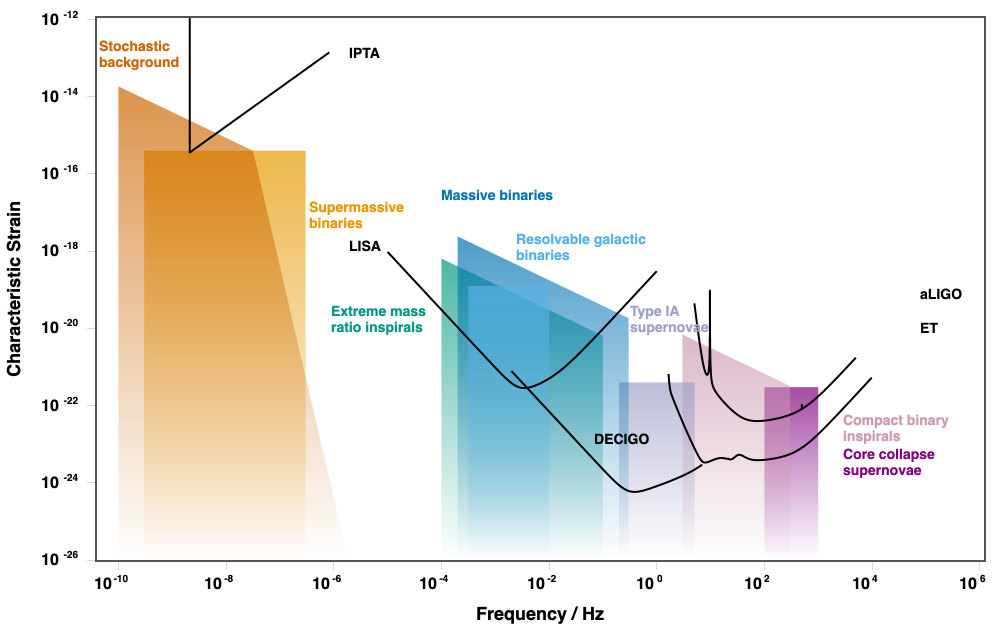}
\caption[Sensitivities of GW detectors]{Sensitivities and frequency bands of the different GW detectors with their main potential sources. Figure by \href{http://rhcole.com/apps/GWplotter/}{Christopher Moore, Robert Cole and Christopher Berry}. }
\label{fig:detectors_sensitivity2}
\end{figure}

\section{GW modeling} \label{sec:GWModeling}

So far, we have concentrated on the detection of GW and remained quite elusive about their formation. In this Section, we will sketch the different theoretical approaches to the formation of GW from compact binaries needed to obtain the template $h(t)$ alluded to in Section~\ref{subsec:data_analysis}. We will begin with the historical quadrupole formula and the first experimental proof of the existence of GW from binary pulsars. Then, we will give a brief account of the post-Newtonian formalism which is currently the most efficient way to obtain a template $h(t)$ for slowly inspiralling binaries. We will finally mention other approaches like the post-Minkowskian expansion or the gravitational self-force programme.

\subsection{The quadrupole formula}

In 1916 Einstein predicted that there would not be dipole gravitational radiation but that a time-varying quadrupole would produce gravitational waves. This is very intuitive in Newtonian physics since the (derivative of the) gravitational dipole of a system of point masses is simply their total momentum: by the center-of-mass theorem, the second derivative of the dipole is zero! The quadrupolar radiation is ultimately related to the spin-2 nature of the graviton. We will rederive the quadrupole formula in Section~\ref{subsec:detected_signal}, Eq.~\eqref{eq:obsgraviton}, so let us instead directly give its result here. Far from a source like a binary system, the gravitational amplitude fluctuation in the TT gauge is given by
\begin{equation}
h_{ij}^\mathrm{TT} = \frac{2G}{r} \Lambda_{ij, kl} \ddot Q^{kl}(t_\mathrm{ret}) \; .
\end{equation}
In this equation, $\Lambda_{ij, kl}$ is the tensor defined in Eq.~\eqref{eq:def_lambda}, $r$ is the distance to the source, $t_\mathrm{ret} = t-r$ is the retarded time and $Q^{kl}$ is the traceless quadrupole moment of the source. For a system of point sources of masses $m_A$ located at coordinates $\mathbf{x}_A$, $A=1, \dots, n$, the quadrupole moment is
\begin{equation}
Q^{kl} = \sum_A m_A \bigg( x_A^k x_A^l - \frac{1}{3} x_A^2 \delta^{kl} \bigg) \; .
\end{equation}
It is a standard exercise to derive the amplitude $h_{ij}^\mathrm{TT}$ for a system of two point masses in circular orbit using the quadrupole formula. Denoting by $\theta$ the angle between the normal to the orbit and the direction of propagation of the GW, one finds that the two polarizations of the GW are
\begin{align}
\begin{split}
h_+(t) &= \frac{4G \mu \omega^2 R^2}{r} \bigg( \frac{1+\cos^2\theta}{2} \bigg) \cos(2 \omega t) \; , \\
h_\times(t) &= \frac{4G \mu \omega^2 R^2}{r}\cos \theta \sin(2 \omega t) \; , \\
\end{split}
\end{align} 
where $\omega$ is the frequency of the orbit, $R$ is the separation of the two masses, and $\mu = m_1 m_2/(m_1+m_2)$ is the reduced mass. In this equation we have included the oscillatory factor in~\eqref{eq:decomp_hijtt} inside $h_+$ and $h_\times$ to insist on the fact that a quadrupole radiates at two times the orbital frequency of the system. With this equation we obtain the order-of-magnitude of the strain mentioned in Section~\ref{subsec:detection_principle}: since $\omega R \sim v \sim 0.3$ in the last stages of a coalescence ($v$ is the typical velocity of the two bodies), and for black holes of masses of order $10 M_\odot$, one obtains roughly
\begin{equation}
h_+ \sim h_\times \sim \frac{1 \mathrm{km}}{r} \; ,
\end{equation}
so that for binaries situated at a few MPc one has $h \sim 10^{-20}$.

The quadrupole formula gives also access to the power dissipated from the system which causes the binaries to spiral closer and closer. We will again derive this dissipated power in Section~\ref{subsec:radiated_gravitons}, Eq.~\eqref{eq:quadrupole_graviton}, so let us give directly the result here:
\begin{equation}
P = \frac{G}{5} \left\langle \dddot Q^{kl} \dddot Q_{kl} \right\rangle \; ,
\end{equation}
where the brackets denote an average over many gravitational wave cycles. Again, for a binary system this translates into
\begin{equation} \label{eq:quadrupoleBinary}
P = \frac{32}{5 G} ( G M_c \omega)^{10/3} \; , \quad M_c = \frac{(m_1 m_2)^{3/5}}{(m_1+m_2)^{1/5}} \; .
\end{equation}
To derive this equation, we have used Kepler's law $\omega^2 = G(m_1+m_2)/R^3$ which is valid at lowest order in the relativistic expansion.
The \textit{chirp mass} $M_c$ is a combination of the masses which naturally appears in the quadrupole formula; since it governs the gravitational dynamics (at lowest order) of binaries, this is the parameter that is the most accurately measured in interferometers (the degeneracy between $m_1$ and $m_2$ which appears in Eq.~\eqref{eq:quadrupoleBinary} is broken by relativistic corrections to this formula).

\subsection{The binary pulsar test} \label{subsec:binary_pulsar}

The first experimental proof of the existence of GW came from observations of pulsars in 1974 by Hulse and Taylor \cite{1981SciAm.245d..74W}. Pulsars are rapidly rotating (with period of order 60 miliseconds), highly magnetized neutron stars. After timing the radio pulses for some time, Hulse and Taylor noticed that there was a systematic variation in the arrival time of the pulses. Sometimes, the pulses were received a little sooner than expected; sometimes, later than expected. These variations changed in a smooth and repetitive manner, with a period of 7.75 hours. They realized that such behavior is predicted if the pulsar were in a binary orbit with another star, later confirmed to be another neutron star. 

\begin{figure}[ht]
\centering
\includegraphics[width=0.6\textwidth]{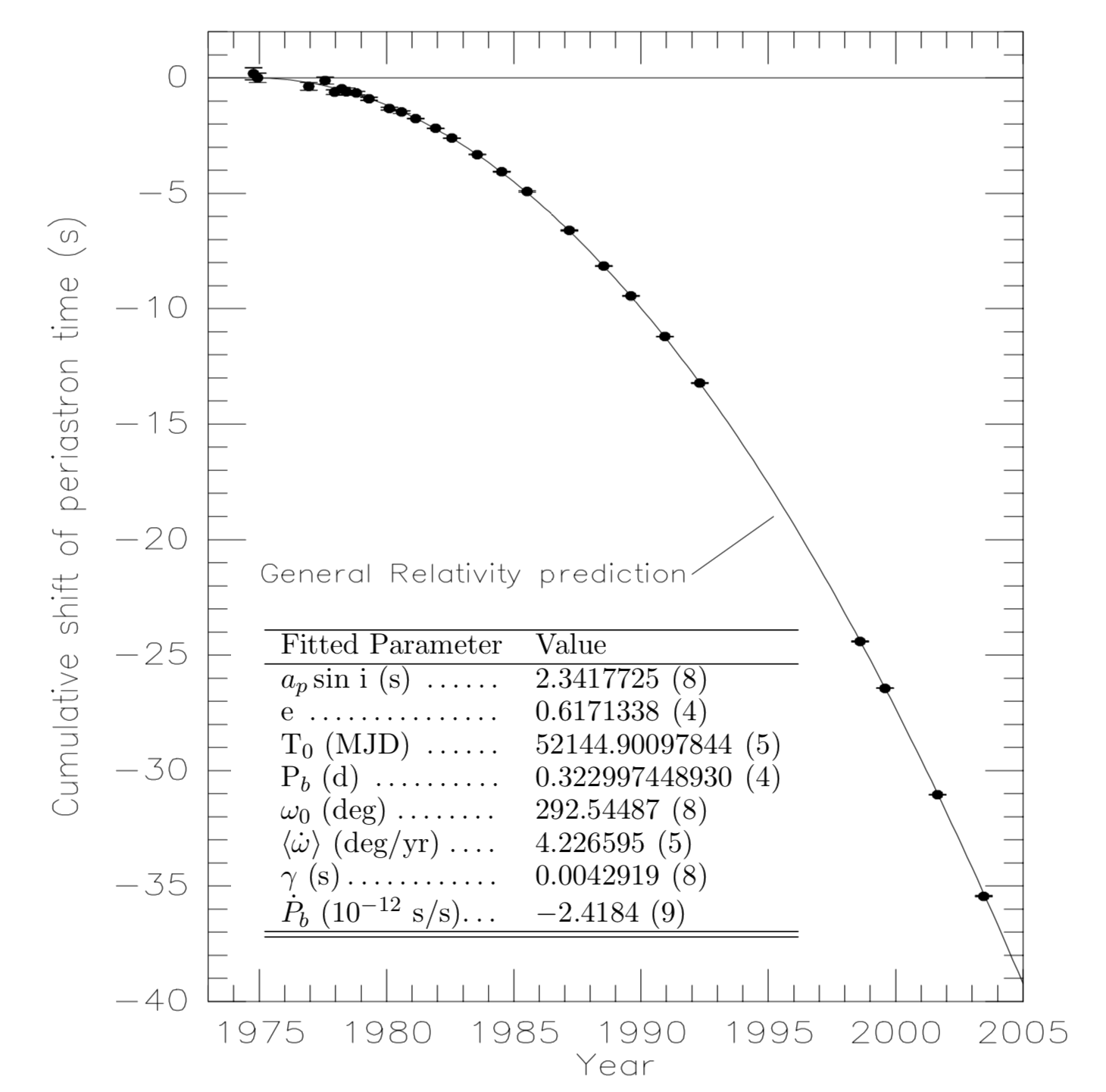}
\caption[Orbital decay of the Hulse-Taylor binary pulsar]{Orbital decay of the Hulse-Taylor binary pulsar in perfect agreement with the quadrupole prediction from GR, together with the measured parameters of the binary system. Adapted from \cite{weisberg2004relativistic} }
\label{fig:hulse_taylor}
\end{figure}

Since its discovery in 1974, the orbital period of the Hulse-Taylor pulsar has been accurately monitored through time. Observations confirm that the motion of the two pulsars is accelerating (so that they lose energy by coming closer together); the precise curve of the period shift due to gravitational radiation is perfectly fitted by the quadrupole formula~\eqref{eq:quadrupoleBinary} (actually, one should also take into account the eccentricity of the orbit, as we will do in Chapter~\ref{Chapter5}, Section~\ref{sec:planar_dynamics}), with a precision of 0.2\%, as illustrated in Figure~\ref{fig:hulse_taylor}. The measurement of the parameters of the binary system is so precise that the uncertainty in the measurement of the gravitational constant $G$ is comparable to the uncertainty in the determination of the masses of the pulsars \cite{weisberg2004relativistic}! The binary pulsar is one of the most stringent test of GR; scalar-tensor theories which generically predict dipole radiation (see Part~\ref{part2}) are strongly constrained by this observation \cite{Freire:2012mg}.

\subsection{The PN formalism} \label{subsec:pn_formalism}

The quadrupole formula was accurately confirmed by the binary pulsar test. However, it not not well-suited to provide an accurate template $h(t)$ for a GW detection: relativistic corrections to the quadrupole should be taken into account if one wants to monitor the waveform to the required accuracy. Let us now show why this is the case.

At the lowest relativistic order, the energy of a binary system is simply its Newtonian energy:
\begin{equation}
E_\mathrm{Newt} = - \frac{G m_1 m_2}{2R} \; .
\end{equation}
We have just seen in Eq.~\eqref{eq:quadrupoleBinary} that this energy drops because of GW emission, so that $R$ (and $\omega$) will slowly vary in time. We make the quasi-circular assumption that $\dot \omega \ll \omega^2$ which is valid up to the last orbits of the binary system. Using the energy balance equation,
\begin{equation} \label{eq:energyBalance}
\mathrm{d}E/\mathrm{d}t = -P \; ,
\end{equation}
one gets an evolution equation for $\omega$,
\begin{equation} \label{eq:omegadot}
\dot \omega = \frac{96}{5} \big(G M_c\big)^{5/3} \omega^{11/3} \; ,
\end{equation}
which shows that $\omega$ increases over time (the binary constituents come closer together). The solution of this equation formally diverges at a finite value of time $t_\mathrm{coal}$,
\begin{equation}
\omega(\tau) = \left( \frac{5}{256 \tau} \right)^{3/8} \big( G M_c \big)^{-5/8} \; ,
\end{equation}
where $\tau = t_\mathrm{coal}-t$ is the time to coalescence. Of course, the divergence is cutoff by the fact that when their separation becomes smaller than a critical distance the two stars or black holes merge. Inverting this relation for $\tau$ we find that for a minimal frequency of $10$ Hz (which is of the order of the lowest frequencies accessible to ground-based interferometers), the total duration of the inspiral phase is 17 min for solar-size objects.

A useful quantity for assessing the sensitivity of detectors to inspiraling binaries is the number of cycles spent in the detector bandwidth. The number of cycles in an interval $\mathrm{d}t$ small compared to the time of variation of $\omega$ is $\mathrm{d} \mathcal{N}_0 = \mathrm{d}\phi / \pi =  \omega \mathrm{d} t / \pi$ (the factor of $2$ accounts for the fact that the GW frequency is two times the binary frequency $\omega$) so that
\begin{equation}
\mathcal{N}_0 = \frac{1}{\pi} \int_{\omega_\mathrm{min}}^{\omega_\mathrm{max}} \mathrm{d}\omega \; \frac{\omega}{\dot \omega} \; ,
\end{equation}
and using Eq.~\eqref{eq:omegadot} one has
\begin{equation} \label{eq:number_cycles}
\mathcal{N}_0 \simeq 1.6 \times 10^4 \left( \frac{10 \mathrm{Hz}}{\omega_\mathrm{min}} \right)^{5/3} \left( \frac{1.2 M_\odot}{M_c} \right)^{5/3} \; ,
\end{equation}
where we have assumed $\omega_\mathrm{min} \ll \omega_\mathrm{max}$. This means that ground-based interferometers can follow the evolution of the signal for thousands of cycles, and space-borne interferometers like LISA can follow it for millions of cycles. However, a detector is sensible to at least one cycle: a difference of one oscillation is clearly visible in the signal. This means that, if there are relativistic corrections to Eq.~\eqref{eq:number_cycles}, we must take them into account as long as they give a correction of more than one cycle! A rough order-of-magnitude estimate is the following: since relativistic corrections to Eq.~\eqref{eq:number_cycles} go as $v^{2n}$ where $n$ is the post-Newtonian order, and that $v \sim 0.3$ close to coalescence, one should go to $n=3$ in order to reach an accuracy of approximately one cycle for ground-based interferometers! This intuition is confirmed by the exact PN computation. Skipping \textit{a lot} of details (going to higher order in Einstein's equations is clearly not as simple as a basic Taylor expansion), at the 3.5PN order the GW phase of the binary system for circular orbits $\phi = \int \mathrm{d}t \omega$ should be replaced by \cite{Blanchet:2013haa}
\begin{align} \label{eq:totalPhase}
\begin{split}
\phi_{3.5 \mathrm{PN}} &= \frac{x^{-5/2}}{32 \nu} \bigg\lbrace 1 + \bigg( \frac{3715}{1008} + \frac{55}{12} \nu \bigg) x - 10 \pi x^{3/2} + \bigg( \frac{15293365}{1016064} + \frac{27145}{1008} \nu + \frac{3085}{144} \nu^2 \bigg) x^2 \\
&+ \bigg( \frac{38645}{1344} - \frac{65}{16} \nu \bigg) \pi x^{5/2} \ln \bigg( \frac{x}{x_0} \bigg) + \bigg[ \frac{12348611926451}{18776862720} - \frac{160}{3} \pi^2 - \frac{1712}{21} \gamma_\mathrm{E} - \frac{856}{21} \ln(16x) \\
&+ \bigg( - \frac{15737765635}{12192768} + \frac{2255}{48} \pi^2 \bigg) \nu + \frac{76055}{6912} \nu^2 - \frac{127825}{5184} \nu^3 \bigg] x^3 \\
&+ \bigg( \frac{77096675}{2032128} + \frac{378515}{12096} \nu - \frac{74045}{6048} \nu^2 \bigg) \pi x^{7/2} + \mathcal{O}(x^4) \bigg\rbrace \; ,
\end{split}
\end{align}
where $\nu=m_1 m_2/(m_1+m_2)^2$ is the symmetric mass ratio, $x=(G (m_1+m_2) \omega)^{2/3}$ is a PN expansion parameter ($x \sim v^2$ in the PN expansion), $\gamma_\mathrm{E}$ is Euler's constant, and $x_0$ is an integration constant that can be fixed by the initial conditions when the wave frequency enters the detector.

\begin{figure}[ht]
\centering
\includegraphics[width=0.8\textwidth]{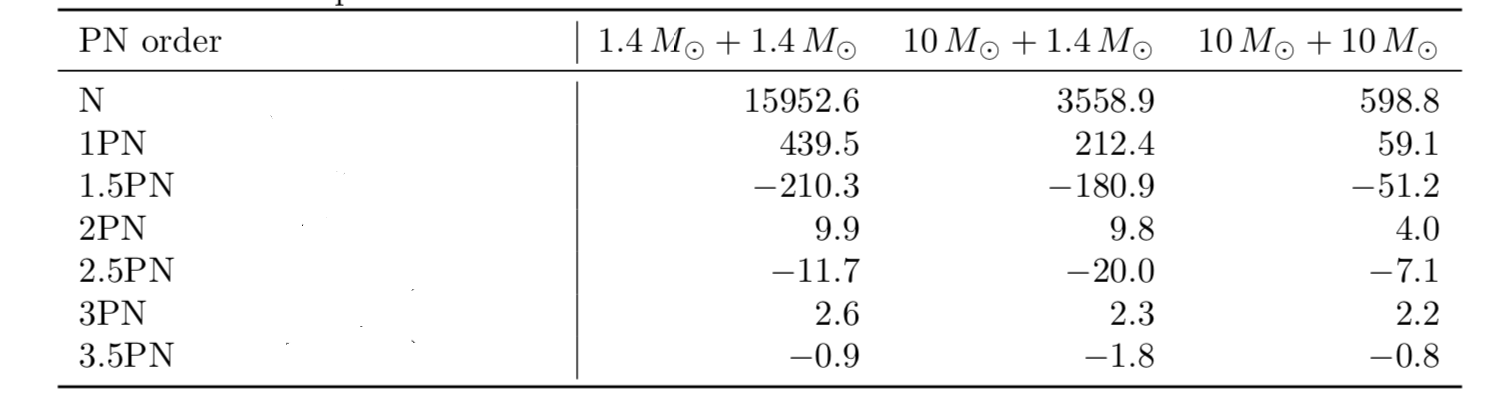}
\caption[Contributions of the different PN order to the total GW phase]{Contributions of the different PN order to the total phase~\eqref{eq:totalPhase} for different mass ratios. The lowest frequency is $f_\mathrm{min} = 10$Hz. Adapted from \cite{Blanchet:2013haa}. }
\label{fig:PN_contributions}
\end{figure}

A detailed presentation of the post-Newtonian expansion within an effective field theory framework will be the subject of Part~\ref{part2}. For the moment, let us just sketch its general idea. As we already emphasized many times, the PN formalism  is based on the expansion in the small parameter $v$, the velocity of the objects forming the binary system. More formally, $v^2$ is the factor between the size of the objects $r_s$ (which for black holes and neutron stars is well approximated by the Schwartzschild radius; in the EFT framework this is the high-energy cutoff of the theory) and the size of the orbit $r$,
\begin{equation}
v^2 \sim \frac{Gm}{r} \sim \frac{r_s}{r}
\end{equation}
Thus, from the PN point of view the expansions in the velocity $v$ and in the gravitational field strength $GM/r$ are formally the same.
At the same time, $v$ relates two IR scales: the orbital size $r$ and the period $T$ — equivalently, the wavelength $\lambda$ of the emitted gravitational waves,
\begin{equation}
v \sim \frac{r}{\lambda}
\end{equation}
What we are ultimately after are the two sides of the balance equation which we used to compute the GW phase, $\mathrm{d}E/\mathrm{d}t = -P$.
Thus in a first approximation, we might think that the problem of computing GW generation from a weakly self-gravitating source, in an expansion in $v$, has two aspects:
\begin{itemize}
\item We must determine the energy of the system to the required order in $v$. This is the 'conservative' aspect of the problem. This energy can be obtained by solving the equation of motion for $h_{\mu \nu}$, Eq.~\eqref{eq:linearizedEinsteinHarmonic}, appropriately generalized to include the nonlinear corrections stemming from the full expansion of the Einstein tensor in Eq.~\eqref{eq:EinsteinEqs}. One can then plug back the metric $h_{\mu \nu}$ in the action to obtain what is known as the Fokker action, i.e an action which depends only on the positions of the two objects we consider and from which one can derive all relevant quantities for the system.
\item Given their motion to desired order, we must compute the dissipated power $P$ which can be done using a multipole expansion. Thus, we cannot limit ourselves to the quadrupole formula but me must include a number of higher multipoles consistent with the order in $v$ which we wish to work.
\end{itemize}

In practice, the situation is far more complicated. The nonlinear structure of the Einstein equation is quite involved so that solving the equations becomes technically quite difficult. Furthermore, divergences appear in the process of obtaining the Fokker action (at lowest order, the metric already diverges at the position of the objects); a consistent treatment of these divergences involves the renormalization procedure. Finally, the two aspects which we highlighted cannot be separated beyond a certain level of approximation. Indeed, the emission of GW costs energy which is drained from the source so, beyond a certain order (namely 2.5PN in the GR case), GW will back-react on the matter sources, influencing their equations of motion. So, a fully-fledged formalism for computing systematically GW production of a self-gravitating source in powers of $v$ is necessarily quite complicated. This will be the subject of Part~\ref{part2}.

\subsection{Other approaches to the two-body problem} \label{subsec:other_approaches}

Before moving on, let us mention a few complementary analytical approaches to the two-body problem which aim at giving a better description of the waveform signal $h(t)$:
\begin{itemize}
\item \textit{Gravitational Self-Force (GSF) \cite{Poisson:2011nh, Quinn_1997, Mino_1997, Thornburg:2011qk}: } The PN formalism was based on an expansion for small velocities and gravitational fields. However, there is another physical situation where a perturbative expansion is relevant: this is when a small object is orbiting a much more massive one, like in the extreme mass ratio inspiral of a solar-size black hole around a supermassive one (this kind of events is among the targets of LISA). Then, one can set up a perturbation theory based on the small mass ratio. This is the GSF programme.

Let us dig a bit deeper into the details of this expansion. For simplicity, we consider a small point-particle object orbiting a massive, spinless \sch black hole. At lowest order in the mass ratio, the conservative part of the dynamics (the lhs in the energy balance equation~\eqref{eq:energyBalance}) is known exactly. This is because we can solve exactly for geodesics in \sch spacetime, so that in \sch coordinates the energy of a point-particle reads
\begin{equation}
E = \mu \frac{r-2GM}{\sqrt{r(r-3GM)}} \; .
\end{equation}
We will rederive this result for more generic spacetimes in Part~\ref{part4}, Section~\ref{sec:background_traj}. On the other hand, the dissipative part of the dynamics (the rhs in Eq.~\eqref{eq:energyBalance}) cannot be known exactly. However, it can be obtained from a perturbative expansion of the Regge-Wheeler (RW) equation which is the equation for small fluctuations of the \sch geometry (again, we refer the reader to Part~\ref{part4} for more details); since a second-order differential equation is much more tractable than full GR, accurate results can be obtained from a high-order PN expansion of the RW equation or from its numerical solution.

When one wants to describe the effects of radiation-reaction on the motion of the small body (so that its trajectory will deviate from a geodesic because it loses energy), though, one faces some difficulties. As already mentioned, the gravitational field is singular at the object location. To find the physical part of the force exerted on the object by radiation-reaction processes is a nontrivial task which the GSF programme addresses. In 1997 the motion of a point mass in a curved background spacetime
was investigated by Mino, Sasaki, and Tanaka \cite{Mino_1997}, who derived an expression for the particle’s acceleration; the same equations of motion were later obtained by
Quinn and Wald \cite{Quinn_1997} using an axiomatic approach. They are now known as the Mino, Sasaki, Tanaka, Quinn and Wald (MiSaTwQuWa) equations. Beyond this first-order computation, the field is still in active development.

\item \textit{Post-Minkowskian (PM) expansion \cite{Antonelli_2019, Cheung:2018wkq, PhysRevLett.122.201603, Bern_2019_2, Damour_2016_2, Damour_2018, damour2019classical}: } The strategy of the PM expansion can be phrased in one line: while the PN formalism relies on a weak-field and small-velocity expansion, the PM method is to expand for weak-field only, keeping the velocity arbitrary. This is particularly adapted to scattering processes where one considers two bodies with high velocities and small impact parameters; one can also relate the scattering observables to quantities related to the two-body problem \cite{Kalin:2019rwq}. However, one can also expect that the PM expansion is in some sense 'more exact' than the PN one in the case of a binary inspiral because it resums an infinite number of terms of the PN expansion (the ones involving a velocity expansion). A comparison between PM and PN Hamiltonians can be found in \cite{Antonelli_2019}.

The computational strategy adopted in the PM expansion is to consider a scattering process and compute observables using amplitudes methods, in a Feynman diagrams expansion similar to the one that we will use in Part~\ref{part2}. This is conceptually very simple, but in practice the gravity vertices entering Feynman diagrams are extremely complicated. Because of these technical difficulties, the PM computations are currently unable to reach the level of accuracy of the PN methods.

\item \textit{Numerical Relativity (NR) \cite{Eisenstein_2019, Grandcl_ment_2009, gourgoulhon200731}: } In the last stages of the binary inspiral, when the Einstein equations become highly nonlinear and all perturbative approximations break down, the only method yielding reliable results is to use numerical simulations. However, due to limited computing power, the evolution of binaries can be monitored only during a few cycles. This is why PN approximations and NR are complementary: the PN method gives access to the waveform during the large number of inspiral cycles, while NR covers the last stages of the dynamics when the two objects collide.

NR codes use the ADM decomposition of the metric (used also in the EFT of DE~\ref{Chapter2}) which allows for a reformulation of the Einstein equations as an initial-value problem \cite{gourgoulhon200731}. This is usually referred to the '3+1' approach because time and space are separated, giving up general covariance which is impractical for numerical computations. The ADM decomposition yields 10 equations: 6 of them are evolution equations which contains derivatives up to first order in time and second order in space. The remaining four equations, containing no time derivatives, are constraint equations. Once the constraints are satisfied initially, mathematically they will remain that way. But for
numerical solutions that is often not the case, especially
when significant non-linearities are present. Small numerical errors can exponentially grow. Keeping the constraints
satisfied at all times has proven essential to reaching stable, convergent solutions of the BH–BH coalescence problem. In that respect, the gauge choice is essential. The 1987 work of Nakamura, Oohara and Kajima, presented \cite{Nakamura:1987zz} a version of ADM that showed much better stability. Later, Shibata and Nakamura \cite{Shibata:1995we} (1995) and Baumgarte and Shapiro \cite{Baumgarte:1998te} (1998) confirmed and extended
those results. These efforts are commonly known as the
BSSNOK approach. It was essential to achieving full 3-
dimensional simulations of BH–BH coalescences and is
in wide use today. 

\item \textit{Effective One-Body (EOB) approach \cite{buonanno_effective_1999, DAMOUR_2008, damour2009effective}: } The EOB formalism is a convenient framework to recast analytical PN computations, so it is not exactly in the same category as the three other approaches already mentioned. In Newtonian mechanics, it is well-known that the two-body problem can be recast as the motion of a point-particle of reduced mass $\mu = m_1 m_2/(m_1+m_2)$ around a central mass $M=m_1+m_2$, thus allowing for an exact solution to the problem. This is obviously not true in GR, but the philosophy of the EOB approach is to make this idea 'as exact as possible'.  The basic claim is that the use of suitable resummation techniques should allow one to describe, by analytical tools, a sufficiently accurate approximation of the entire waveform, from inspiral to ring-down, including the nonperturbative plunge and merger phases. These resummation methods can be thought of using Padé approximants instead of the standard PN Taylor expansion. More precisely, instead of computing the relevant quantities of interest as a series expansion $c_0 + c_1 v + c_2 v^2 + \dots$, one rather uses rational or non-polynomial functions of $v$ such as $\sqrt{1-v^2}$, defined so as to incorporate some of the expected non-perturbative features of the exact solution. This is obtained by relating the PN-expanded dynamics to the motion of a test particle in some external spacetime geometry $\bar g_{\mu \nu}$; the matching is performed in such a way that the two dynamics (the one of the test-particle and the one of the binary problem) coincide up to the required PN accuracy.
 This strategy yields waveforms which compare surprisingly well with numerical simulations \cite{Buonanno:2007pf}.
\end{itemize}

\begin{figure}[ht]
\centering
\includegraphics[width=0.5\textwidth]{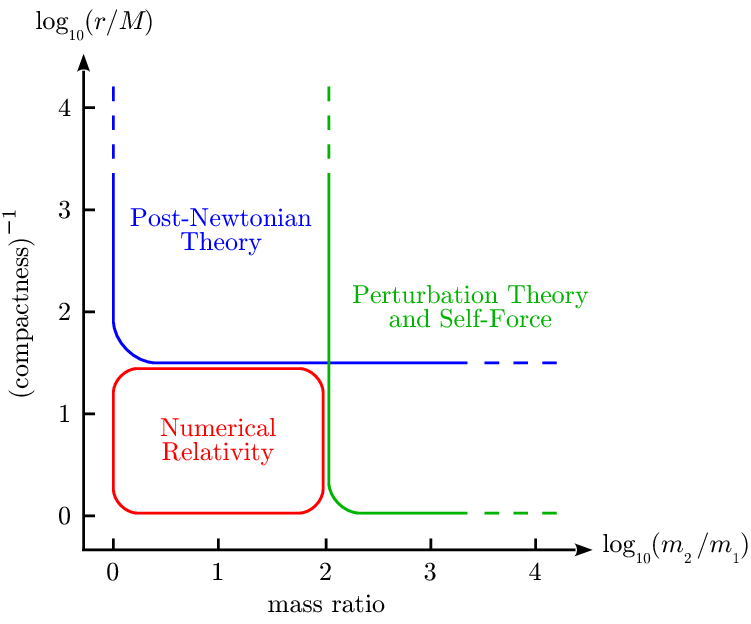}
\caption[The different techniques used to solve the two-body problem]{	Different analytical approximation schemes and numerical techniques are used to model the orbital dynamics and gravitational wave emission from black hole binaries, according to the mass ratio $0 < m_1/m_2 \leq 1$ and the compactness parameter $0 < M/r \lesssim 1$, where $M=m_1+m_2$ is the total mass and $r$ the typical binary separation. Adapted from \cite{Tiec:2014lba} }
\label{fig:methods_2body}
\end{figure}

\section{Tests of GR using GW} \label{sec:tests_GW}

We now come to the Section which is the most relevant to this thesis, that is: what do GW allow us to learn about gravity? Which kinds of tests can we devise which would confirm of falsify GR? Most of this thesis will be devoted to this question, but we will first recap here what is already known on the subject.

\paragraph{Binary pulsars:} To begin with, let us come back on the binary pulsar test of Section~\ref{subsec:binary_pulsar}. As we have said, observations are perfectly consistent with a quadrupolar emission of GW at the $0.2$\% level. However, scalar-tensor theories such as the ones presented in Chapter~\ref{Chapter2} generically predict \textit{dipole} radiation. We will derive this result for BDT theories in Part~\ref{part2}. Thus, scalar-tensor theories are generically disfavored by the binary pulsar test and this places strong constraints on the coupling of the scalar to matter (as we said in Chapter~\ref{Chapter2}, this coupling can be made naturally small by e.g a screening mechanism). 

\paragraph{Speed of gravity: }
The most dramatic constraint on modified gravity theories from GW came from the recent measurement of the speed of GW compared to that of light. On August 17, 2017 the Advanced LIGO and Advanced Virgo gravitational-wave detectors made their first observation of a binary neutron star inspiral~\cite{the_ligo_scientific_collaboration_gw170817:_2017}, GW170817. At the same time, an electromagnetic counterpart was observed in the form of a short gamma-ray burst~\cite{Goldstein_2017}. The difference in the arrival time of the two signals was 1.7s: since the waves had traveled over a distance of approximately $40$ MPc, this allowed to constrain the deviation of the speed of GW $c_\mathrm{GW}$ compared to that of light at the level $\vert c_\mathrm{GW}-1 \vert \lesssim 10^{-15}$ (we recall that we have set $c=1$). However, scalar theories like the Horndeski class generically predict deviations of $c_\mathrm{GW}$ from the speed of light~\cite{Gleyzes:2013ooa,Jimenez:2015bwa,Lombriser:2015sxa,Bettoni:2016mij}; this can be understood from the fact that, since they model cosmological acceleration by a scalar field which acts as a medium spontaneously breaking Lorentz symmetry on a cosmological background,  there
is no a priori reason to expect that gravitational waves, which
are an excitation of this medium, travel at the same speed as
photons.

This feature can easily be seen from the EFT of DE presented in Chapter~\ref{Chapter2}. Indeed, recall that the extrinsic curvature $K_{\mu \nu}$ contains a time derivative of the induced metric from eq.~\eqref{eq:KijADM}. Thus, an operator such as $m_4$ in Eq.~\eqref{eq:ActionEFTQuadratic}, which we rewrite here for clarity,
\begin{equation}
m_4^2(t) \big( \delta K^2 - \delta K^{\mu \nu} \delta K_{\mu \nu} \big) \; ,
\end{equation}
detunes the kinetic term of the graviton from its gradient term and so modifies the speed of propagation of GW. Thus, the GW170817 measurement imposes $m_4^2 = 0$. This is not the end of the story, as one should impose $m_4^2 = 0$ robustly against small variations of the background against which the EFT perturbation are defined. Ref.~\cite{Creminelli:2017sry} showed that it imposes constraints on operators cubic in the perturbations (which we did not write in Eq.~\eqref{eq:ActionEFTQuadratic}), so that the EFT free functions are greatly reduced. Translating this fact in the covariant (beyond) Horndeski approach, they showed that the freedom in the quartic and quintic sector goes from 4 independent terms (listed in Eqs.~\eqref{eq:L_beyond_horndeski}~-~\eqref{eq:L_horndeski}) to only one free function! This kills an impressive list of modified gravity models~\cite{Creminelli:2017sry,Ezquiaga:2017ekz,Baker:2017hug,Sakstein:2017xjx}.  The consequences of this event on the Vainshtein mechanism in scalar-tensor theories  have
been discussed in ~\cite{Crisostomi:2017lbg,Langlois:2017dyl,dima_vainshtein_2018}. 

A possible caveat of this approach is that, as all EFTs, the EFT of DE has a high-energy cutoff which is uncomfortably close to the regime of energy probed by GW events, $\Lambda_\mathrm{cutoff} \sim (\mpl H_0^2)^{1/3} \sim (1000 \mathrm{km} )^{-1}$. One could imagine that the GW speed goes to unity in this high-energy regime, as was argued in~\cite{deRham:2018red}. However, finding an explicit model featuring this GW speed transition at high energy proves to be very difficult.

\paragraph{GW decay into scalars, and dark energy instabilities: } For the same reason that the speed of gravity could differ from unity in a lorentz-breaking medium, one can have interactions between gravitons and the scalar responsible for Dark Energy in modified gravity models. This opens up the possibility that a GW $h$ decay into two scalars $\varphi$, $h \rightarrow \varphi \varphi$. Ref.~\cite{Creminelli:2018xsv} studied the decay rate and found that for typical values of the EFT of DE parameters it leads to a huge suppression of GW signal (essentially, we would not see any GW because they would decay before reaching us). The surviving beyond Horndeski theories contain only the two free functions $G_2$ and $G_3$, i.e.
\begin{equation} \label{eq:Lct1nodecay}
L_{c_\mathrm{GW}=1 \mathrm{, no \; decay}} = f(\varphi) R + G_2(X, \varphi) + G_3(X, \varphi) \square \varphi \; ,
\end{equation}

Finally, the same authors studied another mechanism under which even the $G_3$ function in~\ref{eq:Lct1nodecay} is pathological. Indeed, the scalar that describes dark-energy
fluctuations features ghost and/or gradient instabilities for gravitational-wave amplitudes
that are produced by typical binary systems~\cite{Creminelli:2019kjy}. Ensuring the absence of such an instability means that the cosmological effects of the cubic Horndeski function are negligible, so that the only surviving dark energy theory is the K-Essence $G_2$, which is a dramatic simplification of the freedom of the Horndeski class!

\paragraph{Gravity in the strong-field regime: }

The recent direct detection of GW has enable further tests of GR in the (largely unprobed up to now) strong-field regime~\cite{TheLIGOScientific:2016src}. Indeed, most of the tests of GR carried out up to now were in the low-velocity, weak-field, linear regime. The LIGO/Virgo collaboration have performed several tests on the GW150914 event, a binary black hole merger with observed signal-to-noise ratio of $\sim 24$. Up to now, the main lessons that we have learned from GW data are:
\begin{itemize}
\item \textit{Inspiral–merger–ringdown consistency test: } One can try to evaluate the parameters of the system (masses and spins) from different parts of the waveform, namely its low-frequency and high-frequency parts. These different predictions are in accordance together.
\item \textit{Constraining parameterized deviations from GR inspiral–merger–ringdown waveforms: } The idea of this test is to allow for an independent variation of the PN parameters in order to check if they are compatible with GR. More precisely, one writes a generic formula for the GW phase~\eqref{eq:totalPhase} of the form~\cite{Yunes_2009, Cornish:2011ys}
\begin{equation} \label{eq:variation_PN_coeffs}
\phi = \phi_0 + \frac{x^{-5/2}}{32 \nu} \big(  \hat \varphi_0 + \hat \varphi_1 x^{1/2} + \hat \varphi_2 x + \dots \big)
\end{equation}
and one constrains the parameters $\hat \varphi$ using data. The results up to the $3.5$PN order are shown in Figure~\ref{fig:PNVariationCoeffs}: although the contraints are quite loose (the accuracy on $\hat \varphi$ is $\mathcal{O}(1)$), the data points towards no deviation from GR. Of course, it would be way more satisfying to directly compare data against waveforms in other theories than GR, however waveforms in modified gravity are poorly known up to now. In Part~\ref{part4}, we will advocate for an unifying EFT formalism giving access to waveforms in a large class of modified gravity models.
\begin{figure}[ht]
\centering
\includegraphics[width=0.5\textwidth]{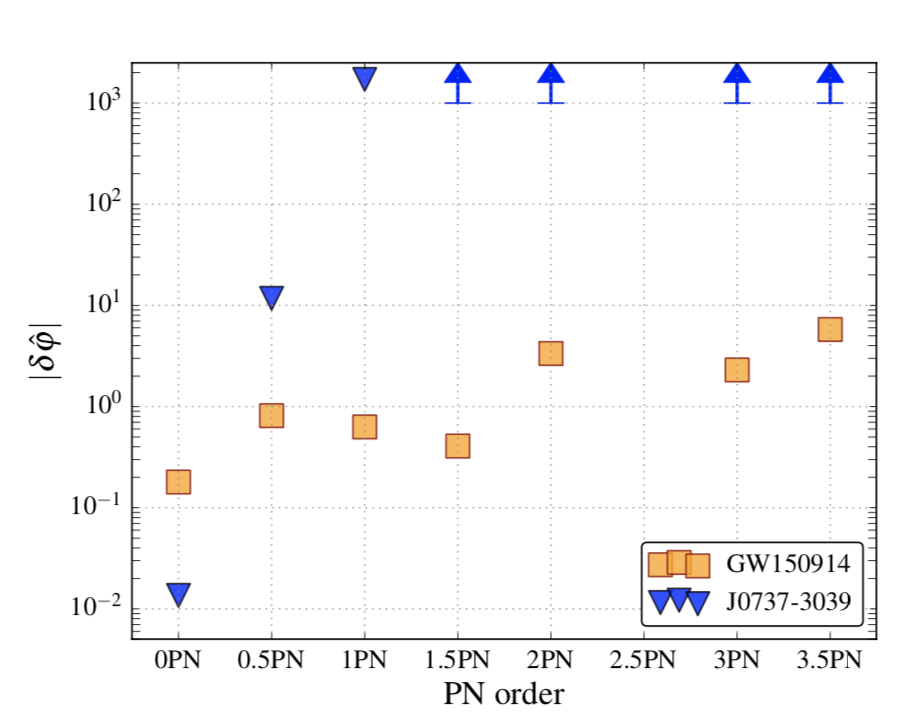}
\caption[Experimental bounds on the PN coefficients]{	90\% upper bounds on the fractional variations of the PN coefficients with respect to their GR value (orange squares). As a comparison, the blue triangles show the 90\% upper bounds extrapolated exclusively from the measured orbital period derivative of the double pulsar J0737-3039. This pulsar measurement is constraining only the lowest PN orders, showing that GW offer a unprecedented test of gravity in the strong-field regime. The deviation of the 2.5PN coefficient is not measurable since it is degenerate with the reference phase. Taken from~\cite{TheLIGOScientific:2016src} }
\label{fig:PNVariationCoeffs}
\end{figure}
\item \textit{Searching for the QNM damping modes: } Just after the merger of two black holes, one expect that the GW signal will look like the one of a perturbed single black hole. The oscillations of black holes are described by the Regge-Wheeler and Teukolski equations (see Part~\ref{part4}); the physical prediction is that the quasi-normal modes (QNMs) of oscillation decay over time (so that black holes are stable objects) with a GW signal of the form
\begin{equation}
h(t) \sim e^{-t/\tau} \cos \omega_0 t \; ,
\end{equation}
where we have only taken into account the mode with the quickest decay (data do not allow yet to distinguish between this fundamental mode and higher harmonics). The values of $\omega_0$ and $\tau$ found by data analysis are consistent with GR. In the future, an interferometer like LISA will allow to measure several QNMs; this is sometimes referred to as 'black hole spectroscopy'~\cite{Berti_2018, Berti_2006}.

\end{itemize}

\paragraph{Supplementary polarizations: } We have seen that in GR there are only two polarizations $h_+$ and $h_\times$ of the GW signal. However, modified gravity theories predict the existence of supplementary polarizations (we will see in Part~\ref{part2} that BDT theories predict one supplementary polarization; in total, there could be up to six different polarizations). This can be seen from Eq.~\eqref{eq:motion_riemann}: the motion of test masses depend on the symmetric rank-2 tensor $R_{i0j0}$ which has six degrees of freedom. In generic theories generalizing GR, all these six DOFs can be present. For example, Einstein-Aether theory generically predicts all six modes~\cite{Jacobson_2004}. For a wave propagating in the $z$ direction, they can be displayed by the matrix~\cite{Will_2014}
\begin{figure}[ht]
\centering
\includegraphics[scale=0.3]{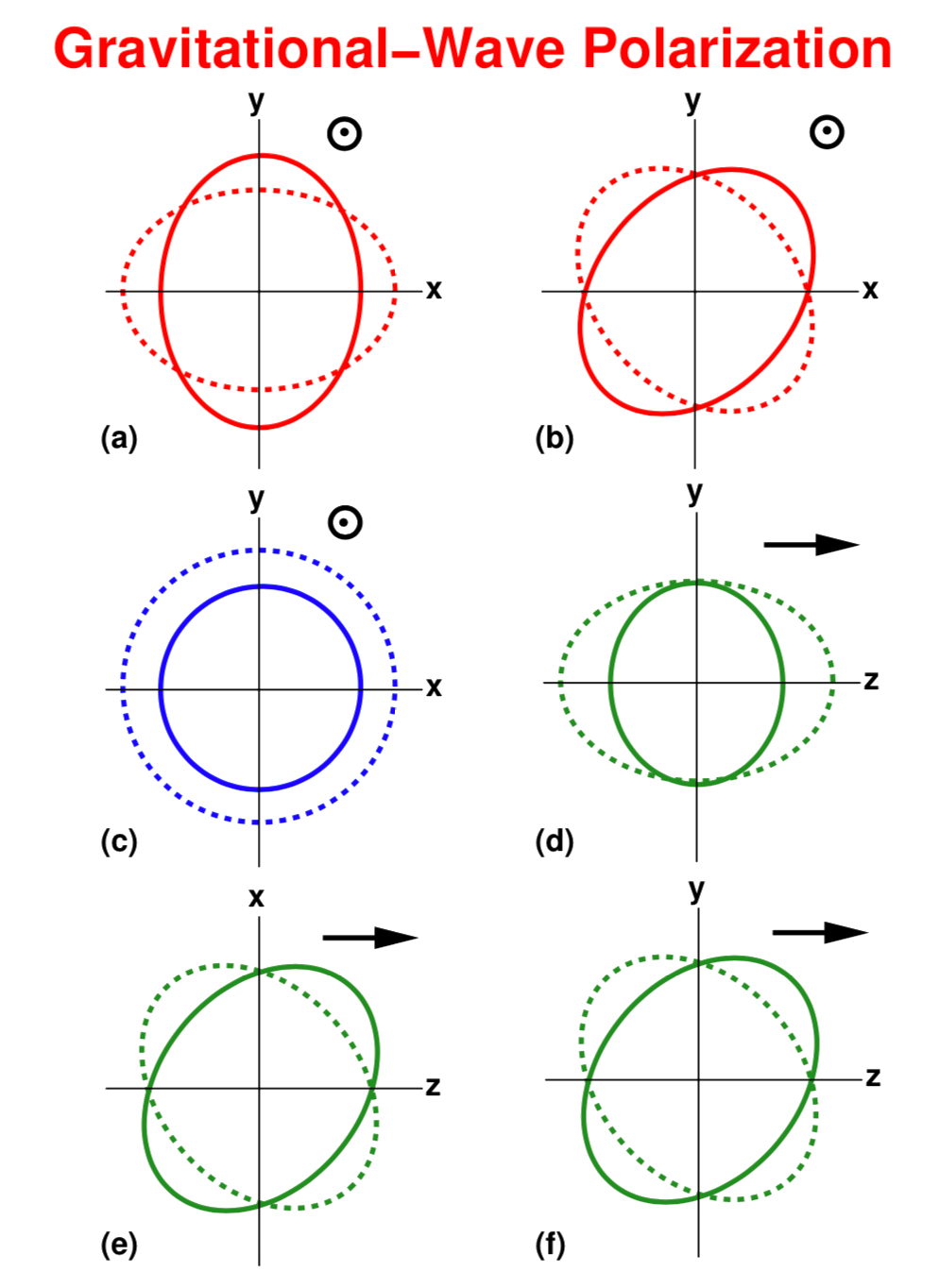}
\caption[Different GW polarizations in a generic gravity theory]{	 The six polarization modes for gravitational waves allowed in a generic metric theory of gravity. We have shown the displacement that each mode induces on a ring of test particles as in Figure~\ref{fig:testmasses}. The wave propagates in the $+z$ direction. There is no displacement out of the plane of the picture. In (a), (b), and (c), the wave propagates out of the plane; in (d), (e), and (f), the wave propagates in the plane. In GR, only (a) and (b) are present; in massless scalar–tensor gravity, (c) is also present. Figure taken from Ref.~\cite{Will_2014} }
\label{fig:6_GW_pola}
\end{figure}
\begin{equation}
h_{ij} =
\begin{pmatrix}
h_\phi + h_+ & h_\times & h_{V1} \\
h_\times & h_\phi - h_+ & h_{V2} \\
h_{V1} & h_{V2} & h_L
\end{pmatrix}_{ij} e^{i\omega (t - z)} \; .
\end{equation}
Three modes ($h_+$, $h_\times$ from GR and $h_\phi$) are transverse to the direction of propagation; the third "breathing" scalar mode is typical of massless scalar-tensor theories, as we will see in Part~\ref{part2}. The three other modes are longitudinal, with one ($h_L$) an axially symmetric stretching mode in the propagation direction, and one quadrupolar mode in each of the two orthogonal planes containing the propagation direction ($h_{V1}$ and $h_{V2}$). Figure~\ref{fig:6_GW_pola} shows the displacements induced on a ring of freely falling test particles by each of these modes.

However, detecting these supplementary polarizations would necessitate to have several GW detectors with different orientations, so that for the LIGO GW150914 this test was not possible. Indeed, with six different orientations one can infer the six components of $R_{i0j0}$ appearing in Eq.~\eqref{eq:motion_riemann} if the direction of the even is known by another measurement, for example by an electromagnetic observation or by time-of-flight measurements between separated detectors. While we are lacking such an array of detectors for the moment, such tests will become possible in the future when several GW detectors will be operating at the same time~\cite{Abbott_2018}.

\part{The two-body problem: an Effective Field Theory viewpoint} \label{part2}


\chapter{The two-body problem in scalar-tensor theories} 

\label{Chapter4} 

In the first part of this thesis, we have given a generic overview of what we currently know on gravity from the three main physical systems in which it plays a fundamental role: the Solar system, cosmology, and gravitational waves (GW). In particular, the emergence of GW astronomy gives a solid motivation for the study of the generation of GW in other theories than General Relativity (GR).
In this Chapter, we will generalize to the well-known Brans-Dicke Type (BDT) theories \footnote{ We presented these theories in Section~\ref{subsec:BDT}; they are called scalar-tensor theories in the literature, but we will refrain to do so for reasons explained in~\ref{subsec:BDT}} the Effective Field Theory (EFT) approach to the two-body problem first advocated by Goldberger and Rothstein in~\cite{goldberger_effective_2006,Goldberger:2007hy}, dubbed Non-Relativistic General Relativity (NRGR) (see also~\cite{Cardoso:2008gn,galley_radiation_2009} and the interesting extension discussed in~\cite{Endlich:2017tqa}). The reader is particularly encouraged to consult the reviews~\cite{Foffa:2013qca,porto_effective_2016,Levi:2018nxp} for a thorough and pedagogical introduction to this formalism, which has become over the years a solid alternative to the traditional post-Newtonian formalism. The perturbative computations of the post-Newtonian approximation have been successfully translated in a series of Feynman diagrams. The current state-of-the art computation in the conservative sector of NRGR is at the 4PN level~\cite{Foffa_2013, Foffa_2017, 2019PhRvD.100b4047F,Foffa_2019}; furthermore NRGR has been generalized to include spin and this yielded results which up to date have not been obtained by any other methods~\cite{Porto_2006, Levi:2015msa, Levi:2020kvb, Levi:2014gsa, Levi:2015ixa, Levi:2016ofk}. In this Chapter, though, we will focus on non-spinning objects.

 This Chapter, based on the JCAP paper~\citeK{Kuntz:2019zef}, introduces a certain number of tools and techniques which will be of great use in the rest of this thesis. Although we will focus on the particular BDT theories for simplicity, most of our results will also hold true in any scalar-tensor theory of gravity so that we will use them again in the next Chapters.
Although we use different methods to obtain them, our results 
overlap with some classic works of Damour and Esposito-Far\`{e}se, where the standard Post-Newtonian formalism is applied to (multifield) BDT theories~\cite{damour_tensor-multi-scalar_1992, Damour:1995kt}\footnote{It should be noted that~\cite{Damour:1995kt} also uses a field theory language (in position space) in the conservative sector of the dynamics. Our approach complements this work by enforcing the power-counting rules directly at the level of the diagrammatic expansion.}. When possible, we compare with the more recent Ref.~\cite{Huang:2018pbu}, where NRGR is used to study the effects of a light axion. . For recent developments on the two-body problem 
in BDT theories see also~\cite{Julie:2017ucp,Julie:2017pkb}, where the conservative dynamics  have been extended to the Effective-One-Body framework at 2PN order, and~\cite{Bernard:2018hta, Bernard:2018ivi} where the equations of motion have been derived  up to 3PN order. See ~\cite{Lang:2015aa,Lang:2013fna} for the tensor and scalar waveform calculations, respectively at 2PN and 1.5PN order.

Note that we use a different notation than~\cite{goldberger_effective_2006}, in that our Planck mass is related to the (bare) Newton constant by $\mpl^2 = {1}/({8 \pi G})$---instead of $m_P^2 = {1}/({32 \pi G})$---and our metric signature is $(- + + +)$. This will make some factor of 4 appear in the graviton propagator and will induce some sign differences.

\section{An EFT approach to the two-body problem in ST theories}

To describe the binary system coupled by gravity and the emitted gravitational wave we use the action
\be
\label{totaction}
S  =  S_{\rm grav} + S_{\rm pp} \;,
\ee
where $S_{\rm grav}$ governs the dynamics of the gravitational and scalar degrees of freedom and is given by
\begin{equation} \label{eq:firstaction}
S_{\rm grav} = \:  \int d^4x \sqrt{-g} \left(\frac{\mpl^2}{2} R - \frac{1}{2}  g^{\mu \nu}\partial_\mu \phi \partial_\nu \phi \right)\, ,
\end{equation}
while $S_{\rm pp}$ is the point-particle action describing the motion of the two inspiralling objects, labelled by $A=1,2$, 
\begin{equation} 
\label{eq:action}
S_{\rm pp} = -\sum _{A=1,2} m_A \int d\tau_A \left[ 1 - \aaa_A \frac{\phi}{\mpl}  - \bbb_A \left(\frac{\phi}{\mpl}\right)^2 +\dots \right] \, ,
\end{equation}
where $d\tau_A^2 = - g_{\mu \nu} d  x_A^\mu d x_A^\nu$ is the proper time of each object. The term in square brackets can be seen as originating from a Taylor expansion of a ``$\phi$-dependent mass'' $m_A(\phi)$~\cite{Eardley1975ApJ}. We now discuss why we restrict to this action and some ways to extend it.

\subsection{The gravitational action}

While trying to extend the NRGR formalism to dark energy/scalar field models we immediately encounter an obstruction. 
It is well known that, to pass Solar System tests, modified gravity theories   display {\it screening} mechanisms that make the scalar interactions weaker in high density environments. For instance, most theories belonging to the Galileon/Horndeski and ``beyond Horndeski" classes  have a rich structure of non-linear terms in their Lagrangians that become more important close to the sources, thereby screening the effects of the scalar fluctuations. We have seen the example of a cubic Galileon in Chapter~\ref{Chapter2}. While it is legitimate to neglect such non-linearities on the largest cosmological scales, they are expected to play a major role in the vicinity of the binaries, causing a breakdown of the perturbative expansion that we use in this Chapter.

Another, related, complication in dealing with dark energy is represented by the spontaneously breaking of Lorentz symmetry. 
A time-evolving background scalar field allows, in the action that governs cosmological perturbations, all sorts of terms that break boosts and time-translations. This is made particularly explicit in the {\it EFT approach} to inflation  and dark energy~\ref{sec:EFTDE}. Let us consider, as an example, the Nambu-Goldstone boson of some broken $U(1)$ symmetry in Minkowski space, 
\begin{equation}
\label{eq:kmoufla}
{\cal L} = - \partial_\mu  \phi \partial^\mu \phi + \frac{(\partial_\mu  \phi \partial^\mu \phi)^2}{\Lambda_*^4} + \dots \, , 
\end{equation}
where $\Lambda_*$ is a mass scale. When this theory is expanded around a homogeneous background configuration $\phi (t)$, which is always a solution in Minkowski, the fluctuations of the field develop a speed of propagation $c_s\neq1$---a very tangible sign of spontaneous breaking of Lorentz symmetry already at quadratic order. 
At the same time, the non-linear terms suppressed by the scale $\Lambda_*$  become important close to the source, giving rise, in this case, to  the ``$k$-mouflage" mechanism~\cite{Babichev:2009ee}, which is a variant of the screening mechanisms discussed above. 

As a full treatment of these  non-linearities is beyond the scope of this Chapter (we will make some steps in this direction in Part~\ref{part3}), we focus on a very standard scalar-tensor action, i.e. the BDT theory in~eq.~\eqref{eq:firstaction}. 
This action is in the so-called ``Einstein frame'' form: possible non minimal couplings between the scalar and the metric fields have been reabsorbed with a field redefinition of the metric and transferred to the matter sector, i.e., to the point-particle action. Moreover, motivated by the dark energy role of the scalar field, we assume it to be effectively massless. The case of a massive axion has been studied in the effective field theory approach in~\cite{Huang:2018pbu}. 

\subsection{The point-particle action} \label{subsec:pp_action}

Let us now motivate the point-particle action ~\eqref{eq:action}. The size of the objects that are orbiting around each other represents our UV scale, so they can be effectively described as point particles. The couplings of a massless spin-2 field are famously constrained by gauge invariance~\cite{Weinberg:1964ew}. As a result, the presence of a scalar provides us with a  richer structure of possible point-particle couplings.

In this Chapter we focus on {\it conformal couplings}, obtained by assuming that matter couples to the gravitational metric multiplied by a general function of the scalar field. Let us point out that there are more generic couplings involving the scalar field. These \textit{disformal} couplings introduced in Section~\ref{subsec:intro_disf} will be studied in Chapter~\ref{Chapter5}.
We have included up to terms quadratic in $\phi$ because, as we will see, higher-order terms become important only at higher order in the Post-Newtonian expansion. Moreover, we have allowed different scalar couplings for different particles because such couplings are not protected against renormalization. Let us explain shortly why this is the case.

For example, let us consider  the actual field theory describing the matter inside the object. Such a ``UV model" might well enjoy a universal scalar coupling of the type 
\begin{equation}
S_{\rm UV} \supset \int d^4x\  \alpha \ \frac{\phi}{M_P} T_{\rm m},
\end{equation}
$T_{\rm m}$ being the trace of the energy momentum tensor of the matter fields. Such a universal coupling is indeed radiatively stable under corrections coming from the sole matter sector (see e.g.~\cite{Hui:2010dn,Nitti:2012ev}). However, the matching into the EFT point-particle action~\eqref{eq:action}  inevitably contains details about the actual shape and density of the body under consideration. For example, as a body becomes more and more self-gravitating, its scalar charge  decreases, down to the point of disappearing when it becomes a black hole (see e.g. the nice discussion in~\cite{Hui:2009kc}). Equivalently, if we started with a universal EFT model~\eqref{eq:action} with some bare mass parameters $m_{\text{bare},A}$ and universal couplings $\alpha_{\text{bare},1} = \alpha_{\text{bare},2} = \alpha_\text{bare}$ (and $\beta_{\text{bare},1} = \beta_{\text{bare},2} = \beta_\text{bare}$),  we can make the corrections to the mass finite by imposing a hard cutoff $\Lambda$ in momentum space, which roughly corresponds to considering a body of size $\Lambda^{-1}$. We get
\begin{equation}
m_A (\Lambda) \ = \  m_{\text{bare},A} + \delta m_A(\Lambda)\, ,
\end{equation}
where $\delta m_A(\Lambda)$ represents the (negative) gravitational energy of the body. The explicit calculation is done in Sec.~\ref{sec2.2}. As we show there, the scalar charge of the body renormalizes in a way that is not universal but actually depends on $\delta m_A(\Lambda)$ (see equation~\eqref{eq:deltaalpha}). 
This result is often stated by saying that a scalar fifth force can satisfy the weak equivalence principle (universality of the free fall for test particles) but not the strong one (universality of the free fall for bodies of non-negligible gravitational self-energy). It is believed that the latter is satisfied only by a purely metric theory as GR~\cite{Will:1993ns}.
It is therefore important to allow different scalar couplings for different objects at the level of the EFT, with the understanding that, in most cases, such a charge is zero for a black hole.

Let us now attempt a systematic discussion of such couplings. Before starting, it is useful to recall the procedure used in GR. As we already mentioned, the cutoff of the EFT is the size of the objects at which the energy scale expansion breaks down. Namely, finite-size effects will enter the EFT as powers of $k r_s$, where $r_s \simeq G m$ is the size of the objects (since we consider compact objects, it is close to their \sch radius) and $k$ the energy scale at which we are probing the EFT. To find the operators associated to such effects (built out of the metric $g_{\mu \nu}$ and the particle's trajectories $x_A^\mu(\lambda_A)$ where $\lambda_A$ is an affine parameter), it is useful to recall the two symmetries under which the point-particle action should be invariant:
\begin{itemize}
\item \textit{General coordinate invariance} $x^\mu \rightarrow \tilde{x}^\mu(x)$
\item \textit{Worldline reparametrization invariance}, i.e changes of the affine parameter $\lambda_A \rightarrow \tilde \lambda_A(\lambda_A)$
\item $SO(3)$ invariance: this guarantees that the compact object is perfectly spherical and spinless.
\end{itemize}

It is straightforward to write down effective Lagrangians that are invariant under these symmetries. To take care of coordinate invariance, we just write down Lagrangians that transform as coordinate scalars constructed from $g_{\mu \nu}$ and $d x_A^\mu / d \lambda_A$. A simple way of ensuring reparametrization invariance is to use the proper time variable
\begin{equation}
d\tau_A^2 = - g_{\mu \nu} \mathrm{d} x_A^\mu \mathrm{d} x_A^\nu
\end{equation}
as the worldline parameter. Since proper time is physical (i.e., measurable) it must be invariant under worldline reparametrizations. At lowest order, the simplest point-particle action that one can write involves the operator
\begin{equation} \label{eq:pp_action_GR}
\mathcal{O}_1 = -m_A \int \mathrm{d} \tau_A \; ,
\end{equation}
where $m_A$ is the particle mass. However, at a higher order in a derivative expansion one can use also the Riemann tensor $R_{\mu \nu \rho \sigma}$. At linear order in the Riemann tensor, there are two nonminimal operators allowed, namely
\begin{equation} \label{eq:lowest_order_correction_pp}
\mathcal{O}_2 = c_R \int \mathrm{d} \tau_A R  \; , \quad \mathcal{O}_3 = c_V \int \mathrm{d} \tau_A R_{\mu \nu} \dot{x}_A^\mu \dot{x}_A^\nu \; .
\end{equation} 
However, these operators are \textit{redundant} in the sense that they can be eliminated because they are proportional to the lowest-order equations of motion in vacuum, $R_{\mu \nu} = 0$. Technically, this comes from the fact that one can make a field redefinition of the metric, $g_{\mu \nu} \rightarrow g_{\mu \nu} + \delta g_{\mu \nu}$ with
\begin{equation}
\delta g_{\mu \nu} = \frac{1}{2 \mpl^2} \int \mathrm{d} \tau_A \frac{\delta^4(x - x_A(\tau_A))}{\sqrt{-g}} \bigg[ - \big( \xi_R - \frac{1}{2} \xi_V \big) g_{\mu \nu} + \xi_V \dot{x}_A^\mu \dot{x}_A^\nu \bigg] \; .
\end{equation}
When plugged into the gravitational action this redefinition induces the shifts $c_{R,V} \rightarrow c_{R,V} + \xi_{R,V}$. Thus by adjusting $\xi_{R,V}$ one can make the coefficients $c_{R,V}$ vanish. Similarly, every operator containing the Ricci tensor could be set to zero, so that all the information is encoded in the Weyl tensor. The curvature tensor is thus decomposed into its electric and magnetic components, namely
\begin{align}
\begin{split}
E_{\mu \nu} &= R_{\mu \alpha \nu \beta} u^{\alpha} u^\beta \; , \\
B_{\mu \nu} &= \frac{1}{2} \epsilon_{\alpha \beta \gamma \mu} R^{\alpha \beta}_{\delta \nu} u^\gamma u^\delta \; .
\end{split}
\end{align}
Thus the lowest-order operators which contribute beyond the point-particle coupling~\eqref{eq:pp_action_GR} are
\begin{equation} \label{eq:pp_nonminimal_weyl}
\mathcal{O}_4 = c_E \int \mathrm{d} \tau_A E_{\mu \nu} E^{\mu \nu}  \; , \quad \mathcal{O}_5 = c_B \int \mathrm{d} \tau_A B_{\mu \nu} B^{\mu \nu} \; .
\end{equation}
The $c_E$ and $c_B$ coefficients are related to the tidal Love numbers of compact objects, which express their deformability under external tidal fields~\cite{Binnington_2009, Damour_2009, Hinderer_2008}
From these expression it is easy to deduce the \textit{effacement theorem} for compact objects: finite-size effects enter at the 5PN order for nonspinning objects. To see this, recall that by splitting $g_{\mu \nu} = \eta_{\mu \nu} + h_{\mu \nu}$ one has the Newtonian scaling $h \sim r_s/r$. Then one can write
\begin{equation}
\frac{\mathcal{O}_4}{\mathcal{O}_1} \sim \frac{c_E \int \mathrm{d} t \; \partial^4 h^2}{m \int \mathrm{d} t \; h} \sim \frac{c_E}{m} \frac{r_s}{r^5}
\end{equation}
The only scale we have to build $c_E$ is the size of the object $r_s$. Thus, the operator $\mathcal{O}_4$ is a $(r_s/r)^5$ correction to the lowest-order operator, i.e a 5PN correction. We can also mention that, beyond the first nonminimal couplings~\eqref{eq:pp_nonminimal_weyl} related to finite-size effects, one can also add operators expressing the absorption of gravitational energy by the horizons of black holes, see e.g.~\cite{Goldberger_2006}.

Let us now come to the case of BDT theories. In a similar fashion, all terms involving the scalar are allowed in the point-particle action: the conformal coupling in Eq.~\eqref{eq:action} is one of them. Furthermore, the trick which we used in GR to set the lowest-order correction~\eqref{eq:lowest_order_correction_pp} to zero is not valid any more. This is because the equations in vacuum of the BDT theory~\eqref{eq:firstaction} is
\begin{align}
\begin{split}
\mpl^2 G_{\mu \nu} &=  \partial_\mu \phi \partial_\nu \phi - \frac{1}{2} g_{\mu \nu} (\partial \phi)^2 \; , \\
\square \phi &= 0 \; ,
\end{split}
\end{align}
where $G_{\mu \nu}$ is the Einstein tensor. Since the spacetime is not Ricci flat \footnote{It would still be the case for a static black hole since in this case the no-hair theorem applies~\cite{hawking1972}, but this would certainly not be true for neutron stars to which the scalar is expect to couple}, one cannot set the lowest-order operators~\eqref{eq:lowest_order_correction_pp} to zero; instead one is only allowed to trade the Ricci tensor for the scalar, so that the first nonminimal finite-size effects operators are
\begin{equation}
\mathcal{O}_6 = c_\phi \int \mathrm{d} \tau_A (\partial \phi)^2 \; , \quad \mathcal{O}_7 = c_D \int \mathrm{d} \tau_A \big( \partial_\mu \phi \dot{x}^\mu \big)^2 \; .
\end{equation}
An interesting remark is that $\mathcal{O}_7$ can also be viewed as arising from a disformal coupling of the scalar to matter. We will investigate in details the effects of such an operator in Chapter~\ref{Chapter5}. On the other hand, the operator $\mathcal{O}_6$ gives rise to the lowest-order finite-size effect in ST theories. By a similar reasoning, it can be found to be a 3PN correction to the lowest-order action. This 'dipolar tide effect' was recently computed in the standard PN formalism in~\cite{Bernard:2019yfz}. It should also be noted that a very similar reasoning has been discussed in Appendix A of Ref.~\cite{Damour_1998}.

\subsection{Outline}

With respect to GR, BDT theories bring new conceptual ideas and we sum up here the main results of this Chapter:
\begin{itemize}
\item While in GR the strong equivalence principle (SEP) prevails which means that strongly self-gravitating bodies fall in a universal way in an external field, in ST theories the SEP is violated and this can be tested using Lunar Laser Ranging (see Chapter~\ref{Chapter1}). We show in Section~\ref{sec2.2} that this SEP violation is intrinsically related to the scalar charge renormalization, which contrary to mass renormalization is not universal. This can directly be seen from the point-particle action~\eqref{eq:action}: the variation of the GR proper time gives the geodesic equation,
\begin{equation}
\delta \bigg( - m_A \int \mathrm{d} \tau_A \bigg) = 0 \quad \Leftrightarrow \quad \ddot x_A^\mu + \Gamma^\mu_{\alpha \beta} \dot x_A^\alpha \dot x_A^\beta = 0 \; ,
\end{equation}
in which the mass $m_A$ (be it bare or renormalized) factors out. On the other hand, the variation of a BDT point-particle action like Eq.~\eqref{eq:action} will generically give trajectories which depend on the scalar coupling $\aaa_A$.

\item The analysis of the conservative sector of the dynamics in Section~\ref{sec:EIH_Lagrangian} shows that at 1PN order all deviations from GR can be encoded in two PPN parameters, $\tilde \gamma$ and $\tilde \beta$, which are constrained by Solar system experiments as discussed in Chapter~\ref{Chapter1}.

\item Scalar-tensor theories generically predict monopole and dipole radiation which are enhanced in the PN expansion with respect to the GR quadrupole. However, the lowest-order monopole does not radiate since it is simply proportional to the total mass of the system $m=m_1+m_2$ which is constant, so that radiation starts at the dipolar order. We will make this statement more precise in Sections~\ref{sec:couplings} and~\ref{sec:dissipative_dynamics}. This could alter the dynamics of binaries.

\item Finally, two other interesting differences with respect to GR is that there is one additional scalar polarization, as already mentioned in Section~\ref{sec:tests_GW}; and that the frequency dependence of a dipolar wave is $\omega$, not $2 \omega$ like in the case of quadrupolar waves. However, these two effects can be detected from the \textit{amplitude} of a GW, which is far less constrained than its phase by GW detectors, so that the modification of the dynamics of binaries mentioned previsouly is of greater phenomenological significance.

\end{itemize}

\section{Non-Relativistic Scalar-Tensor Theories} \label{sec3}

It is straightforward to extend the formalism of Goldberger and Rothstein to the scalar-tensor action~\eqref{totaction}.
We hereby review  the basics of this approach and highlight the novelty represented by the new degree of freedom.

\subsection{Integrating out fluctuating fields}

As explained above, the binary system breaks Lorentz invariance spontaneously. The formalism goes along with this splitting of spacetime into space and time because,  in order to estimate the powers of $v$ that come from different terms in the action and/or from a given Feynman diagram,  we are suggested to split the metric field fluctuation $h_{\mu \nu}$ into a potential part $H_{\mu \nu}$ and a radiative part  $\bar h_{\mu \nu}$, i.e.,
\begin{equation}
g_{\mu \nu} (x)= g^{(0)}_{\mu \nu} (t) + \frac{h_{\mu \nu} (t,\mathbf{ x})}{\mpl}  \;, \qquad h_{\mu \nu} (t,\mathbf{ x})=   H_{\mu \nu} (t,\mathbf{ x})+ \bar h_{\mu \nu}(t,\mathbf{ x}) \;,
\end{equation}
where $g^{(0)}_{\mu \nu} (t)$ is the background metric.
The difference between the potential and radiative parts is in the scaling of their momenta: emitted gravitons  always have the momentum and the frequency of the binary system ${v}/{r}$ (since they should be on-shell), while the spatial momentum of a potential graviton is of the order of the inverse separation between the two components. Denoting the four-momentum of the latter with $k^\mu = (k^0, \mathbf{k})$, one has $k^0 \sim {v}/{r}$ and $\mathbf{k} \sim {1}/{r}$. 

The same separation applies to the scalar field, i.e., 
\begin{equation}
\phi (x)= \phi_0 (t) + \varphi (t,\mathbf{ x})\;, \qquad \varphi (t,\mathbf{ x})= \Phi (t,\mathbf{ x})+ \bar{\varphi}(t,\mathbf{ x})\;, 
\end{equation}
where $\phi_0 (t)$ is the homogeneous time-dependent expectation value of the field.
As we consider systems much smaller than the Hubble radius, we can take the background metric to be the Minkowski metric. Moreover, as we are interested in a dark energy scalar field,  its time variation is of  order Hubble and we can thus neglect it. Therefore, from now on we use 
\be
g^{(0)}_{\mu \nu} = \eta_{\mu \nu} \;, \qquad \phi_0 = \text{const.} \;.
\ee
The constant scalar field VEV   can be reabsorbed in the definition of the masses and scalar charges in eq.~\eqref{eq:action} and can be thus set to zero without loss of generality, $\phi_0=0$.

The effective action is obtained as a two-step path integration, first over  the potential gravitons and scalars, respectively $H_{\mu \nu}$ and $\Phi$, and then over the radiation ones, respectively $\bar h_{\mu \nu}$ and $\bar \varphi$. 
Thus, the first step consists in computing the effective action $ S_{\rm eff}[x_A, \bar{h}_{\mu \nu}, \bar{\varphi}]$, defined by
\be
\label{eq:EFTaction1}
\exp\left({i  S_{\rm eff}[x_A, \bar{h}_{\mu \nu}, \bar{\varphi}]} \right) = \int {\cal D} H_{\mu \nu} {\cal D} \Phi \exp \left({i  S[x_A, {h}_{\mu \nu}, {\varphi}] + i S_{{\rm GF}, H} [H_{\mu \nu}, {\bar h}_{\mu \nu}] } \right) \;,
\ee
where $S_{{\rm GF},H}$ is a gauge-fixing term to the so-called de Donder (or harmonic) gauge, which allows to define the  propagator of $H_{\mu \nu}$. Its expression is given by
\begin{equation}
S_{{\rm GF},H} = - \frac{1}{4} \int d^4x \sqrt{-\bar{g}} \, \bar g^{\mu \nu} \, \Gamma^{(H)}_\mu \Gamma^{(H)}_\nu \;, \qquad \Gamma^{(H)}_\mu \equiv D_\alpha H^\alpha_\mu - \frac{1}{2}D_\mu H^\alpha_\alpha \;,
\label{eq:GFterm} 
\end{equation}
where $\bar g_{\mu \nu} \equiv \eta_{\mu \nu} + \bar{h}_{\mu \nu}/\mpl$ is the  background metric for $H_{\mu \nu}$ and
$D_\mu$ is the covariant derivative compatible with it. It is the path-integral equivalent to the harmonic gauge-fixing of Eq.~\eqref{eq:Harmonic_GF}.
Here we do not consider Faddeev-Popov ghosts because they appear only in loops and we will only compute tree-level diagrams. Indeed, as discussed below loop contributions can be shown to be suppressed with respect to tree level diagrams by the (huge) total angular momentum $L$ of the system~\cite{goldberger_effective_2006}.

The action obtained by this procedure contains the mechanical two-body Lagrangian of the system and the coupling to radiation gravitons. 
For $\bar{h}_{\mu \nu} =0$ and $ \bar{\varphi}=0$ the two body dynamics is conservative and,  to  leading and next to leading order in  $v$  the Lagrangian  reduces   to the Newtonian  and   EIH Lagrangians respectively, extended by the suitable corrections  coming from the scalar fifth force. We compute these in Sec.~\ref{sec:feynman} and~\ref{sec:EIH_Lagrangian}.

The second integration, i.e.~over $\bar{h}_{\mu \nu}$ and $\bar{\varphi}$, gives
\be
\label{eq:finalSeff}
\exp \left(i \hat S_{\rm eff}[x_A] \right) = \int \mathcal{D} \bar{h}_{\mu \nu} \mathcal{D} \bar{\varphi} \exp \left( i  S_{\rm eff}[x_A, \bar{h}_{\mu \nu}, \bar{\varphi}] + i S_{{\rm GF}, \bar{h}}  [{\bar h}_{\mu \nu}] \right)  \;,
\ee
where $S_{{\rm GF},\bar h}$ is the gauge-fixing term for $\bar h_{\mu \nu}$, defined as
\begin{equation}
S_{{\rm GF},\bar h} = - \frac{1}{4} \int d^4x  \, \eta^{\mu \nu} \, \Gamma^{(\bar h)}_\mu \Gamma^{(\bar h)} _\nu \;, \qquad \Gamma^{(\bar h)}_\mu \equiv \partial_\alpha \bar h^\alpha_\mu - \frac{1}{2} \partial_\mu \bar h^\alpha_\alpha \;.
\label{eq:GFterm2}
\end{equation}
We have denoted with a hat the final effective action after the metric and scalar fields have been totally integrated out. As we review in Sec.~\ref{sec:dissipative_dynamics},  $\hat S_{\rm eff}[x_A]$ (more precisely, its imaginary part) contains information about the radiated power into gravitational and scalar waves.

\subsection{Propagators and power counting}
\label{sec3.3}

The fields propagators of the gravitational sector can be obtained from the quadratic  action,
\begin{equation} \label{eq:quadratic_EH}
S^{(2)} = - \frac{1}{8} \int d^4x \left[ -\frac{1}{2}(\partial_\mu h^\alpha_\alpha)^2 + (\partial_\mu h_{\nu \rho})^2 \right]  - \frac12  \int d^4x ( \partial_\mu \varphi )^2 \;,
\end{equation}
where repeated indices are contracted with the Minkowski metric.
In Fourier space it becomes
\begin{equation}
S^{(2)} = - \frac{1}{2} \int \frac{d^4 k}{(2 \pi)^4} k^2 h_{\mu \nu}(k) T^{\mu \nu ; \alpha \beta} h_{\alpha \beta}(-k)  - \frac{1}{2} \int \frac{d^4 k}{(2 \pi)^4} k^2 \varphi(k) \varphi(-k) \;,
\end{equation}
with $k^2 \equiv k_\mu k^\mu$ and $T^{\mu \nu ; \alpha \beta} = \frac{1}{8} \left( \eta^{\mu \alpha} \eta^{\nu \beta} + \eta^{\mu \beta} \eta^{\nu \alpha} - \eta^{\mu \nu} \eta^{\alpha \beta} \right)$.

Let us start discussing the propagators, defined as the inverse quadratic operator. For the metric we have to find the inverse operator of $T^{\mu \nu ; \alpha \beta}$, which is defined as $P_{\mu \nu ; \rho \sigma} T^{\rho \sigma ; \alpha \beta} = I_{\mu \nu}^{\alpha \beta}$ where $I_{\mu \nu}^{\alpha \beta} = \frac{1}{2} (\delta_\mu^\alpha \delta_\nu^\beta + \delta_\mu^\beta \delta_\nu^\alpha)$ is the identity on symmetric two-index tensors. It is straightforward to find 
\be \label{eq:pmunualphabeta}
P_{\mu \nu ; \alpha \beta} = 2 \left( \eta_{\mu \alpha} \eta_{\nu \beta} + \eta_{\mu \beta} \eta_{\nu \alpha} - \eta_{\mu \nu} \eta_{\alpha \beta} \right) \;, 
\ee
where the factor 4 difference with~\cite{goldberger_effective_2006} is due to the different normalization of the Planck mass. 
The propagator for the $h$ field is thus given by 
\begin{equation}
\left\langle T h_{\mu \nu}(x) h_{\alpha \beta}(x') \right\rangle = D_F(x-x') P_{\mu \nu ; \alpha \beta} \;,
\label{eq:propagator}
\end{equation}
where $T$ denotes time ordering and the Feynman propagator $D_F(x-x')$ is given by 
\begin{equation}
D_F(x-x') = \int \frac{d^4 k}{(2 \pi)^4} \frac{-i}{k^2-i\epsilon} e^{-ik(x-x')} \;.
\label{eq:Feyprop}
\end{equation}
The term  $i\epsilon$ is the usual prescription for the contour integral.

\begin{figure}
    \centering
    \subfloat[$\bar{h}_{\mu\nu}$]{  \label{fig:propagartorsa}
%
%
%
\includegraphics{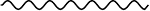}
	} \hspace{1em}
	\subfloat[$H_{\mu\nu}$]{  \label{fig:propagartorsb}
%
%
%
\includegraphics{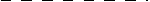}
	} \hspace{1em}
	\subfloat[$\bar{\phi}$]{  \label{fig:propagartorsc}
%
%
%
\includegraphics{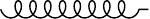}
	} \hspace{1em}
	\subfloat[$\Phi$]{  \label{fig:propagartorsd}
%
%
%
\includegraphics{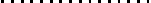}
	} \hspace{1em}
	
    \caption{Representation of the different fields propagators in Feynman diagrams.}
    \label{fig:propagartors}
\end{figure}
At this point it is useful to make a distinction between the propagator of $\bar{h}_{\mu \nu}$ and that of $H_{\mu \nu}$. While for $\bar{h}_{\mu \nu}$ we must use the relativistic propagator  given in eq.~\eqref{eq:propagator}, for $H_{\mu \nu}$ we can take advantage of the fact that its time and space derivatives scale differently with the velocity $v$, $k^0 \sim {v}/{r} \ll |\mathbf{k} | \sim {1}/{r}$. With the partial (only spatial) Fourier decomposition 
\begin{equation}
H_{\mu \nu}(t,\mathbf{x}) = \int \frac{d^3 k}{(2 \pi)^3} H_{\mathbf{k} \, \mu \nu}(t) e^{i \mathbf{k} \cdot \mathbf{x}} \;,
\label{eq:potential_graviton}
\end{equation}
the $v$ power counting becomes more transparent, as we have $\partial_0 H_{\mathbf{k} \, \mu \nu} \sim v \, \mathbf{k}  H_{\mathbf{k}  \, \mu \nu} $.
Therefore, using the expansion 
\be
\frac{-i}{k^2-i\epsilon} = - \frac{i}{\mathbf{k}^2} \left(1 + \frac{k_0^2}{\mathbf{k}^2} + \dots \right) \;,
\ee
(on the right-hand side we have gotten rid of the $i\epsilon$ prescription, which is irrelevant  for off-shell gravitons), one finds the propagator as the lowest-order term in this expansion,
\begin{equation}
\langle T H_{ \mathbf{k}\, \mu \nu}(t) H_{ \mathbf{q}\, \alpha \beta}(t') \rangle = - (2 \pi)^3 \frac{i}{\mathbf{k}^2} \delta^{(3)}(\mathbf{k} + \mathbf{q}) \delta(t-t') P_{\mu \nu ; \alpha \beta}
\label{eq:MixedFourierPropagator} \;.
\end{equation}
Figures~\ref{fig:propagartorsa} and~\ref{fig:propagartorsb} illustrate how  the propagators of $\bar h_{\mu \nu}$ and $H_{\mu \nu}$ are represented in Feynman diagrams.

\begin{figure}
    \centering
    \subfloat[$H_{\mu\nu}$]{
%
%
%
\includegraphics{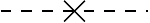}
	} \hspace{1em}
	\subfloat[$\Phi$]{
%
%
%
\includegraphics{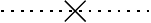}
	} \hspace{1em}

    \caption{Representation of the first velocity correction to the potential propagators.}
    \label{fig:insertion}
\end{figure}

The first correction to the propagator of $H_{\mu \nu}$ is supressed by $v^2$ and reads
\begin{equation}
\langle T H_{\mathbf{k} \, \mu \nu}(t) H_{\mathbf{q} \, \alpha \beta}(t') \rangle_{v^2} = - (2 \pi)^3 \frac{i}{\mathbf{k}^4} \delta^{(3)}(\mathbf{k} + \mathbf{q}) \frac{d^2}{dt dt'}\delta(t-t') P_{\mu \nu ; \alpha \beta} \;,
\label{eq:modH}
\end{equation}
which is represented as an insertion on the propagator, as illustrated in Fig.~\ref{fig:insertion}a.

For the treatment of the scalar field dynamics the procedure is analogous---and without the complication of the indices. We define the partial Fourier transform of $\Phi$ by 
\be
\Phi(t,\mathbf{x}) = \int \frac{d^3 k}{(2 \pi)^3} \Phi_{\mathbf{k} }(t) e^{i \mathbf{k} \cdot \mathbf{x}} \;,
\ee
and for the propagators we obtain
\begin{align}
\langle T \bar{\varphi} (x) \bar{\varphi}(x') \rangle &= D_F(x-x') \; ,  \label{eq:propagatorscalar}\\
\langle T \Phi_\mathbf{k} (t) \Phi_\mathbf{q} (t') \rangle &= - (2 \pi)^3 \frac{i}{\mathbf{k}^2} \delta^{(3)}(\mathbf{k} + \mathbf{q}) \delta(t-t') \;. \label{eq:scalarpot}
\end{align}
We display their Feynman diagram representation in Figs.~\ref{fig:propagartors}c and~\ref{fig:propagartors}d.
The first correction to the propagator of $\Phi_\mathbf{k}$ reads
\be
\langle T \Phi_\mathbf{k} (t) \Phi_\mathbf{q} (t') \rangle_{v^2} = - (2 \pi)^3 \frac{i}{\mathbf{k}^4} \delta^{(3)}(\mathbf{k} + \mathbf{q}) \frac{d^2}{dt dt'}\delta(t-t')  \;,
\label{eq:modPhi}
\ee
and its representation is illustrated in Fig.~\ref{fig:insertion}b.

In order to organize systematically the Feynman diagrams in powers of $v$ and estimate their contributions to the effective action
we need to find the power counting rules of the theory. The power counting rules for the radiating and potential fields can be extracted from their propagators. For a radiation graviton, the propagator in eq.~\eqref{eq:propagator} is given by eq.~\eqref{eq:Feyprop}. Using  $k \sim {v}/{r}$, this scales as $(v/r)^2$, which gives $\bar{h}_{\mu \nu}(x) \sim {v}/{r}$. The same reasoning applies to $\bar \varphi$, which gives $\bar{\varphi}(x) \sim {v}/{r}$.

For the potential graviton we can use the expression of its propagator, eq.~\eqref{eq:MixedFourierPropagator}. Using that the delta function scales as the inverse of its argument, we obtain that the right-hand side of this equation scales as $k^0 / \mathbf{k}^5$, which using that $k^0 \sim {v}/{r}$ and  $\mathbf{k} \sim {1}/{r}$, gives $H_{\mathbf{k} \, \mu \nu}(t) \sim r^2 \sqrt{v}$. The scaling of the scalar can be found in a similar way and gives $\Phi_\mathbf{k} \sim r^2 \sqrt{v}$.

One last subtlety that we need to address to determine the correct power counting of the theory is the presence of the large parameter  ${m}/{\mpl}$ in the point-particle actions, see eq.~\eqref{eq:action}. As discussed in~\cite{goldberger_effective_2006}, this can be resolved  by treating the lowest-order diagrams non-perturbatively, while the next-order diagrams are down by powers of $v$ compared to the leading  ones. 
 In order to make this fact more explicit one 
  introduces the orbital angular momentum associated to the point-particle worldlines,  
\be
L \equiv m v r\;, 
\ee
and  uses the virial relation $v^2 \sim m/(\mpl^2 r)$ to eliminate $m$ and ${m}/{\mpl}$ from the power-counting rules, replacing them by the appropriate combinations of $L$, $v$ and $r$. For instance, one obtains $m/\mpl \sim \sqrt{L v}$.
For the diagrams describing the interactions between the two bodies, the leading  operators scale as $L v^0$ and  must be treated non-perturbatively in accordance 
with the fact that particle worldlines are background non-dynamical fields. Indeed, 
since they
 represent infinitely heavy fields that have been integrated out, they have no associated propagators.

 The large parameter $L$ can be also used to count loop diagrams. Loops that are closed by the particle worldlines are not quantum loops but give tree-level contributions.
 Diagrams with  actual quantum loops (i.e.~not involving particle worldlines) are suppressed by powers of $\hbar/L \ll 1$ and should therefore be discarded.

\subsection{Vertices}

We can now turn to compute the vertices of the action~\eqref{eq:action}. They are of two kinds: the ones generated by the Einstein-Hilbert and scalar field kinetic terms, i.e.~$S_{\rm grav}$, and the ones generated by the point-particle action $S_{{\rm pp}}$.

Let us first discuss the first kind. We will focus on vertices that are needed for our calculations, i.e.~cubic vertices containing only potential fields $H_{\mu \nu}$ and $\Phi$ and vertices that are linear in the radiation fields $\bar h_{\mu \nu}$ and $\bar \varphi$. The Einstein-Hilbert term, together with the gauge-fixing term~\eqref{eq:GFterm}, generates a $H^3$ and a $\bar{h}H^2$  vertex. These have been computed in~\cite{goldberger_effective_2006} (see eqs.~(37) and (45) of that reference) and due to their complexity we do not display them here. The relevant part of the scalar field action  is
\begin{equation}
S_{h\phi^2} = - \frac{1}{2 \mpl} \int d^4x \left( \frac{1}{2} h^\alpha_\alpha \eta^{\mu \nu} - h^{\mu \nu} \right) \partial_\mu \varphi \partial_\nu \varphi \;,
\label{eq:hPhi2Vertex}
\end{equation}
where $h_{\mu \nu} (t, \mathbf{x})= \bar{h}_{\mu \nu}(t, \mathbf{x}) + \int \frac{d^3k}{(2\pi)^3} H_{\mathbf{k}\, \mu \nu}(t) e^{i\mathbf{k} \cdot \mathbf{x}}$  and $ \varphi (t, \mathbf{x}) = \bar{\varphi} (t, \mathbf{x}) + \int \frac{d^3k}{(2\pi)^3} \Phi_\mathbf{k}(t) e^{i\mathbf{k} \cdot \mathbf{x}}$.
This
generates a $H \Phi^2$, a $\bar{h} \Phi^2$ and a $H \bar{\varphi} \Phi$ vertex.

Turning to the point-particle action $S_{ {\rm pp}}$, this can be rewritten, using an affine parameter $\lambda$, as
\begin{equation}
S_{{\rm pp}} = \sum_A m_A \int d \lambda \sqrt{- g_{\mu \nu} \frac{dx_A^\mu}{d\lambda} \frac{dx_A^\nu}{d\lambda} } 
\left[-1 + \aaa_A \frac{\varphi}{\mpl} + \bbb_A \left( \frac{\varphi}{\mpl } \right)^2 \right] \;.
\end{equation}  
We can use reparametrization invariance to set $\lambda$ equal to the local time $t$ of the observer. Using the notation $v_A^\mu(t) =  (1, \mathbf{v}_A)$ and $v_A = |\mathbf{v}_A|$, we arrive at the following expression for $S_{{\rm pp}}$, 
\begin{align}
\begin{split}
S_{{\rm pp}} = \sum_A m_A \int dt & \sqrt{1 - v_A^2 - \frac{h_{\mu \nu}}{\mpl} v_A^\mu v_A^\nu} \left[-1 + \aaa_A \frac{\varphi}{\mpl} + \bbb_A \left( \frac{\varphi}{\mpl} \right)^2 \right] \\
=  \sum_A m_A \int dt& \left[-1 + \frac{v_A^2}{2} + \frac{v_A^4}{8} + O(v^6) \right. \\
& + \frac{h_{00}}{2 \mpl} + \frac{h_{0i}}{\mpl} v_A^i + \frac{h_{ij}}{2 \mpl} v_A^i v_A^j + \frac{h_{00}}{4 \mpl} v_A^2 + O(h v^3) \\
&  + \aaa_A \frac{\varphi}{\mpl} - \aaa_A \frac{\varphi v_A^2}{2 \mpl} + O(\varphi v^3) \\
&  + \left. \frac{h_{00}^2}{8 \mpl^2} + \bbb_A \frac{\varphi^2}{\mpl^2} - \aaa_A \frac{\varphi h_{00}}{2 \mpl^2} + O(h^2v, \varphi^2v, h\varphi v) \right] \;,
\label{eq:pp_Action_expanded}
\end{split}
\end{align}
where in the second equality we have expanded the Lagrangian up to order $v^5$.

To get the vertices from the action, one should multiply by $i$ and specify $h_{\mu \nu}$ and $\varphi$ to radiation  or  potential fields. In order to facilitate the power-counting needed to evaluate the order of a Feynman diagram in the expansion in $v$, Table~\ref{table:power_counting} sums up the power-counting of the different vertices obtained in this section, which will be needed in the following.

\begin{table}
\center
\begin{tabular}{|c|c|}
\hline
Operator & PCR  \\
\hline
$\displaystyle{m_A \int dt v_A^2}$ & $L$ \\
\hline
$\displaystyle{\aaa_A \frac{m_A}{\mpl} \int dt \varphi, \quad \frac{m_A}{2 \mpl}\int dt h_{00}}$ & $\sqrt{L}$ \\
\hline
$\displaystyle{\frac{m_A}{\mpl} \int dt v_A^i h_{0i}}$ & $\sqrt{L} v$ \\
\hline
$\displaystyle{m_A \int dt \frac{v_A^4}{8}}$ & $L v^2$ \\
\hline
$\displaystyle{\frac{m_A}{4 \mpl} \int dt h_{00}v_A^2, \quad \frac{m_A}{2 \mpl} \int dt h_{ij}v_A^i v_A^j, \quad -\aaa_A \frac{m_A}{2 \mpl} \int dt \varphi v_A^2}$ & $\sqrt{L}v^2$ \\
\hline
$\displaystyle{\frac{m_A}{8 \mpl^2} \int dt h_{00}^2, \quad \bbb_A\frac{m_A}{\mpl^2} \int dt \varphi^2, \quad -\aaa_A\frac{m_A}{2 \mpl^2} \int dt \varphi h_{00}}$ & $v^2$ \\
\hline
$h^3$ from $\displaystyle{\frac{\mpl^2}{2} \int d^4x \sqrt{-g}R}$ (see~\cite{goldberger_effective_2006}), $\displaystyle{h \varphi^2}$ from eq.~\eqref{eq:hPhi2Vertex} & $\displaystyle{\frac{v^2}{\sqrt{L}}}$ \\
\hline

\end{tabular}
\caption[Power-counting rules]{Power-counting rules for the vertices obtained by expanding the action, given here for a potential graviton and scalar field. Multiply by $\sqrt{v}$ if needed to replace by a radiation graviton or scalar field.}
\label{table:power_counting}
\end{table}

\subsection{Feynman rules}
\label{sec:feynman}

In this subsection we will give the Feynman rules and, for pedagogical purposes, calculate some simple diagrams. The full set of Feynman rules can be summed up as follows:

\begin{itemize}

\item At a given order in $v$, draw all the diagrams that remain connected when removing the worldlines of the particles, discarding quantum (i.e.~not involving particle worldlines) loop diagrams.

\item For each vertex, multiply the corresponding expression in the Einstein-Hilbert action, in eqs.~\eqref{eq:hPhi2Vertex} and~\eqref{eq:pp_Action_expanded} by $i$ and specify $h_{\mu \nu}$ to $\bar{h}_{\mu \nu}$ or $H_{\mu \nu}$ and $\varphi$ to $\bar{\varphi}$ or $\Phi$, taking into account the associated corresponding power counting rules.

\item Contract all the internal graviton or scalar lines. This gives a combinatorial factor corresponding to the number of Wick contractions. An internal potential graviton line corresponds to multiplying by eq.~\eqref{eq:MixedFourierPropagator}, while an internal radiation one corresponds to a multiplication by eq.~\eqref{eq:propagator}. An internal potential scalar line corresponds to multiplying by eq.~\eqref{eq:scalarpot}, while an internal radiation one corresponds to a multiplication by eq.~\eqref{eq:propagatorscalar}.

\item The combinatorial factor can be obtained from the explicit definition of the effective action, $e^{iS{\rm eff}} = \int {\cal D} h {\cal D} \varphi e^{i(S_0+S_{\rm int})}$ where $S_0$ is the quadratic action and $S_{\rm int}$ contains the vertices.
The rule of thumb is the following: divide by the symmetry factor  of the diagram (coming from the fact that for $n$ vertices, the $1/n!$ of the exponential is not always compensated by the rearrangement of the $(\mathrm{vertex})^n$ term if there are identical vertices) and then multiply by the number of different Wick contractions giving the diagram.

\end{itemize}

Note that if we focus only on the integration over the potential fields so as to obtain $ S_\mathrm{eff}[x_A, \bar{h}, \bar{\varphi}]$, potential gravitons and scalars, respectively $H_{\mu \nu}$ and $\Phi$, can only enter Feynman diagrams as internal lines, while radiation gravitons and scalars, respectively $\bar{h}_{\mu \nu}$ and $\bar{\varphi}$, can only enter Feynman diagrams as external lines, i.e.~they cannot be used as propagators.

Let's now look at the calculation of simple diagrams. The simplest is given in Fig.~\ref{fig:Newt_pot_A}, and represents the Newtonian potential. Using the Feynman rules, we have
\begin{align}
\begin{split}
\left. iS_\mathrm{eff}\right|_{\ref{fig:Newt_pot_A}} &= \left[i \frac{m_1}{2 \mpl} \int dt_1 \int \frac{d^3k_1}{(2\pi)^3} e^{i \mathbf{k}_1 \cdot \mathbf{x}_1(t_1)} \right] \left[i \frac{m_2}{2\mpl} \int dt_2 \int \frac{d^3k_2}{(2\pi)^3} e^{i \mathbf{k}_2 \cdot \mathbf{x}_2(t_2)} \right] \\
& \times \left\langle T H_{00}(t_1, \mathbf{k}_1) H_{00}(t_2, \mathbf{k}_2) \right\rangle \\
&= i P_{00;00} \frac{m_1m_2}{4\mpl^2} \int dt \int \frac{d^3k}{(2\pi)^3} \frac{e^{i \mathbf{k} \cdot (\mathbf{x}_1(t)-\mathbf{x}_2(t) )}}{k^2} \\
&= i \int dt \frac{G m_1 m_2}{r(t)} \;,
\end{split}
\end{align}
where for the last equality we used that
\be
\int \frac{d^3 k}{(2 \pi)^3} \frac{e^{-i \mathbf{k} \cdot \mathbf{x}}}{\mathbf{k}^2} = \frac{1}{4 \pi |\mathbf{x}|} \;,
\ee
and  we have defined $\mathbf{r} \equiv \mathbf{x}_1 - \mathbf{x}_2$ and $r \equiv |\mathbf{r}|$.
An analogous calculation can be done for the scalar interaction given by Fig.~\ref{fig:Newt_pot_b}. This yields
\be
\left. iS_\mathrm{eff}\right|_{\ref{fig:Newt_pot_b}} =  i \int dt \frac{2 G \aaa_1 \aaa_2 m_1 m_2}{r(t) } \;,
\ee
so that the effective gravitational Newton constant between two objects $A$ and $B$ reads
\be
\tilde G_{AB} \equiv G (1+2 \aaa_A \aaa_B ) \;.
\label{eq:Gab}
\ee

A second example with non-trivial symmetry factor is given by Fig.~\ref{subfig:EIH_f} below, i.e.
\be
\begin{split}
\left. iS_\mathrm{eff}\right|_{\ref{subfig:EIH_f}} &= \frac{1}{2!} \left[i \frac{m_1}{8\mpl^2} \int dt_1 \int \frac{d^3k_1}{(2\pi)^3} \frac{d^3k'_1}{(2\pi)^3} e^{i (\mathbf{k}_1 + \mathbf{k}'_1) \cdot \mathbf{x}_1(t_1)} \right] \left[i \frac{m_2}{2\mpl} \int dt_2 \int \frac{d^3k_2}{(2\pi)^3} e^{i \mathbf{k}_2 \cdot \mathbf{x}_2(t_2)} \right] \\
& \times \left[i \frac{m_2}{2\mpl} \int dt'_2 \int \frac{d^3k'_2}{(2\pi)^3} e^{i \mathbf{k}'_2 \cdot \mathbf{x}_2(t'_2)} \right] \left\langle T H_{00}(t_1, \mathbf{k}_1) H_{00}(t_1, \mathbf{k}'_1) H_{00}(t_2, \mathbf{k}_2) H_{00}(t'_2, \mathbf{k}'_2) \right\rangle \\
&= \frac{i m_1 m_2^2 P_{00;00}^2}{2^5\mpl^4} \int dt \left(\int \frac{d^3k}{(2\pi)^3}  \frac{e^{i \mathbf{k} \cdot (\mathbf{x}_1(t)-\mathbf{x}_2(t))}}{k^2} \right)^2 \\
&= i \int dt \frac{m_1 m_2^2 G^2}{2 r^2} \;.
\end{split}
\ee
Now we can move to the complete calculation of the effective action.

\section{Renormalization of  masses and charges}\label{sec2.2}

The action~\eqref{eq:action} contains body-dependent scalar charges $\aaa_A$ and $\bbb_A$. As mentioned in the introduction, even if we assume the validity of the weak equivalence principle---the universality of free falling for test particles---in a scalar-tensor theory such a universality is inevitably spoiled by the gravitational self-energy of massive bodies. It is instructive to see how this happens in the adopted formalism. To this end, in this section, we derive the dependence of the scalar charges on the gravitational self-energy, after computing how the masses of the objects get similarly renormalized. 

\begin{figure}
	\centering
	\subfloat[]{
%
%
%
\includegraphics{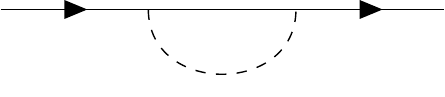}
	} \hspace{1em}
	\subfloat[]{
%
%
%
\includegraphics{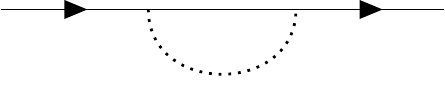}
	} \hspace{1em}
	
\caption{Diagrams contributing to the mass renormalization.}
\label{fig:mass_renormalization}
\end{figure}

Let us consider the point-particle action~\eqref{eq:action} in the static case (i.e.~for $v_A=0$), focussing on a single body.
To simplify the notation, we will omit the index $A$ in this part of the discussion. The action then reads
\be
\label{eq:m}
- m \int d  \tau \left(1 - \alpha \frac{\varphi}{\mpl} - \beta \left( \frac{\varphi}{\mpl} \right)^2 \right)    \;.
\ee
We want to show that the mass $m$ gets renormalized by the contribution of the self-energy of the  object. 
In particular, at lowest order the two diagrams of Fig.~\ref{fig:mass_renormalization} contribute to this action. In the previous derivation, we ignored such diagrams because they are scale-less power-law divergent and, as such, they vanish in dimensional regularization. However, here we will be concerned about the physical significance of such self-energy diagrams, so we choose instead a hard cutoff $\Lambda$ in the momentum integrals, corresponding approximatively to choosing an object of size $r_s \sim 1/\Lambda $. This regularization preserves rotational symmetry. The fact that, on the other hand, it breaks boosts does not concern us too much here as we are considering  objects at rest.

Starting from a bare mass $m_{\text{bare}}$, by including these diagrams we obtain a dressed mass $m (\Lambda)$. In other words, the term in~\eqref{eq:m} that is of lowest order in the fields gets renormalized as
\be
\label{eq:dressedm}
- m_{\text{bare}} \int dt  \quad \to \quad - m (\Lambda) \int dt  \;,
\ee
where 
\begin{equation}
\label{eq:dm}
m (\Lambda) \equiv m_{\text{bare}} + \delta m(\Lambda) \;, \qquad \delta m(\Lambda)  = -  2 \pi  \tilde{G} m_{{\rm bare}}^2 \,  \int^\Lambda \frac{d^3k}{(2\pi)^3} \frac{1}{k^2}  \;, 
\end{equation}
where we have used the Planck mass definition and we have defined
\be
\tilde{G} \equiv G(1+2\aaa_{\rm bare}^2) \;.
\ee 
This (negative) quantity coincides with the gravitational energy of the object, given by the usual expression
\begin{equation}
E = - \frac{\tilde{G}}{2} \int d^3x d^3y \frac{\rho(\mathbf{x}) \rho(\mathbf{y})}{|\mathbf{x} - \mathbf{y}|} \;,
\end{equation}
where $\rho$ is the mass density of the object. Indeed, 
replacing the point-particle density by a regularized version of a delta function, the energy density can be expressed as
\be
\rho(\mathbf{x} )= m_{\rm bare} \int^\Lambda \frac{d^3 k}{(2 \pi)^3} {e^{i \mathbf{k} \cdot \mathbf{x} }} \;,
\ee
and comparing this expression with the second equality in eq.~\eqref{eq:dm}, one obtains
\be
\label{eq:genergy}
E(\Lambda) \equiv   \delta m(\Lambda) \;.
\ee

\begin{figure}
	\centering
	\subfloat[]{
%
%
%
\includegraphics{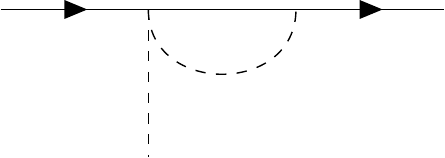}
	} \hspace{1em}
	\subfloat[]{
%
%
%
\includegraphics{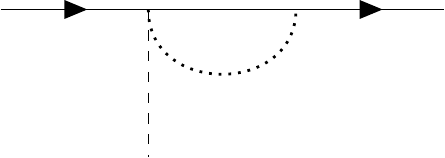}
	} \hspace{1em}
	\subfloat[]{
%
%
%
\includegraphics{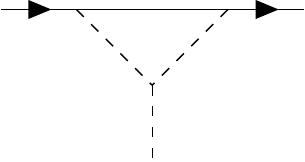}
	} \hspace{1em}
	\subfloat[]{
%
%
%
\includegraphics{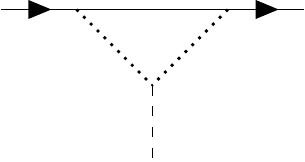}
	}
	
\caption{Diagrams contributing to the mass renormalization of the vertex $\frac{m}{2 \mpl} \int dt h_{00}$.}
\label{fig:mass_renormalization_higher_order}
\end{figure}

Here we have studied the renormalization of the particle mass appearing in the lowest-order vertex~\eqref{eq:m} but, by the equivalence principle, the same mass appears also in 
higher-order operators of the point-particle action~\eqref{eq:action}, such as 
\be
\frac{m}{2 \mpl} \int dt \, h_{00}\;.
\ee 
If the use of our hard-cutoff regulator is  consistent, we should get the same result for the renormalization of this vertex. Indeed, we have  checked that this is the case by calculating the Feynman diagrams shown in Fig.~\ref{fig:mass_renormalization_higher_order}. We will not reproduce this lengthy computation here, as it parallels the one for the scalar coupling that we are going to discuss next, with the complication of the spin-2 vertex. The final result is the same mass renormalization as in the lowest-order vertex~\eqref{eq:m}, i.e.,  
\be
\frac{m_\text{bare}}{2 \mpl} \int dt \, h_{00} \quad \to \frac{m (\Lambda)}{2 \mpl} \int dt \, h_{00} \;,
\ee 
which shows the consistency of the method
\footnote{In calculating the diagrams of Fig.~\ref{fig:mass_renormalization_higher_order}, there appears also terms proportional to $h_{ij} \delta^{ij}$ in the point-particle action. Of course, for a particle at rest such terms should not appear, as in the proper time the only combination involving $h_{\mu \nu}$ is $h_{\mu \nu} v^\mu v^\nu = h_{00}$. We can trace back the appearance of such artifacts from the fact that our regulator breaks Lorentz invariance.}.

We will now address the renormalization of the scalar charges of the objects $A$, appearing in the operator
\be
\label{eq:dressed}
\aaa_A \frac{m_A}{\mpl} \int dt \, \varphi \;.
\ee
In particular, we will show that even if we assume the weak equivalence principle and start with the same bare scalar  charges for the two objects, $\aaa_{\text{bare},1}=\aaa_{\text{bare},2}=\aaa_\text{bare}$, these get renormalized by the higher-order interactions, 
\be
\aaa_\text{bare} \frac{m_{\text{bare},A}}{\mpl} \int dt \, \varphi \qquad \to \qquad  \aaa_A (\Lambda) \frac{m_{A}(\Lambda)}{\mpl} \int dt \, \varphi \;,
\ee
where
\be
\label{eq:renalpha}
\aaa_A (\Lambda) \equiv \aaa_\text{bare} + \delta \aaa_A (\Lambda) \;.
\ee

To study which diagrams contribute to the charge it is convenient to split the scalar field fluctuation into a potential mode, which we will integrate out, and an external source,  i.e.~$\varphi = \Phi + \varphi_\mathrm{ext}$.
After using such a splitting in the action~\eqref{eq:pp_Action_expanded},  the vertices contributing to the renormalization of the operator~\eqref{eq:dressed} are
\begin{equation}
-\aaa_\text{bare} m_{\text{bare},A} \int dt \frac{\varphi_\mathrm{ext}H_{00}}{2 \mpl^2} \;, \qquad 2 \bbb_\text{bare} m_{\text{bare},A} \int dt \frac{\varphi_\mathrm{ext} \Phi}{\mpl^2} \;, 
\end{equation}
where for these bare vertices we have assumed the weak equivalence principle, i.e.~that the scalar couplings $\aaa_\text{bare}$ and $\bbb_\text{bare}$ are common to the two objects. 
\begin{figure}
	\centering
	\subfloat[]{
%
%
		\label{fig:charge_renormalization_a}
\includegraphics{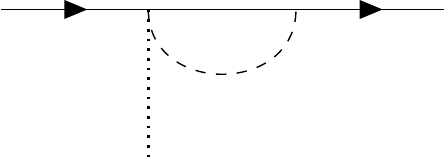}
	} \hspace{1em}
	\subfloat[]{
%
%
		\label{fig:charge_renormalization_b}
\includegraphics{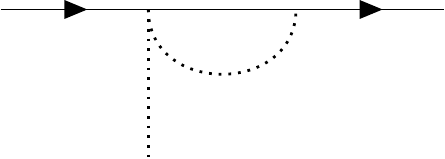}
	} \hspace{1em}
	\subfloat[]{
%
%
		\label{fig:charge_renormalization_c}
\includegraphics{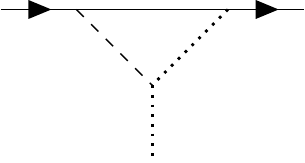}
	}
	
\caption{Diagrams contributing to the charge renormalization.}
\label{fig:charge_renormalization}
\end{figure}
The  corresponding  diagrams are  shown in Fig.~\ref{fig:charge_renormalization_a} and~\ref{fig:charge_renormalization_b}. The correction to the renormalized scalar charge appearing in the vertex $\aaa_A(\Lambda)  \frac{m_A}{\mpl} \int dt \, \varphi_{\rm ext} $
reads
\begin{equation}
\delta \aaa_A (\Lambda)= \left( 2 \aaa  \frac{1-4 \bbb}{1+2 \aaa^2} \right)_{\rm bare} \frac{\delta m_A(\Lambda)}{m_{A}(\Lambda)} \;, 
\label{eq:deltaalpha}
\end{equation}
where for $\delta m_A$ and $m_{A}$ on the right-hand side we have used eqs.~\eqref{eq:dm} and~\eqref{eq:genergy}.
A third vertex, represented in Fig.~\ref{fig:charge_renormalization_c}, comes from the scalar field action~\eqref{eq:hPhi2Vertex} and reads
\begin{equation}
 - \frac{1}{\mpl} \int d^4x \left( \frac{H^\alpha_\alpha}{2} \eta^{\mu \nu} - H^{\mu \nu} \right) \partial_\mu \varphi_\mathrm{ext} \partial_\nu \Phi \;.
\end{equation}
However, the tensorial factor in the  propagator of the potential graviton modes is non-vanishing only for $\mu=0$ and $\nu=0$, which implies that this diagram vanishes in our static case.

Plugging the renormalized values of the scalar couplings~\eqref{eq:renalpha} with eq.~\eqref{eq:deltaalpha} in the expression for the effective Newton constant between two bodies $A$ and $B$ given in eq.~\eqref{eq:Gab} and expanding to leading order in $\delta \aaa$, we obtain
\begin{equation}
\tilde{G}_{AB} \simeq \tilde{G}\left[ 1+ 4\aaa^2 \frac{1 - 4 \bbb}{(1+2 \aaa^2)^2} \left( \frac{E_A}{m_A} + \frac{E_B}{m_B} \right) \right] \;.
\end{equation}
As first realized by Nordtvedt~\cite{Nordtvedt:1968qs}, this implies a violation of the strong Equivalence Principle. This expression agrees with the one derived in Ref.~\cite{damour_tensor-multi-scalar_1992}, which shows that the body-dependent gravitational constant $\tilde{G}_{AB}$ is given by 
\begin{equation}
\tilde{G}_{AB} = \tilde{G}\left[ 1+(4 \tilde \beta - \tilde \gamma - 3) \left( \frac{E_A}{m_A} + \frac{E_B}{m_B} \right) \right] \;,
\end{equation}
where $\tilde \beta$ and $\tilde \gamma$ are the parametrized post-Newtonian (PPN) parameters, given  by\footnote{The PPN parameters are given in eqs.~(4.12b) and (4.12c) of Ref.~\cite{damour_tensor-multi-scalar_1992}, where we  the dictionary between our notation and theirs is $\varphi_{\rm here} = \sqrt{2}M_P \varphi_{\rm there}$, $\alpha^a= -\aaa \sqrt{2}$, $\beta_{ab} = (-4\bbb -2\aaa^2) \delta_{ab}$. We will denote the PPN parameters with a tilde in order not to confuse them with the scalar couplings $\aaa_A$, $\bbb_A$.}
\begin{align}
\tilde \beta &= 1 - 2 \left[ \frac{\aaa ^4+2 \aaa ^2 \bbb }{(1+2 \aaa ^2)^2} \right]_{\rm bare}\;, \\
\tilde \gamma &= 1 - 4 \left[ \frac{\aaa^2}{1+2 \aaa^2} \right]_{\rm bare}\;.
\end{align}
Considerations similar to the one for the renormalization of  $\aaa_A$ can be made for the couplings $\bbb_A$.

Before concluding the section, let us notice that the renormalization of the scalar charges can be also expressed in terms of the so-called ``sensitivity''
of a body to changes in the local value of the effective gravitational constant $G$ due to
changes in the scalar field~\cite{Eardley1975ApJ}. It is explicitely defined by
\be
\label{eq:sensitivity}
s_A \equiv - \frac{d \ln m_A}{d \ln G}  = - \frac{\delta m_A}{m_A}\;,
\ee
where in the last equality we have used eq.~\eqref{eq:dm}. Using eq.~\eqref{eq:deltaalpha}, the sensitivity can be related to the scalar charges by
\be
s_A = -    \left( \frac{1+2 \aaa^2}{1-4 \bbb }\right)_{\rm bare}  \frac{\delta \alpha_A}{2 \aaa_{\rm bare}} \;.
\ee

\section{Conservative dynamics up to $1$PN order}


\label{sec:EIH_Lagrangian}

In Sec.~\ref{sec3} we have shown how to compute the effective action $ S_{\rm eff} [x_A, \bar h_{\mu \nu}, \bar \varphi]$ defined in eq.~\eqref{eq:EFTaction1} by integrating out the potential modes $H_{\mu \nu}$ and $\Phi$. We now focus on the conservative part of this action obtained 
by considering only diagrams without external (and internal) radiation, i.e.~$ S_\mathrm{eff}[x_A, \bar{h}_{\mu \nu}=0, \bar{\varphi}=0]$.

\begin{figure}
	\centering
	\subfloat[]{
%
%
			\label{fig:Newt_pot_A}
%
\includegraphics{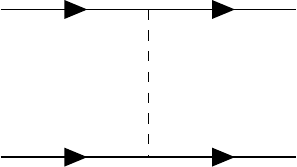}
	} \hspace{1em}
	\subfloat[]{
%
%
		\label{fig:Newt_pot_b}
%
\includegraphics{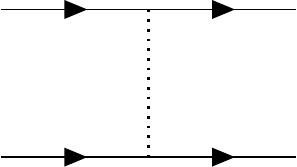}
	}
	
\caption{Feynman diagrams contributing to the $Lv^0$ potential}
\label{fig:Newt_pot}
\end{figure}
At lowest order in $v$, there are only two diagrams contributing to this action, illustrated in Fig.~\ref{fig:Newt_pot}: respectively one graviton  and one scalar exchange. Therefore, the action to order $Lv^0$  is given by 
\begin{equation}
S_{Lv^0} = \int dt \left[\frac{1}{2} m_1 v_1^2 + \frac{1}{2} m_2 v_2^2 + \frac{\tilde G_{12} m_1 m_2}{r}\right] \;.
\end{equation}
The first two terms correspond to the Newtonian kinetic energy of the particles whereas the last term is the effective gravitational potential with the rescaled Newton constant $\tilde{G}_{AB}$ computed in Sec.~\ref{sec:feynman}, see eq.~\eqref{eq:Gab}.

Let us now compute the first relativistic correction to this result. For GR, the corresponding Lagrangian has been calculated for the first time by Einstein, Infeld and Hoffmann~\cite{einstein_gravitational_1938}. It was generalized to   multi-scalar-tensor theories of gravitation by Damour and Esposito-Far\`{e}se ~\cite{damour_tensor-multi-scalar_1992} using a post-Newtonian expansion. In the final equation of this section, see eq.~\eqref{eq:LEIH} below, we will recover the result of~\cite{damour_tensor-multi-scalar_1992} restricted to a single scalar.

\begin{figure}
	\centering
	\subfloat[]{
%
%
%
\includegraphics{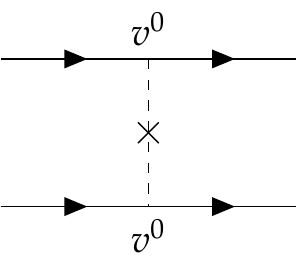}
		\label{subfig:EIH_A}
	} \hspace{1em}
	\subfloat[]{
%
%
%
\includegraphics{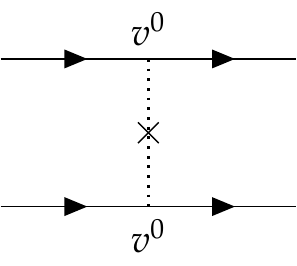}
		\label{subfig:EIH_b}
	} \hspace{1em}
	\subfloat[]{
%
%
%
\includegraphics{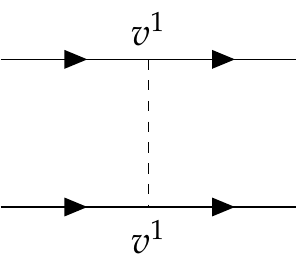}
		\label{subfig:EIH_c}
	} \hspace{1em}
	\subfloat[]{
%
%
%
\includegraphics{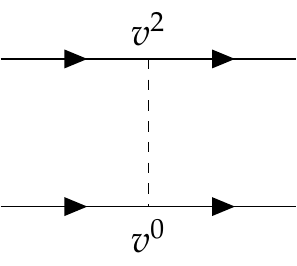}
		\label{subfig:EIH_d}
	} \hspace{1em}
	\subfloat[]{
%
%
%
		\label{subfig:EIH_e}
		\includegraphics{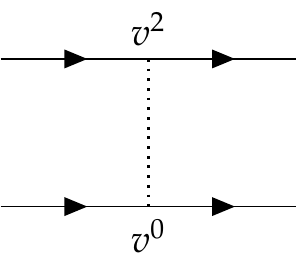}
	} \hspace{1em}
	\subfloat[]{
%
%
%
\includegraphics{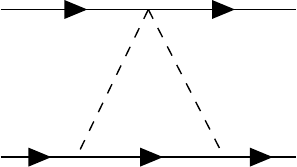}
		\label{subfig:EIH_f}
	} \hspace{1em}
	\subfloat[]{
%
%
%
\includegraphics{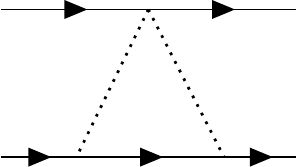}
		\label{subfig:EIH_g}
	} \hspace{1em}
	\subfloat[]{
%
%
%
\includegraphics{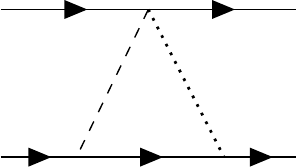}
		\label{subfig:EIH_h}
	} \hspace{1em}
	\subfloat[]{
%
%
%
\includegraphics{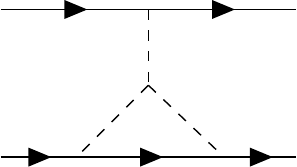}
		\label{subfig:EIH_i}
	} \hspace{1em}
	\subfloat[]{
%
%
%
\includegraphics{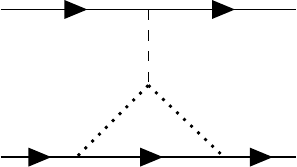}
		\label{subfig:EIH_j}
	} \hspace{1em}
	\subfloat[]{
%
%
%
\includegraphics{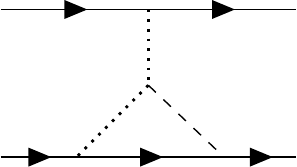}
		\label{subfig:EIH_k}
	} \hspace{1em}
	
\caption[Feynman diagrams contributing to the $Lv^2$ potential.]{Feynman diagrams contributing to the $Lv^2$ potential. Each diagram that is not symmetric under the exchange of particles wordlines should be added with its symmetric counterpart.}
\label{fig:EIH}
\end{figure}
To order $v^2$, the power counting rules dictate that ten diagrams, shown in Fig.~\ref{fig:EIH},  contribute to the potential. Let us see how each of them contributes. (For notational convenience, since we focus on the Lagrangian, we  remove the $i \int dt$ factor in front of each term.)
\begin{itemize}
\item Figures~\ref{subfig:EIH_A} and~\ref{subfig:EIH_b} respectively come from the exchange of a potential graviton and scalar with lowest-order vertex and modified propagators (see eqs.~\eqref{eq:modH} and~\eqref{eq:modPhi}) and yield
\begin{equation}
\frac{\tilde G_{12} m_1 m_2}{2 r} \left( \mathbf{v}_1 \cdot \mathbf{v}_2 - \frac{(\mathbf{v}_1 \cdot \mathbf{r}) (\mathbf{v}_2 \cdot \mathbf{r})}{r^2} \right) \;.
\end{equation}

\item Figure~\ref{subfig:EIH_c} comes from the exchange of a potential graviton with the vertex $h_{0i}v^i$ in the action~\eqref{eq:pp_Action_expanded} and yield
\begin{equation}
-4 \frac{G m_1 m_2}{r} \mathbf{v}_1 \cdot \mathbf{v}_2 \;.
\end{equation}

\item Figures~\ref{subfig:EIH_d} (\ref{subfig:EIH_e})  comes  from the exchange of a potential graviton (scalar) with one of the vertices $hv^2$ ($\phi v^2$) in the action~\eqref{eq:pp_Action_expanded} and yields
\begin{equation}
\frac{G(3-2 \aaa_1 \aaa_2) m_1 m_2}{2 r} (v_1^2 + v_2^2)\;.
\end{equation}

\item  Figures~\ref{subfig:EIH_f},~\ref{subfig:EIH_g} and~\ref{subfig:EIH_h}, respectively coming from the $h^2$, $\phi^2$ and $h \phi$ vertices in the action~\eqref{eq:pp_Action_expanded}, yield
\begin{equation}
\frac{G^2(1 +4 f_{12} -4\aaa_1 \aaa_2) m_1 m_2 (m_1+m_2)}{2 r^2} \;,
\end{equation}
where 
\be
\label{eq:f12}
f_{AB} \equiv \bbb_A \aaa_B^2 + \bbb_B \aaa_A^2 + \kappa_{AB} \left(\bbb_B\aaa_A^2 - \bbb_A \aaa_B^2 \right) \;
\ee
is a symmetric (in the indices $AB$) function built out of $\aaa_A$, $\bbb_A$ and the antisymmetric mass ratio 
\be
\label{eq:kappa}
\kappa_{AB} \equiv \frac{m_A-m_B}{m_A+m_B} \;.
\ee

\item Figure~\ref{subfig:EIH_i} comes from the $H^3$ term in the Einstein-Hilbert action and yields
\begin{equation}
-\frac{G^2 m_1 m_2 (m_1+m_2)}{r^2} \;.
\end{equation}

\item Finally, the last diagrams in Figs.~\ref{subfig:EIH_j} and~\ref{subfig:EIH_k} do not contribute because they are proportional to $\frac{\eta^{\alpha \beta}}{2} P_{00,\alpha \beta} \delta_{ij} - P_{00,ij}$, which vanishes.
\end{itemize}

Gathering all these terms, we can put the action into the following form 
\be
\begin{split}
\label{eq:LEIH}
S_{Lv^2} =& \int dt \;  \bigg\{ \frac{1}{8} \sum_A m_A v_A^4 \\
&+  \frac{\tilde{G}_{12} m_1 m_2}{2 r} \left[ (v_1^2 + v_2^2) - 3 \mathbf{v}_1 \cdot \mathbf{v}_2 -  \frac{(\mathbf{v}_1 \cdot \mathbf{r}) (\mathbf{v}_2 \cdot \mathbf{r})}{r^2} + 2\tilde \gamma_{12} (\mathbf{v}_1 - \mathbf{v}_2)^2 \right] \\
& - \frac{\tilde{G}_{12}^2 m_1 m_2 (m_1+m_2)}{2 r^2} (2 \tilde \beta_{12} -1) \bigg\} \;,
\end{split}
\ee
where as usual  $\tilde{G}_{12} = G (1+2 \aaa_1 \aaa_2)$ is the effective Newton constant and $\tilde \beta_{AB}$ and $\tilde \gamma_{AB}$ are PPN parameters
given here by\footnote{Analgous parameters are defined in the context of multi-scalar-tensor gravity in Ref.~\cite{damour_tensor-multi-scalar_1992}, see eq.~(6.26). The dictionary between our and their notation  is $\alpha^{\rm there}_A = - \aaa^{\rm here}_A \sqrt{2}$, $\beta^{\rm there}_{A} = -4 \bbb^{\rm here}_A -2 (\aaa^{\rm here}_A)^2$.}
\begin{align} \label{eq:ppn_params}
\tilde \gamma_{AB} &= 1 - 4 \frac{\aaa_A \aaa_B}{1+2\aaa_A \aaa_B} \;, \\
\tilde \beta_{AB} &= 1 - 2 \frac{\aaa_A^2\aaa_B^2 + f_{AB}}{(1+2\aaa_A \aaa_B)^2} \;.
\end{align}
For $\aaa_A=0=\bbb_A$ we recover the EIH correction to the Newtonian dynamics originally derived in~\cite{einstein_gravitational_1938} 
and reproduced in the the framework of the NRGR approach in~\cite{goldberger_effective_2006,Kol:2007bc}. 
More generally, the above action agrees with that of Ref.~\cite{damour_tensor-multi-scalar_1992} (see eq.~(3.7) of that reference).\footnote{A similar calculation has been done in~\cite{Huang:2018pbu}  for a massive axion-type field. However, their result disagrees with ours (and with~\cite{damour_tensor-multi-scalar_1992}) in the massless limit. The disagreement may be traced in the calculation of the diagrams in Figs.~\ref{subfig:EIH_j} and~\ref{subfig:EIH_k}. We thank the authors of Ref.~\cite{Huang:2018pbu} for double checking their results and eventually agreeing with us in a private correspondence.}

\section{Couplings to radiative fields}
 \label{sec:couplings} 

In this section we compute the couplings of the radiated fields to the point particles up to 2.5PN order.
In general, we can expand the effective action $S_\mathrm{eff}[x_A, \bar{h}_{\mu \nu}, \bar{\varphi}]$ as
\begin{equation}
S_\mathrm{eff}[x_A,  \bar{h}_{\mu \nu}, \bar{\varphi}] = S_0[x_A] + S_1[x_A,  \bar{h}_{\mu \nu}, \bar{\varphi}] + S_2 [x_A,  \bar{h}_{\mu \nu}, \bar{\varphi}]  + S_{\rm NL} [x_A,  \bar{h}_{\mu \nu}, \bar{\varphi}]  \;.
\end{equation}
The first term of the right-hand side, $S_0$, does not depend on external radiation gravitons. This is the conservative part of the action  that we have computed in Section~\ref{sec:EIH_Lagrangian}  and can be discarded from the following discussion. The next term, $S_1$, is linear in the radiating fields and contains the source that the radiating fields are coupled to. 
On  general grounds, it can be written as 
\begin{equation}
\label{eq:S1}
S_1 = S_{\rm int}^{(h)} + S_{\rm int}^{(\varphi)} \;, \quad S_{\rm int}^{(h)} \equiv -\frac{1}{2\mpl} \int d^4x T^{\mu \nu}(x) \bar{h}_{\mu \nu}(x) \;, \quad S_{\rm int}^{(\varphi)} \equiv \frac{1}{\mpl} \int d^4x J(x) \bar{\varphi}(x)\;, 
\end{equation}
where $T^{\mu \nu}$ and $J$ are respectively the sources for the metric and  the  scalar field radiation fields. In particular, $T^{\mu \nu}$ is the (pseudo) matter  energy-momentum tensor that includes the gravitational self-energy---i.e.~the contributions from the integrated out potential gravitons. It  is conserved in flat spacetime, $\partial_\mu T^{\mu \nu} = 0$, by linear diffeomorphism invariance.

The part quadratic in the radiating fields, $S_2$, provides the kinetic terms of $\bar{h}_{\mu \nu}$ and $\bar \varphi$ while $S_{\rm NL}$ contains higher-order coupling terms. 
The non-linear couplings  in the radiating fields give rise to the so-called \textit{tail effects}~\cite{porto_effective_2016, goldberger_gravitational_2010} and will not be discussed here because they are of order   $1.5$PN higher than the leading order  quadrupole.

Following~\cite{ross_multipole_2012}, to discuss the  couplings to the radiation fields and highlight the power counting in $v$ of the emission process~\cite{goldberger_effective_2006},
we will perform a multipole expansion of the sources of  $\bar{h}_{\mu \nu}$  and $\bar{\varphi}$   at the level of the action.
To simplify the treatment, we will focus here only on the lowest-order coupling but the full derivation can be found in ~\cite{ross_multipole_2012, goldberger_gravitational_2010}.
We will quickly review the graviton case, which has been discussed at length in the literature~\cite{goldberger_gravitational_2010}. We will turn in more details to the scalar case below.

\subsection{Graviton interactions}

\subsubsection{Multipole decomposition} 
Let us first consider the coupling of radiation gravitons with the sources, $S_{\rm int}^{(h)}$. In this subsection we consider a general stress-energy tensor $T^{\mu \nu}$.
In the next subsection we will give the explicit expression of $T^{\mu \nu}$ for our particular physical configuration.

To simplify the calculation and because this case has been studied at length in many references (see e.g.~\cite{goldberger_effective_2006,galley_radiation_2009}\footnote{If one does not chose this gauge and keeps all the components in the discussion, one finds that $\bar{h}_{00}$ couples to the total mass and Newtonian energy of the system while $\bar{h}_{0i}$ couples to the leading-order orbital angular momentum. These are conserved quantities at Newtonian order, which implies that they do not contribute to the radiation emission. 
Moreover, one can find that the quadrupole moment of the stress-energy tensor $I_h^{ij} $ couples to the linearized ``electric-type'' part of the Riemann tensor, $R_{0i0j}$, given by \be
R_{0i0j} \equiv - \frac{1}{2\mpl} \big(\partial_i \dot {\bar h}_{0j} + \partial_j \dot {\bar h}_{0i} - \ddot { \bar{h}}_{ij} - \partial_i \partial_j \bar h_{00} \big)\;,
\ee
whose two-point function is proportional to the projection operator into symmetric and traceless two-index spatial tensors.}), we directly focus on the so-called transverse-traceless gauge, defined by 
\be
\bar h_{0\mu} = 0  \;, \qquad \partial^i \bar h_{ji} =0 \;, \qquad  \bar h^k_k =0 \;.
\ee
Denoting by $\bar{h}_{ij}^{\rm TT}$  the radiated graviton in this gauge, the graviton interaction vertex of eq.~\eqref{eq:S1}  is 
\begin{equation}
\label{eq:gravcoup}
S_{\rm int}^{(h)} = -\frac{1}{2\mpl} \int d^4x T^{ij} \bar{h}_{ij}^{\rm TT} \;.
\end{equation}
Using the equation of motion $\partial_\mu T^{\mu \nu} = 0$, it is straightforward  to rewrite this equation as 
\be
\label{eq:interaction_graviton_multipole}
S_{\rm int}^{(h)} = - \frac{1}{2} \int dt I_h^{ij}  \frac{1}{2\mpl} \ddot { \bar{h}}_{ij}^{\rm TT} \;,  
\ee
where $I_h^{ij}$ is the quadrupole moment of the stress-energy tensor, defined as
\be
\label{eq:Ih}
I_h^{ij} \equiv \int d^3x T^{00} \left( x^i x^j - \frac{1}{3}x^2 \delta^{ij}\right) \;.
\ee

\subsubsection{Quadrupole expression}

As we have just seen in eq.~\eqref{eq:Ih},  to find the gravitational interaction vertex up to quadrupole order we just needed $T^{00}$ to lowest order. After comparison with the full action, eq.~\eqref{eq:pp_Action_expanded}, this is given by 
\begin{equation}
T^{00} = - \sum_A m_A \delta^3(\mathbf{x}-\mathbf{x}_A) \;,
\end{equation}
and the expression of the lowest-order quadrupole is the usual one, i.e.,
\begin{equation}
\label{eq:quadrupole_gr}
I_h^{ij} = - \sum_A m_A  \left( x_A^i x_A^j - \frac{1}{3}x_A^2 \delta^{ij}\right) \;.
\end{equation}
Therefore, at this order the vertex~\eqref{eq:gravcoup} is not modified by the presence of the scalar. However, as we will discuss below, to compute the emitted power we will have to take the third derivative of  the quadrupole moment with respect to time, see eq.~\eqref{eq:quadrupole_graviton}. This involves the acceleration of the two bodies and thus, using the equations of motion,  the modified Newton constant $\tilde{G}_{12} = G (1+2 \aaa_1 \aaa_2)$.
Note that, by using the NRGR power-counting rules explained in Sec.~\ref{sec3}, one finds that the gravitational quadrupole interaction vertex of eq.~\eqref{eq:interaction_graviton_multipole} is of order $\sqrt{L v^5}$.

\subsection{Scalar interactions}

\subsubsection{Multipole decomposition}
\label{subsubsec:multipoledec}

Let us now consider the coupling of radiation scalars with the sources, $S_{\rm int}^{(\varphi)}$.
Including also the  quadratic action of the radiating scalar,
we have
\be
S_{\rm eff} \supset S_2^{(\varphi)} + S_{\rm int}^{(\varphi)}=  \int d^4x  \left( -\frac{1}{2} \eta^{\mu \nu}  \partial_\mu \bar{\varphi} \partial_\nu \bar{\varphi} 
+ \frac{1}{\mpl} J   \bar{\varphi} \right)  \;, 
\ee
which leads to the following equation of motion 
\be
\square \bar{\varphi} = -\frac{J}{\mpl} \;, \qquad \square \equiv \eta^{\mu \nu}  \partial_\mu  \partial_\nu \;.
\ee

Now we want to do a multipole expansion of the scalar field around the center-of-mass $\mathbf{x}_{\rm cm}$, which is defined by\footnote{\label{foot:CM}This definition comes from the invariance of the theory under boosts, which  via Noether theorem gives that the following charge is conserved, 
\begin{equation}
Q^{0i} = \int d^3x \left( T^{00} x^i - T^{0i} t \right) \;.
\end{equation}
Since the total momentum $P^i = \int d^3x T^{0i}$ and energy $E = \int d^3x T^{00}$ are also conserved, we get that the center-of-mass moves with a constant velocity, thus justifying its definition.
Even for standard gravity and point-particle masses, since there are higher-order corrections implied by eq.~\eqref{eq:center_of_mass} the definition $ x_{\rm cm}^i \equiv \sum_A m_A x_A^i/\left( \sum_A m_A \right)$ is valid only at lowest order in the velocity expansion.} 
\begin{equation} \label{eq:def_center_mass}
 \mathbf{x}_{\rm cm} \equiv \frac1{E} \int d^3x \, T^{00} \mathbf{x} \;, \qquad E \equiv  \int d^3 x \, T^{00} \;,
\end{equation}
and can be set at the origin without loss of generality, $\mathbf{x}_{\rm cm}= \mathbf{0}$.
Since we are considering the physical configuration where the radiating scalar field $\bar \varphi$ varies on scales that are much larger than the source term $J$, we can expand the scalar in the interaction $S_{\rm int}^{(\varphi)}$ defined in eq.~\eqref{eq:S1} around the center of mass.  This gives
\begin{equation}
\begin{split}
S_{\rm int}^{(\varphi)} = \int d^4 x \frac{J \bar \varphi }{\mpl}= \int dt \, \int d^3 x \frac{J (t, \mathbf{x})}{\mpl}  \bigg( \bar{\varphi}(t, \mathbf{0}) + x^i \partial_i \bar{\varphi}(t,\mathbf{0}) + \frac{1}{2} x^i x^j \partial_i \partial_j \bar{\varphi}(t, \mathbf{0})  \\
+ \frac{1}{3!} x^i x^j x^k \partial_i \partial_j \partial_k \bar{\varphi}(t, \mathbf{0}) + \ldots \bigg) \;,
\end{split}
\end{equation}
which allows to obtain an expansion of the interactions in terms of the moments of the source, $\int d^3 x {J (t, \mathbf{x}) x^n}$, $n=0,1,\ldots$. 
Recalling that $\partial_i \bar{\varphi} \sim ({v}/{r}) \bar \varphi$, each moment $n$ enters suppressed by $v^n$.

This is not yet organised as a multipole expansion. To achieve this, instead of the moments one should use their irreducible representations under the rotation group. The first term to be modified is the second moment. Instead of $x^i x^j$ we should use 
\be
Q^{ij} \equiv x^i x^j - \frac{1}{3}x^2 \delta^{ij}\;. 
\ee
To compensate the additional term added, one is left with  
\begin{equation}
\frac{1}{6\mpl} \int d^3x \, x^2 J \, \nabla^2 \bar{\varphi} \;.
\end{equation}
We can then use the scalar equation of motion that, up to a contact term renormalizing the point-particle masses, transforms this term into a monopole one. 
Similarly, the  symmetric and traceless tensor associated to the third moment is 
\begin{equation}
Q^{ijk} = x^i x^j x^k - \frac{1}{5} \left(\delta^{ij} x^2 x^k + 2 \; \mathrm{perm} \right) \; ,
\end{equation}
which generates the term 
\begin{equation}
\frac{1}{10\mpl} \int d^3x \, x^2 x^i J \, \partial_i \nabla^2 \bar{\varphi} \;,
\end{equation}
which upon use of the scalar equation of motion contributes to the dipole.
This can go on but we only need these terms for the order we are considering.

Finally, this gives for the interaction (up to order $v^2$),
\begin{align}
\begin{split}
S_{\rm int}^{(\varphi)} =  \frac{1}{\mpl} \int dt \left( I_\varphi \bar{\varphi} + I_\varphi^i \partial_i \bar{\varphi} + \frac{1}{2} I_\varphi^{ij} \partial_i \partial_j \bar{\varphi}  + \ldots \right) \;,
\end{split}
\label{eq:multipole_expansion_scalar}
\end{align}
where 
\be
\label{eq:monodipoquadru}
I_\varphi  \equiv  \int d^3 x \left( J + \frac{1}{6}  \partial_t^2 J  x^2 \right) \;, \quad I_\varphi^i  \equiv \int d^3x \, x^i \left(J + \frac{1}{10} \partial_t^2 J x^2 \right) \;, \quad  I_\varphi^{ij} \equiv \int d^3x J  Q^{ij} \;
\ee
are respectively the scalar monopole, dipole and quadrupole.

\subsubsection{Scalar monopole}

Let us discuss the coupling induced by the monopole, i.e.~the first term on the right-hand side of eq.~\eqref{eq:multipole_expansion_scalar}. At lowest order in $v$, the source is 
\begin{equation}
J_{v^0} = \sum_A \aaa_A m_A \delta^3(\mathbf{x}-\mathbf{x}_A) \;,
\label{eq:J0}
\end{equation}
which translates into a scaling 
\begin{equation}
S_\mathrm{mono}^{(\varphi)} \sim \frac{I_\varphi}{\mpl} \sim \sum_A \aaa_A \sqrt{Lv} \;.
\end{equation}
This could be potentially very constraining if compared to the  radiation graviton, which starts at the quadrupole order of $\sqrt{L v^5}$.
However, this gives a constant coupling and thus,  as we will see in Sec.~\ref{subsec:radiated_scalars}, eq.~\eqref{eq:radiated_power_scalar}, no scalar radiation is emitted.  Therefore, we need to go to higher order.

\begin{figure}
	\centering
	\subfloat[]{
%
%
%
\includegraphics{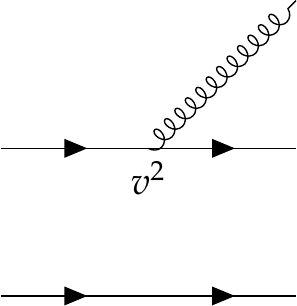}
		\label{subfig:J_v2_A}
	}\hspace{1em}
	\subfloat[]{
%
%
\includegraphics{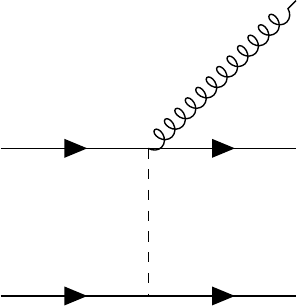}
		\label{subfig:J_v2_b}
	}\hspace{1em}
	\subfloat[]{
%
%
\includegraphics{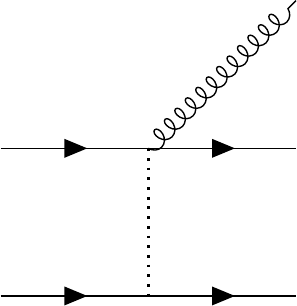}
		\label{subfig:J_v2_c}
	}\hspace{1em}
	\subfloat[]{
%
%
\includegraphics{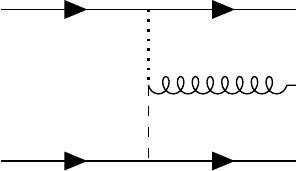}
		\label{subfig:J_v2_d}
	}\hspace{1em}

\caption[Feynman diagrams contributing to the emission of one scalar, at  order $v^2$.]{Feynman diagrams contributing to the emission of one scalar, at  order $v^2$. Diagrams that are not symmetric should be added with their symmetric counterpart.}
\label{fig:J_v2}
\end{figure}

Integrating out potential gravitons and scalars, we find  four diagrams that contribute to $J$ at order $\sqrt{Lv^5}$, all shown in Fig.~\ref{fig:J_v2}. 
The expression for these diagrams are (for convenience we suppress the $i \int dt$ in front of each diagram):
\begin{itemize}

\item Figure~\ref{subfig:J_v2_A}, coming from the $v^2$ term in $\int d\tau \varphi$ (see eq.~\eqref{eq:pp_Action_expanded}),
\begin{equation}
- \sum_A \aaa_A \frac{m_A v_A^2}{2} \frac{\bar{\varphi}(t, \mathbf{x}_A)}{\mpl} \;.
\end{equation}

\item Figure~\ref{subfig:J_v2_b}, coming from the $\bar \varphi H$ term in $\int d\tau \varphi$,\footnote{In~\cite{Huang:2018pbu}, a similar calculation of this diagram (denoted by $7b$ there), for a massive axion-type field, has been reported. We disagree with their result in the massless limit.}
\begin{equation}
- \frac{m_1 m_2 G}{r} \sum_A \aaa_A \frac{\bar{\varphi}(t, \mathbf{x}_A)}{\mpl}\;.
\end{equation}

\item Figure~\ref{subfig:J_v2_c}, coming from the $\bar{\varphi} \Phi$ term in $\int d\tau \varphi^2$,
\begin{equation}
4 \frac{m_1 m_2 G}{r}  \sum_A \bbb_A \aaa_{\bar{A}} \frac{\bar{\varphi}(t, \mathbf{x}_A)}{\mpl}\;,
\end{equation}
where for compactness we have introduced the notation $\aaa_{\bar{A}}$ for the symmetric parameter, i.e.~$\aaa_{\bar{1}} = \aaa_2$ and $\aaa_{\bar{2}}=\aaa_1$.

\item Figure~\ref{subfig:J_v2_d}, coming from the ${\varphi} \Phi H$ term of eq.~\eqref{eq:hPhi2Vertex}. This  vanishes as in the conservative case, because it involves the same projector $\frac{\eta^{\alpha \beta}}{2} P_{00,\alpha \beta} \delta_{ij} - P_{00,ij}$.

\end{itemize}
In conclusion, the complete expression for the coupling $J$ at  order $v^2$ is 
\begin{equation}
J_{v^2} = - \sum_A m_A \aaa_A \frac{v_A^2}{2} \delta^3(\mathbf{x}-\mathbf{x}_A) + \frac{m_1 m_2 G}{r} \sum_A (4 \bbb_A \aaa_{\bar{A}} - \aaa_A) \delta^3(\mathbf{x}-\mathbf{x}_A) \;.
\label{eq:J_v2}
\end{equation}

Let us now discuss the second term in $I_\varphi $, i.e.~$\int d^3x \frac{1}{6} (\partial_t^2 J_{v^0}) x^2$. To calculate it, we use the equations of motion for the point-particles  at lowest order in the velocity expansion, i.e.,
\be
\mathbf{\ddot{x}}_1 = - \frac{\tilde G_{12} m_2}{r^3} \mathbf{r} \;,  \qquad
\mathbf{\ddot{x}}_2 =  \frac{\tilde G_{12} m_1}{r^3} \mathbf{r} \;,
\ee
with the following center-of-mass relations, also valid  at lowest order in the velocity expansion,
\be
\mathbf{x}_1 = \frac{m_2}{m_1+m_2} \mathbf{r} \;, \qquad
\mathbf{x}_2 = - \frac{m_1}{m_1+m_2} \mathbf{r}  \;.
\label{eq:center_of_mass}
\ee

Summing up all contributions and using eq.~\eqref{eq:monodipoquadru}, we finally find
\begin{equation}
I_\varphi = - \frac{1}{6} \sum_A m_A \aaa_A v_A^2 + g_{12} \frac{G m_1 m_2}{r} \;,
\end{equation}
where  $g_{AB}$ is a symmetric combination of the scalar couplings $\aaa_A$, $\bbb_A$ and of the antisymmetric mass ratio $\kappa_{AB}$ defined in eq.~\eqref{eq:kappa}, \begin{equation}
g_{AB} \equiv \aaa_A(4\bbb_B-1)+\aaa_B(4\bbb_A-1) - \frac{1+2\aaa_A\aaa_B}{6}(\aaa_A+\aaa_B+\kappa_{AB}(\aaa_B-\aaa_A)) \;.
\end{equation}

\subsubsection{Scalar dipole}

The fact that scalar-tensor theories generically predict a dipole was first realized by Eardley~\cite{Eardley1975ApJ}. This could induce sizeable deviations from GR because the scalar dipole interaction term is \textit{a priori} of order $\sqrt{Lv^3}$. Using in eq.~\eqref{eq:monodipoquadru} the lowest-order expression for $J$ (eq.~\eqref{eq:J0}), we obtain the lowest-order contribution to the dipole, 
\begin{equation}
I_{\varphi, \rm -1PN}^i = \sum_A \aaa_A m_A x_A^i  \;.
\label{eq:scalar_dipole}
\end{equation}
For equal scalar charges of the two objects,  $\aaa_1 = \aaa_2$,  
the second derivative of $I^i_\varphi$ vanishes due to the conservation of the total momentum and there is no $-$1PN dipole radiation for equal scalar charges.
Thus, for two black holes or two comparable neutron stars, the effect of the dipole is very weak, while for a black hole-neutron star system it is maximal (a black hole has $\aaa_{\rm BH}=0$ in traditional BDT theories due to the no-hair theorem). See~\cite{Huang:2018pbu} for a detailed discussion.

Let us now compute the first-order (1PN) correction to this expression. To do that, we have to take into account the second term in $I^i_\varphi $  in eq.~\eqref{eq:monodipoquadru} coming from the trace part of the octupole, i.e.~$\int d^3x x^i \frac{1}{10} (\partial_t^2 J_{v^0}) x^2$, and the $v^2$ correction to the source from eq.~\eqref{eq:J_v2}. Adding these to the leading-order expression above gives
\begin{align} \label{eq:dipole_full}
\begin{split}
I_\varphi^i &= \sum_A \aaa_A m_A x_A^i + \frac{1}{10} \partial_t^2 \left(\sum_A m_A \aaa_A x_A^2 x_A^i \right) - \sum_A m_A \aaa_A \frac{v_A^2}{2} x_A^i \\
&+ \frac{G m_1 m_2}{r} \sum_A (4 \beta_A \aaa_{\bar{A}} - \aaa_A) x_A^i \;.
\end{split}
\end{align}

\subsubsection{Scalar quadrupole}

The scalar quadrupole interaction vertex is of order $\sqrt{Lv^5}$ and can be straightforwardly computed from eq.~\eqref{eq:monodipoquadru} and the lowest-order expression for $J$. One finds
\begin{equation}
I_\varphi^{ij} =  \sum_A \aaa_A m_A \left(  x_A^i x_A^j - \frac{1}{3}x_A^2 \delta^{ij} \right) \;.
\label{eq:scalar_quadrupole}
\end{equation}

\section{Dissipative dynamics \label{sec:dissipative_dynamics}}

Now that we have a definite expansion for the interaction Lagrangian in terms of multipole moments, we can calculate the power emitted in gravitational waves. As explained below, this can be computed from the imaginary part of the effective action for the two point-like bodies, $\hat S_\mathrm{eff}[x_A]$, obtained by integrating out the radiation fields, see eq.~\eqref{eq:finalSeff}. 

The real part of the effective action generates the coupled equations of motion for the two-body system.
If some energy leaves the system, then $\hat  S_\mathrm{eff}$ contains an imaginary part that is related to the power emitted. To see why this is the case by a simple example, we consider 
a scalar theory with a field $\phi$ coupled to an external source $J$ entering the action as  $\int d^4x J(x) \phi(x)$. 
The effective action obtained by integrating the field $\phi$ is given by the path integral
\begin{equation}\label{pathint}
e^{i S_{\rm eff}[J]}  = \int \mathcal{D}\phi \, e^{iS[\phi, J]} \equiv  Z[J] \;,
\end{equation}
where in the last equality we have defined the generating functional of the Green's functions $Z[J]$.
On the other hand, in the so-called ``in-out'' formalism, $Z$ is also the overlap between initial and final states, i.e.
\begin{equation}
Z[J] = \braket{0_+}{0_-}_J \;.
\end{equation}
From the two equations above, the vacuum transition amplitude between the asymptotic past and future differs from unity if the effective action is not real,
\begin{equation}
|\braket{0_+}{0_-}_J|^2 = e^{-2 \Im[S_{\rm eff}]}  \;.
\label{eq:trans}
\end{equation}
The difference with unity denotes the probability amplitude that particles are lost---or emitted---by the system. 
Expanding the right-hand side of this equation for small $\Im[S_{\rm eff}]$, this can be written as  
\begin{equation}
2 \Im[S_{\rm eff}] = T \int dE d\Omega \frac{d^2 \Gamma}{dEd\Omega} \;,
\label{eq:imaginaryPartAction}
\end{equation}
where $T$ is the  duration of the interaction and $d \Gamma$ is the differential rate for particle emission. The latter can be employed to calculate the radiated power via 
\be
P = \int dE d\Omega E \frac{d^2 \Gamma}{dEd\Omega} \;.
\label{eq:power}
\ee
We will use the two equations above to compute the power radiated into gravitons and scalar particles. 

\subsection{Radiated power}
\subsubsection{Gravitons}
\label{subsec:radiated_gravitons}

\begin{figure} 
	\centering
	\subfloat[]{
%
%
%
\includegraphics{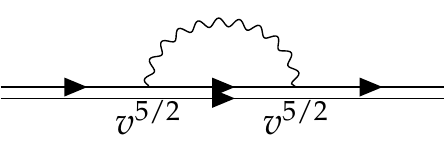}
		\label{subfig:radiation_grav_Seff}
	} \hspace{1em}
	\subfloat[]{
%
%
%
\includegraphics{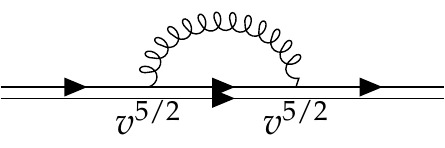}
		\label{subfig:radiation_scalar_Seff}
	} \hspace{1em}
	
\caption{Contribution of the radiation graviton and scalar to the imaginary part of the effective action.}
\label{fig:radiation_Seff}
\end{figure}
Let us first compute the power radiated into gravitons (see e.g.~\cite{goldberger_effective_2006,Kol:2007bc,galley_radiation_2009,Cardoso:2002pa}).  
In the classical approximation the path-integral~\eqref{pathint} is computed at the saddle point of the action,  $\hat S_{\rm eff}[x_A] = S_{\rm eff}[x_A, h_{cl}, \varphi_{cl}]$, and thus decomposes into the two diagrams of~Fig.~\ref{fig:radiation_Seff}: $\hat S_{\rm eff}=\hat S_{\rm eff}^{(h)}+ \hat S_{\rm eff}^{(\varphi)}$. The first term (Fig.~\ref{subfig:radiation_grav_Seff}), contains the interaction vertex of eq.~\eqref{eq:interaction_graviton_multipole}. In particular, using the Feynman rules from this equation we find 
\be
i  \hat S_{\rm eff}^{(h)} = - \frac12 \times \frac{1}{16 \mpl^2} \int dt_1 dt_2   I_h^{ij}(t_1) I_h^{kl}(t_2) \left\langle T \ddot {\bar h}_{ij}^{\rm TT} (t_1, \mathbf{0})   \ddot {\bar h}_{kl}^{\rm TT} (t_2, \mathbf{0}) \right\rangle \;,
\label{eq:Seffgrav}
 \ee
 where we have included the symmetry factor $1/2$ of the diagram.

 To find the propagator for $\bar{h}_{ij}^{\rm TT}$, we can first project $\bar{h}_{ij}$ on the transverse-traceless gauge.
 In terms of the unit vector $\mathbf{ n}$ denoting the direction of  propagation, we have
 \be
 \bar{h}_{ij}^{\rm TT} =  \Lambda_{ij,kl}  ({\mathbf{n}} ) \bar{h}_{kl} \;, 
 \ee 
 where we recall that
 \be
 \Lambda_{ij,kl}  ({\mathbf{n}} ) \equiv  (\delta_{ik}-n_in_k)(\delta_{jl}-n_jn_l)-\frac{1}{2}(\delta_{ij}-n_in_j)(\delta_{kl}-n_kn_l) \;,
 \ee
 is the projector onto transverse-traceless tensors defined in Chapter~\ref{Chapter3}.

Using the  $\bar{h}_{\mu \nu}$ propagator given by eqs.~\eqref{eq:propagator} and~\eqref{eq:Feyprop}, applying  the identities
\be
 \label{eq:ur}
\langle n_i n_j \rangle \ = \ \frac13  \delta_{ij}\;, \qquad \langle n_i n_j n_k n_l \rangle\  = \ \frac1{15}  (\delta_{ij} \delta_{kl} + \delta_{ik} \delta_{jl} + \delta_{il} \delta_{jk} ) \;,
\ee
which follow from the rotational symmetry of the integral, and symmetrizing over the indices $ij$ and $kl$, the expectation value in eq.~\eqref{eq:Seffgrav} can be written as 
 \be
\langle  T \ddot{ \bar{h}}_{ij}^{\rm TT} (t_1) \ddot{ \bar{h}}_{kl}^{\rm TT} (t_2) \rangle = \frac85 \left[ \frac12 (\delta_{ik} \delta_{jl} +\delta_{il} \delta_{jk} ) - \frac13 \delta_{ij} \delta_{kl}  \right]  \int  \frac{d^4 k}{(2\pi)^4}    \frac{-i (k_0)^4 }{k^2 - i \epsilon}  e^{i k_0 (t_1-t_2) } \;,
\ee
where the bracket on the right-hand side contains the projection operator
into symmetric and traceless two-index spatial tensors.
Plugging this expression into eq.~\eqref{eq:Seffgrav}, we find
\be
\hat S_{\rm eff}^{(h)} =  \frac{1}{20 \mpl^2} \int  \frac{d^4 k}{(2\pi)^4}    \frac{(k_0)^4}{k^2 - i \epsilon} | I_h^{ij}(k_0) |^2   \;,
 \ee
 where we have introduced the Fourier transform of the quadrupole moment, 
\be
I_h^{ij}(k_0) = \int dt I_h^{ij} (t) e^{i k_0 t}  \;.
\ee

To extract the imaginary part of the above action, we use the relation for the principal value of a function (denoted by PV). Specifically, we have
\begin{equation}
\frac{1}{k^2-i\epsilon} = \text{PV} \left( \frac{1}{k^2} \right) + i\pi \delta(k^2) \;, 
\label{eq:ImEpsilon}
\end{equation}
where 
\be
\delta(k^2) = \frac{1}{2 |k_0|} \left[ \delta(k_0 - |\mathbf{k}|) + \delta(k_0 + |\mathbf{k}|) \right] \;.
\label{eq:deltarel}
\ee
Using these relations, the imaginary part of the effective action reads
\be
{\Im}[\hat {S}_\mathrm{eff}^{(h)}]  = \frac{1}{20\mpl^2} \int \frac{d^3\mathbf{k}}{(2\pi)^3} \frac{|\mathbf{k}|^4}{2 |\mathbf{k}|}   | I_h^{ij}(|\mathbf{k}|)|^2 =\frac{ G}5 \int_0^\infty  \frac{d\omega \, \omega^5}{2 \pi}   | {I_h^{ij}}(\omega)|^2    \;,
 \ee
where for the second equality we have  integrated over the angles, used $1/\mpl^2 = 8 \pi G$, the on-shell condition $|\mathbf{k}| = \pm k_0$ and defined the emitted frequency as $\omega \equiv | k_0|$.
Comparing with eq.~\eqref{eq:imaginaryPartAction} and applying eq.~\eqref{eq:power}, the expression for the emitted power into gravitons  is 
\be
\begin{split} 
P_g &= \frac{ 2 G}{5  T}  \int_0^\infty \frac{d \omega \, \omega^6}{2 \pi} |{I}_h^{ij} (\omega)|^2 \\
&= \frac{G}{5 T }
 \int_{-\infty}^\infty dt \dddot{I}_h^{ij} (t) \dddot{I}_h^{ij} (t) \equiv \frac{G}{5 }  \big\langle \dddot{I}_h^{ij}\  \dddot{I}_h^{ij} \big\rangle \;,
\label{eq:quadrupole_graviton}
\end{split}
\ee
where in  the second line we have Fourier transformed back the multipoles  to real space and in the last equality we have used the brackets to denote the time average over many gravitational wave cycles.

\subsubsection{Scalars}
\label{subsec:radiated_scalars}

Let us turn now to the power radiated into scalars.
We will now calculate the imaginary part of the effective action $\hat S_{\rm eff}^{(\varphi)}$ obtained by integrating out the radiation scalars. This can be done by computing the self-energy diagram of Fig.~\ref{subfig:radiation_scalar_Seff}, the interaction vertices being the ones of eq.~\eqref{eq:multipole_expansion_scalar}.  
Note that the two vertices in Fig.~\ref{subfig:radiation_scalar_Seff} must be of the same multipole order---if they are not, the remaining indices should be contracted with rotationally invariant tensors, e.g.~$\delta_{ij}$ or $\epsilon_{ijk}$, but such expressions vanish because of the symmetry and the tracelessness of the multipole moments. By applying the multipole expansion derived in Sec.~\ref{subsubsec:multipoledec} and using the Feynman  rules, we get
\be
\begin{split}
i  \hat S_\mathrm{eff}^{(\varphi)} = & - \frac12 \times \frac{1}{\mpl^2} \int dt_1 dt_2 \bigg( I_\varphi(t_1) I_\varphi(t_2) \left\langle T \bar{\varphi}(t_1, \mathbf{0}) \bar{\varphi}(t_2, \mathbf{0}) \right\rangle \\
&+ I_\varphi^i(t_1) I_\varphi^j(t_2) \left\langle T  \partial_i \bar{\varphi}(t_1, \mathbf{0}) \partial_j \bar{\varphi}(t_2, \mathbf{0}) \right\rangle  +  \frac{1}{4} I_\varphi^{ij}(t_1) I_\varphi^{kl}(t_2) \langle T  \partial_i \partial_j \bar{\varphi}(t_1, \mathbf{0}) \partial_k \partial_l \bar{\varphi}(t_2, \mathbf{0}) \rangle \bigg)\;,
 \end{split}
\ee
where we have included again the symmetry factor of $1/2$ for this diagram. By using the expression of the $\bar{\varphi}$ propagator, eq.~\eqref{eq:propagatorscalar}, and the identities~\eqref{eq:ur}, we find
\be
  \hat S_\mathrm{eff}^{(\varphi)} = \frac{1}{2 \mpl^2} \int \frac{d^4 k}{(2 \pi)^4} \frac{1}{k^2 -i \epsilon} \bigg( | I_\varphi (k_0)|^2 + \frac13 |\mathbf{k}|^2   | I_\varphi^i (k_0)|^2 + \frac1{30} |\mathbf{k}|^4   | I_\varphi^{ij} (k_0)|^2  \bigg) \;, 
\ee
where we have introduced the Fourier transforms of the multipole moments, 
\be
I_\varphi(k_0) = \int dt I_\varphi (t) e^{i k_0 t} \;, \quad I_\varphi^i(k_0) = \int dt I_\varphi^i (t) e^{i k_0 t} \;, \quad I_\varphi^{ij}(k_0) = \int dt I_\varphi^{ij} (t) e^{i k_0 t} \;.
\ee

To extract the imaginary part of the above action we use once more eq.~\eqref{eq:ImEpsilon} with~\eqref{eq:deltarel}. By an analogous treatment  to that at the end of Sec.~\ref{subsec:radiated_gravitons}, we find the power emitted  into scalars,
\be 
P_\phi = {2G}\left[  \big\langle \dot{I}_\varphi^2  \big\rangle  + \frac{1}{3} \big\langle  \ddot{I}_\varphi^i \ddot{I}_\varphi^i  \big\rangle + \frac{1}{30} \big\langle \dddot{I}_\varphi^{ij}  \dddot{I}_\varphi^{ij}   \big\rangle \right] \;.
\label{eq:radiated_power_scalar}
\ee
Therefore, beside the quadrupole,  the monopole and the dipole~\cite{Eardley1975ApJ} contribute  as well  to the scalar radiation.

\subsection{Detected signal \label{subsec:detected_signal}}

\begin{figure}
	\centering
	\subfloat[]{
%
%
%
%
\includegraphics{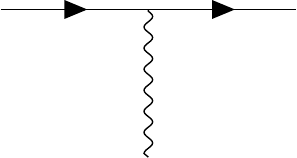}
		\label{subfig:field_isolated_object_graviton}		
	} \hspace{2em}
		\subfloat[]{
%
%
%
%
\includegraphics{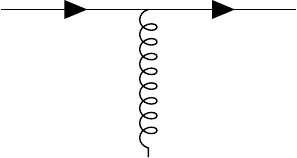}
	              \label{subfig:field_isolated_object_scalar}
} \hspace{1em}

\caption{Feynman diagram giving the radiation field emitted by an object with energy-momentum tensor $T^{\mu \nu}$.}
\label{fig:field_isolated_object}
\end{figure}

Here we compute the  radiation field in gravitons observed at the detector. To simplify the notation we remove the bar over the radiated fields. We need to evaluate the diagram of Fig.~\ref{subfig:field_isolated_object_graviton}---which amounts to find the solution of the equations of motion---but using  a retarded Green's function instead of the Feynman one, so as to enforce the physical nature of the external field. Using the coupling of a radiation graviton to matter directly expanded in multipoles, as found in eq.~\eqref{eq:interaction_graviton_multipole}, in the transverse-traceless gauge this gives \begin{align}
\begin{split}
h_{ij}^{\rm TT}(t, \mathbf{x}) 
= - \frac{i}{\mpl} \Lambda_{ij, kl} \int dt' G_R(t-t', \mathbf{x}) \ddot{I}_h^{kl} (t', \mathbf{0}) \;,
\end{split}
\end{align}
where $ G_R(t-t', \mathbf{x})$ denotes the retarded Green's function between the source located at $(t', \mathbf{0})$ and the observation made at $(t, \mathbf{x})$.
Note that the retarded Green's function is given by a different $i\epsilon$ prescription, which amounts to pick only physical waves modes. In particular, 
\begin{align}
\begin{split}
G_R(t-t', \mathbf{x} - \mathbf{x}') &= \int \frac{d^4k}{(2\pi)^4} \frac{-i}{-(k^0-i\epsilon)^2+\mathbf{k}^2} e^{-ik\cdot (x-x')} \\
&= \frac{i}{4\pi |\mathbf{x}-\mathbf{x}'|} \delta(t'-t-|\mathbf{x}-\mathbf{x}'|) \;.
\end{split}
\end{align}
The second equality comes from the residue theorem. 
Finally, the observed wave (normalized with the Planck mass, so as to agree with the GW literature) is  given by 
\begin{equation}
\frac{h_{ij}^{\rm TT} (t, \mathbf{x})}{\mpl} = \frac{2G}{R} \Lambda_{ij,kl} \ddot{I}_h^{kl}(t_\mathrm{ret}) \;,
\label{eq:obsgraviton}
\end{equation}
where $R $ is the distance to the source, $R = | \mathbf{x}|$, and  $t_\mathrm{ret} = t- R$ is the retarded time. 

The scalar waveform can be found by similar reasoning, evaluating  the diagram of Fig.~\ref{subfig:field_isolated_object_scalar} with  the coupling of a radiation field to matter directly expanded in multipoles, as  in eq.~\eqref{eq:multipole_expansion_scalar}.
Given an on-shell scalar wave propagating in the direction $\mathbf{n}$, we can use   $\partial_i \phi = - n_i \partial_t \phi$ and rewrite these couplings as
\begin{equation}
\hat{S}^{(\varphi)}_\mathrm{int} = \frac{1}{\mpl} \int dt \, \bar \varphi \left( I_\varphi + n_i \dot{I}_\varphi^i + \frac{n_i n_j}{2} \ddot{I}_\varphi^{ij}  \right) \;,
\end{equation}
so that the  observed radiation field into scalars reads 
\be
\varphi (t,\mathbf{x}) = \frac{i}{\mpl} \int dt'  \left( I_\varphi (t', \mathbf{0}) + n_i \dot{I}_\varphi^i (t', \mathbf{0}) + \frac{n_i n_j}{2} \ddot{I}_\varphi^{ij}  (t', \mathbf{0}) \right)  G_R(t-t', \mathbf{x})\;. 
\ee
By a treatment analogous to the one for gravitons, we find the radiated field away from the source,
\begin{equation}
\frac{\varphi(t, \mathbf{x})}{\mpl} = - \frac{2G}{R} \left( \left. I_\varphi + n_i \dot{I}_\varphi^i + \frac{n_i n_j}{2} \ddot{I}_\varphi^{ij}  \right)\right|_{t_\mathrm{ret}} \;.
\label{eq:obsscalar}
\end{equation}

We can now turn to the effect of the gravitational wave passage on the detector. Recall from Chapter~\ref{Chapter3}, eq.~\eqref{eq:motion_riemann}, that the separation $\xi_i$ between two test masses satisfies the geodesic deviation equation
\begin{equation}
\ddot{\xi}_i = - R_{i0j0}    \xi_j \;,
\label{eq:motion_riemann_chap4}
\end{equation}

In standard GR, computing the emitted gravitational wave in the transverse-traceless gauge (so that $h_{00} = h_{0i} = 0$) we have 
\be
R_{i0j0} = -\frac{1}{2 \mpl} \ddot{h}_{ij}^{\rm TT}\;.
\ee 
However, here the detector is non-minimally coupled to $g_{\mu \nu}$. Its scalar charge will generally depend on the local scalar field value (which may be different from the scalar environment of the binary objects) and on the  renormalization effects discessed in Sec.~\ref{sec2.2}.
Defining by $\aaa_{\rm det}$ the  scalar charge of the detector, this can be found by 
\begin{equation}
\int d\tilde{\tau} =  \int d\tau (1-\aaa_{\rm det}\varphi)\;,
\end{equation}
which tells us that, to linear order in the fields,  the physical metric is  
\begin{equation}
\tilde{h}_{\mu \nu} = h_{\mu \nu} - 2\aaa_{\rm det}\varphi \eta_{\mu \nu}\;.
\end{equation}

Using this metric to compute the components of the Riemann tensor in eq.~\eqref{eq:motion_riemann_chap4}, we find
\be
\begin{split}
R_{i0j0} &= \frac{1}{2\mpl} \left( \partial_i \dot{ \tilde{h}}_{0j} + \partial_j \dot{ \tilde{h}}_{0i} - \partial_i \partial_j \tilde{h}_{00} - \ddot{ \tilde{h}}_{ij} \right) \\
&= - \frac{1}{2\mpl} \ddot h_{ij}^{\rm TT} + \aaa_{\rm det} \left( \delta_{ij} \ddot \varphi - \partial_i \partial_j \varphi \right)  = - \frac{1}{2\mpl} \partial_t^2 \left[  h_{ij}^{\rm TT} - 2\aaa_{\rm det} \varphi \left( \delta_{ij} - n_i n_j \right)  \right] \;,
\label{eq:physical_GW}
\end{split}
\ee
where in the last equality we have used again $\partial_i \varphi = - n_i \dot \varphi$. Using the expressions for the observed graviton and scalar waves, eqs.~\eqref{eq:obsgraviton} and~\eqref{eq:obsscalar}, the detector will observe the following metric perturbation,
\be
\frac{h_{ij}^{\rm detector}}{\mpl}  = \frac{2G}{R} \left[  \Lambda_{ij,kl} \ddot I_h^{kl} + 2 \alpha_{\rm det} \left( \delta_{ij} - n_i n_j \right)  \left(  I_\varphi + n_k \dot{I}_\varphi^k + \frac{n_k n_l}{2} \ddot{I}_\varphi^{kl}  \right) \right]_{t_{\rm ret}} \;,
\label{eq:hij_detector}
\ee
For a wave propagating in the direction $ \mathbf{n}$, it is convenient to define the three polarization tensors 
\be
e_{ij}^+  \equiv \mathbf{e}_i \mathbf{e}_j -  \bar {\mathbf{e}}_i \bar {\mathbf{e}}_j \;, \qquad e_{ij}^\times \equiv \mathbf{e}_i \bar {\mathbf{e}}_j + \bar {\mathbf{e}}_i \mathbf{e}_j \;,  \qquad e_{ij}^\phi \equiv \mathbf{e}_i \mathbf{e}_j + \bar {\mathbf{e}}_i \bar {\mathbf{e}}_j  \;, 
\ee
where $\mathbf{e}$ and $ \bar {\mathbf{e}}$ are two unit vectors defining an orthonormal basis with $ \mathbf{n}$..
We can then decompose the metric into these three polarization states, 
\be
\frac{h_{ij}^{\rm detector}}{\mpl}  = \sum_{s= +, \times, \phi} e^s_{ij}( \mathbf{n})  h_s \;,
\ee
where for the two standard transverse-traceless polarizations (also defined in the introduction, Eq.~\eqref{eq:def_eij+x}) we have
\be
\label{eq:hplushcross}
h_{+, \times} = \frac{G}{R}  e_{ij}^{+, \times} ( \mathbf{n})  \ddot I_{h}^{ij}( {t_{\rm ret}}) \;,
\ee
while for the additional scalar polarisation we find
\be
h_\phi =  \frac{4 \alpha_{\rm det} G}{R}  \left(  I_\varphi + n_k \dot{I}_\varphi^k + \frac{n_k n_l}{2} \ddot{I}_\varphi^{kl}  \right) \Big|_{t_{\rm ret}}  \;.
\label{eq:scalar_polarization}
\ee

\subsection{Circular orbits}

We  now  compute  the wave amplitudes emitted by two binary objects in terms of the binary system parameters.
As before, we limit our calculation to the lowest post-Newtonian order. 

As the emission of GW circularizes the orbit, we assume a circular orbit in which the relative coordinate of the system, $\mathbf{r}= \mathbf{x}_1 - \mathbf{x}_2$, has cartesian components parametrized in time as 
\begin{align}
\begin{split}
r_x(t) &= r \cos(\omega t + \pi/2) \;, \\
r_y(t) &= r \sin(\omega t + \pi/2) \;, \\
r_z(t) &= 0 \;.
\end{split}
\end{align}
We  first assume that the frequency of the binary $\omega $ is constant. In the next subsection we will consider its time dependence due to the backreaction of the GW emission on the circular motion. For the following discussion it is convenient to define  the reduced mass of the system $\mu$ and the total mass $M$ as
\be
\mu \equiv \frac{m_1m_2}{M}  \;, \qquad M \equiv m_1+m_2 \;.
\ee

We chose the axis of rotation of the binary system to coincide with the $\mathbf{z}$ axis  while the propagation vector of the GW is  oriented in an arbitrary direction parametrized by the angles $\theta$ and $\phi$, 
\begin{equation}
\mathbf{n} \equiv (\sin \theta \sin \phi, \sin \theta \cos \phi, \cos \theta) \;.
\end{equation}
For the gravitational polarizations $h_+$ and $h_\times$, replacing the above expressions in the quadrupole moment given by eq.~\eqref{eq:quadrupole_gr}, and using this in eq.~\eqref{eq:hplushcross}, one finds (see e.g.~\cite{Maggiore:1900zz})
\begin{align}
\begin{split}
h_+ &= \frac{4 G \mu (\omega r)^2}{R} \left( \frac{1+\cos^2\theta}{2} \right) \cos(2\omega t_\mathrm{ret} +2\phi)\;,\\
h_\times &= \frac{4 G \mu (\omega r)^2}{R} \cos \theta \sin(2\omega t_\mathrm{ret} +2\phi) \;.
\end{split}
\end{align}
By using Kepler's third law to lowest order, i.e.
\be
\label{eq:Kepler}
\omega^2 = \frac{\tilde{G}_{12}M}{r^3} 
\ee
(we remind that $\tilde G_{12} = (1+2 \aaa_1 \aaa_2) G$, see eq.~\eqref{eq:Gab}), we find $\omega r = ( \tilde{G}_{12}M \omega)^{1/3}$.  
Note that this quantity scales as $v$. Using this expression in eq.~\eqref{eq:hgrav} to eliminate $r$,  we can rewrite the scalar waveform as 
\begin{align}
\label{eq:hgrav}
\begin{split}
h_+ &= \frac{4 G \mu}{R} (\tilde{G}_{12}M\omega)^{2/3} \left( \frac{1+\cos^2\theta}{2} \right) \cos(2\omega t_\mathrm{ret} +2\phi) \;, \\
h_\times &= \frac{4 G \mu}{R} (\tilde{G}_{12}M\omega)^{2/3} \cos \theta \sin(2\omega t_\mathrm{ret} +2\phi) \;.
\end{split}
\end{align}

Let us now turn to the scalar polarization, given by eq.~\eqref{eq:scalar_polarization}. 
For circular motion  the monopole term is constant in time and can be discarded. Using the center-of-mass relation~\eqref{eq:center_of_mass} to compute the time derivative of the dipole, $n_k \dot{I}_\varphi^k$ in eq.~\eqref{eq:scalar_dipole}, 
and eliminating the $r$ dependence using eq.~\eqref{eq:Kepler} above, we find the dipolar scalar emission to lowest order,
\begin{align}
\label{dipoleee}
h^{\rm dipole}_\phi = - \frac{4\aaa_{\rm det} G \mu}{R}  (\aaa_1 - \aaa_2) (\tilde{G}_{12}M\omega)^{1/3} \sin \theta \sin(\omega t_\mathrm{ret}+\phi)  \;.
\end{align}
 Similarly, we can compute the second time derivative of the quadrupole moment, ${n_k n_l} \ddot{I}_\varphi^{kl}$  in eq.~\eqref{eq:scalar_dipole}, and find the quadrupolar scalar emission,
\begin{align}
\label{quadrupoleee}
h^{\rm quadrupole}_\phi = - \frac{4\aaa_{\rm det} G \mu}{R} \frac{\aaa_1 m_2 +\aaa_2 m_1}{M}  (\tilde{G}_{12}M\omega)^{2/3} \sin^2 \theta \cos(2\omega t_\mathrm{ret}+2\phi)   \;.
\end{align}

Few comments are in order here. First, notice that  $\aaa_{\rm det}$ is the coupling of the detector to the scalar, while the $\aaa_A$'s are the renormalized couplings of the inspiral objects, which can depend on their masses. Since $\alpha \ll 1$, we expect the scalar amplitude of the GW to be suppressed with respect to the gravitational one.
Second, comparing the powers of the combination $(\tilde{G}_{12}M\omega)^{1/3} \sim v$ in eq.~\eqref{dipoleee} and in eqs.~\eqref{eq:hgrav} and~\eqref{quadrupoleee}   confirms that the dipole is of 0.5PN order less than the gravitational quadrupole, as expected.

\subsection{Frequency dependence}
\label{subsec:frequency_dependance}
Because the number of gravitational wave oscillations within a typical LIGO/Virgo event is very large, gravitational wave detectors are much more sensitive to a phase change rather than a modification of the amplitude. 
For this reason, in this subsection we will compute the frequency dependence of the waveform  from our formalism.

To this aim, we can use the energy balance of the system, i.e.~that the  total power loss is equal to the time derivative of the orbital energy. This reads
\be
\label{energy_balance}
P_g+P_\phi^\mathrm{monopole}+ P_\phi^\mathrm{dipole} + P_\phi^\mathrm{quadrupole} =  - \frac{dE}{dt} \;,
\ee
where $P_g$, $P_\phi^\mathrm{monopole}$, $P_\phi^\mathrm{dipole}$ and $P_\phi^\mathrm{quadrupole}$ are respectively the graviton, and the scalar monopole,  dipole and quadrupole  contributions to the emitted power.
As explained above, since the scalar monopole is constant for circular orbits, its emitted power   vanishes, $P_\phi^\mathrm{monopole}=0$. The orbital energy is given by
\be
E\equiv -\frac{\tilde{G}_{12}m_1m_2}{2r} = - \frac12 (\tilde{G}_{12} M_c \omega )^{2/3} M_c\;,
\ee
where for the last equality we have used again the Kepler's law and we have defined the chirp mass,
\be
M_c \equiv \frac{(m_1 m_2)^{3/5}}{M^{1/5}}= \mu^{3/5} M^{2/5}\;.
\ee

From eq.~\eqref{eq:quadrupole_graviton}, the power emitted  into gravitons reads
\begin{equation}
P_g = \frac{32}{5} G \mu^2 \omega^6 r^4 \;
\end{equation}
and, using again Kepler's law, one can rewrite this  as 
\begin{equation}
P_g = \frac{32}{5\tilde{G}_{12} (1+2\aaa_1 \aaa_2)} (\tilde{G}_{12} M_c \omega)^{10/3} \;.
\end{equation}

One can then proceed analogously for the power emitted into scalars. At lowest order in $v$, the power emitted by the scalar dipole contribution reads
\begin{equation} \label{eq:power_dipole}
P_\phi^\mathrm{dipole} = \frac{2}{3\tilde{G}_{12} (1+2\aaa_1 \aaa_2)} (\aaa_1-\aaa_2)^2 \nu^{2/5} (\tilde{G}_{12} M_c \omega)^{8/3} \;,
\end{equation}
where 
\be
\nu \equiv \frac{m_1m_2}{M^2}
\ee 
is the symmetric mass ratio.

We have derived eq.~\eqref{eq:power_dipole} at lowest order in the velocity expansion. But the quadrupolar power is suppressed by $v^2$ compared to the dipolar one (the Feynman diagrams of Fig.~\ref{fig:radiation_Seff} giving the radiated power involve two interaction vertices that are respectively of order $\sqrt{Lv^3}$ for the dipole and $\sqrt{Lv^5}$ for the quadrupole). We need therefore to compute the dipolar power at next-to-leading order so as to find an expression consistent with the quadrupolar order. 
To simplify the discussion, 
we  discard this correction  here. This approximation can then be used when $\aaa_1-\aaa_2 \lesssim v$, so that the dipole is smaller or of the same order as the quadrupole and its $v^2$ corrections are thus negligible or, alternatively, when $\aaa_1-\aaa_2 \gg v$, in which case the dipole dominates and we can ignore the quadrupolar terms. 
For completeness, we compute the dipolar power at next-to-leading in App.~\ref{app:dipolar}.

Finally, from eq.~\eqref{eq:radiated_power_scalar} one finds that the power emitted by the scalar quadrupole contribution is proportional to that of the gravitational   quadrupole, i.e.,
\begin{equation}\label{eq:scalar_quadrupole_power}
P_\phi^\mathrm{quadrupole} =   \frac{\left( \aaa_1 m_2 +\aaa_2 m_1\right)^2 }{3 M^2}  P_g \;.
\end{equation}

Using these expressions into the left-hand side of the energy balance equation, eq.~\eqref{energy_balance}, we can find a differential equation for the time derivative of the frequency. Following~\cite{Will:1994fb}, it is convenient to define the scalar-tensor chirp mass, 
\be
\tilde M_c^{5/3} \equiv \frac{M_c^{5/3}}{1+2\aaa_1 \aaa_2} \left[1+\frac{(\aaa_1 m_2 + \aaa_2 m_1)^2}{3M^2} \right] \; , 
\ee
and the dipole parameter,
\be
b \equiv \frac{5}{48} (\aaa_1-\aaa_2)^2 \frac{M_c}{\tilde M_c}  \; .
\ee
In terms of these quantities, the evolution equation for $\omega$ reads
\begin{equation} \label{eq:evolution_omega_rescaled}
\dot \omega = \frac{96}{5} (\tilde G_{12} \tilde M_c)^{5/3} \omega^{11/3} \left[1+b \nu^{2/5} (\tilde G_{12} \tilde M_c \omega)^{-2/3} \right] \; .
\end{equation}

We compute the total phase accumulated in the GW detector,  focussing on the quadrupole.
This reads
\begin{equation}
\Phi_{\rm quadrupole} = 2 \int dt \, \omega(t) = 2 \int d\omega \frac{\omega}{\dot \omega} \; ,
\end{equation}
where the factor of two comes from the frequency dependence of the quadrupolar waveform~\eqref{eq:hgrav}. Expanding for small $b (\tilde G_{12} \tilde M_c \omega)^{-2/3} \ll 1$, we can integrate eq.~\eqref{eq:evolution_omega_rescaled} to get
\begin{equation}
\Phi_{\rm quadrupole} = \frac{1}{16} \left[ (\tilde G_{12} \tilde M_c  \pi f)^{-5/3}  - \frac{5}{7} b \nu^{2/5} (\tilde G_{12} \tilde M_c  \pi f)^{-7/3} \right]_{f_\mathrm{in}}^{f_\mathrm{out}} \; ,
\end{equation}
where we used $f \equiv  \omega/ \pi$ to convert the angular frequency of the binary system into the GW frequency emitted by the quadrupole \footnote{Note that there could be a conformal rescaling of the frequency from the time of the GW emission to the one of its detection, due to the cosmological evolution of the field. See the end of Sec.~6 of~\cite{damour_tensor-multi-scalar_1992}. }.
Moreover, $f_\mathrm{in}$ ($f_\mathrm{out}$) denotes the frequency at which the GW signal enters (exits) the  detector. For LIGO/Virgo, we have $f_\mathrm{in} \sim 10\,$Hz $\ll f_\mathrm{out} \sim 1\,$kHz. By requiring that the phase modification is less than $\pi$, we obtain the following approximate bound on the dipole parameter $b$,
\begin{equation}
b \lesssim \frac{112 \pi}{5} \nu^{-2/5} (\tilde G_{12} \tilde M_c \pi f_\mathrm{in})^{7/3}  \simeq 10^{-6}\; .
\end{equation}
Note that the strongest constraint comes from the beginning of the inspiral, when the signal at $f_\mathrm{in} \sim 10$ Hz enters the detector. Our results are in agreement with earlier work by Will~\cite{Will:1994fb}, which uses the sensitivities $s_A$ defined in eq.~\eqref{eq:sensitivity}, instead of the parameters $\alpha_A$. Moreover, the  waveform in BDT gravity has been computed up to 2PN order in~\cite{Sennett:2016klh}.

\section{Concluding remarks}

To finish this Chapter, let us highlight the main differences of BDT theories with respect to GR in the NRGR approach.
First, renormalisation of the scalar charge implies violations of the strong equivalence principle in that the scalar charge of strongly self-gravitating objects like neutron stars depends on their gravitational self-energy.
Second,
the additional power loss in scalar radiation (monopole, dipole and quadrupole) modifies the dynamics of the system,  and so the time evolution of the frequency of the GW,  which ultimately modifies its phase.  Moreover, the dipole radiation, proportional to the {\it difference} of the scalar charges of the two bodies, has the same frequency  $\omega$ as the binary, as opposed to the quadrupole radiation that has frequency $2\omega$. Finally, in the presence of a scalar field, there will be an additional polarization associated to GW (see eq.~\eqref{eq:physical_GW}). It should be noted that these last two effects modify the amplitude of the GW signal, which is far less constrained than the phase by detectors.

The model that we have considered is an important test bench for generalizing the EFT formalism of~\cite{goldberger_effective_2006} to modified gravity, but it also contains obvious limits. Perhaps the most serious one is that its observational signatures will be very hard to detect. The departures from GR that we have just mentioned are proportional to the scalar coupling $\alpha$, which however is constrained to be less than $10^{-2}$ by Solar System tests (see Eq.~\eqref{eq:constraint_gamma}). More realistic models of dark energy, on the other hand, contain non-linearities in the scalar dynamics that are difficult to deal with. These kind of non-linearities will be discussed in Part~\ref{part3}.

\chapter{Disformal inspiralling} \label{Chapter5}

As discussed in part~\ref{part1},  dark energy models generally feature non-linearities that become important in the vicinity of a massive body and can screen the effect of the scalar field. When present, such non-linearities  make our diagrammatic expansion meaningless. In the Feynman perturbative expansion, propagators represent the free part of the Lagrangian, which dominates the dynamics, while interactions are treated perturbatively.
This is no longer the case close to the source.
 
However, we can consider an extension to the models studied in the last Chapter  in which we can trust our usual propagator and where non-linearities show up in a more subtle way. We will now follow the published work \citeK{Brax:2019tcy}.
Consider the {\it disformal coupling} (see Section~\ref{subsec:intro_disf}), relating the Jordan  frame metric $\tilde{g}_{\mu \nu}$ to the Einstein frame one $g_{\mu \nu}$,
\begin{equation}
\label{eq:jordan_frame_metric}
\tilde{g}_{\mu \nu} = A^2(\varphi) g_{\mu \nu} + B(\varphi) \partial_\mu \varphi \partial_\nu \varphi\, .
\end{equation}
In this equation, $A^2(\varphi)$ is the conformal coupling already studied in the last Chapter, and $B(\varphi)$ is the disformal coupling. Expanding for weak-field values $\varphi/\mpl \ll 1$, we get that the point-particle action considered in Eq.~\eqref{eq:action}, namely the proper time of the particle $A$ in the Jordan frame,
\begin{equation}
S_\mathrm{pp, \; A} = - m_A \int \mathrm{d} \tilde \tau_A \; ,
\end{equation}
where $\mathrm{d} \tilde \tau_A^2 = - \tilde g_{\mu \nu} \mathrm{d}x^\mu \mathrm{d}x^\nu$,
 is now given by the expansion
\begin{equation} \label{eq:pp_disf}
S_\mathrm{pp, \; A} = -m_A \int d\tau_A \left(1 - \alpha \frac{\varphi}{\mpl} - \beta \frac{\varphi^2}{\mpl^2}  + \frac{1}{\Lambda^2 \mpl^2} \left( \partial_\mu \varphi \frac{\mathrm{d}x^\mu_A}{\mathrm{d} \tau_A} \right)^2 \right) \; ,
\end{equation}
where we have written the lowest-order (dimensionful) parameter $B$ as $B(\varphi) = - 2 / \mpl^2 \Lambda^2$ (recall from Section~\ref{subsec:intro_disf} that one should have $B < 0$). $\Lambda$ is a new energy scale which characterizes the strength of the disformal interaction. A similar disformal coupling has been studied in theories of dark energy (see e.g.~\cite{Gleyzes:2015pma}), where its natural value  is $\Lambda \sim H_0$. In the above equation, we have focused on \textit{universal} scalar couplings $A^2$ and $B$ since it will be sufficient to capture the essential features of the disformal coupling. As such, there will be no dipole radiation since we have seen in Eq.~\eqref{eq:power_dipole} that the scalar dipole is proportional to the \textit{difference} of scalar charges of two objects. The dominant scalar contributions to the dissipated power will be the monopole and quadrupole, see Eq.~\eqref{eq:scalar_quadrupole_power}.

From Eq.~\eqref{eq:pp_disf}, we see that the disformal coupling
induces at lowest order, on top of other various terms included in our point-particle action~\eqref{eq:pp_Action_expanded}, a vertex
\begin{equation}
- \frac{m_A}{\Lambda^2 \mpl^2} \int \mathrm{d} t (\partial_\mu \varphi v_A^\mu)^2\, ,
\label{eq:disformal_vertex}
\end{equation}
where we recall that $v_A^\mu = (1, \mathbf{v})$.

Let us stress the difference between such a non-linear coupling and those displayed by $k$-mouflage and Galileon-like theories in screened regions. The latter contain non-linear terms in the evolution equation of the scalar field, that show up directly in the spherically symmetric solution of the scalar field configuration generated by a static source. Equivalently, these terms, which become leading close to the source, do not allow to use the standard propagator in a diagrammatic expansion. Here, non linearities are all  hidden in the  coupling to the point-particle. In the  vacuum the field obeys the usual Laplace equation, $\nabla^2 \varphi = 0$, and does not exhibit any transition to a Vainshtein regime at small radii. The standard spherically symmetric/static analysis (equivalently, the one-body diagrams like those, say, of Fig.~\ref{fig:field_isolated_object}) cannot grasp the non-linear dynamical aspects of the disformal model, which are encoded in the velocity dependent {\it two} body diagrams like the one of Fig.~\ref{fig:disformal}.

As such, the disformal coupling will give rise to new terms in the conservative and dissipative dynamics of binary systems.  We will establish the remarkable property that for circular orbits the effects of the disformal coupling vanish up to the seven Post-Newtonian (7PN) order for the conservative part of the dynamics, and 6PN for the dissipative dynamics. Hence only eccentric orbits are sensitive to the presence of the disformal interaction (it will allow us to bound the disformal coupling from the observation of the Hulse-Taylor pulsar in Section~\ref{sec:Hulse_Taylor}). This can be intuitively understood from the fact that the disformal coupling to a point-like object is a time derivative, i.e from Eq.~\eqref{eq:pp_disf} it reads
\begin{equation}
\int d\tau_A \left( \frac{d\varphi}{d\tau_A} \right)^2 \; .
\end{equation}
 As this time derivative vanishes when the  relative distance  $r$ between two orbiting objects is constant for circular orbits, we conclude that the effects of the disformal coupling will only appear when radiation-reaction effects must be taken into account.

Before moving on, we will do some order-of-magnitude estimates. As the scalar field generated by a massive body of mass $m$ in the Newtonian approximation is
\begin{equation}
\varphi \simeq \frac{\alpha m}{4\pi \mpl r}
\end{equation}
where $r$ is the distance to the source,
the condition $\varphi/\mpl \lesssim 1$  translates into $r \gtrsim \tilde r_s$ where
\begin{equation}
\tilde r_s = \alpha \frac{m}{4\pi \mpl^2}
\end{equation}
is the Schwarzschild radius of the source $r_s = 2 G m$  corrected by a factor $\alpha$. As our perturbative treatment is only valid well outside the Schwarzschild radius
of the moving bodies and as $\alpha \ll 1$ from the Cassini bound~\eqref{eq:constraint_gamma}, we conclude that the series expansion in $\varphi/\mpl$ is always valid where the
perturbative treatment of the motion of moving objects can be applied.

By requiring that the disformal factor in eq.~\eqref{eq:jordan_frame_metric} should be less that the conformal one in order to recover Newtonian mechanics, we obtain the following bound
\begin{equation} \label{eq:small_ratio}
\frac{\alpha m}{4\pi \mpl^2 \Lambda r^2} \lesssim 1
\end{equation}
which simply means that our theory is valid in the range $r \gtrsim r_\star$, where $r_\star$ is the nonlinear radius
\begin{equation} \label{eq:nonlinear_radius}
r_\star = \sqrt{\frac{\alpha m}{4 \pi \mpl^2 \Lambda}} \; .
\end{equation}
For distances $r < r_*$, the perturbative calculation that we use in this Chapter will be invalid. One can tune the mass scale $\Lambda$ so that $r_*$ is of the same order as the Schwarzschild radius for an object like the Sun (so that we have interesting departures from GR appearing in the waveform generated by two inspiralling black holes or neutron stars) which gives then the scaling relation
$
r_\star = r_{s, \odot} \sqrt{\frac{m}{m_\odot}}
$
for objects of different masses.
One may worry that this could  give  too large a nonlinear radius for small mass objects, however this is not the case. For example, a body of mass $m=1$ kg would have a nonlinear radius  $r_* = 10^{-12}$ m, consequently usual matter has a nonlinear radius well within its extension.

Finally, let us mention an interesting new development on two-body disformal couplings which goes beyond the scope of the present work.
As already stressed, in this Chapter we consider computations perturbative in the disformal coupling so that the nonlinear radius $r_*$ should be of the order of the Schwarzschild radius of solar-mass objects. However, one can take a totally different path and assume that the nonlinear radius is much bigger, say larger than the solar system size (in cosmological applications where $\Lambda \sim H_0$, $r_*$ is nearly of the parsec size). The perturbative computations which we use in this Chapter would be invalid, and we should evaluate the correction to the energy by other techniques. This ambitious task has been initiated in Ref.~\cite{Davis:2019ltc}: by using a resummation technique, the authors showed that the disformal coupling could give rise to a new screening mechanism. This new effect is still the subject of active research.

\section{Disformal correction to the conservative dynamics}

\subsection{Conservative disformal diagram}

\begin{figure}
	\centering
	\includegraphics{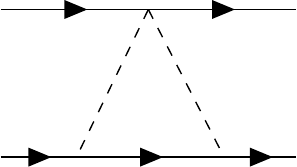}
\caption[Feynman diagram corresponding to the first disformal correction to the conservative dynamics]{Feynman diagram corresponding to the first disformal correction to the conservative dynamics (it should also be included with its symmetric counterpart). The upper vertex is the disformal one in eq.~\eqref{eq:disformal_vertex}.}
\label{fig:disformal}
\end{figure}

We now calculate the diagram given in Figure~\ref{fig:disformal} using the rules of NRGR explained in Section~\ref{sec:feynman}. Using the expression of the proper time given in eq.~\eqref{eq:proper_time}  we find:
\begin{equation}
\mathrm{Fig} \;~\ref{fig:disformal} = - \frac{1}{2} \frac{im_1}{\Lambda^2 \mpl^2} \int dt_1 \frac{(-i)^2\alpha^2m_2^2}{\mpl^2} \int dt_2 dt_3 \left\langle T (\partial_\mu \varphi(t_1, x_1) v_1^\mu)^2 \varphi(t_2, x_2) \varphi(t_3, x_3) \right\rangle
\end{equation}
where the $1/2$ is the symmetry factor of the diagram.
Using the Fourier transform of the field and Wick's contraction we have
\begin{align}
\begin{split}
\mathrm{Fig} \;~\ref{fig:disformal} &= - i \frac{\alpha^2m_1m_2^2}{\Lambda^2\mpl^4} \int dt_1 dt_2 dt_3 \frac{d^3k_1 d^3k_2}{(2\pi)^6} \frac{1}{k_1^2k_2^2} e^{i\mathbf{k}_1 \cdot (\mathbf x_1(t_1) - \mathbf x_2(t_2))} e^{i\mathbf{k}_2 \cdot (\mathbf x_1(t_1) - \mathbf x_2(t_3))} \\
& \times \left[\frac{d}{dt_1}\delta(t_1-t_2) + \delta(t_1-t_2)i\mathbf k_1 \cdot \mathbf v_1 \right] \left[\frac{d}{dt_1}\delta(t_1-t_3) + \delta(t_1-t_3)i\mathbf k_2 \cdot \mathbf v_1 \right]
\end{split}
\end{align}
Using $\frac{d}{dt_1}\delta(t_1-t_2) = - \frac{d}{dt_2}\delta(t_1-t_2)$ and integrating by parts, we have
\begin{align}
\begin{split}
\mathrm{Fig} \;~\ref{fig:disformal} &= i \frac{\alpha^2m_1m_2^2}{\Lambda ^2\mpl^4} \int dt \frac{d^3k_1 d^3k_2}{(2\pi)^6} \frac{1}{k_1^2k_2^2} e^{i\mathbf{k}_1 \cdot (\mathbf x_1 - \mathbf x_2)} e^{i\mathbf{k}_2 \cdot (\mathbf x_1 - \mathbf x_2)} \\
& \times \left[\mathbf k_1 \cdot (\mathbf v_1 - \mathbf v_2) \right] \left[\mathbf k_2 \cdot (\mathbf v_1 - \mathbf v_2) \right]
\end{split}
\end{align}
Finally with the formula $\int \frac{d^3k}{(2\pi)^3} \frac{\mathbf k}{k^2} e^{i \mathbf k \cdot \mathbf r} = i \frac{\mathbf r}{4\pi r^3}$, with $\mathbf r = \mathbf x_1 - \mathbf x_2$ we get
\begin{equation}
\mathrm{Fig} \; ~\ref{fig:disformal} = - i \frac{\alpha^2m_1m_2^2}{16 \pi^2 \Lambda^2\mpl^4} \int dt \frac{(\mathbf n \cdot (\mathbf v_1 - \mathbf v_2))^2}{r^4}
\end{equation}
with $\mathbf n = \mathbf r/r$. Upon using Planck's mass definition $G = 1/8\pi \mpl^2$ and summing over the symmetric diagram, we finally find the formula for the first disformal correction to the two-body Lagrangian given by

\begin{equation}
L_\mathrm{disf} = - \frac{4 \alpha^2 G^2 m_1 m_2 (m_1 + m_2)}{\Lambda^2} \frac{(\mathbf n \cdot (\mathbf v_1 - \mathbf v_2))^2}{r^4}.
\label{eq:disformal_energy}
\end{equation}

This correction recovers the previous result of Ref.~\cite{Brax:2018bow}.


\subsection{Energy for elliptic orbits} \label{sec:planar_dynamics}

In Newtonian mechanics, the two body problem is easily solved and trajectories in the centre of mass frame become elliptical for bound systems with planar trajectories parameterised as
\begin{equation} \label{eq:elliptic_trajectory}
r(\psi) = \frac{a(1-e^2)}{1+e \cos \psi}
\end{equation}
where $a$ is the semi-major axis, $e$ the eccentricity, and $\psi$ is the angle of the trajectory in the plane defined by the motion. We also define $p=a(1-e^2)$.  The conservation of angular momentum leads to the following expression for the time derivative of $\psi$
\begin{equation} \label{eq:psidot}
\dot \psi = \left( \frac{\tilde G M}{p^3} \right)^{1/2} (1+e \cos \psi)^{2}
\end{equation}
where $\tilde G = G(1+2\alpha^2)$ is the renormalised Newton constant that appears in the gravitational force when the leading effect of the conformal interaction is taken into account, i.e. the strength of the Newtonian force is enhanced compared to the purely gravitational case.
Defining by $M=m_1+m_2$ the total mass of the two bodies and $\mu= \frac{m_1 m_2}{m_1+m_2}$ the reduced mass, the energy of the system becomes
\be
E= -\frac{\tilde G M\mu}{2a} \; .
\ee

The disformal interaction produces a correction to this Newtonian energy. From the expression of the disformal Lagrangian~\eqref{eq:disformal_energy}, this correction to the energy is
\be
\delta E= \frac{4\alpha^2 G^2 \mu M^2}{\Lambda^2} \frac{\dot r^2}{r^4} \; .
\ee
where $\dot r = \frac{d r}{dt}$. For elliptical trajectories~\eqref{eq:elliptic_trajectory}, this becomes
\be
\delta E= \frac{4\alpha^2 }{(1+2\alpha^2)^2} \frac{\mu \tilde G^3 M^3}{\Lambda^2 p^5} e^2 \sin ^2 \psi (1+e \cos \psi)^4
\ee
Over one period $T$ of the system, the mean energy correction is defined as
\begin{align}
\begin{split}
\left \langle \delta E \right \rangle &= \frac{1}{T} \int_0^T dt \; \delta E(t) \\
 &= \frac{1}{2\pi} (1-e^2)^{3/2} \int_0^{2\pi} d\psi \frac{\delta E(\psi)}{(1+e\cos \psi)^{2}}
\end{split}
\end{align}
where we have used the time derivative of $\psi$ in eq.~\eqref{eq:psidot} and $T= 2\pi ( a^3 / (\tilde G M) )^{1/2}$  using Kepler's third law. This gives for the correction to the energy due to the disformal interaction
\begin{equation} \label{eq:disformal_energy_elliptic}
\frac{\left \langle \delta E \right \rangle}{E} = - \frac{4\alpha^2}{(1+2\alpha^2)^2} \frac{\tilde G^2 M^2}{\Lambda^2 a^4} \frac{e^2(1+\frac{e^2}{4})}{(1-e^2)^{7/2}} \; .
\end{equation}
A few comments are in order. First, note that the disformal energy~\eqref{eq:disformal_energy_elliptic} vanishes for circular orbits where $e=0$. Unfortunately, circular trajectories are almost always expected for the LIGO/Virgo detector, as the emission of GW circularises the trajectory~\cite{PhysRev.131.435}. However, the space interferometer LISA should be sensitive to trajectories for Extreme Mass Ratio Inspirals, which are expected to be elliptical~\cite{AmaroSeoane:2007aw} and so could provide a constraint on the disformal energy. Second, note that the disformal energy is proportional to the conformal factor $\alpha^2$ which is strongly constrained by the Cassini bound~\eqref{eq:constraint_gamma} as $\alpha^2 \lesssim 10^{-5}$. Lastly, we can rewrite the dimensionless parameter that appears in the ratio of energies in two different suggestive ways; the first one is
\begin{equation}
\frac{\alpha^2 \tilde G^2 M^2}{\Lambda^2 a^4} \sim \frac{\alpha^2 v^4}{\Lambda^2 r^2}
\end{equation}
which simply follows from the virial theorem. This shows that the disformal term induces a 2PN correction to the energy. As the best bound on $\Lambda$ that we will get from the Hulse-Taylor binary pulsar energy loss is $\Lambda \gtrsim 10^{-17}$ eV, the correction to the 2PN Hamiltonian of GR shows up for distances
\begin{equation}
r \sim \frac{\alpha }{\Lambda}  \lesssim 1000 \; \mathrm{km} \;
\end{equation}
We stress that this is not in tension with the LIGO/Virgo observation of GW, as the observed orbits are circular and the disformal energy vanishes in this case (we will be more precise on this statement in Sec.~\ref{sec:circular_disf}).

A second suggestive way to rewrite the dimensionless parameter is
\begin{equation} \label{eq:dimensionless_param}
\frac{\alpha^2 \tilde G^2 M^2}{\Lambda^2 a^4} = \left( \frac{r_\star}{a} \right)^4
\end{equation}
where $r_*$ is the nonlinear radius~\eqref{eq:nonlinear_radius} associated with the total mass $M$. Hence at the limit of validity of the perturbative treatment when $a\simeq r_\star$, this ratio could become close to unity. When the non-linear radius is tuned to be of the order of the Schwarzschild radius for solar mass objects, and close to the merger when the two objects  approach their Schwarzschild radius in relative distance  this parameter will become close to unity and nonperturbative effects should appear, perhaps in the form suggested in Ref.~\cite{Davis:2019ltc}.


\subsection{A theorem for circular orbits} \label{sec:circular_disf}

The first disformal contribution to the relativistic Lagrangian in the case of a circular orbit vanishes. In this part we will show that this vanishing is also valid at higher order in the PN expansion. On the other hand radiation reaction at  2.5PN order renders the two-body orbit inspiralling instead of circular, which will eventually gives rise to  a non-zero contribution of the  disformal coupling to the equations of motion. In this Section we investigate the correction to the two-body Lagrangian which do  not vanish for circular orbits.

Let us first notice that the disformal vertex can be rewritten as
\begin{equation}
- \frac{m_A}{\Lambda^2 \mpl^2} \int d\tau_A \left( \frac{d\varphi(x_A)}{d\tau_A} \right)^2
\end{equation}
where the total derivative of $\varphi$ is taken along the path of the particle $x_A(\tau_A)$. Intuitively, the fact that the disformal coupling does not contribute for circular orbits can be directly seen from this vertex. Indeed, since the scalar field depends on the relative distance $r = |\mathbf x_1 - \mathbf x_2|$ which is constant for a circular orbit, the vertex will vanish because of the presence of a time derivative. Let us now be more precise about this fact.

\begin{figure}
	\centering
	\includegraphics{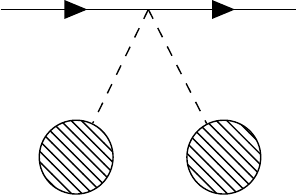}
%
%
%
\caption{Feynman diagram corresponding to the insertion of a disformal vertex with any other arbitrary vertices}
\label{fig:disformal_general}
\end{figure}

 Consider a general Feynman diagram involving a disformal vertex associated to the first particle and any number of other vertices (including possibly disformal ones), as shown in Figure~\ref{fig:disformal_general}. Ignoring the numerical factors, the amplitude of this diagram  can be written as
\begin{equation}
\mathrm{Fig} \;~\ref{fig:disformal_general} = \int d\tau_1 \left\langle T \left(\frac{d\varphi}{d\tau_1} \right)^2 \mathcal{A} \mathcal{B} \right\rangle
\end{equation}
where $\mathcal{A}$ and $\mathcal{B}$ contain the other vertices. We can relate the proper time of each particle to the time of a distant observer \textit{via}
\begin{equation}
d\tau_A = dt \sqrt{1-v_A^2-\frac{h_{\mu \nu}}{\mpl} v^\mu_A v^\nu_A}
\label{eq:proper_time}
\end{equation}
where $v^\mu_A = dx_A^\mu / dt = (1, \mathbf v_A)$ and we have decomposed the metric as $g_{\mu \nu} = \eta_{\mu \nu} + h_{\mu \nu}/\mpl$. There is a  map between $\tau_A$ and $t$, whose precise form will depend on the trajectory of the two particles $\mathbf x_A(t)$.

In the Feynman amplitude, one can contract the two fields coming from the two factors of  $d\varphi/d\tau$ with two fields amongst the ones present in $\mathcal{A}$, $\mathcal{B}$, leading to the following expression
\begin{equation}
\int dt \; H(t) \frac{dF}{dt} \frac{dG}{dt}
\label{eq:general_disformal}
\end{equation}
where $F$, $G$ and $H$ are  functions of time depending on the vertices $\mathcal{A}$ and $\mathcal{B}$. These function can depend on time only through the trajectories of the two particles $\mathbf x_A(t)$  but also on their velocities $\mathbf v_A(t)$, accelerations $\mathbf a_A(t)$ and possibly higher order derivatives. When computing the (conserved) two-body energy from the Lagrangian, one can replace the acceleration and their derivatives using the equations of motion which are of second order, thus retaining a dependence only through positions and velocities.

We will now show that the two functions $F$ and $G$ depend on time only through their relative position and velocities. In the case of a circular motion, these quantities depend very weakly on time, thus rendering the Feynman amplitude heavily suppressed. Let us now derive this statement.

As we did in Section~\ref{subsubsec:multipoledec}, Eq.~\eqref{eq:def_center_mass}, one can define the center-of-mass position \textit{via}
\begin{equation}
\left( \int d^3x T^{00} \right) x_\mathrm{CM}^i = \int d^3x  T^{00} x^i.
\end{equation}
where $T_{\mu\nu}$ is the (pseudo) energy-momentum tensor (including the self-energy of the gravitational field) defined in Eq.~\eqref{eq:S1}. By its very definition, the center-of-mass moves with a constant velocity, $\mathrm{d}^2 \mathbf{x}_\mathrm{CM} / \mathrm{d} t^2 = 0$ (see Footnote~\ref{foot:CM}). In the center-of-mass frame, we further impose
\begin{equation}
\mathbf x_\mathrm{CM} = \frac{d \mathbf x_\mathrm{CM}}{dt} = \mathbf 0
\end{equation}
in order to relate the four unknows $(\mathbf x_A, \; \mathbf v_A)$ to the \textit{relative} coordinates $\mathbf r = \mathbf x_1 - \mathbf x_2$ and $\mathbf v = \mathbf v_1 - \mathbf v_2$. For example, in GR and in the case of the  circular motion of interest  the relation between coordinates and relative coordinates reads
\begin{equation}
M x_1^i = r^i \left[m_2 + 3 \nu \delta m \left( \frac{G M}{r} \right)^2 \right] - \frac{4}{5} \frac{G^2 M^2 \nu \delta m}{r} v^i
\label{eq:CM_PN}
\end{equation}
and similarly for $\mathbf x_2$. Here we have introduced  the mass difference $\delta m = m_1 - m_2$, the symmetric mass ratio $\nu = m_1 m_2/(m_1+m_2)^2$, and we recall that $M=m_1+m_2$ is the total mass. This formula is valid up to 2.5PN order. The precise form of this relation is different for a scalar-tensor theory and has very recently been computed up to 3PN order in Ref~\cite{Bernard:2018ivi} for BDT theories, but the statement that we can always recast the motion in terms of relative coordinates is not modified.

So finally, in the case of a circular motion, the two unknown functions $F$ and $G$ are functions of time through $\mathbf r$ and $\mathbf v$.
A scalar function built out of $\mathbf r$ and $\mathbf v$ must contain only $r$, $\mathbf r \cdot \mathbf v$ and $v$ by $SO(3)$ invariance. For circular motions,  $\dot{r} = \mathbf r \cdot \mathbf v = 0$ and also $\dot{v} = 0$ up to 2.5PN order\footnote{Actually, we have seen in Chapter~\ref{Chapter4} that in a scalar-tensor theory there could be a dipole energy loss term contributing at 1.5PN order to $\dot{r}$. However recall that we made the simplifying assumption of a universal scalar-tensor coupling $\alpha$ so that such a dipole term vanishes.}~\cite{Blanchet:2013haa},
this means that the disformal contribution to the energy is indeed very suppressed.

We just showed that the time derivative of $\varphi$ present in the disformal vertex~\eqref{eq:disformal_vertex} behaves like
\begin{equation}
\frac{d\varphi}{d\tau} \simeq \dot r \partial_r \varphi + \dot v \partial_v \varphi \simeq \frac{\dot r}{r} \varphi + \frac{\dot  v}{v} \varphi
\end{equation}
instead of the usual $\frac{d\varphi}{d\tau} \sim \frac{v}{r} \varphi$ that is expected from the post-Newtonian expansion. From the energy balance between the Newtonian energy and gravitational wave emission one can find that in terms of  velocity power counting $\dot v = \mathcal{O} (v^6)$ and $\dot r= {\cal O}(v^6)$ too. If we then use the power-counting rules of Section~\ref{sec3}, we find that the disformal vertex counts as
\begin{equation}
\frac{m}{\Lambda^2 \mpl^2} \int d\tau \left( \frac{d\varphi}{d\tau} \right)^2 \sim \frac{v^{14}}{\Lambda^2 r^2}
\end{equation}
and the disformal diagram~\ref{fig:disformal} counts as $L \alpha^2 v^{14}/ (\Lambda^2 r^2)$ (where $L$ is the total angular momentum of the system, coming from the two vertices $\alpha m/\mpl \int dt \varphi$). This means that the disformal term is a 7PN effect in the conservative dynamics, so very highly suppressed.


\section{Disformal radiation and back-reaction}
\label{sec:disf}

\subsection{Multipole expansion of the dissipative dynamics}
\label{sec:multi}

\begin{figure}
	\centering
	\includegraphics{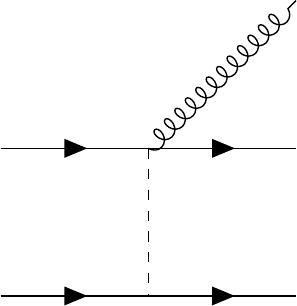}

\caption[Feynman diagram contributing to the emission of one radiation scalar, at  order $v^2$.]{Feynman diagram contributing to the emission of one radiation scalar, at  order $v^2$. It should be added to its symmetric counterpart. The curly line represents the radiated field $\bar \varphi$ while the dotted line represents the potential field $\varphi$. The vertex shown in this diagram is written in eq.~\eqref{eq:dissipative_vertex}}
\label{fig:J_v2_disf}
\end{figure}

The disformal vertex adds one supplementary diagram which contributes to the scalar coupling defined in eq.~\eqref{eq:S1}. Before calculating it, we recall here the expression of the scalar coupling $J$ arising from a scalar conformal coupling up to next-to-leading order from Eq.~\eqref{eq:J_v2},
\begin{align}
\begin{split}
J_{v^0} &= \aaa \big(m_1 \delta^3( \mathbf{x} - \mathbf{x}_1) + (1 \leftrightarrow 2) \big) \\
J_{v^2} &= - \aaa \left( m_1 \frac{v_1^2}{2} \delta^3(\mathbf{x}-\mathbf{x}_1) + (1 \leftrightarrow 2) \right) + \aaa (4 \bbb-1) \frac{G m_1 m_2}{r} \left( \delta^3(\mathbf{x}-\mathbf{x}_1) + (1 \leftrightarrow 2) \right) \; .
\end{split}
\end{align}
From now on, to simplify the expression we will assume that the second-order scalar coupling defined in Eq.~\eqref{eq:pp_disf} is $\beta = 0$. Then, the first disformal contribution to $J$ involves only one Feynman diagram, Figure~\ref{fig:J_v2_disf}. Decomposing the total field into a conservative and a dissipative part as $\varphi = \Phi + \bar \varphi$, this diagram emerges from the vertex coupling $\bar \varphi$ and $\Phi$, which can be written (using eq.~\eqref{eq:disformal_vertex}) as:
\begin{equation} \label{eq:dissipative_vertex}
-2 \frac{m_A}{\Lambda^2 \mpl^2} \int \mathrm{d} t \partial_\mu \bar \varphi v_A^\mu  \partial_\nu \Phi v_A^\nu
\end{equation}
Here as before $\bar \varphi$ is treated as an external field while $\Phi$ is integrated out and enters the Feynman diagrams in the internal lines. Using this expression, the diagram of Figure~\ref{fig:J_v2_disf} can be written as
\begin{equation}
\mathrm{Fig} \;~\ref{fig:J_v2_disf} = - 2i\frac{m_1}{\Lambda^2 \mpl^2} \int \mathrm{d} t_1 i \alpha \frac{m_2}{\mpl} \int \mathrm{d} t_2 \partial_\mu \bar \varphi(t_1, \mathbf x_1) v_1^\mu \left \langle T \partial_\nu \Phi(t_1, \mathbf x_1) v_1^\nu \Phi(t_2, \mathbf x_2) \right \rangle
\end{equation}
Using manipulations similar to the calculation of the previous diagram, one finds that this gives
\begin{equation}
\mathrm{Fig}~\ref{fig:J_v2_disf} =  2i \alpha \frac{m_1 m_2}{\Lambda^2 \mpl^3} \int \mathrm{d} t \partial_\mu \bar \varphi(t_1, \mathbf x_1) v_1^\mu \frac{\mathbf n \cdot (\mathbf v_1 - \mathbf v_2)}{4\pi r^2}
\end{equation}
To this expression one should add the symmetric diagram obtained by exchanging labels 1 and 2. By noticing that $\partial_\mu \bar \varphi(t_1, \mathbf x_1) v_1^\mu = \frac{\mathrm{d}}{\mathrm{d} t} \bar \varphi(x_1)$, one can integrate by parts in order to put this diagram in a form similar to $i\int d^4x J \bar \varphi / \mpl$ as required by the equation~\eqref{eq:S1} defining $J$. By further noticing that $\frac{\mathbf n \cdot (\mathbf v_1 - \mathbf v_2)}{r^2} = - \frac{d}{dt} \frac{1}{r}$, one can write the equation for the disformal contribution to $J$
\begin{equation}
J^\mathrm{disf} =  4 \alpha \frac{Gm_1m_2}{\Lambda^2} \frac{\mathrm{d}^2}{\mathrm{d}t^2} \frac{1}{r} \left( \delta^3(\mathbf{x}-\mathbf{x}_1) + (1 \leftrightarrow 2) \right)
\end{equation}

Notice that, as for the conservative part, this contribution vanishes exactly for a circular orbit.
The lowest-order contribution of the disformal source term $J^\mathrm{disf}$ in the multipole expansion is in the monopole (higher multipoles are further velocity-suppressed). We can consequently write the dipole and quadrupole emission terms using the same equations as before~\eqref{eq:scalar_dipole}~-~\eqref{eq:scalar_quadrupole},
\begin{align}
\begin{split}
I_\varphi^{ij} &= \alpha \left( m_A \left( x^i x^j - \frac{1}{3}x^2 \delta^{ij}\right) + (A \leftrightarrow B) \right) \\
I_\varphi^{i} &= \alpha \left( m_A x_A^i + (A \leftrightarrow B) \right)
\end{split}
\end{align}
and their contribution in the radiated power will be the same as in BDT theories. As we focus on a universal conformal coupling $\alpha$, the contribution of the dipole in the radiated power vanishes; futhermore from Eq.~\eqref{eq:J0} the lowest order contribution to the monopole vanishes also because it is simply $I_{\varphi, \; v^0} = - \alpha (m_1 + m_2)$ which is a constant by conservation of matter. The next contribution to the monopole, including the disformal contribution, starts at the same order as the quadrupole radiation in the velocity expansion and using Eq.~\eqref{eq:monodipoquadru} it can be written as
\begin{align}
\begin{split}
I_{\varphi, \; v^2 + \mathrm{disf}} &= - \frac{\alpha}{6} (m_1 v_1^2 + m_2 v_2^2) - \frac{\alpha}{3}(7+2\alpha^2) \frac{Gm_1m_2}{r} \\
&+ 8 \alpha \frac{Gm_1m_2}{\Lambda^2} \frac{d^2}{dt^2} \frac{1}{r} \; .
\end{split}
\end{align}
where we have used that the acceleration of the first particle is $a_1^i = - \tilde G m_2 r^i / r^3$, and similarly for $a_2^i$.
At this order  $m_1 v_1^2/2 + m_2 v_2^2/2 \simeq G(1+2\alpha^2)m_1m_2 / r$ by conservation of the Newtonian energy hence one can rewrite the monopole term as
\begin{equation} \label{eq:monopole}
I_{\varphi, \; v^2 + \mathrm{disf}} = - 4 \alpha G m_1m_2 \left( \frac{2+\alpha^2}{3r} - \frac{2}{\Lambda^2} \frac{d^2}{dt^2} \frac{1}{r} \right).
\end{equation}
At this point one can easily see that the disformal term is a $v^2/(\Lambda^2 r^2)$ correction to the conformal monopole, i.e a 1PN effect for an elliptic orbit. In the case of a circular orbit, we showed in Sec.~\ref{sec:circular_disf} that one should replace $d/dt \rightarrow v^6/r$ instead of the usual counting $d/dt \rightarrow v/r$, and so the disformal term is a $v^{12}/(\Lambda^2 r^2)$, i.e 6PN, correction to the conformal monopole.

In the next Section we will then use the relation $P_\varphi = 2 G \left \langle \dot I_\varphi^2 \right \rangle$ to find the final expression for the radiated power.

\subsection{Radiated power of elliptic orbits}

In this Section we will calculate the power emitted from the system in an eccentric orbit.
As discussed before, the total emitted power  splits into the power lost into gravitons and the one lost into the scalar field. For the graviton case the power emitted for elliptic orbits is known from the Peter-Mathews formula\cite{PhysRev.131.435}. Here one should use the normalised Newton constant $\tilde G$ instead of $G$ in the quadrupole moment of the source. Using eq.~\eqref{eq:quadrupole_graviton} we obtain for this power
\begin{align}
\begin{split}
P_h &= \frac{32}{5\tilde{G} (1+2\alpha^2)} (\tilde{G} M_c \omega)^{10/3} f_1(e) \;, \\
f_1(e) &= \frac{1}{(1-e^2)^{7/2}}(1+\frac{73}{24}e^2+\frac{37}{96}e^4)
\end{split}
\end{align}
where $M_c = (m_1m_2)^{3/5}/M^{1/5}$ is the chirp mass, $M=m_1 + m_2$ is the total mass of the system, and $\omega = \frac{2\pi}{T}$ is the frequency of the system that satisfies Kepler's third law
\begin{equation}\label{eq:Kepler_disf}
\omega^2 = \frac{\tilde G M}{a^3} \; .
\end{equation}

Let us now discuss the scalar radiation. Since the scalar quadrupole~\eqref{eq:scalar_quadrupole} is proportional to  the gravitational quadrupole, we deduce from eq.~\eqref{eq:radiated_power_scalar} that the scalar quadrupole power loss is
\begin{equation}
P_\varphi^\mathrm{quad} = \frac{\alpha^2}{3} P_h\; .
\end{equation}
As discussed above, the scalar dipole is zero as we focus on a universal scalar coupling $\alpha$. We are left  to calculate the monopole power starting from eq.~\eqref{eq:monopole}. Using the same method as in Sec.~\ref{sec:planar_dynamics}, we can define the average over many gravitational wave cycles appearing in the emitted monopole power~\eqref{eq:radiated_power_scalar} as
\begin{align}
\begin{split}
\left \langle \dot I_\varphi^2 \right \rangle = \frac{1}{2\pi} \int_0^{2\pi} d\psi \frac{(1-e^2)^{3/2}}{(1+e\cos \psi)^2} \dot I_\varphi^2(\psi)
\end{split}
\end{align}
where $T$ is the period of the system, $e$ the excentricity and $\psi$ is the angle along the trajectory defined in eq.~\eqref{eq:elliptic_trajectory}. This integral leads to the monopole power
\begin{align}
\begin{split}
P_\varphi^\mathrm{mono} &= \frac{16}{9\tilde G} \frac{\alpha^2(2+\alpha^2)^2}{(1+2\alpha^2)^3} (\tilde G M_c \omega)^{10/3} \\
&\times \left(f_2(e)+12yf_3(e)+36y^2f_4(e) \right)
\end{split}
\end{align}
where the three ellipticity functions $f_2$, $f_3$ and $f_4$ are given by:
\begin{align}
\begin{split}
f_2(e) &= \frac{e^2}{(1-e^2)^{7/2}} \left(1+\frac{1}{4} e^2\right) \\
f_3(e) &= \frac{e^2}{(1-e^2)^{13/2}} \left(1+\frac{37}{4}e^2+\frac{59}{8}e^4 + \frac{27}{64}e^6 \right) \\
f_4(e) &= \frac{e^2}{(1-e^2)^{19/2}} \left(1+\frac{217}{4}e^2+\frac{1259}{4}e^4+\frac{11815}{32}e^6+\frac{11455}{128}e^8+\frac{1125}{512}e^{10} \right)
\end{split}
\end{align}
and $y$ is the parameter
\begin{equation} \label{eq:def_y}
y = \frac{\tilde G M}{(2+\alpha^2)\Lambda^2a^3} = \frac{1}{2+\alpha^2} \left(\frac{\omega}{\Lambda} \right)^2
\end{equation}
where for the second equality we have used Kepler's third law~\eqref{eq:Kepler_disf}.

The first ellipticity function corresponds to the conformal case without any disformal interaction and agrees with other references~\cite{damour_tensor-multi-scalar_1992}, whereas the other functions contribute only when a disformal interaction is present. We can remark that, in order for the disformal correction to be small, we have to impose $y\lesssim 1$, i.e $\omega \lesssim \Lambda$. This is a scaling slightly different than the one which we obtained in the conservative energy, Eq.~\eqref{eq:dimensionless_param}.
Indeed, one can also relate the parameter $y$ to the nonlinear radius introduced in eq.~\eqref{eq:nonlinear_radius} via
\begin{equation}
y \simeq \frac{1}{\alpha^2} \frac{r_\star}{r_s} \left( \frac{r_*}{a} \right)^3
\end{equation}
in terms of both the nonlinear radius $r_*$ and the Schwarzschild radius $r_s = 2G M$ associated to the total mass $M$. Thus, if $r_* \sim r_s$, the disformal dissipated power is enhanced by a factor $a/(r_* \alpha^2)$ at lowest order with respect to the disformal energy~\eqref{eq:disformal_energy_elliptic}.

By summing up all contributions, we find the final formula for the emitted power:
\begin{align} \label{eq:disformal_power_total}
\begin{split}
P &= \frac{32}{5 (1+2\alpha^2) \tilde G} (\tilde G M_c \omega)^{10/3} \\
& \times \left( (1+\frac{\alpha^2}{3})f_1(e)+\frac{5}{18} \frac{\alpha^2(2+\alpha^2)^2}{(1+2\alpha^2)^2}\left(f_2(e)+12yf_3(e)+36y^2f_4(e) \right) \right) \; .
\end{split}
\end{align}
Notice that as previously stated the disformal contribution vanishes for circular orbits. As for the conservative dynamics, the leading contribution in this case is much suppressed compared to the monopole and quadrupole due to the conformal interaction.

\subsection{Constraint from the Hulse-Taylor pulsar} \label{sec:Hulse_Taylor}

As stated above, the observation of an elliptic inspiral in a GW detector could allow us to put constraints on a disformal interaction. The correction to the total energy~\eqref{eq:disformal_energy_elliptic} and to the dissipated power~\eqref{eq:disformal_power_total} should be consistently used to derive a waveform template in order to perform a matched filter analysis. We can note from eq.~\eqref{eq:def_y} that the strongest constraint would come from systems with high frequencies, i.e from elliptic systems in the LIGO/Virgo band. While the majority of such systems are expected to have circular orbits, it could also be possible to observe an eccentric merger induced by, e.g., Kozai oscillations of a triple system~\cite{Kozai:1962zz, Wen:2002km}.

\begin{figure}
\center \includegraphics[height=.3\textheight]{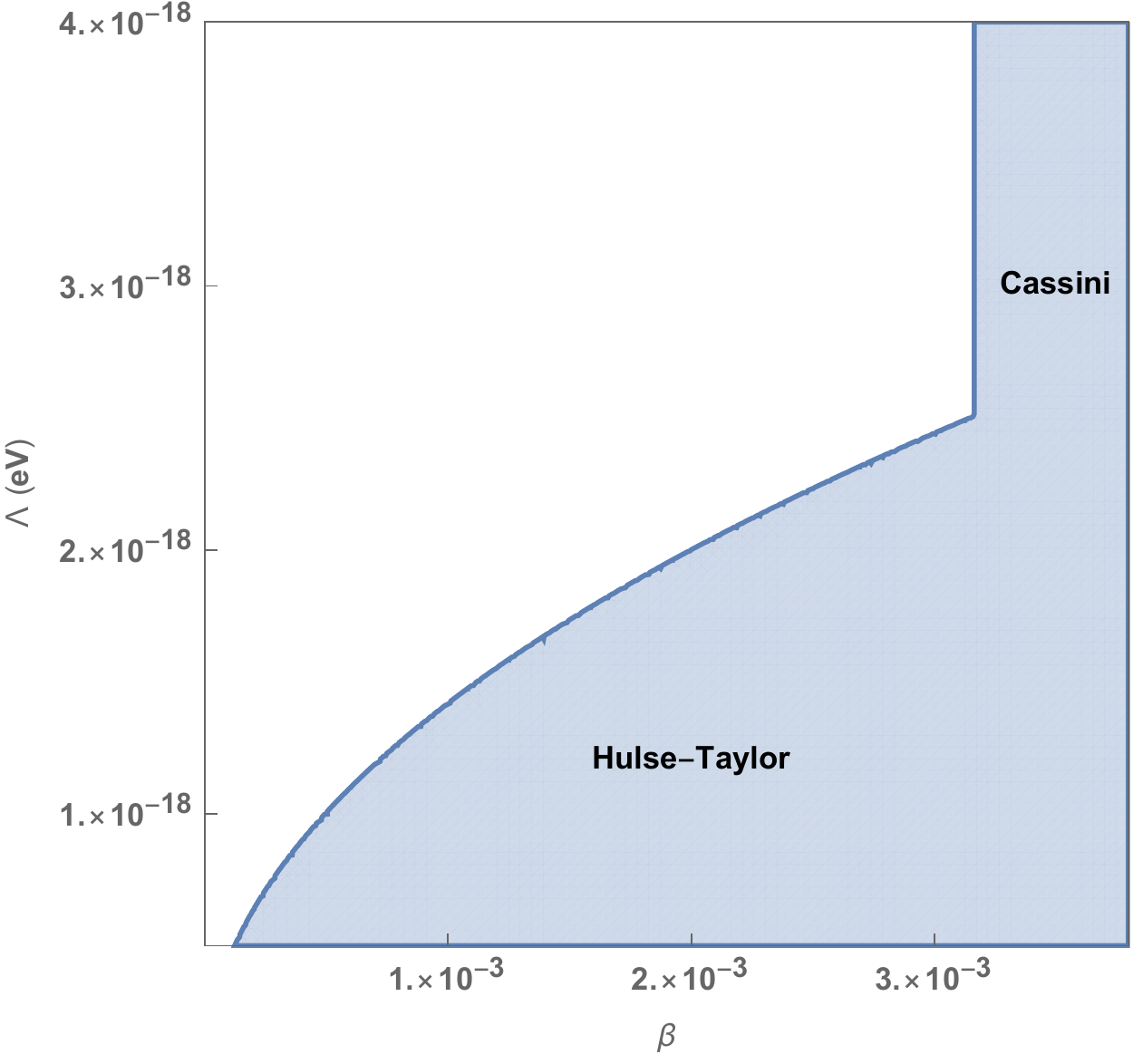}
\caption[Constraints in the $(\alpha, \; \Lambda)$ plane]{Constraints in the $(\alpha, \; \Lambda)$ plane from both the dissipated power of the Hulse-Taylor binary pulsar and the Cassini bound~\eqref{eq:constraint_gamma}. The shaded area is the exclusion region.}
\label{fig:constraints}
\end{figure}

In this Section we will rather focus on the simple constraint coming from the observation of the Hulse-Taylor pulsar B1913+16, which as explained in Section~\ref{subsec:binary_pulsar} is well-known to have an orbital decay consistent with GR at the 0.2 percent level~\cite{Hulse:1974eb, Weisberg:2004hi}. This system does also have a  large eccentricity, $e \simeq 0.6$, which does maximise the disformal effect. We use the parameters inferred from the non-relativistic analysis of arrival time data quoted in Ref.~\cite{Weisberg:2004hi} and displayed in Figure~\ref{fig:hulse_taylor} (we only need the orbital period $P_b = 2\pi /\omega$ and the eccentricity $e$). By simply requiring that the dissipated power in eq.~\eqref{eq:disformal_power_total} should not be corrected by more than $0.2$ percent by the disformal effect, we find the following bound on $\Lambda$
\begin{equation} \label{eq:bound_lambda}
\Lambda \gtrsim 3 \times 10^{-18} \; \mathrm{eV}
\end{equation}
where we have taken the conformal coupling to be at the upper limit of the Cassini bound~\eqref{eq:constraint_gamma}, $\alpha^2 \simeq 10^{-5}$. This bound is comparable with the one from torsion pendulum experiments~\cite{Brax:2014d}. In Figure~\ref{fig:constraints} we plot the constraints from both the Cassini bound and the Hulse-Taylor pulsar in the $(\alpha, \; \Lambda)$ plane.

\section{The GW speed bound} \label{sec:disformal_speed}

We will finish this Chapter by commenting on a different kind of constraint on the disformal coupling, namely the gravitational wave speed constraint that we mentioned in Section~\ref{sec:tests_GW}. It could seem surprising at first sight that such a constraint applies on the theory which we consider in this Chapter: the gravitational action is still the Einstein-Hilbert term~\eqref{eq:firstaction} so that gravitons propagate luminally. However, if one adds a disformal coupling then 'light does not propagate luminally': contrary to a conformal coupling, the disformal coupling~\eqref{eq:jordan_frame_metric} relates two metrics with different lightcones.

First, let us evaluate the speed difference of gravitons and photons on a cosmological background. There, $\partial_\mu \phi = (\dot \phi, \mathbf{0})$ and the time evolution of the scalar is cosmological, $\dot \phi \sim \mpl H_0$. Then, the order-of-magnitude of the disformal term in Eq.~\eqref{eq:jordan_frame_metric} is
\begin{equation}
\frac{1}{\mpl^2 \Lambda^2} \partial_\mu \phi \partial_\nu \phi \sim \left( \frac{H_0}{\Lambda} \right)^2 \sim 10^{-30} \; ,
\end{equation}
where we have taken for $\Lambda$ the lowest bound allowed from Eq.~\eqref{eq:bound_lambda}. Thus, the fractional deviation of the speed of photons from unity is way below the actual bound $\vert c_\mathrm{GW}- c_\mathrm{photons} \vert \lesssim 10^{-15}$ mentioned in Section~\ref{sec:tests_GW}.

Another, more promising, constraint from the GW170817 event comes from the time-of-flight difference caused by propagation of photons close to the neutron stars. There, the scalar field takes its Newtonian value at lowest order, $\phi / \mpl \simeq \alpha G M_\odot /r$ as an order-of-magnitude, so that the disformal term is approximately
\begin{equation}
\frac{1}{\mpl^2 \Lambda^2} \partial_\mu \phi \partial_\nu \phi \sim \left( \frac{\alpha G M_\odot }{\Lambda r^2} \right)^2 \; .
\end{equation}
To compute the time-of-flight from this equation, one should integrate it along the path of the photons. Taking the photon to start at approximately the radius of the neutron stars, close to the \sch radius, gives a time difference
\begin{equation}
\Delta t \simeq \alpha^2 \frac{1}{\Lambda^2 G M_\odot} \; .
\end{equation}
This time should be less than approximately $1$s, so that the lower bound on $\Lambda$ that we get using this constraint is $\Lambda \gtrsim 10^{-14}$ eV (in other words, the nonlinear radius $r_*$ is close to the \sch radius for solar-mass objects). As such, the constraint is better than the one obtained from binary pulsars~\eqref{eq:bound_lambda}. However, one should keep in mind that the bounds obtained in this Chapter are assuming a perturbative treatment of the disformal coupling; as mentioned in the introduction, a first step towards nonperturbative computations involving a disformal coupling can be found in~\cite{Davis:2019ltc}.

\chapter{A resummation technique} \label{Chapter6}

Before finishing Part~\ref{part2} and moving on to the study of the two-body problem in scalar theories equipped with a screening mechanism, we will devote a chapter to a particular resummation technique within the framework of NRGR. In this Chapter based on the paper~\citeK{Kuntz:2020gan}, we will concentrate on GR (since our resummation technique was not in use already in the case of GR) so that it will be mainly independent of the other parts, although it uses the NRGR formalism. The extension to ST theories is the subject of ongoing work. We will take a slightly different route than in the two last Chapters and apply post-Minkoskian (PM) ideas to the two-body problem.
 Compared to the PN formalism, the PM philosophy (born from the field of scattering amplitudes~\cite{2017arXiv170803872C}) is to keep the velocity arbitrary while still expanding observables for weak fields. This procedure is well adapted to scattering processes in which velocities may be high but such that one still restricts to large values of the impact parameter. The scattering angle is known at the 3PM order~\cite{PhysRevLett.122.201603} and can be translated in conservative Hamiltonians for binary systems~\cite{Cheung:2018wkq} allowing for a useful cross-check with the PN formalism within their overlapping domain of validity. Even if the PM computations are far from reaching the same perturbative order as the PN ones, one can argue that they provide a somewhat \textit{more exact} result at the same order since they are exact to all orders in velocity~\cite{Antonelli_2019}. In this Chapter, we will adopt the PM philosophy in the sense that our results will be nonperturbative in some part of the dynamics, while still being perturbative in the other part. This is achieved by a simplification of the Feynman rules of NRGR, such that no perturbative assumption is made on the matter sector. Let us now be more precise about this statement.

In Chapter~\ref{Chapter4} we have seen that the vertices in the Feynman diagrams of the NRGR approach are of two types: the \textit{bulk} nonlinearities originating from the Einstein-Hilbert action, and the \textit{worldline} couplings coming from the matter action describing the two point-particles. In this Chapter, we point out that the introduction of two worldline parameters - or einbeins - allows to drastically simplify the worldline couplings, leaving an action containing only a \textit{linear} coupling of the graviton to the source. The number of Feynman diagrams needed to evaluate the action at each perturbative order is thus greatly reduced, so that the only computational obstacle in the two-body problem is entirely contained in the Einstein-Hilbert action. Such a procedure is the perfect analogue of going from the Nambu-Goto to the Polyakov action in string theory~\cite{Polyakov}. Concerning the two-body problem, similar worldline parameters were also introduced in~\cite{Galley_2013} in the ultra-relativistic limit of NRGR  and in~\cite{Davis:2019ltc} to resum a series of Feynman diagrams when considering disformally coupled scalar fields (we already mentioned their approach in Chapter~\ref{Chapter5}).

We show that such a simplification allows to drastically reduce the number of Feynman diagrams needed to get the conservative Lagrangian at a given PN order. Next, we investigate in more details the physical properties of the worldline parameters which we introduced. In the case of circular orbits, these are known as the redshift variables~\cite{Detweiler:2008ft} as they represent the redshift of a photon emitted close to the point-particles and detected at large distance from the system. We find that these variables obey a fifth-order polynomial equation whose properties are examined in both the static and circular orbit case. We show that this equation does not admit solutions for close enough binaries, so that it allows to define an 'effective two-body horizon' ; more precisely, we find that for a critical separation no circular orbit can exist at all for the two-body problem. This is a two-body generalization of the well-known Innermost Circular Orbit (ICO) \footnote{Defined as the smallest possible circular orbit ; not to be confused with the Innermost \textit{stable} Circular Orbit or ISCO } of the \sch geometry. This result could shed light on nonperturbatives properties of the two-body motion.

\section{Integrating out gravity}

We begin by directly introducing the worldline parameters. Recall that in the NRGR formalism the matter action is constituted of point-particles $A=1,2$,
\begin{equation} \label{eq:pp_action}
S_{m,A} = - m_A \int \mathrm{d}t \sqrt{-g_{\mu \nu} v_A^\mu v_A^\nu} \; , \quad v_A^\mu = \frac{\mathrm{d} x_A^\mu}{\mathrm{d} t} \;  .
\end{equation}
and it is expanded for small deviations from flat space and small velocities, giving rise to an infinite series of vertices (the lowest-order ones are written in Eq.~\eqref{eq:pp_Action_expanded} in the case of BDT theories). Similarly, the expansion of the Einstein-Hilbert action gives rise to the quadratic term written in Eq.~\eqref{eq:quadratic_EH}, which defines the graviton propagator~\eqref{eq:propagator}. Notice that the real part of the Feynman propagator, relevant to the conservative dynamics which we will focus on in this Chapter, can be computed exactly (without using a slow-motion approximation) using $1/(k^2-i\epsilon) = PV(1/k^2) + i \pi \delta(k^2)$ and reads
\begin{equation}  \label{eq:Feyprop_exact}
\Re i D_F(x_1-x_2) = \frac{1}{8 \pi \vert \mathbf{x}_1 - \mathbf{x}_2 \vert} \left( \delta\left(t_1 - t_2 - \vert \mathbf{x}_1 - \mathbf{x}_2 \vert \right) +  \delta\left(t_1 - t_2 + \vert \mathbf{x}_1 - \mathbf{x}_2 \vert \right) \right) \; .
\end{equation}

 The essential point of this Chapter is that the nonlinearities associated to the point-particle action (we will refer to them as \textit{worldline} nonlinearities) can be computed exactly, by introducing two auxiliary parameters (we will give their physical interpretation in Section~\ref{sec:circular}). Let us rewrite each point-particle action as
\begin{equation} \label{eq:pp_action_e}
S_{m,A} = - \frac{m_A}{2} \int \mathrm{d}t \left[ e_A - \frac{g_{\mu \nu} v_A^\mu v_A^\nu}{e_A} \right] \; .
\end{equation}
Variation with respect to the einbein gives $e_A = \sqrt{-g_{\mu \nu} v_A^\mu v_A^\nu}$
which yields back the original point-particle action~\ref{eq:pp_action}. Instead, we will keep the $e_A$ undetermined from now on and integrate out the gravitational field. The crucial improvement that this procedure yields is that the point-particle vertex is now \textit{linear} in the gravitational field, allowing for an \textit{exact} computation of the effective action: all worldline couplings are now linear, so that Feynman diagrams like the one in Figure~\ref{subfig:EIH_f} do not exist! Of course, the bulk action originating from the Einstein-Hilbert term is not simply given by its quadratic term, but there are also cubic vertices giving rise to Feynman diagrams such as the one in Figure~\ref{subfig:EIH_i}, which we will not consider for the time being.
These vertices are suppressed by 1PN order in the PN expansion, so that strictly speaking our computation will be of 0PN order. On the other hand our result should generalize the 1PM results~\cite{Cheung:2018wkq}, since the 1PM order consists in taking the full graviton propagator~\eqref{eq:Feyprop_exact} with a linearized source term~\cite{Foffa:2013gja}.

\begin{figure}
	\centering
	\subfloat[]{
%
%
\includegraphics{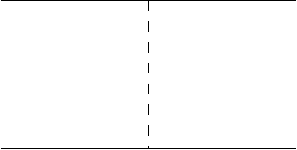}
	} \hspace{1em}
	
\caption[The only Feynman diagram arising from the linear coupling~\ref{eq:matter_coupling_Lagrange}]{The only Feynman diagram arising from the linear coupling~\ref{eq:matter_coupling_Lagrange}. The dotted line represents an insertion of the full graviton propagator~\ref{eq:Feyprop_exact}, while the vertex contain the worldline parameters from the second line of Eq.~\eqref{eq:matter_coupling_Lagrange} }
\label{fig:one_feynman}
\end{figure}

  We rewrite the point-particle action as
\begin{align}
S_{m,A} &= - \frac{m_A}{2} \int \mathrm{d}t \left[e_A + \frac{1-v_A^2}{e_A} \right] \\
&+ \frac{m_A}{2 \mpl} \int \frac{\mathrm{d}t}{e_A(t)} h_{\mu \nu} v_A^\mu v_A^\nu \; , \label{eq:matter_coupling_Lagrange}
\end{align}
where the first line does not depend on the gravitational field. The couplings of the graviton to the point-particle give rise to \textit{only} one Feynman diagram, represented in Figure~\ref{fig:one_feynman}. It is easily computed as
\begin{equation}
\left. i S_\mathrm{eff} \right\vert _{\ref{fig:one_feynman}} = \frac{i m_1}{2 \mpl} \frac{i m_2}{2 \mpl} \int \frac{\mathrm{d} t_1}{e_1(t_1)} \frac{\mathrm{d} t_2}{e_2(t_2)} P_{\mu \nu; \alpha \beta} v_1^\mu v_1^\nu v_2^\alpha v_2^\beta D_F \left( t_1-t_2, \mathbf{x}_1(t_1) - \mathbf{x}_2(t_2) \right) \; ,
\end{equation}
where the tensor $P_{\mu \nu; \alpha \beta}$ has been defined in Eq.~\eqref{eq:pmunualphabeta}.

We will ignore retardation effects as a first step, and show after how to include them. This means that we set $t=t'$ in the Feynman propagator~\ref{eq:Feyprop_exact}, giving the final effective action where the gravitational field has been effectively removed
\begin{equation} \label{eq:my_Lagrangian}
S_\mathrm{eff} = \int \mathrm{d} t \left[ - \frac{m_1}{2} \left(e_1 + \frac{1-v_1^2}{e_1} \right) - \frac{m_2}{2} \left(e_2 + \frac{1-v_2^2}{e_2} \right) + \frac{\lambda G m_1 m_2 }{e_1 e_2 r} \right] \; ,
\end{equation}
where as before $r = \vert  \mathbf{x}_1 - \mathbf{x}_2 \vert$ and $\lambda$ is a combination of the two velocities,
\begin{equation}
\lambda = 1 + v_1^2 + v_2^2 - 4 \mathbf{v}_1 \cdot \mathbf{v}_2 - v_1^2 v_2^2 + 2 (\mathbf{v}_1 \cdot \mathbf{v}_2)^2 \; . 
\end{equation}

 It seems hard to believe that such a simple Lagrangian contains infinitely many orders of the two-body post-Newtonian expansion. However, we will show in Section~\ref{sec:PN_expansion} that it is indeed the case. In the following, we will sometimes refer to the first two terms of this equation as the "kinetic" Lagrangian, and the last one as the "interaction" Lagrangian, even if the separation between kinetic and potential energy is only valid in the Newtonian limit.

It is straightforward to generalize the precedent result to include retardation effects. The first delta function in the Feynman propagator fixes the time $t_2$ to be the retarded time, defined by the equation
\begin{equation}
t_2^R = t_1 - \left \vert \mathbf{x}_1(t_1) - \mathbf{x}_2(t_2^R) \right \vert \; ,
\end{equation}
and the second delta function is just a relabeling $1 \leftrightarrow 2$. Taking into account the Jacobian in the delta function, we obtain the action
\begin{align}
\begin{split} \label{eq:my_eq_retarded}
S_\mathrm{eff} &= \int \mathrm{d} t \left[ - \frac{m_1}{2} \left(e_1(t) + \frac{1-v_1^2}{e_1(t)} \right) - \frac{m_2}{2} \left(e_2(t) + \frac{1-v_2^2}{e_2(t)} \right) \right. \\
& + \left. \frac{\lambda(t, t^R) G m_1 m_2 }{2 e_1(t) e_2(t^R) \left[ \left \vert \mathbf{x}_1(t) - \mathbf{x}_2(t^R) \right \vert - \mathbf{v}_2(t^R) \cdot (\mathbf{x}_1(t) - \mathbf{x}_2(t^R))\right]} + (1 \leftrightarrow 2) \right] \; ,
\end{split}
\end{align}
where we have highlighted the fact that in $\lambda$, the velocity of the second particle should be evaluated at retarded time.
This Lagrangian shows causal propagation from particle $2$ to particle $1$ and vice-versa. The symmetric term $(1 \leftrightarrow 2)$ can also be rewritten as a dependence on advanced time, as in the Feynman-Wheeler absorber theory~\cite{Wheeler:1945aa}.

This Lagrangian has a number of interesting properties. First, it is exactly Poincaré invariant, as we have integrated out gravitons without breaking Poincaré invariance. At the level of the effective Lagrangian, while the invariance under spacetime translations and space rotations is obvious, it is not immediately clear that $L$ is also invariant under Lorentz boosts. But remember that the combination of velocities in $\lambda$ comes from the Lorentz invariant contraction $P_{\mu \nu; \alpha \beta} v_1^\mu v_1^\nu v_2^\alpha v_2^\beta$, while the Liénard-Wiechert dependence on the positions is also Lorentz invariant. Contrast this with the usual post-Newtonian Lagrangian, which is Lorentz invariant only to next order in the PN expansion.

This exact Poincaré invariance implies the conservation of the ten usual quantities: momentum, angular momentum, energy and center-of-mass theorem. We will use these conserved quantities to derive the equations of motion of circular orbits in Section~\ref{sec:circular}.

Another particularity of the Lagrangian~\eqref{eq:my_eq_retarded} (which in fact is just a consequence of its Poincaré invariance) is that it is conservative, i.e it represents a system which does not dissipate any energy in the form of gravitational waves. Of course, such a system is perfectly unphysical, but conservative Lagrangians are nonetheless useful because, as highlighted in the last Chapters, there is a clean separation between conservative and dissipative dynamics in most approaches to the two-body problem~\cite{Blanchet:2013haa}. Some properties of the full dynamics are due to its conservative part (such as the existence of an innermost circular orbit), and other properties emerge from the dissipative part (such as the adiabatic inspiral of the two point-particles). In this Chapter we will only focus on conservative dynamics.

\section{Post-Newtonian expansion} \label{sec:PN_expansion}

\subsection{Instantaneous dynamics} \label{subsec:instantaneous}

Let us now focus on the instantaneous Lagrangian~\ref{eq:my_Lagrangian} and show how we recover the standard post-Newtonian expansion. To get the effective two-body dynamics from the Lagrangian~\ref{eq:my_Lagrangian}, one should integrate out the two auxiliary parameters. This gives the two equations
\begin{align} \label{eq:system_e1e2}
\begin{split}
e_1^2 &= 1 - v_1^2 - \frac{2 \lambda G m_2}{e_2 r} \; , \\
e_2^2 &= 1 - v_2^2 - \frac{2 \lambda G m_1}{e_1 r} \; ,
\end{split}
\end{align}
which together imply a fifth-order equation for e.g $e_1$,
\begin{equation}
(e_1^2 - 1 + v_1^2)^2 \left( e_1(1-v_2^2) - \frac{2 \lambda G m_1}{r} \right) - \frac{4 \lambda^2 G^2 m_2^2 e_1 }{r^2} = 0  \; ,
\end{equation}
with labels interchanged for $e_2$. By applying the post-Newtonian scaling $1/r = \mathcal{O}(v^2) \ll 1$ it is easy to perturbatively solve this equation. There are four branches of solutions and we select the one whose PN expansion is consistent. We find
\begin{align}
\begin{split} \label{eq:PN_sol_e1}
e_1 &= 1 - \frac{1}{2} \left(v_1^2 + \frac{2 G m_2}{r} \right) \\
&- \frac{1}{8} \left(v_1^4 +  \frac{G m_2}{r}( 12 v_1^2 + 12 v_2^2 - 32 \mathbf{v}_1 \cdot \mathbf{v}_2 ) + \frac{4 G^2 m_2(m_2 + 2 m_1)}{r^2}  \right) + \mathcal{O}(v^6) \; ,
\end{split}
\end{align}
which gives the two-body Lagrangian up to order $\mathcal{O}(v^6)$,
\begin{align}
\begin{split}
L &= -(m_1+m_2) + \frac{1}{2} \left(m_1 v_1^2 + m_2 v_2^2 + \frac{2 G m_1 m_2}{r} \right) \\
&+ \frac{1}{8} \left( m_1 v_1^4 + m_2 v_2^4 + 4 \frac{G m_1 m_2}{r} \left(3 v_1^2 + 3v_2^2 - 8 \mathbf{v}_1 \cdot \mathbf{v}_2  \right) + \frac{4 G^2 m_1 m_2(m_1 + m_2)}{r^2} \right) + \mathcal{O}(v^6) \; .
\end{split}
\end{align}

In the $\mathcal{O}(v^2)$ term we recognize the usual Newtonian potential. The next order should give the 1PN or Einstein-Infeld-Hoffmann Lagrangian, given in Eq.~\eqref{eq:LEIH} for BDT theories (GR is recovered by setting the PPN parameters $\gamma_{AB}$ and $\beta_{AB}$ to one). However, recall that we do not yet consider \textit{bulk} nonlinearities, as well as propagators corrections coming from retarded effects. This means that our expression should only recover diagrams~\ref{subfig:EIH_c},~\ref{subfig:EIH_d} and~\ref{subfig:EIH_f}, and it is indeed the case. It is remarkable that by a single linear diagram one can get directly the diagrams with nonlinear wordline fields insertions such as diagram~\ref{subfig:EIH_f}. Once propagator insertions are included, to obtain the full 1PN Lagrangian we will only miss one diagram which is the one with the cubic graviton vertex given by Figure~\ref{subfig:EIH_i}.

Going further, one may want to check if this property holds also at the 2PN order. Continuing the procedure outlined before, we find
\begin{align}
\begin{split}\label{eq:L2PN}
L_{2PN} &= \frac{1}{16} \left(m_1 v_1^6 + m_2 v_2^6 + 2 \frac{G m_1 m_2}{r} \left(7 v_1^4 + 7 v_2^4 + 2 v_1^2 v_2^2 - 16 \mathbf{v}_1 \cdot \mathbf{v}_2 (v_1^2 + v_2^2)  + 16 (\mathbf{v}_1 \cdot \mathbf{v}_2)^2 \right)  \right. \\
&+ \left. 4 \frac{G^2 m_1 m_2}{r^2} \left(v_1^2(6m_1+7m_2) + v_2^2(6m_2+7m_1) - 16 \mathbf{v}_1 \cdot \mathbf{v}_2 (m_1+m_2)  \right) + 8 \frac{G^3 m_1 m_2(m_1+m_2)^2}{r^3}  \right) \; .
\end{split}
\end{align}

The 2PN conservative Lagrangian in the NRGR formalism has been derived in Ref.~\cite{Gilmore_2008}. However, when comparing our results one should be careful about the fact that Ref.~\cite{Gilmore_2008} uses a different parametrization of the metric, namely the Kol-Smolkin variables~\cite{Kol:2007rx} (similar but not equivalent to the ADM variables),
\begin{equation}
g_{\mu \nu} = \begin{pmatrix}
e^{2 \phi / \mpl} \vspace{1em} & - e^{2 \phi / \mpl} A_j / \mpl \vspace{1em} \\
- e^{2 \phi / \mpl} A_i / \mpl \hspace{2em} & - e^{-2 \phi / \mpl} \gamma_{ij} + e^{2 \phi / \mpl} A_i A_j / \mpl^2 
\end{pmatrix} \; ,
\end{equation}
where the metric excitations are described in terms of one scalar $\phi$, one vector $A_i$ and one tensor $\gamma_{ij}$. Thus, our perturbation $h_{\mu \nu}$ contains in fact an infinite number of powers of $\phi$. This means that our linearized Einstein-Hilbert action~\ref{eq:quadratic_EH} contains an infinite number of bulk vertices in the Kol-Smolkin variables (but not all of them). This makes a direct comparison with Ref.~\cite{Gilmore_2008} difficult at the nonlinear level. 

In the following we will separate the radial and velocity dependence of the different terms at a given PN level. This means that at 2PN order, there are four classes of terms entering eq.~\eqref{eq:L2PN}: $\mathcal{O}(v^6)$ (trivial since it just comes from the expansion of $\sqrt{1-v^2}$), $\mathcal{O}(v^4/r)$, $\mathcal{O}(v^2/r^2)$ and $\mathcal{O}(1/r^3)$. The diagrams giving the $\mathcal{O}(v^4/r)$ terms do not involve any bulk vertices, so that we can directly compare our results to those of Ref.~\cite{Gilmore_2008} which are contained in diagrams $a$, $d$ and $f$ of this reference. Indeed, the sum of these diagrams exactly gives the first line of eq.~\eqref{eq:L2PN}.

\begin{figure}
	\centering
	\subfloat[]{ \label{subfig:2PN_a}
%
%
%
%
\includegraphics{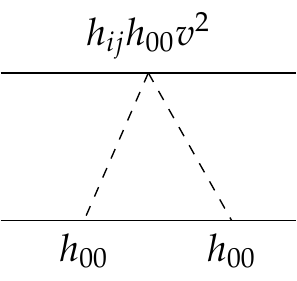}
	} \hspace{1em}
	\subfloat[]{ \label{subfig:2PN_b}
%
%
%
%
\includegraphics{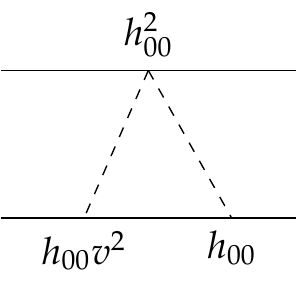}
	} \hspace{1em}
	\subfloat[]{ \label{subfig:2PN_c}
%
%
%
%
\includegraphics{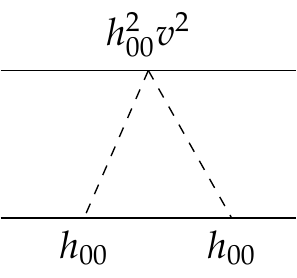}
	} \hspace{1em}
	\subfloat[]{ \label{subfig:2PN_d}
%
%
%
%
\includegraphics{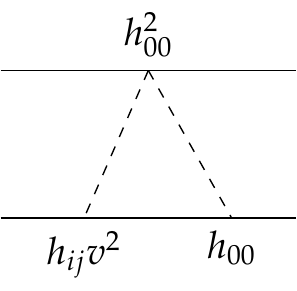}
	} \hspace{1em}
	\subfloat[]{ \label{subfig:2PN_e}
%
%
%
%
\includegraphics{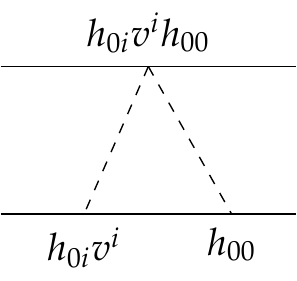}
	} \hspace{1em}
	\subfloat[]{ \label{subfig:2PN_f}
%
%
%
%
\includegraphics{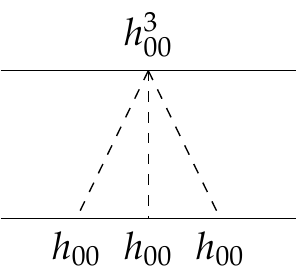}
	} \hspace{1em}
	\subfloat[]{ \label{subfig:2PN_g}
%
%
%
%
\includegraphics{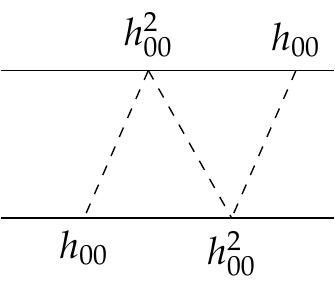}
	} \hspace{1em}

\caption[The seven Feynman diagrams needed at 2PN order.]{The seven Feynman diagrams needed at 2PN order. Each vertex is labeled with the corresponding term it refers to in the expansion of the square root in the matter action~\ref{eq:pp_action}. Each diagram which is not symmetric under the exchange $1 \leftrightarrow 2$ should be added with its symmetric counterpart.}
\label{fig:Feyn_2PN}
\end{figure}

Concerning the second line of eq.~\eqref{eq:L2PN}, we will rather evaluate ourselves the diagrams needed without using the Kol-Smolkin variables for the reason explained above. The diagrams needed to be evaluated are represented in Figure~\ref{fig:Feyn_2PN}: since they do not involve bulk nonlinearities, they are rather straightforward to evaluate using the rules of NRGR. We find

\begin{align}
\mathrm{Fig} \;~\ref{subfig:2PN_a} &= i \frac{G^2 m_1 m_2^2}{r^2} v_1^2 + (1 \leftrightarrow 2) \; , \\
\mathrm{Fig} \;~\ref{subfig:2PN_b} &= i \frac{G^2 m_1 m_2^2}{2 r^2} v_2^2 + (1 \leftrightarrow 2) \; ,  \\
\mathrm{Fig} \;~\ref{subfig:2PN_c} &= i \frac{3 G^2 m_1 m_2^2}{4 r^2} v_1^2 + (1 \leftrightarrow 2) \; ,  \\
\mathrm{Fig} \;~\ref{subfig:2PN_d} &= i \frac{G^2 m_1 m_2^2}{r^2} v_2^2 + (1 \leftrightarrow 2) \; ,  \\
\mathrm{Fig} \;~\ref{subfig:2PN_e} &= - 4 i \frac{G^2 m_1 m_2^2}{r^2} \mathbf{v}_1 \cdot \mathbf{v}_2 + (1 \leftrightarrow 2) \; ,  \\
\mathrm{Fig} \;~\ref{subfig:2PN_f} &= i \frac{G^3 m_1 m_2^3}{2 r^3} + (1 \leftrightarrow 2) \; ,   \\
\mathrm{Fig} \;~\ref{subfig:2PN_g} &= i \frac{G^3 m_1^2 m_2^2}{r^3} \; . 
\end{align}

Summing all these different contributions, we recover exactly the second line of the 2PN Lagrangian~\eqref{eq:L2PN}. However, we also discover that our resummation technique is somewhat not so interesting as one could imagine in a first time: indeed, all of the diagrams in Figure~\ref{fig:Feyn_2PN} are quite easy to compute compared to diagrams involving bulk vertices like the one in Figure~\ref{subfig:EIH_i}. This means that the true complexity of the GR two-body problem is not in the (point-particle) matter sector, but in the gravitational action.

\subsection{Retardation effects}


In this Section we will include the retardation effects in the 1PN and 2PN Lagrangians and show that we also recover known results (in the NRGR formalism, these are taken into account by modifications of the propagator as described in Section~\ref{sec3.3}). To do so, it is convenient to rewrite the Feynman propagator as
\begin{equation}
D_F(x_1-x_2) = \frac{- i}{4 \pi} \sum_n \frac{(-1)^n}{(2n)!} \vert \mathbf{x}_1 - \mathbf{x}_2 \vert^{2n-1} \frac{\mathrm{d}^{2n}}{\mathrm{d}t_1^n \mathrm{d}t_2^n} \delta(t_1-t_2) \; ,
\end{equation}
where we have expanded the delta functions in term of $\vert \mathbf{x}_1 - \mathbf{x}_2 \vert$. This shows that the interacting action can be rewritten as
\begin{equation} \label{eq:Sint_powerSeries}
S_\mathrm{int} = \sum_n \frac{(-1)^n}{(2n)!} \int \mathrm{d} t \frac{\mathrm{d}^{2n}}{\mathrm{d}t_1^n \mathrm{d}t_2^n} \left[ \frac{G m_1 m_2 \lambda(t_1, t_2) \vert \mathbf{x}_1(t_1) - \mathbf{x}_2(t_2) \vert^{2n-1} }{e_1(t_1) e_2(t_2)} \right]_{t_1 = t_2 = t} \; .
\end{equation}
Note that it is \textit{not} a total derivative as it may seem at first glance. To this interacting action one should also add the 'kinetic' action in the first line of Eq.~\eqref{eq:my_eq_retarded}.

At the 1PN level, one can simply set $e_1 = e_2 = \lambda = 1$ and compute the $n=1$ term in eq.~\eqref{eq:Sint_powerSeries} (even if $e_1$ or $e_2$ are modified by retardation effects, this dependence will cancel at this order because the 'kinetic' Lagrangian involves the combination $e_\alpha + 1/e_\alpha$). This yields
\begin{equation}
L_{1PN, \mathrm{retarded}} = \frac{G m_1 m_2}{2 r} \left( \mathbf{v}_1 \cdot \mathbf{v}_2 - (\mathbf{v}_1 \cdot \mathbf{n}) (\mathbf{v}_2 \cdot \mathbf{n}) \right) \; ,
\end{equation}
where $\mathbf{n} = \mathbf{x}_1 - \mathbf{x}_2$. This is exactly diagram~\ref{subfig:EIH_A}.

At the 2PN level, modifications to the einbeins will need to be taken into account. To this aim, it is convenient to rewrite the interacting action using integration by parts as
\begin{align}
\begin{split} \label{eq:Sint_powerSeries_IPP}
S_\mathrm{int} &= \sum_n \frac{(-1)^n}{(2n)!} \sum_{j,k=0}^n (-1)^{j+k} {n \choose j} {n \choose k} \int \mathrm{d} t \frac{1}{e_1(t) e_2(t)} \\
& \times \frac{\mathrm{d}^{j+k}}{\mathrm{d} t^{j+k}} \left[ \left. \frac{\mathrm{d}^{2n-j-k}}{\mathrm{d} t_1^{n-j} \mathrm{d} t_2^{n-k} } \left[ G m_1 m_2 \lambda(t_1, t_2) \vert \mathbf{x}_1(t_1) - \mathbf{x}_2(t_2) \vert^{2n-1} \right] \right\vert_{t_1 = t_2 = t} \right] \; ,
\end{split}
\end{align}
which allows for a direct variation of the action with respect to $e_1$ and $e_2$.

There are now three contributions from retardation effects to the effective action at the 2PN order: the $n=2$ term in eq.~\eqref{eq:Sint_powerSeries} taking $e_1 = e_2 = \lambda = 1$, the $n=1$ term in the same equation with the 1PN correction to $e_1$, $e_2$ and $\lambda$, and the 2PN term in the kinetic Lagrangian coming from the 1PN correction to $e_1$ and $e_2$. We consider here only corrections to the $\mathcal{O}(v^4/r)$ sector of the Lagrangian to allow for a direct comparison with Ref.~\cite{Gilmore_2008}, as explained in the previous Section. Taking them all into account recovers exactly diagrams $b$, $c$ and $e$ of this reference. 

\subsection{Finite-size effects}

A great improvement of our approach compared to standard NRGR is that finite-size effect have a more direct graphical interpretation: they are the only non-minimal couplings to the point-particles worldlines. As shown in Section~\ref{subsec:pp_action}, these effects are known to arise at the 5PN order for nonspinning sources and consequently necessitate the knowledge of the PN dynamics up to this very high order. At lowest order, the expansion of the finite-size effect operator in Eq.~\eqref{eq:pp_nonminimal_weyl} is
\begin{equation} \label{eq:CE}
\frac{C_E}{\mpl^2} \int \mathrm{d}t (\partial_i \partial_j h_{00})^2 \; ,
\end{equation}
We choose here to treat these kind of operators perturbatively, while still staying nonperturbative in the lowest-order point-particle coupling $-m \int \mathrm{d} \tau$. Hence a simple Feynman diagram gives the lowest-order contribution of finite-size effects to the effective Lagrangian~\eqref{eq:my_eq_retarded}:
\begin{figure}
	\centering
	\subfloat[]{
%
%
%
%
\includegraphics{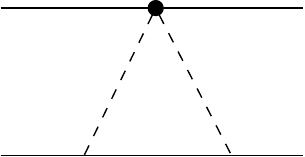}
	} \hspace{1em}

\caption[Leading order contribution to the finite-size effects.]{Leading order contribution to the finite-size effects. The dotted line represents an insertion of the nonminimal coupling operator~\ref{eq:CE}. The diagram should be added with its symmetric counterpart.}
\label{fig:Feyn_finite-size}
\end{figure}
\begin{equation}
\mathrm{Figure} \;~\ref{fig:Feyn_finite-size} = 24 i \frac{G^2 C_{E,1} m_2^2}{e_2^2 r^6} + (1 \leftrightarrow 2) \; .
\end{equation}


\section{Static potential} \label{sec:static_pot}

Having confidently established the perturbative validity of our Lagrangian~\eqref{eq:my_eq_retarded}, we will now analyze in more details the non-perturbative effects it implies. By non-perturbative we mean that we do not resort to any post-Newtonian expansion beyond the first approximation of ignoring bulk nonlinearities. This approach is similar in spirit to Ref.~\cite{Antonelli_2019}, who choose to consider PM hamiltonians as exact in order to gauge the improvement of the PM approximation when comparing to a PN expansion. Of course, there is no physically conceivable situation where the bulk nonlinearities would be subdominant while not being in a post-Newtonian approximation scheme. However, as stated in the Introduction, our results will be in some sense 'more exact' than the post-Newtonian ones. We will see that we will be able to define an 'effective two-body horizon', a quantity whose very existence cannot be recasted in the standard post-Newtonian formalism.

 To get some insight, we will begin by analyzing the static potential between the two point-particles. This means that we will set $v_1 = v_2 = 0$ in all our preceding formulas and imagine that an external operator pinpoints the two masses at their location so that they do not move (of course, such a procedure is perfectly unphysical; we will deal with measurable quantities in Section~\ref{sec:circular}). The potential energy of the system is simply the opposite of the Lagrangian and is given by
\begin{equation}
V(r) =  \frac{m_1}{2} \left( e_1 + \frac{1}{e_1} \right) + \frac{m_2}{2} \left( e_2 + \frac{1}{e_2} \right) - \frac{G m_1 m_2}{e_1 e_2 r} \; ,
\end{equation}
and $e_1, e_2$ obey the quintic equation
\begin{equation} \label{eq:quintic_static}
f_1(e_1, r) \equiv (e_1^2 - 1)^2 \left( e_1 - \frac{2 G m_1}{r} \right) - \frac{4 G^2 m_2^2 e_1 }{r^2} = 0 \; ,
\end{equation}
with labels interchanged for $e_2$. 

\begin{figure}
\center \includegraphics[width=.6\columnwidth]{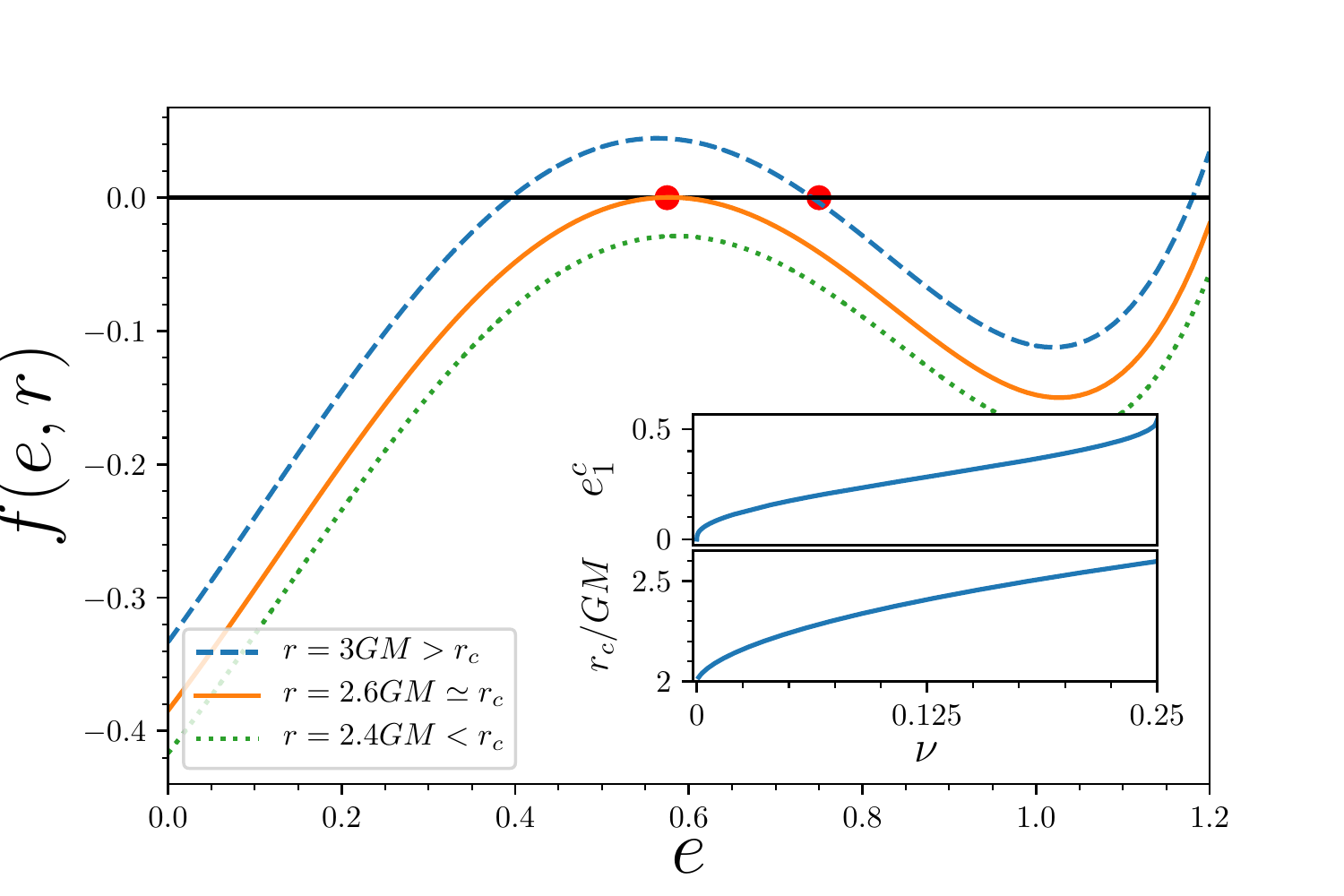}
\caption[Polynomial equation~\eqref{eq:quintic_static} in the equal-mass case $m_1 = m_2 = M/2$, for different values of $r$]{\textit{Main plot}: Polynomial equation~\eqref{eq:quintic_static} in the equal-mass case $m_1 = m_2 = M/2$, for different values of $r$. The red dot represents the solution with the correct post-Newtonian behavior for $r \rightarrow \infty$. For $r$ smaller than the critical radius $r_c$ this solution becomes complex-valued.
\textit{Subplot}: Critical radius in units of $GM$ and critical parameter $e_1$, as a function of the symmetric mass ratio $\nu = m_1 m_2/M^2$. It is assumed that $m_1 < m_2$, so that in the test-mass limit $m_1 \rightarrow 0$, $r_c = 2G M = 2Gm_2$ and $e_1^c = 0$; in the equal-mass case, $r_c = 3 \sqrt{3} G M /2 \simeq 2.6 G M$ and $e_1^c = 1/\sqrt{3} \simeq 0.58$ }
\label{fig:quintic_static}
\end{figure}

The polynomial equation~\eqref{eq:quintic_static} has five roots and only one has the correct post-Newtonian expansion as in~\eqref{eq:PN_sol_e1}.
In Figure~\ref{fig:quintic_static} we have plotted the quintic equation~\eqref{eq:quintic_static} in the equal-mass case as a function of $e_1$ and for different values of $r$. It appears that for a certain critical value of the radius $r_c$, the root $e_1(r)$ with the correct post-Newtonian behavior cease to exist (more precisely, this root becomes complex). This critical radius is defined as the point where
\begin{equation}
f_1(e_1, r) = \frac{\partial f_1}{\partial e_1} = 0 \; ,
\end{equation} 
and is also plotted as a function of the symmetric mass ratio $\nu = m_1 m_2/(m_1+m_2)^2$ in Figure~\ref{fig:quintic_static} (an exact expression for $r_c$ does exist but is not so illuminating). Note that due to the symmetry of the equations, the critical radius is the same if one instead considers the polynomial equation on $e_2$.

To get an understanding of the phenomenon at play, let us consider the test-mass ratio $m_1 \rightarrow 0$. In this case the critical radius becomes $r=2Gm_2$. But we also know the exact solution for $e_1$ from the original point-particle action~\eqref{eq:pp_action_e}: it is $e_1^2 = -\bar g_{00}$  where $\bar g_{\mu \nu}$ is the \sch metric in harmonic coordinates,
\begin{equation} \label{eq:g00_harmonic}
\bar g_{00} = - \frac{r-Gm_2}{r+Gm_2} \; .
\end{equation}
More precisely, since we are considering a one-graviton exchange in the Lagrangian~\eqref{eq:my_eq_retarded} we should rather take the linearized \sch metric so that $e_1^2 \simeq 1 - 2Gm_2/r$.
 It is then clear that in this case the critical radius corresponds to the point-particle becoming lightlike $e_1=0$, i.e the critical radius is reached when the point-particle is on the horizon of the massive particle $m_2$.
In this sense, our equations define an effective horizon for two interacting point-particles: for $r < r_c$, our static assumption is necessarily invalid since there are no solution to the equations, i.e the two particles are forced to move. We will translate this statement in a measurable gauge-invariant quantity in the Section~\ref{sec:circular}.

Of course, the true horizon of the \sch metric in harmonic coordinates is situated at $r=Gm_2$ and not at $r=2Gm_2$.
 We expect that taking into account cubic and higher vertices would give a more accurate estimate of the location of the horizon. This is indeed the case when approximating the rational fraction~\eqref{eq:g00_harmonic} by polynomials of increasing degree in order to find its zero.

It is interesting to note that the disappearance of a solution of the quintic equation~\eqref{eq:quintic_static} is an intrinsically nonperturbative phenomenon. Indeed, if one tried to solve the quintic equation for $e_1$~\eqref{eq:quintic_static} perturbatively as in Section~\ref{sec:PN_expansion}, one would get e.g
\begin{equation} \label{eq:PN_static}
e_1 = 1 - \frac{G m_2}{r} - \frac{G^2m_2(m_2+2m_1)}{2r^2} - \frac{G^3 m_2(m_2^2 + 4m_1 m_2 + 3 m_1^2)}{2r^3} + \mathcal{O}\left( \frac{1}{r^4} \right) \; ,
\end{equation}
and this solution exists for all $r$. Likewise, one could solve equation~\eqref{eq:quintic_static} perturbatively in the mass ratio $m_1/m_2$ to get
\begin{equation}
e_1 = \sqrt{1 - \frac{2Gm_2}{r}} - \frac{G^2 m_1 m_2}{r(r-2Gm_2)} - \frac{3 G^3 m_1^2 m_2(r-Gm_2)}{2r^{3/2}(r-2Gm_2)^{5/2}} + \mathcal{O}(m_1^3) \; .
\end{equation}
The point is that, as the lowest-order solution presents a singular behavior only at $r=2Gm_2$, the same will be true of any perturbative order beyond the leading one, and so the solution for $e_1$ can never cease to exist for any $r$ greater than $2Gm_2$ contrary to the behavior identified in Figure~\ref{fig:quintic_static}.

Another feature of the critical point is the characteristic behavior of $e$ near to this point. One can define a 'critical exponent' $\gamma$ by the scaling $e_1 - e_1^c \propto (r - r_c)^\gamma$ close to the critical point, with $\gamma = 1/2$. This is because the vanishing of the derivative of the polynomial equation $\partial f / \partial e_1 = 0$ at the critical point imposes that for $e_1$ close to $e_1^c$ and $r$ close to $r_c$,
\begin{equation}
0 = f_1(e_1, r) \simeq  (r-r_c) \frac{\partial f_1}{\partial r} + \frac{1}{2} (e_1-e_1^c)^2 \frac{\partial^2 f_1}{\partial e_1^2} \; ,
\end{equation}
such that $e_1 - e_1^c \propto (r - r_c)^{1/2}$. This generalizes to any mass ratio the well-known behavior in the test-mass limit, Eq.~\eqref{eq:g00_harmonic}.

We now move on to the case of circular orbits.

\section{Circular orbits} \label{sec:circular}

In this Section, we will derive the perturbative and nonperturbative properties of the conservative Lagrangian~\eqref{eq:my_eq_retarded} (which we simply denote by $L$) in the case of an exactly circular two-body orbit of frequency $\omega$ (we recall that we concentrate on conservative dynamics). In this setting, a physical interpretation of the auxiliary parameters $e_1, e_2$ has been given by Detweiler~\cite{Detweiler:2008ft}. Namely, these are the redshift of photons emitted near to the point-particles and detected by an observer situated far from the system on its axis of rotation, and as such are gauge invariant within the physically reasonable class of gauges defined in~\cite{Blanchet:2011aha}. In the following, we will mainly be interested by this \textit{redshift observable}, as well as by the energy of circular orbits, considered as functions of the frequency of the orbit.

\subsection{Equations of motion}

In order to derive the equations of motion of e.g the first particle, we choose to rewrite the $(1 \leftrightarrow 2)$ term in~\eqref{eq:my_eq_retarded} as a dependence on advanced time for the particle 2, so that the variables concerning the first particle are always evaluated at time $t$:
\begin{align}
\begin{split} \label{eq:Lagrangian_advanced}
L &=  - \frac{m_1}{2} \left(e_1(t) + \frac{1-v_1^2}{e_1(t)} \right) - \frac{m_2}{2} \left(e_2(t) + \frac{1-v_2^2}{e_2(t)} \right) 
 + \frac{G m_1 m_2}{2 e_1(t)} \left[  \frac{\lambda(t, t^R) }{ e_2(t^R) \tilde r^R} 
 + \frac{\lambda(t, t^A) }{ e_2(t^A) \tilde r^A} \right] \; ,
\end{split}
\end{align}
where
\begin{align} \label{eq:def_retarded_advanced}
\begin{split}
t^R &= t - \left \vert \mathbf{x}_1(t) - \mathbf{x}_2(t^R) \right \vert \; , \\
t^A &= t + \left \vert \mathbf{x}_1(t) - \mathbf{x}_2(t^A) \right \vert \; , \\
\tilde r^R &= \left \vert \mathbf{x}_1(t) - \mathbf{x}_2(t^R) \right \vert - \mathbf{v}_2(t^R) \cdot (\mathbf{x}_1(t) - \mathbf{x}_2(t^R)) \; , \\
\tilde r^A &= \left \vert \mathbf{x}_1(t) - \mathbf{x}_2(t^A) \right \vert + \mathbf{v}_2(t^A) \cdot (\mathbf{x}_1(t) - \mathbf{x}_2(t^A)) \; .
\end{split}
\end{align}

As stated below Eq.~\eqref{eq:my_eq_retarded}, exact Poincaré invariance of the Lagrangian implies the conservation of the ten usual quantities. In particular, conservation of angular momentum restricts the motion to a two-dimensional plane.
We thus parameterize the circular trajectories according to
\begin{equation} \label{eq:trajectory_x1x2}
\mathbf{x}_1(t) = R_1 (\cos \omega t, \sin \omega t)^T \; , \quad \mathbf{x}_2(t) = - R_2 (\cos \omega t, \sin \omega t)^T \; ,
\end{equation}
where $\omega$ is the frequency of the circular orbit. This ansatz solves both the equations of motion and the center-of-mass theorem (see below); indeed, because of the time-symmetric (non-dissipative) character of the equations, the momentum of each particle will be aligned with the common axis of the velocities.

By defining the two (positive) variables
\begin{equation}
u^R = t - t^R \; , \quad u^A = t^A - t \; ,
\end{equation}
and plugging the trajectory~\eqref{eq:trajectory_x1x2} into the definitions of the retarded and advanced times~\eqref{eq:def_retarded_advanced}, we find that $u^R= u^A \equiv u$ satisfy the same equation,
\begin{equation} \label{eq:retarded}
u = \sqrt{R_1^2 + R_2^2 + 2R_1 R_2 \cos \omega u} \; .
\end{equation}

In principle there could be multiple solutions to this equation, but we can focus on the one continuously related to the nonrelativistic solution at large distances ($\omega \rightarrow 0$) where $u = R_1 + R_2$.


Then, the equality $u^R= u^A $ implies that $\tilde r^R$ and $\tilde r^A$ take the common value
\begin{equation}
\tilde r^R = \tilde r^A \equiv \tilde r = u + \omega R_1 R_2 \sin \omega u \; ,
\end{equation}
and similarly for $\lambda$,
\begin{equation}
\lambda(t, t^R) = \lambda(t, t^A) \equiv \tilde \lambda = 1 + \omega^2(R_1^2 + R_2^2 + 4 R_1 R_2 \cos \omega u) + R_1^2 R_2^2 \omega^4 \cos 2 \omega u \; .
\end{equation}

We are now in position to compute the polynomial equations on $e_i$. To this aim, we will rather use the Lagrangian expanded in a power series~\eqref{eq:Sint_powerSeries_IPP}. Minimization with respect to $e_1$ gives
\begin{align}
\begin{split}
e_1^2 &= 1 - v_1^2 - \frac{2Gm_2}{e_2} \bigg( \sum_n \frac{(-1)^n}{(2n)!} \sum_{j,k=0}^n (-1)^{j+k} {n \choose j} {n \choose k} \\
& \times  \frac{\mathrm{d}^{j+k}}{\mathrm{d} t^{j+k}} \left[ \left. \frac{\mathrm{d}^{2n-j-k}}{\mathrm{d} t_1^{n-j} \mathrm{d} t_2^{n-k} } \left[ \lambda(t_1, t_2) \vert \mathbf{x}_1(t_1) - \mathbf{x}_2(t_2) \vert^{2n-1} \right] \right\vert_{t_1 = t_2 = t} \right] \bigg) \; .
\end{split}
\end{align}
However, since the $(t_1,t_2)$ dependence in $\lambda(t_1, t_2)$ and $\vert \mathbf{x}_1(t_1) - \mathbf{x}_2(t_2) \vert$ is only contained in terms proportional to $\cos \omega(t_1-t_2)$, this equation simplifies drastically. Indeed, by setting $t_1=t_2=t$ the expressions become constant in time, so the derivation with respect to $t$ selects the term $j=k=0$ only. The equation simplifies to
\begin{equation}
e_1^2 = 1 - v_1^2 - \frac{2Gm_2}{e_2} \left( \sum_n \frac{(-1)^n}{(2n)!} \frac{\mathrm{d}^{2n}}{\mathrm{d} t_1^{n} \mathrm{d} t_2^{n} } \left[ \lambda(t_1, t_2) \vert \mathbf{x}_1(t_1) - \mathbf{x}_2(t_2) \vert^{2n-1} \right] \right) \; .
\end{equation}
One easily recognizes the usual expansion in term of retarded and advanced times (trading the derivatives on $t_1$ to derivatives on $t_2$ by using the time dependence $\propto \cos \omega(t_1-t_2)$) , and so the equation is
\begin{equation}
e_1^2 = 1 - v_1^2 - \frac{Gm_2}{e_2} \left( \frac{\lambda(t,t^R)}{\tilde r^R} + \frac{\lambda(t,t^A)}{\tilde r^A} \right) \; .
\end{equation}
With our previous notations, this becomes
\begin{equation}
e_1^2 = 1 - R_1^2 \omega^2 - \frac{2 \tilde \lambda G m_2}{\tilde r e_2}  \; .
\end{equation}
The equation on $e_2$ can simply be found by interchanging $1 \leftrightarrow 2$. The point of this derivation was to show that one can indeed take $e_1$ and $e_2$ to be a constant (independent of time) in all the equations of motion.


Finally, the last equation that we are after is the equation of motion for one of the point-particles (say $m_1$), which will give the generalization of Kepler's law:
\begin{equation} \label{eq:EOM_base}
\frac{d \mathbf{p}_1}{dt} = \frac{\partial L}{\partial \mathbf{x}_1} \; ,
\end{equation}
where 
\begin{equation}
\mathbf{p}_1 = \frac{\partial L}{\partial \mathbf{v}_1} = \frac{m_1}{e_1} \mathbf{v}_1 + \frac{G m_1 m_2}{2 e_1 e_2 \tilde r} \left( \frac{\partial \lambda(t,t^R)}{\partial \mathbf{v}_1} + \frac{\partial \lambda(t,t^A)}{\partial \mathbf{v}_1} \right) \; .
\end{equation}
which upon using the ansatz~\eqref{eq:trajectory_x1x2} on $\mathbf{x}_1, \mathbf{x}_2$ gives
\begin{align}
\begin{split}
\mathbf{p}_1 &= \bigg[ \frac{m_1}{e_1} R_1 \omega + \frac{G m_1 m_2}{e_1 e_2 \tilde r} \left( 2(1-R_2^2 \omega^2) R_1 \omega + 4 \cos \omega u (1+R_1 R_2 \omega^2 \cos \omega u) R_2 \omega \right) \bigg] \\
& \times (-\sin \omega t, \cos \omega t)^T  \; .
\end{split}
\end{align}
Thus, as advertised before, the momentum is aligned with the common direction of the velocities.

 To compute $\partial L / \partial \mathbf{x}_1$, one should be careful to the fact that $t^R$ and $t^A$ depend on $\mathbf{x}_1$, so that 
\begin{equation}
\frac{\partial L}{\partial \mathbf{x}_1} = \frac{G m_1 m_2}{2 e_1 e_2 \tilde r} \left( \frac{\partial \lambda^R}{\partial t^R} \frac{\partial t^R}{\partial \mathbf{x}_1} - \frac{\tilde \lambda}{\tilde r} \frac{\partial \tilde r^R}{\partial \mathbf{x}_1} + (R \leftrightarrow A) \right) \; ,
\end{equation}
where $\lambda^R = \lambda(t, t^R)$. From the definitions of the advanced and retarded times~\eqref{eq:def_retarded_advanced} one gets
\begin{align}
\frac{\partial t^R}{\partial \mathbf{x}_1} &= - \frac{\mathbf{r}^R}{\tilde r} \; , \quad  \frac{\partial t^A}{\partial \mathbf{x}_1} =  \frac{\mathbf{r}^A}{\tilde r} \; , \\
\frac{\partial \tilde r^R}{\partial \mathbf{x}_1} &= - \mathbf{v}_2^R + \frac{1 - v_2^2 + \mathbf{a}_2^R \cdot \mathbf{r}^R}{\tilde r} \mathbf{r}^R \; , \quad \frac{\partial \tilde r^A}{\partial \mathbf{x}_1} =  \mathbf{v}_2^A + \frac{1 - v_2^2 + \mathbf{a}_2^A \cdot \mathbf{r}^A}{\tilde r} \mathbf{r}^A \; ,
\end{align}
where $\mathbf{r}^R = \mathbf{x}_1(t) - \mathbf{x}_2(t^R)$, and generically a superscript $R$ denotes evaluation at retarded time (the same being true for $A$). With our parameterization, one gets
\begin{align}
\frac{\partial \lambda^R}{\partial t^R} \frac{\partial t^R}{\partial \mathbf{x}_1} + (R \leftrightarrow A) &= -8 R_1 R_2 \omega^3 \sin \omega u (1 + R_1 R_2 \omega^2 \cos \omega u) \frac{R_1 + R_2 \cos \omega u}{\tilde r} (\cos \omega t, \sin \omega t)^T \\
\frac{\partial \tilde r^R}{\partial \mathbf{x}_1} + (R \leftrightarrow A) &= 2 \left[ R_2 \omega \sin \omega u + (1 + R_1 R_2 \omega^2 \cos \omega u) \frac{R_1 + R_2 \cos \omega u}{\tilde r}  \right] (\cos \omega t, \sin \omega t)^T \; .
\end{align}
Finally, the projection of the equation of motion~\eqref{eq:EOM_base} gives the generalized Kepler law,
\begin{align}
\begin{split} \label{eq:EOM_Kepler}
R_1 \omega^2 &= \frac{G m_2}{e_2 \tilde r^2} \left\lbrace \tilde \lambda \left[ R_2 \omega \sin \omega u + (1 + R_1 R_2 \omega^2 \cos \omega u) \frac{R_1 + R_2 \cos \omega u}{\tilde r}  \right] \right. \\
&+ \left. \vphantom{\frac{R_1}{\tilde r}} 4 R_2 \omega^2 (1 + R_1 R_2 \omega^2 \cos \omega u) (R_1^2 \omega \sin \omega u - u \cos \omega u) + 2 (R_2^2 \omega^2 - 1) R_1 \tilde r \omega^2 \right \rbrace \; .
\end{split}
\end{align}

It is easily checked that in the nonrelativistic case $e_2=1, \omega \rightarrow 0, R_1 = m_2/(m_1+m_2)r$ (where $r = R_1 + R_2$), one recovers the usual Kepler law, $\omega^2 r^3 = G (m_1+m_2)$. We now have all the equations needed to solve for the two-body motion, namely: the equation one the retarded time $u$, the definitions of $\tilde r$ and $\tilde \lambda$, the equations of motion and the coupled equations on $e_1,e_2$, which we all rewrite here for convenience:
\begin{align}
\begin{split} \label{eq:system_circular}
u &= \sqrt{R_1^2 + R_2^2 + 2R_1 R_2 \cos \omega u} \; , \\
\tilde r &= u + \omega R_1 R_2 \sin \omega u \; , \\
\tilde \lambda &= 1 + \omega^2(R_1^2 + R_2^2 + 4 R_1 R_2 \cos \omega u) + R_1^2 R_2^2 \omega^4 \cos 2 \omega u \; , \\
R_1 \omega^2 &= \frac{G m_2}{e_2 \tilde r^2} \left\lbrace \tilde \lambda \left[ R_2 \omega \sin \omega u + (1 + R_1 R_2 \omega^2 \cos \omega u) \frac{R_1 + R_2 \cos \omega u}{\tilde r}  \right] \right. \\
&+ \left. \vphantom{\frac{R_1}{\tilde r}} 4 R_2 \omega^2 (1 + R_1 R_2 \omega^2 \cos \omega u) (R_1^2 \omega \sin \omega u - u \cos \omega u) + 2 (R_2^2 \omega^2 - 1) R_1 \tilde r \omega^2 \right \rbrace \; ,
 \\
e_1^2 &= 1 - R_1^2 \omega^2 - \frac{2 \tilde \lambda G m_2}{\tilde r e_2}  \; ,
\end{split}
\end{align}
where we did not write the two other equations on the second point-particle arising from a $1 \leftrightarrow 2$ permutation of the last two equations.

\subsection{Conserved energy}

Once these equations are solved, we can easily obtain the conserved energy as the Hamiltonian of the system. A naive guess for $H$ would be
\begin{equation} \label{eq:guess_H}
H = \mathbf{p}_1 \cdot \mathbf{v}_1 + \mathbf{p}_2 \cdot \mathbf{v}_2 - L \; .
\end{equation}
However, because of the presence of retarded and advanced times, the application of Noether's theorem to time translations is not so straightforward, and in fact we will see that Eq.~\eqref{eq:guess_H} is actually incomplete. We will now look in more details at the boundary term in the variation of the action needed for Noether's theorem. We rewrite the action in its particle-symmetric form as
\begin{align}
\begin{split} \label{eq:action_symmetric_noether}
S &= \int_{t_-}^{t_+} \mathrm{d} t \left[ - \frac{m_1}{2} \left(e_1 + \frac{1-v_1^2}{e_1} \right) - \frac{m_2}{2} \left(e_2 + \frac{1-v_2^2}{e_2} \right) +  L_2^R + L_1^R \right] \; ,
\end{split}
\end{align}
where we have introduced boundaries for the integration on the time variable, and the retarded Lagrangians are defined as
\begin{align}
\begin{split}
L_2^R &= \frac{\lambda(t, t_2^R) G m_1 m_2 }{2 e_1 e_2(t_2^R) \left[ \left \vert \mathbf{x}_1 - \mathbf{x}_2(t_2^R) \right \vert - \mathbf{v}_2(t_2^R) \cdot (\mathbf{x}_1 - \mathbf{x}_2(t_2^R))\right]} \; , \\
L_1^R &=  \frac{\lambda(t_1^R, t) G m_1 m_2 }{2 e_1(t_1^R) e_2 \left[ \left \vert \mathbf{x}_1(t_1^R) - \mathbf{x}_2 \right \vert + \mathbf{v}_1(t_1^R) \cdot (\mathbf{x}_1(t_1^R) - \mathbf{x}_2)\right]}  \; , \\
t_1^R &= t - \left \vert \mathbf{x}_1(t_1^R) - \mathbf{x}_2 \right \vert \; , \\
t_2^R &= t - \left \vert \mathbf{x}_1 - \mathbf{x}_2(t_2^R) \right \vert \; . \\
\end{split}
\end{align}

To avoid cluttering notation, in this equation and from now on it is implicit that each variable is evaluated at $t$ when we do not write its argument. Let us consider the variation of this action under an arbitrary transformation $\mathbf{x}_\alpha \rightarrow \mathbf{x}_\alpha + \delta \mathbf{x}_\alpha$. To this aim one should rewrite the action so that it contains only the appropriate variables evaluated at their present time. For the first particle, this can be achieved by the change of variable $t' = t_1^R$ whose Jacobian is
\begin{equation} \label{eq:jacobian}
\frac{\mathrm{d}t}{\left \vert \mathbf{x}_1(t_1^R) - \mathbf{x}_2 \right \vert + \mathbf{v}_1(t_1^R) \cdot (\mathbf{x}_1(t_1^R) - \mathbf{x}_2)} = \frac{\mathrm{d}t'}{\left \vert \mathbf{x}_1(t') - \mathbf{x}_2(t_2^A) \right \vert + \mathbf{v}_2(t_2^A) \cdot (\mathbf{x}_1(t') - \mathbf{x}_2(t_2^A))} \; ,
\end{equation}
with obvious notations for the advanced time of the second particle. This has the effect of transforming $L_1^R$ in $L_2^A$ where the positions of the first particle is evaluated at present time while the second one at advanced time.

 Due to this change of variable, the boundaries change from $t_- \rightarrow t_+$ to $t_{-,1}^R \rightarrow t_{+,1}^R$. Since the boundaries now contain the positions of the particles, we should take them into account when varying the action. The variation of $t_{+,1}^R$ with respect to $\mathbf{x}_1$ is
\begin{equation}
\delta t_{+,1}^R = - \frac{\delta \mathbf{x}_1(t_{+,1}^R) \cdot \big( \mathbf{x}_1(t_{+,1}^R) - \mathbf{x}_2(t_+) \big)}{\left \vert \mathbf{x}_1(t_{+,1}^R) - \mathbf{x}_2(t_+) \right \vert + \mathbf{v}_1(t_{+,1}^R) \cdot (\mathbf{x}_1(t_{+,1}^R) - \mathbf{x}_2(t_+))} \; ,
\end{equation}
and similarly for $t_-$. We can then write the total variation of the action as
\begin{align}
\begin{split}
\delta S &= \int_{t_-}^{t_+} \mathrm{d}t \left\lbrace \frac{\partial L_2^R}{\partial \mathbf{x}_1} \delta \mathbf{x}_1 + \left( \frac{m_1}{e_1} \mathbf{v}_1 + \frac{\partial L_2^R}{\partial \mathbf{v}_1}  \right) \delta \mathbf{v}_1  \right\rbrace \\
&+ \int_{t_{-,1}^R}^{t_{+,1}^R} \mathrm{d}t \left\lbrace \frac{\partial L_2^A}{\partial \mathbf{x}_1} \delta \mathbf{x}_1 + \frac{\partial L_2^A}{\partial \mathbf{v}_1} \delta \mathbf{v}_1   - \frac{\mathrm{d}}{\mathrm{d}t} \left[ L_2^A \frac{\delta \mathbf{x}_1 \cdot \big( \mathbf{x}_1 - \mathbf{x}_2(t_2^A) \big)}{\left \vert \mathbf{x}_1 - \mathbf{x}_2(t_2^A) \right \vert + \mathbf{v}_1 \cdot (\mathbf{x}_1 - \mathbf{x}_2(t_2^A))} \right]  \right\rbrace + (1 \leftrightarrow 2) \; .
\end{split}
\end{align}
Upon integration by parts, and using the equations of motion, $\delta S$ reduces to a boundary term
\begin{align}
\begin{split} \label{eq:deltaS}
\delta S = \int_{t_-}^{t_+} \mathrm{d}t \frac{\mathrm{d}}{\mathrm{d}t} \left[ \mathbf{p}_1 \cdot \delta \mathbf{x}_1 - L_2^A \frac{\delta \mathbf{x}_1(t_1^R) \cdot \big( \mathbf{x}_1(t_1^R) - \mathbf{x}_2 \big)}{\left \vert \mathbf{x}_1(t_1^R) - \mathbf{x}_2 \right \vert + \mathbf{v}_1(t_1^R) \cdot (\mathbf{x}_1(t_1^R) - \mathbf{x}_2)} - F_1  \right] + (1 \leftrightarrow 2) \; ,
\end{split}
\end{align}
where we have used that, when the equations of motion are satisfied, the Jacobian of the change of variable in Eq~\eqref{eq:jacobian} is trivial, $\mathrm{d}t = \mathrm{d}t'$. In this equation, the function $F_1$ is defined by
\begin{equation}
F_1(t) = \int_{t_1^R}^t \mathrm{d}t' \left[ \frac{\partial L_2^A}{\partial \mathbf{x}_1} \delta \mathbf{x}_1 + \frac{\partial L_2^A}{\partial \mathbf{v}_1} \delta \mathbf{v}_1 \right] \; .
\end{equation}

We are now ready to derive Nother's theorem. For $\delta \mathbf{x}_1 = \delta \mathbf{x}_2 = \epsilon \mathbf{n}$ with $\epsilon$ a small parameter and $\mathbf{n}$ a constant direction (expressing the invariance of the action under space translations), by setting $\delta S = 0$ we obtain an equation expressing the conservation of total momentum. It can easily be checked to be redundant with the equations of motion~\eqref{eq:EOM_Kepler}, and the same is true for the center-of-mass theorem originating from the invariance of the action under boosts.

On the other hand, for $\delta \mathbf{x}_\alpha = \epsilon \mathbf{v}_\alpha$ ($\alpha=1,2$), the variation of the action is a total derivative, $\delta S = \int_{t_-}^{t_+} \mathrm{d}t \; \mathrm{d}L/\mathrm{d}t$ where $L$ is the total Lagrangian. Setting this quantity equal to the one we just computed~\eqref{eq:deltaS}, we can express the conservation of the total energy of the system which reads
\begin{align}
\begin{split} \label{eq:energy_circular}
E &= \frac{m_1}{2} \left(e_1+\frac{1+R_1^2 \omega^2}{e_1} \right) + \frac{m_2}{2} \left(e_2+\frac{1+R_2^2 \omega^2}{e_2} \right) - \frac{Gm_1m_2}{e_1e_2 \tilde r} \bigg[ 1 - \omega^2\left(R_1^2 + R_2^2 + 4 R_1 R_2 \cos \omega u \right)  \\
&- 3 R_1^2 R_2^2 \omega^4 \cos 2 \omega u + 4 \frac{u^2}{\tilde r} R_1 R_2 \omega^3 \sin \omega u \left(1+R_1 R_2 \omega^2 \cos \omega u\right)  \\
  &\quad + \left. \frac{\tilde \lambda}{\tilde r} \left( R_1 R_2 \omega \sin \omega u + \frac{u}{\tilde r} R_1 R_2 \omega (\omega u \cos \omega u - \sin \omega u) \right)   \right] \; .
\end{split}
\end{align}

A useful check on the validity of our computation consists in the expansion of the energy as a function of $\omega$ in the post-Newtonian regime, using the scaling $\omega R_\alpha \sim v$, $Gm/R_\alpha \sim v^2$ where $\alpha=1,2$. Finding a perturbative solution of the system of equations~\eqref{eq:system_circular} is straightforward. We introduce the standard post-Newtonian parameter
\begin{equation}
x = \left( G M \omega \right)^{2/3} \; ,
\end{equation}
where $M = m_1 + m_2$ is the total mass, in terms of which the total energy $\mathcal{E} = E - m_1 - m_2$  is
\begin{align}
\begin{split}
\mathcal{E} &= - \frac{\mu x}{2} \left( 1-\frac{x}{12}  (\nu +17)-\frac{x^2}{24}  \left(\nu ^2-209
   \nu +145\right) \right. \\
   & - \left.\frac{5 x^3}{5184}  \left(7 \nu ^3+6810 \nu ^2-22593
   \nu +23591\right) + \mathcal{O}(x^4) \right) \; ,
\end{split}
\end{align}
where as usual $\mu = m_1 m_2/(m_1+m_2)$ and $\nu = \mu / (m_1+m_2)$. Although each term in this expansion would need to be corrected to get the correct PN result (our computations are technically of 0PN order), we can observe that the $\nu^n x^n$ coefficient is correct at each order. This was already noticed in Refs.~\cite{Kalin:2019rwq,Foffa:2013gja}: the 1PM energy correctly captures the $\nu^n x^n$ coefficient, and since our computation generalizes the 1PM results we happily recover this fact.

\subsection{Innermost circular orbit} \label{sec:ICO-CICO}

Let us start this Section by considering a point-particle of mass $\mu$ in a circular orbit around a Schwarzschild black hole of mass $M$. It is well-known that the Schwarzschild solution possesses an Innermost Stable Circular Orbit (ISCO) situated at $R = 6 GM$, where $R$ is a gauge-invariant distance defined by
\begin{equation} \label{eq:GI_distance}
R = \left(\frac{GM}{\omega^2} \right)^{1/3} \; .
\end{equation}
This ISCO is situated at the minimum of the energy of the point-particle, which is
\begin{equation}
E_\mathrm{pp} = \mu \left[ \frac{1-2x}{\sqrt{1-3x}} - 1 \right] \; .
\end{equation}
Another feature of Schwarzschild geometry is the existence of a last circular orbit at $R = 3GM$, under which no circular orbits (even unstable) can exist at all. This locus corresponds to the four-velocity $v^\mu$ of the point-particle becoming lightlike thus justifying its name of 'light-ring'.

In the two-body case, various definitions exist for the Innermost Circular Orbit (ICO). One can stick to the minimum of the energy as given by equation~\eqref{eq:energy_circular}, however there is no notion of stability in this definition. An analysis of the stability of circular orbits in the post-Newtonian framework can be found in Ref.~\cite{Blanchet:2013haa}. In particular, this work suggests that all circular orbits may be stable in the equal-mass case (we recall here that we concentrate on the conservative part of the dynamics, neglecting dissipation which would make the two point-particles fall into each other in a relatively short time).

\begin{figure}
\center \includegraphics[width=.6\columnwidth]{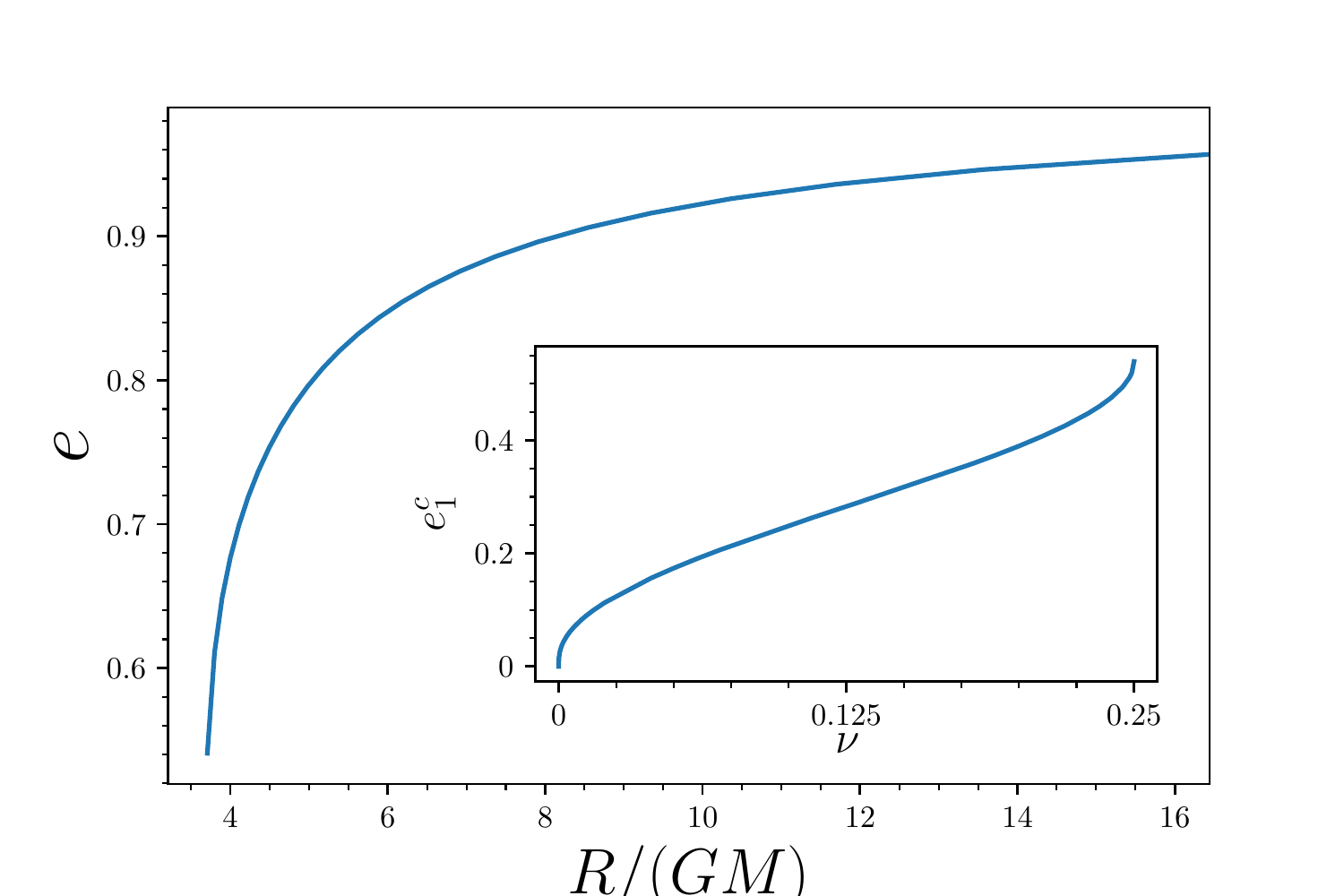}
\caption[Redshift function $e_1=e_2=e$ in the equal-mass case $m_1=m_2$]{\textit{Main plot}: Redshift function $e_1=e_2=e$ in the equal-mass case $m_1=m_2$, plotted as a function of the gauge-invariant distance~\eqref{eq:GI_distance}. The derivative of $e$ at the critical radius is infinite.
\textit{Subplot}: Critical redshift $e_1^c$ for different symmetric mass ratios $\nu$. In the test-mass limit, $e_1^c = 0$;  in the equal-mass case, $e_1^c \simeq 0.54$.}
\label{fig:ecrit_circular}
\end{figure}

We now add a third (and, we believe, more suited to the name) definition of the innermost circular orbit to the two mentioned above, which we name Critical Innermost Circular Orbit (CICO). It is defined by the point where the redshift functions $e_1, e_2$ become complex-valued \footnote{due to the symmetry of our equations, $e_1$ and $e_2$ become complex-valued at the same value of the frequency $\omega$}: no circular orbit can exist at all beyond the CICO. Moreover, for two particles approaching the CICO an observer situated on the axis of rotation of the binary system would see an abrupt change in the redshift of photons emitted near to the point-particles (i.e., an abrupt change in the functions $e_i$). This is due to the critical behavior observed in Section~\ref{sec:static_pot}: at the critical point $(e_c, \omega_c)$, both the polynomial equation on $e_i$ and its derivative vanishes, so that one has the scaling $e - e_c \propto (\omega - \omega_c)^{1/2}$ and the derivative of $e_i$ at the critical point is infinite. This is illustrated in Figure~\ref{fig:ecrit_circular}.

\begin{figure}
\center \includegraphics[width=.6\columnwidth]{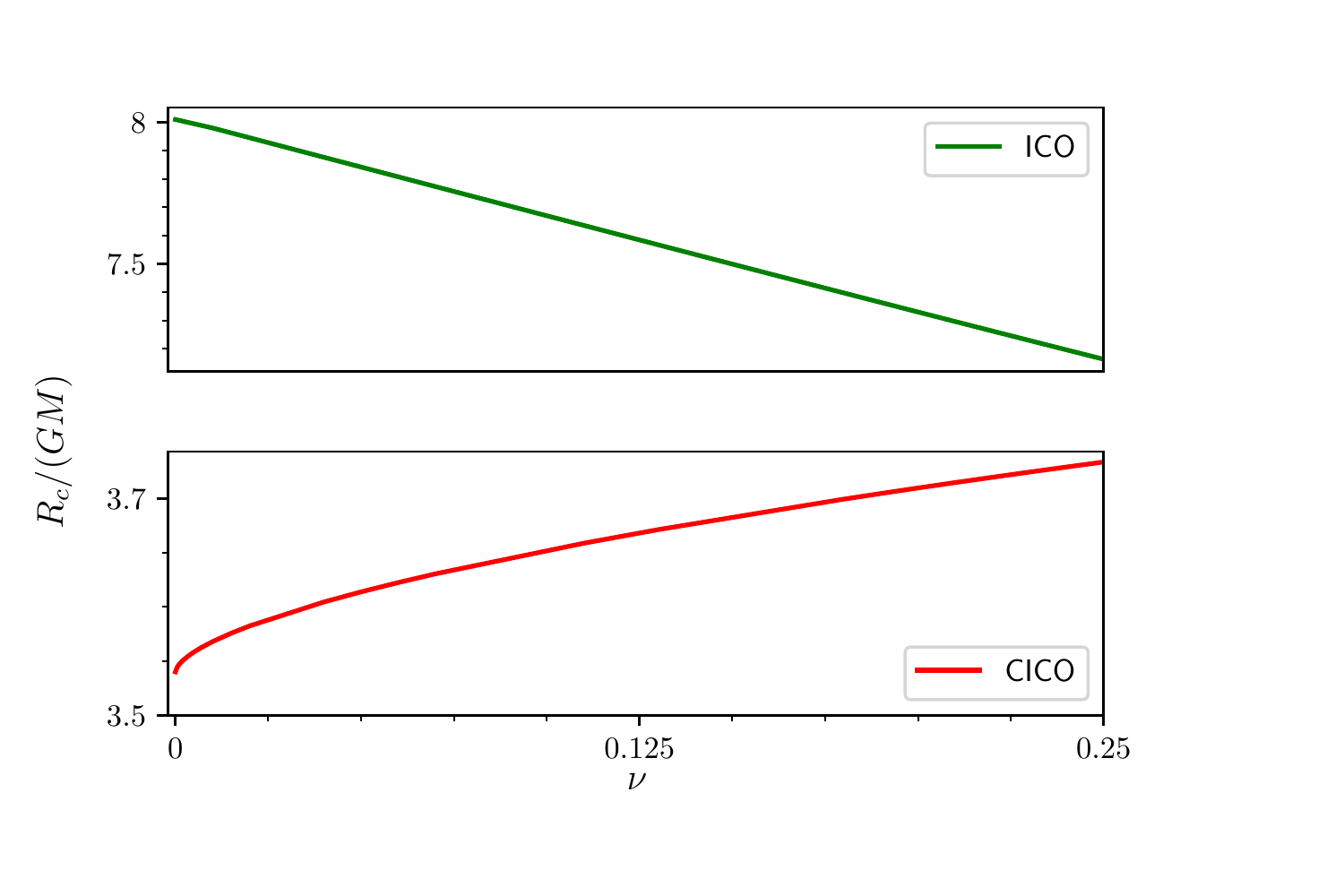}
\caption[Location of the ICO and CICO]{Location of the ICO and CICO (translated in the gauge-invariant distance~\eqref{eq:GI_distance}) for different symmetric mass ratios $\nu$. The ICO is almost linear in $\nu$. In the test-mass limit, $R^\mathrm{ICO} \simeq 8.01 GM$ and $R^\mathrm{CICO} \simeq 3.54 GM$; in the equal-mass case, $R^\mathrm{ICO} \simeq 7.16 GM$ and $R^\mathrm{CICO} \simeq 3.73 GM$. } 
\label{fig:ico_cico}
\end{figure}

Obtaining the value of the ICO by numerically solving the system of equations~\eqref{eq:system_circular} is straightforward. In Figure~\ref{fig:ico_cico} we have plotted the location of the ICO (given by its 'standard' definition, i.e the minimum of the energy~\eqref{eq:energy_circular}) and the CICO as a function of the symmetric mass ratio $\nu$. Note that their value in the test-mass limit $\nu \rightarrow 0$ is not the correct Schwarzschild result since our Lagrangian corresponds to a single graviton exchange. Indeed, our results in the test-mass limit can be recovered by considering the linearized Schwarzschild metric in harmonic coordinates,
\begin{equation}
\mathrm{d}s^2 = - \left(1 - \frac{2GM}{r} \right) \mathrm{d} t^2 + \left(1+\frac{2GM}{r} \right) \left( \mathrm{d} r^2 + r^2 \mathrm{d}\Omega^2 \right) \; .
\end{equation}
In this test-mass limit, the CICO is determined by the equation $g_{\mu \nu} v^\mu v^\nu = 0$, while the ICO is the minimum of the test-mass energy. It is straightforward to derive Kepler's law and the point-particle energy in these coordinates,
\begin{equation}
\omega^2 = \frac{GM}{r^2(r+GM)} \; , \quad E = (r-2GM) \sqrt{\frac{r+GM}{r(r^2-2GMr-4G^2M^2)}} \; .
\end{equation}
Then one readily finds that in the test-mass limit of our linearized approximation the ICO and CICO are situated at
\begin{equation}
R^\mathrm{ICO} \simeq 8.01 GM \; , \quad R^\mathrm{CICO} \simeq 3.54 GM \; .
\end{equation}
Thus our approximation overestimates the critical radius $R^\mathrm{CICO} = 3GM$ in the test-mass limit; this could be improved by taking into account higher-order nonlinear vertices in the Einstein-Hilbert action.
Note that the 1PN and 2PN results in the test-mass limit predict respectively $R^\mathrm{ICO, 1PN} \simeq 1.5 GM$ and $R^\mathrm{ICO, 2PN} \simeq 4.02 GM$.

\section{Conclusions}

In this Chapter, we have shown that the introduction of two einbeins allows for a drastic simplification of the Feynman rules of NRGR; diagrams of increasing complexity are simply recovered from the expansion of a polynomial equation. We thus expect our result to be particularly relevant for the computation of higher order PN dynamics, although as we emphasised in Section~\ref{subsec:instantaneous} our resummation technique only applies to the simplest diagrams.

Furthermore, we have shown that the polynomial equations obeyed by the redshift variables allows one to define an horizon for two interacting point-particles in GR. For circular orbits, the standard PN solution for the worldline parameters becomes complex-valued for small enough separations so that no circular orbit can exist at all beyond this critical distance. More generally, the disappearance of PN solutions for close enough binaries points towards an inadequacy of the PN parameterization in this strong-field regime.

There are multiple avenues for extending and improving our results. Apart from the inclusion of higher PM orders and the extension to BDT theories, it would also be interesting to explore the synergies of our resummation with the Effective One-Body (EOB)~\cite{buonanno_effective_1999} formalism. Indeed, while the EOB philosophy is to recast the two-body motion as the one of a point-particle in an effective metric, our resummation is somewhat two-body in essence: worldline nonlinearities do not contribute to the field of an isolated object, so that the Feynman diagrams included in our resummation contain only genuine two-body effects. On the other hand, our treatment misses the one-body dynamics (which is fully contained in the bulk nonlinearities), so that an EOB approach would be complementary.

\part{The Vainshtein mechanism} \label{part3}

\chapter{The Two-body Vainshtein potential} \label{Chapter7}

In Sections~\ref{subsec:KEssence}~-~\ref{subsec:Galileons} we have presented two cosmological theories, Galileons and K-Essence, which feature a screening mechanism, namely the Vainshtein mechanism for Galileons and the K-Mouflage one for K-Essence. Since the essential features of these two screening mechanisms are almost identical, in this Chapter we will actually refer to both of them when speaking of Vainshtein screening.
The essence of this mechanism, which solves the vDVZ discontinuity of massive gravity~\cite{1970NuPhB..22..397V, 1970JETPL..12..312Z}, is that the nonlinearities present in the scalar action dominate at small scales so that one does not recover a Newtonian behavior of the field; generically, it is found that the field goes as $\varphi \sim r^{n}$ with $n > -1$ so that it is subdominant compared to the Newtonian gravitational potential $\Phi \sim r^{-1}$. For definiteness, in this Chapter (based on the PRD paper~\citeK{Kuntz:2019plo}) we will consider a K-Mouflage theory defined by the scalar action
\begin{equation}
S = \int \mathrm{d}^4x \sqrt{-g} \left[ - \frac{(\partial \varphi)^2}{2} - \frac{1}{4\Lambda^4} (\partial \varphi)^4 + \beta \frac{\varphi T}{\mpl} \right] \; ,
\label{eq:k-mouflage}
\end{equation}
where the scalar couples to the trace of the energy-momentum tensor $T$ with a universal coupling $\beta$, and $\Lambda$ is an energy scale which (if one wants that the scalar actively participate in dark energy) is related to the Hubble rate by $\Lambda^2 = \mpl H_0$~\cite{Barreira_2015}.
Note that we have concentrated on the scalar part of the action, the gravitational action being the one of standard GR. Furthermore,
let us stress that the chosen form for the interaction (with a negative sign in front of the quartic term, which is needed in order to have a well-behaved screening) has a speed of sound around a cosmological background greater than one. Ref~\cite{Barreira_2015} discusses the conditions needed in order to have a viable k-Mouflage theory. Finally, from an EFT viewpoint a particular K-Mouflage theory such as the one chosen in our action \eqref{eq:k-mouflage} could seem quite difficult to justify: contrary to the Galileons in which there exists a non-renormalization theorem, quantum loops could generate any kind of operators such as the quartic one considered in Eq. \eqref{eq:k-mouflage} and which dominate the action in the Vainshtein regime (we will quickly recall this fact below). However, it has been argued in Ref.~\cite{deRham:2014wfa} that K-Mouflage theories could be stable under quantum corrections.
Despite all these issues, we chose the theory \eqref{eq:k-mouflage} to test our method because it is one of the simplest settings in which there is Vainshtein screening and the numerical implementation with finite elements is easier since its (weak form) equation of motion involves only first derivatives.

Let us recall the basics of the Vainshtein screening mechanism.
 The equation of motion for the scalar which follows from the action \eqref{eq:k-mouflage} is
\begin{equation} \label{eq:eom_kmouflage}
\frac{1}{\sqrt{-g}} \partial_\mu \bigg[ \sqrt{-g} \partial^\mu \varphi \bigg( 1 + \frac{(\partial \varphi)^2}{\Lambda^4} \bigg) \bigg] = -  \beta \frac{T}{\mpl} \; .
\end{equation}
Let us consider the Sun as a single point-particle at rest, $T = - M_\odot \delta^3(\mathbf{x})$. Then, expanding around flat space $g_{\mu \nu} = \eta_{\mu \nu}$ one finds using Gauss' law that the scalar obeys the equation
\begin{equation} \label{eq:single_pp_kmouflage}
\varphi' \bigg( 1 + \frac{(\varphi')^2}{\Lambda^4} \bigg) = \frac{\beta M_\odot}{4 \pi \mpl r^2} \; .
\end{equation}
The essential point of the Vainshtein mechanism is that, in the small-scale regime $r \rightarrow 0$, the nonlinear term in the lhs of Eq. \eqref{eq:single_pp_kmouflage} dominates the equation, so that the field is approximated as
\begin{equation} \label{eq:ss_field_single_pp}
\varphi \simeq 3 \left( \frac{\Lambda^4 \beta M_\odot}{4 \pi \mpl} \right)^{1/3} r^{1/3} \; .
\end{equation}
Introducing the nonlinear scale or Vainshtein radius
\begin{equation}
r_* = \sqrt{\frac{\beta M_\odot}{4 \pi \mpl \Lambda^2}} \; ,
\end{equation}
the ratio of $\varphi$ to the Newtonian potential $\phi_N = M_\odot / (4 \pi \mpl r)$ is
\begin{equation}
\frac{\varphi}{\phi_N} = \beta  \left(\frac{r}{r_*}\right)^{4/3} \; ,
\end{equation}
so that for $r \ll r_*$ the field is very suppressed. Choosing $\Lambda^2 = \mpl H_0$ for the scalar to give rise to an accelerated expansion , one has $r_* \sim \beta^{1/2} \times 0.1$ Pc for the sun so that Vainshtein screening is indeed very effective! However, the main difference with the BDT theories presented in Chapter~\ref{Chapter4} is that, even if very suppressed, the 'fifth force' induces a supplementary perihelion precession which can be used to constrain the theory since the perihelion data from the Solar system planets are very accurate \footnote{In the BDT theories of Chapter~\ref{Chapter4}, the modification to the Newtonian potential took the form of a renormalization of Newton's constant since the scalar fifth force also goes as $1/r$. Contrary to the Vainshtein case, this has no observable consequences. }

To be more precise, the supplementary perihelion angle over one revolution of a planet, say Mercury, is proportional to the ratio of fifth forces which can be obtained from the field by $F_\varphi \sim \beta m \varphi /\mpl$, $F_N \sim m \phi_N / \mpl$ where $m$ is the mass of the planet, so that the perihelion angle is
\begin{equation} \label{eq:ratio_fifth_forces}
\Delta \theta \sim \beta^2  \left(\frac{r}{r_*}\right)^{4/3} \; .
\end{equation}

The Solar system data constrains the rate of precession of the different plants perihelions, $\Delta \bar \omega \sim \Delta \theta / T$ where $T$ is the period of the planet. Since $T \propto r^{3/2}$ by Kepler's law and $ \Delta \theta \propto r^{4/3}$, we see that (since the constraint on $\Delta \bar \omega$ is similar for most of the planets of the Solar system, $\Delta \bar \omega \lesssim 0.5$ milliarcsecond/century) the most constraining measurement is the one from the closest planet to the Sun, i.e Mercury (this scaling would be different for a Galileon-3, see~\cite{Iorio:2012pv, andrews_galileon_2013, dvali_accelerated_2003}). Plugging in the numerical values, this translates in a constraint on $\Delta \theta$,
\begin{equation} \label{eq:perihelion_constraint}
\Delta \theta \lesssim 10^{-12} \; ,
\end{equation}
or in other words $\beta$ should not be larger than (do not forget that there is a factor of $\beta$ inside the Vainshtein radius $r_*$)
\begin{equation}
\beta \lesssim 10^{-5} \; .
\end{equation}

The perihelion constraint is so accurate that it strongly disfavors the K-Mouflage screening we are considering! This order-of-magnitude estimate is confirmed by a more precise calculation~\cite{Barreira_2015}.  The same analysis on a Galileon-3 yields $\beta \lesssim 0.3$~\cite{Iorio:2012pv}, which shows that Galileon screening is more effective. Nonetheless, as we already stressed before, we will stick with a K-Mouflage action \eqref{eq:k-mouflage} since its numerical implementation is simpler. Our results could easily be generalized to Galileon theories.

Another, related viewpoint on Vainshtein screening is to see it as a suppression of scalar fluctuations around massive sources. Let us assume that there are small perturbations in the energy-momentum tensor, $T = \bar T + \delta T$ where $\bar T = - m \delta^3(\mathbf{x})$.
 Splitting the field as $\varphi = \bar \varphi + \delta \varphi$ where $\bar \varphi$ is the spherically symmetric field given in Eq. \eqref{eq:ss_field_single_pp} and $\delta \varphi$ is a small fluctuation, the quadratic action for $\delta \varphi$ is 
\begin{equation}
S_\mathrm{quad} = \int \mathrm{d}^4x \left[ - \frac{(\partial \delta \varphi)^2}{2} \bigg( 1 + \frac{(\bar{\varphi}')^2}{\Lambda^4} \bigg) -\frac{(\bar{\varphi}')^2}{\Lambda^4} \big( \partial_r \delta \varphi \big)^2 + \beta \frac{\delta \varphi \delta T}{\mpl} \right] \; ,
\end{equation}
where the term linear in $\delta \varphi$ is zero because $\bar \varphi$ satisfies the equations of motion. Then in the Vainshtein regime one has $Z \equiv (\bar{\varphi}')^2 / \Lambda^4 \gg 1$ so that the kinetic term for $\delta \varphi$ is multiplied by a huge number. Canonically normalizing the fluctuations $\psi = \sqrt{Z} \varphi$, one finds that the coupling of $\psi$ to matter is suppressed by a factor $\sqrt{Z}$: this is the Vainshtein mechanism. Such an approach will be adopted in Chapter~\ref{Chapter8} when considering perturbations caused by 'small', solar-size black holes orbiting around a supermassive one.

In both approaches discussed above, one can compute the Vainshtein effect using the fact that the Sun provides the dominant part of the field so that one can use spherical symmetry to simplify the equations. However, little is known on the Vainshtein mechanism beyond this simple spherically-symmetric one-body case. With an eye on the constraints coming from GW observations, it would be very interesting to investigate Vainshtein screening between two comparable-mass bodies. 
Indeed,
recent work on the radiation of binary systems in Galileon theories, both theoretical~\cite{chu_retarded_2013, deRham:2012fg, deRham:2012fw} and numerical~\cite{dar_scalar_2018}, have shown that the radiation itself is screened with powers of $\lambda/r_*$, where $\lambda$ is the wavelength of the emitted radiation which is greater by an amount $1/v$ than the size of the system $r$ (here $v$ is the typical velocity of the bodies). The theoretical calculations, which are performed assuming a background field generated by a central mass $M=m_1+m_2$, are better suited for a small mass ratio inspiral. We will follow this approach in the next Chapter; however for the moment we will be more concerned about the comparable-mass case which has never been studied in details yet.

In this Chapter we present a first step towards this direction, namely an approximation to the non-relativistic conserved energy of two bodies. To do so, we will first show that the perturbative tools we employed in Chapter~\ref{Chapter4} are adapted to the calculation of this energy outside the screening regime, when the distance between the two bodies is large and nonlinearities are only a small perturbation of the quadratic term. While this is not the relevant regime of interest in which we are interested for Solar system constraints, this will allow us to get some insight on the form of the energy. Notably, we will be able to use tools borrowed from the Effective One-Body formalism (EOB formalism, see Section~\ref{subsec:other_approaches}) to resum a part of the nonlinear corrections (note that the EOB formalism has been   generalized to BDT theories in~\cite{Julie:2017ucp, Julie:2017pkb}).

We will then postulate that the EOB relations can also be used in the screening region where nonlinearities dominate the action.
Since these relations map the two-body problem into a spherically symmetric one-body problem, it allows to recover the energy as a function of the known exact field solution. The difficulty is then to identify the parameters of this deformed exact solution. Outside the screening radius, our perturbative (Feynman) expansion allows us to identify the EOB parameters. However, the perturbative expansion breaks down for objects separated by less than the screening radius.

It is however easy to identify the EOB parameters in the screening region using a simple numerical simulation, and the rest of this Chapter will be devoted to this task. Numerical simulations of screened theories in the quasi-static limit have mostly focused on N-body simulations for cosmological applications~\cite{Barreira:2013b, Li:2011vk, Li:2013nua, Khoury:2009tk, 2009PhRvD..80l3003S, Schmidt:2009sg}, which is not our concern here. We thus implement our own code using the Finite element solver Fenics~\cite{ans20553}. We first obtain the energy outside the screening radius in order to confirm our preliminary results. We then concentrate on the two-body energy in the fully screened situation which is relevant for astrophysical systems, and for arbitrary mass ratios. The most important result of this Chapter is the final mass ratio dependance of the two-body energy, and it is shown in Figure~\ref{fig:b0_num}.

Finally, we will present an important application of our results concerning the two-body Vainshtein potential for arbitrary mass ratio: we will show that the dependence of the energy on the mass ratio implies a violation of the Weak Equivalence Principle (WEP). Indeed, in a Vainshtein screened theory two bodies of comparable mass would not fall in the same way as a test-particle in an external gravitational field. This has direct implications on the orbit of the Moon which is measured with great accuracy by Lunar Laser Ranging (LLR, see Section~\ref{subsec:expe_ppn}). We will see how this translates in a bound on the parameters of the theory.

\section{Two-body energy outside the screening radius} \label{sec:outside}

Let us now apply the perturbative Feynman expansion introduced in Chapter~\ref{Chapter4} to the theory \eqref{eq:k-mouflage},
 We will only consider static point-sources,
 \begin{equation}
T = -m_1 \delta^3(\mathbf{x} - \mathbf{x}_1) - m_2 \delta^3(\mathbf{x} - \mathbf{x}_2) \; .
\end{equation} 
  so we will ignore time from now on and focus only on the potential energy $E$. In order to have simpler expression in what follows and also to compare directly our results to the numerical simulation, we will introduce rescaled variables
\begin{equation} \label{eq:rescaling}
\tilde{\varphi} = \frac{\varphi}{\Lambda^2}, \quad \tilde{m} = \frac{\beta m}{4\pi M_P \Lambda^2} \; ,
\end{equation}
so that the action of the system writes as
\begin{equation}
\frac{S}{\Lambda^4} = \int dt d^3x \left[ - \frac{1}{2} (\nabla \tilde{\varphi})^2 - \frac{1}{4} (\nabla \tilde{\varphi})^4 + \tilde{\varphi} \tilde{T} \right] \; ,
\label{eq:k_mouflage_rescaled}
\end{equation}
where $\tilde{T} = -4\pi \tilde{m}_1 \delta^3(\mathbf{x} - \mathbf{x}_1) - 4 \pi \tilde{m}_2 \delta^3(\mathbf{x} - \mathbf{x}_2)$, and we will ignore the tilde from now on. Be careful that now the dimension of $m$ is (in natural units) $-2$ and the dimension of $\varphi$ is $-1$.

\subsection{Spherically symmetric case}

Consider the action \eqref{eq:k_mouflage_rescaled} with a single point-like source $T = -4\pi M \delta^3(\mathbf{x})$. The equation of motion \eqref{eq:eom_kmouflage} in our rescaled units reads
\begin{equation}
\partial_i \left(\partial_i \varphi + (\partial \varphi)^2 \partial_i \varphi \right) = - T \; .
\end{equation}

Using spherical symmetry and integrating over a sphere, one can reduce it to a single ordinary differential equation
\begin{equation}
\varphi' + (\varphi')^3 = \frac{M}{r^2} \; .
\label{eq:cubic_eq}
\end{equation}

This can be solved exactly so that, for a field vanishing at infinity,
\begin{equation}\label{eq:hypergeom}
\varphi_M(r) = - \frac{M}{r} {}_3F_2 \left(\frac{1}{4}, \frac{1}{3}, \frac{2}{3} ; \frac{5}{4}, \frac{3}{2} ;- \frac{27M^2}{4r^4} \right) \; ,
\end{equation}
where ${}_pF_q$ is the generalized hypergeometric function.

The solution has two regimes separated by the nonlinear scale $r_*$
\begin{align}
\begin{split}
\varphi_M &= - \frac{M}{r} + \frac{M^3}{5 r^5} + \dots, \quad r > r_* \; , \\
&= C + 3 \left(  M r \right)^{1/3} + \dots, \quad r<r_* \; ,
\end{split}
\label{eq:spherical_symmetry}
\end{align}
where $C \simeq -3.7 \sqrt{M}$ is a constant of integration (we chose the constant such that the field vanishes at infinity, so it cannot also vanish in zero), and the nonlinear scale is given by
\begin{equation}
r_* = \left( \frac{27}{4} \right)^{1/4} \sqrt{M} \; ,
\end{equation}
and corresponds to the radius of convergence of the two series written above.

These two regimes can be seen as expansions in $r_*/r$ and $r/r_*$ respectively (in fact, it is even possible to reformulate the initial action with additional fields in order to make the screened regime appear from the beginning, see~\cite{gabadadze_classical_2012}). We have expanded up to next-to-leading order in the $r>r_*$ case because it will prove useful in the following. We have shown in the introduction that $r_* \sim 0.1$ Pc for the sun, thus rendering the next order in the $r<r_*$ case very subdominant concerning Solar System experiments.

\subsection{Two-body problem} \label{sec:two_body}

Let's now take a two-body source $T = -4 \pi m_1 \delta^3(\mathbf{x} - \mathbf{x}_1) - 4 \pi m_2 \delta^3(\mathbf{x} - \mathbf{x}_2)$. The salient feature of the two-body problem in screened theories is that one can not compute the energy by superposing one-body energies, as one usually does in Newtonian gravity. Said differently, one can not recast the problem in terms of a simple ODE, and instead one should solve a nonlinear PDE of 2 variables, thus explaining the lack of analytical results. Nonetheless, the problem is well formulated outside the screening radius where the nonlinear term can be treated as an interaction. Let us now imagine to have two bodies outside of their respective screening radius and calculate the first nonlinear correction to the potential energy of the two objects. To this aim, we will use Feynman rules derived from the action \eqref{eq:k_mouflage_rescaled}, as explained in Section~\ref{sec:feynman}.
In particular, we recall that the non-relativistic propagator of the field obtained from the quadratic term in the action~\ref{eq:k_mouflage_rescaled} is given by
\begin{equation}
\left\langle T \varphi_\mathbf{k}(t) \varphi_\mathbf{q}(t') \right\rangle = -i (2\pi)^3 \delta^3(\mathbf{k} + \mathbf{q}) \delta(t-t') \frac{1}{k^2} \; ,
\end{equation}
where our Fourier convention is $\varphi = \int_\mathbf{k} \varphi_\mathbf{k} e^{i \mathbf{k} \cdot \mathbf{r}}$, and $\int_\mathbf{k}$ means $\int \frac{\mathrm{d}^3k}{(2\pi)^3}$.

On the other hand, the new nonlinear term gives a new interaction vertex
\begin{equation}
\frac{-i}{4} \int \mathrm{d} t \int_{\mathbf{k}_1, \; \mathbf{k}_2, \; \mathbf{k}_3, \; \mathbf{k}_4} (2\pi)^3 \delta^{(3)}( \mathbf{k}_1 + \mathbf{k}_2 + \mathbf{k}_3 + \mathbf{k}_4) \; \big( \mathbf{k}_1 \cdot \mathbf{k}_2 \big) \; \big( \mathbf{k}_3 \cdot \mathbf{k}_4 \big) \varphi_{\mathbf{k}_1} \; \varphi_{\mathbf{k}_2} \; \varphi_{\mathbf{k}_3} \; \varphi_{\mathbf{k}_4} \; .
\end{equation}

\begin{figure}[h]
	\centering

\includegraphics{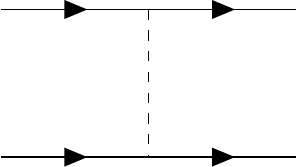}
	
\caption[Feynman diagram contributing to the Newtonian potential]{Feynman diagram contributing to the Newtonian potential. External sources are represented as straight lines and scalars as dotted lines}
\label{fig:Newt_pot_chap7}
\end{figure}

In the classical saddle-point action $S_\mathrm{cl}$, at lowest order there is the well-known one-scalar exchange of Fig~\ref{fig:Newt_pot_chap7} (we already computed it in Section~\ref{sec:feynman}) that gives rise to the Newtonian potential between the two sources. In our units, it is (with $r = |\mathbf{x}_1 - \mathbf{x}_2|$)
\begin{equation}
\frac{E_\mathrm{Newt}}{4\pi \Lambda^4} = -\frac{m_1 m_2}{r} \; ,
\label{eq:Newtonian_energy}
\end{equation}
As a side note, remember that the quantity on the lhs of Eq. \eqref{eq:k_mouflage_rescaled} is $S/ \Lambda^4$, so we are rather computing the rescaled energy $E/\Lambda^4$ with the above Feynman rules.
 For later convenience we will now use a rescaled energy defined by
\begin{equation}
\tilde{E} = \frac{E}{\Lambda^4} \; ,
\end{equation}
and forget the tilde from now on.

\begin{figure}
	\centering
	
	\subfloat[]{

			\label{subfig:correction_kmouflage_4a}

\includegraphics{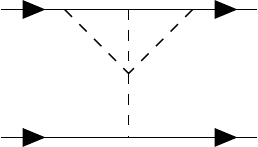}
	}
	\subfloat[]{

			\label{subfig:correction_kmouflage_4b}

\includegraphics{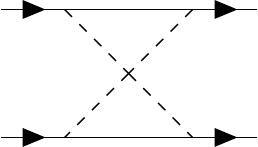}
	}

\caption[Feynman diagrams contributing to the first nonlinear correction in a K-mouflage theory]{Feynman diagrams contributing to the first nonlinear correction in a K-mouflage theory. The first one should be added with its symmetric counterpart.}
\label{fig:correction_kmouflage}
\end{figure}

At next order in perturbation theory, we have to calculate the two diagrams of Figure~\ref{fig:correction_kmouflage}. The first one can be put in the following form
\begin{align}
\begin{split}
\mathrm{Fig}~\ref{subfig:correction_kmouflage_4a} &= i \frac{(4\pi)^4}{3} m_1^3m_2 \int_{\mathbf{K}, \mathbf{k}_1, \mathbf{k}_2} e^{i \mathbf{K} \cdot \mathbf{r}} \frac{1}{\mathbf{k}_1^2 \mathbf{k}_2^2 \mathbf{K}^2 (\mathbf{K}+\mathbf{k}_1 + \mathbf{k}_2)^2} \\
& \times \left( \mathbf{k}_1 \cdot \mathbf{k}_2 (\mathbf{K}+\mathbf{k}_1 + \mathbf{k}_2) \cdot \mathbf{K} + 2 \; \mathrm{perm}  \right) \; ,
\end{split}
\end{align}
where $\mathbf{r} = \mathbf{x}_1 - \mathbf{x}_2$. The remaining integrals over momentum should be computed with dimensional regularization. To this aim a set of useful integrals in $d=3-2\epsilon$ are given in Appendix~\ref{sec:useful_integrals}.  One can simplify the calculation with the following observation: after having integrated $k_1$ and $k_2$, one is left with an integral over $K$ which, for dimensional reasons (the first correction to the two-body potential is proportional to $r^{-5}$, see \eqref{eq:spherical_symmetry}), is of the form
\begin{equation}
\int_{\mathbf{K}} \mathbf{K}^2 e^{i \mathbf{K} \cdot \mathbf{r}} \; ,
\end{equation}
(there could also be a factor of $\epsilon$ in the power of $\mathbf{K}^2$, but it does not change the validity of the argument). Then using formula \eqref{eq:intK}, one can see that there is a pole of the Gamma function in the \textit{denominator}. Consequently, this integral will vanish in dim. reg (it will be proportional to $\epsilon$) unless there is also a pole of the Gamma function in the numerator. Keeping the only pole that appear in the numerator, and repeatedly using the formulaes given in Appendix~\ref{sec:useful_integrals}, one can find that
\begin{equation}
\mathrm{Fig}~\ref{subfig:correction_kmouflage_4a} = -4\pi i \frac{m_1^3m_2}{5 r^5} \; .
\end{equation}

One can use the same machinery to calculate the second diagram~\ref{subfig:correction_kmouflage_4b}. However, in this case there is no pole of the Gamma function at the numerator, and consequently this diagram vanishes in dim. reg. Including the symmetric counterpart of~\ref{subfig:correction_kmouflage_4a}, we finally obtain the formula for the first correction to the two-body energy,
\begin{equation}
\frac{E}{4\pi} = - \frac{m_1 m_2}{r} + \frac{m_1m_2(m_1^2+m_2^2)}{5 r^5} + \dots \; .
\label{eq:2_body_energy}
\end{equation}

\subsection{Resumming the test-mass diagrams} \label{sec:resum_TM}

The result of Feynman integrals concerning diagram~\ref{subfig:correction_kmouflage_4a} should come as no surprise. Indeed, as we will now explain, this diagram is equivalent to the corresponding energy of a point-particle mass in case of one of the masses goes to zero. In the following, we will assume without loss of generality that $m_1 < m_2$.

Consider the limit $m_1 \rightarrow 0$. Then the two-body energy should reduce to the energy of a point-particle $m_1$ in the external field generated by $m_2$, which is (the energy being obtained from the field $\varphi_\mathrm{cl}$ of eq. \eqref{eq:spherical_symmetry} via $E_\mathrm{pp} = - \int d^3x \varphi_\mathrm{cl} T$)
\begin{equation}
\frac{E_\mathrm{pp}}{4\pi} = - \frac{m_1 m_2}{r} + \frac{m_1m_2^3}{5 r^5} + \dots
\end{equation}

From there we see that the result of diagram~\ref{subfig:correction_kmouflage_4a}, which ultimately gives the numerical prefactor in front of the first-order correction to the two-body energy, could not have been otherwise. Had the diagram of figure~\ref{subfig:correction_kmouflage_4b} been nonzero, its value would not have been fixed by this observation, because it is proportional to $m_1^2 m_2^2$, which vanish (compared to the test-mass diagram proportional to $m_1$) in the test-mass limit.

\begin{figure}
	\centering

\includegraphics{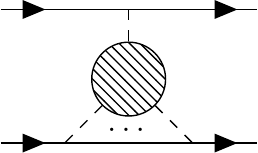}
	
\caption[Test-mass diagram with a single coupling to the first particle $m_1$ and $P$ couplings to the second particle $m_2$]{Test-mass diagram with a single coupling to the first particle $m_1$ and $P$ couplings to the second particle $m_2$. The number of internal vertices (not involving particles worldlines) is $N$.}
\label{fig:test_mass}
\end{figure}

Based on this observation, we propose to resum a particular class of diagram which share the same property at any order in the nonlinear expansion, which we call test-mass diagrams. They are given by a single coupling to the mass $m_1$ and any number of couplings to $m_2$, as illustrated in Figure~\ref{fig:test_mass}. Denoting by $N$ the expansion order (i.e, the number of vertex corresponding to the insertion of the nonlinear operator in the diagrams) and by $P$ the number of $m_2$ mass insertions, we have that there are $1+4N+P$ field insertions in this diagram. Since there are no loops and the diagram should be connected once we remove the particles wordlines, there are $N-1$ 'internal' propagators, i.e propagators that connect two nonlinear vertices. Finally, the total number of propagators is $P+1+(N-1)$. Since the total number of field insertions is two times the number of propagators, we get the relation
\begin{equation}
P=2N+1 \; ,
\end{equation}
so that at order $N$ the mass coefficient of this graph (and its symmetric counterpart) is $m_1m_2(m_1^{2N}+m_2^{2N})$.

One can even argue that, at a given perturbation order $N$, this test-mass diagram is the leading one away from the test-mass limit. Indeed, the contribution of other kinds of diagrams, with more $m_1$ insertions (we found the only other second-order diagram to vanish, but we see no reason why it should be the case at higher orders), would be of the form $(m_1m_2)^q(m_1^{2(N+1-q)} + m_2^{2(N+1-q)})$ where $2 \leq q \leq N+1$. For $m_2 > m_1$ and in the large $N$ limit, the ratio of this quantity to the test-mass diagram is
\begin{equation}
\left( \frac{m_1}{m_2} \right)^{q-1} \; ,
\end{equation}
which is less than one. However, this is not accounting for the fact that there can be a large number of these other diagrams, which can make them count as much as the test-mass one. Anyway, we shall content ourselves with having understood at least a part of the nonlinear energy.

Now the exact energy of a point-particle $m_1$ in the field generated by $m_2$ is
\begin{equation}
\frac{E_{pp}}{4 \pi} = m_1 \sum_{N \geq 0} \alpha_N \frac{m_2^{2N+1}}{r^{4N+1}} \; ,
\end{equation}
where $\alpha_N$ are numerical coefficients that can be easily found by solving eq. \eqref{eq:cubic_eq}. By writing in a identical way the contribution of the test-mass graphs to the two-body energy,
\begin{equation}
\frac{E}{4\pi} = \sum_{N \geq 0} \beta_N \frac{m_1m_2(m_1^{2N}+m_2^{2N})}{r^{4N+1}} \; ,
\label{eq:2_body_energy_resummed}
\end{equation}
we see that in order to have the good test-mass limit one should impose $\beta_N = \alpha_N$ for $N \geq 1$, and $\beta_0 = \frac{\alpha_0}{2} = - \frac{1}{2}$.
Denoting by $\varphi_m$ the (exact) spherically symmetric field generated by a body of mass $m$ in eq. \eqref{eq:spherical_symmetry}, one finally finds for the contribution of test-mass graphs to the two-body energy
\begin{equation}
\frac{E}{4\pi} = \frac{m_1 m_2}{r} + m_1 \varphi_{m_2}(r) + m_2 \varphi_{m_1}(r) \; .
\label{eq:energy_outside_exact}
\end{equation}

This result can be intuitively understood as being the symmetric sum of one-body energies, plus a compensating term that ensures that the Newtonian limit $r \rightarrow \infty$ (where $\varphi_m \sim -m/r$) is correct. We will compare this analytical resummed energy to the numerical solution in Sec.~\ref{sec:results}.

\section{Effective One-Body approach}
\label{sec:EOB}

\subsection{Energy map outside}

Very much like in GR, where the motion of a two-body system (expanded in powers of $r_s/r$, where $r_s$ is the Schwartzchild radius of the combined mass $M=m_1+m_2$) can be recast in the motion of a test-mass in a modified external Schwartzchild metric~\cite{buonanno_effective_1999}, the above formula for the two-body energy can be expressed into the energy of such a point-particle in a modified external field. Of course, what will now play the role of the Schwartzchild radius is the nonlinear radius $r_*$. Let us define, in top of the reduced mass $\mu$ and the total mass $M$ which are the two masses naturally associated to the effective problem, the mass ratio $x$ as
\begin{align}
\begin{split}
\mu &= \frac{m_1m_2}{m_1+m_2} \; , \\
M &= m_1+m_2 \; , \\
x &= \frac{m_1}{m_1+m_2} \; .
\end{split}
\end{align}

The kinetic energy of the two objects is easy to rewrite in terms of an effective kinetic energy since we have the well-known relation
\begin{equation}
\frac{1}{2} m_1 v_1^2 + \frac{1}{2}m_2 v_2^2 = \frac{1}{2} \mu v^2 \; ,
\end{equation}
where $\mathbf{v} = \mathbf{v}_1 - \mathbf{v}_2$, and we have set the center-of-mass to the origin of coordinates (this center-of-mass definition would be modified by relativistic corrections, but we do not consider these in this treatment).

As for the approximate potential energy that we obtained above \eqref{eq:2_body_energy_resummed}, it can be rewritten in terms of the effective parameters as
\begin{equation}
\frac{E}{4\pi} = \mu \sum_{N \geq 0} \beta_N \frac{M^{2N+1} \left(x^{2N}+(1-x)^{2N}\right)}{r^{4N+1}} \; .
\end{equation}

We now see that, outside the nonlinear radius, the motion of a two-body system separated by $r = |\mathbf{x}_1 - \mathbf{x}_2|$ can be identified with the motion of a test-particle of mass $\mu$ (at a distance $r$ from the origin) in a modified external field created by $M$, whose modified coefficients in the nonlinear expansion are given by
\begin{align}
\begin{split}
\tilde{\alpha}_N &= \alpha_N \left(x^{2N}+(1-x)^{2N}\right), \; N>0 \;, \\
\tilde{\alpha}_0 &= \alpha_0 \; .
\end{split}
\label{eq:modified_coeffs}
\end{align}

What is the nonlinear radius associated to the effective problem ? The perturbative expansion \eqref{eq:modified_coeffs} that we wrote above breaks down at the nonlinear radius associated to the biggest of the two masses, i.e $r_* = M(1-x)$. However this result is really tied to the test-mass resummation, and calculating more precisely the two-body energy could change it. It is nonetheless natural to assume that the nonlinear radius of the full two-body problem is the one associated to the total mass $M$, i.e.,
\begin{equation}
r_* = \left( \frac{27}{4} \right)^{1/4} \sqrt{M} \; ,
\end{equation}
knowing that the real two-body Vainshtein radius (defined as the radius of convergence of the energy expansion for $r \rightarrow \infty$) could differ by numerical factors dependant on $x$.

Alternativaly, we can formulate the equivalent one-body problem in a different way which will prove useful when investigating the behavior inside the nonlinear radius. Using the energy of a test-mass $\mu$ in an external field generated by $M=m_1+m_2$,
\begin{align} \label{eq:pp_energy}
\begin{split}
\frac{E_\mathrm{tm}}{4 \pi} &= \mu \varphi_M(r) \\
 &= \mu \sum_{N \geq 0} \alpha_N \frac{M^{2N+1}}{r^{4N+1}} \; ,
\end{split}
\end{align}
we can build an \textit{energy map} between the real energy $E$ and the effective test-mass energy $E_\mathrm{tm}$ as follows
\begin{align} \label{eq:energy_map_outside}
\begin{split}
\frac{E}{E_\mathrm{tm}} &= f\left(\frac{E_\mathrm{tm}}{E_N}-1\right) \\
&= a_0 + a_1 \left(\frac{E_\mathrm{tm}}{E_N}-1\right) + a_2 \left(\frac{E_\mathrm{tm}}{E_N}-1\right)^2 + \dots
\end{split}
\end{align}

Here $E_N = - \mu M/r$ is the Newtonian reference energy, and the function $f$ has been Taylor expanded for small values of the dimensionless ratio $E_\mathrm{tm}/E_N-1$. Indeed, from eq. \eqref{eq:spherical_symmetry} this ratio can be expanded outside the nonlinear radius as
\begin{equation}
\frac{E_\mathrm{tm}}{E_N}-1 = - \frac{M^2}{5r^4} + \dots
\end{equation}

To construct such an energy map, one can choose each value $a_N$ such that each coefficient in front of $(M^2/r^4)^N$ of eq. \eqref{eq:energy_map_outside} matches. Since the small ratio $E_\mathrm{tm}/E_N-1$ is chosen such that $a_N \left(E_\mathrm{tm}/E_N-1 \right)^N$ contributes only at order $(M^2/r^4)^N$ or higher, this procedure yields an unambiguous value for $a_N$ for all $N$.

This energy map proves very useful because it allows to resum the nonlinear behavior into the small parameter $E_\mathrm{tm}/E_N-1$. We refer the reader to Ref.~\cite{buonanno_effective_1999} for its derivation in the context of the Post-Newtonian formalism.

\subsection{Energy map inside}

Having understood the behavior of the energy outside the nonlinear radius, we would like now to generalize to astrophysical situations of interest where the two bodies lie deep within their nonlinear radius. The above formula \eqref{eq:energy_outside_exact} for the two-body energy, even if it resums part of the nonlinear corrections, has no chance to be valid inside $r_*$ because it contains the Newtonian reference energy \eqref{eq:Newtonian_energy} valid only at large radius. However, one can still obtain the coefficients of an energy map from numerical simulations. In App.~\ref{sec:matching} we adopt an analytical approach that attempts to relate the two energy maps by a matching condition at the nonlinear radius. While we will argue that this part should yield a qualitative result concerning the two-body energy, we will see when comparing to the numerical simulation that even this qualitative result does not compare well to the real two-body energy. Further improvement is needed in order to obtain a sensible analytical result.

In the following, we will \textit{assume} that the two-body energy for $r<r_*$ can be related, in a spirit similar to the one for $r>r_*$, to the energy of a test-mass $\mu$ in an external field generated by the total mass $M$ through an energy map. We have no possibility to calculate directly the modified coefficients as we did in the last Section, but we can obtain them numerically as we will do in Sec.~\ref{sec:numerical}.

Inside the nonlinear radius, the energy map should take the form
\begin{align} \label{eq:energy_map_inside}
\begin{split}
\frac{E'}{E'_\mathrm{tm}} &= g\left(\frac{E'_\mathrm{tm}}{E'_\mathrm{ref}}-1\right) \\
&= b_0 + b_1 \left(\frac{E'_\mathrm{tm}}{E'_\mathrm{ref}}-1\right) + b_2 \left(\frac{E'_\mathrm{tm}}{E'_\mathrm{ref}}-1\right)^2 + \dots
\end{split}
\end{align}

A few comments are required here. First, in order to avoid the appearance of an unphysical (mass-dependant) constant in the energy (which we calculated before by assuming its vanishing at infinity, so it cannot also vanish in zero), we chose to write the energy map using the $r$-derivative of the energy $E'$. Second, we use a reference energy level equal to the small-$r$ value of the energy since the Newtonian energy is irrelevant inside the Vainshtein radius,
\begin{equation}
\frac{E'_\mathrm{ref}}{4\pi} = \mu \left( \frac{M}{r^2} \right)^{1/3} \; .
\end{equation}

Finally, note that the small parameter $E'_\mathrm{tm}/E'_\mathrm{ref} - 1$ is this time expanded as
\begin{align}
\begin{split}
\frac{E'_\mathrm{tm}}{E'_\mathrm{ref}}-1 &= \sum_{N \geq 1} \gamma_N \left( \frac{r^2}{M} \right)^{2N/3} \\
&= - \frac{1}{3} \left( \frac{r^2}{M} \right)^{2/3} + \dots \; ,
\end{split}
\end{align}
where the coefficients $\gamma_N$ can be found by solving exactly the nonlinear equation \eqref{eq:cubic_eq} (we will not need their exact expression here).

In the test-mass limit, the real two-body energy should be approximated by the point-particle energy, and consequently $b_0 = 1$ and $b_1$, $\dots$ $b_N=0$. Now, away from the test-mass limit, we are only interested by the behavior of $b_0$ as a function of the mass ratio $x$, since we recall that the next order is further Vainshtein suppressed and thus irrelevant for astrophysical systems. We will now turn to a numerical implementation that will allow us to obtain the coefficient $b_0$.


\section{Numerical simulation} \label{sec:numerical}

In this Section, we will directly solve the nonlinear PDE in the two-body case using a finite element solver and obtain the two-body energy in order to compare it to the analytical result.

\subsection{Setup} \label{sec:setup_chap7}

Using the action \eqref{eq:k_mouflage_rescaled}, one has the following equation of motion for the scalar field
\begin{equation}
\nabla \cdot \left( \nabla \varphi + (\nabla \varphi)^2 \nabla \varphi \right) = -T \; ,
\end{equation}
where we recall that
\begin{equation}
T = -4\pi m_1 \delta^3(\mathbf{x} - \mathbf{x}_1) - 4 \pi m_2 \delta^3(\mathbf{x} - \mathbf{x}_2) \; .
\end{equation}


To numerically solve this equation, a finite element (FEM) solver \footnote{We use the Python FEM solver fenics~\cite{ans20553}} is well adapted to the problem, since the PDE can easily be put into a weak form: for any test function $v$ that vanish on the boundary of the integration domain,
\begin{equation} \label{eq:weak_equation}
\int d^3x \left( (1+ (\nabla \varphi)^2) \nabla \varphi \cdot \nabla v - Tv \right) = 0 \; .
\end{equation}

FEM solvers solve the weak form equation by decomposing the unknown $\varphi$ on some basis functions $\psi_j$, here chosen to be the continuous Lagrange polynomials of second order on the grid chosen to discretize the problem. The solver then finds the coefficients $c_j$ of this decomposition $\varphi = \sum_j c_j \psi_j$ by evaluating the weak form equation \eqref{eq:weak_equation} with the test function $v$ being one of the basis functions $\psi_i$. This produces a matrix equation for the unknown vector of coefficients $(c_j)_{j \geq 0}$ that is solved by an efficient sparse LU decomposition. The non-linear term is dealt with Newton iterations, i.e. by setting $\varphi = \varphi_0 + \varphi_1$ with $\varphi_0$ a function that is close to the solution sought after, linearizing over $\varphi_1$ then solving for it, and finally iterate the procedure until a desired convergence threshold has been reached.

In cylindrical coordinates (so that the problem becomes effectively two-dimensional), we choose the two bodies to lie along the $z$ axis at positions $+a/2$ and $-a/2$. We regularize the delta-functions by replacing them with Gaussians,
\begin{align}
\begin{split}
\delta^3(\mathbf{x}) = \frac{\delta(r) \delta(z)}{2\pi r} 
= \frac{1}{2\pi^2 \sigma^2 r} e^{-\frac{r^2+z^2}{2\sigma^2}} \; ,
\end{split}
\end{align}
where we have taken care of the fact that the $r$ variable goes from $0$ to $\infty$ while $z$ goes from $- \infty$ to $\infty$.

There are two scales involved in this problem, the separation between the two bodies that we denote by $a$ and the nonlinear scale $r_* = \sqrt{M}$. We choose for the domain of integration the half-disk defined by $r>0$, $R \leq R_\mathrm{max}$ where $R = \sqrt{r^2+z^2}$. The domain is automatically discretized by the FEM solver with a resolution of approximately $64 \times 64$ points, with a manual refinement of the grid near the two bodies. The boundary conditions are chosen such that $\partial_r \varphi (0,z) = 0$ (as required by the symmetry of the problem) and $\varphi(r,z) = - \frac{m_1+m_2}{R}$ for $R=R_\mathrm{max}$. The second boundary condition corresponds to recovering both spherical symmetry and the Newtonian-like behavior of the field far from the two bodies, where we know the exact solution which is given by eq. \eqref{eq:spherical_symmetry}. For it to be consistent, we must also ensure $r_* \ll R_\mathrm{max}$.

Once we get the field solution, the energy can be computed as
\begin{align}
\begin{split}
E = \int d^3x \left( \frac{(\nabla \varphi)^2}{2} +  \frac{(\nabla \varphi)^4}{4} \right) 
+ 4\pi m_1 \varphi(\mathbf{x}_1) + 4\pi m_2 \varphi(\mathbf{x}_2) \; .
\label{eq:numerical_energy}
\end{split}
\end{align}

A remarkable fact to be noted is that the self-energy contribution $4\pi m_1 \varphi(\mathbf{x}_1) + 4\pi m_2 \varphi(\mathbf{x}_2)$ is not divergent, contrary to the Newtonian case. This is due to the fact that the field goes to a constant as $\varphi \sim |\mathbf{r} - \mathbf{x}_\alpha|^{1/3}$ close to the source $\alpha$, instead of diverging as $1/|\mathbf{r} - \mathbf{x}_\alpha|$. The same goes for the integral over all space. We consequently do not need to renormalize the energy.

Finally, after having compared the behavior of the energy for $r \gtrsim r_*$ to the theoretical predictions, we will need the energy in the realistic $r \ll r_*$ case. In this setup, we can ignore the quadratic term in the equation of motion \eqref{eq:weak_equation}. This corresponds to take an infinite Vainshtein radius. Correspondingly, we have to change the exterior boundary condition to $\varphi(r,z) = 3((m_1+m_2)R)^{1/3}$ for $R=R_\mathrm{max}$ because the field is screened over all space. The energy is computed ignoring also the quadratic term, but there is a subtelty in its definition because the integral is formally divergent as $R_\mathrm{max}^{1/3}$ (a consequence of the good UV and bad IR behavior of the field). This divergence is the same as the one associated to a single particle of mass $M=m_1+m_2$ sitting at the origin, and so to renormalize the theory we susbtract to the energy the contribution of such a field which is
\begin{equation}
3\pi ((m_1+m_2)R_\mathrm{max})^{1/3} \; .
\end{equation}



\subsection{Numerical tests} \label{sec:num_tests}

In order to assess the validity of our numerical scheme, we have performed two numerical tests. Before presenting them, let us first discuss the choice of the numerical parameters $R_\mathrm{max}$ and $\sigma$. A finite choice of $R_\mathrm{max}$ brings a correction of the order $a/R_\mathrm{max}$ to the field (where we recall that $a$ is the separation between the two bodies), and so we choose $R_\mathrm{max}=50a$  in our simulation.

Concerning the effect of $\sigma$, let us consider a single point-particle of mass $m_2$ in the $r \ll r_*$ case with boundary condition $\varphi(r,z) = 3(m_2 R)^{1/3}$ for $R=R_\mathrm{max}$. In this case, if the particle was really pointlike, the field would vanish at the origin. This is no longer the case if the particle has a finite extent $\sigma$. Rather, the central value of the field is
\begin{equation} \label{eq:field_sigma}
\varphi(0,0) \sim (m_2 \sigma)^{1/3} \; ,
\end{equation}
simply by dimensional analysis. But this could potentially be problematic in the calculation of the two-body energy which requires the evaluation of the field at the particle location, see eq. \eqref{eq:numerical_energy}. The contribution of this term in the total two-body energy is
\begin{align}
\begin{split}\label{eq:field_center}
m_2 \varphi(\mathbf{x}_2) &\sim E_\mathrm{tm} \\ 
&\sim \frac{m_1m_2}{m_1+m_2} ((m_1+m_2)a)^{1/3} \; ,
\end{split}
\end{align}
this equality being true up to a numerical factor.

In order to avoid an unphysical dependance on $\sigma$ in the energy, we have to tune $\sigma$ in order that the term in eq. \eqref{eq:field_sigma} is much smaller than the term in eq. \eqref{eq:field_center}. This gives the following bound:
\begin{equation} \label{eq:condition_sigma}
\sigma \ll \frac{x^3}{1-x} a \; ,
\end{equation}
where we recall that $x = m_1/(m_1+m_2)$ is the mass ratio. We choose $\sigma = 10^{-6} a$, which is consistent with the minimal value of $x$, $x_\mathrm{min}=0.03$ for $a=1$, that we will use.

\begin{figure}
\centering
\includegraphics[width=0.7\columnwidth]{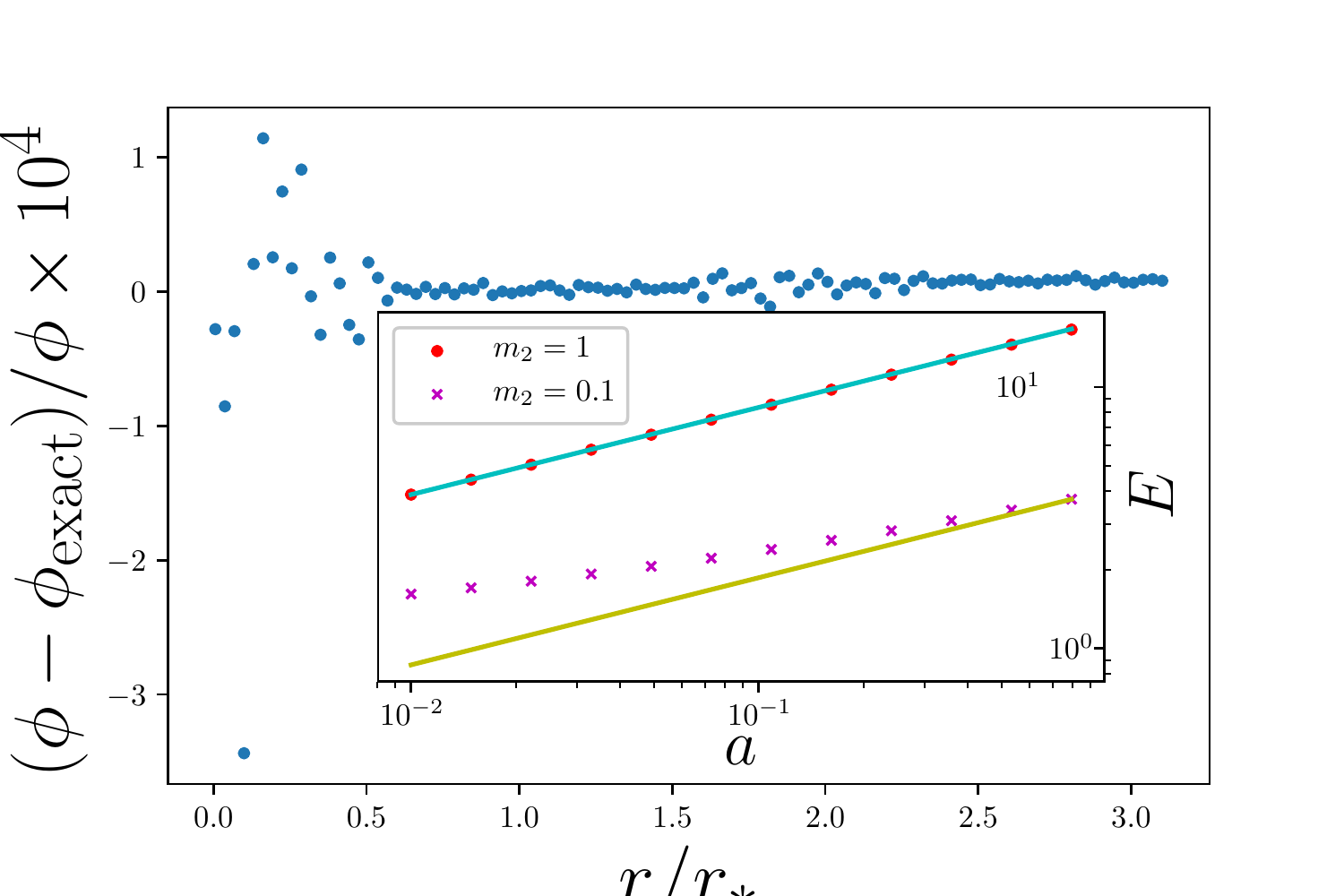}
\caption[Checks of the numerical code.]{Checks of the numerical code. The main plot is the fractional difference of the numerical field solution in the spherically symmetric case of a single particle of mass $M = 1$, compared to the exact solution of eq. \eqref{eq:hypergeom}.
The subplot is the two-body energy defined by eq. \eqref{eq:numerical_energy} in the $r \ll r_*$ case, keeping $m_1=1$ and for two values of $m_2$. The parameter $\sigma$ is taken to be $\sigma = 10^{-4}$. The upper points, with parameter $m_2 = 1$, show good agremment between the expected power-law behavior $E \propto a^{1/3}$ (continuous cyan curve) and the numerical result. The agreement between the two is a good check of the validity of our code. The bottom points, with $m_2 = 0.1$, show a deviation from the simple power-law behavior (continuous yellow curve) due to the fact that condition \eqref{eq:condition_sigma} is not satisfied any more.}
\label{fig:spher_sym}
\end{figure}

Having chosen a value for the numerical parameters, we have first checked that, for a single particle at the origin of the coordinates, the field solution corresponds to the exact solution given in eq. \eqref{eq:hypergeom}. The result is plotted in Figure~\ref{fig:spher_sym}, where we see that there is perfect agreement between the theoretical and numerical values.

The second nontrivial check that we have performed is to verify that the two-body energy in the fully screened regime (i.e, neglecting the quadratic operator as explained in Sec.~\ref{sec:setup_chap7}) indeed varies as $E \propto a^{1/3}$. In Figure~\ref{fig:spher_sym} we have plotted the numerical value of the two-body energy in the $r \ll r_*$ case in logarithmic plot, from which we immediately confirm the expected behavior.


\subsection{Results} \label{sec:results}

\begin{figure}
\centering
\includegraphics[width=0.7\columnwidth]{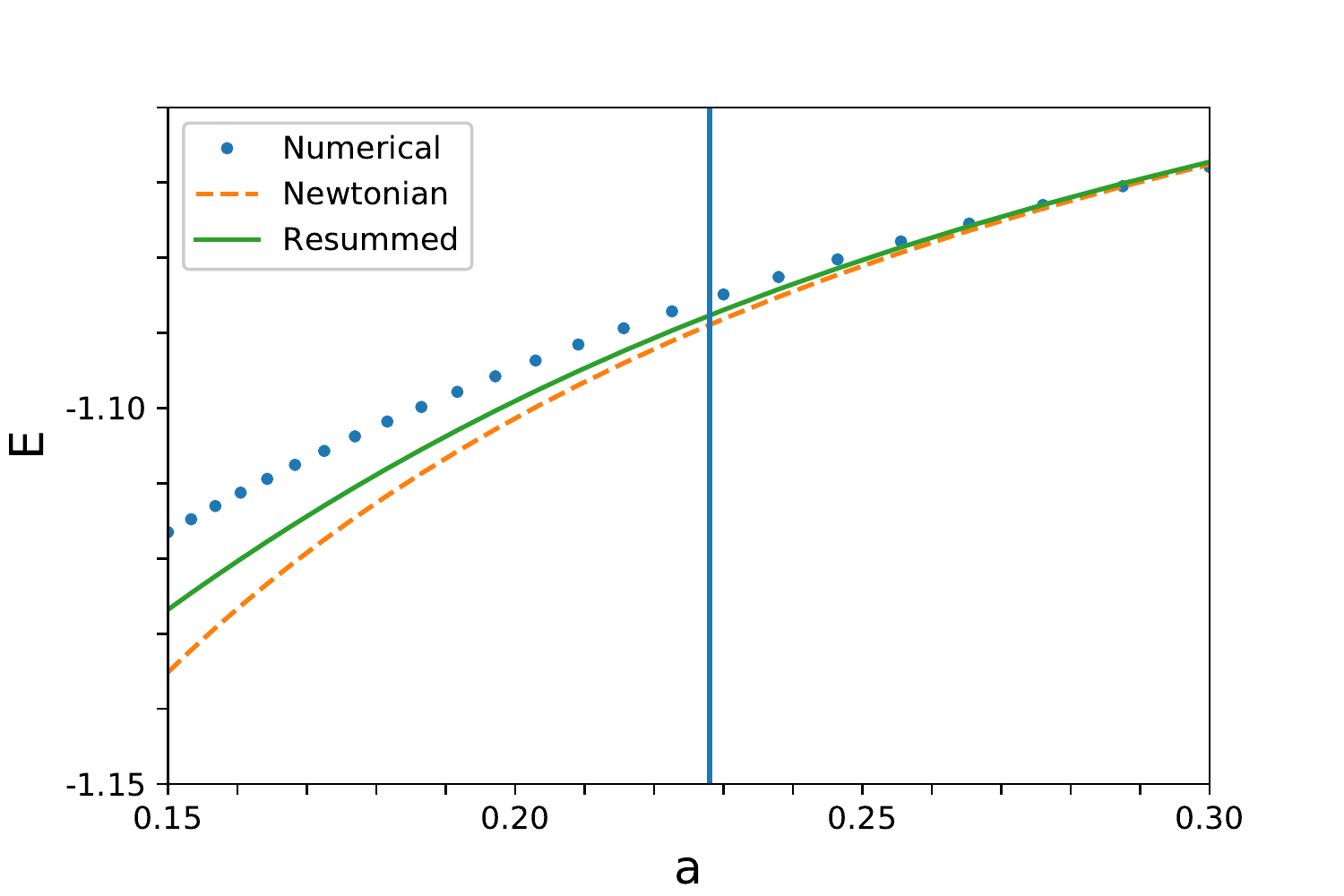}
\caption[Plot of the numerical solution to the energy]{Plot of the numerical solution to the energy close to the Vainshtein radius as a function of the point-masses spacing $a$, for parameters $m_1=m_2=10^{-2}$. The energy is normalized to its absolute value at infinity \eqref{eq:E_infinity}. The numerical solution corresponds to the blue filled circles, the Newtonian potential $-m_1m_2/r+E_\infty$ to the dashed line and the resummed solution \eqref{eq:energy_outside_exact} (shifted  with respect to $E_\infty$) to the solid curve. The vertical bar is the location of the Vainshtein radius associated to the total mass $M=m_1+m_2$.}
\label{fig:energy_outside}
\end{figure}

Figure~\ref{fig:energy_outside} presents the numerical two-body in the $r \gtrsim r_*$ regime against different theoretical predictions for equal masses. The value of the energy when the charges are taken infinitely far apart is not zero: it can be easily calculated as two times the value of the energy when we plug the exact one-body field solution \eqref{eq:hypergeom} into the action. This yields
\begin{equation}\label{eq:E_infinity}
E(\infty) \simeq -31 (m_1^{3/2}+m_2^{3/2}) \; .
\end{equation}

We can see that the numerical value of the energy for large separation matches very well this result, thus providing again a strong check of the validity of our code. The resummed energy provides an improvement over the simple Newtonian potential, although the gain is $\sim 50$ \% at the nonlinear radius.

\begin{figure}
\centering
\includegraphics[width=0.7\columnwidth]{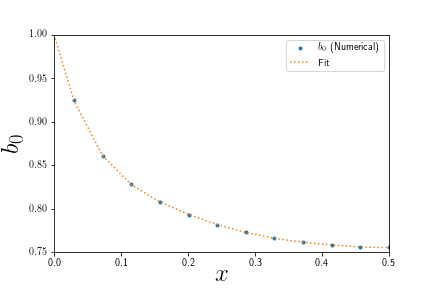}
\caption[Plot of the numerical solution for the coefficient $b_0$]{Plot of the numerical solution for the coefficient $b_0$ defined in eq. \eqref{eq:b0_num}, for a spacing between point-particles of $a=1$.}
\label{fig:b0_num}
\end{figure}

Concerning the fully screened situation $r \ll r_*$, Figure~\ref{fig:b0_num} presents the effective parameter $b_0$ as a function of the mass ratio $x$, where $b_0$ is defined as
\begin{align} \label{eq:b0_num}
\begin{split}
b_0 &= \frac{E}{E_\mathrm{tm}} \; , \\
\frac{E_\mathrm{tm}}{4\pi} &= 3 \mu (Ma)^{1/3} \; ,
\end{split}
\end{align}
where we chose $a=1$ to obtain our results ($b_0$ does not depend on $a$, as emphasised in Sec.~\ref{sec:num_tests}) and the numerical energy $E$ is computed by keeping only the nonlinear term in the action, as explained in Sec.~\ref{sec:setup_chap7}. This is the most important result of this work, since it presents the two-body energy in astrophysically relevant situations and for arbitrary mass ratios. We can see that $b_0 \simeq 0.75$ in the equal-mass case, which means that screening is a bit more efficient than when there is a large mass hierarchy (for which $b_0 = 1$). For convenience, a fit to $b_0$ with a sixth-order polynomial gives

\begin{align}
\begin{split}
b_0(x) &= 1 -3.17 x + 23.7 x^2 - 105.89 x^3 \\
 &+ 266.48 x^4 - 347.46 x^5 + 182.64 x^6
\end{split}
\end{align}

The dependance of the two-body energy on the mass ratio implies a direct violation of the Weak Equivalence Principle (WEP). In the next Section we will explore the consequences of this result on the orbit of the Moon, which is measured with a great accuracy by Lunar Laser Ranging.

\section{Violation of the Weak Equivalence Principle} \label{sec:WEP}

In this Section we will show how our results imply an Equivalence Principle violation that would be visible on the Moon orbit. We will make heavy use of the following mass ratios
\begin{align}
\begin{split}
x_\mathrm{SE} = \frac{m_\oplus}{m_\odot + m_\oplus} &\simeq 3 \times 10^{-6} \\
x_\mathrm{SM} = \frac{m_M}{m_\odot + m_M} &\simeq 3 \times 10^{-8} \\
x_\mathrm{EM} = \frac{m_M}{m_\oplus + m_M} &\simeq 10^{-2} \; ,
\end{split}
\end{align}
where $m_\odot$ is the Sun mass, $m_\oplus$ is the Earth mass and $m_M$ is the Moon mass. 

\subsection{Three-body system, and finite size corrections} \label{sec:three-body}

In this part we will examine the applicability of our results to a three body system like the one formed by the Sun, the Earth and the Moon (herafter, SEM system). As explained in the introduction of this Chapter, fifth forces generated by the scalar interaction are expected to induce a supplementary perihelion precession that would be visible on planetary orbits~\cite{PhysRevLett.61.1159}. But the lunar perihelion precession (when considering its motion around the Earth) cannot be computed by ignoring the scalar field generated by the Sun, and using a perturbative treatment Ref.~\cite{andrews_galileon_2013} reached the conclusion that the perturbations blows up once distances hierarchies are taken into account. We will recover their result using a different approach. See also~\cite{hiramatsu_equivalence_2013} for an interesting numerical work on the violation of the equivalence principle in the DGP model.


To start with, let us reformulate the action of a two-body system in a different way. Labelling the two objects as $1$ and $2$, we split the scalar field according to
\begin{equation}
\varphi = \varphi_2(r_2) + \psi \; ,
\end{equation}
where $r_2 = \vert \mathbf{x} - \mathbf{x}_2 \vert$ is the distance to the body $2$, and $\varphi_2$ is the spherically symmetric field~\eqref{eq:spherical_symmetry} generated by the same body. In the fully screened situation that is of interest to us, this fields writes as 
\begin{equation}
\varphi_2(r_2) = 3 (m_2 r_2)^{1/3} \; .
\end{equation}

If $m_2$ is much greater than the other mass, we can think of $\psi$ being a fluctuation on top of the dominant field generated by the mass $m_2$ as we did in the introduction of this Chapter, but for now let's keep the discussion general and not assume any mass hierarchy (we still assume $m_1 \leq m_2$ by symmetry). Then by inserting this decomposition in the action~\eqref{eq:k_mouflage_rescaled} (where we ignored the quadratic term that is subdominant on small scales), we get the following action
\begin{align}
\begin{split} \label{eq:expanded_action}
S &= S[\varphi_2] + \Lambda^4 \int dt d^3x \left[ -\frac{1}{2} (\nabla \varphi_2)^2 (\nabla \psi)^2 - (\nabla \varphi_2 \cdot \nabla \psi)^2  \right. \\
&- \left. (\nabla \psi)^2 \nabla \varphi_2 \cdot \nabla \psi - \frac{1}{4} (\nabla \psi)^4 + \psi T_1 \right] \; ,
\end{split}
\end{align}
where $S[\varphi_2]$ is the original action~\eqref{eq:k_mouflage_rescaled} applied to $\varphi_2$, and $T_1 = -4\pi m_1 \delta^3(\mathbf{x} - \mathbf{x}_1)$ is the source term corresponding to the object $1$. The term linear in $\psi$ vanishes because of the equations of motion for $\varphi_2$.

We now ask the question: close to the object $1$, what is the behavior of the fluctuation $\psi$ ? In other words, is there an operator that dominates the action for $\psi$ in~\eqref{eq:expanded_action} ? It seems natural to assume that close enough to the source $1$, we recover the behavior $\psi \sim (m_1 r_1)^{1/3}$, which means that the last nonlinear operator in eq.~\eqref{eq:expanded_action} dominates. Let us assume this is the case and derive the condition on $r_1$ for this to be true.

If the term $(\nabla \psi)^4$ dominates in the action, then we recover the same spherically symmetric action as for the second object, and we consequently obtain
\begin{equation}
\psi \simeq 3 (m_1 r_1)^{1/3} \; .
\end{equation}

Let us now make the ratio between the operator $(\nabla \psi)^4$ and another one in the action, say the term cubic in $\psi$. Using Cauchy-Schwarz inequality, we get
\begin{equation} \label{eq:PX_condition_background}
\frac{(\nabla \psi)^2 \nabla \varphi_2 \cdot \nabla \psi}{(\nabla \psi)^4} \leq \frac{\vert \nabla \varphi_2 \vert}{\vert \nabla \psi \vert} \simeq \left( \frac{m_2}{m_1} \left(\frac{r_1}{r_2} \right)^2 \right)^{1/3} \; .
\end{equation}

Using the same scaling, one can show that the term quadratic in $\psi$ is similarly suppressed with respect to the term cubic in $\psi$. This result is quite important. It means that around the first body, the field can be well approximated simply by taking the linear superposition $\varphi_2 + \psi$ of two spherically symmetric solutions. This is true up to the maximal distance to the first body
\begin{equation}
r_1^\mathrm{max} = r_2 \sqrt{\frac{m_1}{m_2}} \;,
\end{equation}
which depends on the mass ratio. Since the Vainshtein radius of a massive body is $r_{*, \alpha} = \sqrt{m_\alpha}$, this equation can also be interpreted as
\begin{equation}
\frac{r_1^\mathrm{max}}{r_{*, 1}} = \frac{r_2}{r_{*,2}}
\end{equation}
which simply means that the distance to each body is measured in units of its own Vainshtein radius.

This has several consequences. First, it means that we can ignore the finite-size of the bodies and treat them as point-particles as long as their radius is less than $r_1^\mathrm{max}$. In this particular theory, this is true both for the Sun-Earth system and for the Earth-Moon system if we take the numerical values of their respective radius and masses.

Second, it means that we can simply add the fifth forces felt by a satellite that orbits sufficiently close to a planet itslef orbiting around its star. Unfortunately, for the SEM system and the particular screening theory considered here this is not true, as
\begin{equation}
\left( \frac{m_\odot}{m_\oplus} \left(\frac{r_1}{r_2} \right)^2 \right)^{1/3} \simeq 1 \; ,
\end{equation}
where we have taken $r_1$ to be the Earth-Moon distance and $r_2$ to be the Sun-Earth distance. The end result is that we expect the Lunar perihelion precession rate to be corrected by an amount depending on the masses and distances hierarchies.

For another type of Vainshtein screening such as the Galileon-3~\cite{nicolis_galileon_2009}, the same reasoning shows that we can superpose the nonlinear solutions provided
\begin{equation} \label{eq:galileon_condition_background}
\left( \frac{m_2}{m_1} \left(\frac{r_1}{r_2} \right)^3 \right)^{1/2} \ll 1 \; ,
\end{equation}
which is true at the 10\% level for the SEM system. The same conclusion was also reached in~\cite{Nicolis:2004qq} with similar arguments. This shows that in the case of a Galileon-3, the Lunar perihelion precession can be calculated by simply ignoring the Sun \footnote{Note that we can also apply this reasoning to the Sun embedded inside our own galaxy. In this case, $m_2/m_1 = m_\mathrm{Gal}/m_\odot \simeq 10^{12}$, $r_1/r_2 = r_{\mathrm{SE}}/r_\mathrm{Gal} \simeq 6 \times 10^{-10}$ where $m_\mathrm{Gal}$ is the galactic mass, $r_\mathrm{Gal}$ is the distance to the Galactic centre, and we have made the simplifying assumption that the mass of the Galaxy is concentrated at its centre. Eqs~\ref{eq:PX_condition_background} and~\ref{eq:galileon_condition_background} then show that, for both Galileons and $P(X)$, we can indeed neglect the background field generated by the galaxy for the inner planets of the Solar System. I am grateful to Philippe Brax for pointing this out.} Since the Earth-Moon mass ratio is $x \simeq 10^{-2}$, the calculation can be carried out in the test-mass approximation and yields an interesting constraint on the size of a Galileon-3 operator~\cite{dvali_accelerated_2003}.


\subsection{Weak Equivalence Principle violation}

The fact that the two-body energy of two massive particles is not the one of a reduced mass $\mu$ in an external field implies a violation of the Weak Equivalence Principle, as we will now show. In this Section we will assume a general Vainshtein screening mechanism that gives rise to a fifth force (not necessarily the specific $P(X)$ theory that we studied in the rest of the Chapter). If we first neglect the moon, the total (gravitational and scalar) interacting Lagrangian of the Earth and the Sun writes as
\begin{equation} \label{eq:total_2_body_lagrangian}
L_{SE} = \frac{G m_\odot m_\oplus}{r_{SE}} \left(1 + \alpha(x_{SE}) \left(\frac{r_{SE}}{r_*} \right)^n \right) \; .
\end{equation}
Here $r_{SE} = \vert \mathbf{y}_S - \mathbf{y}_E \vert$ is the Earth-Sun distance, $x_{SE} = m_\oplus/(m_\odot + m_\oplus)$ is the Earth-Sun mass ratio, $r_*$ is the Vainshtein radius of the Sun, $n$ is an exponent that depends on the type of screening considered, and $\alpha$ is an unknown function of the coefficient $x_{SE}$ that can be found numerically as in Sec.~\ref{sec:numerical}. 
 In order to avoid confusion between positions and mass ratios, we denote in this Section the positions by the letter $\mathbf{y}$. This type of Lagrangian is common to all theories endorsed with Vainshtein screening, with the expressions of $n$, $r_*$ and $\alpha$ different among theories. For example, $n = 3/2$ for a Galileon-3, and for the $P(X)$ theory considered above in this Chapter one has $n = 4/3$ with a function $\alpha$ given by
\begin{equation}
\alpha(x) = \beta^2 b_0(x)\; ,
\end{equation}
where we have taken into account the scalar charge $\beta$ present in the ratio of fifth forces in Eq.~\eqref{eq:ratio_fifth_forces}.

We will make the supplementary assumption that we can get the total force felt by the Moon by simply adding the fifth forces of the Earth and the Sun. As we showed in Sec.~\ref{sec:three-body}, this is the case for a Galileon-3 but not for the particular $P(X)$ example examined in the rest of the Chapter. A complete treatment of this case would necessitate further work. With this assumption we can write the total (non-relativistic) interaction Lagrangian of this three-body system (to get the total Lagrangian from this, one should also add the kinetic energies from the three bodies)
\begin{align} \label{eq:total_3_body_lagrangian}
\begin{split}
L_\mathrm{int} &= \frac{G m_\odot m_\oplus}{r_{SE}} \left(1 + \alpha(x_{SE}) \left(\frac{r_{SE}}{r_*} \right)^n \right) \\
&+ \frac{G m_\odot m_M}{r_{SM}} \left(1 + \alpha(x_{SM}) \left(\frac{r_{SM}}{r_*} \right)^n \right) \\
&+ \frac{G m_\oplus m_M}{r_{EM}} \left(1 + \alpha(x_{EM}) \left(\frac{r_{EM}}{\tilde r_*} \right)^n \right) \; ,
\end{split}
\end{align}
where $S$ designates the Sun, $E$ the Earth, $M$ the Moon, and each line of this equation is the two-body Lagrangian of eq.~\eqref{eq:total_2_body_lagrangian} adapted to each pair of bodies. The last line of this equation contains the Earth Vainshtein radius $\tilde r_*$. We will now derive the fifth force incidence on the Lunar motion along the lines of Ref.~\cite{Damour:1995gi}.

The first line of eq.~\eqref{eq:total_3_body_lagrangian} gives rise to the Earth anomalous perihelion precession, and the third line to the Lunar anomalous perihelion precession. These effects are already discussed in other References~\cite{dvali_accelerated_2003, Iorio:2012pv, andrews_galileon_2013} and we will not comment on them. From now on, we will ignore the last line of eq.~\eqref{eq:total_3_body_lagrangian} which does not give rise to the leading order equivalence principle violation that we are going to derive. Let us expand the distances to the Sun around the Earth-Moon center-of-mass, which is defined with the usual expression
\begin{equation}
(m_\oplus + m_M) \mathbf{Y} = m_\oplus \mathbf{y}_E + m_M \mathbf{y}_M \; ,
\end{equation}
where $\mathbf{y}_E$ and $\mathbf{y}_M$ are the positions of the Earth and the Moon respectively. Then the distances to the Sun can be expressed by
\begin{align}
\begin{split}
r_{SE} &= \vert \mathbf{y}_S - \mathbf{Y} - x_{SM} \mathbf{y}_{EM}  \vert \\
r_{SM} &= \vert \mathbf{y}_S - \mathbf{Y} + x_{SE} \mathbf{y}_{EM}  \vert \; ,
\end{split}
\end{align}
where $\mathbf{y}_S$ is the Sun position. Expanding the total Lagrangian to first order in $r_{EM}$, one finds
\begin{align}
\begin{split}
L_\mathrm{int} &= \frac{G m_\odot (m_\oplus + m_M)}{r} \left(1 + \left[ (1-x_{EM}) \alpha(x_{SE}) + x_{EM} \alpha(x_{SM}) \right] \left(\frac{r}{r_*} \right)^n \right) \\
&- G m_\odot m_\oplus x_{EM} r_{EM}^i \frac{\partial}{\partial r^i} \left[ \frac{1}{r} \left(1+\alpha(x_{SE}) \left(\frac{r}{r_*} \right)^n  \right) \right] \\
&+ G m_\odot m_M (1-x_{EM}) r_{EM}^i \frac{\partial}{\partial r^i} \left[ \frac{1}{r} \left(1+\alpha(x_{SM}) \left(\frac{r}{r_*} \right)^n  \right) \right] \;,
\end{split}
\end{align}
where $r = \vert \mathbf{y}_S - \mathbf{Y} \vert$ is the center-of-mass distance to the Sun, and as discussed above we have dropped the third line of eq.~\eqref{eq:total_3_body_lagrangian}. Upon introducing the reduced mass $\mu_{EM} = m_\oplus m_M / (m_\oplus + m_M)$, one finds the following expression
\begin{align} \label{eq:reduced_3_body_lagrangian}
\begin{split}
L_\mathrm{int} &= \frac{G m_\odot (m_\oplus + m_M)}{r} \left(1 + \left[ (1-x_{EM}) \alpha(x_{SE}) + x_{EM} \alpha(x_{SM}) \right] \left(\frac{r}{r_*} \right)^n \right) \\
&+ G \mu_{EM} m_\odot \frac{r_{EM}^i r_i}{r^3} \left( (1-n) (\alpha(x_{SE})-\alpha(x_{SM})) \left(\frac{r}{r_*} \right)^n \right) \; .
\end{split}
\end{align}

As discussed in Ref.~\cite{Damour:1995gi}, there are two physical effects stemming from this Lagrangian \footnote{In fact Ref.~\cite{Damour:1995gi} discusses three physical consequences, but the third one is not apparent at the expansion order we work with, and is expected not to be measureable.}. The first line of eq.~\eqref{eq:reduced_3_body_lagrangian} implies that the gravitational constant involved in the motion of the Earth-Moon system around the Sun is not the same than the one involved in the motion of this system around its barycenter (third line of eq.~\eqref{eq:total_3_body_lagrangian}). However, this effect has practically no observable consequences.

The second physical effect, on which we will concentrate, is the perturbation on the lunar orbit implied by the second line of eq.~\eqref{eq:reduced_3_body_lagrangian}. Nordtvedt~\cite{PhysRev.170.1186} showed in 1968, in the context of BDT theories where the strong equivalence principle is violated (see Section~\ref{sec2.2}), that this term implies a modulation of the Lunar orbit with amplitude
\begin{equation}
\delta r_{EM} \simeq 3 \times 10^{12} \vert \delta_\oplus - \delta_M \vert \; \mathrm{cm} \; .
\end{equation}
In BDT theories, $ \vert \delta_\oplus - \delta_M \vert$ is the fractional variation of Newton's constant due to the gravitational self-energy of each body; said equivalently, Newton's constant appearing in the gravitational attraction between two bodies $A$ and $B$ is
\begin{equation}
G_{AB} = G(1+\delta_A + \delta_B) \; ,
\end{equation}
this expression being derived in Chapter~\ref{Chapter4}.

Although the physical origin is quite different, the equivalence principle violation considered here gives rise to the same term in the second line of the Lagrangian~\eqref{eq:reduced_3_body_lagrangian} than the strong equivalence principle violation of BDT theories. The parameter $\delta_\oplus - \delta_M$ is replaced by the following quantity
\begin{align}
\begin{split}
\delta_\oplus - \delta_M &\rightarrow (1-n) (\alpha(x_{SE})-\alpha(x_{SM})) \left(\frac{r}{r_*} \right)^n \\
& \simeq (1-n) \alpha_1 x_{SE} \left(\frac{r}{r_*} \right)^n \; .
\end{split}
\end{align}
In the second line we have expanded $\alpha$ to first order in the mass ratios, $\alpha(x) \simeq \alpha_0 + \alpha_1 x$ and neglected the Moon mass ratio compared to the Earth mass ratio, $x_{SM} \ll x_{SE}$.

Current LLR data give the constraint $\vert \delta_\oplus - \delta_M \vert \lesssim 10^{-13}$~\cite{Williams:2012nc}, which by ignoring the $\mathcal{O}(1)$ factor of $(1-n)$ yield the constraint
\begin{equation}
\alpha_1 x_{SE} \left(\frac{r}{r_*} \right)^n \lesssim 10^{-13} \; .
\end{equation}

Let us contrast this with the anomalous perihelion constraint of the Earth, which from the introduction of this Chapter constrains the fifth force exerted in the Earth by the Sun to be smaller than
\begin{equation}
\alpha_0 \left(\frac{r}{r_*} \right)^n \lesssim 10^{-11} \; .
\end{equation}

On the one hand, we gain two orders of magnitude by using Lunar Laser Ranging data, but on the other hand the equivalence principle violation is suppressed by the mass ratio $x_{SE} \simeq 10^{-6}$ with respect to the perihelion bound, resulting in a looser constraint if we assume that $\alpha_0 \sim \alpha_1$.

\section{Conclusions} \label{sec:conclusion}

Let us recap here the main results of this Chapter, in which we have 
 analyzed for the first time the two-body potential energy of pointlike objects in Vainshtein screened theories for arbitrary mass ratios. One the one hand, from outside the nonlinear radius, the problem is amenable to a perturbative treatment which we use to resum a class of Feynman graphs. We derive an Effective One-Body energy map which relates the two-body energy to the one of a test particle in an external field. On the other hand, we conjecture the existence of such an energy map inside the nonlinear radius where the nonlinear screening term dominates the action. We have tried to get the analytical behavior of this expansion by a matching procedure at the nonlinear scale. Improving the accuracy of this matching procedure would necessitate to know the exterior potential energy with a higher accuracy. This could be done by calculating Feynman diagrams with \textit{two} insertions of the nonlinear operator, instead of only one as we did in Sec.~\ref{sec:two_body}. This would necessitate more involved computations that we leave for future work.

At the same time, we have performed a numerical simulation in order to get the most relevant effective coefficient $b_0$ corresponding to the ratio between the real two-body energy and the energy of a test-mass in an external field. In the test-mass limit, $b_0 \simeq 1$ while in the equal-mass limit, we find $b_0 \simeq 0.75$ in the particular $P(X)$ theory that we considered. This means that Vainshtein screening is active even in the fully nonlinear situation where we do not assume one mass to be smaller than the other, with departure from simple order-of-magnitude estimates encoded into a simple coefficient $b_0$ which can be found with a numerical simulation. While we have focused on a particular model exhibiting nonlinearities, we expect such a feature to be valid in any model endorsed with Vainshtein screening such as Galileons.

As we showed in Sec.~\ref{sec:WEP}, the fact that the two-body energy differs from the test-mass one implies a violation of the weak equivalence principle, as the Earth and the Moon would not fall the same way towards the Sun. We used this fact to bound the size of the coupling parameters of any theory relying on Vainshtein screening to hide the effects of a fifth force in the Solar system. Although the final constraint is looser than the one obtained from perihelion precession for a Galileon-3, the methodology employed is quite general and we will use it in the next Chapter
 to investigate on the case of two neutron stars or black holes in their inspiral phase.

\chapter{Extreme mass ratio inspiral in the cubic Galileon} \label{Chapter8}

In the last Chapter, we have focused on the static two-body energy in a theory with Vainshtein screening. Although the problem is strongly coupled so that we cannot solve it with perturbation theory, we have obtained a numerical estimate of the two-body energy. We now want to consider the case of two inspiralling objects, say neutron stars or black holes. In this case the numerical algorithm would necessitate to be more elaborated: see e.g \cite{dar_scalar_2018} for a numerical implementation of a fully dynamical setting in the cubic Galileon. However, there is a regime in which analytical computations are reliable: this is the case of an Extreme Mass Ratio Inspiral (EMRI) of a solar-size BH onto a supermassive one. Supermassive BHs typically lie at the center of galaxies; the inspiral of solar-size BH around a supermassive one is expected to produce GW detectable by the future space-borne interferometer LISA (see Section~\ref{subsec:current_planned_inter}), and can lead to an interesting phenomenology such as Kozai-Lidov oscillations~\cite{Randall:2019sab,Randall:2017jop,Randall:2018nud}. Such a system could thus allow for a perturbative treatment (the small BH move in the background field generated by the supermassive one) and is observationally relevant, which makes it very interesting as a probe of the Vainshtein mechanism.

On the conservative side, the problem is trivially solved since one falls back on the energy of a point-particle in a background Vainshtein field (see e.g the introduction of Chapter \ref{Chapter7} where we have explained how this supplementary fifth force can be constrained with perihelion data). On the other hand, the dissipative dynamics of Galileons in the extreme mass ratio case can be found in Refs. \cite{deRham:2012fg,deRham:2012fw}. 

In this Chapter based on the JCAP paper~\citeK{Brax:2020ujo}, we will consider a cubic Galileon theory and follow a route similar to these References, with two notable exceptions. First, while Refs. \cite{deRham:2012fg,deRham:2012fw} have focused on binary pulsars data, we will be more interested in the phase of GW emitted from a highly asymmetric binary system of BHs. The second difference concerns the coupling of the scalar to the objects which we model as point-particles. Indeed, pulsars are rapidly rotating neutron stars which we expect to be coupled to the scalar field \textit{via} the usual matter coupling. On the other hand, it is well-known that scalar fields are often trivial around BHs, i.e the scalar charge of a BH is zero in many different theories. This is the content of the no-hair theorem, to which we will dedicate Chapter \ref{Chapter9}. While referring the reader to this Chapter for more details, let us simply state here that
 there are many ways out of this restriction. Indeed the  time-dependence of a scalar field at spatial infinity induces scalar hair around BHs~\cite{Jacobson_1999}, which can seem to be kind of obvious since the field \textit{cannot} be nontrivial. This time-dependence could be due to cosmological boundary conditions or to the environment in which the BHs are located~\cite{Horbatsch_2012}. Concerning the Galileon, we will review in the next Chapter, Section~\ref{sec:BH_hair_timedep}, how hair can be  induced by an asymptotic timelike gradient of the scalar, showing that it can even lead to the generation of large scalar couplings. For the moment though, all what we need to know is that BH couple to the scalar with an \textit{effective} coupling $\beta_\mathrm{eff}$ which is a free parameter of our problem. In the cubic Galileon case, this effective coupling could even be quite large if one considers the timelike gradient of the field to be sourced by some dynamical process in the local environment of the system.
 
Even if one circumvents the no-hair theorem, the fact that the Vainshtein suppression is so huge (the fifth force of the cubic Galileon was suppressed by 16 orders of magnitude in the Solar system !) could cast doubts on the detectability of such a signal. However, we have seen that scalar-tensor theories generically predict dipole radiation which is of greater mangitude than the GR quadrupole; furthermore, GW detectors such as LISA can monitor the inspiral of compact objects during a large number of GW cycles, thereby greatly enhancing the potentiality of detecting any deviation from GR. It is thus worth investigating  the detectability of a Galileon field by LISA.

\section{Perturbative scheme} \label{sec:setup_chap8}

Our starting point will be the action of a cubic Galileon which is known from Section~\ref{subsec:Galileons} to exhibit the Vainshtein screening mechanism:
\begin{equation} \label{eq:base_action}
S = \frac{1}{2} \int \mathrm{d}^4 x \sqrt{-g} \left[ \mpl^2 R - (\partial \varphi)^2 - \frac{1}{\Lambda^3} (\partial \varphi)^2 \square \varphi \right] + S_m[\tilde g_{\mu \nu}, \psi_i] \; ,
\end{equation}
where $\Lambda$ is the energy scale of the cubic Galileon interaction, and we denote the scalar by $\varphi$ to avoid confusion with the angular variable $\phi$. In order to give a rough estimate for $\Lambda$, if the Galileon is supposed to contribute to the accelerated expansion of our universe one should impose $\Lambda^3 \sim H^2 \mpl$. Nevertheless, we will keep this scale arbitrary and assume that a cosmological constant lies behind the acceleration. The metric $\tilde g_{\mu \nu}$ is as usual the Jordan frame metric, which couples to matter. We take it to be simply conformally related to the Einstein frame metric, $\tilde g_{\mu \nu} = A^2(\varphi) g_{\mu \nu}$ where $A(\varphi) = e^{\beta \varphi / \mpl}$ is the (universal) coupling factor. 

We have seen in Section~\ref{sec:tests_GW} that the cubic Galileon does not change the speed of gravity and so it is not constrained by the GW speed bound (this bound essentially removes any quartic or quintic Galileon of any cosmological model). However, the instability bound mentioned in the same Section~\ref{sec:tests_GW} indeed applies to the cubic Galileon, and this severely limits its cosmological significance (this is confirmed by other cosmological constraints~\cite{Bellini_2013}); furthermore,
 as we have seen in the introduction of the last Chapter, the perihelion constraint provides a relevant bound on the size of a cubic Galileon operator. In this Chapter, we will analyze independently which kind of constraints we can obtain on the cubic Galileon from GW emitted by the system which we consider.
 
This system is composed of a supermassive BH of mass $m_0 \sim 10^6 M_\odot$ orbited by a 'small' BH of mass $m_1 \sim 50 M_\odot$, which as discussed in Section~\ref{subsec:current_planned_inter} can be observed through its GW emission in the LISA window. For simplicity, we will assume the orbit of the small object to be circular of radius $r$, although the eccentricity of the orbits could grow to significant values in this kind of configurations~\cite{Randall:2019sab,Randall:2017jop,Randall:2018nud}. Similarly, we consider a \sch BH for simplicity but an interesting generalization of this work could consist in including the spin of the BHs.

Let us consider the central supermassive BH first. In the cubic Galileon, the no-hair theorem discussed in Chapter \ref{Chapter9} prevents the formation of any scalar field around a static, spherically symmetric BH. However, this holds true only for a static configuration. Let us assume that the scalar has some temporal gradient at spatial infinity, i.e
 \begin{equation}
 \varphi = q t + \varphi_0(r)
 \end{equation}
This time-dependence can arise if one considers dynamical effects in the vicinity of the system. In fact, this is a quite natural assumption to make since supermassive BH are situated at the center of galaxies where several dynamical processes can happen. Then, following the work of Ref. \cite{Babichev_2013} which we will recall in Chapter~\ref{Chapter9}, Section~\ref{sec:BH_hair_timedep}, it can be shown that $\varphi_0(r)$ couples to the BH with an effective scalar charge. More precisely, 
\begin{equation}
\label{eq:phi_SS}
\varphi_0' =  \left( \frac{\beta_\mathrm{eff} m_0 \Lambda^3 }{8 \pi \mpl r} \right)^{1/2}
 =  \frac{\beta_\mathrm{eff} m_0}{8 \pi \mpl r_*^2} \left( \frac{r_*}{r} \right)^{1/2} \;,  \quad r_*^3 = \frac{\beta_\mathrm{eff} m_0}{8 \pi \mpl \Lambda^3} \; , \quad  \beta_\mathrm{eff} = \frac{q^2}{2 \Lambda^3 \mpl} \; .
\end{equation}
In fact, we already gave this expression in Section~\ref{subsec:Galileons} when explaining the basics of the Vainshtein mechanism: this is the field of a cubic Galileon coupling to the matter \textit{via} $\beta_\mathrm{eff}$.  The effective scalar charge will be a free parameter of our problem: if no deviation from GR is detected in the signal, one can see the final result that we will obtain as constraining \textit{any} temporal gradient of a cubic Galileon field.


Let us now enter into the details of the perturbative scheme that we will follow. The hierarchy $m_1 \ll m_0$ suggests the field decomposition $\varphi = q t + \varphi_0 + \psi$ where $\varphi_0$ is given in Eq.~\eqref{eq:phi_SS} (and by $r$ we mean the distance to the center-of-mass of the system), and $\psi$ is generated by the small perturbation that $m_1$ creates. We similarly split $g_{\mu \nu} = \eta_{\mu \nu} + h_{\mu \nu} / \mpl$\footnote{A more sensible way to do would be to split $g_{\mu \nu} = g^{(0)}_{\mu \nu} + h_{\mu \nu} / \mpl$ where $g^{(0)}_{\mu \nu}$ is the \sch metric. However, as we will argue below this would correct our result only at a higher post-Newtonian order. }. Expanding to quadratic order in the fields and integrating by parts the terms with higher derivatives, we get
\begin{align}
\begin{split}\label{eq:expansion_action}
S^\mathrm{quad} &= \int \mathrm{d}^4 x \; \frac{1}{4} \left[ - \partial_\mu h_{\alpha \beta} P^{\alpha \beta \gamma \delta} \partial^\mu h_{\gamma \delta} + \left(\partial_\alpha h^\alpha_\nu - \frac{1}{2} \partial_\nu h \right)^2 \right] - \frac{1}{2} (\partial \psi)^2 - \frac{1}{2 \Lambda^3} \bigg[ 2 \square \varphi_0 (\partial \psi)^2  \\
& - 2 \partial_\mu \partial_\nu \varphi_0 \partial^\mu \psi \partial^\nu \psi + \frac{1}{\mpl} \partial_\mu \varphi_0 \partial_\nu \varphi_0 \partial_\alpha \psi \partial^\alpha h^{\mu \nu}
- \frac{1}{\mpl} 2 \partial^\mu \varphi_0 \partial^\nu \varphi_0 \partial_\mu \psi \left(\partial_\alpha h^\alpha_\nu - \frac{1}{2} \partial_\nu h \right) \bigg] + S_m  \; ,
\end{split}
\end{align}
where $P^{\alpha \beta \gamma \delta} = \eta^{\alpha \gamma} \eta^{\beta \delta}/2 - \eta^{\alpha \beta} \eta^{\gamma \delta}/4$, $h$ is the trace of $h_{\mu \nu}$, and we got rid of the part of the action linear in $\psi$ with the equations of motion. This equation present a mixing terms $h \psi$ between the graviton and the scalar; following~\cite{Babichev_2013}, one can disentangle the two degrees of freedom with the change of variable
\begin{equation} \label{eq:change_var}
\bar h_{\mu \nu} =  h_{\mu \nu} + \frac{2}{\mpl \Lambda^3} \left[ \partial_\mu \varphi_0 \partial_\nu \varphi_0 - \frac{1}{2} \eta_{\mu \nu} (\partial \varphi_0)^2 \right] \psi \; .
\end{equation}
Now Eq.~\eqref{eq:expansion_action} takes the following form
\begin{equation}\label{eq:quadratic_action}
S^\mathrm{quad} =  \int \mathrm{d}^4 x \; \frac{1}{4} \left[ - \partial_\mu \bar h_{\alpha \beta} P^{\alpha \beta \gamma \delta} \partial^\mu \bar h_{\gamma \delta} + \left(\partial_\alpha \bar h^\alpha_\nu - \frac{1}{2} \partial_\nu \bar h \right)^2 \right] - \frac{1}{2} \mathcal{G}^{\mu \nu} \partial_\mu \psi \partial_\nu \psi + S_m \; ,
\end{equation}
where the effective metric $\mathcal{G}^{\mu \nu}$ in which $\psi$ propagates is
\begin{equation} \label{eq:gmunu}
\mathcal{G}^{\mu \nu} = \eta^{\mu \nu} \left[ 1 + \frac{2}{\Lambda^3} \square \varphi_0 - \frac{1}{2 \mpl^2 \Lambda^6} (\partial \varphi_0)^4 \right] - \frac{2}{\Lambda^3} \partial^\mu \partial^\nu \varphi_0 + \frac{2}{\mpl^2 \Lambda^6} (\partial \varphi_0)^2 \partial^\mu \varphi_0 \partial^\nu \varphi_0 \; ,
\end{equation}
Let us simplify this expression. By definition, the Vainshtein regime corresponds to $\square \varphi_0 / \Lambda^3 \gg 1$. On the other hand, for a field given in Eq.~\eqref{eq:phi_SS} one can write
\begin{equation}
\frac{(\partial \varphi_0)^4 / (\mpl^2 \Lambda^6)}{\square \varphi_0 / \Lambda^3} \sim \left( \frac{\beta_\mathrm{eff} G m_0}{r} \right)^2 \left( \frac{r}{r_*} \right)^3 \ll 1 \; .
\end{equation}
Thus, the terms quartic in $\varphi_0$ in Eq.~\eqref{eq:gmunu} are negligible in the Vainshtein regime. The effective metric then simplifies to
\begin{equation}
\mathcal{G}^{\mu \nu} = \frac{2}{\Lambda^3} \square \varphi_0 \eta^{\mu \nu} - \frac{2}{\Lambda^3} \partial^\mu \partial^\nu \varphi_0 \; .
\end{equation}
Before moving on to the calculation of the Green's function (i.e, the inverse operator of $\mathcal{G}^{\mu \nu}$), let us comment on the matter action. We model the two BHs as point-particles,
\begin{equation}
S_m = - m_0 \int \mathrm{d}t \;  \sqrt{-g_{\mu \nu} v_0^\mu v_0^\nu} - m_1 \int \mathrm{d}t \; \sqrt{-g_{\mu \nu} v_1^\mu v_1^\nu} \; ,
\end{equation}
where $v_0^\mu = \mathrm{d}x_0^\mu / \mathrm{d}t$ is the velocity of the massive BH and similarly $v_1^\mu$ is the velocity of the small BH. In this equation, the scalar is not coupled to the BHs as one would expect from the no-hair theorem. However, the change of variable in Eq.~\eqref{eq:change_var} induces an \textit{effective} coupling of the scalar to the BHs. To lowest post-Newtonian order, one can write
\begin{equation}
S_m = - m_0 \int \mathrm{d}t \;  \sqrt{-\bar g_{\mu \nu} v_0^\mu v_0^\nu} \left( 1 +  \beta_\mathrm{eff} \frac{\psi}{\mpl} \right) - m_1 \int \mathrm{d}t \; \sqrt{-\bar g_{\mu \nu} v_1^\mu v_1^\nu} \left( 1 +  \beta_\mathrm{eff} \frac{\psi}{\mpl} \right) \; ,
\end{equation}
where $\bar g_{\mu \nu} = \eta_{\mu \nu} + \bar h_{\mu \nu} / \mpl$.

\section{Galileon propagation: the Green's function} \label{sec:Green}
In this section we exactly compute the propagation of the Galileon in the geometrical  background configuration, following the computations of the previous subsection. For this purpose we essentially need to compute the Green's function of the second order action in perturbations. This will enable us to compute the power dissipated in scalar radiation.

As the gravitational and scalar fluctuations are now decoupled at lowest post-Newtonian order,
we concentrate on the scalar part of the action.
Rewriting the quadratic action~\eqref{eq:quadratic_action} for $\psi$ in spherical coordinates, we find
\begin{equation}
S =  \int d^4x \frac{1}{2} \left[ K_{t} (\partial_t \psi)^2 - K_{r} (\partial_r \psi)^2 - K_{\Omega} (\partial_\Omega \psi)^2 \right] +  \frac{\beta_\mathrm{eff}}{\mpl} \psi T \; ,
\end{equation}
where the kinetic, angular and radial factors read
\begin{align}
K_t = 3 \left( \frac{r_*}{r} \right)^{3/2}, \qquad
K_r = 4 \left( \frac{r_*}{r} \right)^{3/2} \qquad \text{and} \qquad
K_\Omega = \left( \frac{r_*}{r} \right)^{3/2} \; ,
\end{align}
and $T$ is (to lowest PN order)
\begin{equation}
T = - m_0 \delta^{(3)} (\mathbf{x} - \mathbf{x}_0) - m_1 \delta^{(3)} (\mathbf{x} - \mathbf{x}_1) \; .
\end{equation}

 The Green's function associated to this quadratic operator is defined by
\begin{equation}
\left[- K_t(r) \partial_t^2 + \frac{1}{r^2} \partial_r \left(r^2 K_r \partial_r \right) + \frac{K_\Omega}{r^2} \nabla_\Omega^2 \right] G(x, x') = \delta^4(x - x')
\end{equation}
where $\nabla_\Omega^2 = \partial_\theta^2 + 1/\sin^2 \theta \partial_\phi^2$. It is worth mentioning  the following special boundary condition. We will take the field to vanish at the origin of coordinates where the massive object lies. Since the field goes as $r^{1/2}$ at the origin, this is consistent, contrary to the Newtonian problem where the field goes as $1/r$ but vanishes at infinity. We can obtain the field as
\begin{equation} \label{eq:field_green_fct}
\psi(x) = \int d^4x' G(x, x') \left( - \frac{\beta_\mathrm{eff} T(x')}{\mpl} \right)
\end{equation}

We will follow  Refs.~\cite{andrews_galileon_2013,deRham:2012fg,chu_retarded_2013} in order to calculate the Green's function.
%
 By introducing the rescaled variable $\vec u = \frac{\vec r}{r_*}$, we rewrite the previous equation as
\begin{equation}
\left( -3 r_*^2 \partial_t^2 + 4 \partial_u^2 + \frac{2}{u} \partial_u + \frac{\nabla_\Omega^2}{u^2} \right) G(\vec u, t ; \vec u', t') = \frac{\delta^3(\vec u - \vec u') \delta(t-t')}{r_*}\;.
\end{equation}
We further decompose the Green function in a Fourier and spherical harmonics basis as
\begin{equation} \label{eq:green_fourier}
G(\vec u, t ; \vec u', t') = \int \frac{d\omega}{2\pi} e^{-i \omega (t-t')} \sum_{l=0}^{\infty} R_{lm \omega}(u, u') \sum_{m=-l}^l Y_l^m[\theta, \phi] \bar Y_l^m[\theta', \phi']\;.
\end{equation}
Using the resolution of the identity
\begin{equation}
\sum_{l,m} Y_l^m[\theta, \phi] \bar Y_l^m[\theta', \phi'] = \delta( \cos(\theta) - \cos(\theta') ) \delta(\phi - \phi')
\end{equation}
one easily finds the equation for the mode function $R_{lm}$
\begin{equation} \label{eq:diffeq_R}
\left( \partial_u^2 + \frac{1}{2u} \partial_u + \frac{3}{4} \tilde \omega^2 - \frac{l(l+1)}{4 u^2} \right) R_{lm \omega}(u,u') = \frac{\delta(u - u')}{4 r_* \sqrt{u}}
\end{equation}
where $\tilde \omega = r_* \omega$.

The general continuous solution of eq. \eqref{eq:diffeq_R} is
\begin{equation}
R(u,u') = A R^1(u) R^1(u') + B R^2(u) R^2(u') + C R^1(u_<) R^2(u_>) + D R^1(u_>) R^2(u_<)
\end{equation}
where we have omitted the index $l,m, \omega$ for clarity and the constants $A,B,C,D$ are to be fixed by the normalization of the mode functions and the boundary conditions. Here we have introduced the notation $u_> = \mathrm{max}(u, u')$ and $u_< = \mathrm{min}(u, u')$, and the homogeneous solutions are given by the Bessel functions
\begin{align}
\begin{split}
R^1(u) &= \mathcal{N} u^{1/4} J_\nu \left(\sqrt{\frac{3}{4}} u \tilde \omega \right) \\
R^2(u) &= \mathcal{N} u^{1/4} J_{-\nu}\left(\sqrt{\frac{3}{4}} u \tilde \omega \right)\;,
\end{split}
\end{align}
where $\mathcal{N}$ is a normalization constant and $\nu = (2l+1)/4$. Integrating eq. \eqref{eq:diffeq_R} for $u$ close to $u'$, we get
\begin{equation}
(D-C)W = \frac{1}{4r_* \sqrt{u}}
\end{equation}
where
\begin{equation}
W = R^{1'} R^2 - R^1 R^{2'} = \frac{2 \mathcal{N}^2 \sin(\nu \pi)}{\pi \sqrt{u}}
\end{equation}
is the Wronskian of the two homogeneous solutions. We choose
\begin{equation}
\mathcal{N} = \sqrt{\frac{\pi}{8r_* \sin(\nu \pi)}}
\end{equation}
such that $D-C = 1$.

We next determine the constants from the boundary conditions. We require the flux to be purely outgoing at infinity, which corresponds to taking the retarded Green's function. On the other hand, the boundary condition at the origin can be fixed by the following observation. Consider the field produced by a variation $m_0 \rightarrow m_0 + \delta m_0$ of the central mass. From eq. \eqref{eq:phi_SS} (with a field vanishing at the origin), it is
\begin{equation}
\delta \pi_0(r) = \frac{\beta_\mathrm{eff} \delta m_0}{8\pi \mpl r_*} \left( \frac{r}{r_*} \right)^{1/2}
\end{equation}
On the other hand, from eq. \eqref{eq:field_green_fct}, we have
\begin{align}
\begin{split}
\delta \pi_0(t, \vec x) &= - \frac{\beta_\mathrm{eff}}{\mpl} \int d^4x' G( x, x') T( x') \\
&= \frac{\beta_\mathrm{eff}  \delta m_0}{\mpl} \int dt' G(\vec r, t ; \vec 0, t')\;.
\end{split}
\end{align}
By equating these two equations, we find the boundary condition at the origin
\begin{equation}\label{eq:BC_origin}
\mathrm{lim}_{\omega \rightarrow 0} \sum_l \frac{2l+1}{4\pi} P_l(\cos(\theta)) R_{lm\omega}(u,0) = \frac{\sqrt{u}}{8\pi r_*}\;,
\end{equation}
where $P_l$ represent the Legendre polynomials, and we have used the following identities
\begin{equation}
Y_l^m(0,\phi) = \sqrt{\frac{2l+1}{4\pi}} \delta_{m0} \; , \quad Y_l^0(\theta, \phi) = \sqrt{\frac{2l+1}{4\pi}} P_l(\cos(\theta))\;.
\end{equation}

Let us examine the asymptotic behaviour for $l>0$ first. From the behaviour of the Bessel's functions at the origin
\begin{equation}
J_\nu(z) \sim \frac{1}{\Gamma(\nu+1)} \left( \frac{z}{2} \right)^\nu
\end{equation}
we immediately deduce that $B=D=0$ for $l \geq 1$ in order for the Green's function to be continuous at the origin. The solution is now
\begin{equation}
R(u,u') = A R^1(u) R^1(u') - R^1(u_<) R^2(u_>)\;.
\end{equation}
The constant $A$ is fixed by requiring the flux to be outgoing at infinity. Indeed, by rewriting the Bessel functions in terms of the two Hankel functions
\begin{align}
\begin{split}
J_\nu &= \frac{H_\nu^{(1)} +H_\nu^{(2)} }{2} \\
J_{-\nu} &= \frac{1}{2} \left( H_\nu^{(1)}(1-i \tan(\nu \pi)) - H_\nu^{(2)}(1+i \tan(\nu \pi)) \right)
\end{split}
\end{align}
and using the asymptotic behaviour at infinity
\begin{align} \label{eq:asymptotic_bessel}
\begin{split}
H_\nu^{(1)}(z) &\sim \sqrt{\frac{2}{\pi z}} e^{i\left(z-\nu \frac{\pi}{2} - \frac{\pi}{4} \right)} \\
H_\nu^{(2)}(z) &\sim \sqrt{\frac{2}{\pi z}} e^{-i\left(z-\nu \frac{\pi}{2} - \frac{\pi}{4} \right)}
\end{split}
\end{align}
the condition that the flux is purely outgoing imposes
\begin{equation}
A = - (1+i \tan(\nu \pi))
\end{equation}

Let us now examine the $l=0$ case. In this case, $R^2$ is not divergent any more at the origin but takes a finite value,
\begin{equation}
R^2(0) = \frac{1}{\Gamma(3/4)} \left( \frac{1}{8 \sqrt{3} \tilde \omega} \right)^{1/4} \sqrt{\frac{\pi}{r_*}}\;.
\end{equation}
The solution when one of the points is taken to be the origin is
\begin{equation}
R(u,0) = R^2(0) \left( B R^2(u) + D R^1(u) \right)\;.
\end{equation}
We can now use eq. \eqref{eq:BC_origin} by noticing that $R_{lm \omega}(u,0) = 0$ for $l > 0$, which gives
\begin{equation}
\mathrm{lim}_{\omega \rightarrow 0} R(u,0) = \frac{\sqrt{u}}{2 r_*}\;.
\end{equation}
This imposes $D = 1$ (so $C=0$). The $B$ coefficient multiplies a power-law divergent term which we simply subtract as it does not depend on the variables $u, u'$.  $B$ is  left undetermined here.
To find $A$ and $B$, let us rewrite the solution when one of the endpoints is taken to infinity, say $u'$,
\begin{equation}
R(u,u') = A R^1(u) R^1(u') + R^2(u)(B R^2(u') + R^1(u')) \; ,
\end{equation}
so that in order to have a purely outgoing flux, one should impose
\begin{align}
\begin{split}
A &= 0 \\
B &= \frac{1}{1+i}\;.
\end{split}
\end{align}
In conclusion, the mode functions are
\begin{equation} \label{eq:mode_functions}
\begin{array}{rclc}
R(u,u') &=& R^1(u_>) R^2(u_<) + \frac{1}{1+i} R^2(u) R^2(u') & \text{for }l=0 \\
& & & \\
R(u,u') &=& - R^1(u_<) R^2(u_>) - (1+i \tan(\nu \pi)) R^1(u) R^1(u') & \text{for }l>0  \;.
\end{array}
\end{equation}
In this way the Green's function \eqref{eq:green_fourier} is completely characterised.

\section{Dissipative dynamics} \label{sec:diss_power}

In this section we  compute the dissipated power in scalar radiation due to the presence of the cubic Galileon. The power emitted in the tensor sector will follow the usual quadrupole formula and hence we will focus on the scalar sector.

\subsection{Energy-momentum tensor}
The total energy-momentum tensor splits into gravitational and scalar contributions,
\begin{equation}
T_{\mu \nu} = T^\varphi_{\mu \nu} + T^g_{\mu \nu} \; ,
\end{equation}
where $T^g_{\mu \nu}$ is the usual Landau-Lifschitz pseudo-tensor, and $T^\varphi_{\mu \nu} = - 2 / \sqrt{-g} \delta S_\varphi/\delta g^{\mu \nu}$, where $S_\varphi$ is the scalar part of the action. Far from matter sources the total energy-momentum tensor is conserved, which allows to find the power lost into radiation by integrating it over a distant sphere of radius $\mathcal{R}$ centered on the system,
\begin{equation}
P_\varphi = \int d^2 S T_{0i}n^i = \mathcal{R}^2 \int d^2 \Omega T_{0r} \; ,
\end{equation}
where $n^i$ is the outward pointing vector of the sphere. The Landau-Lifschitz pseudo-tensor will give rise to the usual quadrupole formula at lowest order in the post-Newtonian expansion, so there remains only to find the scalar dissipated power. The scalar energy-momentum tensor calculated from the action \eqref{eq:base_action} reads
\begin{align}
\begin{split} \label{eq:EMT}
T^\varphi_{\mu \nu} &= \partial_\mu \varphi \partial_\nu \varphi - \frac{1}{2} g_{\mu \nu} (\partial \varphi)^2 + \frac{1}{\Lambda^3} \left( \partial_\mu \varphi \partial_\nu \varphi \square \varphi + \frac{1}{2} g_{\mu \nu} \partial_\alpha \big( (\partial \varphi)^2 \big ) \partial^\alpha \varphi \right. \\
&- \left. \frac{1}{2} \big [ \partial_\mu \big( (\partial \varphi)^2 \big) \partial_\nu \varphi + \mathrm{sym} \big ]  \right) \; .
\end{split}
\end{align}
Splitting the field $\varphi = \varphi_0 + \psi$ as in Eq. \eqref{eq:expansion_action}, one can collect the terms quadratic in $\psi$ in the energy-momentum tensor. The linear terms average to zero in time in the dissipated power. As emphasised in Section~\ref{sec:setup_chap8}, the dominant terms will be the quadratic ones coming from the Galileon term. By neglecting angular and time total derivatives, which, once again, will average to zero in the dissipated power, one finds the $0r$ part of the scalar energy-momentum tensor
\begin{equation}
T^\varphi_{0r} = \frac{4  \varphi_0'}{\Lambda^3 \mathcal{R}} \partial_t \psi \partial_r \psi.
\end{equation}
For a wave travelling far from the massive objects, one has $\partial_r \psi = - \partial_t \psi / c_r$ where $c_r = \sqrt{3}/2$ is the radial propagation speed. This gives
\begin{equation} \label{eq:power_chap8}
P_\varphi = \frac{8}{\sqrt{3}} \mathcal{R}^{1/2} r_*^{3/2} \int d^2 \Omega (\partial_t \psi)^2\;.
\end{equation}

\subsection{Dissipated power}

Let us now find the power radiated at infinity, which reduces to finding $\psi(\vec x,t)$ at large distance from the source. Let us recall that the origin of our coordinates is the center-of-mass of the binary system so that at lowest order in the PN expansion
\begin{equation} \label{eq:CM_relations}
\mathbf{x}_1 = \frac{m_0}{m_1 + m_0} \mathbf{r} \; , \quad \mathbf{x}_0 = - \frac{m_1}{m_1 + m_0} \mathbf{r}
\end{equation}
and $\mathbf{r} = \mathbf{x}_1 - \mathbf{x}_0$.
 By using eqs. \eqref{eq:field_green_fct} and \eqref{eq:green_fourier}, one finds
\begin{align}
\begin{split} \label{eq:field}
\psi(\vec x, t) = \frac{\beta_\mathrm{eff}}{\mpl} \int dt' \frac{d\omega}{2\pi} e^{-i \omega(t-t')} \sum_{l,m} Y_l^m(\theta, \phi)
 \left[ m_0 R_{l\omega}(\mathcal{R}, \vert \mathbf{x}_0 \vert) \bar Y_l^m (\pi/2, \Omega t') + (0 \leftrightarrow 1) \right] \; ,
\end{split}
\end{align}
where $\mathcal{R}, \theta, \phi$ are the coordinates of the distant sphere of integration, and the trajectory lies in the $\pi/2$ plane. We have denoted by $\Omega t$ the angle of the central BH in its trajectory, the angle of $m_1$ being found by adding a phase of $\pi$ (the two BHs are on opposite sides of the circle).
Simplifying the time and frequency integrals, using  using eq. \eqref{eq:power_chap8} and integrating over a sphere, we find the dissipated power,
\begin{equation}
P_\varphi = \frac{8}{\sqrt{3}} \mathcal{R}^{1/2} r_*^{3/2} \frac{\beta_\mathrm{eff}^2}{\mpl^2} \sum_{l,m} s_{lm}^2 m^2 \Omega^2  \big[ m_0  R_{l\omega}(\mathcal{R}, x_0) - m_1  R_{l\omega}(\mathcal{R}, x_1) \big]^2 \; ,
\end{equation}
where $\omega = m \Omega$ and $s_{lm} = Y_l^m(\pi/2, 0)$. 
We immediately see that, as expected for a circular trajectory, there is no contribution of the monopole $l=m=0$. 
Tidying up a bit the expression for the mode functions $R_{l\omega}$, we find
\begin{equation} \label{eq:power_beautiful}
P_\varphi = \frac{\pi \beta_\mathrm{eff}^2}{3 r_*^{3/2} \mpl^2} \sum_{l,m} \frac{s_{lm}^2}{\Gamma((2l+5)/4)^2} (m \Omega)^{l+3/2} \left( \frac{\sqrt{3}}{4} \right)^{l+1/2} \big[ m_0 x_0^{(l+1)/2}  - m_1 x_1^{(l+1)/2} \big]^2 \; ,
\end{equation}
where we  have expanded the Bessel function for small arguments,
\begin{equation}
J_\nu(z) \sim \frac{1}{\Gamma(\nu+1)} \left( \frac{z}{2} \right)^\nu \; .
\end{equation}

From Eq.~\eqref{eq:power_beautiful} we can deduce that each multipole is suppressed by a factor $r \Omega \sim v$ with respect to the preceding one in the PN expansion. Furthermore, using the center-of-mass relations~\eqref{eq:CM_relations}  we can deduce that the contribution of the dipole $l=1$ is zero, as expected since we consider a universal scalar charge. In order to be consistent one should also compute the relativistic correction to the dipole (coming from e.g relativistic corrections to the trace of the energy-momentum tensor $T$) which should be of the same order than the quadrupole in the PN expansion. However, we can use the results of Ref. \cite{deRham:2012fw} who showed that the dipole receives a further suppression in the cubic Galileon, so that the lowest-order scalar radiation is given by the quadrupole term only, i.e
\begin{equation} \label{eq:power_quadrupole}
P_\varphi = \mathcal{A} \beta_\mathrm{eff}^2 \frac{m_1^2}{\mpl^2} \frac{(\Omega r)^3}{(\Omega r_*)^{3/2}} \Omega^2 \; ,
\end{equation}
where we have expanded for $m_1 \ll m_0$, we have defined $r = \vert \mathbf{x}_1 - \mathbf{x}_0 \vert$ and the numerical factor is $\mathcal{A} = 5\cdot 3^{5/4} \cdot \sqrt{2} / (64 \Gamma(9/4)^2) \simeq 0.34$. Comparing the order-of-magnitude with the GR quadrupole, $P_\mathrm{GR} \sim (m_1/m_0)^2 v^{10}/G$ where $v$ is the velocity of the small BH, we find
\begin{equation}
\frac{P_\varphi}{P_\mathrm{GR}} \sim \beta_\mathrm{eff}^2 \frac{1}{v (\Omega r_*)^{3/2}} \sim \beta_\mathrm{eff}^2 v^{-5/2} \left( \frac{r}{r_*} \right)^{3/2} \; .
\end{equation}
Thus, compared to the static Vainshtein suppression, there is an enhancement of the signal by a factor $v^{-5/2}$.

\section{Inspirals and scalar correction to the phase} \label{sec:inspiral}

Here we compute the Galileon scalar correction to the GW phase recorded in a detector such as LISA.  This will allow us to derive a lower bound on the effective coupling $\beta_\mathrm{eff}$ in order to be able to detect a Galileon correction to the GW phase. We still assume that the GR quadrupole dominates the power loss.

Using Kepler's third law, the Newtonian energy of the system is
\begin{equation}
E = - \frac{m_1 v^2}{2} \; ,
\end{equation}
where $v = (Gm_0 \Omega_0)^{1/3}$ is the velocity of the small BH. The time evolution of the binary system is given by the balance equation
\begin{equation} \label{eq:balance}
\frac{dE}{dt}= -P_{GR}- P_\varphi \; ,
\end{equation}
where $P_\varphi$ is the quadrupolar scalar power loss in Eq. \eqref{eq:power_quadrupole} and $P_\mathrm{GR}$ is the GR quadrupolar power loss given by
\begin{equation}
P_\mathrm{GR} = \frac{32}{5G} \left( \frac{m_1}{m_0} \right)^2 v^{10}
\end{equation}
Let us introduce the dimensionless constant $C$ by writing $P_\varphi = C P_\mathrm{GR} v^{-11/2}$, i.e.
\begin{equation}
C = \frac{5 \pi \mathcal{A} \beta_\mathrm{eff}^2}{4} \left( \frac{Gm_0}{r_*} \right)^{3/2} \; .
\end{equation}
The balance equation \eqref{eq:balance} then takes the form
\begin{equation}
\frac{dv}{dt} = \frac{32 m_1}{5 G m_0^2} v^9 \left( 1 + C v^{-11/2} \right) \; .
\end{equation}

The number of observable GW cycles of the binary system in the detector is $2 \Phi(t)$ where
\begin{equation}
\Phi(t)= \int_{t_\mathrm{in}}^t \Omega dt= \frac{1}{Gm_0} \int_{v_\mathrm{in}}^v dv v^3 \left(\frac{dt}{dv}\right) \; ,
\end{equation}
where $t_\mathrm{in}$ is the initial time at which the signal enters in the GW detector, and $v_\mathrm{in}=v(t_\mathrm{in})$.

Assuming a final velocity $v_\mathrm{out} \sim 1 \gg v_\mathrm{in}$ when the small BH plunges into the central BH, the total accumulated phase during the inspiral is $\Phi = \Phi_\mathrm{GR} + \Delta \Phi$ where $\Phi_\mathrm{GR}$ is the usual GR phase,
\begin{equation}
\Phi_\mathrm{GR} = \frac{1}{32} \frac{m_0}{m_1} v_\mathrm{in}^{-5} \; ,
\end{equation}
and $\Delta \Phi$ is the correction due to the scalar field,
\begin{equation}
\Delta \Phi = - \frac{10 C}{672}  \frac{m_0}{m_1} v_\mathrm{in}^{-21/2} \; .
\end{equation}
Let us display a more 'user-friendly' version of this equation. The physical parameters of our system are $\Lambda$, $m_0 \sim 10^6 M_\odot$, $m_1 \sim 50 M_\odot$ and the lowest frequency accessible in LISA (which will determine the beginning of the observation of the signal) is of order $\Omega_\mathrm{in} \sim 10^{-3}$ Hz, so that
\begin{equation} \label{eq:delta_phi_EMRI}
\Delta \Phi \simeq 3.5 \times 10^{-7} \beta_\mathrm{eff}^{3/2} \left( \frac{\Lambda}{10^{-12} \mathrm{eV}} \right)^{3/2} \; \left( \frac{m_1}{50 M_\odot} \right)^{-1} \; \left( \frac{m_0}{10^6 M_\odot} \right)^{-3/2} \; \left( \frac{\Omega_\mathrm{in}}{10^{-3} \mathrm{Hz}} \right)^{-21/6} \; ,
\end{equation}

The precision achieved on the phase of GW observatories is at the level of $\Delta \Phi \sim 0.1$~\cite{Babak_2017}, so that an effective scalar charge $\beta_\mathrm{eff} \gtrsim 10^4$ would induce observable modifications to the GW phase. Note that $\Delta \Phi$ is quite sensitive to the minimal frequency $\Omega_\mathrm{in}$ at which the signal is detected in the interferometer, so that a lower $\Omega_\mathrm{in}$ would greatly increase the detectability of such an event. Likewise, a value of $\Lambda \sim 10^{-8}$ eV would allow for a detection with $\beta_\mathrm{eff} \sim 1$. This equation concludes this Chapter.

\part{Testing the no-hair theorem with gravitational waves} \label{part4}

\chapter{Black holes and scalar hair} \label{Chapter9}

The final part of this thesis is dedicated to no-hair theorems and their possible tests with GW. The very essence of no-hair theorems is that BHs in equilibrium are quite simple objects. Indeed, in GR they can be characterized by three numbers: their mass $M$, angular momentum $J$, and electric charge $Q$. Contrast this with stars, which could be very different even if they possess the same mass, angular momentum and charge (for example, two stars do not necessarily share the same internal composition). At the root of the no-hair theorem are the uniqueness theorems~\cite{Chru_ciel_2012} stating that all higher multipole moments are determined by only $M$, $J$ and $Q$. This led to the conjecture that the outcome of gravitational collapse of any kind of matter is a Kerr-Newman BH: no other physical parameter can describe a BH. This is the no-hair \textit{conjecture}; trying to prove it by assuming different kinds of matter leads to the no-hair \textit{theorems}.



Scalar fields are one of the simplest types of 'matter' (in the Einstein frame, the contribution of the scalar to the equations of motion can always be recast on the form of an energy-momentum tensor) and, as we have argued in Chapter~\ref{Chapter2}, they can be used in many models of dark matter or dark energy. The question which immediately arises is: what is the fundamental difference between scalar fields and electromagnetism, such that the former vanish around BHs while the latter can endow the BHs with a charge? In the first Section of this Chapter, we will recall this fundamental statement which is at the heart of the no-hair theorems for scalars. Then, we will show how the argument can be adapted to the Horndeski class of Lagrangians. The proofs are always quite simple and elegant, which suggests that the no-hair theorem is indeed a very restrictive statement.

In spite of this, as usual in physics a theorem is as good as its assumptions, which, even if quite general, can fail to be true under some circumstances. We will dedicate the next Section to an exploration of all the possible ways to evade the no-hair theorems; it will appear that in several physically significant cases scalar hair should be expected to be present around BH.
This Chapter relies heavily on the excellent review~\cite{herdeiro2015asymptotically}, to which we refer the reader for more details. It will
 serve as an introduction to the next Chapter, where we will apply EFT ideas to BHs endowed with scalar hair. There, we will explore the possibility of detecting the presence of scalar hair with GW data: we will be able to provide waveform templates for \textit{any} theory predicting the existence of scalar hair. This opens up interesting avenues for testing gravity with GW.

\section{The no-hair theorems} \label{sec:no_hair}

\subsection{Electro-vacuum vs scalar vacuum}

In this subsection, we will highlight the main difference between vectors and scalar fields around \sch BHs. Let us consider two types of action, the first one being the action of a BDT theory in the Einstein frame, 
\begin{equation} \label{eq:scalar_action}
S_{\varphi} = \:  \int \mathrm{d}^4x \sqrt{-g} \left(\frac{\mpl^2}{2} R - \frac{1}{2}  \partial^\mu \varphi \partial_\mu \varphi \right)\, ,
\end{equation}
and the second one being an electromagnetic field minimally coupled to gravity,
\begin{equation} \label{eq:firstaction_chap9}
S_{\rm em} = \:  \int d^4x \sqrt{-g} \left(\frac{\mpl^2}{2} R - \frac{1}{4}  F_{\mu \nu} F^{\mu \nu} \right)\, ,
\end{equation}
In both case, the \sch metric with mass $M$ is a solution of the theory, with $\partial_\mu \varphi = 0$ and $F_{\mu \nu} = 0$ respectively. Let us now consider respectively a test vector and scalar field on this \sch background. Due to the symmetries of spacetime, the vector field writes as $A_\mu = (A_0(r), \mathbf{0})$, while the scalar is $\varphi(r)$. The equation of motion for the vector imposes
\begin{equation}
\nabla_\mu F^{\mu \nu} = 0 \quad  \Rightarrow \quad A_0 = - \frac{Q_E}{r} \; ,
\end{equation}
where $Q_E$ is the dimensionless electric charge of the BH (we have set $4 \pi \epsilon_0 = 1$). This solution sources an energy-momentum tensor which is everywhere regular,
\begin{equation}
T_{\mu \nu}^\mathrm{E} = F_{\mu \alpha} F_\nu^\alpha - \frac{1}{4} g_{\mu \nu} F_{\alpha \beta}F^{\alpha \beta} \; .
\end{equation}
Moreover, by making the electric field backreact on the metric via the Einstein equations $\mpl^2 G_{\mu \nu} = T_{\mu \nu}^\mathrm{E}$, one obtains the Reissner-Nordström BH solution:
\begin{align} \label{eq:reissner_nordtstrom}
\mathrm{d}s^2 &= - \left( 1 - \frac{2GM}{r} + \frac{GQ_\mathrm{E}^2}{r^2} \right) \mathrm{d}t^2 + \left( 1 - \frac{2GM}{r} + \frac{GQ_\mathrm{E}^2}{r^2} \right)^{-1} \mathrm{d}r^2 + r^2 \mathrm{d} \Omega^2 \; , \\
A_0 &= - \frac{Q_\mathrm{E}}{r}
\end{align}
Remarkably, the solution for the electromagnetic field is the same than in the test-field approximation (i.e, it is linear).

On the other hand, the Klein-Gordon equation for the scalar is much less well-behaved,
\begin{equation} \label{eq:phi_ss_KG}
\square \varphi = 0 \quad \Rightarrow \quad \partial_r \varphi = \frac{Q_\mathrm{S}}{r^2} \left( 1 - \frac{2G M}{r} \right)^{-1} \quad \Rightarrow \quad \varphi(r) = \frac{Q_\mathrm{S}}{2G M} \ln \left( \frac{2G M}{r} - 1 \right) \; ,
\end{equation}
where $Q_\mathrm{S}$ is the dimensionless scalar charge. The scalar field diverges logarithmically at the horizon; the same is true for the scalar energy-momentum tensor:
\begin{equation}
T_{\mu \nu}^\mathrm{S} = \partial_\mu \varphi \partial_\nu \varphi - \frac{1}{2} g_{\mu \nu} \partial_\alpha \varphi \partial^\alpha \varphi \;,
\end{equation}
Indeed, a spacetime scalar such as the trace of the energy-momentum tensor diverges as
\begin{equation}
T^\mathrm{S} = \frac{Q_\mathrm{S}^2}{r^4} \left( \frac{2GM}{r} - 1 \right)^{-1}
\end{equation}
This proves that
no regular (on and outside a horizon), spherically symmetric and static solution of
a BH with scalar hair exists, connecting continuously to the \sch solution.

\subsection{The theorem} \label{subsec:hawking_th}

We may also be interested in the case of a rotating BH, under which the spherically symmetric assumption of the last Subsection breaks down. In 1972, the proof was given by both Hawking~\cite{hawking1972} and Bekenstein~\cite{Bekenstein:1971hc} (following earlier works~\cite{DeLaCruz:1970kk},~\cite{Penney:1968zz}) that, under some assumptions which we will highlight below,  \textit{any} static and axially symmetric, asymptotically flat BH spacetime cannot support scalar hair. We will now give the essence of this proof, in a modern version given by Sotiriou and Faraoni~\cite{Sotiriou_2012}.

Consider the BDT action with an arbitrary potential,
\begin{equation} \label{eq:scalar_action_potential}
S_{\varphi} = \:  \int \mathrm{d}^4x \sqrt{-g} \left(\frac{\mpl^2}{2} R - \frac{1}{2}  \partial^\mu \varphi \partial_\nu \varphi - V(\varphi) \right)\, .
\end{equation}
The scalar obeys the Klein-Gordon equation,
\begin{equation} \label{eq:KGEq}
\square \varphi - V'(\varphi) = 0 \; ,
\end{equation}
where the d'Alembertian is taken in curved space, $\square = \nabla_\mu \nabla^\mu$. The staticity and the axial symmetry implies that the spacetime possesses two killing vectors, $\partial_t$ and $\partial_\phi$. Let us assume that the scalar inherits this spacetime symmetries so that $\partial_t \varphi = \partial_\phi \varphi = 0$. Then, multiplying the Klein-Gordon equation by $V'(\varphi)$ and integrating gives
\begin{equation}
\int \mathrm{d}^4 x \sqrt{-g} \big( V'(\varphi) \square \varphi - V'(\varphi)^2 \big) = 0 \; .
\end{equation}
Now, integrating the first term by parts gives
\begin{equation}
\int \mathrm{d}^4 x \sqrt{-g} \big( - V''(\varphi) \partial_\mu \varphi \partial^\mu \varphi - V'(\varphi)^2 \big) + \int_\mathcal{H} \mathrm{d}^3 \sigma V'(\varphi) n^\mu \partial_\mu \varphi = 0 \; ,
\end{equation}
where the boundary term is computed on the horizon (the boundary term at infinity vanishes provided the scalar vanishes sufficiently fast at infinity). But, since the event horizon of a stationary, asymptotically flat spacetime is a Killing horizon, the normal $n^\mu$ to $\mathcal{H}$ is a linear combination of the Killing vector fields. Thus, $n^\mu \partial_\mu \varphi = 0$ and we conclude that
\begin{equation}
\int \mathrm{d}^4 x \sqrt{-g} \big(  V''(\varphi) \partial_\mu \varphi \partial^\mu \varphi + V'(\varphi)^2 \big) = 0 \; .
\end{equation}
Under the assumption that $V''(\varphi) \geq 0$ (which is required for stability of solutions to the Klein-Gordon equation \eqref{eq:KGEq}), and since $\partial_\mu \varphi$ is spacelike or zero so that $\partial_\mu \varphi \partial^\mu \varphi \geq 0$, one is forced to have $\partial_\mu \varphi = 0$ over all spacetime, i.e the scalar is trivial. Note the remarkable fact that the theorem did not use Einstein's equations.

\subsection{Generalizing to Galileons and Horndeski} \label{sec:galileon_no_hair}

Another no-hair theorem for shift-symmetric Galileon theories (and which also applies to beyond Horndeski) has been proposed by Hui and Nicolis~\cite{Hui_2013}. Let us sketch the (very simple) proof here. Assuming a spherically symmetric and static BH, we choose a gauge where the line element is
\begin{equation}
\mathrm{d}s^2 = - f(r) \mathrm{d}t^2 + \frac{1}{f(r)} \mathrm{d}r^2 + \rho^2(r) \mathrm{d}\Omega^2 \; .
\end{equation}
The proof proceeds in four steps:
\begin{enumerate}
\item \textit{The Galileon/Horndeski EOM is a current conservation equation: } Indeed, shift symmetry $\varphi \rightarrow \varphi + C$ allows to define a Noether current associated with this symmetry,
\begin{equation} \label{eq:def_current}
J^\mu = \frac{1}{\sqrt{-g}} \frac{\delta S[\varphi]}{\delta(\partial_\mu \phi)} \; .
\end{equation}
The equation of motion of the scalar is then equivalent to the current conservation equation $\nabla_\mu J^\mu = 0$, i.e
\begin{equation} \label{eq:current_conservation}
\frac{1}{\sqrt{-g}} \partial_\mu \big( \sqrt{-g} J^\mu \big) = 0 \; .
\end{equation}
Moreover, the symmetries of spacetime are such that the only nontrivial component of $J^\mu$ is $J^r$.

\item \textit{$J^r$ vanishes at the horizon: } Indeed, the norm of the current $J^\mu$ is $J^\mu J_\mu = (J^r)^2/f$ which would be infinite at the horizon $f(r_h)=0$ unless $J^r=0$. Note that this requirement of imposing the finiteness of $J^\mu J_\mu$ is actually not respected by the Gauss-Bonnet hair which we will present in the next Section. In this case, the authors argue that this divergence does not lead to any pathology in the theory.

\item \textit{$J^r$ vanishes everywhere:} The current conservation equation \eqref{eq:current_conservation} can be simplified in this spherically symmetric case to give
\begin{equation}
\rho^2 J^r = \mathrm{Const} \; ,
\end{equation}
and since $\rho^2$ is expected to be finite at the horizon (it measures the area of constant-$r$ hypersurfaces), we reach the conclusion that $J^r = 0$ for all $r$.

\item \textit{$J^r = 0$ implies $\varphi = 0$: } This is the part of the theorem which is the easiest to break, as we will see in Section~\ref{sec:evading_no_hair}. Hui and Nicolis argue that, since in Galileon theories the current takes the form
\begin{equation} \label{eq:jr_galileon}
J^r = f \cdot \varphi' \cdot F( \varphi', g, g', g'') \; ,
\end{equation}
(see Eq. \eqref{eq:current_BH} below), where $F$ is a polynomial of $\varphi'$ whose coefficients depend on the metric and its derivatives. The crucial observation on $F$ is that it asymptotes to a nonzero constant when $\varphi'$ goes to zero at spatial infinity.  The reason is simply that in the weak $\varphi$ limit, the
action is well approximated by its quadratic terms and
the shift-current reduces simply to $J^\mu \simeq \partial^\mu \varphi$ up to an
overall constant which defines $\varphi$’s normalization. But, by continuity, if $\varphi'$ starts to deviate a bit from zero then the current in Eq. \eqref{eq:jr_galileon} would be nonzero which contradicts point $3$. We should then have $\varphi'=0$ over all spacetime, i.e the field is trivial.

\end{enumerate}

\section{Evading the no-hair theorems} \label{sec:evading_no_hair}

In the previous Section we have showed that, under quite general assumptions, no scalar 'hair' can exist around BH. Now we will reconsider independently each of the assumptions of Section~\ref{sec:no_hair} and exhibit situations in which scalar hair can form when they are violated. Generically, scalar hair can be of two types: \textit{primary} hair associated to an independent scalar charge, and \textit{secondary} hair where the scalar charge is not independent of the mass of the BH. The examples we will present in this Section will belong to both categories.

\subsection{Time-dependence at spatial infinity} \label{sec:BH_hair_timedep}

One of the simplest possibility of having scalar hair around a BH is to break asymptotic flatness or stationarity. For example, let us add a time-dependent term to the BDT solution found in \eqref{eq:phi_ss_KG} so that~\cite{Jacobson_1999}
\begin{equation} \label{eq:ST_with_hair}
\varphi(t,r) = \frac{Q_\mathrm{S}}{2GM} \left[ \frac{t}{2GM} + \ln \left( \frac{2GM}{r} - 1 \right) \right] \; .
\end{equation}
Then the scalar still solves $\square \varphi = 0$ on the \sch background. Moreover, this solution is regular on the horizon because the divergence in $t$ precisely cancels the divergence in $r$ when going close to the horizon. This can be seen from the fact that the advanced time coordinate $v = t + \tilde r$ is regular on the horizon, where $\tilde r$ is the tortoise coordinate,
\begin{equation}
\tilde r = r + 2 GM \ln \left( \frac{r}{2GM} - 1 \right) \; .
\end{equation}
Indeed, the scalar can be expressed as
\begin{equation}
\varphi = \frac{Q_\mathrm{S}}{(2GM)^2} \left[ v - r - 2GM \ln \left(\frac{r}{2GM} \right) \right] \; ,
\end{equation}
which is regular on the horizon. Here, $Q_\mathrm{S}/(2GM)^2$ represents the asymptotic value $\dot \varphi_\infty$ of the derivative of the scalar at large distances. For example, considering cosmological boundary conditions with scaling $\dot \varphi_\infty \sim H_0 \mpl$ would impose that closer to the BH,
\begin{equation}
\frac{\varphi}{\mpl} \sim \frac{Q_\mathrm{S}}{GM \mpl} \sim \frac{H_0}{\mpl} \frac{M}{\mpl} \; .
\end{equation}
We thus see that, even for a supermassive BH the factor $\frac{M}{\mpl}$ cannot counterbalance the extremely small number $ \frac{H_0}{\mpl} \sim 10^{-60}$ so that cosmological scalar hair are quite negligible. However, this is not the end of the story since the asymptotic gradient $\dot \varphi_\infty$ could also be generated by some dynamical process far from the BH (in fact, for BHs embedded in a galactical environment, we can even say that this hypothesis is the most natural than one can make). \eqref{eq:ST_with_hair} then shows that \textit{any} BH embedded in a galactical environment will have scalar hair. This was used in~\cite{Horbatsch_2012} to put constraints on the scalar coupling to BHs using observations of a BH binary system at the center of a galaxy. Numerical simulations for this type of hair can be found in~\cite{Berti_2013}.

This approach can also be adopted in the Galileon/Horndeski case. Let us review in details the case of a cubic Galileon whose action is given in Eq. \eqref{eq:base_action}. 
BHs in the presence of a cubic Galileon interaction have been studied extensively~\cite{Babichev:2016fbg,Babichev:2015rva,Babichev:2013cya,Babichev_2013}. In particular, Ref.~\cite{Babichev:2016fbg} showed that hairy solutions do exist once we impose cosmological boundary conditions. The scalar charge of massive objects in a cubic Galileon was studied in~\cite{Babichev_2013} where it was shown that, even starting from a negligible bare coupling of the scalar to matter $\beta \sim 0$, an order-one \textit{effective} scalar charge $\beta_\mathrm{eff}$ emerges from  the  cosmological boundary condition. Let us see in more details how this effect arises.

As discussed in Section~\ref{sec:galileon_no_hair}, since the theory is shift symmetric the scalar EOM takes the form of a current conservation $\nabla_\mu J^\mu = 0$ with~\cite{Babichev_2013, Babichev:2016fbg}
\begin{equation}
J^\mu = \partial^\mu \varphi + \frac{1}{\Lambda^3} \square \varphi \partial^\mu \varphi - \frac{1}{2 \Lambda^3} \nabla^\mu \left( (\partial_\nu \varphi)^2  \right) \; .
\end{equation}
We consider a vacuum spherically symmetric solution of the field equations with an ansatz for the scalar field
\begin{equation}
\varphi = q t + \bar \varphi(r)\; ,
\end{equation}
where $q= \mpl t_{\rm scalar}^{-1}$ is the time derivative of the field, and a static and spherically symmetric ansatz for the metric is chosen
\begin{equation}
\mathrm{d}s^2 = - e^{\nu(r)} \mathrm{d}t^2 + e^{\lambda(r)} \mathrm{d}r^2 + r^2 \mathrm{d}\Omega^2 \; .
\end{equation}
Notice that the value of the phenomenological parameter $q$ is arbitrary here: as emphasized before, the asymptotic boundary condition for the scalar can be due to cosmology or to dynamical processes in the environment of the BH.
The $(tr)$ component of the metric equations is then equivalent to $J^r = 0$. We further make the assumption that the scalar is a test  field, i.e. we neglect its backreaction on the metric. We seek for solutions perturbatively close to the Schwarzschild one. With this supplementary assumption, Ref.~\cite{Babichev_2013} then showed that the scalar field solution to the $J^r = 0$ equation is
\begin{equation} \label{eq:sol_phiprime_BEF}
\varphi' = - \frac{1}{4} \Lambda^3 r \left( 1 - \sqrt{1 + \frac{8 M q^2}{8 \pi \mpl^2 r^3 \Lambda^6}} \right) \left[ 1 + \mathcal{O}\left( \frac{2GM}{r} \right) \right] \; ,
\end{equation}
where $M$ is the ADM mass of the BH, and the solution has been expanded outside the Schwarzschild radius $r_s = 2GM$.
 We can define an effective scalar charge and its associated Vainshtein radius,
\begin{equation} \label{eq:def_vainshtein_radius}
\beta_\mathrm{eff} = \frac{q^2}{2 \Lambda^3 \mpl}, \quad r_V^3 = \frac{\beta_\mathrm{eff} M}{8 \pi \mpl \Lambda^3} \; ,
\end{equation}
such that for $r \ll r_V$ the solution reads
\begin{align} \label{eq:phi_SS_chap9}
\varphi' =  \left( \frac{\beta_\mathrm{eff} M \Lambda^3 }{8 \pi \mpl r} \right)^{1/2}
 =  \frac{\beta_\mathrm{eff} M}{8 \pi \mpl r_V^2} \left( \frac{r_V}{r} \right)^{1/2} \;.
\end{align}
This solution is exactly the field generated by a massive body coupled with a Jordan frame metric $A(\varphi) = e^{\beta_\mathrm{eff} \varphi / \mpl}$. We can consequently model BHs with a point-particle action with coupling $\beta_\mathrm{eff}$ as
\begin{equation}
S_m = - M \int dt e^{\beta_\mathrm{eff} \varphi / \mpl} \sqrt{- g_{\mu \nu} v^\mu v^\nu} \; ,
\end{equation}
where $v^\mu = \frac{dx^\mu}{dt}$ is the four-velocity of the BH. This result has been extensively used in Chapter~\ref{Chapter8}.

It is important to notice that for cosmological boundary conditions, $q \sim H \mpl$ and for $\Lambda$ related to the dark energy scale, $\Lambda^3 \sim H^2 \mpl$, then the effective scalar charge is close to unity. The associated Vainshtein radius is, for an object of solar mass, of order of a kiloparsec. This leads to a Vainshtein suppression of the fifth force
\begin{equation}
\frac{\varphi}{\phi_N} \sim \beta_\mathrm{eff} \left( \frac{r}{r_V} \right)^{3/2}\;,
\end{equation}
where $\phi_N = M / (4 \pi \mpl r)$ is the Newtonian potential. However, if $q$ is given by some dynamical process is the vicinity of the BH, then the scalar coupling could grow to appreciable values $\beta_\mathrm{eff} \gg 1$. This would enhance the detectability of the Galileon field imprint on GW discussed in Chapter~\ref{Chapter8}.


\subsection{Non-minimally coupled scalars}

The theorem discussed in Section~\ref{subsec:hawking_th} applies to BDT theories in the Einstein frame where the scalar is minimally coupled to gravity. Since going in the Jordan frame is just a matter of field redefinition, we are led to think that this theorem also holds in the Jordan frame. However, this fails to be true for a particular class of BH in which the conformal transformation needed to go from the Einstein frame to the Jordan frame is singular. Consider the Bocharova–Bronnikov–Melnikov –Bekenstein (BBMB) BH solution~\cite{Bekenstein:1975ts, Bekenstein:1974sf} of conformal scalar-vacuum, which has the action
\begin{equation}
S = \frac{1}{2} \int \mathrm{d}^4 x \sqrt{-g} \left[ \mpl^2 R \left( 1 - \frac{\varphi^2}{3 \mpl^2} \right) - \partial_\mu \varphi \partial^\mu \varphi \right] \; .
\end{equation}
Then a hairy BH solution exists with
\begin{align}
\mathrm{d}s^2 &= - \left( 1 - \frac{GM}{r} \right)^2 \mathrm{d}t^2 + \left( 1 - \frac{GM}{r} \right)^{-2} \mathrm{d}r^2 + r^2 \mathrm{d}\Omega^2 \\
\frac{\varphi}{\mpl} &= \frac{\sqrt{3} GM}{r - GM} \; .
\end{align}
This is a one-parameter family of solutions, whose parameter is $M$, the total mass. The geometry coincides with the one of an extremal Reissner-Nordström BH, i.e. Eq. \eqref{eq:reissner_nordtstrom}  with $\vert Q_\mathrm{E} \vert = \sqrt{G} M$. In particular it has a regular horizon situated at $r=GM$ and hence it is a BH. Moreover, the scalar field diverges at the horizon, even though the geometry is regular at this locus.
Why does the no-hair discussed in Section~\ref{subsec:hawking_th} not apply to this BH? The conformal transformation needed to go from the Jordan frame to the Einstein frame is
\begin{equation}
A^2(\varphi) = 1 - \frac{\varphi^2}{3 \mpl^2} = \frac{r(r-2GM)}{(r-GM)^2} \; .
\end{equation}
Thus the conformal transformation becomes singular at $r=2M$ and one cannot easily go from the Jordan frame to the Einstein frame. However, this kind of solution has been shown to be unstable against linear perturbations so that it cannot be formed in astrophysical processes~\cite{Bronnikov:1978mx}.


\subsection{Superradiance and scalar clouds}

Superradiance is a phenomenon which amplifies waves (be it scalar, vector or tensor waves) when they hit a rotating object with absorbing boundary conditions~\cite{brito2015superradiance}. When considering a scalar wave impinging a BH, if the scalar has a mass it can trigger an instability named 'BH bomb'~\cite{Press:1972zz, cardoso2004black}; more generally, long-lived modes of the scalar can be excited around the BH which could lead to observable signatures~\cite{Baumann_2019, Boskovic:2018lkj}. Let us quickly review the formation process of this 'scalar cloud'.

Consider a simple BDT action with a \textit{complex} scalar field,
\begin{equation}
S = \frac{1}{2} \int \mathrm{d}^4 x \sqrt{-g} \left[ \mpl^2 R - \partial_\mu \varphi^* \partial^\mu \varphi - \mu^2 \varphi^* \varphi  \right] \; .
\end{equation}
Using a test-field analysis, let us study the modes of the scalar field on a Kerr background. The EOM for the scalar is simply the Klein-Gordon equation, $\square \varphi = \mu^2 \varphi$, in which we plug the separable ansatz
\begin{equation}
\varphi = e^{-i \omega t} e^{i m \phi} S_{lm}(\theta) R_{lm}(r) \; ,
\end{equation}
where $\omega$ is the frequency, $m, l \in \mathbb{Z}$ such that $-l\leq m \leq l$, $S_{lm}(\theta)$ are the spheroidal harmonics and $R_{lm}(r)$ obeys a second-order differential equation (we will use a similar decomposition when deriving the Regge-Wheeler equation in the next Chapter). Imposing purely absorbing boundary conditions at the horizon and purely outgoing waves at infinity, one generically expects a complex frequency $\omega = \omega_\mathrm{R} + i \omega_\mathrm{I}$ with $\omega_\mathrm{I} <0$ signalling a decay of the scalar (stable mode), while $\omega_\mathrm{I} >0$ would indicate the presence of an unstable mode.

In the \sch case, it turns out that all scalar modes are stable. However, around a Kerr BH one can have $\omega_\mathrm{I} >0$ under the condition that $\vert \omega \vert < m \Omega_H$ where  $\Omega_H$ is the angular velocity of the horizon. This is the superradiance phenomenon: the wave is amplified at the horizon of the BH, and since the mass term provides a potential barrier at infinity the scalar mode grows in time. At the threshold of superradiance, i.e when $\vert \omega \vert = m \Omega_H$, Hod observed that there are bound states with real frequency~\cite{Hod_2012, Hod_2013}. These scalar bound states are regular on and outside the horizon and, due to the complex nature of the scalar field, they source a $t, \phi$-independent energy-momentum tensor. These have been called scalar clouds around Kerr BHs~\cite{Herdeiro_2014}. These solutions can be extended to a full non-linear solution of BDT theories; these are asymptotically flat, regular on and outside an event horizon rotating BH solutions with primary scalar hair.

\begin{figure}[ht]
\centering
\includegraphics[width=0.8\textwidth]{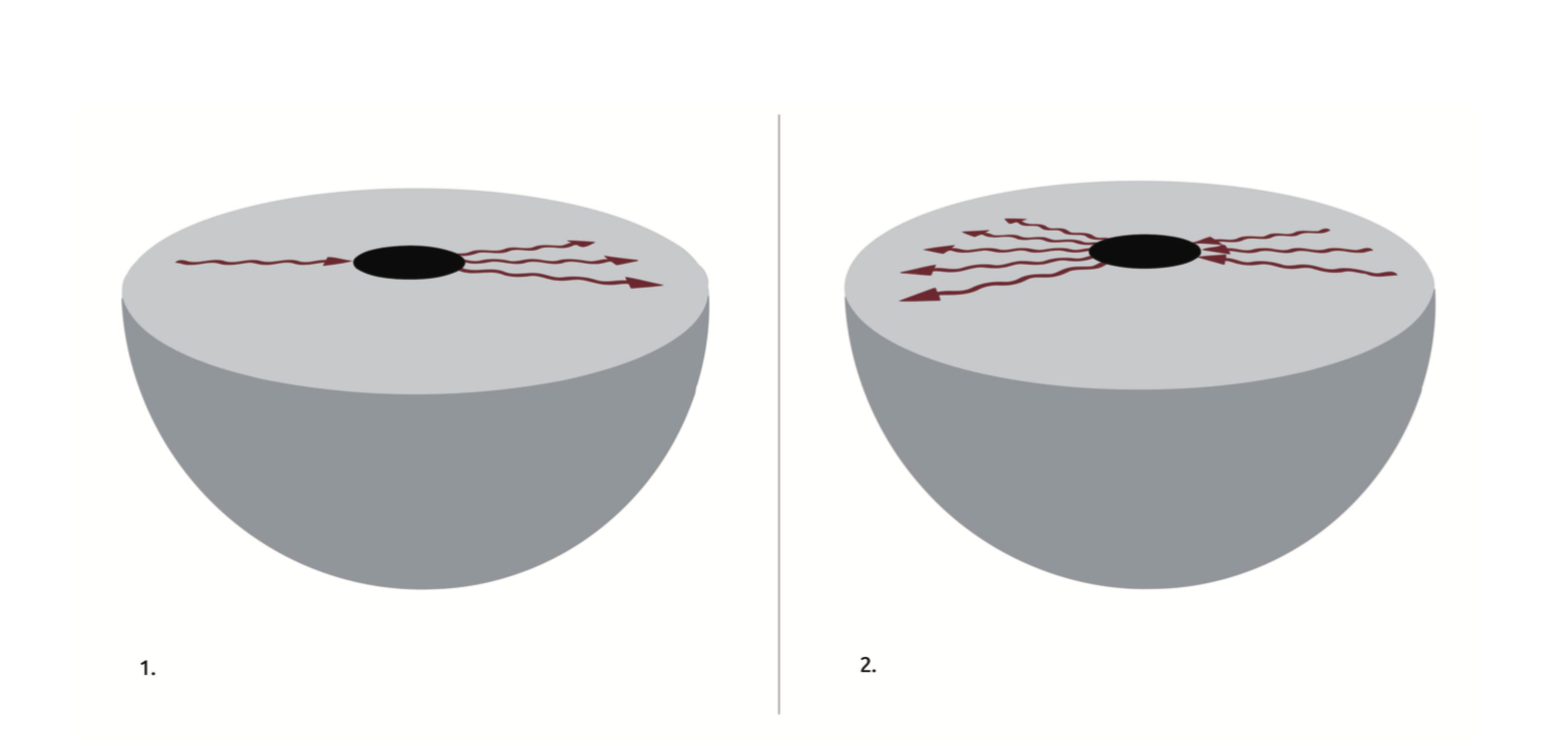}
\caption[Illustration of the 'BH bomb' process]{ Illustration of the 'BH bomb' process: a scalar wave is amplified at the horizon and trapped at infinity since the mass term acts as a potential barrier. Figure taken from~\cite{brito2015superradiance}}
\label{fig:BH_bomb}
\end{figure}

\subsection{Evading the Galileon no-hair theorem} \label{subsec:evading_galileon}

We now come to a Section which is of greater interest in relation with this thesis since it will correspond to the domain of applicability of the next Chapter. We will show that there can indeed exist hairy solution in the Horndeski class of theories without any time-dependence of the scalar. These solutions rely on a loophole in the last step of the proof presented in~\ref{sec:galileon_no_hair}. Let us give the expression of the current defined in \eqref{eq:def_current} in the beyond Horndeski class ~\cite{Babichev_2017}
\begin{align}
\begin{split} \label{eq:current_BH}
J^r &= 2 f \varphi' G_{2X} + f \frac{rh'+4h}{rh} X G_{3X} - 4f \varphi'  \frac{fh - h + rfh'}{r^2 h} G_{4X} - 8 f^2 \varphi' \frac{h+rh'}{r^2h} X G_{4XX} \\
&-f h' \frac{1-3f}{r^2 h} X G_{5X} + 2 \frac{h' f^2}{r^2 h} X^2 G_{5XX} - 4 f^2 \varphi' \frac{h+rh'}{r^2h} X (2 F_4 + X F_{4X}) \\
&+ 3 \frac{f^2 h'}{r h} X^2 (5 F_5 + 2XF_{5X}) \; ,
\end{split}
\end{align}
where as usual $X = \partial^\mu \varphi \partial_\mu \varphi  = g^{rr} \varphi'^2 $, and the functions $G_i$, $F_i$ are defined in Section~\ref{subsec:Horndeski} (they depend on $X$ only since the theory is shift symmetric). We have used the following form of the background metric,
\begin{equation}
\mathrm{d}s^2 = - h(r) \mathrm{d}t^2 + \frac{1}{f(r)} \mathrm{d}r^2 + r^2 \mathrm{d}\Omega^2 \; ,
\end{equation}
(this is a different choice than the one adopted in Section~\ref{sec:galileon_no_hair} but it does not change the validity of the argument). Recall from Section~\ref{sec:galileon_no_hair} that we should have $J^r=0$ for all $r$ in order for the norm of the current to be finite at the horizon. Since we also assume that a kinetic term is present in the action, $G_2 \supseteq X$, the only way to avoid the trivial solution $\phi'=0$ is to make one of the $G_i$ or $F_i$ piece in \eqref{eq:current_BH} independent of $\phi'$. Thus, $\phi'$ appearing in the kinetic term will be forced to take a nontrivial value from the condition $J^r=0$. Ref.~\cite{Babichev_2017} showed that it can occur if one chooses the beyond-Horndeski functions to be one among these forms,
\begin{align}
\begin{split}
G_2 &\supseteq \sqrt{\vert X \vert} \; , \quad G_3 \supseteq \ln \vert X \vert \; , \quad G_4 \supseteq \sqrt{\vert X \vert} \; , \quad  G_5 \supseteq \ln \vert X \vert \; , \\
F_4 &\supseteq \vert X \vert^{3/2} \; , \quad F_5 \supseteq \vert X \vert^{-2} \; .
\end{split}
\end{align}
If any of these terms is present in the action, then $\phi'$ will be non-trivial (although this does not guarantee by itself the existence of a BH solution). Ref.~\cite{Babichev_2017} gives examples of hairy BH solutions in the $G_4$ and $F_4$ cases. From this perspective, the presence of scalar hair seems to be a quite generic phenomenon. However, in a subsequent study \cite{Saravani_2019} it has been shown that all of these hair, except the $G_5$ one, do not lead to solutions smoothly connected to Minkowski spacetime (this conclusion is further supported by Ref. \cite{Creminelli:2020lxn} which appeared during the redaction of this thesis).

Let us finish by commenting on the $G_5$ case. It is well known (see Section~\ref{subsec:GB_CS}) that $G_5 \propto \ln \vert X \vert$ corresponds to a coupling of the scalar to the Gauss-Bonnet invariant, i.e a term in the action of the type
\begin{equation}
\phi \big( R^{\mu \nu \rho \sigma} R_{\mu \nu \rho \sigma} - 4 R^{\mu \nu} R_{\mu \nu} + R^2 \big)\; ,
\end{equation}
This type of term is known to generate scalar hair and has been studied by several authors~\cite{Sotiriou_2014, Silva:2017uqg, Silva:2018qhn, Antoniou_2018}. These solutions could even lead to 'spontaneous BH scalarization', i.e situations where hairy solution are dynamically preferred over GR solutions. However, it appears that such solutions cannot be consistent with $J^r=0$, i.e they violate the assumption of the finiteness of the current at the horizon. On the other hand, the curvature invariants are finite and perturbations of the scalar are stable so that it is argued that the requirement of the finiteness of the current is unphysical \cite{Creminelli:2020lxn}.

Before moving on, let us insist again on a crucial point concerning BH hair in Horndeski theories. One may be tempted to think that, since we have mentioned in Section~\ref{sec:tests_GW} that the freedom in Horndeski functions has been immensely simplified to the sole function $G_2$, these hairy BH solutions are not phenomenologically viable. However, the constraints discussed in Section~\ref{sec:tests_GW} apply for a \textit{cosmological} scaling of the Horndeski functions; if the new energy scales involved in the $G_i$, $F_i$ functions are higher so that their cosmological relevance is negligible, then all the freedom of the beyond-Horndeski is still allowed. This has been showed in particular for the $G_5$ term in Section~\ref{subsec:GB_CS}.

\section{An EFT for scalar hair}

\subsection{Quadratic action}

The last Section made clear that one can expect scalar hair to be present around BH in several theories, and it is often the most natural choice to make if one is seeking departures from GR. Since we are opening a new era of observational evidence of BHs with GW, we are finally in a position where the absence or presence of scalar hair is a feature that can be tested experimentally rather than ruled out by no-go theorems based on a set of assumptions. In that respect, the EFT approach advocated in Chapter~\ref{Chapter2} for dark energy provides a very handful formalism: we will see that we will be able to encode deviations from GR in a set of coefficients which can be measured with GW observations. This Section will be dedicated to a detailed presentation of the results of Ref.~\cite{Franciolini:2018uyq} since we will make an heavy use of them in the next Chapter. In short, the authors of Ref.~\cite{Franciolini:2018uyq} studied the imprints of scalar hair on the Quasi-Normal Modes (QNMs) of BHs. These QNMs describe the gravitational perturbations of BHs; they should be measurable at the very end of a GW signal from a binary merger, when the two BHs have collided and the geometry approaches the one of a single, perturbed BH~\cite{Berti_2016}. Currently, one is only able to extract the dominant QNM mode from data as discussed in Section~\ref{sec:tests_GW}, but as the accuracy of the GW detectors improves we will be able to see other modes in the QNM data. Thus, one will be able to test if these modes really correspond to Kerr BHs or if they differ in some way because of the presence of hair around the BH.

Let us assume that one of the mechanisms discussed in Section~\ref{sec:evading_no_hair} gives a nontrivial scalar profile $\bar \varphi(r)$ around a \sch BH\footnote{Note that one can also construct an effective theory assuming that the no-hair theorem holds and that the background solution is exactly \sch, but leaving open the possibility that perturbation differ from GR. This has been carried out in~\cite{Tattersall_2018_2, Tattersall_2018_3}}. For our purposes it will not matter whether this is a primary or secondary hair, i.e. whether or not such profile is associated with an additional conserved charge. The situation is then very similar to the EFT of DE, where the field took a timelike VEV $\bar \varphi(t)$ when acting as DE. To study perturbations of the scalar hair $\varphi = \bar \varphi + \delta \varphi$, one can again go to the unitary gauge by performing a radial diffeomorphism such that the perturbation vanishes $\delta \varphi = 0$ (notice that this essential point requires the background value of the scalar to be nontrivial, $\bar \varphi' \neq 0$). Therefore, one can construct an effective action in the very same spirit as in Section~\ref{subsec:construction_EFTDE} by using the fact that the background scalar provides a preferred \textit{radial} foliation of spacetime. One will be left with an effective action which will be invariant under time- and angular-diffeomorphisms, but not under radial ones so that arbitrary functions of the radius are allowed. The most generic action of such type is, up to quadratic order in perturbations and second order in derivatives~\cite{Franciolini:2018uyq},
\begin{align}
& S =  \int \mathrm{d}^4x \, \sqrt{-g} \bigg[
\frac{1}{2}M^2_1(r) R -\Lambda(r) - f(r)g^{rr} - \alpha(r)\bar K^\mu_{\ \nu} K^\nu_{\ \mu} \nonumber
\\
&	+ M_2^4(r)(\delta g^{rr})^2
	+M_3^3(r) \delta g^{rr }\delta K  + M_4^2(r) \bar K^\mu_{\ \nu} \delta K^\nu_{\ \mu} \delta g^{rr }\nonumber
\\[2mm]
&	+ M_5^2(r)(\partial_r\delta g^{rr})^2
	+M_6^2(r) (\partial_r\delta g^{rr})\delta K  + M_7(r)\bar K^\mu_{\ \nu} \delta K^\nu_{\ \mu}  (\partial_r\delta g^{rr})
	+ M_8^2(r)(\partial_\mu \delta g^{rr})^2 \label{eq: effective action for perturbations}
\\[2mm]
&	+M_9^2(r)(\delta K)^2 + M_{10}^2(r)\delta K^\mu_{\ \nu}\delta K^\nu_{\ \mu}
+ M_{11}(r) \bar K^\mu_{\ \nu} \delta K^\nu_{\ \mu} \delta K 
+ M_{12}(r)  \bar K^\mu_{\ \nu} \delta K^\nu_{\ \rho} \delta K^\rho_{\ \mu}  \nonumber
\\
&	+ \lambda(r) \bar K^\mu_{\ \nu} \bar K^\nu_{\ \rho} \delta K^\rho_{\ \mu} \delta K 
+ M_{13}^2(r) \delta g^{rr }\,  \delta\!\! \  ^{(3)}\!R
+ M_{14}(r)  \bar K^\mu_{\ \nu}\,  \delta\!\! \ ^{(3)}\!R^\nu_{\ \mu} \delta g^{rr } + \ldots
\bigg] \, , \nonumber 
\end{align}
In this equation (which is very similar to~\eqref{eq:ActionEFTQuadratic}, with some differences which we will highlight below), $K_{\mu \nu}$ and $^{(3)}R$ are respectively the extrinsic and intrinsic curvatures of constant-$r$ hypersurfaces. At first sight, the freedom seems much higher than in the cosmological case~\eqref{eq:ActionEFTQuadratic} since there are many more free functions. This comes from the fact that the action has been expanded around a spherically symmetric background \footnote{One of the three unknown functions in Eq.~\eqref{eq:ss_background} can be fixed by a background radial diffeomorphism, however to allow for an easier comparison with different theories we leave $a$, $b$ and $c$ arbitrary for the moment },
\begin{equation}\label{eq:ss_background} 
\mathrm{d} s^2 = \bar g_{\mu\nu} \mathrm{d} x^\mu \mathrm{d}  x^\nu = -a^2(r)\mathrm{d} t^2 + \frac{\mathrm{d} r^2}{b^2(r)} + c^2(r)\left(\mathrm{d}\theta^2+\sin^2\theta\mathrm{d} \phi^2 \right) \, 
\end{equation} 
instead of a FLRW one. Indeed, in the cosmological case the background is highly symmetric so that any background ADM quantity can be expressed in term of the background metric~\cite{Cheung:2007st}. This is no longer true in this less symmetric case, see below for details. The net result is that there is no proper way to define a covariant perturbation tensor, however this turns out not to be a limitation of the formalism.

At background level, only the first line of Eq.~\eqref{eq: effective action for perturbations} (the \textit{tadpole} operators) contribute. Compared to the cosmological case, there is one additional operator $\bar K^\mu_{\ \nu} K^\nu_{\ \mu}$ which, contrary to the FLRW case, cannot be rewritten in terms of the other tadpole terms. The action is expressed in the Jordan frame so that there is an arbitrary   function $M_1^2(r)$ in front of the Ricci scalar. As usual, one can set $M_1^2(r) = \mpl^2$ with a conformal transformation; in this Jordan frame, matter would be nonminimally coupled to the metric.

The other tadpole operators appearing in~\eqref{eq: effective action for perturbations} are not arbitrary, but are instead fixed by the requirement that the background metric be a solution to Einstein's equations. More specifically, a spherically symmetric background metric of the form 
\eqref{eq:ss_background}
will satisfy Einstein's equations only if the coefficients $\Lambda(r), f(r)$ and $\alpha(r)$ appearing in the first line of~\eqref{eq: effective action for perturbations} obey the following \emph{tadpole conditions}:
\begin{align} 
& f(r) 
=  \left(\frac{a'c'}{ac}-\frac{b'c'}{bc} - \frac{c''}{c}  \right)M^2_1 + \frac{1}{2}\left(\frac{a'}{a}-\frac{b'}{b}\right)(M^2_1)' - \frac{1}{2}(M^2_1)'' \label{fm} 
\\
& \qquad \qquad\qquad\qquad\qquad\qquad  - \left( \frac{a'^2}{2a^2}-\frac{a'b'}{2ab}-\frac{a'c'}{ac} + \frac{c'^2}{c^2} - \frac{a''}{2a}\right)\alpha + \frac{a'}{2a}\alpha' \, , \nonumber
 \\[2mm]
&\Lambda(r) 
= - b^2\left(\frac{c''}{c} + \frac{a'c'}{ac}+ \frac{b'c'}{bc}+\frac{c'^2}{c^2}-\frac{1}{b^2c^2}  \right)M^2_1 - b^2\left(\frac{a'}{2a}+\frac{b'}{2b}+\frac{2c'}{c} \right) (M^2_1)'\label{lm} 
\\ 
& \qquad \qquad\qquad\qquad\qquad\qquad  - \frac{b^2}{2}(M^2_1)'' - b^2\left( \frac{a'^2}{2a^2}-\frac{a'b'}{2ab}-\frac{a'c'}{ac} + \frac{c'^2}{c^2} - \frac{a''}{2a}\right)\alpha + \frac{b^2a'}{2a}\alpha'
 \, , \nonumber 
 \\[2mm]
&\left(\frac{a'}{a}-\frac{c'}{c} \right)(M^2_1+\alpha)'  + \left( \frac{a''}{a} - \frac{c''}{c} + \frac{a'b'}{ab} + \frac{a'c'}{ac}- \frac{b'c'}{bc}-\frac{c'^2}{c^2}\right)(M^2_1+\alpha) + \frac{M^2_1}{b^2c^2} = 0\label{thirdtadpcondition}  .  
\end{align}%
Thus, one can think of $\Lambda(r), f(r)$ and $\alpha(r)$ as being completely specified once the background metric and the function $M_1(r)$ are given. All other functions appearing in~\eqref{eq: effective action for perturbations} (which contribute to the quadratic action) are in principle arbitrary and must be constrained by observations.\footnote{As we have seen in Section~\ref{subsec:evading_galileon}, static hairy BH only exists for specific choices of the scalar action. Whether or not this is encoded by additional, hidden constraints on these arbitrary functions---akin to a swampland conjecture for hairy black holes---remains an interesting open question.} 

At this point, we should stress an important difference between the effective action we are using here and the one first derived in~\cite{Franciolini:2018uyq}: in Eq.~\eqref{eq: effective action for perturbations}, $\bar K^\mu_{\ \nu}$ and $\delta K^\mu_{\ \nu}$ always appear with one upper and one lower index. This was not the case in the effective action used in~\cite{Franciolini:2018uyq}. In particular, the tadpole equations appearing in~\cite{Franciolini:2018uyq} imply the specific index structure index structure
$\bar K^{\mu \nu}K_{\mu \nu}$ for the tadpole term proportional to $\alpha(r)$.  As a result our tadpole conditions~\eqref{fm}--\eqref{thirdtadpcondition} differ from the ones quoted in~\cite{Franciolini:2018uyq}.  From a conceptual viewpoint the two effective actions are completely equivalent, as they correspond to choosing a different basis of operators in the Lagrangian. From a technical viewpoint, however, the choice we are making here turns out the be more convenient for two reasons.  First, the transformation properties of $\alpha(r)$ under conformal redefinitions of the metric are much simpler when the third tadpole is defined as in Eq.~\eqref{eq: effective action for perturbations}.  Second, matching an explicit scalar-tensor theory onto the effective action~\eqref{eq: effective action for perturbations} is much simpler. This is in part due to the fact that the perturbed and background induced metrics with mixed indices have identical components.

The example of the extrinsic curvature tensor will serve as a clarification on this point.  In order to raise and lower the indices of the unperturbed  extrinsic curvature tensor $\bar K^\mu_{\ \nu}$, it looks natural to use the background metric. However, when dealing also with perturbations, this is far from being the most convenient option. A quantity such as $\delta K^\mu_{\ \nu} = K^\mu_{\ \nu} - \bar K^\mu_{\ \nu}$ would end up transforming in some hybrid cumbersome way. In practice, when trying to translate a general theory in the EFT language, one has to expand in perturbations terms such as $K_{\mu \nu} K^{\nu \rho} K_\rho^{\ \mu}$, and would like to be able to raise and lower the indices in some definite standard way at any step of the process. In the case of a spatially flat Friedmann Robertson Walker (FRW) universe~\cite{Piazza:2013coa}, the extrinsic curvature of the constant time hyper-surfaces evaluates $\bar K^\mu_{\ \nu} = {\rm diag}(0, H, H, H) = H h^\mu_{\ \nu}$, $H$ being the Hubble parameter and $h^\mu_{\ \nu}$ the induced (full, \emph{i.e.} containing the perturbations!) three-dimensional metric. One can thus \emph{define} the perturbations as the fully covariant tensor $\delta K_{\mu \nu} \equiv K_{\mu \nu} - H h_{\mu \nu}$ and raise and lower the indices accordingly~\cite{Cheung:2007st}. 
 
In the present less symmetric case, there is no natural way to define a covariant perturbation tensor $\delta K^\mu_{\ \nu}$. By looking at the  metric in the form
\eqref{eq:ss_background}
one finds
\begin{equation}
 \bar K^\mu_{\ \nu} \ = \ b(r) \cdot {\rm diag}\left(\frac{a'(r)}{a(r)},\ 0, \ \frac{c'(r)}{c(r)}, \ \frac{c'(r)}{c(r)}\right)\, .
\end{equation}

In the absence of a covariant perturbation tensor, it is misleading to manipulate terms containing background and perturbation quantities. One solution is to just abstain to do so, by always contracting extrinsic curvature tensors with an index up and an index down, without the need of ever lowering and raising indices. For example, when expanding in perturbations, the cubic extrinsic curvature term previously mentioned should here be written as 
\begin{align}
K^{\mu}_{\  \nu}\,  K^\nu_{\  \rho}\,  K^\rho_{\ \mu} \ & = \ (\bar K^{\mu}_{\  \nu} + \delta K^{\mu}_{\  \nu}) \, (\bar K^\nu_{\  \rho} + \delta K^\nu_{\  \rho})\,  (\bar K^\rho_{\ \mu} + \delta K^\rho_{\ \mu})\\ \nonumber
& = \ - 2 \bar K^{\mu}_{\  \nu}\, \bar K^\nu_{\  \rho}\, \bar K^\rho_{\ \mu} + 3 \bar K^{\mu}_{\  \nu}\, \bar K^\nu_{\  \rho}\, K^\rho_{\ \mu} + 3
\bar K^{\mu}_{\  \nu}\, \delta K^\nu_{\  \rho}\, \delta K^\rho_{\ \mu} + \delta K^{\mu}_{\  \nu}\, \delta K^\nu_{\  \rho}\, \delta K^\rho_{\ \mu} \, ,
\end{align}
\emph{i.e.} with each term having one index up and one index down. 
A similar reasoning applies to the induced intrinsic curvature, $\  ^{(3)}\!\bar R^\mu_{\ \nu} = {\rm diag}\left(0, 0, c^{-2}(r), c^{-2}(r)\right)$. Contractions involving more indices, such as those involving covariant derivatives of $K^\mu_{\ \nu}$, will require more care, but they appear at higher order in the derivative expansion and can be overlooked at this time. 

With the caution required by the issues just discussed, one can follow the construction of~\cite{Franciolini:2018uyq} 
and show that the only terms that contribute to the quadratic action for metric perturbations up to second order in derivatives are~\eqref{eq: effective action for perturbations}.


The effective action for perturbations in Eq.~\eqref{eq: effective action for perturbations} was used in~\cite{Franciolini:2018uyq} to constrain departures from the quasi-normal mode spectrum of Schwarzschild black holes in General Relativity. This EFT framework was also used in~\cite{Franciolini:2018aad} to argue for the existence of stable wormhole solutions in scalar-tensor theories. In the next Chapter we will develop a third application of this formalism by studying the modifications to the waveform produced by a binary inspiral with extreme mass ratio. However before to move on let us recall how one can obtain a Regge-Wheeler equation starting from the EFT action~\eqref{eq: effective action for perturbations} and how it allows to compute the QNM spectrum of BHs with scalar hair.


\subsection{QNM spectrum in the odd sector}

When looking at perturbations of a spherically symmetric BH, metric fluctuations can be classified in scalars ($\delta g_{00}$, $\delta g_{0r}$, $\delta g_{rr}$), vectors ($\delta g_{0i}$, $\delta g_{r i}$) and tensor ($\delta g_{ij}$) where $i,j = \theta, \phi$. Each vector and tensor can be further split into even and odd components: an odd perturbation changes sign under parity $(\theta, \phi) \rightarrow (\pi - \theta, \phi + \pi)$, while an even perturbation do not. Let us focus for now on the odd perturbations (the odd sector, in GR as well as in scalar-tensor theories, is much simpler than the even one), which
are parametrized by 3 functions $h_0$, $h_1$ and $h_2$ as follows~\cite{Regge:1957td}:
\begin{eqnarray} \label{odd metric perturbations}
\delta g_{\mu\nu}^{\rm odd} = \left(
\begin{array}{ccc}
0 & 0 & \epsilon^k {}_j \nabla_k h_0\\
0 & 0 & \epsilon^k {}_j \nabla_k h_1\\
\epsilon^k {}_i \nabla_k h_0 \,\,\,\, &  \epsilon^k {}_i \nabla_k 
                                    h_1 & \,\,\,\, {1\over 2} (\epsilon_i {}^k
                                           \nabla_k \nabla_j +
                                           \epsilon_j {}^k \nabla_k
                                           \nabla_i ) h_2
\end{array}\right) ,
\end{eqnarray}
where
\begin{align} \label{harmonics decomposition odd perturbations}
	h_n (t,r,\theta, \phi) = \sum_{\ell =2}^{\infty} \sum_{m = - \ell}^\ell h_n^{\ell m} (t,r) Y_{\ell m} (\theta, \phi), \qquad \qquad n=1,2,3, 
\end{align}
and $\epsilon_{ij}$ and $\nabla_i$ are respectively the Levi-Civita tensor and covariant derivative on the 2-sphere---see {\it e.g.} Sec. 3 of~\cite{Franciolini:2018uyq} for their explicit expressions. Odd perturbations with angular momentum number $\ell = 0, 1$ have been omitted in Eq.~\eqref{harmonics decomposition odd perturbations} because they do not propagate, partly due to an enhanced gauge invariance.  This can be easily checked by deriving the quadratic action for such modes, and is consistent with the fact that the additional scalar degree of freedom belongs to the even sector. 

The only operators in the effective action~\eqref{eq: effective action for perturbations} that contribute to the odd sector are~\cite{Franciolini:2018uyq}:
\begin{align}
& S =  \int \mathrm{d}^4x \, \sqrt{-g} \bigg[
\frac{1}{2}M^2_1(r) R -\Lambda(r) - f(r)g^{rr} - \alpha(r)\bar K^\mu_{\ \nu} K^\nu_{\ \mu} \nonumber
\\
& \qquad \qquad \qquad \qquad \qquad + M_{10}^2(r)\delta K^\mu_{\ \nu}\delta K^\nu_{\ \mu}  + M_{12}(r)  \bar K^\mu_{\ \nu} \delta K^\nu_{\ \rho} \delta K^\rho_{\ \mu} \bigg]. \label{S odd}
\end{align}
Indeed, the quantities $\delta g^{rr}$, $\delta K$ and $\delta^{(3)} R$ appearing in the quadratic action~\eqref{eq: effective action for perturbations} contain only even-type perturbations, so that the only terms containing odd-type perturbations are $M_{10}$ and $M_{12}$.

From this action it is possible to derive an equation of motion for the perturbations. One of the metric functions in the odd sector~\eqref{odd metric perturbations} is fixed by a gauge choice (the Regge-Wheeler gauge fixes $h_2=0$), another one is fixed by a constraint equation so that the EOM for the only degree of freedom left is of the Regge-Wheeler type~\cite{Regge:1957td,Chandrasekhar:1985kt}:
\begin{equation} \label{eq:RW_without_source}
\frac{d^2 \Psi}{d \tilde r^2} +  V(\tilde r) \Psi = 0 \; ,
\end{equation}
where the form of $\Psi$ (related to $h_0$, $h_1$), $\tilde r$ and $V$ is given in~\cite{Franciolini:2018uyq}. In the next Chapter, we will derive again this equation with a source term on the right-hand side corresponding to the presence of a small point-particle sourcing the perturbations. Imposing to the solutions of this Schrödinger-type equation purely ingoing boundary conditions at the horizon (when the tortoise coodinate $\tilde r \rightarrow \infty$) and purely outgoing flux at infinity, one finds that only a discrete set of frequencies solve the problem: these are the QNMs, representing the frequencies of oscillation of a BH. Schutz and Will~\cite{Iyer:1986vv} showed that the quasi-normal spectrum associated with the equation~\eqref{eq:RW_with_source} can be approximated analytically using the WKB method, with a precision reaching the percent level. Their main result is the
relation
\begin{equation} \label{eq:WKB}
\left. \frac{V}{\big(2 \partial_{\tilde r} V \big)^{1/2}} \right\vert_{\tilde r = \tilde r_*} = - i \left( n + \frac{1}{2} \right) \; ,
\end{equation}
where $n \in \mathbb{N}$ and $\tilde r_*$ is the position of the light-ring, i.e the maximum of $-V$. Since $V$ depends on $\omega$, this equation implicitely defines the QNM frequencies. If we further assume that deviations from GR are small (as is evidenced by the first GW observations), one can evaluate the small shift in the QNM frequency $ \omega_\mathrm{QNM} = \omega_\mathrm{QNM, \; GR} + \delta \omega$ from Eq.~\eqref{eq:WKB}. This shift will depend on derivatives of the EFT functions $a$, $b$, $c$ (defined in~\eqref{eq:ss_background}), $M_{10}$ and $M_{12}$ at the light-ring $r_{*, \; GR}$. This is the so-called the \textit{light-ring expansion}: knowing the EFT functions and a few of their derivatives at the light ring (see~\cite{Franciolini:2018uyq} for a more precise statement) allows to reconstruct the QNMs of hairy BHs. Measurement of several QNM frequencies by e.g LISA~\cite{Baibhav_2019} could allow to constrain the parameters of this expansion.

Finally, let us mention that this procedure is perfectly transportable to the even sector, albeit this is technically much more involved. In the next Chapter, we will also focus on the odd sector for simplicity and leave a detailed study of the even sector for further work.

\chapter{Extreme mass ratio inspiral with scalar hair} \label{Chapter10}

In the last few Chapters we have analyzed how a GW signal could be altered in some specific scalar-tensor theories. Of course, it is clear that studying such modifications on a model-by-model basis is impractical and ultimately very inefficient. A unifying formalism, akin to the PPN formalism or to the EFT of DE respectively presented in Chapters~\ref{Chapter1} and~\ref{Chapter2}, would greatly reduce the amount of work needed to compare theory with experiment.
In the last Chapter of this thesis (based on the JCAP paper~\citeK{Kuntz:2020yow}), we will develop an effective formalism adapted to GW generation in the extreme mass ratio limit. We will focus on the inspiral phase of the dynamics, since the very large number of cycles recorded in a detector such as LISA will allow for an exquisite measure of all the parameters of the system \cite{Babak_2017}.

As we already mentioned in Section~\ref{sec:tests_GW}, a simple approach to the question could be e.g to allow for independent variations of the PN coefficients in the GW phase, Eq.~\eqref{eq:variation_PN_coeffs}. This approach has been advocated in the Parametrized post-Einsteinian (ppE) formalism~\cite{Yunes_2009, Cornish:2011ys, Chatziioannou:2012rf}, which aims at devising a generic framework encompassing the inspiral, merger and ringdown signals of binaries emitting GW. However, there remains some major points that this formalism does not address. First, matching to an explicit modified gravity model still requires to repeatedly derive the same equations on a model-by-model basis~\cite{Tahura_2018}; studying the two-body dynamics in modified theories of gravity proves to be an herculean task even in simplest of setups, as the metric and the putative supplementary fields quickly get non-linearly coupled down the PN expansion \cite{Yagi:2011xp}.  Second, allowing for a new coefficient at each PN order gives way too many independent parameters when fitting to data. One could try to vary the PN coefficients one by one, as has been done in the LIGO analysis presented in Section~\ref{sec:tests_GW}, but then the formalism is not related to any particular theory since each modified gravity theory generically predicts a modification of \textit{all} the PN parameters. Finally, the ppE formalism is generically useful for assessing the compatibility of a given signal with GR \textit{once it has been detected with a GR template}, but cannot replace a modeled search of a signal truly different from GR in data. In this respect, a fully-fledged waveform in the simplest scalar-tensor setting, namely BDT theories, is only known to 1PN order~\cite{Lang:2015aa, Bernard:2018hta, Bernard:2018ivi}. This level of  accuracy would need to be further improved for a meaningful comparison with data, which generally requires the energy flux to be known at least up to 3PN order~\cite{Blanchet:2013haa}. The formalism which we will present in this Chapter will address all of these points.

Let us now be more precise on the middle-of-the-road approach that we will adopt. We will concentrate on scalar-tensor theories and consider a black hole endowed with a scalar charge or hair. As we have seen in the last Chapter, the no-hair theorem which forbids the appearance of scalar charge for BHs in the simplest scalar-tensor theories can fail to be true under some circumstances. In this Chapter we adopt a pragmatic approach: since scalar hair would be among the strongest scalar-tensor modifications to GR that one can devise, it is one of the first signatures that one should aim to detect or constrain~\cite{Horbatsch_2012,Berti_2013,Franciolini:2018uyq}.
We will consider an Extreme Mass Ratio Inspiral (EMRI) of a solar-size BH onto a supermassive one. The EFT approach described in the last Chapter thus allows to write the most generic scalar-tensor Lagrangian quadratic in the perturbations caused by the small BH. One major advantage of this approach is that it doesn't require as an input the microscopic Lagrangian of the scalar-tensor theory. As a consequence, it bypasses potential ambiguities associated with field redefinitions of the scalar (and, in a particular frame, also with conformal transformations of the metric)~\cite{Piazza:2013coa}.

As we have seen in Chapter~\ref{Chapter9}, linear perturbations around spherically symmetric solutions can be classified into even and odd parity modes. The even sector in particular is responsible for some of the most interesting features of GW emission in scalar tensor theories. On the one hand, it is singlehandedly responsible for the leading quadrupole emission (this is the case also in pure GR). Indeed, one byproduct of our computation is that the lowest order radiation in the odd sector is of 1PN order. On the other hand, it contains fluctuations of the scalar field itself, and thus the dipolar component of radiation.  Despite all this, in this Chapter we will focus our attention on the \emph{odd parity sector}. This is mainly done for reasons of technical simplicity, since this sector contains a single degree of freedom and is described by an effective Lagrangian that includes only a handful of operators. The general formalism that we develop here will be extended to the even parity sector in a future work.

The equations for perturbations of static black hole solutions can be cast into the standard Regge-Wheeler form~\eqref{eq:RW_without_source}, characterized by an effective potential $V(r)$ taking values outside the event horizon of the black hole. While quasinormal modes are sensitive to the entire shape of $V$, the PN approximation that we implement here only requires a limited number of terms in a $1/r$ expansion of the potential. In Sec.~\ref{sec-3} we \emph{define} our background metric and the EFT operators by means of such an expansion, and impose  the constraints arising from the tadpole equations to find relations among the coefficients. In Sec.~\ref{sec:examples}, we illustrate our formalism by providing a few examples of covariant theories in the unitary gauge and working out the coefficients of their PN expansion.

To describe EMRI systems we use a point-like source term representing the small mass in circular orbit around the large black hole. We find that the relevant coupling with the odd sector is of the conformal type, and that finite size ({\it i.e.} higher derivative) corrections are negligible with respect to PN ones. This allows us in Sec.~\ref{sec:RW} to write the Regge-Wheeler equation in the presence of a source and express the effective potential $V(r)$ in terms of the PN parameters of the EFT Lagrangian. 

Finally, Sec.~\ref{sec:perturbative_sol_RW} contains the main results of this Chapter.  Following~\cite{Poisson:1993aa,Sasaki:1994aa}, we  solve the Regge-Wheeler equation in powers $v$. We will work with an accuracy of $\mathcal{O}(v^5)$ beyond the lowest order solution. Since, as emphasized before, the odd sector itself is suppressed by 1PN order with respect to the usual quadrupole formula,  
by calculating the flux at infinity we will obtain an analytic expression for the dissipated power up to 3.5PN order ({\it i.e.}, up to $\mathcal{O}(v^7)$ beyond the leading GR quadrupole), which is the minimal required accuracy for waveform templates.\footnote{Of course, the neglected mass ratio of the system, even if very small in EMRI ($10^{-5}$), will need to be taken into account for an accurate waveform template. Taking into account this mass ratio in a second-order self-force calculation would be an interesting direction in which to extend this work.}

\vspace{.3cm}

\noindent{\it Conventions:} In most of this Chapter and contrary to the convention adopted in this thesis, we will work in units such that $G=1$. Thus, mass and length will have the same dimension.

\section{Sourced odd sector in the PN limit} \label{sec-3}

\subsection{Effective action for odd perturbations with a point-like source}

It is convenient to classify perturbations around a spherically symmetric background according to their transformation properties under parity. Because the action \eqref{eq: effective action for perturbations} discussed in Chapter \ref{Chapter9} does not contain parity violating terms, even and odd modes decouple from each other at linear level and can be studied separately. In this Chapter we restrict our attention to the odd sector of perturbations, which only includes one propagating degree of freedom. The general EFT action~\eqref{eq: effective action for perturbations} is considerably simplified by such a restriction, as only the small subset of operators showed in Eq.~\eqref{S odd} contribute. As shown in the next Section, these terms contribute to the standard Regge-Wheeler equation with two radial and two temporal derivatives. Terms with higher derivatives are not necessarily negligible from the point of view of the PN expansion. However, as they can be \emph{naively} associated with ghost-like instabilities, they must represent small perturbative corrections in the effective Lagrangian. We do not include them here because we assume that they are suppressed by some high scale. 

We now need to supplement this action with a term that describes the coupling with a point-like source. The point-particle action should be invariant under the same symmetries as the bulk action, and for our purposes it will be sufficient to consider only the leading term in its derivative expansion, {\it i.e.}
\begin{align} \label{S source}
	S_{\rm source} = - \int \mathrm{d}\tau \; \mu(r), \qquad\qquad\quad \mathrm{d}\tau^2 = - g_{\mu \nu}\mathrm{d}x^\mu \mathrm{d}x^\nu \; .
\end{align}
In fact, under very reasonable assumptions, higher derivative corrections to the point-particle action above are more suppressed compared to the PN corrections we are interested in. We will discuss this more explicitly in Sec \eqref{sec: finite size} after introducing the PN parametrization of our EFT coefficients.

In order to simplify the subsequent analysis, it is helpful to perform a conformal redefinition of the metric to set $M_1(r) \equiv \mpl$. This can always be achieved provided the arbitrary functions of $r$ appearing in Eqs. \eqref{S odd} and \eqref{S source} are appropriately redefined. We will also extract an overall factor of $(8 \pi)^{-1}$ from all the bulk coefficients, so that out final action reads
 \begin{align}
\begin{split} \label{eq:action-again}
S_\mathrm{odd} = \frac{1}{8\pi} \int \mathrm{d}^4x& \sqrt{-g} \left[\frac{R}{2} - \Lambda(r) -  f(r) g^{rr} - \alpha(r)\bar K^\mu_{\ \nu} K^\nu_{\ \mu}  \right. \\
&  \quad \left. + M_{10}^2(r) \delta K^\mu_{\ \nu}\delta K^\nu_{\ \mu}  +  M_{12}(r)  \bar K^\mu_{\ \nu} \delta K^\nu_{\ \rho} \delta K^\rho_{\ \mu}+ \dots \right] \ - \ \int \mathrm{d}\tau \mu(r) \; .
\end{split}
\end{align}
The last step will allow us to adopt units where $G=1$---which is particularly convenient since we'll be working in the PN regime---without introducing factors of $8 \pi$ in our equations. The fact that our EFT coefficients differ by a factor of $8\pi$ compared to those in~\cite{Franciolini:2018uyq} means that our Einstein equations with $G=1$ should agree with those in~\cite{Franciolini:2018uyq} with $M_{\rm Pl} = 1$.

 In this section, we will first review such a reduced action, and then introduce a parametrization of the EFT coefficients that is appropriate for the PN regime we are ultimately interested in.

\subsection{Parametrization of the EFT coefficients} \label{sec:param}

Our next step is to parametrize the background metric components in Eq.~\eqref{eq:ss_background}, as well as the arbitrary functions of $r$ appearing in the (Einstein-frame) EFT action~\eqref{eq:action-again}. As discussed in Chapter~\ref{Chapter9}, this action (without the source term) was first used in~\cite{Franciolini:2018uyq} to study black hole quasi-normal modes. In that case, the quasi-normal frequencies were found to depend on the values of the EFT coefficients, the metric components, and their derivatives close to the horizon, {\it i.e.} at $r \sim 2M$. Here, in contrast, we will be interested in the PN regime where $r \gg 2M$.

From now on we choose to work with a radial coordinate such that $c(r) = r$ or, equivalently, such that the surface area of the 2-spheres is $4 \pi r^2$. Then, we can expand the remaining components of the background metric in the PN regime as follows 
\begin{subequations}\label{eq:param_a_b}
\begin{align} 
a^2(r) &= 1 - \frac{2M}{r} + a_2 \left(\frac{2M}{r} \right)^2 + a_3 \left(\frac{2M}{r} \right)^3 + \mathcal{O} \left( \frac{M}{r} \right)^4 \; , \\
b^2(r) &= 1 -   \frac{2M}{r} + b_2 \left(\frac{2M}{r} \right)^2 + b_3 \left(\frac{2M}{r} \right)^3 + \mathcal{O} \left( \frac{M}{r} \right)^4 \; .
\end{align}
\end{subequations}
Here, $M$ is the ADM mass of the black hole, which is defined by the first term in the series expansion of $a(r)$. The coefficients $a_i$ and $b_i$  parametrize instead possible deviations from GR. 
We will see later on that terms up to $1/r^3$ are needed to calculate the waveform at $\mathcal{O}(v^5)$ beyond leading order.

One might think that the first few coefficients in the expansion~\eqref{eq:param_a_b} are already tightly constrained by solar system tests. However, the constraints that apply to the Sun do not necessarily apply to other objects. This is particularly evident in theories where the scalar charge is an independent parameter in addition to the mass---in which case the scalar hair is usually called \emph{primary}. Moreover, even in theories with only \emph{secondary} hair,  Birkhoff's theorem does not necessarily apply, and there might exist more than one branch of solutions characterized by different scalar charge/mass ratios.

The reader might also have noticed that $a$ and $b$ are equal up to ${\cal O}(1/r^2)$. Alternatively, a more general expansion for $b$ could  be considered, of the type 
\begin{equation}
 b^2(r) = 1 -  (1+b_1) \frac{2M}{r} + \dots\, .
\end{equation}
In App.~B of the article associated with this Chapter~\cite{Kuntz:2020yow}, we show that $b_1\neq0$ inevitably leads to violations of the null energy condition (NEC). Theories that violate the NEC are prone to developing instabilities in the scalar sector, although counterexamples also exist (e.g.~\cite{Creminelli:2010ba,Piazza:2013pua}). Considering $b_1\neq0$, while considerably complicating the equations, does not seem to lead to any instability in the present case. We suspect however that this might be because we are considering the odd parity sector which does not include scalar field perturbations. Be that as it may, in this Chapter we specialize to the case where $b_1=0$ and expand the metric coefficients according to~\eqref{eq:param_a_b}. A deeper analysis of this issue would require studying the even sector, and is left for future work.

We now proceed with an analogous expansion for the background coefficients appearing in the EFT action,
\begin{align} \label{exp-1}
\Lambda(r) & =  \frac{2M}{r^3} \left[\Lambda_3 + \Lambda_4 \frac{2M}{r}  + \Lambda_5 \left( \frac{2M}{r} \right)^2 + \dots  \right] \, , \\ \label{exp-2}
f(r) & =  \frac{2M}{r^3} \left[f_3 + f_4 \frac{2M}{r} + f_5 \left(\frac{2M}{r} \right)^2 + \ \dots \ \, \right] \, , \\
\alpha(r) & =  \frac{2M}{r} \left[\alpha_1 + \alpha_2 \frac{2M}{r}  + \alpha_3 \left(\frac{2M}{r} \right)^2 + \ \dots \  \  \label{exp-3} \right] \, , 
\end{align}
Upon use of the  tadpole equations~\eqref{fm}-\eqref{thirdtadpcondition}, the above coefficients can be calculated as polynomials of the metric ones, up to an integration constant that we can choose to be the parameter~$\alpha_1$:

\begin{align}
\Lambda_3  =& -  \alpha_1\, , \qquad \label{3-tadpole1}
\Lambda_4  = \frac{12 a_2 -3 \alpha_1}{4} \, , \\ \label{3-tadpole2}
\Lambda_5  =& \frac{-13 {\alpha_1}+20 {\alpha_1} a_2+24 a_2+30 a_3
-4 {\alpha_1} b_2-8 b_2+6 b_3}{8}\\ \label{3-tadpole3}
 f_3  =& - \alpha_1 \, , \qquad 
 f_4  = \frac{ -7 \alpha_1+4 a_2  + 4 b_2}{4} \, , \\ \label{3-tadpole4}
 f_5 = & \frac{ -27 {\alpha_1}+20 {\alpha_1} {a_2}  
 +4 {\alpha_1} b_2 
+24 {a_2}+6 a_3+8 {b_2}+14 {b_3} }{8}\, ,  \\\label{3-tadpole5}
\alpha_2 =& \frac{6  \alpha_1 -8 a_2}{4} \; , \qquad
\alpha_3 = \frac{\alpha_1(18  - 4  b_2 -12 a_2) - 28 a_2 -18 a_3+4 b_2-2 b_3}{8} . 
 \end{align}

Up to second order in derivatives, the bulk part of the action \eqref{eq:action-again} contains two unknown functions of the radius: $M_{10}(r)$ and $M_{12}(r)$. We will expand also these functions in inverse powers of the radius. A non-zero asymptotic value of $M_{10} (r)$ would induce a non-luminal GW speed at large distances from the black hole. This can be easily seen from the fact that the operator $\delta K_{\mu \nu}$ detunes the radial derivative of $g_{\mu \nu}$ from its temporal derivative (an analogous phenomenon occurs in the EFT of dark energy \cite{Gleyzes:2013ooa}).  Since the recent detection of a nearly coincident GW and electromagnetic signal constrains deviations from an exactly luminal speed of gravity to be smaller than $ \mathcal{O}(10^{-15})$ (see Section~\ref{sec:tests_GW}), we will assume that $M_{10}^2$ starts at order $1/r$ in the PN expansion. It has been argued that the parameters of the EFT of Dark Energy may not be constrained by this measurement~\cite{deRham:2018red}, because the energy scale at which such theory is defined is very different from the typical GW frequency of a LIGO signal. However, our EFT is precisely designed to study GWs and this bound is particularly relevant. A non-zero asymptotic value of $M_{12}$, on the other hand, does not induce a different speed of gravity in flat space. Therefore, we will parameterize our two EFT coefficients as follows:
\begin{align} \label{M10 expansion}
M_{10}^2 &= \gamma_1 \frac{2M}{r} + \gamma_2 \left( \frac{2M}{r} \right)^2 + \mathcal{O} \left( \frac{M}{r} \right)^3 \; ,  \\
\frac{M_{12}}{M} &= \lambda_0 + \lambda_1 \frac{2M}{r} + \mathcal{O} \left( \frac{M}{r} \right)^2 \; , \label{M12 expansion}
\end{align}
where we have expanded each function at the desired order for our calculations as will be clear in Section \ref{sec:perturbative_sol_RW}. Note that, in units such that $G=1$, $M_{10}$ is dimensionless while $M_{12}$ has mass (or length) dimensions. We have normalized $M_{12}$ by the mass of the central black hole because this choice will simplify later equations. However, it should be kept in mind that $M_{12}$ can also depend on microscopic scales that enter the action of any given scalar-tensor theory (scalar Gauss-Bonnet is an example of this, see Section \ref{sec:examples}). Further observational constraints on the size of the expansion coefficients $\gamma_1$, $\lambda_0$ etc. are discussed in the next Subsection.

Finally, there remains an unknown function in the matter sector introduced by the conformal redefinition of the metric to go in the Einstein frame, $\mu(r)$. This conformal redefinition depends on the background scalar field which is itself expanded in a PN series. We can therefore write an expansion of $\mu$:
\begin{equation}
\mu(r) = \mu_0 \left[1 + \mu_1 \frac{2M}{r} + \mu_2 \left(\frac{2M}{r} \right)^2 + \mu_3 \left(\frac{2M}{r} \right)^3 + \mathcal{O} \left( \frac{M}{r} \right)^4  \right] \; , \label{mu expansion}
\end{equation}
where $\mu_0$ is the asymptotic ADM mass of the point-particle. This is reminiscent of the expansion of the mass of an object in terms of the so-called \textit{sensitivities} in BDT theories. Indeed, we will show in App. \ref{app:matching_BD} how the coefficients $\mu_i$ are related to the sensitivities in these theories.

\subsection{Observational constraints from time delays} \label{app:obs}

From eq. \eqref{PN expansion tortoise coordinate}  we could infer that there is an additional Shapiro time delay induced by the parameters $\lambda_0$ and $\gamma_1$ on a GW as compared to a photon. As the recent GW170817 event has remarkably set a stringent bound on the speed of gravitons compared to photons (as discussed in Section~\ref{sec:tests_GW}), let us see the consequences of this measurement in our formalism (even if the GW170817 event concerns two neutrons stars of comparable mass where our perturbative treatment is expected to break down, we are just concerned here by orders of magnitude). Such a violation of the equivalence principle was already used in \cite{Abbott_2017} to constrain the difference $\left \vert \gamma_\mathrm{GW} - \gamma_\mathrm{EM} \right \vert$ between the Eddington parameters of GW and photons respectively.

For an external observer, a photon traveling on a radial geodesic of the metric \eqref{eq:ss_background} has an apparent ``speed''
\footnote{The photon speed is unchanged by the conformal transformation in the matter sector of eq. \eqref{eq:action-again}. However, one should be careful to the fact that the mass appearing in Eq.~\eqref{eq:drdt_phot} is not the physical mass that one would measure with a gravitational experiment. As suggested by the form of Kepler's law \eqref{eq:kepler_chap10} derived later in this Chapter, this physical mass is related to $M$ \textit{via} $M_\mathrm{phys} = M (1-2 \mu_1)$.  }
\begin{equation} \label{eq:drdt_phot}
\frac{dr}{dt} = a b \simeq 1 - 2 \frac{M}{r} + \mathcal{O}\left( \frac{M}{r} \right)^2 \; .
\end{equation}
If the photon starts its trajectory from near the horizon of the central black hole and ends it in a detector on Earth, the Shapiro time delay (defined as the delay between the time of flight of this photon and the time of an equivalent photon traveling in a Minkowski spacetime \cite{Shapiro:1964aa}) is
\begin{equation}
\Delta t_\mathrm{photon} = - 2 M \log \left( \frac{d}{r_s} \right) \; .
\end{equation}
where $d$ is the distance of the black hole to the Earth, and $r_s$ is its Schwarzschild radius. Note that this time is finite while we would expect a true signal emitted near to the horizon to be infinitely redshifted: this is due to the first-order approximation in $1/r$ which we are using. On the other hand, a GW is a solution to the RW equation \eqref{eq:RW_with_source} and its wavelike properties are associated to the coordinate $\tilde r$. From Eq. \eqref{PN expansion tortoise coordinate} the delay for a graviton then reads
\begin{equation}
\Delta t_\mathrm{graviton} = - (2+\lambda_0+2\gamma_1) M \log \left( \frac{d}{r_s} \right) \; .
\end{equation}
For nonzero $\lambda_0$, $\gamma_1$ there is a difference in the time of arrival of photons and gravitons. However, the order of magnitude of this time difference is (in natural units) the Schwarzschild radius of the central object. This is way below the $1.7$s time difference measured in the GW170817 event for a solar mass object, even if log-enhanced by the ratio $d/r_s$; however this could become relevant for a supermassive black hole. We conclude that the current bound on the GW speed is not currently constraining the parameters of our expansion.




\subsection{Finite-size corrections to the point-particle action} \label{sec: finite size}

Let us now return to an issue we alluded to earlier, namely the possibility of including higher derivative corrections in the point-particle action \eqref{S source}. As we already pointed out, the point-particle action should be invariant under the same symmetries as the bulk part of the action. This means that, in unitary gauge, all possible terms invariant under residual diffeomorphisms are allowed. These terms will once again be organized in a derivative expansion, and encode the fact that the object under consideration is not truly point-like. 

For example, all the terms\footnote{Notice that the Lagrangian \eqref{finite size Lagrangian} should in principle contain also powers of $u^r$. However, $u^r =0$ for the circular orbits we are interested in, and therefore we have omitted such dependence at the outset for simplicity.} that could in principle contribute a linear coupling with at most one derivative on the metric perturbation are
\begin{align}
	S &= \int d \tau \Big\{ \mu (r) + c_1 (r)  \delta g^{rr} + c_2 (r) \partial^r \delta g^{rr} + c_3 (r) u^\mu \nabla_\mu \delta g^{rr} + c_4 (r) \delta K \nonumber \\
	& \qquad \qquad \qquad \qquad \qquad \qquad \qquad \qquad  + c_5 (r) \bar K^\mu{}_\nu \delta K^\nu{}_\mu + c_6 (r) u_\mu u^\nu \delta K^\mu{}_\nu + \cdots \Big\} . \label{finite size Lagrangian}
\end{align}
It turns out that only the very last term in this Lagrangian yields an additional linear coupling with odd perturbations. Using the asymptotic expansions \eqref{eq:param_a_b} and \eqref{mu expansion}, as well as the definition of the angular momentum $L$ in \eqref{eq:def_E_L}, we find that for a circular orbit of radius $r_0$ this last term is schematically of the form
\begin{align}
	u_\mu u^\nu \delta K^\mu{}_\nu \sim \frac{L}{\mu_0 r_0^3} \times \left\{ \begin{array}{c} h_0 \\ h_1 \end{array} \right\}
\end{align}
at leading order in a $1/r_0$ expansion. Thus, assuming that the dimensionless quantity $c_6 (r_0) \sim \mathcal{O}(1)$, as would expected on naturalness grounds, we find that this last term is suppressed compared to the leading interaction \eqref{leading interaction with source} by a factor of $(r_0 \mu_0)^{-1}$. In turn, this correction is negligible compared to the PN corrections we will consider below provided 
\begin{align}
	\frac{1}{r_0 \mu_0} \ll \frac{G M}{r_0} \qquad \Longrightarrow \qquad G M \mu_0 \gg 1 ,
\end{align}
which is certainly satisfied for ``macroscopic'' masses $M$ and $\mu_0$. Thus, in what follows we will neglect finite-size corrections to the point-particle action.

\section{Examples} \label{sec:examples}

To see how the formalism described in the previous sections works in practice, in what follows we write some covariant theories with scalar hair in unitary gauge and spell out the values of the coupling functions of action~\eqref{eq:action-again} in these cases.

\subsection{Brans-Dicke type theories}

As repeatedly emphasized in this thesis, scalar-tensor theories \emph{\`a la} Brans Dicke are the simplest playground for modifications of gravity. Black holes cannot display scalar charges in these theories as a consequence of a known and early example no-hair theorem discussed in Section~\ref{subsec:hawking_th}.
However, a neutron star is expected to develop a nonzero scalar charge which can be further enhanced by spontaneous and dynamical scalarization \cite{Damour:1993aa, Barausse_2013}. Of course, we do not expect the waveforms that we will obtain by matching our EFT to a BDT theory to be relevant in the case of a binary neutron star inspiral where the two components are approximately of the same mass; however, our results could provide a useful cross-check with the existing PN literature \cite{Lang:2015aa, Bernard:2018hta, Bernard:2018ivi}. Indeed, our waveform should be recovered in the extreme mass ratio limit of these references.
 Futhermore, as was pointed out in~\cite{Cardoso_2013}, a black hole will develop a nonzero scalar charge in realistic astrophysical situations when it is surrounded by matter.

Incidentally,  theories of the Brans-Dicke type provide a nice example of the power of the formalism outlined in the previous sections. For historical reasons, they are often introduced by means of the rather inconvenient action
\begin{equation} \label{eq:bd-action}
S = \frac{1}{16 \pi} \int \mathrm{d}^4x \sqrt{-g} \left[ \phi R - \frac{\omega(\phi)}{\phi} g^{\mu \nu} \partial_\mu \phi \partial_\nu \phi \right] + S_m  \; ,
\end{equation}
which is of the form presented in Chapter \ref{Chapter1}, Eq.~\eqref{eq:action_BD}. 
However, the precise coefficient in front of the Ricci scalar and the functional form of $\omega (\phi)$ are not unambiguously defined, since they can be changed by conformal transformations of the metric and scalar field redefinitions.
On the contrary, by working in the Einstein frame and in the unitary gauge, our EFT~\eqref{eq:action-again} is free from these ambiguities. 

With an appropriate field redefinition of the metric (\emph{i.e.} by going to the Einstein frame) and of the scalar field, the action~\eqref{eq:bd-action} takes the form (see App.~\ref{app:matching_BD} for details and for connection with more standard notation) 
\begin{align} \label{eq:bd_action_EinsteinFrame}
S = \frac{1}{16 \pi} \int \mathrm{d}^4x \sqrt{- \tilde g} \left[ \tilde R -   \tilde g^{\mu \nu} \partial_\mu \varphi \partial_\nu \varphi \right] 
 - \sum_A \int \mathrm{d} t \; m_A(\varphi) \sqrt{-  \tilde g_{\mu \nu} v^\mu_A v^\nu_A} \; ,
\end{align}
where the index $A$ refers to the different point-particle objects,  $v^\mu_A = d x^\mu_A / dt$ is the velocity of each object, and $m_A(\varphi)$ is a field-dependent mass. 
From now on we will drop the tildes for notational convenience. 

For a background field $\bar \varphi(r)$ around a single object of mass $m(\varphi)$, the EFT coefficients in the action \eqref{eq:action-again} are easily identified. Only $f$ and $\mu$ are nonzero and they read
\begin{align}
\begin{split} \label{eq:EFT_coeffs_ST}
f(r) &= \frac{ \bar \varphi'^2(r)}{2} \; , \\
\mu(r) &= m(\bar \varphi(r)) \; .
\end{split}
\end{align}
 In order to obtain a PN expansion of these functions, the only remaining task is to find the background value $\bar \varphi$. By varying the action with respect to $g_{\mu \nu}$ and $\varphi$ one finds the following equations of motion in vacuum
\begin{align}
\begin{split}
&R_{\mu \nu} = \partial_\mu \varphi \partial_\nu \varphi \; , \\
&\partial_\mu \left(\sqrt{-g} g^{\mu \nu}  \partial_\nu \varphi  \right) = 0 \; .
\end{split}
\end{align}
Although a exact solution of these equations is known in the so-called Just coordinates \cite{damour_tensor-multi-scalar_1992}, we find it more convenient to solve them perturbatively for large values of $r$ in the standard coordinate system of Eq.~\eqref{eq:ss_background} with $c(r) = r$. Plugging in our spherically symmetric ansatz for the metric \eqref{eq:ss_background} and the background field $\bar \varphi(r)$, we find the following system of equations for the three unknown functions $a$, $b$ and $\bar \varphi$:
\begin{align}
\begin{split} \label{eq:system_ST}
 a'' a  + \frac{a' b' a}{b} + \frac{2}{r} a a' &= 0 \; , \\
\frac{a''}{a} + \frac{a' b'}{a b}  + \frac{2b'}{rb} &= - \bar \varphi'^2 \; , \\
r \left( \frac{a' b^2}{a} + b b' \right) + b^2 - 1 &= 0 \; , \\
\partial_r \left(r^2 a b  \bar \varphi'  \right) &= 0 \; .
\end{split}
\end{align}
 The last equation is immediately integrated to find
\begin{equation} \label{eq:def_scalar_charge}
\bar \varphi = \varphi_0 - \int \mathrm{d}r \; \frac{q M}{r^2 a b} \; ,
\end{equation}
where $\varphi_0$ is the asymptotic value of the field, $M = m(\varphi_0)$ and $q$ is called the \textit{scalar charge} of the object. While for a test-mass the scalar charge can be easily deduced from the coupling function $\omega(\phi)$, for a compact object like a neutron star, instead, one should resort to a specific neutron star model encapsulating short-distance physics. In this case, $q$ can grow to appreciable values due to spontaneous scalarization \cite{Damour:1993aa}.

In order to avoid these complications, we choose to adopt $q$ as an independent parameter. Note that in this way we are being more general than post-Newtonian dynamics which perturbatively relates the scalar charge $q$ to the two free functions of the theory $\omega(\phi)$ and $m_A(\phi)$ in a weak-field approximation. In contrast, we do not rely on any weak-field approximation to evaluate the scalar charge $q$. For completeness, we include the computation of the scalar charge in the weak-field approximation in App. \ref{app:matching_BD}.

 The remaining task is now to perturbatively solve the three first equations of \eqref{eq:system_ST}. We parametrize $a$ and $b$ as in eq. \eqref{eq:param_a_b}.  It is easy to show that, to be consistent with the order of expansion we are dealing with, one should expand the two first equations to $\mathcal{O}(1/r^6)$ and the third one to $\mathcal{O}(1/r^4)$ (excluded). We obtain
\begin{align}
\begin{split}
a_2 = 0, \quad b_2 = \frac{q^2}{8}, \quad b_3 = \frac{q^2}{16}, \quad a_3 = \frac{q^2}{48}\, .
\end{split}
\end{align}
The above equations allow to find also the expansion of $\bar \varphi$
\begin{equation} \label{eq:expansion_barphi}
\bar \varphi = \varphi_0 +\frac{M q}{r} +\frac{M^2 q}{r^2} -\frac{M^3 q \left(q^2-16\right)}{12r^3} + \dots \; .
\end{equation}
But this is not really needed because the coefficients of the effective action are now given through the tadpole equations. In particular, by setting $\alpha_1=0$ we find by using~\eqref{3-tadpole3} and~\eqref{3-tadpole4} that
\begin{equation} \label{eq:f_bd}
f_4 \ = \ \frac{q^2}{8}, \qquad f_5 \ = \ \frac{q^2}{4}\, .
\end{equation}
As expected, the other background coefficients $\Lambda$ and $\alpha$ vanish. To sum up our discussion, in BDT theories the only independent parameters of our expansion are the scalar charge $q$ (which depends on the short-distance physics model of neutron star considered) and the PN expansion of the mass function $\mu(r)$. Without any further restriction on the form of $\mu(r)$, a new independent parameter appears at each PN order which is certainly problematic from the point of view of data analysis as pointed out in the introduction. Thus, we expect that phenomenologically viable BDT waveforms can only be obtained if one restricts to a particular form of the function $\mu(r)$.

While referring to App.~\ref{app:matching_BD} for more connections with the existing literature, here we point out that the energy flux for brans-dicke type scalar-tensor theories is known up to the 1PN order \cite{Lang:2015aa} (although the 2PN order is underway \cite{Bernard:2018hta, Bernard:2018ivi}). As already mentioned, when the symmetric mass ratio $\nu = m_1 m_2/(m_1+m_2)^2$ is negligible, we should recover the results of Lang \cite{Lang:2015aa}. However, since we are still missing the even part of the spectrum, we are not able to do this comparison for the time being. We hope to come back to it in a near future.

\subsection{Gauss-Bonnet}

The linear Gauss-Bonnet (GB) scalar tensor model is defined by the following action
\begin{equation} \label{eq:GB-action}
S = \frac{M_P^2}{2} \int d^4 x \sqrt{-g}\left(R - \partial^\mu \varphi \partial_\mu \varphi  +  2 \bar \alpha  \, \varphi \, {\cal G}\right)\, ,
\end{equation}
where ${\cal G}$ is the GB total derivative term,
\begin{equation}
{\cal G} \ =\  R^{\mu \nu \rho \sigma}R_{\mu \nu \rho \sigma} - 4 R^{\mu \nu} R_{\mu \nu} + R^2 \, .
\end{equation}
Following the conventions of~\cite{Sotiriou_2014}, the scalar $\varphi$ is dimensionless here, and the coupling of the GB term has been denoted with a bar to distinguish it from the coefficient in the EFT action~\eqref{eq: effective action for perturbations}. Note that EMRI in the closely related Einstein-dilaton-Gauss-Bonnet and Chern-Simons theories were respectively considered in~\cite{2016PhRvD..94j4024B} and~\cite{Pani_2011}.


It is a rather cumbersome task to write the action of this theory in the EFT form~\eqref{eq: effective action for perturbations}. In what follows we describe the basic steps that we made in order to obtain this. First, it proves useful to exploit the equivalence of GB gravity and Horndeski theory~\cite{Kobayashi:2011nu}, according to which the GB term in~\eqref{eq:GB-action} is equivalent, up to boundary terms, to the Horndeski-5 action 
\begin{equation} 
 \varphi \, {\cal G} \ \simeq \ G_5(X) G_{\mu \nu} \varphi^{; \mu \nu} + \frac13 G'_{5}(X)(\Box\varphi^3 - 3 \Box \varphi \varphi_{; \mu \nu} \varphi^{;\mu \nu} + 2  \varphi_{; \mu \nu} \varphi^{;\nu \rho} \varphi_{; \rho}^{;\mu})
 \end{equation}
with $G_5(X) = - 4 \bar \alpha \ln(|X|)$. In the above, $X \equiv \partial_\mu \varphi \partial^\mu \varphi$ and $G_{\mu \nu}$ is the Einstein tensor. At this point, we can try adapt the vocabulary of~\cite{Gleyzes:2013ooa} to the present spherically symmetric case in order to write the Horndeski-5 Lagrangian in the EFT form 
~\eqref{eq:GB-action}. To start with, let us focus on the coefficient $M_1^2$  multiplying the Einstein Hilbert term. For the theory at hands it is relatively easy to obtain 
\begin{equation} \label{eq:M1}
M_1^2(r) \ = \ M_P^2\ \left[1 - 8 \bar\alpha b\left(b \varphi' \right)'\right]\, .
\end{equation}

The authors of~\cite{Sotiriou_2014} have studied a spherically symmetric black hole solution of the above theory and given the asymptotic behavior at large $r$ of the metric coefficients and of the scalar itself, 
\begin{align} 
a^2(r)&= 1 -\frac{2M}{r} + \frac{M P^2}{6 r^3} +\frac{M^2 P^2+24 \bar \alpha M P }{3 r^4} + {\cal O}(r^{-5})\, , \\ 
b^2(r)&= 1 -\frac{2M}{r} +\frac{P^2}{2 r^2} +\frac{M P^2}{2 r^3} + \frac{48 \bar \alpha M P + 2 M^2 P^2}{3 r^4} + {\cal O}(r^{-5})\, , \\ \label{metricso-3}
c^2(r) & = r^2 \\
\varphi(r)& = \frac{P}{r} +\frac{M P}{r^2} +\frac{16 M^2 P-P^3}{12 r^3} +\frac{6 M^3 P -12 \bar \alpha M^2-M P^3}{3 r^4}+ {\cal O}(r^{-5}) \, .
\end{align}
In the above, $P$ is the scalar charge of the black hole, which however is not independent of its mass~\cite{Sotiriou_2014}, and~\eqref{metricso-3} is just our standard gauge fixing condition. 

The above expressions can now be used inside~\eqref{thirdtadpcondition} in order to get an expansion for $\alpha(r)$. Then, by using~\eqref{fm} and~\eqref{lm} we can solve for the remaining tadpole coefficients, $f(r)$ and $\Lambda(r)$. In summary, the effective action coefficients read
\begin{align} \label{M1q-ex}
M_1^2(r) & = 1 - \frac{16 \bar \alpha P}{r^3} - \frac{8 \bar \alpha M P}{r^4} - \frac{4 \bar\alpha(4 M^2 P + P^3)}{r^5} + {\cal O}(r^{-6}) \\ \label{fr-ex}
f(r) & = \frac{P^2}{2 r^4} +\frac{96 \bar \alpha P+2 M P^2}{r^5}+  \frac{240 \bar \alpha M P+24 M^2 P^2-P^4}{4 r^6}+{\cal O}(r^{-7}) \, ,\\ \label{La-ex}
\Lambda(r) & = -\frac{48 \bar \alpha P}{r^5} + \frac{44 \bar \alpha M P}{r^6} + {\cal O}(r^{-7})\, ,\\ \label{alpha-ex}
\alpha(r) & = \frac{24  \bar \alpha P}{r^3} + \frac{8  \bar \alpha M P}{r^4} + {\cal O}(r^{-5})\, ,
\end{align}

This is not the end of the story however, because all the quantities above are ``Jordan frame'' quantities. We can always perform a conformal transformation of the metric tensor and bring the action to the form~\eqref{eq:action-again}, with no radius-dependent coefficient multiplying the Einstein Hilbert term. This is achieved by the field redefinition $g_{\mu \nu}^{\rm (J)}(x) \rightarrow g_{\mu \nu}^{\rm (E)}(x) = g_{\mu \nu}^{\rm (J)}(x) M_1^2(r)$, where $J$ and $E$ stand for Jordan- and Einstein- frames. One can check that equation~\eqref{thirdtadpcondition} is covariant under such a conformal transformation, provided that $\alpha$ transforms as $\alpha^{(J)}(r)\rightarrow \alpha^{(E)}(r) = \alpha^{(J)}(r)/ M_1^2(r)$. Of course, the radial gauge choice $r^2 \equiv g_{\theta\theta} = c(r)$ cannot hold after the conformal transformation. Once we conformally transform to the Einstein frame, we can introduce an ``Einstein frame radius'' $r^{(E)} =  g_{\theta\theta} ^{(E)}$ so that the radial gauge choice still holds in the new frame.

However, from the expansion~\eqref{M1q-ex}, it is clear that the conformal factor $M_1^2$ is different from unity only at a relatively high order in the $1/r$ expansion, which make the Einstein- and Jordan- frame radii differ from each other only to relative order $r^{-3}$. 
As a result, the expressions of the coefficients $f$, $\Lambda$ and $\alpha$ given in~\eqref{fr-ex},~\eqref{La-ex} and~\eqref{alpha-ex} are already valid also in the Einstein frame. 

Finally, by explicitly translating the Horndeski 5 action into the EFT formalism we get the two quadratic operators that are relevant for odd perturbations, 
\begin{align} 
M_{10}^2(r) &=  8 \bar \alpha b^2 \varphi'\left(\frac{a'}{2 a} -  \frac{b'}{b} +  \frac{c'}{c} -  \frac{\varphi''}{\varphi'}\right) \\ \nonumber
& = -\frac{24 \bar \alpha P}{r^3} -\frac{12 \bar \alpha M P}{r^4} -\frac{6 \bar \alpha \left(4 M^2 P+P^3\right)}{r^5} + {\cal O}(r^{-6})\, , \\[2mm]
M_{12}(r) & = - 8 \bar \alpha b \varphi' \\ \nonumber
&= \frac{8 \bar \alpha P}{r^2} +\frac{8 \bar \alpha M P}{r^3} + \frac{12 \bar \alpha M^2 P}{r^4}+{\cal O}(r^{-5})\, .
\end{align}

\section{Circular orbits and the sourced Regge-Wheeler equation} \label{sec:RW}

In the extreme mass ratio regime, GWs emitted during the inspiral phase can be thought of as arising from perturbations of the heavy companion generated by a point-like source~\cite{Sasaki:1994aa, Poisson:1993aa}. We can then constrain the presence of a scalar hair for the heavy companion by using the effective theory introduced in the previous section. To this end, we will first discuss some aspects of trajectories in a generic spherically symmetric background, and then derive the Regge-Wheeler equation with a source term.

\subsection{Background trajectories} \label{sec:background_traj}

Before turning our attention to the dynamics of linear perturbations, we will pause for a moment to discuss some features of circular orbits in generic spherically-symmetric backgrounds. The results derived in this section will be used later on to simplify the linear coupling between perturbations and point-like source. Since the metric \eqref{eq:ss_background} is symmetric around $\theta = \pi/2$, we will consider trajectories that are restricted to this plane. Remember that we are working with a radial coordinate such that $c(r) = r$.

The equations of motion for the point-particle are found by varying the background action,
\begin{equation} \label{eq:matter_action}
\bar S_m = - \int \mathrm{d} \lambda \; \mu(r) \sqrt{\bar g_{\mu \nu} \frac{dx^\mu}{d\lambda} \frac{dx^\nu}{d\lambda}} \; ,
\end{equation}
where we have reintroduced an affine parameter $\lambda$ along the trajectory for convenience.  Denoting $T \equiv \mu(r) \sqrt{\bar{g}_{\mu \nu} \dot{x}^\mu \dot x^\nu}$ and $\dot{x}^\mu \equiv dx^\mu/d\lambda$, the equations of motion take the form
\begin{equation}
\frac{d}{d\lambda} \frac{\partial T}{\partial \dot x^\mu} = \frac{\partial T}{\partial x^\mu} \; .
\end{equation}
Choosing now the affine parameter $\lambda$ to be equal to the background proper time $\tau$, and using the fact that $T$ does not depend explicitly on $t$ nor on $\phi$, we obtain the two equations
\begin{align}
\frac{d}{d\tau} \left(\mu(r) a^2(r) \frac{dt}{d\tau} \right) = 0 \qquad \qquad 
\qquad \frac{d}{d\tau} \left(\mu(r) r^2 \frac{d\phi}{d\tau} \right) = 0 \; .
\end{align}
The conserved quantities in parentheses are respectively the energy and the angular momentum of the point-particle:
\begin{equation} \label{eq:def_E_L}
E = \mu(r) a^2(r) \frac{dt}{d\tau}, \qquad\qquad\qquad L = \mu(r) r^2 \frac{d\phi}{d\tau} \; .
\end{equation}
Using $d \tau^2 = - \bar{g}_{\mu \nu} dx^\mu dx^\nu$, one finds that the radial component of the equations of motion can be written using the conserved quantities above as
\begin{equation}\label{1st order eq circular orbits}
\left( \frac{dr}{d\tau} \right)^2 = b^2 \left[ \frac{E^2}{\mu^2 a^2} - \left(1+\frac{L^2}{\mu^2 r^2} \right)  \right] \; .
\end{equation}
Deriving now Eq. \eqref{1st order eq circular orbits} with respect to $\tau$, we find the second order equation
\begin{equation}  \label{2nd order eq circular orbits}
2 \frac{d^2r}{d\tau^2} = \frac{d}{dr}\left[ \frac{E^2 b^2}{\mu^2 a^2} - b^2 \left(1+\frac{L^2}{\mu^2 r^2} \right)  \right] \; .
\end{equation}

In the particular case of a circular trajectory, for which $dr/d\tau = d^2r/d\tau^2 = 0$, we can solve Eqs. \eqref{1st order eq circular orbits} and \eqref{2nd order eq circular orbits} to find the energy and the angular momentum of the particle as a function of the background parameters evaluated at the radius of the orbit:
\begin{align} \label{eq:L_E_background}
L = \mu r \; \sqrt{\frac{r(a \mu)'}{\mu(a-r a')}} , \qquad \qquad \qquad E = \mu a \; \sqrt{\frac{\mu a+r a \mu'}{\mu(a-r a')}} \; ,
\end{align}
where $( \,\, )'\equiv d/dr ( \,\, )$. It is easy to check that these results reduce to the usual expressions for angular momentum and energy of a non-relativistic point particle when $\mu(r) = m$, $a(r)^2 = 1 - 2 M/r$, and $M/r = v^2 \ll 1$  . It is also interesting to notice that, generically, there will be corrections to Kepler's law (we will come back to this point in Section \ref{sec:last_part}). Indeed, from Eqs. \eqref{eq:def_E_L} and the definition of the angular frequency $\Omega = d\phi/dt$ we find the relation
\begin{equation} \label{Omega eq.}
\Omega = \frac{d\phi}{dt} = \frac{L a^2}{E r_0^2} \; ,
\end{equation}
with $r_0$ the radius of the orbit. In GR, it is easily checked that $\Omega^2 r_0^3 = M$ for a Schwarzschild solution in coordinates such that $4 \pi r^2$ is the area of the invariant 2-spheres. In general, this will no longer be true in the presence of a scalar hair.

\subsection{Sourced Regge-Wheeler equation} \label{sec:RW_with_source}

We are finally in a position to derive the linearized equation for odd metric perturbations sourced by a test particle. In what follows, we will work in Regge-Wheeler gauge by setting $h_2 =0$ (see eq.~\ref{odd metric perturbations}). Perturbations with different values of $\ell$ and $|m|$ decouple at linear level due to rotational invariance. Therefore, we will focus on a single $(\ell, |m|)$ sector and suppress the angular momentum labels whenever possible to simplify the notation. By expanding the bulk part of the effective action \eqref{eq:action-again} up to quadratic order in perturbations we find 
\begin{align} \label{bulk action 1}
	S_\mathrm{bulk} &= \frac{1}{8\pi} \int \mathrm{d} t \mathrm{d} r \left[u_1 |h_0|^2 + u_2 |h_1|^2 + u_3( |\dot h_1|^2 - 2 \dot h_1^* h_0' + |h_0'|^2 + 2u_4\dot h_1^* h_0) + \text{c.c.} \right],
\end{align}
with the understanding that for $m=0$ one should add an overall factor of $1/2$ to avoid overcounting.
The real functions $u_i(r)$ were calculated in~\cite{Franciolini:2018uyq}, and are reproduced here in Appendix~\ref{app:def_complicated_functions} for completeness. 

The bulk action should be supplemented with the point-particle action expanded up to linear order in perturbations. Considering again a circular trajectory in the $\theta = \pi /2$ plane with radius $r_0$ and  angular frequency $\Omega$, and choosing the affine parameter so that $\lambda = t$, we find the following expression for the source action

\begin{align} \label{leading interaction with source}
	S_\mathrm{source} &= - \frac{s_{\ell m} L}{r_0^2} \int \mathrm{d} t \; h_0 (t,r_0) e^{i m \Omega t} + \text{c.c} ,
\end{align}
where $L$ is the angular momentum defined in Eq. \eqref{eq:def_E_L}, and the coupling constant $s_{\ell m}$ is given by
\begin{align} \label{PPP}
	s_{\ell m} =\, \sqrt{\frac{2\ell +1}{4\pi} \frac{(\ell-m)!}{(\ell +m)!}} \, P_\ell^{m+1} (0) \ ,
\end{align}
with $P_\ell^{m+1}$ an associated Legendre polynomial.

The perturbation $h_0$ never appears with a time derivative  in the total action $S=S_{\rm bulk} + S_{\rm source}$. Therefore, it can be integrated out by solving its equation of motion, which is just a constraint equation. This equation is however a second-order ordinary differential equation in the $r$ variable. To overcome this difficulty, we will follow Refs.~\cite{Franciolini:2018uyq, De_Felice_2011} and introduce an auxiliary variable $q(t,r)$ defined by
\begin{equation} \label{eq:def_q}
q = \dot h_1 - h_0' + u_4 h_0 \; ,
\end{equation}
and rewrite the total action as
\begin{align}
S &= \frac{1}{8\pi} \int \mathrm{d} t \mathrm{d} r \biggr\lbrace (u_1 - \partial_r(u_3 u_4) - u_3 u_4^2) |h_0|^2 + u_2 |h_1|^2 +  u_3 q^*\left[2 (\dot h_1 - h_0' + u_4 h_0) - q \right] \biggr\rbrace  \nonumber \\
& \qquad \qquad \qquad \qquad  \qquad \qquad \qquad \qquad \qquad \ - \frac{s_{\ell m} L}{r_0^2} \int \mathrm{d} t \; h_0 e^{i m \Omega t} + \text{c.c.} \; . \label{total action 1}
\end{align}
It is easy to show that the bulk part of the action is equivalent to the one in Eq.~\eqref{bulk action 1} after integrating out~$q$. Varying instead with respect to $h_0^*$ and $h_1^*$ one obtains the algebraic constraints
\begin{equation}
h_0 = \frac{\partial_r(u_3 q) + u_3 u_4 q}{\partial_r(u_3 u_4) + u_4^2 u_3 - u_1} + A(r) \delta(r-r_0) e^{im \Omega t} \; ,\qquad \qquad \quad h_1 = \frac{u_3}{u_2} \dot q \; . \label{sols 0 h1}
\end{equation}
Once again $r_0$ is the radius of the circular orbit, whereas $A(r)$ is defined by
\begin{equation}
A(r) =  \frac{8 \pi s_{\ell m} L}{2r^2(u_1-\partial_r(u_3 u_4) - u_4^2 u_3)} \; .
\end{equation}
Plugging the solutions \eqref{sols 0 h1} for $h_0$ and $h_1$ into the action \eqref{total action 1}, we obtain (up to an irrelevant divergent constant) the effective action
\begin{align}
S &= \frac{1}{8\pi} \int \mathrm{d} t \mathrm{d} r \left[ \mathcal{G}_{00} \dot q^2 + \mathcal{G}_{rr} q'^2 + \mathcal{G}_{qq} q^2 \right] \nonumber  \; \\
& \qquad \qquad \qquad \qquad \quad + \frac{1}{4\pi} \int \mathrm{d}t \mathrm{d}r \; u_3 \left [ (u_4 A - A')\delta(r-r_0) - A \delta'(r-r_0) \right ] q \, e^{i m \Omega t} , \label{q action}
\end{align}
where the explicit expressions of the $\mathcal{G}$ functions are given in Appendix \ref{app:def_complicated_functions}. By varying the quadratic action above with respect to $q$ we obtain a second order equation for the only propagating degree of freedom in the odd sector: 
\begin{equation}
- \frac{\mathcal{G}_{rr}}{\mathcal{G}_{00}} q'' - \frac{\partial_r \mathcal{G}_{rr}}{\mathcal{G}_{00}} q' + \left(\omega^2 + \frac{\mathcal{G}_{qq}}{\mathcal{G}_{00}} \right) q = -  \frac{u_3}{\mathcal{G}_{00}} \left [ (u_4 A - A')\delta(r-r_0) - A \delta'(r-r_0) \right ] 2 \pi \delta(\omega - m \Omega) \; , \label{eq. odd perts 1}
\end{equation}
where we have transformed to Fourier space (in time) with the convention $h(\omega) = \int \mathrm{d} t \; e^{i \omega t} h(t)$. This equation can be recast in a Schr\"odinger-like form by rescaling $Q$ and redefining the radial coordinate as follows:
\begin{subequations}
\begin{align}
 \Psi &= ( \vert \mathcal{G}_{rr} \mathcal{G}_{00} \vert)^{1/4} q \label{Psi def} \\
 \frac{d}{d\tilde r} &= \sqrt{\left \vert \frac{\mathcal{G}_{rr}}{\mathcal{G}_{00}} \right \vert} \frac{d}{dr} \; \label{eq:def_tortoise_coord} ,
\end{align}
\end{subequations}
Then, Eq. \eqref{eq. odd perts 1} reduces to the Regge-Wheeler equation
\begin{equation} \label{eq:RW_with_source}
\frac{d^2 \Psi}{d \tilde r^2} + \left[\omega^2 - V(\tilde r)\right] \Psi = J(\tilde r) \; ,
\end{equation}
with
\begin{align}\label{eq:source_term}
J &= 2 \pi \delta(\omega - m \Omega) u_3 \frac{ \vert \mathcal{G}_{rr} \mathcal{G}_{00} \vert^{1/4}}{\mathcal{G}_{00}} \left[ (A'-u_4A) \delta(r-r_0) + A \delta'(r-r_0)\right] \; , \\
V &= \frac{5 \mathcal{G}_{00}'{}^2
   \mathcal{G}_{rr}{}^2-\mathcal{G}_{00}{}^2 \left(16
   \mathcal{G}_{qq}
   \mathcal{G}_{rr}-\mathcal{G}_{rr}'{}^2+4
   \mathcal{G}_{rr}
   \mathcal{G}_{rr}''\right)-2 \mathcal{G}_{00}
   \mathcal{G}_{rr}
   \left(\mathcal{G}_{00}'
   \mathcal{G}_{rr}'+2
   \mathcal{G}_{00}''
   \mathcal{G}_{rr}\right)
   }{1
   6 \mathcal{G}_{00}{}^3
   \mathcal{G}_{rr}} \ . \label{eq:def_RW_potential}
\end{align}

\section{Power emitted in the PN regime} \label{sec:perturbative_sol_RW}

In this section we derive the main result of this Chapter: a PN expansion of the power emitted in the odd sector by an extreme mass-ratio binary in the presence of a scalar hair. Our approach will be similar to the one first developed in~\cite{Poisson:1993aa} and then used in~\cite{Sasaki:1994aa, Tagoshi_1994} to calculate the power dissipated by a particle in circular orbit around a Schwarzschild black hole  up to the 4PN order. We organize our discussion in several steps: first, we show how the power emitted is related to the form of $\Psi$ far away from the source (Sec. \ref{sec:GW_fluw}); second, we review how the asymptotic form of $\Psi$ is related to the homogeneous solution $\Psi^\mathrm{in}$ satisfying ingoing boundary conditions at the horizon (Sec. \ref{subsec:setup}); third, we calculate $\Psi^\mathrm{in}$ in a PN expansion (Sec. \ref{sec-6.2}); and finally, we assemble all our results to obtain a PN expansion of the power emitted (Sec. \ref{sec:last_part}).

\subsection{Dissipated power from asymptotic solution} \label{sec:GW_fluw}

We will start by expressing the energy flux leaving the binary system in terms of the variable $\Psi$ defined by Eqs. \eqref{eq:def_q} and \eqref{Psi def}. To this end, we will make the simplifying assumption that our quadratic action for perturbations reduces to the one we would derive from GR at large distances. This assumption is certainly well supported by present observations---{\it e.g.}, by the current bounds on the luminal propagation of speed of GWs~\cite{Creminelli:2017sry}---and it can be translated into a bound on the asymptotic behavior of the coefficients in the effective action \eqref{eq:action-again}: both $M_{10}(r)$ and $M_{12}(r)/r$ must vanish at large~$r$.  As we will see in Sec.~\ref{sec:examples}, this assumption is also easily satisfied by known black hole solutions with scalar hair.

We can now combine the explicit expressions provided in App.~\ref{app:def_complicated_functions} together with the PN expansions \eqref{eq:param_a_b} to show that at large $r$
\begin{equation} \label{estimates}
u_4 \to \frac{2}{r}, \qquad\qquad\quad   ( \vert \mathcal{G}_{rr} \mathcal{G}_{00} \vert)^{1/4} \to \sqrt{\frac{\ell(\ell+1)}{(\ell-1)(\ell+2)}} \frac{r}{2} \; ,
\end{equation}
and therefore that our variable $\Psi$ reduces asymptotically to 
\begin{equation}
	\Psi \to \frac{r}{(\ell+2)(\ell -1)} \, [ \dot h_1 - r^2 \partial_r(h_0/r^2)] \ .
\end{equation}
Up to an overall multiplicative factor that only depends on $\ell$, this  result is equal to the usual Regge-Wheeler variable, see {\it e.g.} ~\cite{Nagar:2005ea}. Thus, we can immediately borrow standard results to relate the asymptotic form of $\Psi$ to the usual $+$ and $\times$ polarizations in flat space. Keeping track of the aforementioned overall factor, we have~\cite{Nagar:2005ea}
\begin{equation} \label{h plus cross to Psi}
h_+ - i h_\times = \frac{2 i}{r} \; _{-2} Y_{\ell m} (\theta, \phi) \Psi \; ,
\end{equation}
where $ \; _{-2} Y_{\ell m}$ are the $s=-2$ spin-weighted spherical harmonics. Finally, because the asymptotic action for $h_+$ and $h_\times$ is the same as in GR, so will be the asymptotic form of the stress-energy tensor of perturbations. Thus, the total instantaneous power emitted takes the usual form in terms of $h_{+,\times}$~\cite{Maggiore:1900zz}:
\begin{align} \label{power emitted general}
	P \ = \ \lim_{r \to \infty} \int d\Omega \, \frac{r^2}{16 \pi} ( \dot h_+^2 +  \dot h_\times^2) \ = \ \lim_{r \to \infty} \ \frac{1}{4\pi} \ \sum_{\ell \geq 2} \sum_{m=-\ell}^\ell \left \vert \frac{d \Psi}{dt} \right \vert^2  ,
\end{align}
where in the last step we used the orthonormality of spin-weighted spherical harmonics and assumed that $h_+ - i h_\times$ is a linear superposition of modes with all possible values of $\ell, m$.


We will use this result later in this section to calculate the total power emitted in the odd sector in a PN expansion. For now, we just point out that the corresponding expression in the even sector would be more complicated because it would need to include also the power emitted by the additional  scalar mode. We leave a full investigation of the even sector for future work.

\subsection{Asymptotic solution from homogenous solution} \label{subsec:setup}

Equation \eqref{power emitted general} means that the power emitted is completely determined by the asymptotic form of the solution $\Psi$ to the inhomogeneous equation \eqref{eq:RW_with_source}. Following~\cite{Sasaki:1994aa}, we will now show how the latter can in turn be expressed in terms of one particular solution to the associated homogeneous equation. 

As long as the ``potential'' $V(\tilde r)$ vanishes at the horizon ($\tilde r \to - \infty$) and at spatial infinity ($\tilde r \to + \infty$), $\Psi$ will asymptotically approach a linear combination of complex exponentials $e^{\pm i \omega \tilde r}$ (notice that $J(\tilde r)$ vanishes at both ends). On physical grounds, we will impose ingoing boundary conditions at the horizon---corresponding to the fact that no classical signal can leave the horizon---and outgoing boundary conditions at spatial infinity---since we are not interested in gravitational radiation produced by other, far-away sources. Using the Green's functions method, we can express such a solution as follows: 
\begin{align} \label{Green function solution}
\Psi(\tilde r) &= \int_{-\infty}^\infty d\tilde r' G(\tilde r, \tilde r') J(\tilde r') \; . 
\end{align}
The Green's function $G$ with the correct boundary conditions is in turn equal to
\begin{align} \label{Feynman Green's function}
	G(\tilde r,\tilde r') &= \frac{1}{W}  \left[ \theta(\tilde r-\tilde r') \Psi^\mathrm{out}(\tilde r) \Psi^\mathrm{in}(\tilde r') + \theta(\tilde r'-\tilde r) \Psi^\mathrm{out}(\tilde r') \Psi^\mathrm{in}(\tilde r) \right]\; , 
\end{align}
with $W = \Psi^\mathrm{in} \partial_{\tilde{r}} \Psi^\mathrm{out} - \Psi^\mathrm{out} \partial_{\tilde{r}} \Psi^\mathrm{in}$ the Wronskian,
$\theta$ the Heaviside function, and $\Psi^\mathrm{out}$ and $\Psi^\mathrm{in}$ solutions to the homogeneous Regge-Wheeler equation subject to ingoing and outgoing boundary conditions respectively:
\begin{subequations}\label{eq:BC}
\begin{align}  \label{Psi in bc}
\Psi^\mathrm{in} &= \left\{
\begin{array}{ll}
C e^{-i \omega \tilde{r}}, & \qquad \mathrm{for} \; \tilde{r} \rightarrow - \infty \\
A^\mathrm{in} e^{-i \omega \tilde{r}} + A^\mathrm{out} e^{i \omega \tilde{r}}, & \qquad \mathrm{for} \; \tilde r \rightarrow +\infty
\end{array} \right. \\
\Psi^\mathrm{out} &= \left\{
\begin{array}{ll}
B^\mathrm{in} e^{-i \omega \tilde{r}} + B^\mathrm{out} e^{i \omega \tilde{r}}, & \qquad \mathrm{for} \; \tilde{r} \to - \infty \\
e^{i \omega \tilde{r}}, & \qquad \mathrm{for} \; \tilde r \rightarrow +\infty \\
\end{array} \right. .
\end{align}
\end{subequations}
Notice that we have fixed the normalization of $\Psi^\mathrm{out}$ at $\tilde r \rightarrow + \infty$, but have kept the normalization of $\Psi^\mathrm{in}$ arbitrary for later convenience. 

The constancy of the Wronskian gives
\begin{equation}
W = 2 i \omega A^\mathrm{in} = 2 i \omega C B^\mathrm{out} \; .
\end{equation}
In addition, the same argument applied to the other two linearly independent solutions $\Psi^\mathrm{in}$ and $\bar \Psi^\mathrm{in}$ (where the overbar denotes complex conjugation) and the same for $\Psi^\mathrm{out}$ gives
\begin{equation} \label{eq:ain_aout}
\vert A^\mathrm{in} \vert^2 - \vert A^\mathrm{out} \vert^2 = \vert C \vert ^2, \quad \vert B^\mathrm{out} \vert^2 - \vert B^\mathrm{in} \vert^2 = 1 \; .
\end{equation}

Taking now the limit $\tilde r \rightarrow + \infty$ of \eqref{Green function solution}, we obtain an expression for $\Psi$ far away from the emission region:
\begin{equation} \label{eq:solution_RW_with_source}
\Psi(\tilde r) \rightarrow \frac{e^{i \omega \tilde r}}{2i\omega A^\mathrm{in}} \left[ \int_{-\infty}^\infty d\tilde r' \, \Psi^\mathrm{in}(\tilde r') J(\tilde r') \right]  \; .
\end{equation}
Thus, we see that the amplitude of the emitted wave is completely determined by the ingoing-wave solution of the homogeneous Regge-Wheeler equation and its related coefficient $A^\mathrm{in}$. Of course, $\Psi^{\rm in}$ and therefore $A^{\rm in}$ are determined up to an overall multiplicative constant, but this ambiguity does not affect Eq. \eqref{eq:solution_RW_with_source}, which only depends on the ratio $\Psi^{\rm in}/A^\mathrm{in}$.


\subsection{PN expansion of the homogeneous solution} \label{sec-6.2}

Let us now calculate $\Psi^{\rm in}$ and $A^{\rm in}$ by solving the homogeneous Regge-Wheeler equation in a PN scheme with the ingoing boundary conditions \eqref{Psi in bc}. Given how we parametrized the various functions entering the Regge-Wheeler equation in Section \ref{sec:param}, we will find it more convenient to work in terms of the coordinate $r$ rather than the tortoise coordinate $\tilde r$. Even better, we will introduce a dimensionless coordinate $z \equiv \omega r$, so that our homogeneous equation becomes:
\begin{equation} \label{eq:RW}
 \frac{dz}{d\tilde r} \frac{d}{dz} \left( \frac{dz}{d\tilde r}  \frac{d \Psi^\mathrm{in}}{d z } \right) + (\omega^2 -V) \Psi^\mathrm{in} = 0 \; .
\end{equation}
The advantage of working with a dimensionless coordinate is that the two scales entering this equation, $M$ and $\omega$, can only appear in the dimensionless combination $\epsilon \equiv  2 M \omega$ (remember, our units are such that $G = 1$). Moreover, the source $J$ in \eqref{eq:source_term} is non-zero only for $\omega = m \Omega$, and in this regime we have 
\begin{align}
	\epsilon &\sim 2M \Omega = \frac{2M}{r_0} \times r_0 \Omega = v^3 \ll 1 \; , \\
	z_0 &\sim \Omega r_0 \sim v
\end{align}
where $r_0$ is once again the radius of the circular orbit. This suggests that we solve Eq. \eqref{eq:RW} in perturbation theory by expanding in powers of $\epsilon$ (but keeping $z$ arbitrary for now). More precisely, we will write our equation~as 
\begin{equation}
E_0[\Psi^\mathrm{in}] + \epsilon E_1[\Psi^\mathrm{in}] + \epsilon^2 E_2[\Psi^\mathrm{in}] + \mathcal{O}(\epsilon^3) = 0 \; ,
\end{equation}
and look for a perturbative solution of the form
\begin{equation} \label{eq:expansion_psi}
\Psi^\mathrm{in} = \Psi^\mathrm{in}_0 + \epsilon \Psi^\mathrm{in}_1 + \epsilon^2 \Psi^\mathrm{in}_2 + \mathcal{O}(\epsilon^3) \; ,
\end{equation}
satisfying ingoing boundary conditions at the horizon. Similarly, we expand $A^\mathrm{in}$ in powers of $\epsilon$ as $A^\mathrm{in} = A^\mathrm{in}_0 + \epsilon A^\mathrm{in}_1 + \epsilon^2 A^\mathrm{in}_2 + \mathcal{O}(\epsilon^3)$. The equations that the different orders $\Psi^\mathrm{in}_i$ solve are
\begin{align}
\begin{split} \label{eq:diffeq_expansion}
E_0[\Psi_0] &= 0 \; , \\
E_0[\Psi_1] &= - E_1[\Psi_0] \; , \\
E_0[\Psi_2] &= - E_1[\Psi_1] - E_2[\Psi_0] \; .
\end{split}
\end{align}
We thus have to solve the equation $E_0$ with inhomogeneous term given by the lowest-order solution.
We will see later on that working up to $\mathcal{O}(\epsilon^2)$ included is sufficient to calculate the waveform up to $\mathcal{O}(v^5)$ beyond the leading order result. 

In order to expand Eq. \eqref{eq:RW} in powers of $\epsilon$, we will use the following results, which can be derived from those in App.~\ref{app:def_complicated_functions}:
\begin{align}
\frac{d\tilde r}{dr} &= 1 + \kappa_1 \frac{\epsilon}{z} + \kappa_2 \frac{\epsilon^2}{z^2} + \mathcal{O} \left( \epsilon^3 \right) \; , \label{PN expansion tortoise coordinate}\\
V(z)   &= \frac{\omega^2}{z^2} \bigg(  \ell(\ell+1) + \kappa_3 \frac{\epsilon}{z} + \kappa_4 \frac{\epsilon^2}{z^2} + \mathcal{O} \left( \epsilon^3 \right) \bigg) \; ,
\end{align}
where
\begin{subequations}\label{eq:def_kappa}
\begin{align} 
\kappa_1 &= 1 + \gamma_1 + \frac{\lambda_0}{2} \; , \\
\kappa_2 &= \frac{1}{8(\ell -1)(\ell + 2)}  \big[  8 a_2 - 8 b_2 - (2 \alpha_1 + 4 \gamma_1 + \lambda_0)^2 \nonumber \\
 & +(\ell^2 + \ell - 2) (8 - 4 a_2 - 4 b_2 + 8 \gamma_1 + 12 \gamma_1^2 + 8 \gamma_2 + 
 2 \lambda_0 + 12 \gamma_1 \lambda_0 + 3 \lambda_0^2 + 
 4 \lambda_1)  \big]  \vphantom{\frac{1}{2}} \; , \\
 \kappa_3 &= -6 + 4 \alpha_1 + 5 \gamma_1 - (\ell^2 + \ell - 3) \bigg( 1 + \frac{\lambda_0}{2} \bigg)  \; ,
\end{align}
\begin{align}
 \kappa_4 &=  \frac{1}{ 16(\ell -1)(\ell + 2)} \bigg\lbrace \ell (\ell+1) \big[ \ell( \ell+1) \big( 16 a_2 - 4 \lambda_0 (-4 + 4 \gamma_1 + \lambda_0) - 8 \lambda_1 \big) \nonumber \vphantom{\frac{1}{2}} \\
&-336 a_2 - 8 \alpha_1 \lambda_0 + 76 \gamma_1 \lambda_0 + 
 11 \lambda_0^2 + 
 4 \big( 12 + 52 b_2 + 4 (-4 + \alpha_1) \alpha_1 - 62 \gamma_1 \nonumber \vphantom{\frac{1}{2}} \\
 & + 
    24 \alpha_1 \gamma_1 + 35 \gamma_1^2 + 36 \gamma_2  - 
    23 \lambda_0 + 16 \lambda_1 \big) \big] + 8 \big[ 54 a_2 - 30 b_2 + 7 \alpha_1^2 + 
   62 \gamma_1 \nonumber \vphantom{\frac{1}{2}} \\
   & + (\gamma_1 + \lambda_0) (9 \gamma_1 + 
      2 \lambda_0) + \alpha_1 (16 + 20 \gamma_1 + 13 \lambda_0) - 
   3 (4 + 12 \gamma_2 - 5 \lambda_0 + 4 \lambda_1) \big] \vphantom{\frac{1}{2}} \bigg\rbrace \; .
\end{align}
\end{subequations}
In the GR limit, where all our EFT parameters are set to zero, we recover the usual tortoise coordinate and Regge-Wheeler potential. Note that a similar PN parametrization of the Regge-Wheeler potential was discussed in \cite{2019PhRvD..99j4077C, 2019PhRvD.100d4061M}. However, here we are able to express the parameters $\kappa_i$ in terms of more fundamental ones.

\subsubsection{Zeroth-order solution: $\Psi^\mathrm{in}_0$} \label{subsec:psi0}

At lowest order in perturbation theory, the equation we need to solve is 
\begin{align}
	E_0[\Psi_0^\mathrm{in}] \equiv \Psi_0^\mathrm{in}\hphantom{}''(z) + \left(1-\frac{\ell(\ell+1)}{z^2}\right) \Psi_0^\mathrm{in}(z) = 0 .
\end{align}
The equation for $\Psi_0^{\rm in}$ is the same as in GR and there is no dependence on modified gravity parameters. This simple equation admits the two independent solutions $z j_\ell(z)$ and $z y_\ell(z)$, where $j_\ell$ and $y_\ell$ are spherical Bessel functions. The ingoing boundary condition should be imposed at the horizon $z_* \sim M \omega \sim \epsilon$. At zeroth order in our perturbative expansion, consistency requires that we set $z_* = 0$.  Our solutions scale like $z j_\ell(z) \sim z^{\ell+1}$ and $z y_\ell(z)\sim z^{-\ell}$ for small $z$,  and therefore regularity at the horizon singles out the solution
\begin{align}
	\Psi_0^\mathrm{in}(z) = z j_\ell(z) .
\end{align}
We can now calculate the zeroth order contribution to the coefficient $A^{\rm in}$ using the well-known asymptotic form of the spherical Bessel function $j_\ell$, which gives:
\begin{align}
	\Psi_0^\mathrm{in}(z) \ \stackrel{z \to \infty}{\longrightarrow} \ \frac{i}{2} \left( e^{-i z} e^{i \ell \pi/2} - e^{i z} e^{-i \ell \pi/2} \right) .
\end{align}
Using the fact that $z = \omega r \simeq \omega (\tilde r+ \mathcal \varphi)$ asymptotically, with $\varphi$ an arbitrary integration constant, we conclude that $A^{\rm in}_0 = i^{\ell+1} e^{i \varphi} / 2 $. Thus,  we find that $A^{\mathrm{in}}$ is determined up to an arbitrary phase, in agreement with the remarks of Ref.~\cite{Sasaki:1994aa}. This phase has no physical meaning, as evidenced by the fact that the power emitted ultimately depends on $|A^{\rm in}|^2$.

We now discuss an important issue concerning the inner boundary condition at the horizon. When choosing the lowest-order solution to be the spherical Bessel $z j_\ell$, we have made no use of the fact that the flux should be purely ingoing at the horizon: it could as well be a mixture of ingoing and outgoing modes, for the moment being regularity fixes the solution uniquely. When does the fact that the object is a BH, and not a very compact neutron star for example, makes a difference? There are two quantities in the asymptotic behavior of $\Psi$, Eq.~\eqref{eq:solution_RW_with_source}, where the presence of the horizon could show up: the RW function $\Psi^\mathrm{in}$ could be modified at the source location $z_0$ where $J$ has support, and the coefficient $A^\mathrm{in}$ could be modified.

Let us discuss the coefficient $A^\mathrm{in}$ first. From the behavior $\Psi^\mathrm{in} \propto  \mathcal{O}(z^{\ell+1})$ at $z=0$, we deduce that $C=\mathcal{O}(\epsilon^{\ell+1})$ where $C$ is the inner boundary condition defined in eq. \eqref{eq:BC}. On the other hand, $A^\mathrm{in}$ and $A^\mathrm{out}$ are of order unity since $z j_\ell \sim \mathcal{O}(1)$ for $z \rightarrow \infty$. Then the difference between the two is from eq. \eqref{eq:ain_aout}
\begin{equation}
\vert A^\mathrm{in} \vert - \vert A^\mathrm{out} \vert = \frac{\vert C \vert^2}{\vert A^\mathrm{in} \vert + \vert A^\mathrm{out} \vert} = \mathcal{O}(\epsilon^{2\ell+2}) \; .
\end{equation}
This shows that the difference between the ingoing and the outgoing flux is of order $\mathcal{O}(\epsilon^{2\ell+2})$, i.e. the presence of an horizon will show in $A^\mathrm{in}$ at the very high order $\mathcal{O}(v^{18})$ since $\ell \geq 2$. This is consistent with the result of Sasaki \cite{Sasaki:1994aa}.

On the other hand, the modification to $\Psi^\mathrm{in}$ at the particle's location $z_0$ shows up at a lowest PN order, as we now show. To see this, one can remark that at each order in $\epsilon$ the solution $\Psi^\mathrm{in}_i$ is a solution of the Bessel equation $E_0$ with inhomogeneous terms, see Eq.~\eqref{eq:diffeq_expansion}. Thus, at higher orders in $\epsilon$ the freedom to impose the ingoing boundary condition is related to the freedom of choosing the coefficients of the solution to the homogeneous equation $E_0$, i.e to choose the coefficients of $z j_\ell$ and $z y_\ell$ in $\Psi^\mathrm{in}_i$. The coefficient in front of $z j_\ell$ at all orders is just a normalization factor of $\Psi^\mathrm{in}_0 = z j_\ell$ so that we can set it to one. On the other hand, the lowest order at which the other Bessel function $z y_\ell$ is allowed to appear represents the order at which one can tune a parameter to enforce the ingoing boundary condition. Let us denote by $n$ the order in $\epsilon$ at which one is allowed to tune the coefficient of the homogeneous solution $z y_\ell$ in the expansion~\eqref{eq:expansion_psi}. In order to preserve the lowest-order behavior $\Psi^\mathrm{in} \sim \epsilon^{\ell+1}$ at the horizon, one should have
\begin{equation}
\epsilon^{\ell+1} \simeq \epsilon^n \times \epsilon y_\ell(\epsilon) \sim \epsilon^{n-\ell} \; ,
\end{equation}
since $z y_\ell \sim \epsilon^{-\ell}$ for $z \rightarrow \epsilon$. Thus, since $\ell \geq 2$ one has $n = 5$. Now, taking into account this supplementary term in $\Psi^\mathrm{in}$ for $\ell = 2$ one has
\begin{equation}
\left. \Psi^\mathrm{in}(z) \right\vert_{\ell = 2} \sim z^3 \bigg[ 1 + \dots + \epsilon^5 z y_\ell(z) + \dots \bigg] \sim z^3 \bigg[ 1 + \dots + \frac{\epsilon^5}{z^5} + \dots \bigg] \; ,
\end{equation}
where we have expanded for $z \ll 1$.
For a particle in circular orbit where $\epsilon \sim v^3$ and $z_0 \sim v$, the influence of the inner boundary condition on $\Psi^\mathrm{in}$ is of order $v^{10}$. Note that this point was not discussed by Sasaki~\cite{Sasaki:1994aa}, which led him to the wrong estimate of the influence of an horizon at the order $v^{18}$ in the outgoing flux. However, the correct PN order $v^{10}$ can be found e.g in Ref.~\cite{Mino:1997bx}.


One final comment we should make is that Eq.~\eqref{power emitted general} only accounts for the power dissipated in GWs at \textit{infinity}. However, a fraction of the GWs emitted will also be absorbed by the black hole horizon. Fortunately, this effect is again of higher PN order compared to the accuracy considered in this Chapter. This can be seen from the near-horizon behavior of the full solution $\Psi$ in Eq.~\eqref{Green function solution},
\begin{equation} \label{eq:solution_RW_with_source_horizon}
\Psi(\tilde r) \xrightarrow[\tilde r \to - \infty]{} \frac{C e^{-i \omega \tilde r}}{2i\omega A^\mathrm{in}} \left[ \int_{-\infty}^\infty d\tilde r' \, \Psi^\mathrm{out}(\tilde r') J(\tilde r') \right]  \; .
\end{equation}
On the one hand, we have that $C/ A^{\rm in} \sim \Psi_0^\mathrm{in}(\epsilon) /A^{\rm in}_0 \sim \epsilon^{\ell+1}$; on the other hand, $\Psi^\mathrm{out}$ must be (at lowest order) a combination of spherical Bessel functions $z j_\ell(z)$ and $z y_\ell(z)$ in order to enforce purely outgoing boundary conditions at infinity. Thus, $\Psi^\mathrm{out} (z_0) \sim z_0^{-\ell}$ for $z_0 \ll 1$, and therefore the ratio of the solution to the Regge-Wheeler equation at the two boundaries is
\begin{equation}
\frac{\Psi(\tilde r \rightarrow - \infty)}{\Psi(\tilde r \rightarrow  \infty)} \sim v^{\ell+2} \; .
\end{equation}
Since $\ell \geq 2$ and the power is proportional to $\vert \Psi \vert^2$, this means that the power dissipated at the horizon is suppressed by $v^8$ compared to the power emitted at infinity. This is consistent with the results derived by Poisson and Sasaki in pure GR~\cite{Poisson_1995}.

\subsubsection{First-order solution: $\Psi^\mathrm{in}_1$} \label{subsec:psi1}

Let us know turn our attention to the first order correction to the solution above. We now need to solve an inhomogeneous equation for $\Psi_1^{\rm in}$ of the form $E_0 (\Psi_1^{\rm in}) = - E_1 (\Psi_0^{\rm in})$, where the explicit form of the source term is
\begin{align}
	E_1[\Psi_0^\mathrm{in}] &= -2 \frac{\kappa_1}{z} \Psi_0^\mathrm{in}\hphantom{}''(z) + \frac{\kappa_1}{z^2} \Psi_0^\mathrm{in}\hphantom{}'(z)  - \frac{\kappa_3}{z^3}  \Psi_0^\mathrm{in}(z) \; ,
\end{align}
where the $\kappa_i$'s are defined in Eq.~\eqref{eq:def_kappa}.
%
%

Once again, the first order solution can be found using the Green's functions method, which yields
\begin{align} \label{Psi 1}
	\Psi_1^{\rm in}(z) = - \int_0^\infty d z' G_0 (z, z') E_1 (\Psi_0^{\rm in}(z')),
\end{align}
with $G_0$ a Green's function of the differential equation $E_0$ which satisfy the appropriate boundary conditions. 
We want to make sure that lowest order solution $\Psi_0^{\rm in}$ is regular at the horizon, and we want to make sure not to spoil this. For this reason, we will choose $G_0$ in such a way that $\Psi_1^{\rm in}(z=0) =0$. This is accomplished by the Green function
\begin{align}
	 G_0 (z, z') = \theta (z-z') \left[ z y_\ell(z)  z' j_\ell(z') - z j_\ell(z) z' y_\ell(z') \right] ,
\end{align}
where we used the fact that the Wronskian of $E_0$ is $W = z j_\ell \partial_z (z y_\ell) - z y_\ell \partial_z (z j_\ell) = 	1$. The Green function $G_0$ is the analog of a retarded Green's function (with time replaced by the radial coordinate $z$), unlike the one in Eq. \eqref{Feynman Green's function}, which is closer in spirit to a Feynman's Green function. 

Taking now the large-$z$ limit of \eqref{Psi 1}, we can read off the coefficient of the exponential $e^{-i \omega \tilde r}$ to extract the first order correction to $A^{\rm in}$.
Using the Hankel functions:
\begin{equation}
j_\ell = \frac{h_\ell^{(1)} + h_\ell^{(2)}}{2}, \quad y_\ell = \frac{h_\ell^{(1)} - h_\ell^{(2)}}{2i} \; ,
\end{equation}
and the symptotic behavior for $z \rightarrow \infty$
\begin{equation}
h_\ell^{(1)} \sim (-i)^{\ell+1} \frac{e^{iz}}{z}, \quad h_\ell^{(2)} \sim i^{\ell+1} \frac{e^{-iz}}{z} \; ,
\end{equation}
we find the following behavior:
\begin{equation}
\Psi^\mathrm{in} \sim \frac{i^{\ell+1}}{2} e^{-iz} (1 + \epsilon(\mathcal{A}(z)-i\mathcal{B}(z))) + \text{outgoing} \; ,
\end{equation}
where by outgoing we mean a term proportional to $e^{iz}$, and we have defined the integrals
\begin{align}
\begin{split} \label{eq:integrals_psiin}
\mathcal{A}(z) &= \int_0^z dz' z' y_\ell(z') E_1[\Psi_0] \; , \\
\mathcal{B}(z) &= \int_0^z dz' z' j_\ell(z') E_1[\Psi_0] \; ,
\end{split}
\end{align}
On the other hand, from the boundary conditions \eqref{eq:BC} and the fact that the tortoise coordinate writes as (from eq. \eqref{PN expansion tortoise coordinate})
\begin{equation}
\tilde z = \varphi + z + \epsilon \kappa_1  \log(z) + \mathcal{O}(\epsilon^2) \; ,
\end{equation}
where $\varphi$ is the unknown integration constant already discussed above in Section~\ref{subsec:psi0}, we get that the boundary condition for $\Psi$ is
\begin{equation}
\Psi \sim A^{\mathrm{in}} e^{i\varphi} e^{-iz} \left(1 - i \epsilon \kappa_1 \log(z)  \right) + \text{ outgoing} \; ,
\end{equation}
from which we immediately identify (with a slight abuse of notation which we will explain below)
\begin{equation}
A^{\mathrm{in}} = \frac{i^{\ell+1}}{2} e^{-i\varphi} \left(1+\epsilon\left[\mathcal{A}(z) + i\left(-\mathcal{B}(z) + \kappa_1 \log(z) \right) \right]  \right) \; .
\end{equation}
Of course the whole expression of $A^{\mathrm{in}}$ should not depend on $z$, so the log term should compensate in the asymptotic expressions of $\mathcal{A}$ and $\mathcal{B}$. This will be a useful check of our calculation.
Extracting the asymptotic behavior of $\mathcal{A}$ and $\mathcal{B}$, the final result is
%
\begin{align} \label{A^in_1}
A^{\rm in} &=  \frac{i^{\ell+1}e^{i \varphi}}{2} \bigg\lbrace 1 + \frac{\epsilon}{2}  \bigg[ \frac{\lambda_0 + 4 (\alpha_1 + \gamma_1 - 1)}{\ell(\ell+1)} - 1 - \frac{\lambda_0}{2} + (2 + 2 \gamma_1 + \lambda_0) (1 + \psi_\ell - \ln(2))
\bigg] \bigg\rbrace \;  , 
\end{align}
where $\psi_\ell$ is the digamma function defined by
\begin{equation}
\psi_\ell = \sum_{k=1}^{\ell-1} \frac{1}{k} - \gamma \; ,
\end{equation}
and $\gamma = 0.577...$ is Euler's constant. 

The careful reader may worry about the fact that our result \eqref{A^in_1} does not seem to reduce to the GR result (see {\it e.g.} Eq. (4.6) of Ref. \cite{Sasaki:1994aa}) when all our EFT parameters vanish. This discrepancy is however due to a different choice of normalization and phase for $\Psi$. As stated in Ref. \cite{Sasaki:1994aa}, only the difference between the value of $A^{\rm in}_1$ for different $\ell$'s is physically meaningful, and indeed we have checked that such difference correctly reproduces the GR result in the appropriate limit.

\subsubsection{Second-order solution: $\Psi^\mathrm{in}_2$} \label{subsec:psi2}

To know the full solution $\Psi$ at order $\mathcal{O}(v^5)$, one needs also to compute the second-order solution $\Psi^\mathrm{in}_2$. This can be seen from the expansion of $\Psi^\mathrm{in}$
in powers of $\epsilon$ and for $z=z_0 \ll 1$ which yields an expression of the form            
\begin{equation}
\frac{\Psi^\mathrm{in} (z_0)}{z_0^{\ell+1}} \sim  [1 + \mathcal{O}(z_0^2)] + \epsilon \left[ \frac{1}{z_0} + \mathcal{O}(z_0) \right] + \epsilon^2 \left[ \frac{1}{z_0^2} + \mathcal{O}(1) \right] +\epsilon^3 \left[ \frac{1}{z_0^3} + \mathcal{O}(1/z_0) \right] + \mathcal{O}(\epsilon^4) \; ,
\end{equation}
and so there is a part of the $\mathcal{O}(v^5)$ correction to $\Psi^\mathrm{in}$ which comes from the $\Psi^\mathrm{in}_2$ contribution $(\epsilon / z_0)^2 \sim v^4$.
Fortunately, we need only the expansion of $\Psi^\mathrm{in}_2$ at small values of $z$, because for large $z$ the contribution to $A^{\mathrm{in}}$ will be of order $\mathcal{O}(\epsilon^2) = \mathcal{O}(v^6)$. Likewise,
the contribution cubic in $\epsilon$ that we are neglecting would give a leading correction that scales like $(\epsilon / z_0)^3 \sim v^6$.

Extending our first order analysis to higher orders is conceptually straightfoward. In particular, the equation for the second order correction $\Psi_2^{\rm in}$ is of the form $E_0[\Psi^\mathrm{in}_2] = - E_1[\Psi^\mathrm{in}_1] - E_2[\Psi^\mathrm{in}_0]$, with 
\begin{align}
	E_2[\Psi^\mathrm{in}_0] =  \frac{3 \kappa_1^2 - 2 \kappa_2 }{z^2} \Psi_0^\mathrm{in}\hphantom{}''(z) - \frac{3 \kappa_1^2 - 2 \kappa_2}{z^3} \Psi_0^\mathrm{in}\hphantom{}'(z)  - \frac{\kappa_4}{z^4}  \Psi_0^\mathrm{in}(z) \; ,
\end{align}
Following the same logic we adopted to derive the first order solution, we conclude immediately that 
\begin{align}
	\Psi_2^{\rm in}(z) = - \int_0^\infty d z' G_0 (z, z') [E_1 (\Psi_1^{\rm in}(z')) + E_2 (\Psi_0^{\rm in}(z'))].
\end{align}
From this expression we can easily obtain the small-$z$ behavior of $\Psi^\mathrm{in}_2$:
\begin{equation}
\Psi^\mathrm{in}_2 \sim \mathcal{C}_2 z^{\ell-1} + \mathcal{O}(z^{\ell+1})
\end{equation}  
At the order we are interested in, only the $\ell=2$ term will contribute to the dissipated power (since from eq. \eqref{eq:power_chap10} the power is proportional to $\Psi^2$) and we find
\begin{align}
\mathcal{C}_2 &= \frac{1}{2880} \big(552 a_2 - 360 b_2 + 76 \alpha_1^2 + 
   8 \gamma_1 (12 + 53 \gamma_1) + 176 \gamma_1 \lambda_0 + 
   25 \lambda_0^2 \\
   & + \alpha_1 (-96 + 400 \gamma_1 + 
      76 \lambda_0) - 96 (5 \gamma_2 + \lambda_1)) \; .
\end{align}

As a sanity check, we have also verified that our generic solution for $\mathcal{C}_2$ in pure GR and for different values of $\ell$ corresponds to the expansion given in Appendix A of Ref. \cite{Tagoshi_1994}.

\subsection{PN expansion of the dissipated power}\label{sec:last_part}

We are now finally in a position to calculate the power emitted by combining all the results we have derived so far in this section. Combining the explicit expression \eqref{eq:source_term} for the source term $J$ with the asymptotic form of $\Psi^{\rm in}$ in Eq. \eqref{eq:solution_RW_with_source}, Fourier transforming back from frequency to time, and then plugging the result into the formula  \eqref{power emitted general} for the power emitted, we find
\begin{align}
	P = \frac{1}{16 \pi} \sum_{\ell \geq 2} \sum_{m=-\ell}^\ell \left\vert \left(\frac{A'}{A} - u_4 - \frac{d}{dr} \right) \left( \frac{\Psi^{\rm in}}{A^{\rm in}} A \, u_3 \frac{ \vert \mathcal{G}_{rr} \mathcal{G}_{00} \vert^{1/4}}{\mathcal{G}_{00}}  \frac{d \tilde r}{dr}\right) \right\vert^2_{r=r_0} .
\end{align}
Thus, we see that the power emitted depends  on $\Psi^{\rm in}$, the background metric coefficients, and all other EFT coefficients evaluated only at the radius $r_0$ of the orbit. This means in particular that we can expand all coefficients in powers of $2M /r_0 \sim v^2$, and $\Psi^{\rm in}(z_0)$ (which itself is an expansion in $\epsilon \sim v^3$) in powers of $z_0 \equiv m \Omega r_0 = m v$. When taken all together, these expansions will yield a PN approximation for the power $P$.

In order to carry out these expansions systematically, we need to take into account the fact that Kepler's law---and thus the relation between $r_0$ and $v$---is modified in the presence of a scalar hair. In fact, the velocity is equal to 
\begin{equation}
v = \Omega r_0 = \frac{L a^2}{E r_0} \; ,
\end{equation}
where in the last step we used eq. \eqref{Omega eq.}, while the energy $E$ and angular momentum $L$ are defined in Eq.  \eqref{eq:L_E_background}. Expanding $E$ and $L$ in inverse powers of $r_0$, we can invert this equation to express $2M/r_0$ in terms of $v$:
\begin{align} \label{eq:kepler_chap10}
	\frac{2M}{r_0} = \frac{ v^2}{1-2\mu_1} \left[1 + \frac{4a_2-6\mu_1+8\mu_2}{(1-2\mu_1)^2} v^2 + \mathcal{O}(v^4) \right].
\end{align}
This implies that the parameter $\epsilon = 2 M m \Omega = (2 M / r_0) m v $ also admits an expansion in powers of $v$. Incidentally, the fact that the leading term on the righthand side depends on the parameter $\mu_1$ can be viewed as a renormalization of Newton's constant---a common feature in scalar-tensor theories \cite{damour_tensor-multi-scalar_1992, Kuntz:2019zef}.


Our PN expansion for the power emitted in the odd sector can be cast in the form   
\begin{align}
\begin{split}\label{eq:power_chap10}
P &= P_N v^2 \Big[ p_{0} \;  + p_{2} \; v^2 + p_{4} \; v^4 + \mathcal{O}(v^6) \Big]  \; ,
\end{split}
\end{align}
where we have denoted by $P_N$ the standard quadrupole energy loss,
\begin{equation}
P_N = \frac{32}{5} \left( \frac{\mu_0}{M} \right)^2 v^{10} \; ,
\end{equation}
and the coefficients $p_i$ in Eq. \eqref{eq:power_chap10} are reported in App.~\ref{power-app}. This is the main result of this Chapter: our EFT approach provides a model-independent parametrization of how the coefficients $p_i$ can differ from their GR value in the presence of a scalar hair.  A departure from GR would lead to changes in the phase of the waveform. 

An interesting byproduct of our result is that we can turn off all our EFT coefficients to obtain the power emitted in the odd sector in GR. To our knowledge, this is the first time that the contribution of the odd sector to the luminosity formula for a circular orbit in GR has been appeared in the literature (earlier work was based on the Teukolsky equation and yielded results that included both even and odd sector).  In particular, our result shows that the odd sector contribution is already of 1PN order---{\it i.e.} suppressed by $v ^2$ compared to the quadrupole expression, which comes from the even sector. Thus, we have determined the total power emitted in the odd sector up to 3.5PN order.

Before finishing this Chapter, let us stress again one important point. Given our final equation~\eqref{eq:power_chap10} for the dissipated power, one may be tempted to ask what is the advantage of our formalism compared to e.g letting $p_0$, $p_1$ and $p_2$ free in an actual template, as this is done in the parametrized post-Einsteinian (ppE) formalism \cite{Yunes_2009}.  However, our approach is much more valuable because we are able to provide a dictionary between the expansion parameters and the fundamental theories, as this is done in Section \ref{sec:examples}. Once a particular theory is chosen, say with one free modified gravity parameter, then our EFT provides a one-parameter family of templates which can be used to draw much more conclusive tests of GR than the ppE formalism can provide due to its inherent freedom of parametrization.
 We think that this method can give rise to interesting and efficient modeled searches in a large class of modified gravity theories.

{
\ihead{Concluding remarks}
\addchaptertocentry{Concluding remarks}
\chapter*{Concluding remarks}

This work aimed at characterizing gravity by contrasting GR with one of its most simple alternatives, namely scalar-tensor theories. In this thesis, we have focused on the two-body problem and we have highlighted the aspects in which observables could differ from their GR values in these alternative theories. To do so, we have often adopted an EFT viewpoint on the two-body problem: this reasoning allows for a straightforward comparison of theories with experiments, while at the same time pointing towards the direction in which we could hope to measure new effects. Since the recent detection of gravitational waves open a new era in astronomy, we have mainly focused on the strong-field regime of gravity when gravitational waves are emitted by binary systems of neutron stars or black holes.

The first part of the thesis consists in a generic introduction on the use of effective field theories in gravity, cosmology and the two-body problem in GR. Chapter~\ref{Chapter1} presents the PPN formalism as one of the first historical example of the application of EFT ideas to gravity. This formalism allows a straightforward classification of observable deviations from GR in the weak-field regime of the Solar system; it serves as a guideline for the reasonings that we have adopted in the strong-field regime throughout the thesis. Chapter~\ref{Chapter2} is a presentation of the basics aspects of cosmology, its current issues and the modified theories of gravity which can address these issues. It is not directly related to the two-body problem but it offers a strong motivation for the theories that we have studied in the two-body regime in the remaining of the thesis. Moreover, we have introduced in this Chapter the EFT of dark energy which is another important example of application of EFT ideas to gravity; this formalism has been extensively used in Chapters~\ref{Chapter9} and~\ref{Chapter10}. Finally, Chapter~\ref{Chapter3} contains an introduction to gravitational waves: how to describe them in linearized GR, the interferometers and how they operate to detect GW, a brief presentation of the tools used to solve the two-body problem in the strong-field regime of GR, and the current tests of GR using experimental GW data. The rest of the manuscript primarily deals with new tests of gravity using GW.

The second part of the thesis presents an EFT formalism designed for high-order, accurate computations in the two-body problem, namely Non-Relativistic General Relativity (NRGR). In Chapter~\ref{Chapter4}, we have adapted this formalism to Brans-Dicke type theories. We have shown that this Non-Relativistic Scalar-Tensor formalism allows to shed a new light on well-known properties of scalar-tensor theories: finite-size effects, violation of the strong equivalence principle, dipolar radiation. In Chapter~\ref{Chapter5}, we have extended this formalism to include a disformal coupling of the scalar field. We have proven that there is no contribution of such a disformal coupling in the case of circular orbits up to a very high PN order, and we have then used data from the Hulse-Taylor pulsar whose orbit is quite eccentric to constrain the magnitude of this coupling. Finally, in Chapter~\ref{Chapter6} we have turned back to GR only and have introduced a resummation technique within the NRGR formalism. This technique allows for a nice simplification of the two-body computations; moreover, the new parameters introduced define an effective two-body horizon. We have generalized the notion of innermost circular orbit to the two-body problem, showing that no circular orbit can exist at all under a critical radius which we have given as a function of the masses of the system. 

The third part of the thesis deals with the Vainshtein screening mechanism which is often used in cosmology as an example of theory in which there could be large deviations from GR on cosmological scales while still recovering GR on Solar system scale where the tests are very stringent. In Chapter~\ref{Chapter7}, after a short introduction on the Vainshtein mechanism, we have defined the two-body problem in this kind of theories and we have found the two-body potential in the static case. The high nonlinearity of the theory implies a violation of the equivalence principle which we derive in the Sun-Earth-Moon system. This sets a new bound on the magnitude of Vainsthein screening. Chapter~\ref{Chapter8} presents the effect of Vainshtein screening on the motion of black holes when the no-hair theorem is violated. We have computed the modification to the GW phase coming from this modification of gravity in the case where one black hole is much smaller than the other one.

The last part of the manuscript is dedicated to the no-hair theorems and their observable effects. Chapter~\ref{Chapter9} presents the no-hair theorems as well as many interesting physical situations in which they do not apply and consequently we should expect large deviations from GR for the motion of black holes. We have also introduced an EFT formalism inspired from the EFT of dark energy and designed to study quasi-normal modes of hairy black holes. In Chapter~\ref{Chapter10}, we have used this new EFT in another regime, namely GW emission by a very asymmetric system composed of a small black hole and a supermassive one. We have shown that such an EFT allows a straightforward computation of the waveform in a large class of scalar-tensor theories, thus giving access to modeled searches beyond GR.

We are confident that the methodology employed in this thesis can lead to fruitful new developments. Let us highlight here some of the most promising directions in which we think that this work could be continued.
\begin{enumerate}
\item Spinning objects and finite-size effects are quite under-developed topics in the modified gravity landscape. It could be interesting to adapt the EFT formalism developed in Chapter~\ref{Chapter4} to spinning black holes or neutron stars. Spin degrees of freedom are described in the EFT approach by new worldline operators~\cite{Porto_2006, Levi:2018nxp}; their generalization to the scalar-tensor case should be quite straightforward.
\item Higher-order computations in the GR two-body problem with the kind of resummation techniques discussed in Chapter~\ref{Chapter6} are certainly worth exploring. For example, it could be interesting to get a more accurate approximation of the CICO defined in Section~\ref{sec:ICO-CICO} and to compare it to numerical simulations, or to investigate on the properties of the worldline parameters in the extreme mass ratio case.
\item It would be interesting to reconsider the no-hair theorems within the formalism developed in Chapters~\ref{Chapter9} and~\ref{Chapter10}: could the no-hair arguments be translated in constraints on the free functions appearing in the EFT action~\eqref{eq: effective action for perturbations}?
\item  Last but not least, the last Chapter of this thesis should be completed as soon as possible with the even sector of the dynamics. Once this is achieved, implementing the EFT parametrization in a waveform template accessible to all the community would pave the way towards modeled searches of gravitational wave signals truly differing from GR, thus potentially opening the door to new discoveries in our Universe.
\end{enumerate}

}




\appendix 


\chapter{Dipolar dissipated power at next-to-leading order} \label{app:dipolar}

In order to have  the dipolar radiated power at the same order as the quadrupolar one in the final part of Chapter~\ref{Chapter4}, we  derive here the first-order correction to the dipolar power. To this aim, we will use the full next-to-leading order dipole in eq. \eqref{eq:dipole_full}. However, we must  correct the center-of-mass relations  in eq.~\eqref{eq:center_of_mass}  by terms higher order in the velocity. 
In particular, using the centre-of-mass definition in eq.~\eqref{eq:def_center_mass} with the $00$-component of the stress-energy tensor given by
\begin{equation}
T^{00} = \sum_A m_A \left(1+\frac{v_A^2}{2} - \frac{\tilde{G}_{12} m_{\bar{A}}}{2r} \right) \delta^3(\mathbf{x} - \mathbf{x}_A ) \; ,
\end{equation}
the center-of-mass relations become
\begin{equation} \label{eq:CM_NLO}
\mathbf{x}_1 = \frac{\mu_2}{\mu_1+\mu_2} \mathbf{r}, \qquad \mathbf{x}_2 = - \frac{\mu_1}{\mu_1+\mu_2} \mathbf{r} \; ,
\end{equation}
with
\begin{equation}
\mu_A \equiv m_A \left(1+\frac{v_A^2}{2} - \frac{\tilde{G}_{12} m_{\bar{A}}}{2r} \right) \; .
\end{equation}

The generalized equations of motion can be found by varying the EIH lagrangian \eqref{eq:LEIH}. In the case of a circular motion,
\begin{equation}
v^2 \equiv (\mathbf{v}_1 - \mathbf{v}_2)^2 = \frac{\tilde{G}_{12}(m_1+m_2)}{r} \; ,
\end{equation}
and using the centre-of-mass relations  above we find, for the acceleration up to next-to-leading order,
\begin{align} \label{eq:EOM_NLO}
\begin{split}
\frac{d^2 \mathbf{r}}{dt^2} &= - \frac{\tilde{G}_{12}M}{r^3} \mathbf{r} \bigg\{1+\frac{\tilde{G}_{12}M}{r} \left[ \nu - \frac{3+2 \aaa_1 \aaa_2}{1+2 \aaa_1 \aaa_2} \right.  \\
& + 4  \left. \frac{ \aaa_1^2 \aaa_2^2 + \beta_1 \aaa^2_2(1+\kappa_{12}) + \beta_2 \aaa_1^2(1-\kappa_{12})}{(1+2 \aaa_1 \aaa_2)^2} \right] \bigg\} \; ,
\end{split}
\end{align}
where we recall  the notation: $M \equiv m_1 + m_2$,  $\nu \equiv \frac{m_1 m_2}{M^2}$ and $ \kappa_{12} \equiv \frac{m_1-m_2}{M}$.

The scalar radiated power by the dipole is given by (see eq.~\eqref{eq:radiated_power_scalar})
\begin{equation}
P_\phi^\mathrm{dipole} = \frac{2G_N}{3} \big\langle  \ddot{I}_\varphi^i \ddot{I}_\varphi^i  \big\rangle \; .
\end{equation}
Inserting the dipole expression up to next-to-leading order in eq.~\eqref{eq:dipole_full} and using eqs.~\eqref{eq:CM_NLO} and \eqref{eq:EOM_NLO}  derived above we finally find, for the dipolar power up to next-to-leading order,
\begin{align}
\begin{split}
P_\phi^\mathrm{dipole} &= \frac{2  (\aaa_1-\aaa_2)^2}{3\tilde{G}_{12} (1+2\aaa_1 \aaa_2)} \nu^{2/5} (\tilde{G}_{12} M_c \omega)^{8/3} \\
& + \frac{4 (\aaa_1-\aaa_2)}{15 \tilde{G}_{12} (1+2\aaa_1 \aaa_2)^3} (\tilde{G}_{12} M_c \omega)^{10/3}  
 \left[17 (\aaa_1-\aaa_2) + (\aaa_1-\aaa_2) \aaa_1 \aaa_2 (2 \aaa_1 \aaa_2-43) \right. \\
&+ \kappa_{12} (\aaa_1+\aaa_2)(1+2\aaa_1 \aaa_2)(4+3\aaa_1 \aaa_2) -2 \kappa_{12}^2 (\aaa_1-\aaa_2) (1+2\aaa_1 \aaa_2)^2 \\
& - \left. 10 \beta_2 \aaa_1(1+\kappa_{12})(1-2\aaa_1(\aaa_1-2\aaa_2)) + 10 \beta_1 \aaa_2(1-\kappa_{12})(1-2\aaa_2(\aaa_2-2\aaa_1))  \right] \; .
\end{split}
\end{align}
After some algebra, it is possible to show that this result agrees with eq. (6.44) of Ref.~\cite{damour_tensor-multi-scalar_1992}. On the other hand, it differs from eq.~(55) of Ref.~\cite{Huang:2018pbu}.

%
%
%
%
%
%
%
%
%

\chapter{Matching interior and exterior solutions}
\label{sec:matching}

\begin{figure}
\centering
\includegraphics[width=0.7\columnwidth]{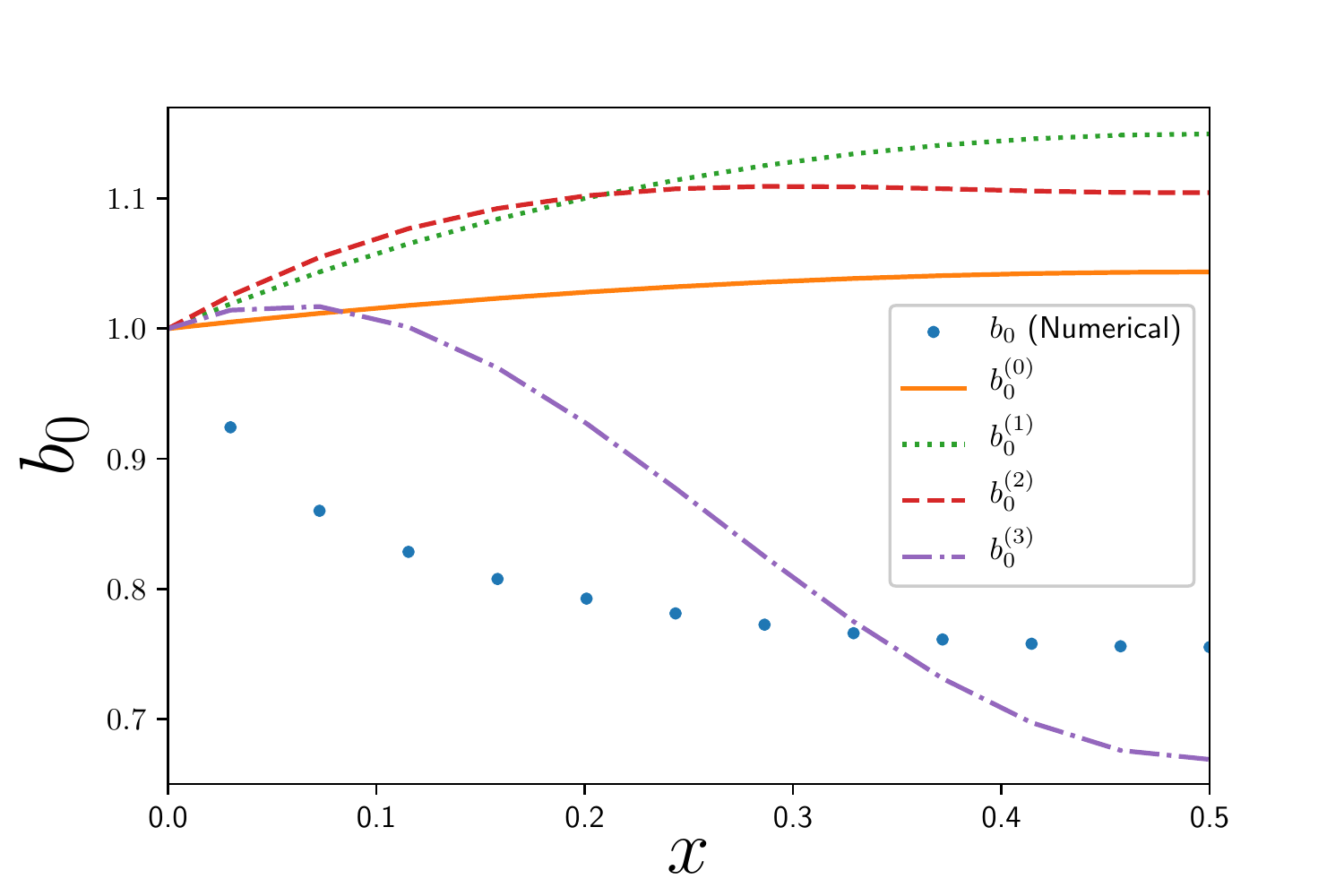}
\caption[Plot of the first four predictions for the coefficient $b_0$]{Plot of the first four predictions for the coefficient $b_0$ defined in eq. \eqref{eq:energy_map_inside} using the procedure defined in the main text. For comparison, the numerical coefficient $b_0$ is also plotted.}
\label{fig:b0}
\end{figure}

\begin{figure}
\centering
\includegraphics[width=0.7\columnwidth]{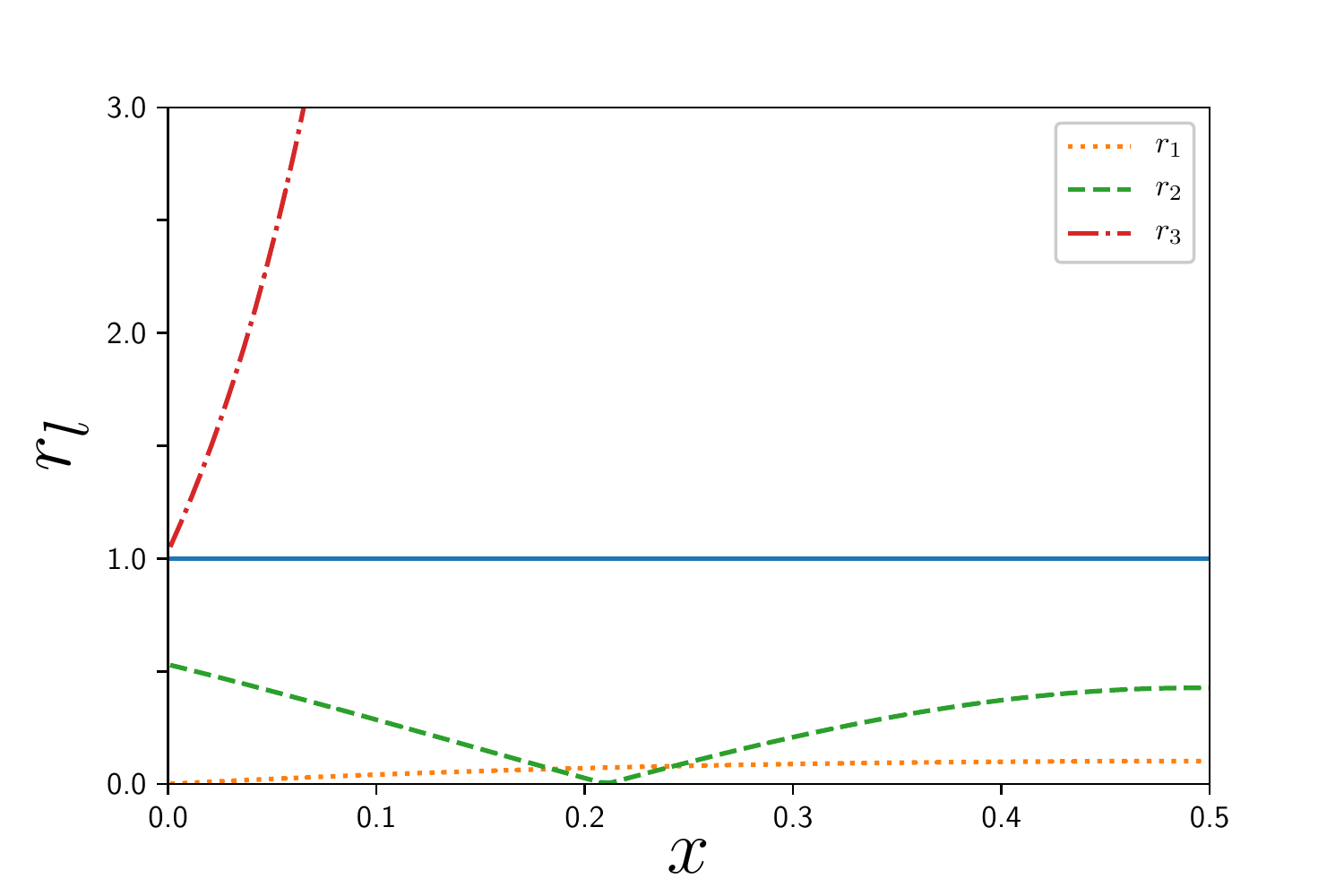}
\caption[Plot of the coefficient $r_l$]{Plot of the coefficient $r_l$ defined in eq. \eqref{eq:rl}. The horizontal bar at $r = 1$ is the maximum authorized value for $r_l$. If $r_l > 1$, then the corresponding term in the asymptotic series \eqref{eq:asymptotic_series} is increasing. }
\label{fig:rl}
\end{figure}

In this Appendix related to Chapter \ref{Chapter7}, we will attempt to obtain a qualitative analytical behavior of the energy as a function of the mass ratio inside the nonlinear radius. We can use the formula for the energy outside the screening radius that we found in Sec. \ref{sec:resum_TM} as a limiting boundary condition for the energy map inside the screening radius. Since the energy map partly resums the nonlinear behavior, it should allow to get a qualitative description of the coefficient $b_0$.

Using the energy derived in eq. \eqref{eq:energy_outside_exact}, one finds that the left-hand side of the energy map \eqref{eq:energy_map_inside} writes as
\begin{align}
\begin{split}
\frac{E'}{E'_\mathrm{tm}} &= \frac{1}{\varphi'_M} \left[ -\frac{M}{r^2} + \frac{\varphi'_{M(1-x)}}{1-x} + \frac{\varphi'_{Mx}}{x} \right] \\
&\equiv F(r,x) \; ,
\end{split}
\end{align}
for $r$ close to $r_*$. We recall that $x$ is the mass ratio $x = m_1/M \leq 1/2$ and that $\varphi_m$ is the spherically symmetric field  \eqref{eq:spherical_symmetry} generated by a mass $m$. Note that, as $\frac{\varphi'_{Mx}}{x} \rightarrow \frac{M}{r^2}$ as $x \rightarrow 0$, this expression has the good test-mass limit, and in this limit we recover that $b_0 = 1$ and $b_1, \; \dots \; b_N = 0$.

If the energy map really resums a part of the nonlinear dynamics, then it should give an asymptotic expansion near to the Vainshtein radius, as suggested by the relative smallness of the expansion parameter at $r_*$,
\begin{equation}
\left.\frac{E'_\mathrm{tm}}{E'_\mathrm{ref}}-1 \right|_{r_*} \simeq - 0.53 \; ,
\end{equation}
where we recall that $r_* = \left( 27/4 \right)^{1/4} \sqrt{M}$.

Consequently, we can try to find a first approximation for the value of $b_0$ by neglecting higher-order contributions in the right-hand side of eq. \eqref{eq:energy_map_inside}. This gives for $b_0$
\begin{equation}
b_0^{(0)}(x) \simeq  F(r_*, x) \; .
\end{equation}

This first prediction for $b_0$ is plotted in Figure \ref{fig:b0}.

One can try to improve the result by matching not only the numerical value of the energy map \eqref{eq:energy_map_inside} at $r=r_*$, but also its derivatives with respect to $r$. This will allow us to extract $b_1$, $b_2$... while at the same time refining our prediction on $b_0$. More precisely, matching up to the $k^\mathrm{th}$ derivative will allow us to determine $k+1$ coefficients in the energy map, say $b_0^{(k)} , \; b_1^{(k)}, \; \dots \; b_k^{(k)}$.

When should we stop this procedure ? Obviously it cannot be pushed to arbitrary order because the boundary condition on the energy is not sufficiently precise. A sensible criterion can be found by thinking in terms of asymptotic series. Let us write the $k^\mathrm{th}$ prediction for $b_0$ as
\begin{equation} \label{eq:asymptotic_series}
b_0^{(k)} = \sum_{l=0}^{k} b_0^{(l)} - b_0^{(l-1)} \; ,
\end{equation}
where we have defined $b_0^{(-1)} = 0$.

In this equation we have highlighted the fact that each step of the iteration brings out an additional factor of $b_0^{(l)} - b_0^{(l-1)}$ to $b_0$. Then the asymptotic series criterion requires us to cut the sum at the term at which $|b_0^{(l)} - b_0^{(l-1)}|$ is minimized in order to get the best precision on $b_0$. In Figure \ref{fig:rl} we have plotted the ratio
\begin{equation} \label{eq:rl}
r_l = \left| \frac{b_0^{(l)} - b_0^{(l-1)}}{b_0^{(l-1)} - b_0^{(l-2)}} \right| \; ,
\end{equation}
for $l=1, \; 2, \; 3$. The minimal term of the series is the one for which $r_l < 1$ and $r_{l+1} > 1$. We can see directly that, while the two first iterations of the procedure (i.e., matching up to the second derivative) seem to bring an improvement in the determination of $b_0$, the third one gives a large correction.

The end result of this Section is shown in Figure \ref{fig:b0}, where we plot the first four predictions $b_0^{(0)}, \; b_0^{(1)}, \; b_0^{(2)}$ and $b_0^{(3)}$ for $b_0$. Of course, an immediate drawback of this method is that we are not able to assess quantitatively the precision of the estimation of this coefficient ; moreover, it is apparent from Figure \ref{fig:b0} that even the qualitative behavior is not correct. A refinement of the method is needed but we leave this for future work.

\chapter{Useful integrals}
\label{sec:useful_integrals}

Here we give a number of integrals which we heavily used in this thesis in the Feynman diagrams calculations.

\begin{equation}
\int \frac{d^dk}{(2 \pi)^d} \frac{1}{(\mathbf{k}^2)^a ((\mathbf{k} + \mathbf{K})^2)^b} = \frac{1}{(4\pi)^{d/2}} \frac{\Gamma(a+b-d/2) \Gamma(d/2-a) \Gamma(d/2-b)}{\Gamma(a) \Gamma(b) \Gamma(d-a-b)} (\mathbf{K}^2)^{d/2-a-b}
\end{equation}

\begin{equation}
\int \frac{d^dk}{(2 \pi)^d} \frac{k^i}{(\mathbf{k}^2)^a ((\mathbf{k} + \mathbf{K})^2)^b} = \frac{1}{(4\pi)^{d/2}} \frac{\Gamma(a+b-d/2) \Gamma(d/2-a+1) \Gamma(d/2-b)}{\Gamma(a) \Gamma(b) \Gamma(d-a-b+1)} (\mathbf{K}^2)^{d/2-a-b} K^i
\end{equation}

\begin{align}
\begin{split}
\int \frac{d^dk}{(2 \pi)^d} \frac{k^i k^j}{(\mathbf{k}^2)^a ((\mathbf{k} + \mathbf{K})^2)^b} &= \frac{1}{(4\pi)^{d/2}} \frac{\Gamma(a+b-d/2-1) \Gamma(d/2-a+1) \Gamma(d/2-b)}{\Gamma(a) \Gamma(b) \Gamma(d-a-b+2)} (\mathbf{K}^2)^{d/2-a-b} \\
& \times \left( \frac{d/2-b}{2} \mathbf{K}^2 \delta^{ij} + (a+b-d/2-1)(d/2-a+1)K^i K^j \right)
\end{split}
\end{align}

\begin{equation}
\int \frac{d^dk}{(2 \pi)^d} (\mathbf{k}^2)^\alpha e^{i \mathbf{k} \mathbf{r}} = \frac{2^{2\alpha-1}}{\pi^{(d-1)/2}} \frac{\Gamma(\alpha + d/2)}{\Gamma(d/2) \Gamma((3-d)/2-\alpha)} (\mathbf{r}^2)^{-\alpha -d/2}
\label{eq:intK}
\end{equation}

\chapter{Matching to Brans-Dicke type theories} \label{app:matching_BD}

In this Appendix, we will connect the results for Brans-Dicke type theories obtained in Chapter~\ref{Chapter10} with the existing PN literature \cite{Lang:2015aa, Bernard:2018hta, Bernard:2018ivi}. In these references, the two fundamental quantities are the Brans-Dicke coupling function $\omega(\phi)$ and the field-dependent mass $m_A(\phi)$, defined by the Brans-Dicke like action (in the Jordan frame)
\begin{equation} \label{eq:bd_action_app}
S = \frac{1}{16 \pi} \int \mathrm{d}^4x \sqrt{-g} \left[ \phi R - \frac{\omega(\phi)}{\phi} g^{\mu \nu} \partial_\mu \phi \partial_\nu \phi \right] - \sum_A \int \mathrm{d} t \; m_A(\phi) \sqrt{- g_{\mu \nu} v^\mu_A v^\nu_A} \; ,
\end{equation}
where the index $A$ refers to the different point-particle objects,  $v^\mu_A = d x^\mu_A / dt$ is the velocity of each object, and $m_A(\phi)$ is a field-dependent mass. 
The field dependence of $m_A(\phi)$ is expected since local physics now depends on the environment and so of the value of $\phi$ at the object location \cite{Eardley1975ApJ}. For light objects, we do not expect any dependence of $m_A$ on $\phi$, while for strongly self-gravitating objects like neutron stars we cannot ignore this dependence. This phenomenon is known as spontaneous scalarization~\cite{Damour:1993aa}. Following \cite{Mirshekari_2013, Bernard:2018hta} the coupling function $\omega(\phi)$ is parameterized in a weak-field expansion as
\begin{equation} \label{eq:expansion_omega}
\omega(\phi) = \frac{1-4\zeta}{2 \zeta} + \lambda_1 \frac{1-\zeta}{\zeta^2} \left( \frac{\phi}{\phi_0} - 1 \right) +  \frac{\lambda_2}{2} \frac{1-\zeta}{\zeta^3} \left( \frac{\phi}{\phi_0} - 1 \right)^2 + \dots \; ,
\end{equation}
where $\phi_0$ is the asymptotic value of the scalar, and the mass function $m_A(\phi)$ as
\begin{equation} \label{eq:expansion_mA}
\log m_A(\phi) = \log m_A^{(0)} + s_A \log \frac{\phi}{\phi_0} + \frac{s_A'}{2} \left( \log  \frac{\phi}{\phi_0} \right)^2 + \frac{s_A''}{6} \left( \log  \frac{\phi}{\phi_0} \right)^3 + \dots
\end{equation}

We can express the Jordan frame action~\eqref{eq:bd_action_app} in our formalism by making the conformal field redefinition
\begin{equation} \label{eq:change_variable_ST}
\tilde g_{\mu \nu} = \phi g_{\mu \nu}, \quad \phi = e^\lambda, \quad \varphi = \int d\lambda \left( \omega(e^\lambda) + \frac{3}{2} \right)^{1/2}  \; ,
\end{equation}
where the relation between $\varphi$ and $\lambda$ depends on the exact form of the function $\omega(\phi)$. It brings the action in the following Einstein frame form already displayed in Eq.~\eqref{eq:bd_action_EinsteinFrame},
\begin{align}
\begin{split} \label{eq:bd_action_Einstein_app}
S &= \frac{1}{16 \pi} \int \mathrm{d}^4x \sqrt{-\tilde g} \left[ \tilde R -  \tilde g^{\mu \nu} \partial_\mu \varphi \partial_\nu \varphi \right] \\
& - \sum_A \int \mathrm{d} t \; m_A(e^{\lambda(\varphi)}) \sqrt{- e^{- \lambda(\varphi)}  \tilde g_{\mu \nu} v^\mu_A v^\nu_A} \; .
\end{split}
\end{align}
From now on we will drop the tildes for simplicity. The equations of motion for such an action are given in Eq.~\eqref{eq:system_ST}. However, in the main text the scalar charge defined in Eq.~\eqref{eq:def_scalar_charge} was a free parameter related to the short-distance physics coupling the scalar to the neutron star. In the PN literature, this charge can be related to the functions $\omega$ and $m_A$ in the weak-field approximation. This can be done by taking into account the point-particle source term in the scalar EOM. Thus, the field equation for $\varphi$ becomes
\begin{equation}
\frac{1}{\sqrt{-g}} \partial_\mu \left(\sqrt{-g} g^{\mu \nu}  \partial_\nu \varphi  \right) = 8 \pi \sqrt{-g_{00}} \frac{d\lambda}{d\varphi} \frac{d}{d\lambda} \left(m_A(e^\lambda) e^{-\lambda/2} \right) \delta^3(\mathbf{x}) \; ,
\end{equation}
where we recall that $\lambda$ is defined by eq. \eqref{eq:change_variable_ST}, and we have taken the point-particle to sit at the origin of the coordinates. In spherical coordinates, this is rewritten as
\begin{equation} \label{eq:field_eq_bd_charge}
\partial_r \left(r^2 a b  \bar \varphi'  \right) = 2 \frac{a}{\sqrt{\omega(e^\lambda) + 3/2}} \frac{d}{d\lambda} \left(m_A(e^\lambda) e^{-\lambda/2} \right) \delta(r) \; .
\end{equation}
The left-hand side contains the scalar charge of the body, while the right-hand side depends on the fields $a$ and $\lambda$ evaluated at the location of the point-particle (formally divergent in our approach where the cutoff scale corresponding to the inverse size of the neutron star is sent to infinity). We use a weak-field expansion in which $\lambda = \lambda_0 + \mathcal{O}(m/r)$, $a=1+\mathcal{O}(m/r)$ at the location of the point-particle (note that such expansion is not accurate for neutron stars where $m/r \sim \mathcal{O}(1)$: this is why we chose to consider the scalar charge as a free parameter in the main text). Integrating Eq. \eqref{eq:field_eq_bd_charge}, this yields
\begin{equation} \label{eq:def_scalar_charge_app}
\bar \varphi = \varphi_0 - \int \mathrm{d}r \; \frac{Q}{r^2 a b} \; ,
\end{equation}
where
\begin{equation}
Q = \sqrt{\frac{2 \zeta}{1 - \zeta}} m_A^{(0)} e^{-\lambda_0/2} (1-2s_A) \; ,
\end{equation}
where $\zeta$ is related to the asymptotic value of $\omega$ through Eq.~\eqref{eq:expansion_omega}, and the sensitivity $s_A$ is defined in Eq.~\eqref{eq:expansion_mA}. This is in agreement with the conventional PN literature \cite{damour_tensor-multi-scalar_1992, Damour:1993aa}. In the main text, we have made use of the reduced scalar charge $q = Q / m_A^{(0)}$.

Since the tadpole function $f$ was already given in Eq.~\eqref{eq:f_bd}, the only remaining task of this Appendix is to find the expansion of the mass function $\mu(r)$, defined in Eq.~\eqref{mu expansion}. From Eq.~\eqref{eq:bd_action_Einstein_app}, $\mu(r)$ is given by
\begin{equation}
\mu(r) = m_A\big(e^{\lambda(r)} \big) e^{- \lambda(r) / 2}
\end{equation}
Thus, we have to find the expansion of
 the function $\lambda$. 
By using that, from the very definition of $\lambda$ in Eq.~\eqref{eq:change_variable_ST},
\begin{equation} 
\frac{\mathrm{d} \lambda}{\mathrm{d} r} = \frac{1}{\sqrt{\omega(e^\lambda) + 3/2}} \frac{\mathrm{d} \varphi}{\mathrm{d} r} \; ,
\end{equation}
one can parametrically solve for $\lambda$ in a $1/r$ expansion using the expansion of $\varphi$ given in Eq.~\eqref{eq:expansion_barphi}. This finally translates into the coefficients of $\mu(r)$,
\begin{align}
\begin{split}
\mu_0 &= m_A^{(0)} e^{-\lambda_0/2} \; , \\
\mu_1 &= \tilde q \left(s_A-\frac{1}{2} \right) \; , \\
\mu_2 &= \frac{\tilde q}{8 \zeta} \left( \zeta (4s_A - 2 + \tilde q(1-4s_A+4s_A^2+4s_A')) + 2\tilde q \lambda_1(1-2s_A) \right) \; , \\
\mu_3 &= \frac{\tilde q}{48 \zeta^2} \left[ 8 \zeta ^2 (2 s_A-1)
 +6 \zeta  \tilde q \left(2 \lambda_1+\zeta 
   \left(4 s_A^2-4 s_A+4 s_A'+1\right)-4 \lambda_1
   s_A\right) \right. \\
    &+ \tilde q^2 \left\lbrace-\zeta  \left(2 \lambda_1+24 \lambda_1 s_A^2-16
   \lambda_1 s_A+2 s_A+24 \lambda_1 s_A'-1\right) \right. \\
   &+\left. \left. 2\zeta ^2 \left(4 s_A^3-6 s_A^2+12 s_A s_A'+4 s_A-6
   s_A'+4 s_A''-1\right)+4 (2 s_A-1) \left(4 \lambda_1^2-\lambda_2\right)\right\rbrace \right] \; ,
\end{split}
\end{align}
where $\tilde q = q \sqrt{\frac{\zeta}{2 -2\zeta}}$.

\chapter{Various functions and coefficients} \label{app:def_complicated_functions}

\section{Coefficients in the Lagrangian of odd modes}

In this Appendix we give the expressions of the functions $u_i$ and $\mathcal{G}_i$ respectively defined in the main text by Eq. \eqref{bulk action 1} and Eq. \eqref{q action}

\begin{subequations}
\begin{align}\label{u and v}
u_1 &= - \frac{\ell(\ell+1)}{4 a^4 c^3} \bigg[a c \left(a \left(b \left(\alpha  c^2 a''-3 a \alpha +a b M_{12}'-a (\alpha +2) c
   c''-a c \alpha '-2 a\right)-a (\alpha +2) c b'\right) \right. \nonumber \\
   & \left. +a c a' \left(\alpha  c b'+b
   \left(5 \alpha +2 b M_{12}'+c \alpha '+2\right)\right)+b c^2 \left(a'\right)^2
   \left(b M_{12}'-2 \alpha \right)\right) \nonumber \\
   &+2 a c M_{10}^2 \left(c \left(a \left(b c
   a''+a b'+a b c''\right)+a a' \left(c b'+3 b\right)-2 b c \left(a'\right)^2\right)-a^2
   b\right) \nonumber \\
   &+b M_{12} \left(c a'+a\right) \left(c \left(2 a \left(b c a''+a b'+a b
   c''\right)+a a' \left(2 c b'+3 b\right)-3 b c \left(a'\right)^2\right)-2 a^2
   b\right) \nonumber \\
   &+4 a^2 b c^2 M_{10} M_{10}' \left(c a'+a\right) - \frac{a^3 c}{b} (\ell - 1)(\ell+2) \bigg] \\
u_2  &= \frac{\ell (\ell+1) (\ell-1) (\ell+2) a b }{4 c^3}\left[2 b  c' M_{12}-c \left(1-2 M_{10}^2\right)\right]\, ,  \\
u_3  &= -\frac{\ell (\ell+1) b }{4
   a^2 c}\left[ \partial_r (a c) b M_{12} - a c \left(1-2 M_{10}^2\right)\right]\, ,  \\
u_4  &= \frac{1}{a b c
   \left[ \partial_r (a c) b M_{12} - a c \left(1-2 M_{10}^2\right)\right]}
   \bigg[ c^2 \left(b a'
   \left(b M_{12}
   a'+a \left(\alpha +2
   M_{10}^2-2
   \right)\right)+2
    a^2
   b'\right) \nonumber \\
   &+a b c c'
   \left(2 b
   M_{12}
   a'-a \left(\alpha -2
   M_{10}^2+2
   \right)\right)+a^2
   b^2 M_{12}
   c'^2 \bigg] \; .
\end{align}
\end{subequations}
The functions $\mathcal{G}_{00},\mathcal{G}_{rr}$ and $\mathcal{G}_{qq}$ that appear in the action \eqref{q action} are defined in terms of the above functions as follows~\cite{Franciolini:2018uyq}:
\begin{subequations}
\begin{align}
\mathcal{G}_{00} =& -\frac{u_3^2}{u_2}\, ,\\
 \mathcal{G}_{rr} =& \frac{u_3^2}{u_4 u_3' +u_3 \left(u_4'+u_4^2\right)-u_1}\, , \\
\mathcal{G}_{qq} =& \frac{(\mathcal{G}_{rr} )^2}{u_3^3} \left[u_3' \left(2 u_3' u_4'-u_1'\right)+u_3 \left(-u_4 u_1'-u_3'' u_4'+u_3' u_4''+u_4 u_3'
   u_4'\right) + u_1 u_3''\right .
   \\
  & \qquad \quad \qquad \qquad \qquad \left . +2 u_4 u_1 u_3'+u_1 u_3 \left(3 u_4'+u_4^2\right)-u_1^2+u_3^2 \left(u_4 u_4''-2
   u_4'^2\right)\right]\, . \nonumber 
\end{align}
\end{subequations}

\section{Power emitted} \label{power-app}
Here we report the expressions of the coefficients appearing in the main result of Chapter \ref{Chapter10}, Eq.~\eqref{eq:power_chap10}:

\begin{align}
p_0 &= \frac{1}{36(1-2 \mu_1)^2}, \\
p_2 &= \frac{1}{504 (1-2 \mu_1)^4} \big( 23 + 112 a_2 - 140 \gamma_1 - 35 \lambda_0 - 260 \mu_1 + 
 280 \gamma_1 \mu_1 + 70 \lambda_0 \mu_1 + 92 \mu_1^2 \nonumber \\
 & + 
 28 \alpha_1 ( 2 \mu_1 - 1) + 224 \mu_2 \big), \\
p_4 &= \frac{1}{181440 (1-2 \mu_1)^6} \bigg( -33817+ 47040 \alpha_1^2 \mu_1^2-47040 \alpha_1^2 \mu_1+11760
   \alpha_1^2 \nonumber \\
   &-389760
   \alpha_1 \gamma_1 \mu_1+97440 \alpha_1 \gamma_1+97440 \alpha_1 \lambda_0 \mu_1^2-97440 \alpha_1
   \lambda_0 \mu_1+24360 \alpha_1 \lambda_0 \nonumber \\
   &+43200
   \alpha_1 \mu_1^3-427680 \alpha_1 \mu_1^2+483840
   \alpha_1 \mu_1 \mu_2+213840 \alpha_1 \mu_1-241920 \alpha_1 \mu_2 \nonumber \\
   &-5400 \alpha_1-80640 a_2^2+1440
   a_2 \big(84 \alpha_1 (2 \mu_1-1)+420 \gamma_1 (2
   \mu_1-1)+210 \lambda_0 \mu_1-105 \lambda_0 \nonumber \\
   &+184
   \mu_1^2+40 \mu_1-224 \mu_2+18\big)+60480 a_3 (2
   \mu_1-1)+80640 b_2 \mu_1^2-80640 b_2 \mu_1 \nonumber \\
   &+20160
   b_2+772800 \gamma_1^2 \mu_1^2-772800 \gamma_1^2 \mu_1+193200 \gamma_1^2+386400 \gamma_1 \lambda_0 \mu_1^2-386400 \gamma_1 \lambda_0 \mu_1 \nonumber \\
   &+96600 \gamma_1
   \lambda_0+1066560 \gamma_1 \mu_1^3-3414240 \gamma_1
   \mu_1^2+2419200 \gamma_1 \mu_1 \mu_2+1707120
   \gamma_1 \mu_1 \nonumber \\
   &-1209600 \gamma_1 \mu_2-133320
   \gamma_1-161280 \gamma_2 \mu_1^2+161280 \gamma_2
   \mu_1-40320 \gamma_2+48300 \lambda_0^2 \mu_1^2 \nonumber \\
   &-48300
   \lambda_0^2 \mu_1+12075 \lambda_0^2+283440 \lambda_0
   \mu_1^3-878760 \lambda_0 \mu_1^2+604800 \lambda_0
   \mu_1 \mu_2+439380 \lambda_0 \mu_1 \nonumber \\
   &-302400 \lambda_0 \mu_2-35430 \lambda_0-40320 \lambda_1 \mu_1^2+40320
   \lambda_1 \mu_1-10080 \lambda_1+117488 \mu_1^4 \nonumber \\
   &-182176
   \mu_1^3+449280 \mu_1^2 \mu_2-222648 \mu_1^2-167040
   \mu_1 \mu_2+241920 \mu_1 \mu_3+119096 \mu_1 \nonumber \\
   &-322560 \mu_2^2+213120 \mu_2-120960 \mu_3 +389760 \alpha_1 \gamma_1 \mu_1^2 \bigg).
\end{align}


\bibliographystyle{utphys}
\bibliography{these.bib}


\end{document}